%% file: main.tex
\documentclass[12pt,twoside]{mitthesis}
\usepackage{lgrind}
\usepackage{graphicx}
\begin{document}
\include{cover}
\pagestyle{plain}
\include{contents}

\include{introduction}

\include{Exp}

\include{Analysis}
\include{Analysis_2}

\include{impacts}

\appendix
\include{app0}
\include{app1}
\include{app2}
\include{app3}
\include{app4}
\include{app5}
\include{biblio}
\end{document}

%% file: cover.tex
%
%
%
%
%
%
%
\title{High Precision Measurement of the Proton Elastic Form Factor Ratio at Low $Q^2$}

\author{Xiaohui Zhan}
\department{Department of Physics}
\degree{Doctor of Philosophy}
\degreemonth{January}
\degreeyear{2010}
\thesisdate{January 25, 2010}


\supervisor{William Bertozzi}{Professor of Physics}
\supervisor{Shalev Gilad}{Principle Research Scientist}

\chairman{Thomas J. Greytak}{Associate Department Head for Education}

\maketitle



\cleardoublepage
\setcounter{savepage}{\thepage}
\begin{abstractpage}
\input{abstract}

\end{abstractpage}


\cleardoublepage

\section*{Acknowledgments}

This work could not have be completed without the people who have supported and helped me along this long journey. I am extremely grateful for their care and considerations along these years for which I don't have many opportunities to show my gratitude in front of them.

First, I would like to thank my advisor, Prof. William Bertozzi, for giving me the opportunity to start the graduate study at MIT, and for his continuing guidance, attention and support throughout these years. I won't forget the ``hard moments'' he gave me during the preparation of the part III exam the same as the encouragement when I was frustrated. He helped me to understand how to become a physicist and at the mean time a happy person in life. I also would like to thank my another advisor Dr. Shalev Gilad for his valuable advices and suggestions during the whole analysis and the encouragement throughout my study and research at Jefferson Lab. Without their support, I would not complete the thesis experiment and finish the degree.

I would like to thank my academic advisor Prof. Bernd Surrow for his careful guidance in my graduate courses and the discussions for the future career. Many thanks to my thesis committee members: Prof. William Donnelly and Prof. Iain Stewart for their valuables comments and suggestions to this thesis.

Although it's only been less than two years since I join the E08-007 collaboration, I had a wonderful experience and learned a lot in working with the spokespersons, post-docs and former graduate students. Individually, I sincerely appreciate Prof. Ronald Gilman for his guidance and support before and during the running of the experiment, and his helpful suggestions and discussions for the analysis afterwards. I would like to thank Dr. Douglas Higinbotham for being my mentor at Jefferson Lab and giving valuable advices in resolving different problems I encountered along the way. I would like to thank Dr. Guy Ron, for providing the first hand experimental running and analysis experience, the experiment would not run so smoothly without his effort. I would like to thank Dr. John Arrington for the valuable comments and discussions on the analysis and providing the form factors global fits. I also would like to thank Prof. Steffen Strauch, Prof. Eliazer Piasetzky, Prof. Adam Sarty, Dr. Jackie Glister and Dr. Mike Paolone for their guidance and inspiring discussions through the whole analysis process. In addition, I would like to thank the group former post-docs Dr. Nikos Sparveris and Dr. Bryan Moffit for their generous help on the experimental setup and assistance through the experiment. This work could not be done without the contribution from any one of them.

I would like to thank the Hall A staff members and the entire the Hall A collaboration for their commitment and shift efforts for this experiment. I would also like to thank the Jefferson Lab accelerator crew for delivering high quality beam for this experiment.

In the earlier days at Jefferson Lab, I worked with the saGDH/polarized $^3$He group. It was very special to me since that's when I completed my first analysis assignment and learned quite some knowledge about the target system. I would like to thank Dr. Jian-Ping Chen for his supervision and guidance when I started the research in Jefferson Lab without any experience, his passion and rigorous attitude for physics have served as a model for me. I also would like to thank the former graduate students Vince Sulkosky, Jaideep Singh, Ameya Kolarkar, Patricia Solvignon and Aidan Kelleher for their patience and generous support on various things. I would like to thank the $^3$He lab/transversity fellow students: Chiranjib Dutta, Joe Katich, and Huan Yao for the great experience we had worked together.

Although I started my graduate life at MIT, I spent the last four years at Jefferson Lab. I am lucky to have friendships at both cities which made my graduate study and research an enjoyable experience. I would like to thank them for their support and encouragement: Bryan Moffit, Vince Sulkosky, Bo Zhao, Kalyan Allada, Lulin Yuan, Fatiha Benmokhtar, Ya Li, Jianxun Yan, Linyan Zhu, Ameya Kolarkar, Xin Qian, Zhihong Ye, Andrew Puckett, Peter Monaghan, Jin Huang, Navaphon Muangma, Kai Pan, Wen Feng, Wei Li, Feng Zhou,. Especially I would like to thank the group post-doc Vince Sulkosky for his effort in reading and correcting my thesis draft.

And Finally, I would like to show my deepest appreciation to my parents. I would not be anywhere without them and there is no words could ever match the love and support they gave since I was born. I also want to thank my fiance Yi Qiang, for the endless support over the past 8 years, and loving me for who I am.


%% file: abstract.tex
%
%
%
Experiment E08-007 measured the proton elastic form factor ratio
$\mu_pG_E/G_M$ in the range of $Q^2=0.3-0.7(\mathrm{GeV}/c)^2$ by recoil polarimetry. Data were taken in 2008 at the Thomas Jefferson National Accelerator Facility in Virginia,
USA. A 1.2 GeV polarized electron beam was scattered off a cryogenic
hydrogen target. The recoil proton was detected in the left HRS in coincidence with the elasticly scattered electrons tagged  by the BigBite spectrometer. The proton polarization was measured by the focal plane polarimeter (FPP).

 In this low $Q^2$ region, previous measurement from Jefferson Lab Hall A (LEDEX) along with various fits and calculations indicate substantial deviations of the ratio from unity. For this new measurement, the proposed statistical uncertainty ($<1\%$) was achieved. These new results are a few percent lower than expected from previous world data and fits, which indicate a smaller $G_{Ep}$ at this region. Beyond the intrinsic interest in nucleon structure, the new results also have implications in determining the proton Zemach radius and the strangeness form factors from parity violation experiments.

%% file: contents.tex
\tableofcontents
\newpage
\listoffigures
\newpage
\listoftables

%% file: introduction.tex
\chapter{Introduction}

When the proton and the neutron were discovered in 1919 and 1931
respectively, they were believed to be Dirac particles, just like
the electron. They were expected to be point-like and to have a
Dirac magnetic moment, expressed by:
\begin{equation}
\mu_D=\frac{q}{mc}|\vec s|
\end{equation}
where \textsl{q}, \textsl{m}, and \textsl{s} are the electric charge,
mass and spin of the particle respectively. However, later
measurements of these nucleons magnetic moments revealed the existence
of the nucleon substructure. The first direct evidence that the proton has an
internal structure came from the measurement of its anomalous magnetic
moment 70 years ago by O. Stern~\cite{Stern},
\begin{equation}
\mu_p = 2.79\mu_B,
\end{equation}
where $\mu_B$ is the Bohr magneton. The first measurement of the
charge radius of the proton by Hofstadter \textsl{et al.}~\cite{Hofs,
 Hofs1} yielded a value of 0.8fm, which is quite close to the modern value.

Starting from 1950s, electron scattering experiments were used to
unravel the nucleon internal structure. Through the measurements of electromagnetic form
factors and nucleon structure functions in elastic and deep
inelastic lepton-nucleon scattering, it's commonly accepted
that in a simplistic picture, a nucleon is composed of three valence
quarks interacting with each other through the strong force. The strong interaction theory, Quantum Chromodynamics (QCD) can make rigorous
predictions when the four-momentum transfer squared, $Q^2$, is very
large and the quarks become asymptotically free. However, predicting
nucleon form factors in the non-perturbative regime is difficult due to the dominance of the soft
scattering processes. As a consequence there
are several phenomenological models which attempt to explain the data in
this domain, and precise measurements of the nucleon form factors are
desired to constrain and test these models.

In the one-photon-exchange (OPE) approximation, the $ep$
elastic scattering cross section formalism is well known and can be
parameterized by two form factors, $G_E$ and $G_M$ which are functions of $Q^2$. At low momentum
transfer, the form factors can be interpreted as the fourier transform
of the nucleon charge and magnetic densities. Earlier experiments measured the cross section of
the $ep$ elastic scattering which contains information about the
internal structure responsible for the deviation from the scattering
off point-like particles. However, after four decades of effort, there
were still large kinematic regions where only very limited
measurements of the form factors were possible, since the cross section of the unpolarized electron scattering is only sensitive to a specific
combination of the form factors and the lack of a free neutron target.

In the last two decades, advances in the technology of intense
polarized electron beams, polarized targets and polarimetry have
ushered in a new generation of electron scattering experiments which rely on
spin degrees of freedom. Compared to the cross section measurement,
the polarization techniques have several distinct advantages. First,
they have increased sensitivity to a small amplitude of interest by measuring an interference term. Second, spin-dependent experiments involve the measurement of
polarizations or helicity asymmetries, and these quantities are independent
of the cross section normalization, since most of the helicity independent systematic
uncertainties can be canceled by measuring a ratio of polarization
observable.

The first experiment to measure the recoil proton polarization observable in $ep$ elastic scattering was done at SLAC by Alguard {\it et al.}~\cite{alguard}, but the impact of the results was severely limited by the low statistics. Followed by that, the proton form factor measurements using recoil polarimetry were carried out at MIT-Bates~\cite{Mil, bar99} and MAMI~\cite{mami, mami1}. Due to limited statistics and kinematics coverage, the ratio values were in agreement within uncertainties with the unpolarized measurements. More recent measurements of the proton form factor ratio $\mu_pG_E/G_M$ using recoil polarimetry at Jefferson Lab~\cite{punj,Mark,Gayou}, which
have much better precision at high $Q^2$, deviated dramatically from
the unpolarized data. This has prompted intense theoretical and experimental
activities to resolve the discrepancy. The validity of analyzing data
in the OPE approximation has been questioned, and two-photon-exchange
(TWE) processes are now considered as an significant correction to the unpolarized data
and mostly account for the discrepancy at high $Q^2$~\cite{john:TPE}.

While extending our knowledge at higher momentum transfer region is an ongoing endeavor, the proton form factor ratio behavior at low $Q^2$ has also become the subject of considerable interest, especially, when potential discrepancy was observed from the most recent high precision measurements for $Q^2<1~\mathrm{GeV}^2$. BLAST~\cite{BLAST} did the first proton form factor ratio measurement via
beam-target asymmetry at $Q^2$ values of 0.15 to 0.65 $\mathrm{Gev}^2$, and the results are consistent with 1 in this region. LEDEX~\cite{LEDEX}, which used the recoil polarimetry technique, observed a substantial deviation from unity at $Q^2\sim 0.35~\mathrm{GeV}^2$ . However, the data quality of LEDEX was compromised due to the low beam polarization and background issues~\cite{guy_thesis}. Hence, it was necessary to carry out a new high precision measurement to either confirm to refute the deviations at low momentum transfers. Beyond
the intrinsic interest in the nucleon structure, an improved proton
form factor ratio also impacts other high precision
measurements such as parity violation experiments
(HAPPEX)~\cite{happex, happex2}, deeply virtual Compton scattering (DVCS)~\cite{Ji_dvcs, Ji_dvcs_2}, and also determination of other physical quantities such as the proton Zemach radius.

This thesis presents the analysis and results of experiment E08-007, which was conducted in 2008 at Jefferson Lab Hall A. In this experiment, the proton form factor ratio $\mu_pG_E/G_M$ was measured
at $Q^2= 0.3-0.7 \mathrm{GeV^2}$ using recoil polarimetry.
\section{Definitions and Formalism}
\subsection{Exclusive electron scattering}
\begin{figure}
  \begin{center}
    \includegraphics[angle=0, width=0.45\textwidth]{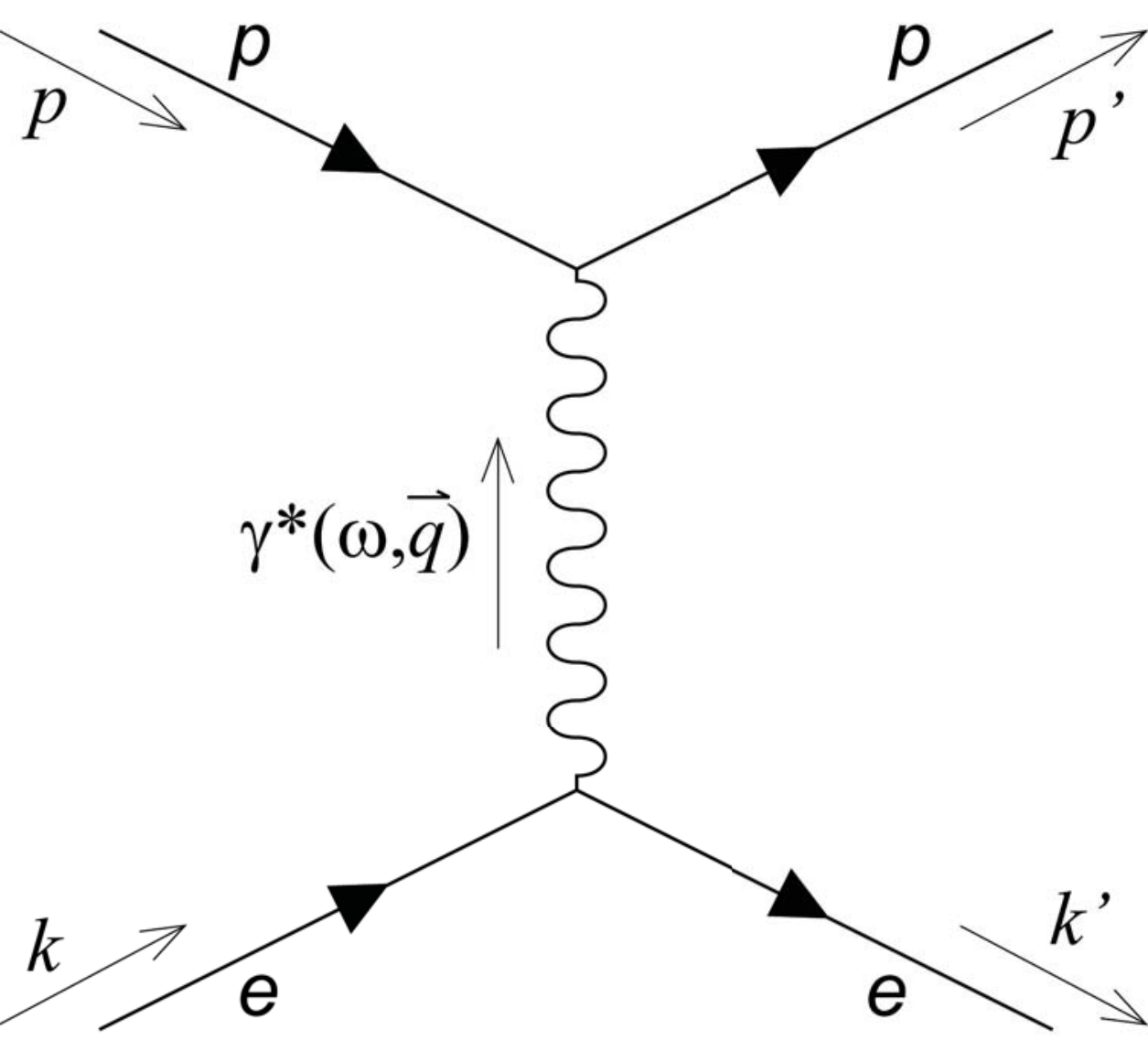}
    \caption{The leading order diagram of $ep$ elastic scattering.}
    \label{fig:ep_lead}
  \end{center}
\end{figure}

When scattered off a nuclear target, the electron exchanges virtual
photons with the nucleus, which probes the electromagnetic
structure of the nucleus. The electromagnetic coupling is small
enough $(\alpha=1/137)$ that it is valid to only consider the leading
order. For the elastic scattering reaction off a proton,
$e(k)+P(p)\to e(k')+P(p')$, the leading order diagram is shown in
Fig.~\ref{fig:ep_lead}.
Initial and final electrons have four-momenta $k=(E,\vec k)$
and $k' = (E',\vec {k'})$ respectively, and the initial and final
protons $p=(E_p,\vec p)$ and $p'=(E'_p,\vec {p'})$. The
virtual photon has four-momentum $q=(\omega,\vec q)$, and the
Lorentz-invariant four-momentum transfer squared $Q^2$ is defined as:
\begin{equation}
Q^2=-q^2=-(\omega^2-\vec q^2)=-(k-k')^2\sim
4EE'\sin^2\frac{\theta_e}{2},
\end{equation}
where the last expression is valid in the Lab frame by neglecting the
electron mass. The amplitude of $Q^2$ is associated with the scale that the electromagnetic probe is sensitive to.

For exclusive elastic scattering, the recoil proton is also detected,
so that $Q^2$ can be defined from the proton momenta:
\begin{equation}
Q^2= -(p'-p)^2=-[(E'_p-E_p)^2-(\vec {p'}-\vec p)^2].
\label{equ:kine}
\end{equation}
In the Lab frame,the initial proton is at rest, and Eq.~\ref{equ:kine} becomes:
\begin{equation}
Q^2=-(E'^2+m_p^2-2E'_pm_p-\vec p^2)=-(2m_p^2-2m_pE'_p)=2m_pT_p,
\end{equation}
where $m_p$ is the proton mass and $T_p=E'_p-m_p$ is the kinetic
energy of the final proton in the Lab frame.
\subsection{Formalism}
One of the advantages of the electromagnetic probe lies in the fact
that the leptonic vertex $e(k)\to e(k')+\gamma^*(q)$ is fully described
by the theory of the electromagnetic interaction, Quantum
ElectroDynamics (QED), and the information related to the unknown electromagnetic properties of the nucleon are contained by only the hadronic vertex
$\gamma^*(q)+P(p)\to P(p')$. From the Feynman
diagram in Fig.~\ref{fig:ep_lead}, the amplitude for $ep$ elastic
scattering can be written as:
\begin{eqnarray}
i\mathcal{M}&=&[ie\bar{v}
(p')\Gamma^{\mu}(p',p)v(p)]\frac{-ig_{\mu\nu}}{q^2}[ie\bar{u}(k')\gamma^{\nu}u(k)]\\
&=&\frac{-i}{q^2}[ie\bar{v}(p')\Gamma^{\mu}(p',p)v(p)][ie\bar{u}(k')\gamma_{\mu}u(k)],
\label{eq:mag}
\end{eqnarray}
where $\gamma^{\mu},\mu=0,1,2,3$ with the 0-th component as the time
component, are the Dirac $4\times 4$ matrices in the chiral representation:
\begin{equation}
\gamma_0= \left(\begin{array}{cc}
0 & 1 \\
1 & 0
\end{array}\right),
\vec{\gamma}= \left(\begin{array}{cc}
0 & \vec{\sigma} \\
-\vec{\sigma} & 0
\label{eq:gamma}
\end{array}\right)
\end{equation},
and $\vec{\sigma}$ is the set of standard Pauli matrices:
\begin{equation}
\sigma_1 = \left(\begin{array}{cc}
0 & 1 \\
1 & 0
\end{array}\right),
\sigma_2 = \left(\begin{array}{cc}
0 & -i\\
i & 0
\end{array}\right),
\sigma_3 = \left(\begin{array}{cc}
1 & 0 \\
0 & -1
\end{array}\right).
\end{equation}
$u(k)$ and $\bar{u}(k')$ are the Dirac spinors for the initial and final
electron, and $v(p)$, $\bar{v}(p')$ are the Dirac four-spinors for the
initial and recoil proton respectively. In particular, the proton spinors enter in
the plane-wave solution for a spin 1/2 particle
$\psi(x)=v(p)e^{-ip\cdot x}$ which satisfies the Dirac equation:
\begin{equation}
(-i\gamma^{\mu}\partial_{\mu}-m)\psi(x)=0,
\end{equation}
and one can write:
\begin{equation}
v(p)=\left(\begin{array}{c}
\sqrt{p \cdot \sigma} \chi\\
\sqrt{p \cdot \bar{\sigma}} \chi
\label{eq:v(p)}
\end{array}\right)
\end{equation}
with $\sigma^{\mu}\equiv
(1,\vec{\sigma}),\bar{\sigma}\equiv(1,-\vec{\sigma})$ and $\chi$ is a
normalized two-spinor, such that
\begin{equation}
\chi^{\dagger}\chi=1.
\end{equation}
While the leptonic current $j_{\mu}=ie\bar{\mu}(k')\gamma_{\mu}u(k)$ is
fully described by QED, the hadronic current
$\mathcal{J}^{\mu}=ie\bar{v}(p')\Gamma^{\mu}v(p)$ involves the factor
$\Gamma^{\mu}$, which contains the information about the internal
electromagnetic structure of the proton. In general $\Gamma^{\mu}$ is
some expression that involves $p,p',\gamma^{\mu}$ and constants such as
the proton mass $m$, the electric charge $e$. Since the hadronic current
transforms as a vector, $\Gamma^{\mu}$ must be a linear combination of
these vectors, where the coefficients can only be function of
$Q^2$. It is convenient to write the current in the following way:
\begin{equation}
\mathcal{J}^{\mu}=ie\bar{v}(p')\Gamma^{\mu}v(p)=ie\bar{v}(p')[\gamma^{\mu}F_1(q^2)+\frac{i\sigma^{\mu\nu}q_{\nu}}{2m}\kappa
F_2(q^2)]v(p),
\end{equation}
where
$\sigma^{\mu\nu}=\frac{i}{2}[\gamma^{\mu},\gamma^{\nu}]$,~$\kappa\simeq 1.793$
is the proton anomalous magnetic moment and $F_{1,2}(Q^2)$ are the proton elastic form factors. They
contain the information about the electromagnetic structure of the
proton.

\subsection{Nucleon Form Factors}

$F_1(Q^2)$ and $F_2(Q^2)$ are distinguished according to their helicity
$(\vec{\sigma}\cdot \vec{p}/|\vec{p}|)$ characteristics, the projection of electron
intrinsic spin $\vec{\sigma}$ along its direction of motion
$\vec{p}/|\vec{p}|$. $F_1(Q^2)$ is the
Dirac form factor; it represents the helicity-preserving part of the
scattering. On the other hand, the Pauli form factor $F_2(Q^2)$ represents the
helicity-flipping part. $F_1$ and $F_2$ are defined in a similar
way for the neutron. The form factors are normalized according to
their static properties at $Q^2=0$. For the proton:
\begin{equation}
F_{1p}(0)=1, F_{2p}(0)=1,
\end{equation}
and for the neutron:
\begin{equation}
F_{1n}(0)=0, F_{2n}(0)=1.
\end{equation}
For reasons that will soon become obvious, it is more convenient to
use the Sachs form factors~\cite{sachs}: $G_E(Q^2)$ and $G_M(Q^2)$, which
are defined as:
\begin{eqnarray}
G_E&=&F_1-\tau\kappa F_2{}
\nonumber\\
{}G_M&=&F_1+\kappa F_2,
\label{eq:sachs}
\end{eqnarray}
where $\tau=\frac{Q^2}{4m^2}$ is a kinematic factor. The Sachs form
factors also have particular values at $Q^2=0$ according to the static
properties of the corresponding nucleon:
\begin{eqnarray}
G_{Ep}(0)=1,G_{Mp}(0)=\mu_p\\
G_{En}(0)=0,G_{Mn}(0)=\mu_n,
\end{eqnarray}
where $\mu_p=2.79$ and $\mu_n=-1.91$ in units of nuclear magneton.

\subsection{Hadronic Current in the Breit Frame}
In the Breit frame, which is defined as the frame where the initial
and final nucleon momenta are equal and opposite, the hadronic current
has a simplified interpretation. A definition of variables in the Breit frame, which are noted with a subscript $B$, is elaborated in Appendix A. Using the Gordon
identity~\cite{Gordon}
\begin{equation}
\bar{v}(p')\gamma^{\mu}v(p)=\bar{v}(p')[\frac{p'^{\mu}+p^{\mu}}{2m}+\frac{i\sigma^{\mu\nu}q_{\nu}}{2m}]v(p)
\end{equation}
similarly, we can write:
\begin{equation}
\bar{v}(p')\Gamma^{\mu}v(p) = \bar{v}(p')[(F_1+\kappa
F_2)\gamma^{\mu}-\frac{(p+p')^{\mu}}{2m}\kappa F_2] v(p).
\end{equation}
In the Breit frame, the explicit expression of the hadronic current
$\mathcal{J}=(\mathcal{J}^0,\mathcal{\vec{J}})$ is simplified:
\begin{eqnarray}
\mathcal{J}^0 &=& ie\bar{v}(p')[(F_1+\kappa
F_2)\gamma^0-\frac{E_{pB}}{m}\kappa F_2]v(p)\\
\mathcal{\vec{J}}&=& ie(F_1+\kappa F_2)\bar{v}(p')\vec{\gamma}v(p),
\end{eqnarray}
where $E_{pB}$ is the
Using $\bar{v}(p')=v^\dagger(p')\gamma^0$, the time component
$\mathcal{J}^0$ can be expressed by:
\begin{equation}
\mathcal{J}^0=ie[(F_1+\kappa F_2)v^\dagger(p')v(p)-\kappa F_2\frac{E_{pB}}{m}v^\dagger(p')\gamma^0v(p)].
\end{equation}
By the definition of $v(p)$ and $\gamma^0$ in Eqs.~\ref{eq:v(p)}
and~\ref{eq:gamma}, we now have:
\begin{eqnarray}
\mathcal{J}^0&=&ie(F_1+\kappa F_2)\chi '^\dagger\left(\sqrt{p'\cdot
  \sigma},\sqrt{p'\cdot \bar{\sigma}}\right)\left(\begin{array}{c}
\sqrt{p\cdot \sigma}\\
\sqrt{p\cdot \bar{\sigma}}
\end{array}\right)\chi{}
\nonumber\\
& & {}-ie\kappa F_2\frac{E_{pB}}{m}\chi '\left(\sqrt{p'\cdot
\sigma}, \sqrt{p'\cdot\bar{\sigma}}\right)\left(\begin{array}{cc}
0 & 1\\
1 & 0
\end{array}\right)
\left(\begin{array}{c}
\sqrt{p\cdot \sigma}\\
\sqrt{p\cdot \bar{\sigma}}
\end{array}\right)\chi.
\end{eqnarray}
Then, by the expressions:
\begin{eqnarray}
\sqrt{p'\cdot \sigma}\sqrt{p\cdot
  \sigma}=\sqrt{p'\cdot\bar{\sigma}}\sqrt{p\cdot\bar{\sigma}}&=&m\\
\sqrt{p'\cdot \sigma}\sqrt{p\cdot\bar{\sigma}}+\sqrt{p'\cdot
  \bar{\sigma}}\sqrt{p\cdot \sigma}& =&2E_{pB}\\
\tau=\frac{Q^2}{4m^2}=\frac{\vec{q_B}^2}{4m^2}&=&\frac{E_{pB}^2-m^2}{m^2},
\end{eqnarray}
we can finally get the simple relation:
\begin{equation}
\mathcal{J}^0=ie2m\chi '^\dagger\chi(F_1-\tau\kappa F_2)=ie2m\chi
'^\dagger\chi G_E.
\end{equation}
The vector current $\mathcal{\vec{J}}$ can also be expressed in a
similar way in the Breit frame:
\begin{equation}
\mathcal{\vec{J}}=-e\chi '^\dagger(\vec{\sigma}\times\vec{q_B})\chi
(F_1+\kappa F_2)=ie\chi '^\dagger(\vec{\sigma}\times\vec{q_B})\chi G_M.
\end{equation}
Therefore, in the Breit frame, the electric form factor $G_E$ is
directly related to the electric part of the interaction between the
virtual photon and the nucleon, and the magnetic form factor $G_M$ is
related to the magnetic part of this interaction. The electric and magnetic form factors can be associated
with the Fourier transforms of the charge and magnetic current
densities in this frame in the non-relativistic limit. The Fourier transforms can be expanded in powers of $q^2$:
\begin{eqnarray}
G_{E,M}(Q^2)&=&\int\rho(\vec{r})e^{i\vec{q}\cdot\vec{r}}d^3\vec{r}\\
&=&\int \rho(\vec{r})d^3\vec{r}-\frac{\vec{q}^2}{6}\int \rho(\vec{r})\vec{r}^2d^3\vec{r}+\dots
\end{eqnarray}
Notice that the first integral gives the total electric or magnetic
charge, and the second integral defines the RMS electric or magnetic
radii of the nucleon. However, the Breit frame is a mathematical
abstraction, and for different $Q^2$ value, the Breit frame experiences
relativistic effect which is essentially a Lorentz contraction of the
nucleon along the direction of motion. This results in a non-spherical
distribution for the charge densities, and complicates the Fourier
transform interpretation of the form factors in the rest frame.

\section{Form Factor Measurements}
\subsection{Rosenbluth Cross Section}
The differential cross section for $ep$ scattering in the lab frame can be written as:
\begin{equation}
d\sigma=\frac{(2\pi^4|\mathcal{M}|^2)}{4(k\cdot p)}\delta^4(k+p-k'-p')\frac{d^3\vec{k'}}{(2\pi^3)2E'}\frac{d^3\vec{p'}}{(2\pi^3)2E'_p},
\end{equation}
where we have neglected the electron mass, and $\mathcal{M}$ is the
amplitude defined in Eq.~\ref{eq:mag}. Integrating over $\vec{k'}$ and
$\vec{p'}$ gives:
\begin{equation}
\frac{d\sigma}{d\Omega_e}=\frac{|\mathcal{M}|^2}{64\pi^2}\frac{1}{m^2}\left(\frac{E'_p}{E_p}\right)^2,
\end{equation}
where $\Omega_e$ is the solid angle in which the electron is
scattered, and $|\mathcal{M}|^2$ has the form:
\begin{equation}
|\mathcal{M}|^2=[\mathcal{J}^\mu\frac{-i}{q^2}j_\mu][\mathcal{J^\nu}\frac{-i}{q^2}j_\nu]^\ast=\left(\frac{1}{q^2}\right)^2[\mathcal{J}^\mu\mathcal{J}^{\nu\ast}][j_\mu
j_\nu^\ast]=\left(\frac{e^2}{q^2}\right)^2W^{\mu\nu}L_{\mu\nu}.
\end{equation}
The hadronic and leptonic tensors are defined respectively as:
\begin{eqnarray}
W^{\mu\nu}&=&\frac{1}{e^2}\mathcal{J}^\mu\mathcal{J}^{\nu\ast}\\
L_{\mu\nu}&=&\frac{1}{e^2}j_\mu j_\nu^\ast.
\end{eqnarray}
For unpolarized cross section, $W^{\mu\nu}$ and $L_{\mu\nu}$ are
averaged over the incident particle spin states, and summed over the
final particle spin states. Since the contraction of these two
tensors is a Lorentz invariant, they can be calculated in any frame,
as long as they are both calculated in the same frame.

In the single-photon exchange (Born) approximation, the formula for
the differential cross section of electron scattering off nucleons is
 given by~\cite{cross}:
\begin{equation}
\frac{d\sigma}{d\Omega_e}=\left(\frac{d\sigma}{d\Omega}\right)_{Mott}\frac{E'}{E}\big\{F_1^2(Q^2)+2(F_1(Q^2)+F_2(Q^2))^2\tan^2\frac{\theta_e}{2}\big\},
\label{eq:cx}
\end{equation}
where
\begin{equation}
\left(\frac{d\sigma}{d\Omega}\right)_{Mott}=\left(\frac{e^2}{2E}\right)^2\left(\frac{\cos^2\frac{\theta_e}{2}}{\sin^4\frac{\theta_e}{2}}\right)
\end{equation}
is the Mott cross section for the scattering of a spin-1/2 electron
from a spinless, point-like target, and $\frac{E}{E'}$ is the recoil
factor. This is the most general form for the electron elastic scattering
cross section. Using Eq.~\ref{eq:sachs}, we can rewrite Eq.~\ref{eq:cx} without the
interference term:
\begin{equation}
\frac{d\sigma}{d\Omega_e}=\frac{\alpha^2}{Q^2}\left(\frac{E'}{E}\right)^2\big[2\tau G_M^2+\frac{\cot^2\frac{\theta}{2}}{1+\tau}(G_E^2+\tau
G_M^2)\big],
\end{equation}
where $\alpha=e^2/4\pi\sim1/137$ is the electromagnetic fine structure constant, and
this expression is known as the {\it Rosenbluth formula}.
\subsubsection{Rosenbluth Separation}
The Rosenbluth cross section has two contributions: the electric term
$G_E$ and the magnetic term $G_M$. As noted earlier, there is no
interference term, so that the two contributions can be separated. We define the {\it reduced cross section} as:
\begin{equation}
\sigma_{red}=\frac{d\sigma}{d\Omega}\frac{\varepsilon(1+\tau)}{\frac{d\sigma}{d\Omega}_{Mott}} = \tau G_M^2+\varepsilon G_E^2,
\label{eq:red}
\end{equation}
where $\varepsilon=(1+2(1+\tau)\tan^2(\theta_e/2))^{-1}$ is the virtual photon
polarization parameter. The quantity $\varepsilon$ can be changed at a given $Q^2$, by changing the incident
electron beam energy and the scattering angle. Therefore, at a fixed
$Q^2$ by varying $\epsilon$, one can measure the elastic scattering
cross section and separate the two form factors using a linear fit to
the cross section. The slope is equal to $G_E^2$ and the
intercept is equal to $\tau G_M^2$.

This method has been extensively used in the past 40 years to measure
the elastic form factors and proved to be a very powerful method to
measure the proton and the neutron magnetic form factors up to large
$Q^2$. However, there are practical limitations. First, the
neutron electric form factor is normalized to the
static electric charge of the neutron, which is 0, and the cross section
is completely dominated by the magnetic form factor. For the proton,
the magnetic term will also dominate at large $Q^2$, since the factor
$\tau=\frac{Q^2}{4m^2}$ increases quadratically as $Q^2$ increases. As an example, at
$Q^2=2\mathrm{GeV}^2$ the magnetic term contributes about 95$\%$ of
the total cross section. On the other hand, in the low $Q^2$ region, the
magnetic term is suppressed for the same reason and the electric term
becomes dominant. Besides the difficulties from the physics side,
the precision of Rosenbluth separation is also limited by the cross
section measurements due to a widely different kinematic settings in order
to cover a wide range of $\varepsilon$. Systematic errors are introduced by the inconsistent acceptance, luminosity, detector efficiency between different kinematics.

\subsection{Polarization Transfer Measurements}
In 1974, Akhiezer and Rekalo~\cite{Akhie} discussed the interest of
measuring an interference term of the form $G_EG_M$ by measuring the
transverse component of the recoiling proton polarization in the
reaction $\vec{e}+p\to e'+\vec{p'}$. Thus, one could obtain $G_E$ in the presence of
a dominating $G_M$ at large $Q^2$. Instead of directly measuring the separate
form factors, the ratio $G_E/G_M$ can be accessed by measuring the
polarization of the recoil nucleon. The virtue of the polarization
transfer technique is that it is sensitive only to the ratio
$G_E/G_M$ and does not suffer from the dramatically
reduced sensitivity to a small component. Another way of measuring the
interference term would be to measure the asymmetries in the
scattering of a polarized beam off a polarized target.

By measuring the polarization $P_{\hat{u}}$ of the recoil nucleon
along a unit vector $\hat{u}$, we measure a preferential
orientation of the spin along $\hat{u}$. In this case, when we
average over initial proton spin states and sum over final proton spin
states, the completeness relation
\begin{equation}
\sum_{s=1,2}\chi^s\chi^{\dagger s}=1
\end{equation}
no longer holds. Instead we have to use:
\begin{equation}
\sum_{s=1,2}\chi '^s\chi '^{\dagger s}=1+\vec{\sigma}\cdot \hat{u}
\end{equation}
so that the hadronic tensor becomes:
\begin{equation}
W^{\mu\nu}=\frac{1}{2}Tr[\mathcal{F}^\mu(1+\vec{\sigma}\cdot \hat{u})\mathcal{F}^{\nu\dagger}]=W^{\mu\nu}_u+W^{\mu\nu}_p,
\end{equation}
where $W^{\mu\nu}_u$ is the unpolarized hadronic tensor and
$W^{\mu\nu}_p$ is the polarized one:
\begin{equation}
W^{\mu\nu}_p=\frac{1}{2}Tr[\mathcal{F}^\mu\mathcal{F}^{\nu\dagger}\vec{\sigma}\cdot\hat{u}].
\end{equation}
For recoil proton polarization measurements, a longitudinally polarized
beam is required. The polarization of the beam is defined as:
\begin{equation}
h=\frac{N^+-N^-}{N^++N^-},
\end{equation}
where $N^+$ and $N^-$ are the number of electrons with their spin
parallel and anti-parallel to their momentum respectively. Therefore,
with a polarized electron beam, the leptonic tensor is modified and a
new $\gamma$ matrix is introduced:
\begin{equation}
\gamma_5=i\gamma_0\gamma_1\gamma_2\gamma_3=
\left(\begin{array}{cc}
-1 & 0\\
0 & 1
\end{array}\right)
\end{equation}
The operator:
\begin{equation}
\frac{1-\gamma_5}{2}=
\left(\begin{array}{cc}
1&0\\
0&0
\end{array}\right).
\end{equation}
projects the spin along the momentum in a preferential direction. If
the beam polarization is $h$, and by further neglecting the electron mass, the leptonic tensor can be written as:
\begin{eqnarray}
L_{\mu\nu}&=&\frac{1}{2}Tr[(\gamma\cdot
k'-m_e)\gamma_\mu(1-h\gamma_5)(\gamma\cdot k-m_e)\gamma_\nu]{}
\nonumber\\
&=&{}2k_\mu k'_\nu+2k_\nu k'_\mu-2g_{\mu\nu}k\cdot
k'+2ih\epsilon_{\mu\nu\alpha\beta}k_\alpha k'_\beta{}
\nonumber\\
&=&{}L_{\mu\nu}^u+L_{\mu\nu}^p,
\label{eq:lep}
\end{eqnarray}
where $\epsilon_{\mu\nu\alpha\beta}$ is the Levi-Civita symbol. It is
0 if any two indices are identical, -1 under an even number of
permutations and +1 under an odd number of permutations. Note that
$L_{\mu\nu}^p$ is anti-symmetrical.

In order to get the polarized amplitude, we contract the leptonic and
the hadronic tensors:
\begin{equation}
W^{\mu\nu}L_{\mu\nu}=W^{\mu\nu}_uL^u_{\mu\nu}+W^{\mu\nu}_uL^p_{\mu\nu}+W_p^{\mu\nu}L^u_{\mu\nu}+W_p^{\mu\nu}L^p_{\mu\nu}
\end{equation}
where
\begin{itemize}
\item  $W_u^{\mu\nu}L_{\mu\nu}^u$ is the amplitude squared of the
  unpolarized process.
\item  $W_u^{\mu\nu}L^p_{\mu\nu}=0$ because it is the product of a
  symmetrical and an anti-symmetrical tensors.
\item  $W_p^{\mu\nu}L^u_{\mu\nu}$ is the {\it induced
    polarization}, it represents the polarization state of the recoil
  proton after scattering with an unpolarized beam off an unpolarized
  target.
\item  $W_p^{\mu\nu}L^p_{\mu\nu}$ is the {\it transferred
    polarization}, it represents the polarization state of the recoil
  proton after scattering with a polarized beam.
\end{itemize}
The recoil polarization along the vector $\hat{u}$ are given by:
\begin{eqnarray}
P_{\hat{u}}^{ind}&=&\frac{W_p^{\mu\nu}L^u_{\mu\nu}}{W_u^{\mu\nu}L^u_{\mu\nu}}{}
\nonumber\\
{}hP_{\hat{u}}^{transf}&=&\frac{W_p^{\mu\nu}L^p_{\mu\nu}}{W_u^{\mu\nu}L^u_{\mu\nu}}.
\end{eqnarray}
With the equations above, we can write the amplitude as:
\begin{equation}
W^{\mu\nu}L_{\mu\nu}=W_u^{\mu\nu}L^u_{\mu\nu}(1+P^{ind}_{\hat{u}}+hP^{transf}_{\hat{u}}),
\end{equation}
where $h$ is the polarization of the beam.

First, assume we measure the polarization along the 1-direction, and
we can derive each term of the hadronic tensor:
\begin{equation}
W^{\mu\nu}_{p,1}=\frac{1}{2}Tr[\mathcal{F}^\mu\mathcal{F}^{\nu\dagger}\sigma^1].
\end{equation}
Using $\sigma^1\sigma^2=i\sigma^3$, $\sigma^3\sigma^1=i\sigma^2$ and
$\sigma^2\sigma^3=i\sigma^1$, we have:
\begin{eqnarray}
\mathcal{F}^{0\dagger}\sigma^1&=&2mG_E\sigma^1\\
\mathcal{F}^{1\dagger}\sigma^1&=&-\sqrt{Q^2}G_M\sigma^3\\
\mathcal{F}^{2\dagger}\sigma^1&=&i\sqrt{Q^2}G_M\\
\mathcal{F}^{3\dagger}\sigma^1&=&0.
\end{eqnarray}
The $\vec{\sigma}$ matrices have the
trace properties:
\begin{eqnarray}
Tr[\gamma_\mu\gamma_\nu]&=&4g_{\mu\nu}{}
\nonumber\\
{}Tr[\gamma_\mu\gamma_\nu\gamma_\rho\gamma_\sigma]&=&4(g_{\mu\nu}g_{\rho\sigma}-g_{\mu\rho}
g_{\nu\sigma}+g_{\mu\sigma}g_{\nu\rho}),
\end{eqnarray}
where $g_{\mu\nu}$ is the Minkowski metric. The only non-zero terms arising are:
\begin{eqnarray}
W^{02}_{p,1}&=&i\sqrt{Q^2}2mG_EG_M{}
\nonumber\\
{}W^{20}_{p,1}&=&-i\sqrt{Q^2}2mG_EG_M.
\end{eqnarray}
We note here that the polarized tensor is anti-symmetrical, hence,
when it multiplied by the unpolarized leptonic tensor, the terms will
vanish, which applies for all the components.

The corresponding polarized terms of the leptonic
tensor in the Breit frame are anti-symmetrical, and obeys:
\begin{equation}
L^p_{02}=-L^p_{20}.
\end{equation}
According to Eq.~\ref{eq:lep},
\begin{eqnarray}
L^p_{02}&=&2ih\epsilon_{02\alpha\beta}k_\alpha k'_\beta{}
\nonumber\\
{}&=&2ih(k_{1B}k'_{3B}-k_{3B}k'_{1B})=-ihQ^2\cot\frac{\theta_B}{2}.
\end{eqnarray}
By contracting the hadronic tensor and the leptonic tensor, we get the
transferred polarization amplitude:
\begin{equation}
W^{\mu\nu}_{p,1}L^p_{\mu\nu}=4hmQ^2\sqrt{Q^2}\cot\frac{\theta_B}{2}G_EG_M.
\end{equation}
Therefore, measuring the 1-component, or transverse component of the
recoil proton polarization, gives access to the interference term
$G_EG_M$, which is inaccessible from an unpolarized cross section measurement.

The derivation for the 2-component is exactly identical. It involves
the terms $W_{p,2}^{01}$ and $L_{01}^p$ of the tensors, in particular:
\begin{equation}
L_{01}^p=2ih\epsilon_{01\alpha\beta}k_{\alpha B}k'_{\beta B}=2ih(k_{3B}k'_{2B}-k_{2B}k'_{3B})=0
\end{equation}
since $k_{2B}=k'_{2B}=0$. Therefore, {\bf in the Born approximation, there is
no normal component to the transferred polarization in elastic scattering}.

The same derivation applies to the longitudinal, 3-component, and we
can obtain:
\begin{equation}
W_{p,3}^{\mu\nu}L_{\mu\nu}^p=-4hQ^2\sqrt{Q^2}\frac{G_M^2}{\sin\frac{\theta_B}{2}},
\end{equation}
hence, the measurement of the longitudinal component of the recoil proton
polarization is a measurement of the magnetic form factor $G_M^2$.

We can now change the notation of the transferred polarization
components by $1\to y, 2\to x, 3\to z$. By applying the transformation
from the Breit frame to the Lab frame as defined in Appendix A, we have:
\begin{eqnarray}
\sigma_{red}P_x&=&0{}
\nonumber\\
{}\sigma_{red}P_y&=&-2\varepsilon\sqrt{\tau(1+\tau)}\tan\frac{\theta_e}{2}G_EG_M{}
\nonumber\\
{}\sigma_{red}P_z&=&\varepsilon\frac{E+E'}{m}\sqrt{\tau(1+\tau)}\tan^2\frac{\theta_e}{2}G^2_M,
\end{eqnarray}
where $\sigma_{red}=\varepsilon G^2_E+\tau G_M^2$ is the reduced cross
section as defined in Eq.~\ref{eq:red}. From this equation, we can see
that the ratio of the form factors $G_E/G_M$ can be extracted by a
simultaneous measurement of the transverse and longitudinal components
of the polarization of the recoil proton:
\begin{equation}
\frac{G_E}{G_M}=-\frac{P_y}{P_z}\frac{E+E'}{2m}\tan\frac{\theta_e}{2}.
\end{equation}
Compared to the cross section measurement, this method offers several
experimental advantages:
\begin{itemize}
\item Only a single measurement is required at each $Q^2$, and this
  greatly reduces the systematic error associated with the settings of
  the spectrometer and beam energy changes.
\item Since it's a polarization ratio measurement, it is not sensitive
  to the knowledge of helicity independent factors, such as the detector efficiency, beam
  polarization and the analyzing power of the polarimeter.
\item The measurement of the interference term $G_EG_M$ provides a much
  more accurate characterization of the electric form factor.
\item There is no need to measure the absolute cross section,
  therefore, the associated systematic uncertainties are usually much smaller.
\end{itemize}
With so many advantages, the polarization measurements cannot
provide absolute measurements of either form factor by
themselves. However, when coupled with cross section measurements, they allow
for a precise extraction of both form factors, even in regions where
the cross section is sensitive to only one of the form factors.

\section{World Data}

Proton and neutron form factors have been measured for over 50 years
at different electron accelerator facilities around the world. Earlier cross
section measurements (Rosenbluth separation) at low $Q^2$ found that
the form factors, except $G_{En}$, were in approximate agreement with the
diploe form:
\begin{equation}
\frac{G_{Mp}}{\mu_p}\simeq G_{Ep}\simeq\frac{G_{Mn}}{\mu_n}\simeq G_D
\end{equation}
where:
\begin{equation}
G_D=(1+\frac{Q^2}{0.71\mathrm{GeV}^2})^{-2}.
\end{equation}
This implies that the charge and magnetization distributions would
be well described by an exponential distribution, corresponding to the
Fourier transform of the dipole form.

\begin{figure}
  \begin{center}
    \includegraphics[angle=0, width=0.80\textwidth]{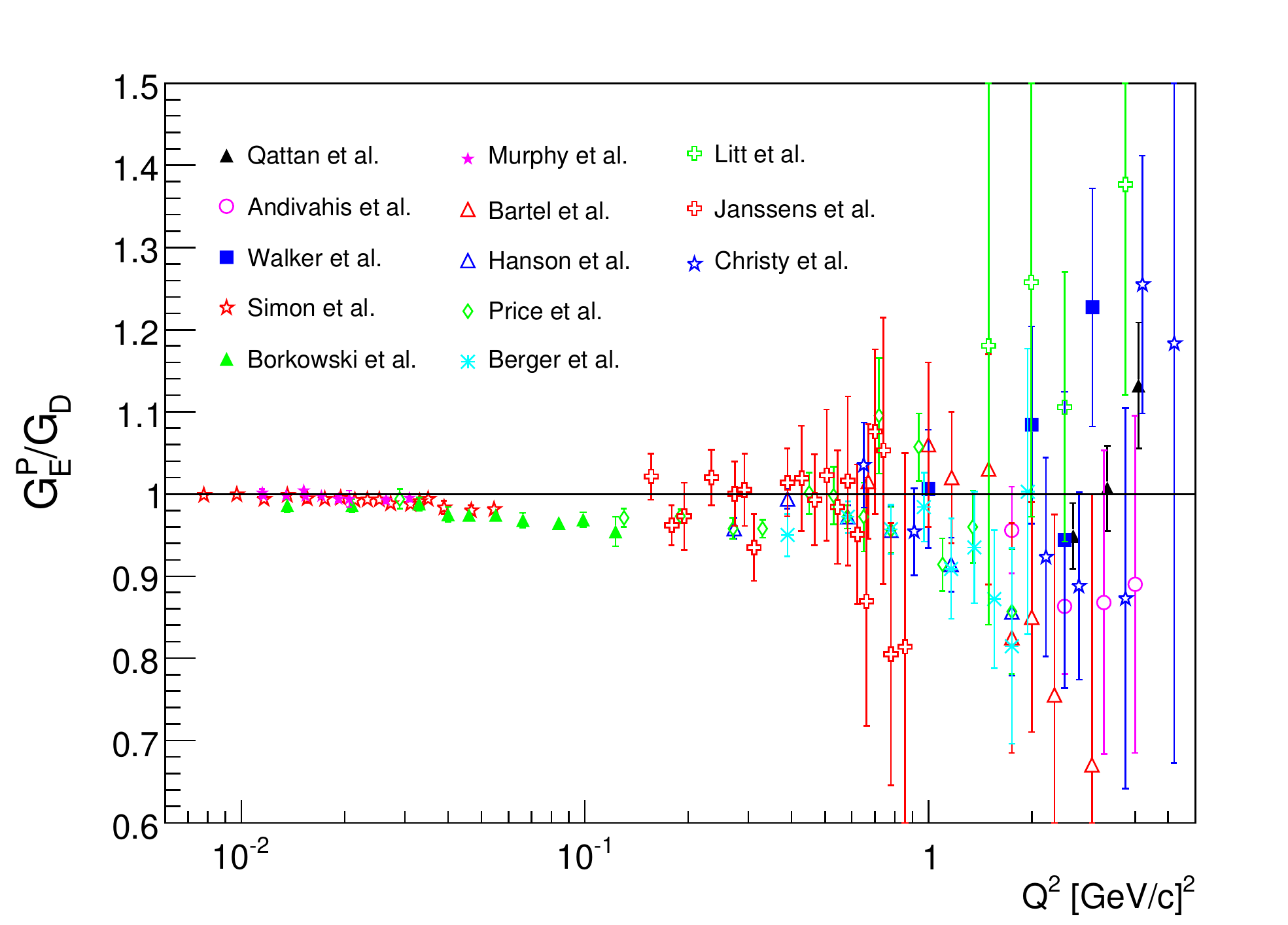}
    \caption{World data of $G_{Ep}$ from unpolarized
      measurements~\cite{qattan,andi,walker,simon,borkow,murphy,bartel,hanson,price,berger,litt,janss,christy},
      using the Rosenbluth method, normalized to the dipole parameterization.}
    \label{fig:gep}
  \end{center}
\end{figure}

\begin{figure}
  \begin{center}
    \includegraphics[angle=0, width=0.80\textwidth]{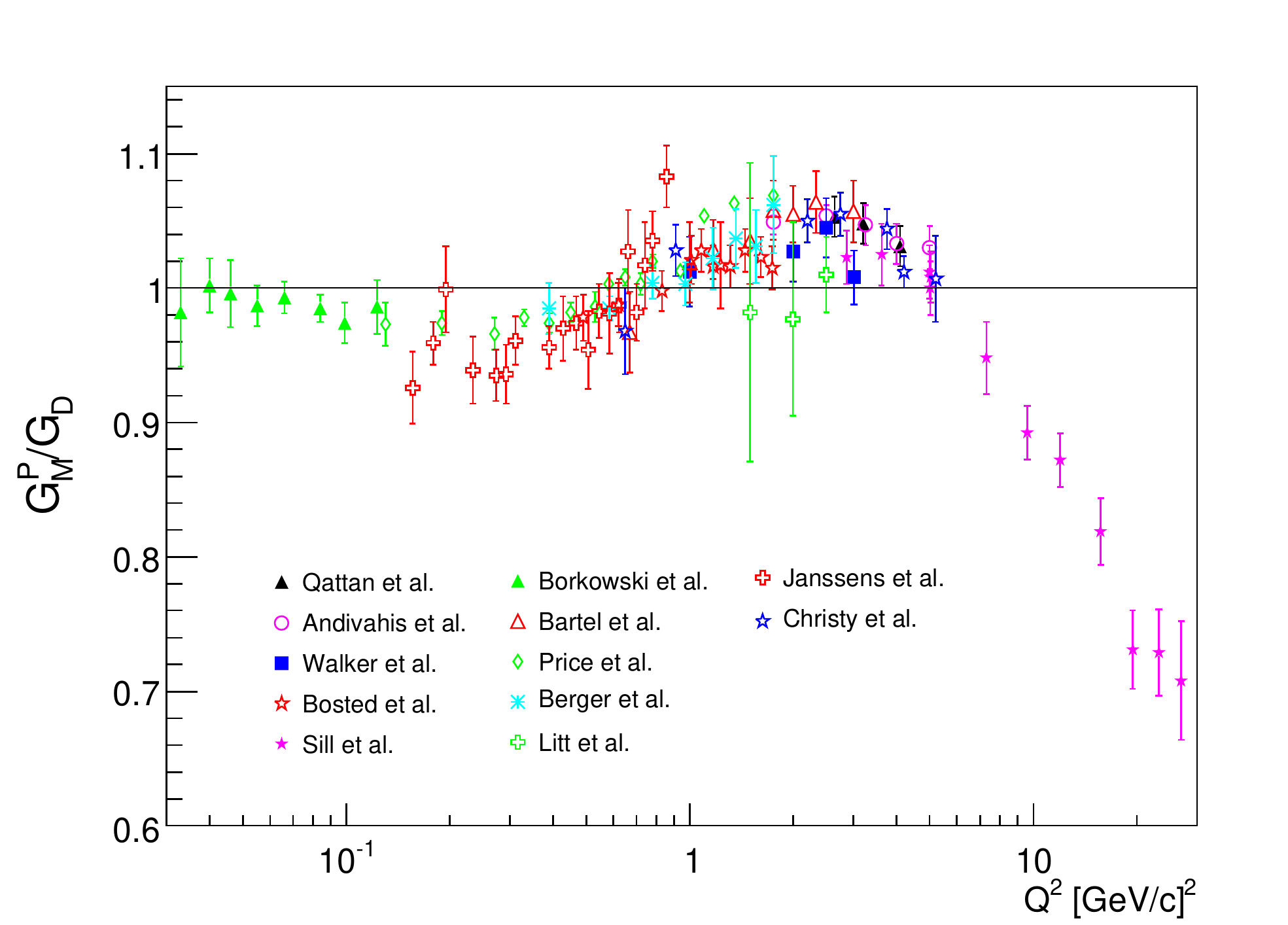}
    \caption{World data of $G_{Mp}$ from unpolarized
      measurements~\cite{qattan,walker,bosted_gm,sill,borkow,bartel,price,berger,litt,janss,christy},
      using the Rosenbluth method, normalized to the dipole parameterization.}
    \label{fig:gmp}
  \end{center}
\end{figure}

\begin{figure}
  \begin{center}
    \includegraphics[angle=0, width=.85\textwidth]{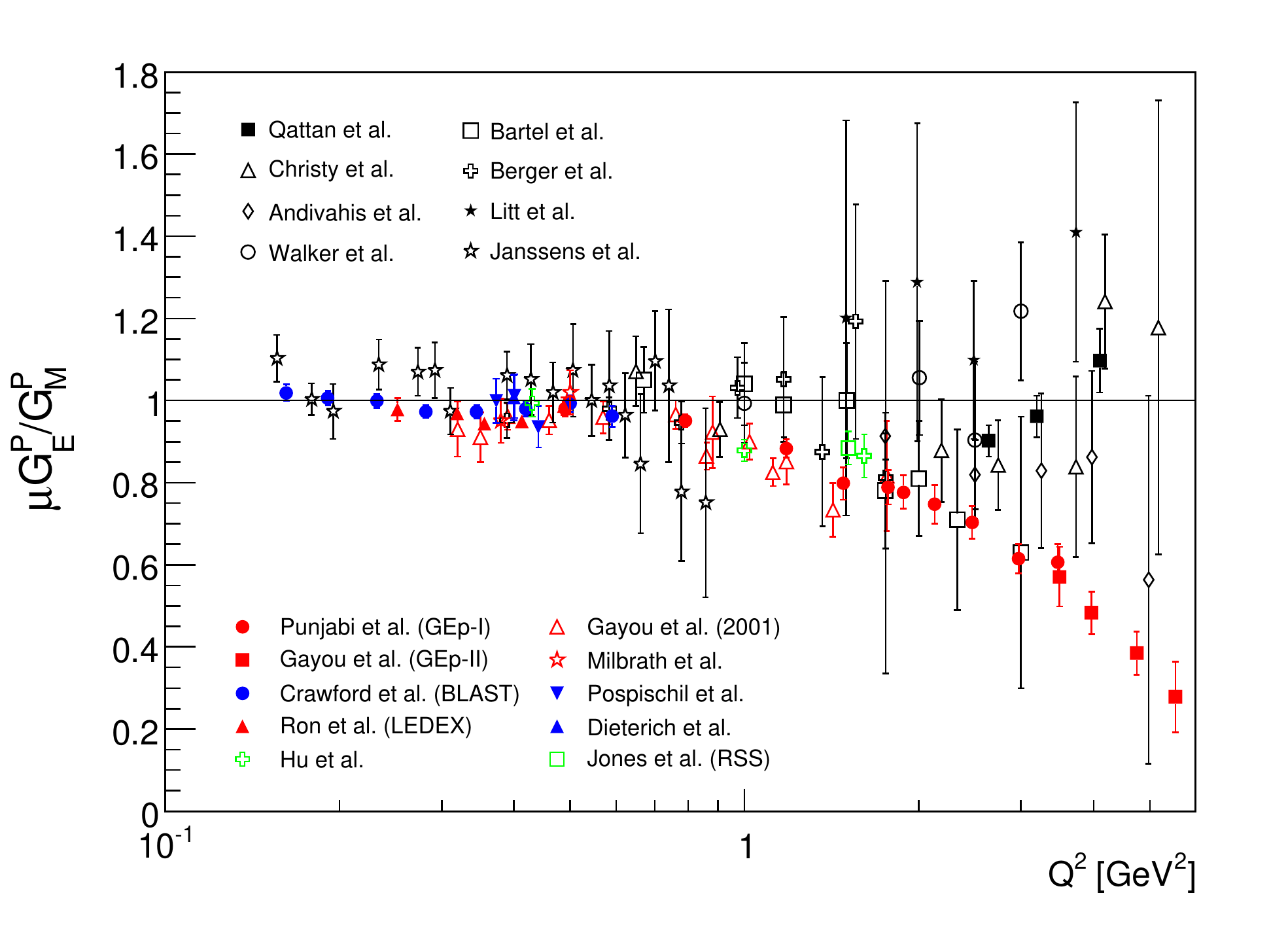}
    \caption{World data of the ratio $\mu_pG_{Ep}/G_{Mp}$ from
      unpolarized measurements (black symbols) using the Rosenbluth
      method and from polarization experiments (colored
      symbols) ~\cite{punj,Mark,Gayou,BLAST,LEDEX,hu,dieterich,posp}.}
    \label{fig:ratio}
  \end{center}
\end{figure}

Figs.~\ref{fig:gep} and \ref{fig:gmp} give a summary of the world data on the separated
proton form factors for unpolarized measurement using the Rosenbluth
separation method. It is clear that the extractions of $G_{Ep}$ from
Rosenbluth separation are of limited precision at high $Q^2$, and for
$G_{Mp}$, the data follow the dipole shape reasonably well up to $Q^2\sim7\mathrm{GeV}^2$ but show a large
deviation from this behavior at higher $Q^2$. Fig.~\ref{fig:ratio} shows the ratio
$\mu_p G_{Ep}/G_{Mp}$ from Rosenbluth separation. Earlier results
generally indicated that the form factor ratio stays around 1 but with
poor quality.

The polarization transfer technique was used for the first time by
Milbrath {\it et al.}~\cite{Mil} at the MIT-Bates facility at
$Q^2$ values of 0.38 and 0.50 $\mathrm{GeV}^2$. A
follow-up measurement was performed at the MAMI facility~\cite{mami},
for $Q^2 = 0.4\mathrm{GeV}^2$. A greater impact of the
polarization transfer measurement was made by two recent
experiments~\cite{Mark,punj,Gayou}, at Jefferson Lab Hall
A as shown in Fig.~\ref{fig:ratio}. The most striking feature
of the data is the sharp, practically linear decline as $Q^2$ increases.

\begin{figure}
  \begin{center}
    \includegraphics[angle=0, width=0.6\textwidth]{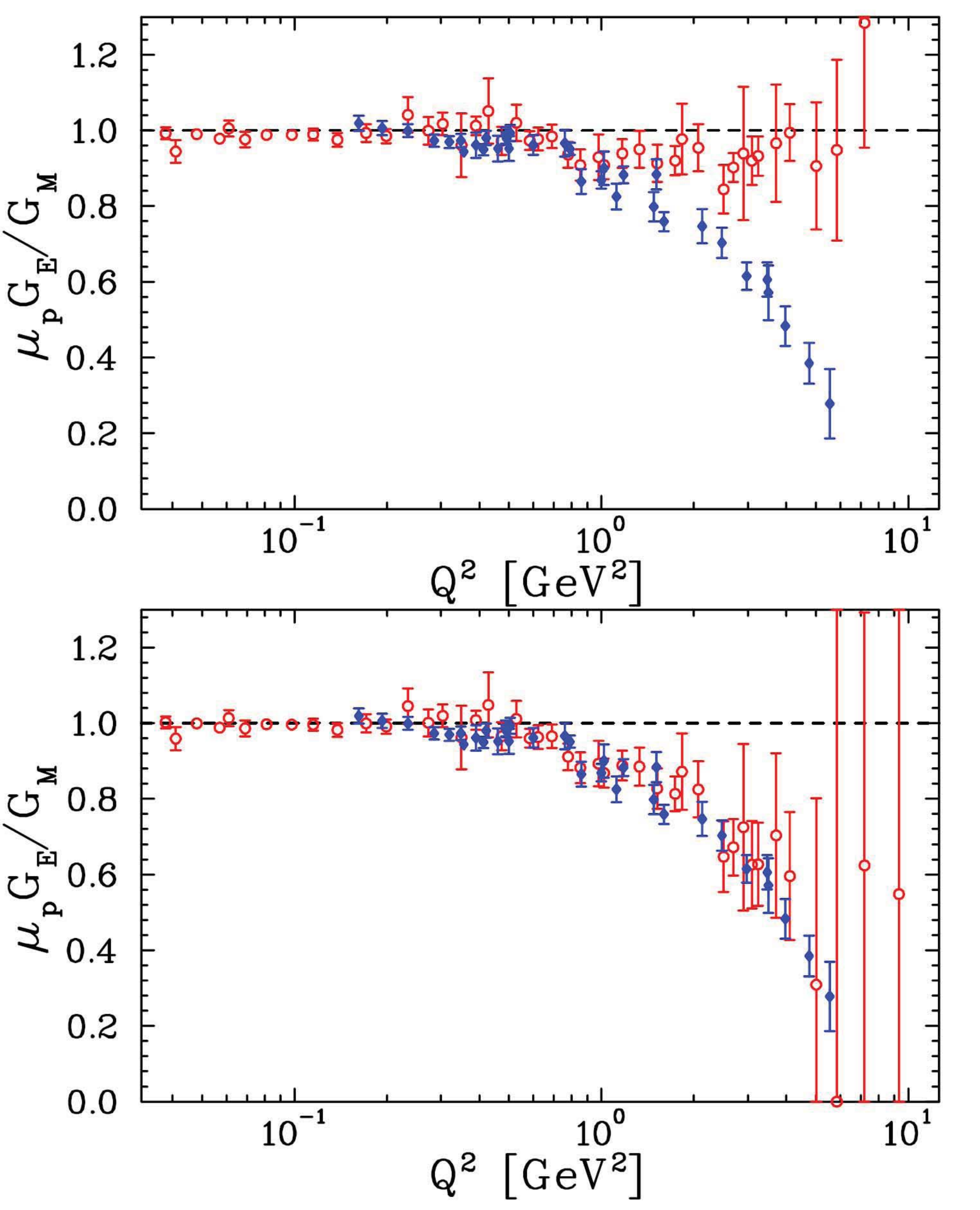}
    \caption{Ratio $\mu_pG_{Ep}/G_{Mp}$ extracted from
      polarization transfer (filled diamonds) and Rosenbluth method
      (open circles). The top (bottom) figures show Rosenbluth method
      data without (with) TPE corrections applied to the cross sections. Figures from~\cite{john:TPE}.}
    \label{fig:tpe}
  \end{center}
\end{figure}

In order to resolve the discrepancy between the results of the form
factor ratio from the two experimental techniques, an $\varepsilon$
dependent modification of the cross section is necessary. More recently, two-photon-exchange (TPE) contribution
 is considered as the main origin of this discrepancy. A number of recent theoretical studies of
TPE in elastic scattering have been performed~\cite{guichon,blunden,Chen,afana,blunden_melnitch,kondra,jain,borigy,Carlson}. These
indicate that TPE effects give rise to a strong angular-dependent
correction to the elastic cross section, which can lead to large
corrections to the extracted ratio. In fact, the results of
quantitative calculations based both on hadronic intermediate states
and on generalized parton distributions, provide strong evidence that
TPE effects can account for most of the difference between the
polarized and unpolarized data sets. Fig.~\ref{fig:tpe} shows a
comparison of the Rosenbluth data and the polarization data from the global
analysis~\cite{john:TPE}. The TPE correction brings the high $Q^2$
$\mu_pG_E/G_M$ points from unpolarized measurements into decent agreement with
the polarization transfer measurement data.

\section{Models and Global Fits}
While the world experimental data have been quite fruitful for the
nucleon electromagnetic form factors, significant theoretical progress
has also been made in recent years in understanding the nucleon
electromagnetic structure from the underlying theory of QCD. As the
theory of the strong interaction, QCD has been extremely well tested in
the high-energy region, i.e., in the perturbative QCD (pQCD)
regime. Ideally, one should be able to calculate the nucleon
electromagnetic form factors directly in pQCD regime to confront the
data. Unfortunately, it's impossible to solve QCD analytically in the
confinement regime where the available world experimental data are
located. Lattice QCD calculations based on first principles of QCD, on
the other hand, have shown much promise in this field, and is developing
rapidly. While pQCD give prediction for the nucleon form factors in
the perturbative region, QCD effective theories such as the chiral
perturbation theory can in principle provide reliable prediction in
the very low energy region. In between the low energy region and the
pQCD regime, various QCD-inspired models and other phenomenology
models exist. Thus, precision data in all experimentally accessible
regions is crucial in testing these predictions. There are some recent reviews~\cite{gao_sum, john_rev, charles_rev} that provide a nice summary on these models and predictions.

The newly developed Generalized Parton distributions (GPDs)~\cite{Ji_dvcs,Ji_dvcs_2,Rady,Rady_2,Rady_3}, which can be accessed through deeply virtual Compton scattering and deeply
virtual meson production, connect the nucleon form factors and the
nucleon structure functions probed in the deep inelastic scattering
experiments. The GPDs provide new insights into the structure of the
nucleon, and possibly provide a complete map of the nucleon wave-function.

The rest of the section will give a brief
discussion of various theoretical approaches used to calculate the
nucleon electromagnetic form factors.

\subsubsection{Scaling and pQCD}
In contract to the QED dynamics of the
leptonic probe, the QCD running coupling constant at 1-loop order is:
\begin{equation}
\alpha_s(Q^2)=\frac{\alpha_s(0)}{1+\frac{\alpha_s(0)}{16\pi^2}(11-\frac{2}{3}N_f)\mathrm{ln}(\frac{Q^2}{\Lambda^2})},
\end{equation}
where the string coupling constant $\alpha_s\to 0$ as the inter-quark
distance $\to$ 0. Thus, one can solve QCD using the perturbation
method in the limit of $Q^2\to \infty$. As illustrated in Fig.~\ref{fig:pqcd}, in pQCD picture, the large momentum of the virtual photon resolves the three leading quarks of the nucleon, and the momentum is transferred between the quarks through two successive gluon exchanges. Brodsky and Farrar~\cite{brodsky} proposed the following scaling law for the proton Dirac ($F_1$) and
Pauli form factor ($F_2$) at large momentum transfers based on dimensional
analysis:
\begin{equation}
F_1\propto (Q^2)^{-2},\:F_2\sim\frac{F_1}{Q^2}
\end{equation}
This prediction is a natural consequence of hadron helicity
conservation. Hadron helicity conservation arises from the vector
coupling nature of the quark-gluon interaction, the quark helicity
conservation at high energies, and assumption of zero quark
orbital angular momentum state in the nucleon.
\begin{figure}
  \begin{center}
    \includegraphics[angle=0, width=0.55\textwidth]{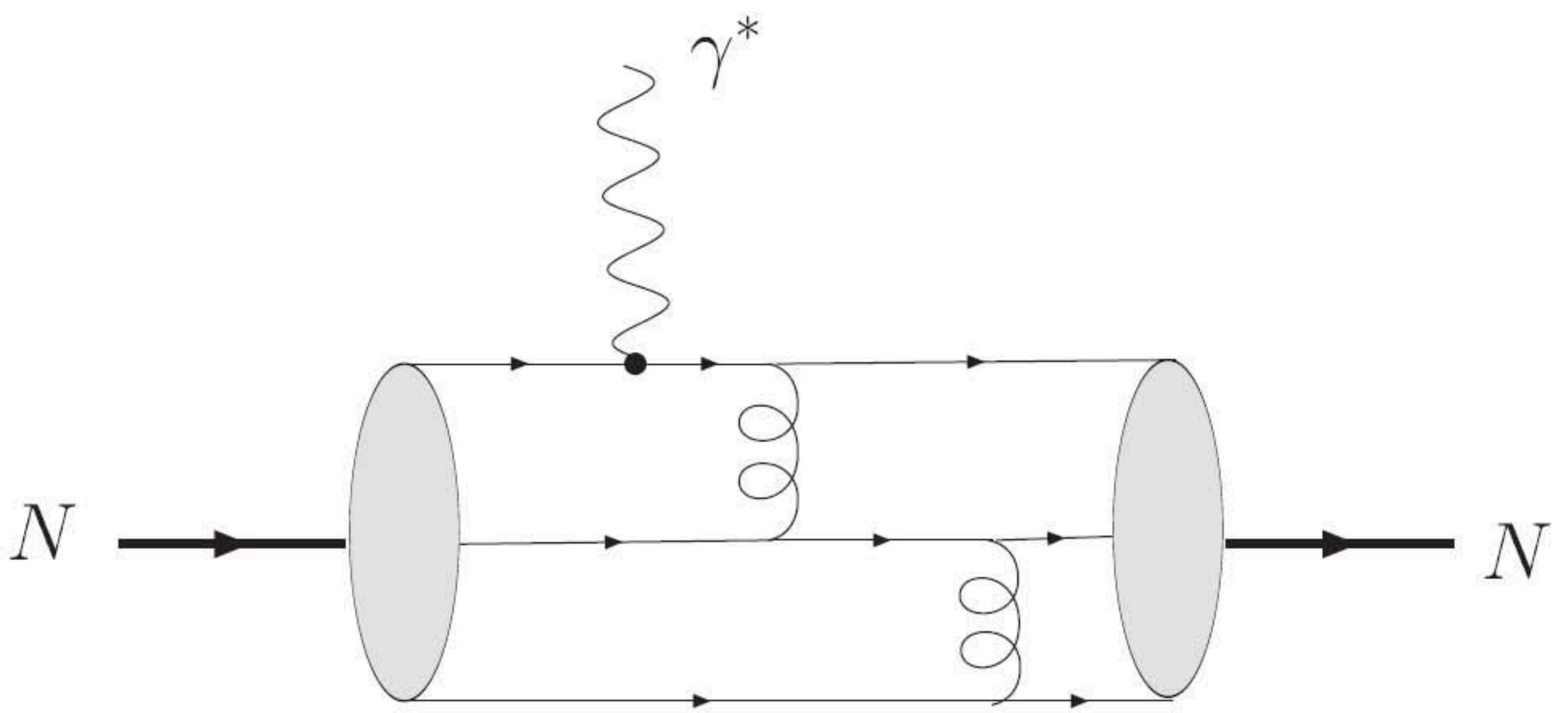}
    \caption{Pertubative QCD picture for the nucleon EM form factors.}
    \label{fig:pqcd}
  \end{center}
\end{figure}
In terms of the Sach's form factors $G_{Ep}$ and $G_{Mp}$, the scaling
result predicts: $\frac{G_{Ep}}{G_{Mp}}\to$ constant at large
$Q^2$. Such scaling results were confirmed in a short-distance pQCD
analysis carried out by Brodsky and Lepage~\cite{lepage}. Considering the
proton magnetic form factor at large $Q^2$ in the Breit frame, the
initial proton is moving in the $z$ direction and is struck by a highly
virtual photon carrying a large transverse momentum, $q_\perp^2=Q^2$.
The form factor corresponds to the amplitude that the composite proton
absorbs the virtual photon and stays intact. Thus, the form factor
becomes the product of the following three probability amplitudes:
\begin{itemize}
\item the amplitude for finding the valence $|qqq>$ state in the
  incoming proton.
\item the amplitude for this quark state to scatter from the incoming
  photon producing the final three-quark state with colinear momenta.
\item the amplitude for the final three-quark state to reform a proton.
\end{itemize}
Based on this picture, Brodsky and Lepage obtained the following
result within their short-distance pQCD analysis~\cite{lepage}:
\begin{eqnarray}
G_M(Q^2)&=&\frac{32\pi^2}{9}\frac{\alpha_s^2(Q^2)}{Q^4}\sum_{n,m}b_{nm}(\mathrm{ln}\frac{Q^2}{\Lambda^2})^{-\gamma_n-\gamma_m}[1+\mathcal{O}(\alpha_s(Q^2),m^2/Q^2)]{}
\nonumber\\
{}&\to&\frac{32\pi^2}{9}C^2\frac{\alpha_s^2(Q^2)}{Q^4}(\mathrm{ln}\frac{Q^2}{\Lambda^2})^{-4/3\beta}(-e_\parallel),
\end{eqnarray}
where $\alpha_s(Q^2)$ and $\Lambda$ are the QCD strong coupling
constant and scale parameter respectively, $b_{nm}$ and $\gamma_{m,n}$ are QCD
anomalous dimensions and constants, and $e_\parallel\:(-e_\parallel)$
is the mean total charge of quarks with helicity parallel (anti-parallel) to the nucleon's
helicity. For protons and neutrons, the mean total charge is given by:
\begin{equation}
e_\parallel^p=1,-e_\parallel^p=0,e_\parallel^n=-e_\parallel^n=-1/3,
\end{equation}
and based on the fully symmetric flavor-helicity wave function. For the
proton electric form factor, one obtains similar results for the $Q^2$
dependence in the $Q^2\to\infty$ limit, and the short-distance pQCD
analysis predicts the same scaling law as the dimensional analysis for
the proton form factors: $\frac{G_{Ep}}{G_{Mp}}\to$ constant and
$\frac{Q^2F_2}{F_1}\to$ constant.

Recently, Belitsky, Ji and Yuan~\cite{belitsky} performed a pQCD analysis of
the nucleon's Pauli form factor $F_2$ in the asymptotically large
$Q^2$ limit. They found that the leading contribution to $F_2$ goes
like $1/Q^6$, which is consistent with the scaling result obtained by
Brodsky and Farrar~\cite{brodsky}. Fig.~\ref{fig:pqcd} shows data on the
scaled proton Dirac and Pauli form factor ratio $\frac{Q^2F_2}{F_1}$ from
Jefferson Lab as a function of $Q^2$ together with various
predictions. While the short-distance pQCD analysis~\cite{lepage} predicts a
constant behavior for the $\frac{Q^2F_2}{F_1}$ in the $Q^2\to\infty$,
the data are in better agreement with the $\frac{QF_2}{F_1}$ scaling
behavior. The data could imply that the asymptotic pQCD scaling region
has not been reached so far or that hadron helicity is not conserved in the
experimentally tested regime. However, Brodsky, Hwang and
Hill~\cite{brodsky_2} were able to fit the Jefferson Lab data using a form
consistent with pQCD analysis and hadron helicity conservation by
taking into account higher twist contributions. Ralston and
Jain~\cite{ralston} argue that the $\frac{QF_2}{F_1}$ scaling behavior is
expected from pQCD when one takes into account contributions to the
proton quark wave function from states with non-zero orbital angular
momentum.
\begin{figure}
  \begin{center}
    \includegraphics[angle=0, width=0.75\textwidth]{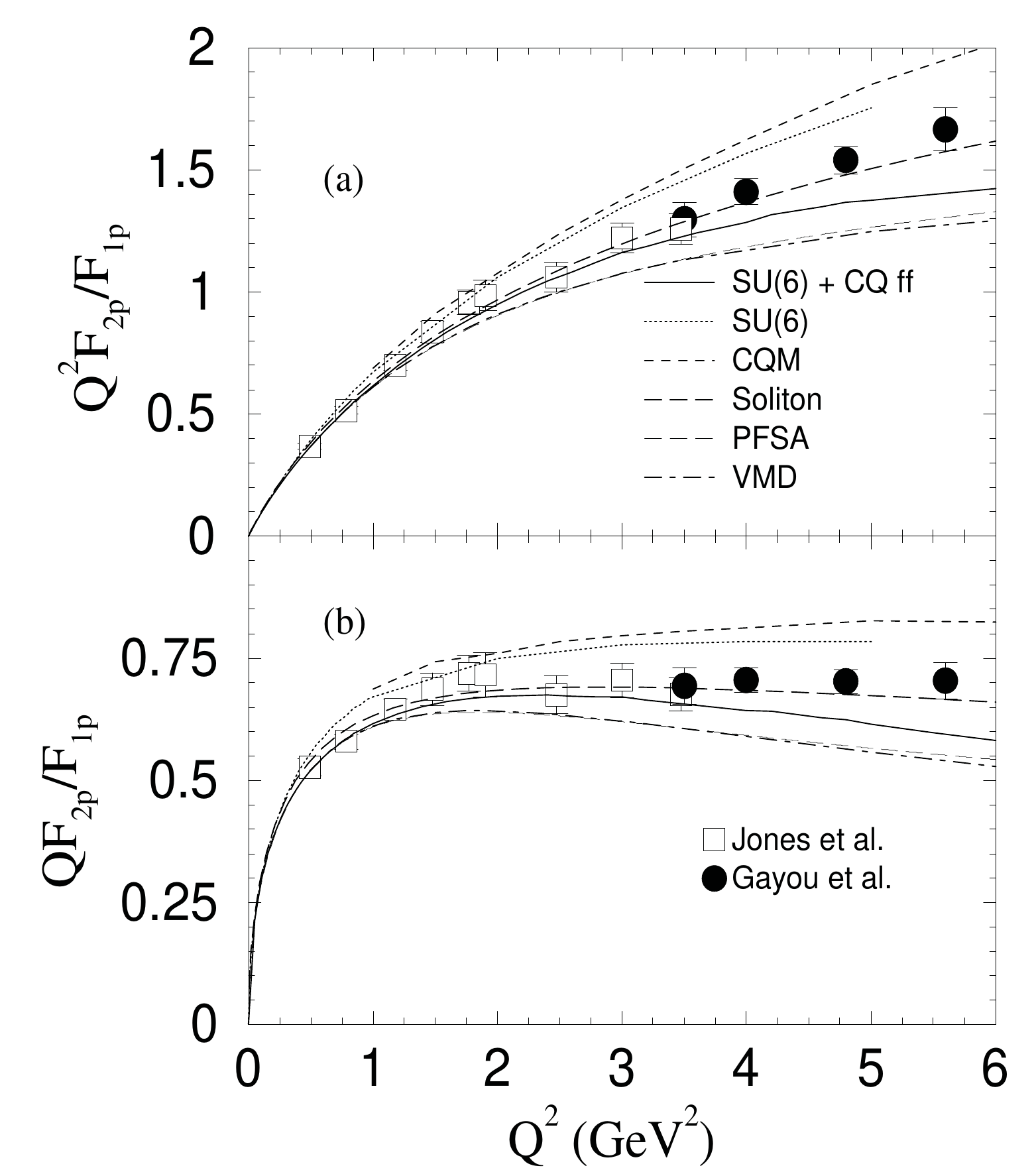}
    \caption{The scaled proton Dirac and Pauli form factor ratio:
      $\frac{Q^2F_2}{F_1}$ (upper panel) and $\frac{QF_2}{F_1}$ (lower
      panel) as a function of $Q^2$ in $\mathrm{GeV}^2$. The data are
      from~\cite{Mark,Gayou}. Shown with statistical uncertainties
      only. The dash-dotted curve is a new fit based on vector meson
      dominance model (VMD) by Lomon~\cite{lomon}. The thin long dashed
    curve is a point-form spectator approximation (PFSA) prediction of
  the Goldstone boson exchange constituent quark model
  (CQM)~\cite{wagen}. The solid and the dotted curves are the CQM
  calculations by Cardarelli and Simula~\cite{card} including SU(6)
  symmetry breaking with and without constituent quark form factors,
  repectively. The long dashed curve is a relativistic chiral soliton
  model calculation~\cite{holz}. The dashed curve is a relativistic CQM by Frank,
  Jennings, and Miller~\cite{frank}. Figure from~\cite{gao_sum}.}
    \label{fig:pqcd}
  \end{center}
\end{figure}
Miller~\cite{Miller} recently used light front dynamics in modeling the
nucleon as a relativistic system of three bound constituent quarks
surrounded by a cloud of pions. While the pion cloud is important for
understanding the nucleon structure at low momentum transfer,
particularly in understanding the neutron electric form factor, quark
effects are expected to dominate at large momentum transfers. The model was able to
reproduce the observed constant behavior of $\frac{QF_2}{F_1}$ as a
function of $Q^2$ and the $\frac{QF_2}{F_1}$ is predicted to be a
constant up to a $Q^2$ value of 20 $\mathrm{GeV}^2$.

 \subsubsection{Lattice QCD Calculations}
An analytical approach in solving QCD at low momentum transfers is
prevented due to the non-perturbative nature of QCD at large
distance. However, important conceptual and technical progress has
been made over the last decade in solving QCD on the lattice. In general, lattice QCD calculations are a discretized version of QCD formulated in terms of path integrals on a space-time lattice~\cite{wil74} with the bare quark masses and the coupling constant as the only parameters. The parameters commonly defined in lattice calculations are:
\begin{itemize}
\item lattice spacing $a$: separate calculation at several values of $a$ is required in order to extrapolate results at finite lattice spacing $a$ to $a=0$ by continuum theory.
\item spatial length of the box $L$: as lattice calculations are performed for a finite lattice size, one must define a box size large enough to fit the hadrons inside, and this requires to increase the number of sites as one decreases $a$.
\item pion mass $m_{\pi}$: to keep finite volume effects small, one must have a box size much larger than the Compton wavelength of the pion. Present lattice QCD calculations take $Lm_{\pi}\ge 5$.
\end{itemize}
State-of-the-art lattice calculations for nucleon structure studies use $a\le 0.1$ fm and $L\sim3$ fm, and the pion mass down to a few hundred MeV. These results are connected with the physical world by extrapolation down to the physical quark masses ($m_q$ is proportional to $m^2_{\pi}$ for small quark masses). As the computational costs of such calculations increase like $m_{\pi}^{-9}$, it was only until very recently that pion mass values below 350 MeV~\cite{ber00,far04} have been reached.

\begin{figure}
  \begin{center}
    \includegraphics[angle=0, width=0.4\textwidth]{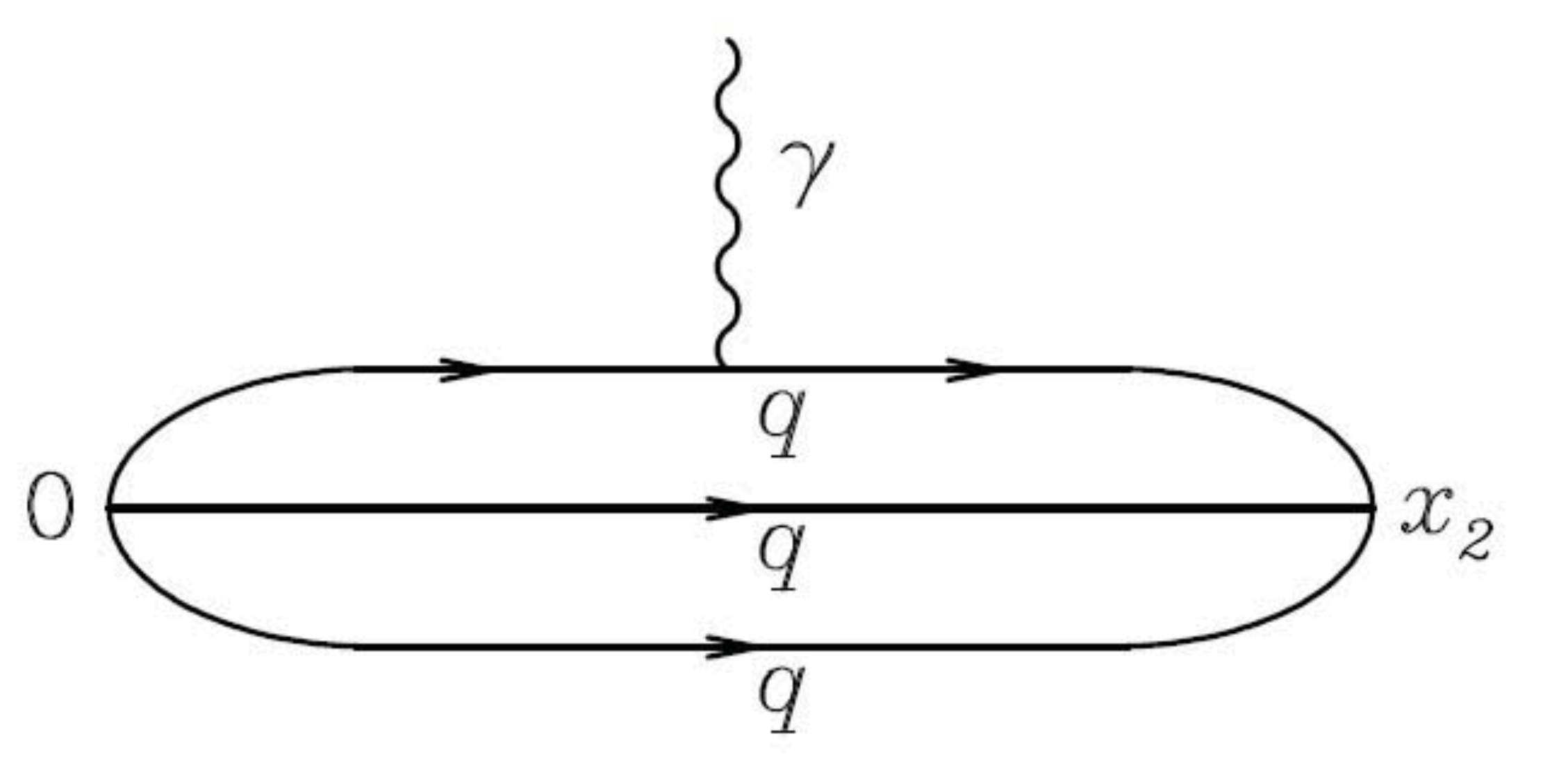}
    \includegraphics[angle=0, width=0.4\textwidth]{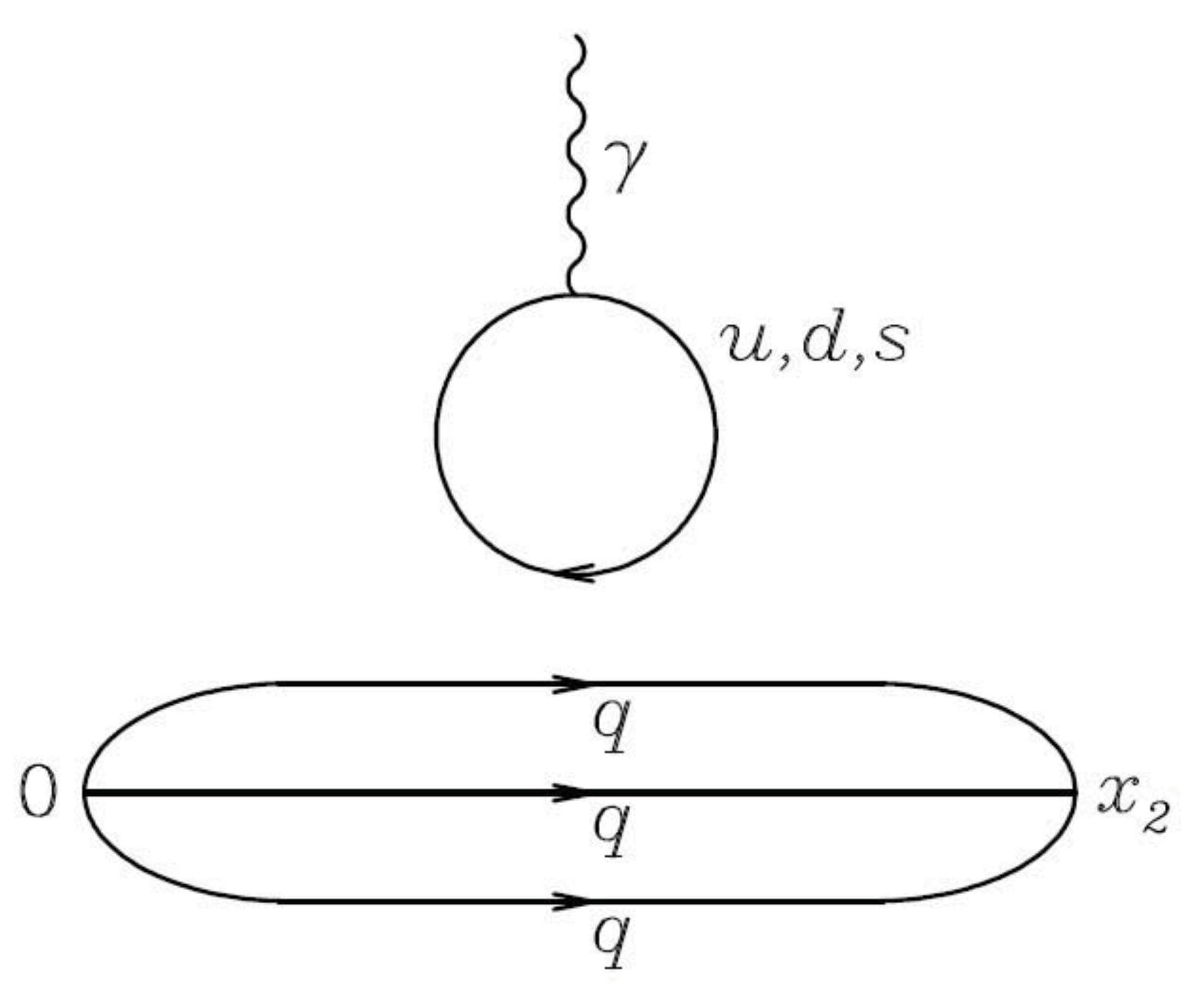}
    \caption{Diagrams illustrating the two topologically different contributions when calculating nucleon EM form factors in lattice QCD~\cite{boi06}.}
    \label{fig:dia_lattice}
  \end{center}
\end{figure}

\begin{figure}
  \begin{center}
    \includegraphics[angle=0, width=0.45\textwidth]{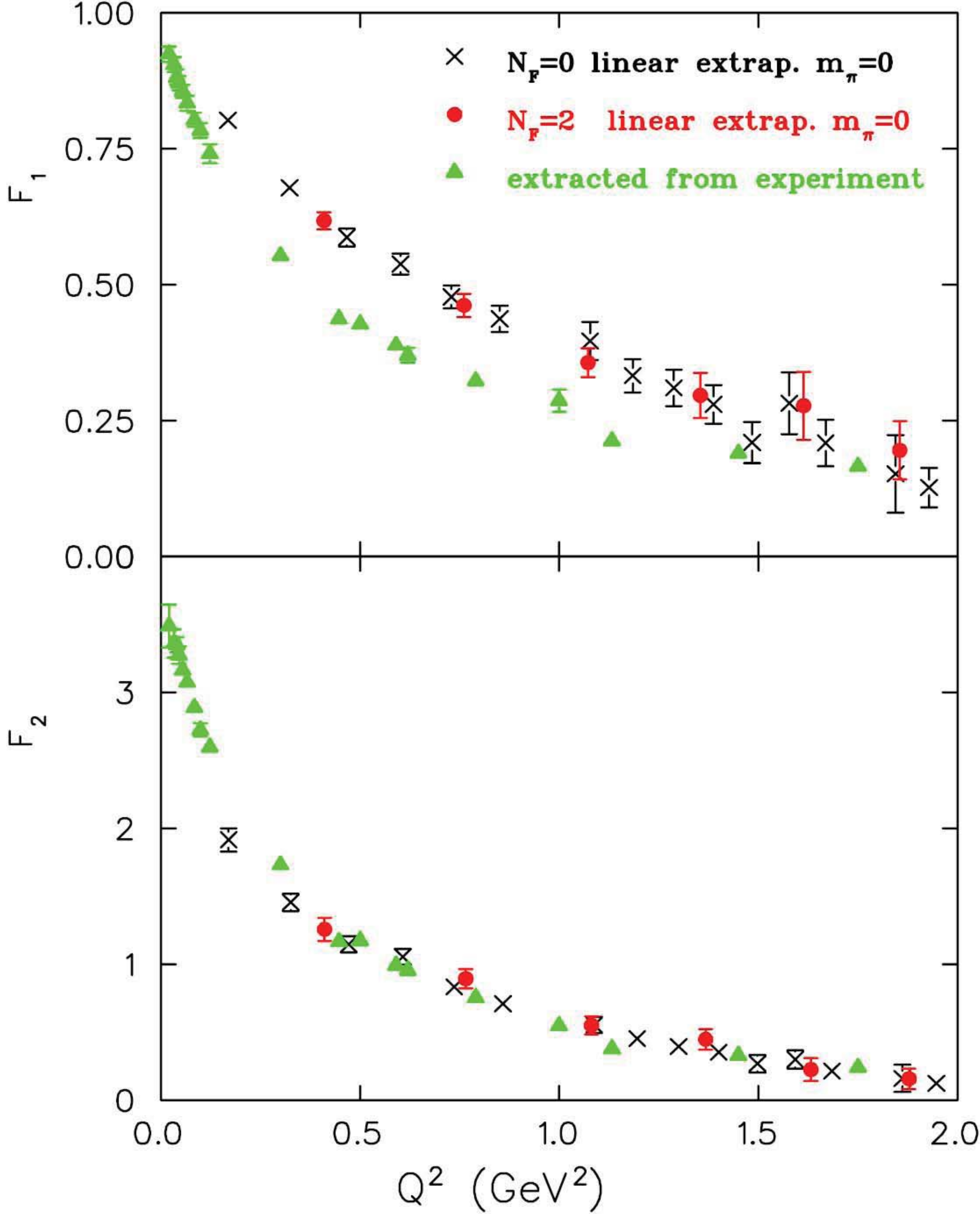}
    \includegraphics[angle=0, width=0.45\textwidth]{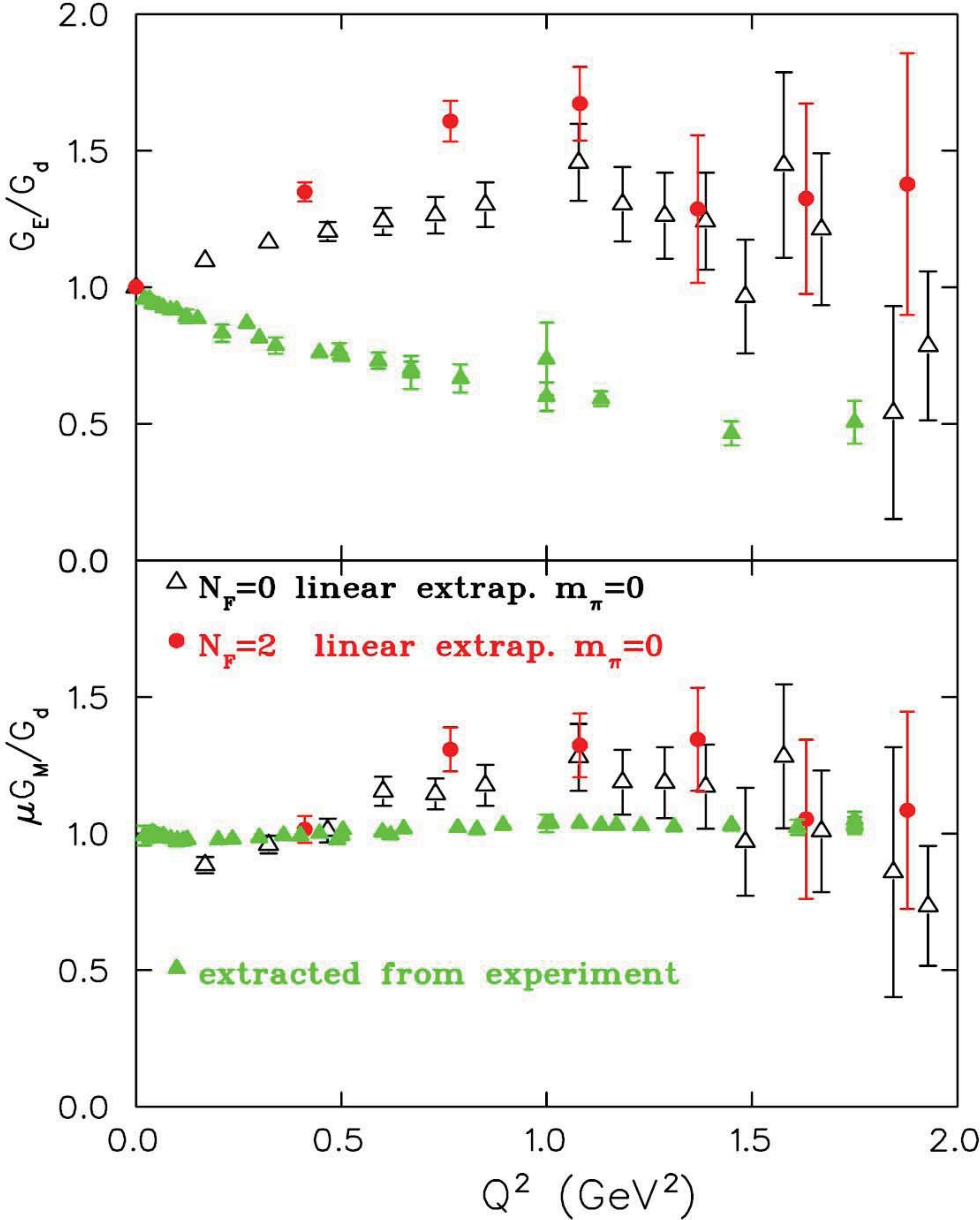}
    \caption{Lattice QCD results from the Nicosia-MIT group~\cite{ale06a} for the isovector form factors $F^V_1$ (upper left) and $F^V_2$ (lower left) as a function of $Q^2$. Both the quenched results ($N_F=0$) and unquenched lattice results with two dynamical Wilson fermions ($N_F=2$) are shown for three different pion mass values. The right panels show the results for $G^V_E$ (upper right) and $G^V_M$ (lower right), divided by the standard dipole form factor, as a function of $Q^2$ in the chiral limit. The filled triangles show the experimental results for the isovector form factors extracted from the experimental data for the proton and neutron form factors. Figure from~\cite{ale06a}.}
    \label{fig:mit_lattice}
  \end{center}
\end{figure}

Also, most of the lattice results obtained so far were carried out in the so-called quenched approximation in which the quark loop contributions, i.e. the sea quark contributions, are suppressed. As illustrated in Fig.~\ref{fig:dia_lattice}, the disconnected diagram (right panel) involves a coupling to a $q\bar q$ loop, thus, it requires a numerically more intensive calculation and is neglected in most lattice studies. The Nicosia-MIT group~\cite{ale06a} has performed a high-statistics calculation of nucleon isovector EM form factors, both in the quenched approximation and in full QCD, using two dynamical Wilson fermions. The largest $Q^2$ value accessible is around $Q^2\simeq 2~\mathrm{GeV}^2$. When comparing with experiments, the Nicosia-MIT group uses a linear fit in $m^2_{\pi}$. As shown in Fig.~\ref{fig:mit_lattice}, one can see that both the quenched and unquenched lattice results of~\cite{ale06a} largely overestimate the data for $F^V_1$. For $F^V_2$, one observes a stronger quark mass dependence, bringing the lattice results closer to experiment with decreasing $m_{\pi}$.

\begin{figure}
  \begin{center}
    \includegraphics[angle=0, width=0.95\textwidth]{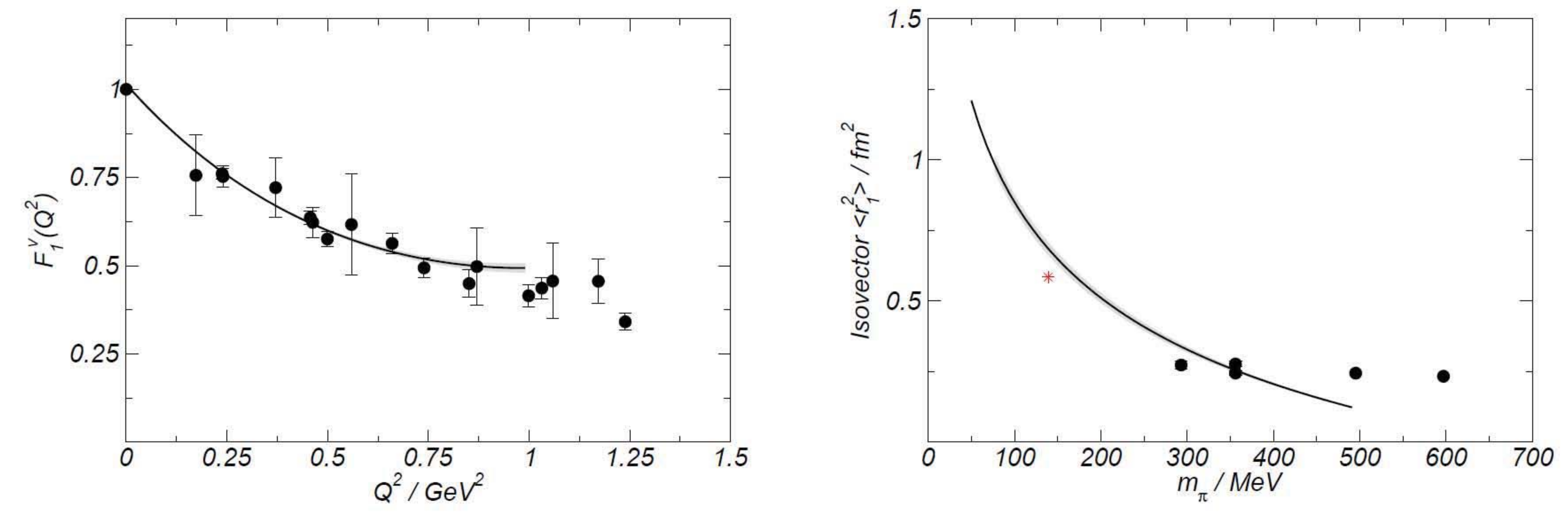}
    \caption{Isovector form factor $F^V_1(Q^2)$ lattice data with best fit small scale expansion (SSE) at $m_{\pi} = 292.99$ MeV (left panel). The line in the right-hand panel shows the resulting Dirac radii, $\langle r_1^2\rangle$. Also shown as the data points are the Dirac radii obtained from dipole fits to the form factors at different pion masses. Figure from~\cite{schroers09}.}
    \label{fig:lattice_new}
  \end{center}
\end{figure}

The lattice calculations at present are still severely limited by available
computing power. Hence, the uncertainties in extrapolating lattice results
to the physical quark mass are rather large, particularly with the
naive linear extrapolation in quark mass. Thus, the challenge is to
find an accurate and reliable way of extracting the lattice results to
the physical quark mass. The extrapolation methods which incorporate
the model independent constraints of chiral symmetry~\cite{detmold,lein},
especially the leading non-analytic (LNA) behavior of chiral
perturbation theory~\cite{hackett} and the heavy quark limit~\cite{shifman} are
exciting development in these years. Recently, the LHPC collaboration~\cite{schroers09} calculated new high-statistics results using a mixed action of domain wall valence quarks on an improved staggered sea, and performed chiral fits to both vector and axial form factors. Through the comparison with the experimental data (see Fig.~\ref{fig:lattice_new}), they found that a combination of chiral fits and lattice data is promising with the current generation of lattice calculations.

\subsubsection{Vector Menson Dominance (VMD) Model}
In the low $Q^2$ region, several effective models have been developed to describe the nucleon properties. Most of them are semi-phenomenological, which means that they require experimental data as inputs and thus have little predictive power. Usually each model is valid in a limited $Q^2$ range.
\begin{figure}
  \begin{center}
    \includegraphics[angle=0, width=0.35\textwidth]{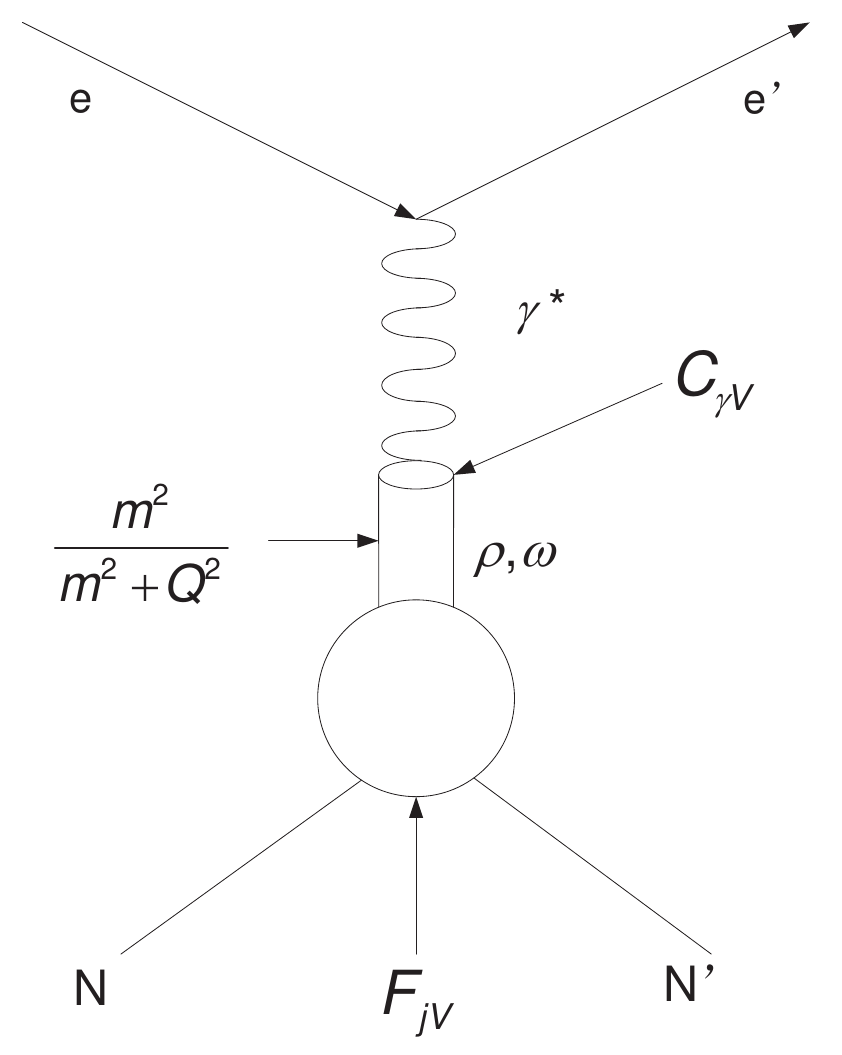}
    \caption{Photon-nucleon coupling in the VMD picture.}
    \label{fig:vmd}
  \end{center}
\end{figure}
One of the earlier attempts to describe the proton form factors is a
semiphenomenological fit introduced by Iachello {\it et al.}~\cite{iache}. It is based on a model that the scattering amplitude is written as an intrinsic form factor
of a bare nucleon multiplied by an amplitude derived from the
interaction with the virtual photon via vector meson dominance
(VMD). As shown in Fig.~\ref{fig:vmd}, the nucleon form factors are expressed
in terms of photon-meson coupling strengths $C_{\gamma V}$ and
meson-nucleon vertex form factors $F_{jV}$:
\begin{equation}
F_j^{is,iv}(Q^2)=\sum_i\frac{m_i^2C_{\gamma Vi}}{m_i^2+Q^2}F_{jVi}(Q^2),
\end{equation}
where the sum is over vector mesons of mass $m_i$ and $is$ and $iv$
correspond to the isoscalar and isovector electromagnetic currents
respectively. The form factors are then given by:
\begin{equation}
2F_{jp}=F^{is}_j+F^{iv}_j;\:2F_{jn}=F^{is}_j-F^{iv}_j,
\end{equation}
where $j=1,2$ and $p$ and $n$ denote the proton and neutron
respectively.

\begin{figure}
  \begin{center}
    \includegraphics[angle=0, width=0.75\textwidth]{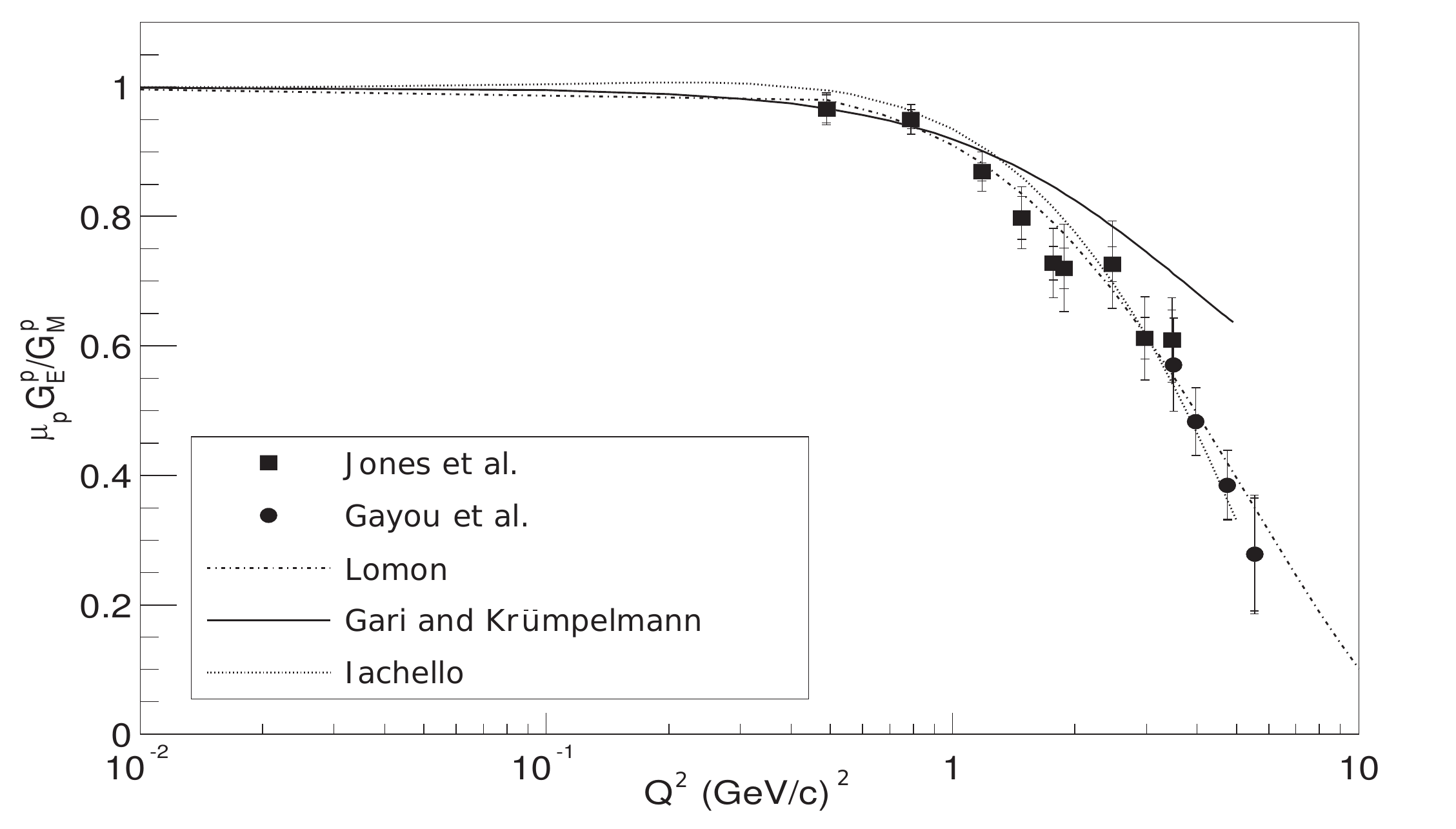}
    \caption{The proton form factor ratio $\mu_pG_{Ep}/G_{Mp}$ from
      Jefferson Lab Hall A together with calculations from various VMD
    models.}
    \label{fig:vmd1}
  \end{center}
\end{figure}

\begin{figure}
  \begin{center}
    \includegraphics[angle=0, width=0.75\textwidth]{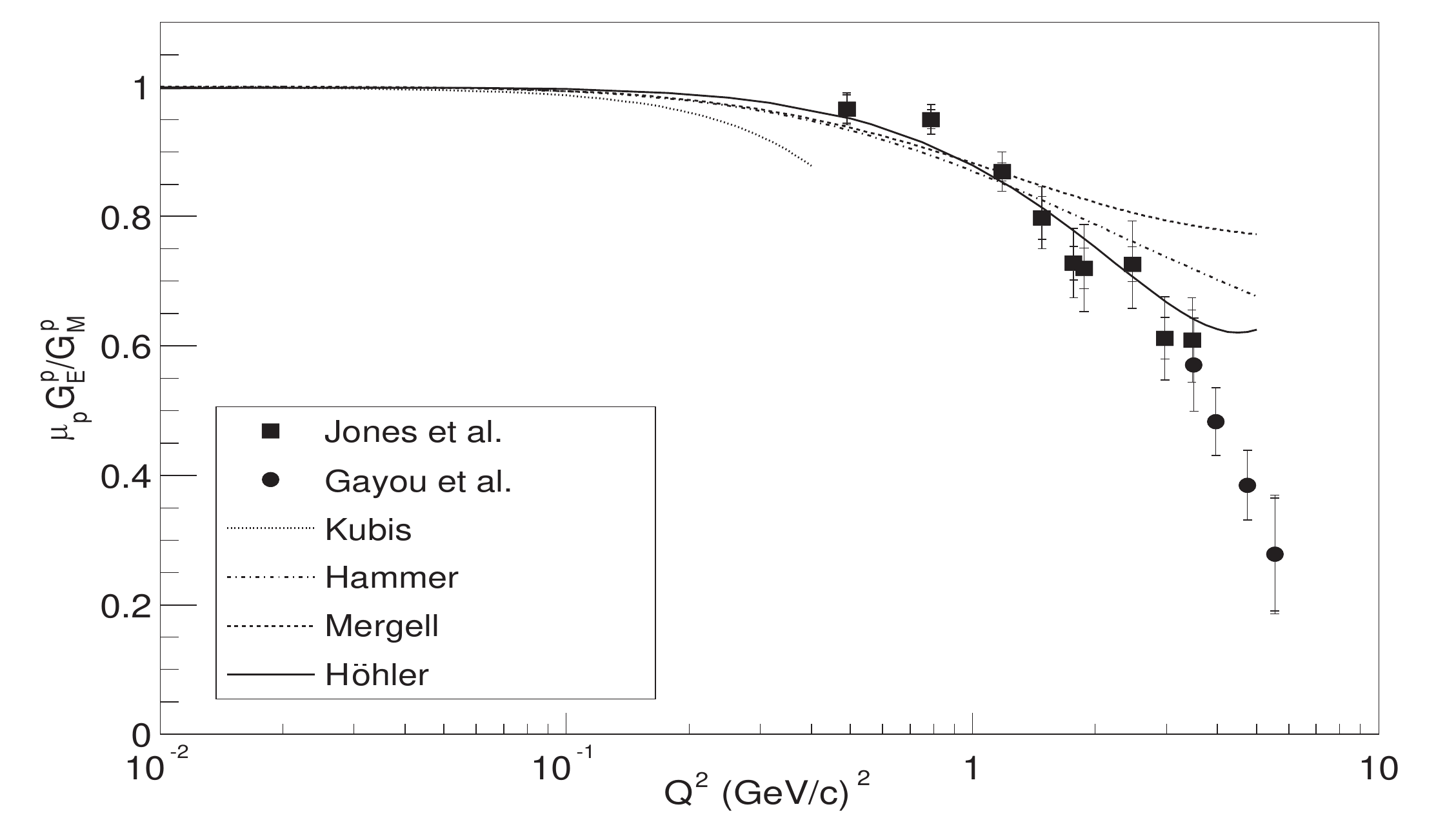}
    \caption{The proton form factor ratio $\mu_pG_{Ep}/G_{Mp}$ from
      Jefferson Lab Hall A together with calculations from dispersion
      theory fits. Figure from~\cite{gao_sum}}
    \label{fig:vmd2}
  \end{center}
\end{figure}

Various forms of the intrinsic bare nucleon form factor have been used:
dipole, monopole, eikonal. However, since this function is
multiplicative, it cancels out in the ratio $G_E/G_M$. The VMD
amplitude was written in terms of parameters fit to data. Gari and
Kr$\ddot \mathrm{u}$mpelmann~\cite{gari_1} extended the basic VMD model with
an additional term to include quark dynamics at large $Q^2$ via
pQCD. Lomon updated this model~\cite{lomon} by including the width of the
$\rho$ meson and additional higher mass vector meson exchanges. The model
has been further extended~\cite{lomon_2} to include the $\omega'$(1419)
isoscalar vector meson pole in order to describe the Jefferson Lab
proton form factor ratio data at high $Q^2$. Fig.~\ref{fig:vmd1} shows the proton form
factor ratio data as a function of $Q^2$ together with predictions
from various VMD models discussed above. While these models have
limited predictive power due to the tunable parameters, once the high $Q^2$ data have fixed the
parameters, the approach to low $Q^2$ can be constrained. However,
one can obviously see that these calculations are still different in
the low $Q^2$ range.
H$\ddot\mathrm{o}$hler~\cite{hohler} fit the $e-N$ scattering data with a
dispersion ansatz, and the contributions from
$\rho,\:\omega,\:\phi,\:\rho'$ and $\omega'$ were included and
parameterized. The proton form factor ratio is obtained and is in good
agreement with the Jefferson Lab data up to $Q^2\approx 3~\mathrm{GeV}^2$ as shown in Fig.~\ref{fig:vmd2}.

In recent years, these VMD relation approaches have been extended to include
chiral perturbation theory~\cite{mergell,hammer,meiss_1,meiss_2,kubis}. Mergell {\it et al.}~\cite{mergell}
obtained a best fit that gave an rms proton radius near 0.85 fm, which is close
 to the accepted value of 0.86 fm. However, simultaneously fitting the neutron data did not yield better results.
 Hammer {\it et al.}~\cite{hammer}
included the available data in the time-like region in the fit to
determine the model parameters. The later work by Kubis~\cite{kubis} was restricted to the low $Q^2$ domain of $0 - 0.4~\mathrm{GeV}^2$ and used
the accepted proton RMS radius of 0.86 fm as a constraint. The
comparison between data and the different models are shown in
Fig.~\ref{fig:vmd2}. It is not a surprise to find that these models failed
to describe the high $Q^2$ data when their region of validity was
claimed to be for $Q^2\le0.4~\mathrm{GeV}^2$.

Recently, an updated dispersion-theoretical analysis~\cite{hammer_2} describes the nucleon form factors
through the inclusion of additional unphysical isovector and
isoscalar poles whose masses and widths are fit parameters to
the form factors. The parametrization of the spectral functions
includes constraints from unitarity, pQCD, and recent measurements of
the neutron charge radius.
\begin{figure}
  \begin{center}
    \includegraphics[angle=0, width=0.95\textwidth]{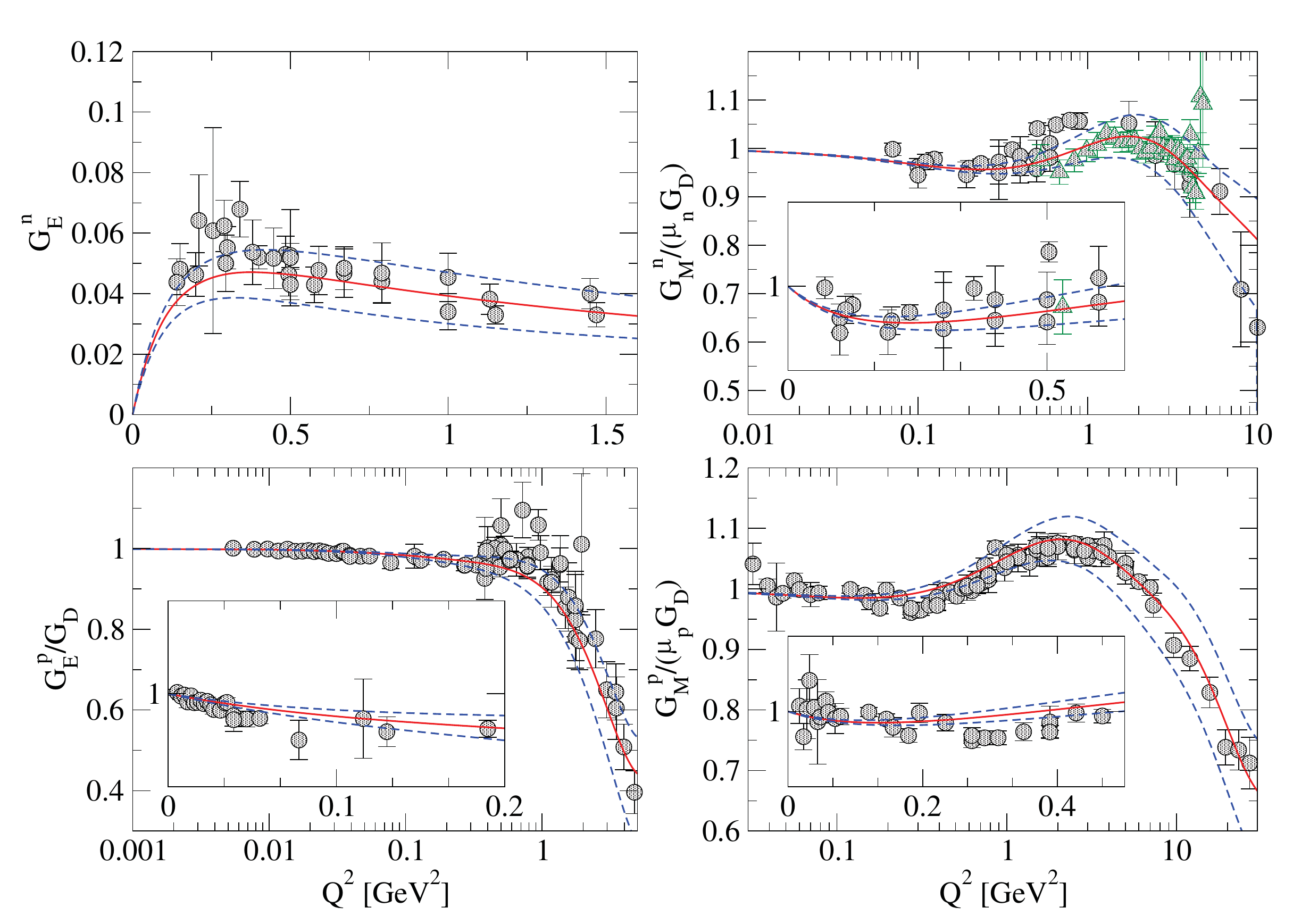}
    \caption{The nucleon electromagnetic form factors for space-like
      momentum transfer with the explicit pQCD continuum. The
      solid line gives the fit~\cite{belu_2} together with the world
      data (circles) including the JLab/CLAS data for $G_{Mn}$ (triangles), while the dashed lines indicate the
      error band. Figure from~\cite{belu_2}.}
    \label{fig:ff_vmd4}
  \end{center}
\end{figure}
Belushkin {\it et al.}~\cite{belu_2} updated the analysis by including
contributions from the $\rho\pi$ and $K\bar{K}$ isoscalar continua as
independent inputs, in addition to the $2\pi$ continuum. The $2\pi$
continuum is evaluated using the latest experimental data for the pion
time-like form factor~\cite{belu_2}. The $K\bar{K}$ continuum is obtained
from an analytic continuation of $KN$ scattering data~\cite{hammer_3}.
World data were analyzed in both space-like and time-like regions, and
the fits were in general agreement with the data. Fig.~\ref{fig:ff_vmd4} shows the results
for space-like momentum transfers compared to the published world
data, which includes preliminary CLAS data.

\subsubsection{Constituent Quark Models}
In the constituent quark model (CQM), the nucleon is described as the
ground state of a three-quark system in a confining potential. In this picture,
the ground state baryon, which is composed of the three
lightest quarks ($u$, $d$, $s$), is described by $SU(6)$ flavor wave functions
and an antisymmetric color wave function. This non-relativistic model, despite its simplicity,
gives a relatively good description of baryon static properties, such as nucleon magnetic moments and the charge and magnetic radii.

However, to calculate electromagnetic form factors in the high-$Q^2
 (1-10~\mathrm{GeV}^2)$ region, relativistic effects need to be considered.
 Relativistic constituent quark models (RCQM) are
based on relativistic quantum mechanics as opposed to quantum field
theory. The goal is to formulate a mechanics where the Hamiltonian
acts on a suitable Hilbert space, similar to the non-relativistic
case. For any relativistic quantum theory, it must respect Poincar$\acute{\mathrm{e}}$
invariance. There are three classes of
hamiltonian quantum dynamics which satisfy Poincar$\acute{e}$
invariance~\cite{dirac}: the instant form, light-front form, and point form.

In the instant form, the Einstein mass relation $p_\mu p^\mu=m^2$
takes the form:
\begin{equation}
p^0=\pm \sqrt{\vec{p}^2+m^2}
\end{equation}
which has two solutions for $p^0$, thus allowing quark-antiquark pair
creation and annihilation in the vacuum, and it makes the theory
complicated. In this case, the generators of the Poincar$\acute{\mathrm{e}}$
group are the energy of the system, whereas, the rotations do not
contain interactions. This allows states of good angular momentum to be
easily constructed.

In the point form, where the dynamical variables
refer to the physical conditions on some three-dimensional surface
rather than an instant, boosts and rotations are
dynamical. It has the angular momenta and Lorentz boosts the same as the
free case, but has complications in dealing with all four
momentum components.

In the light-front dynamics, the space-time variables $x$ and $t$ are
transformed to $x^\pm = \frac{1}{\sqrt{2}}(t\pm x)$ with corresponding
canonical momenta $p^\pm$. This system has the advantages of a simple
Hamiltonian without negative energies, the ability to separate the
center of mass from the relative motion of particles, and boosts which
are independent of the interactions.

Several theorists have calculated the proton electric and magnetic
form factors using various versions of CQM. Chung and
Coester~\cite{chung}, Aznauryan~\cite{aznau}, and Schlumpf~\cite{schlumpf} all used RCQM to calculate nucleon form factors in the $Q^2$
range of $0-6~\mathrm{Gev}^2$. Both groups were able to reproduce the available data on
$F_{1p}$ and $F_{2p}$ between $Q^2$ from 2 to 4 $\mathrm{GeV}^2$. The calculation by Schlumpf is in good
agreement with the unpolarized data, showing a rise in the ratio
$\mu_pG_{Ep}/G_{Mp}$, but fails to reproduce the polarized data from Jefferson Lab.

More recent calculations have been made using the CQM in light front
dynamics (LFCQM)~\cite{card, card_2}. This approach uses a one-body current
operator with phenomenological form factors for the CQMs and
light-front wave functions which are eigenvectors of a mass operator
. The $SU(6)$
symmetry breaking effects with and without the constituent quark form
factor are also included. These calculations are able to describe the
trend of the high-$Q^2$ polarized data.

Previously, Frank, Jennings and Miller~\cite{frank} considered medium modifications
in real nuclei and calculated the proton form factors in CQM. Their
results for the free proton are in reasonable agreement with the polarization data
and predict a change in sign of $G_{Ep}$ at slightly higher
$Q^2$.

A relativistic quark model (RQM) calculated by Li~\cite{li} requires symmetry
in the center-of-mass frame. By adding additional terms to the
baryon wave function, which are generated by the $SU(6)$ symmetry
requirements, it represents the inclusion of the sea quarks. The result
of this calculation originally preceded the publication of the
polarized data from Jefferson Lab, and the model has good agreement
with the data.

\begin{figure}
  \begin{center}
    \includegraphics[angle=0, width=0.7\textwidth]{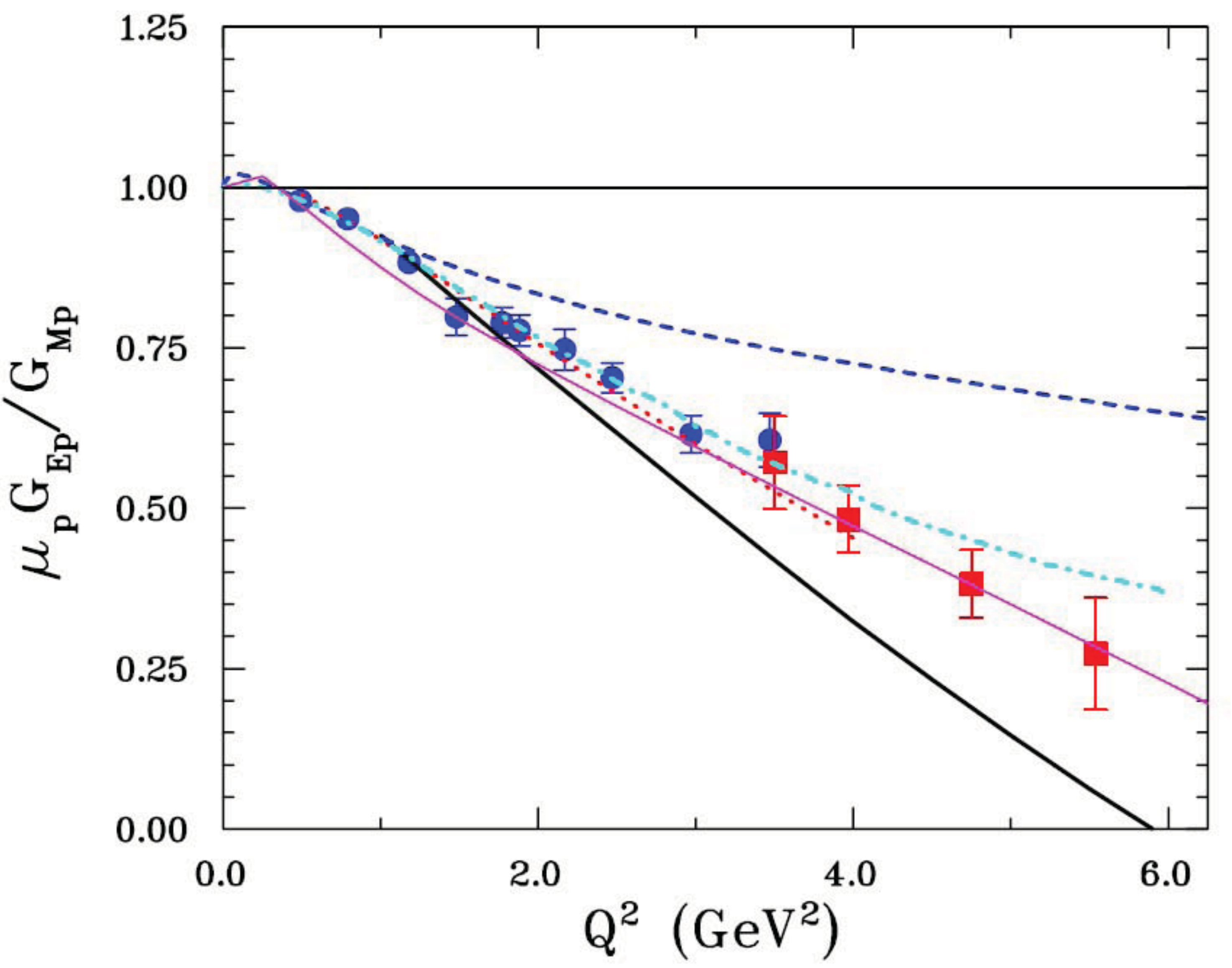}
    \caption{Comparison of various relativistic CQM calculations with the data for $\mu_pG_{Ep}/G_{Mp}$. Dotted curve: front form calculation of Chung and Coester~\cite{chung} with point-like constituent quarks; thick solid curve: front form calculation of Frank {\it et al.}~\cite{frank}; dashed curve: point form calculation of Boffi {\it et al.}~\cite{bof01} in the Goldstone boson exchange model with point-like constituent quarks; thin solid curve: covariant spectator model of Gross and Agbakpe~\cite{gro06}. Figure from~\cite{charles_rev}.}
    \label{fig:cqm}
  \end{center}
\end{figure}

A variant of the CQM model is the diquark model of Kroll {\it et
  al.}. Two of the constituent quarks are tightly-bound into a spin-0
or 1 diquark with a phenomenological form factor which allows the
diquark to behave as free quarks at high $Q^2$. When an electron scatters
from the spin-1 diquark, helicity-flip amplitudes are generated. Ma,
Qing, and Schmidet~\cite{ma,ma_2} performed calculations of a quark
spectato-diquark model using the light-cone formalism. They also
describe the available data well. Recently,
Wagenbrunn Boffi {\it et al.}~\cite{wagen} calculated the
neutron and proton electromagnetic form factors for the first time
using the Goldstone-boson-exchange constituent quark model. The
calculations are performed in a covariant frame work using the
point-form approach to relativistic quantum mechanics, and is in good
agreement with the form factors from polarized data. The comparison between various CQM models and the data are shown in Fig.~\ref{fig:cqm}.

\begin{figure}
  \begin{center}
    \includegraphics[angle=0, width=0.7\textwidth]{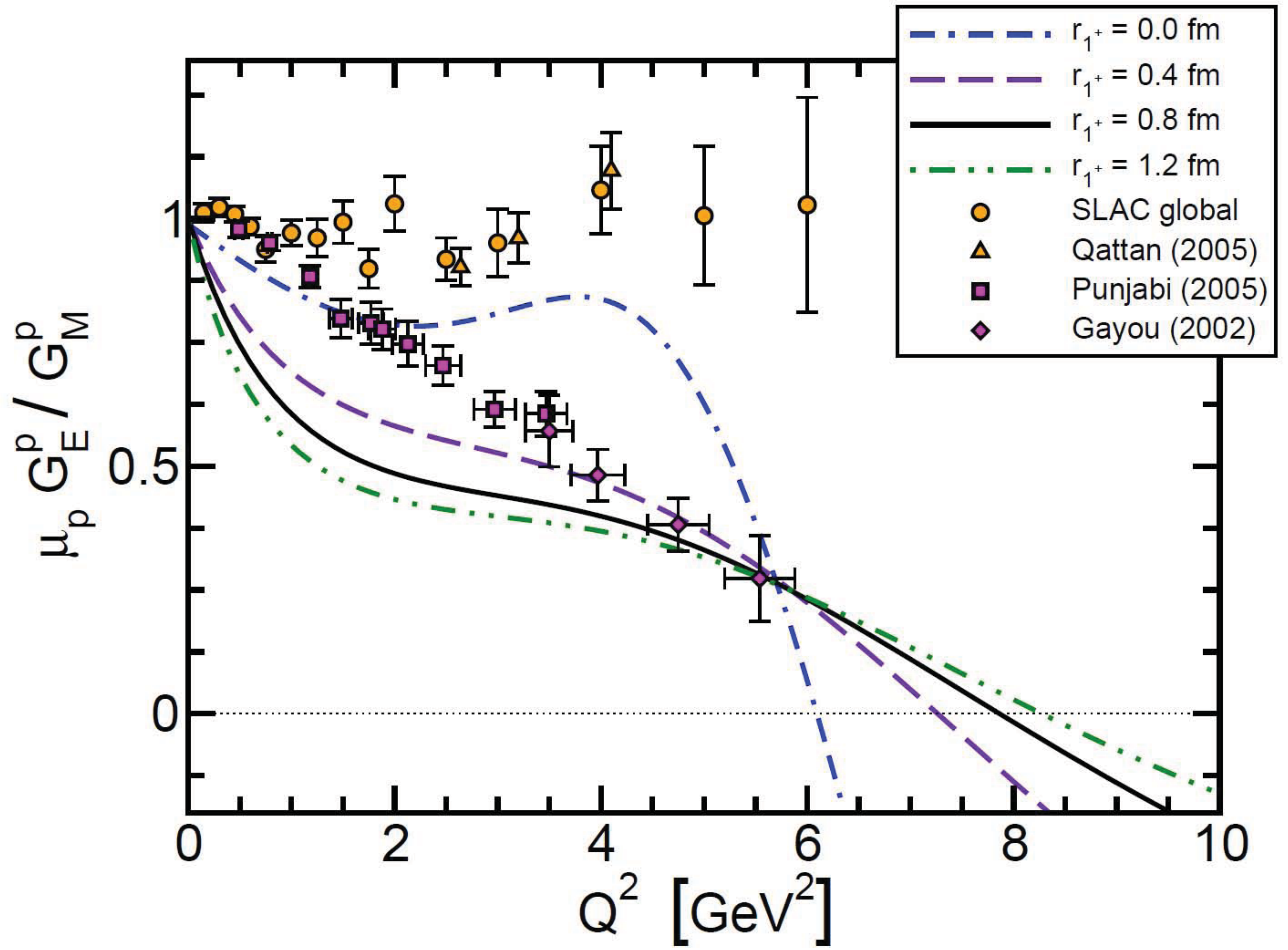}
    \caption{Result for the proton form factor ratio $\mu_p G_{E}^p/G_M^p$ computed with four different diquark radii, $r_{1+}$. Figure from~\cite{diquark}.}
    \label{fig:diquark}
  \end{center}
\end{figure}

Recently, Cl$\ddot \mathrm{e}$t {\it et al.}~\cite{diquark} calculated the form factors contributed by a dressed-quark core. It is defined by the solution of a Pioncar$\acute{\mathrm{a}}$ covariant Faddeev equation, in which dressed-quarks provide the elementary degree of freedom and the correlations between them are expressed via diquarks. The nucleon-photon vertex only has the diquark charge radius as the free parameter. The calculation of the proton Sach's form factor ratio through this model is compared with the experimental data as shown in Fig.~\ref{fig:diquark}.

Irrespective of the diquark radius, however, the proton's electric form factor possesses a zero and the magnetic form factor is positive definite. For $Q^2 <3~\mathrm{GeV^2}$, the result of the calculation lies below experiment, which can likely be attributed to the omission of pseudoscalar-meson-cloud contributions.

\subsubsection{Pion Cloud Models}
As the lightest hadrons, pions dominate the long-distance behavior of hadron wave functions and yield characteristic signatures in the low momentum transfer behavior of hadronic form factors. Therefore, a natural way to qualitatively improve the CQMs is to include the pionic degrees of freedom~\cite{man84}.

In the early MIT Bag Model, the nucleon is described as three quark fields
confined in a potential that maintains them within a finite sphere of
radius $R$. The introduction of the pion cloud~\cite{lu,lu_2} improves the static properties of the nucleon by restoring chiral symmetry and also provides a convenient connection to $\pi N$ and $NN$
scattering. To extend the calculation to larger $Q^2$, Miller performed a light-front cloudy bag model calculation~\cite{Miller}, which give a relatively good global account of the data both at low and larger $Q^2$.

\subsubsection{Chiral Perturbation Theory}
At low momentum region, the nucleon form factors can also be studied within chiral perturbation theory ($\chi$PT) expansions based on chiral Lagrangians with pion and nucleon fields. In $\chi$PT, the short-distance physics is parameterized in terms of low-energy-constants (LECs), which ideally can be determined by matching to QCD; but in practice, they are fit to experimental data or estimated using resonance saturation. In the calculation of the nucleon form factors, the LECs can be obtained from the nucleon static properties, such as the charge radii and the anomalous magnetic moments. Once these LECs are determined, the $Q^2$-dependence of the form factors can be predicted.

\begin{figure}
  \begin{center}
    \includegraphics[angle=0, width=0.85\textwidth]{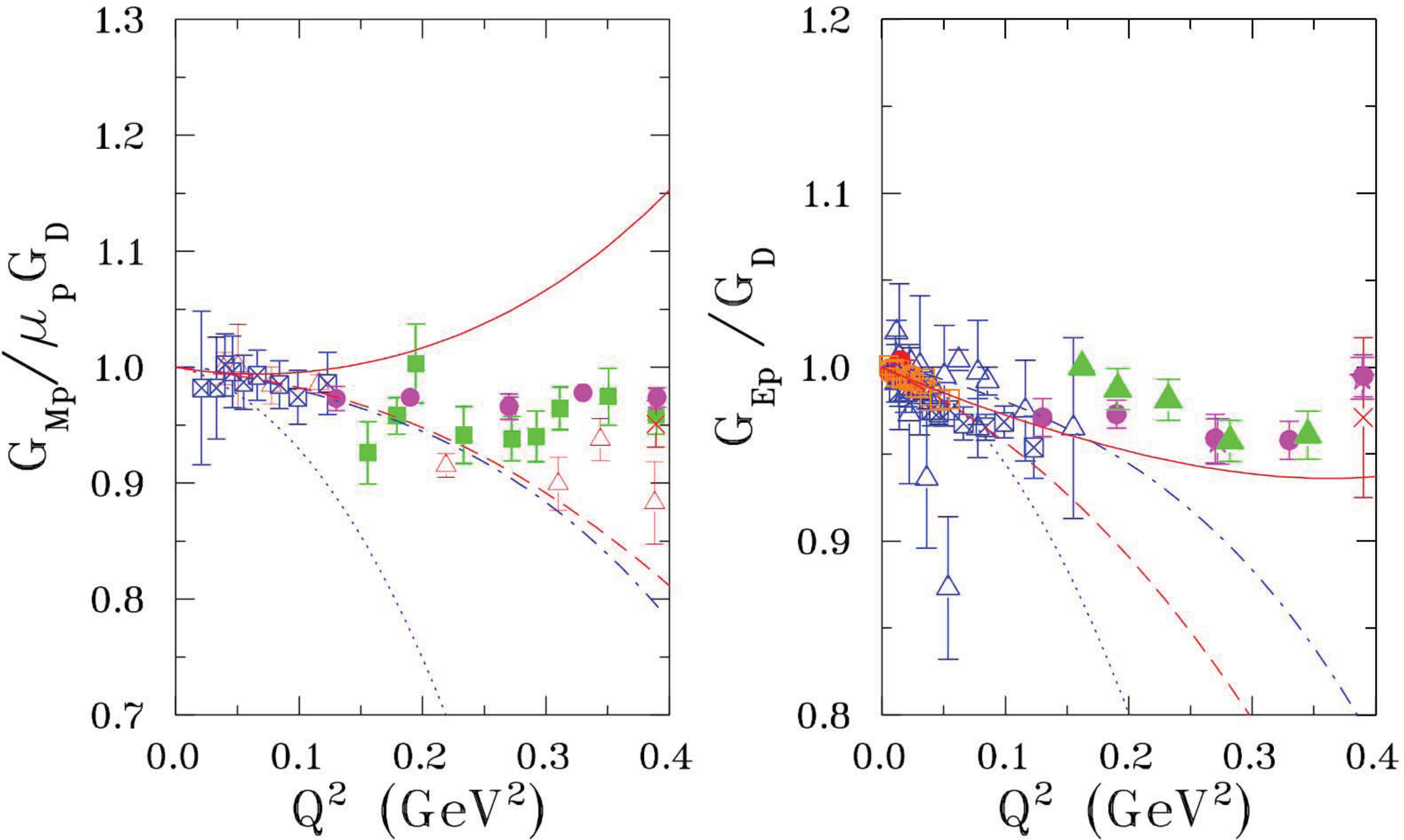}
    \caption{The proton form factors in the relativistic baryon $\chi$PT of~\cite{kub01} (IR scheme) and~\cite{schi05} (EOMS scheme). The results of~\cite{kub01} including vector mesons are shown to third (dashed curves) and fourth (solid curves) orders. The results of~\cite{schi05} to fourth order are displayed both without vector mesons (dotted curves) and when including vector mesons (dashed-dotted curves). Figure from~\cite{charles_rev}.}
    \label{fig:chiral}
  \end{center}
\end{figure}

The calculation of the nucleon EM form factors involves a simultaneous expansion in soft scales: $Q^2$ and $m_{\pi}$, which are understood to be small relative to the chiral symmetry breaking scale $\Lambda_{\chi SB}\sim 1$ GeV. Several expansion schemes have been developed in the literature. Early calculations of the nucleon from factors in the small scale expansion (SSE)~\cite{hem97} have been performed in~\cite{ber98}. Since such an approach is based on a heavy baryon expansion it is limited to $Q^2$ values much below 0.2 $\mathrm{GeV}^2$. Subsequently, several calculations of the nucleon form factors have been performed in manifestly Lorentz invariant $\chi$PT. Kubis and Meissner~\cite{kub01} performed a calculation in relativistic baryon $\chi$PT, employing the infrared regularization (IR) scheme. Schindler~\cite{schi05} also performed a calculation employing the extended on-mass-shell (EOMS) renormalization scheme. Both groups found that when only pion and nucleon degrees of freedom are included, one cannot well describe the data over a significant range of $Q^2$. On the other hand, it was found that the vector meson pole diagrams play an important role, which also confirms the findings of VMD models and dispersion theory mentioned earlier. The corresponding results in both IR and EOMS schemes are shown in Fig.~\ref{fig:chiral}.

\subsubsection{\label{sec:den}Nucleon Charge and Magnetization Densities}
Although models of nucleon structure can calculate the form factor directly, it is desirable to relate form factors to spatial densities because our intuition tends to be grounded more firmly in space than momentum transfer. The interpretation of the form factors $G_E$ and $G_M$ has the simplest interpretation in the nucleon Breit frame where the energy transfer vanishes, and the charge and magnetization densities can be written as:
\begin{eqnarray}
\rho_{ch}^{NR}(r) = \frac{2}{\pi}\int_0^{\infty} dQ Q^2j_0(Qr)G_E{Q^2}\\
\mu\rho_{m}^{NR}(r) = \frac{2}{\pi}\int_0^{\infty} dQ Q^2j_0(Qr)G_M{Q^2}.
\end{eqnarray}
However, this naive inversion is only valid when it ignores the variation of the Breit frame with $Q^2$, also known as the non-relativistic (NR) limit. For the nucleon, when the form factors are measured for $Q^2$ values much larger than $M^2$, one needs to take the effect of relativity into account. Kelly~\cite{kelly} provided a relativistic prescription to relate the Sachs form factors to the nucleon charge and magnetization densities, which accounts for the Lorentz contraction of the densities in the Breit frame relative to the rest frame.

If we start from the spherical charge $\rho_{ch}(r)$ and magnetization densities $\rho_m(r)$ in the nucleon rest frame which are normalized according to the static properties:
\begin{eqnarray}
\int_0^{\infty} dr r^2\rho_{ch}(r) &=& Z\\
\int_0^{\infty} dr r^2\rho_m(r) &=& 1,
\end{eqnarray}
the Fourier-Bessel transforms of the intrinsic densities are defined as:
\begin{equation}
\tilde\rho(k)=\int_0^{\infty}dr r^2j_0(kr)\rho(r),
\end{equation}
where k is the spatial frequency (or wave number), and $\tilde\rho(k)$ is described as the intrinsic form factor. If one can find the connection between the Sachs form factor and the intrinsic form factors, the intrinsic density is obtained simply by inverting the Fourier transform:
\begin{equation}
\rho(r) = \frac{2}{\pi}\int_0^{\infty}dk k^2j_0(kr)\tilde\rho(k).
\end{equation}
In the non-relativistic limit, $k\to Q$ and $\tilde\rho(Q)\to G(Q^2)$, where $G(Q^2)$ is the appropriate Sachs form factor. However, this naive inversion causes unphysical cusps at the origin for the common dipole and Galster parameterizations. Licht and Pagnamenta~\cite{licht} attributed these failures to the replacement of the intrinsic spatial frequency $k$ with the momentum transfer $Q$ and demonstrated that the density softens, when a Lorentz boost from the the Breit frame with momentum $q_B = Q$ to the rest frame is applied. Consequently, the spacial frequency is replaced by:
\begin{equation}
k^2 = \frac{Q^2}{1+\tau},
\label{eq:k}
\end{equation}
where $\tau = Q^2/4M^2_N$, and a measurement with Breit-frame momentum transfer $q_B = Q$ probes a reduced spatial frequency $k$ in the rest frame.

\begin{figure}
  \begin{center}
    \includegraphics[angle=0, width=0.50\textwidth]{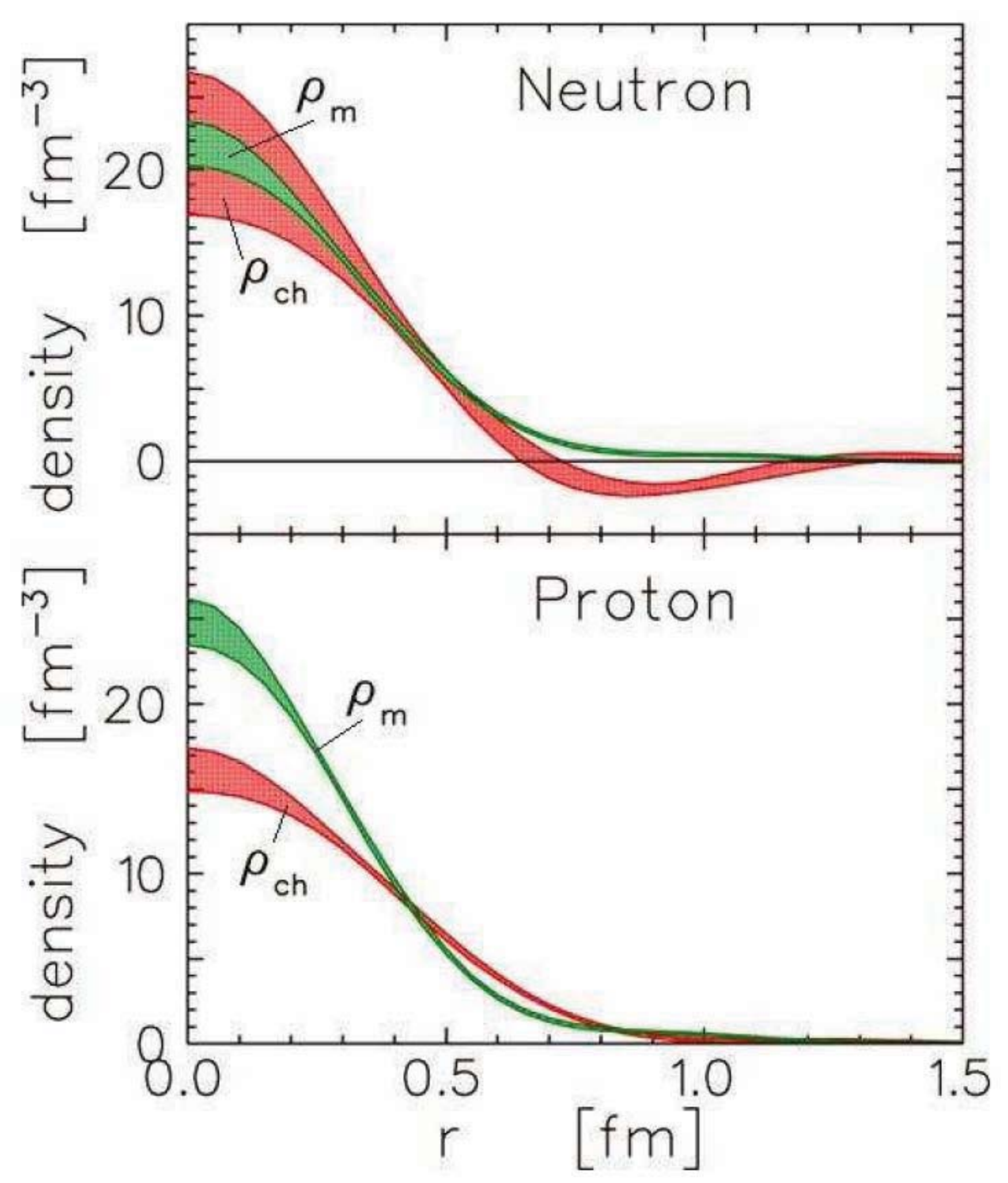}
    \caption{Comparison between charge and magnetization densities for the proton and neutron. Figure from~\cite{charles_rev}}
    \label{fig:kelly}
  \end{center}
\end{figure}

Unfortunately, unique relativistic relationships between the Sachs form factors measured by finite $Q^2$ and the static charge and magnetization densities in the rest frame do not exist. The fundamental problem is that electron scattering measures transition matrix elements between states of a composite system that have different momenta, and the transition densities between such states are different from the static densities in the rest frame. Several models have employed similar relativistic prescriptions, which can be written in the following form:
\begin{eqnarray}
\tilde\rho_{ch}(k) & = & G_E(Q^2)(1+\tau)^{\lambda_E}\\
\mu\tilde\rho_m(k) & = & G_M(Q^2)(1+\tau)^{\lambda_M}
\end{eqnarray}
where $k$ and $Q^2$ are related as in Eq.~\ref{eq:k} and $\lambda$ is a model-dependent constant. One can see that the accessible spatial frequency is limited to $k\le2M_N$ determined by the nucleon Compton wavelength. To account for an asymptotic $1/Q^4$ form factor behavior, Kelly followed the choice $\lambda_E = \lambda_M = 2$, and he employed linear expansions in complete sets of basis functions to minimize the model dependence. Fig.~\ref{fig:kelly} shows the charge and magnetization densities for neutron and proton from his analysis as determined from fits of the world data.

The low $Q^2$ behavior of the form factors also play an important role in defining the transition radii obtained from integral moments of the underlying density. The integral moments are defined by:
\begin{equation}
M_{\alpha} = \int_0^{\infty} dr r^{2+\alpha}\rho(r).
\end{equation}
While the lowest nonvanishing moment is free of discrete ambiguities, the higher moments depend upon $\lambda$. For example, the proton radius retains a small dependence upon $\lambda$,
\begin{equation}
<r^2>_{\lambda,p} = -6\frac{dG(0)}{dQ^2}|_{Q^2\to0}-\frac{3\lambda}{2m^2_p},
\end{equation}

Recently, Miller~\cite{miller_ff} proposed a model independent analysis in the infinite-momentum-frame (IMF). In this frame, the charge density $\rho(\mathrm {\bf b})$ in the transverse plane is in fact a two-dimensional Fourier transform of the $F_1$ form factor:
 \begin{equation}
 \rho(\mathrm{\bf b}) \equiv \sum_q e_q\int dx q(x,\mathrm {\bf b}) = \int \frac{d^2q}{(2\pi)^2}F_1(Q^2=q^2)e^{iq\cdot \mathrm {\bf b}}.
 \end{equation}

 In contrast with earlier expectations, from this analysis, the neutron charge density is negative at the center, and the proton's central $d$ quark charge density is larger than that of the $u$ quark by about $30\%$.

\subsubsection{Global Fits}
As the most basic quantities, nucleon electromagnetic form factors are
needed for various calculations in nuclear physics. Hence, a simple parametrization which accurately represents the
data over a wide range of $Q^2$ and has reasonable behavior for both
$Q^2\to 0$ and $Q^2\to\infty$ would be convenient.

\begin{figure}
  \begin{center}
    \includegraphics[angle=0, width=0.9\textwidth]{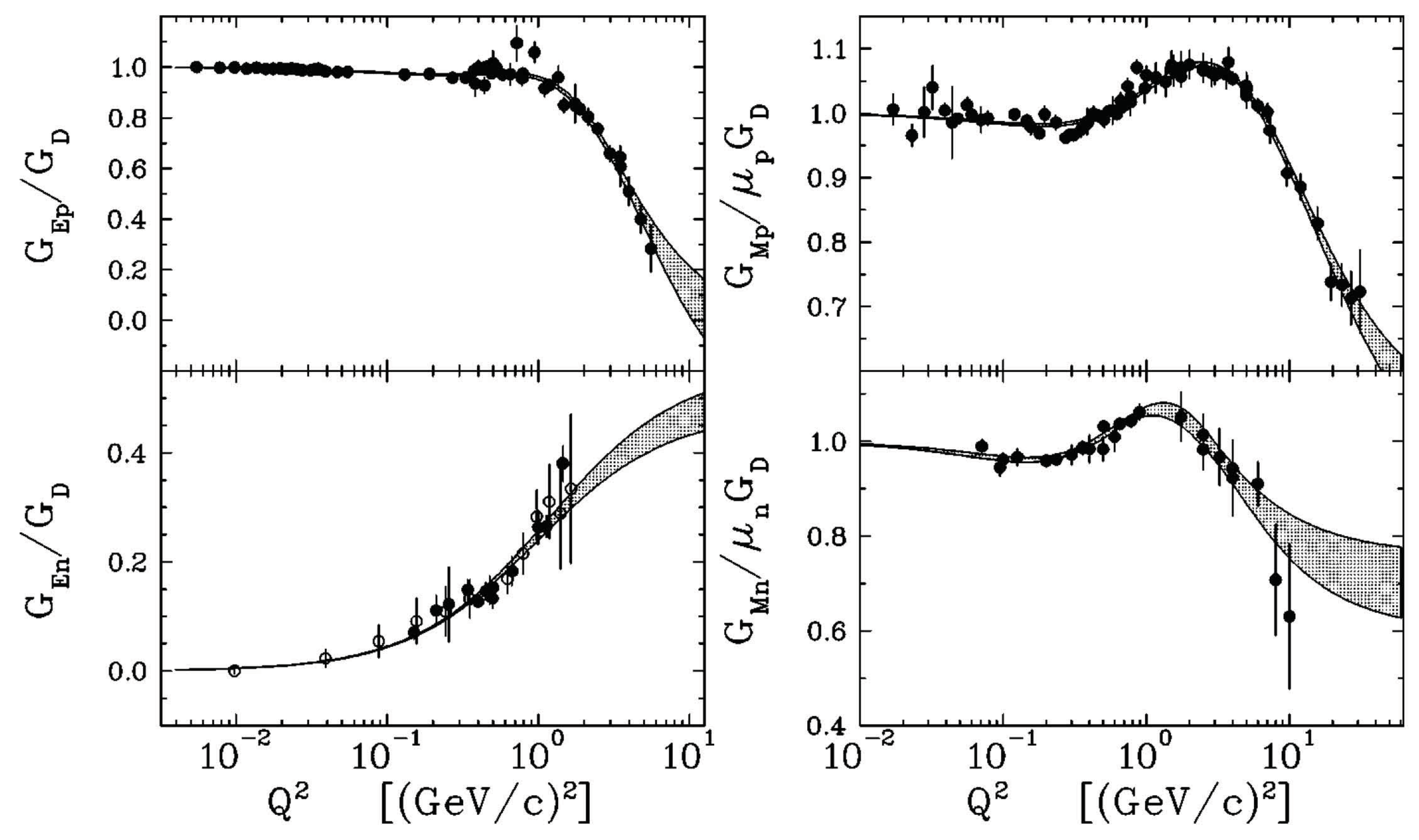}
    \caption{Kelly's fits~\cite{kelly} to nucleon electromagnetic form factors. The error bands were of the fits. Figure from~\cite{kelly}.}
    \label{fig:fit_kelly}
  \end{center}
\end{figure}

For reasonable behavior at low $Q^2$, the power-series
representation should involve only even powers of $Q$. At high $Q^2$,
dimensional scaling rules require $G\propto Q^{-4}$. However, at present the most
common parameterizations violate one or both of these conditions. Often
the reciprocal of a polynomial in $Q$~\cite{bosted,brash,arrington_2} is used, but this method has
difficulty in determining the RMS radius due to the unphysical odd powers of
$Q$. Recently, Kelly~\cite{kelly} proposed a much simpler parametrization which
takes the form:
\begin{equation}
G(Q^2)\propto\frac{\sum_{k=0}^na_k\tau^k}{1+\sum_{k=1}^{n+2}b_k\tau^k},
\end{equation}
where both numerator and denominator are polynomials in
$\tau=Q^2/4m_p^2$ and the degree of the denominator is larger
than that of the numerator to ensure the $\propto Q^{-4}$ for large
$Q^2$. Good fits by this form require only four parameters each for
$G_{Ep},G_{Mp}$ and $G_{Mn}$, and only two for $G_{En}$. Fig.~\ref{fig:fit_kelly}
shows the results of the parametrization.

\begin{figure}
  \begin{center}
    \includegraphics[angle=0, width=0.9\textwidth]{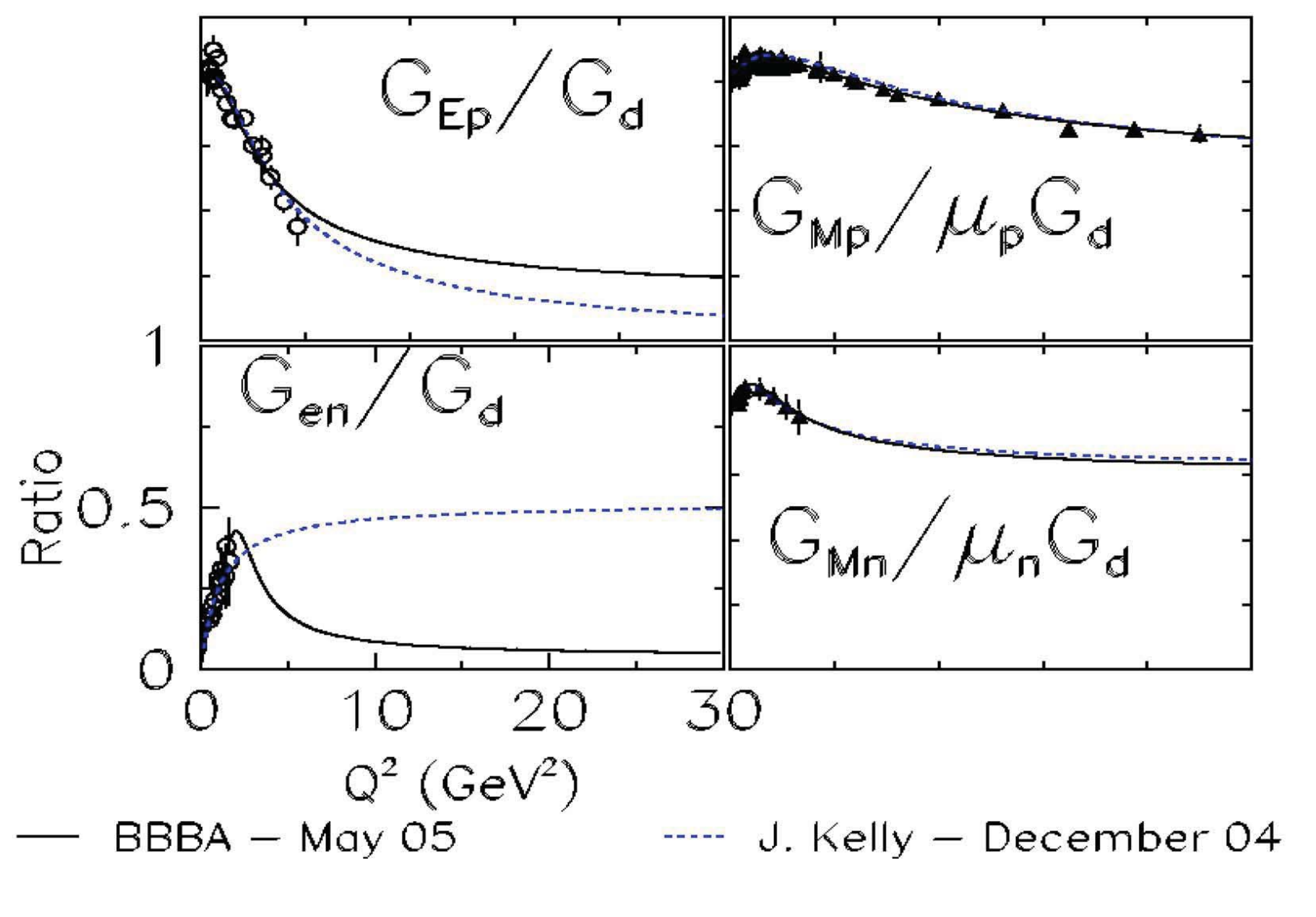}
    \caption{The parametrization of Bradford {\it et al.} compared with Kelly's, together with
      world data. Figure is from~\cite{bradford}.}
    \label{fig:fit_brad}
  \end{center}
\end{figure}

\begin{figure}
  \begin{center}
    \includegraphics[angle=0, width=0.7\textwidth]{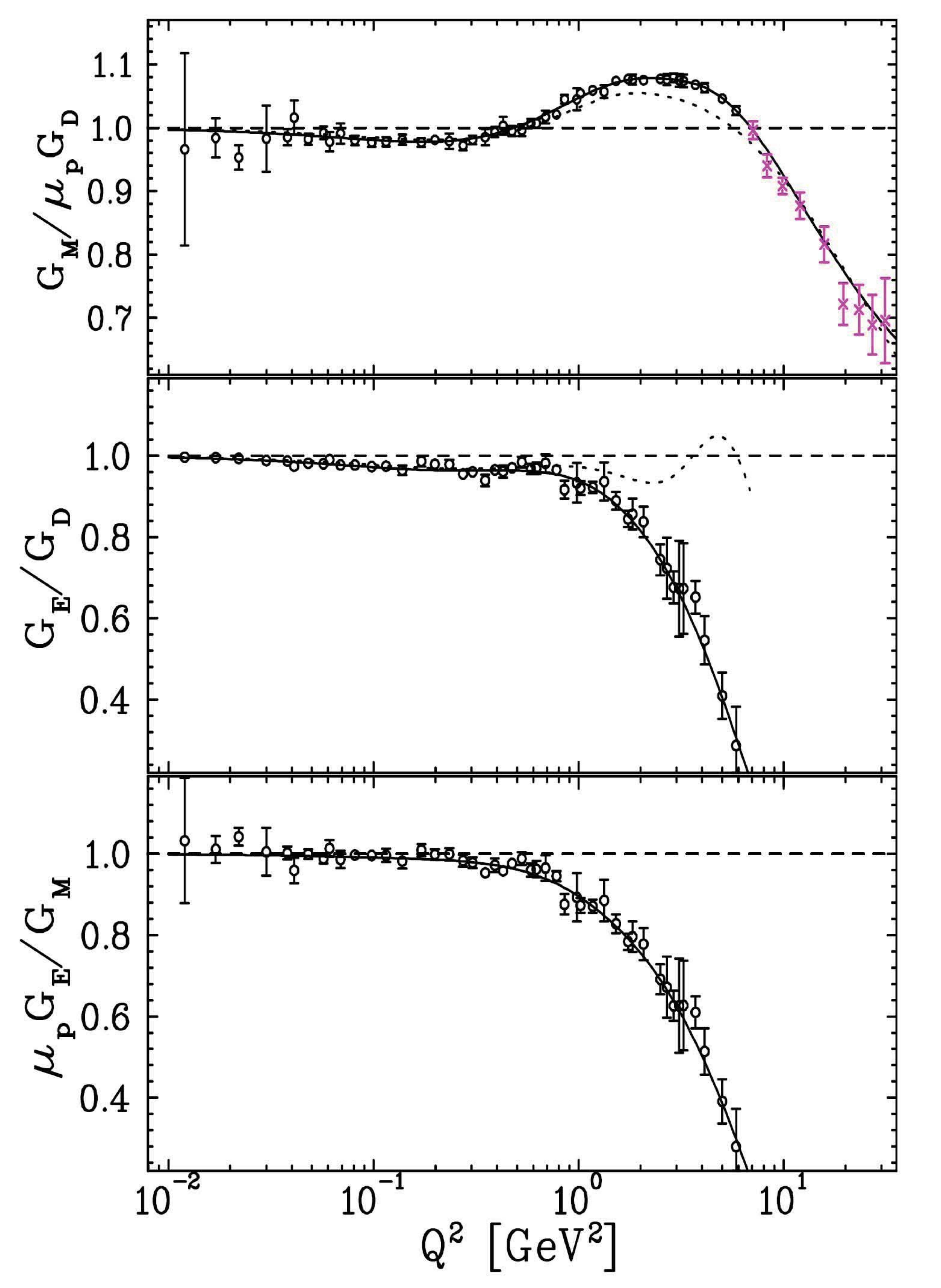}
    \caption{Extracted values of $G_E$ and $G_M$ from the global
      analysis. The open circles are the results of the combined
      analysis of the cross section data and polarization
      measurements. The solid lines are the fits to TPE-corrected cross
      section and polarization data. The dotted curves show the results of taking $G_E$ and $G_M$ from a fit to the TPE-uncorrected reduced cross section. Figure from~\cite{john:TPE}.}
    \label{fig:fit_john}
  \end{center}
\end{figure}

Bradford {\it et al.}~\cite{bradford} did another parametrization that uses the
same functional but with two additional constraints. The first
constraint comes from local duality, and a second constraint is based on
QCD sum rules including a further application of duality. The constraints
were implemented by scaling the high $Q^2$ data of $G_{Mp}$ and then
adding these scaled points to the data sets for $G_{En}$ and $G_{Mn}$
during the fits. Fig.~\ref{fig:fit_brad} shows the new parameterizations.
Arrington and Sick~\cite{arrington_sick} performed a fit of the world data at very low momentum
transfer by a Continued Fraction (CF) expansion:
\begin{equation}
G_{CF}(Q)=\frac{1}{1+\frac{b_1Q^2}{1+\frac{b_2Q^2}{1+\cdots}}}.
\end{equation}
This expansion is suitable for the lower momentum transfers, and extends up to
$Q=\sqrt{Q^2}\approx 0.8~\mathrm{GeV}/c$. The analysis included the
effect of Coulomb distortion and the Two-Photon-Exchange (TPE) exchange
beyond Coulomb distortion, which includes only the exchange of an
additional soft photon.
Later on, Arrington {\it et al.}~\cite{john:TPE} performed a global analysis of
the world data. The analysis combined the corrected Rosenbluth cross
section and polarized data, and this is the first extraction of $G_{E}$ and
$G_M$ including the explicit TPE correction. Fig.~\ref{fig:fit_john} shows this
global analysis compared with the world data.

\begin{figure}
  \begin{center}
    \includegraphics[angle=0, width=0.9\textwidth]{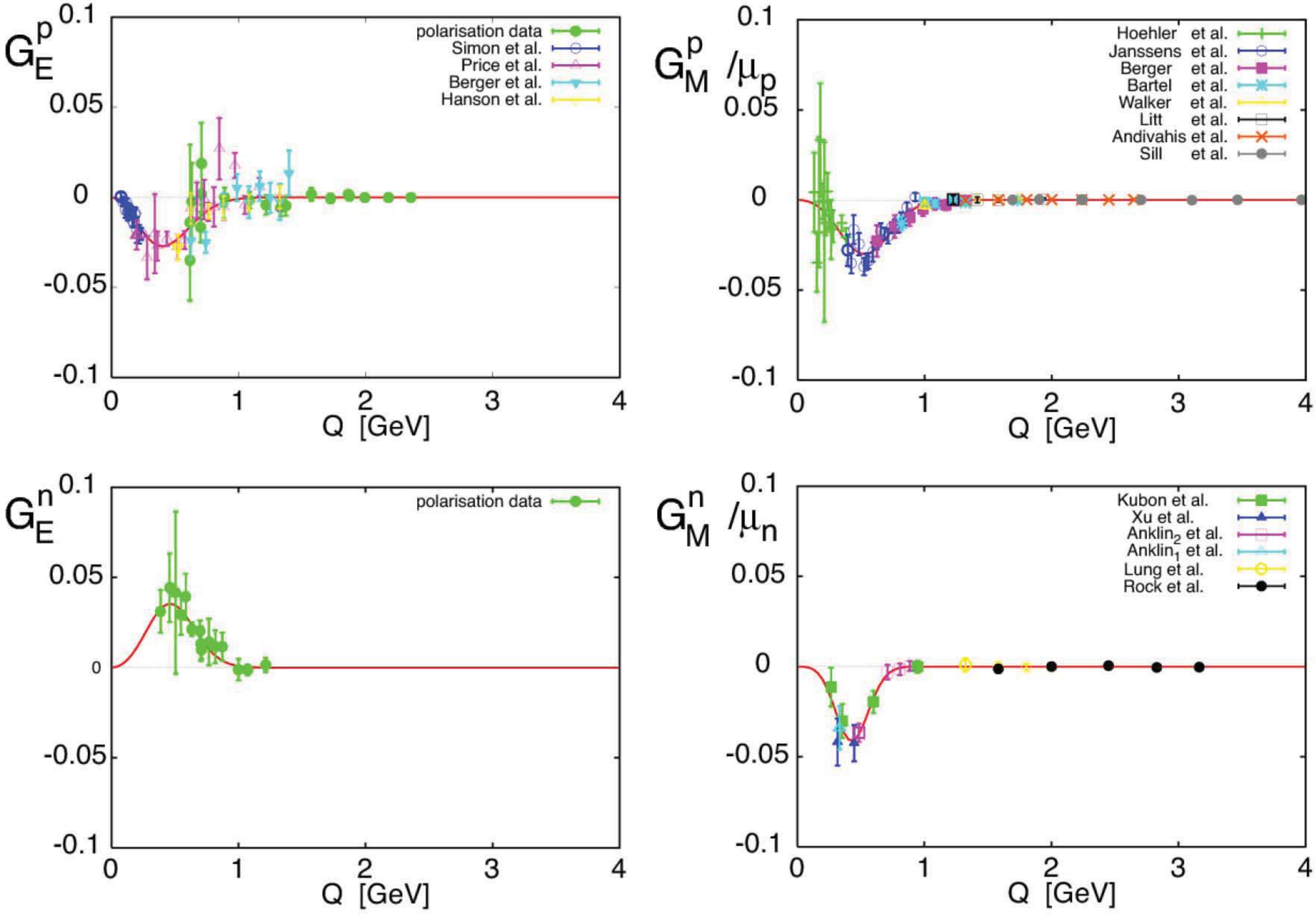}
    \caption{The difference between the measure nucleon form factors and the 2-components phenomenological fit of~\cite{fried} for all four form factors.}
    \label{fig:fried_ffs}
  \end{center}
\end{figure}

In 2003, Friedrich and Walcher performed various phenomenological fits~\cite{fried} at low $Q^2$ with the ``bump and dip'' structure on the base of the smooth 2-dipole form. Shown in Fig.~\ref{fig:fried_ffs} are the difference between the measured nucleon form factors at that time and the smooth part of their phenomenological ansatz. It is found that all four form factors exhibit similar structure at low momentum transfer region, which they interpreted as an effect of the pion cloud around a bare nucleon. They found a very long-range contribution to the charge
distribution in the Breit frame extending out to about 2 fm which could arise from the pion cloud. With the hint of the existence of the ``bump and dip'' structure, their analysis reinvigorated the interest in investigating the form factor behavior in the low $Q^2$ region.

\section{Measurements at Low $Q^2$}
While at high $Q^2$, it is generally accepted that the proton form factor ratio $\mu_pG_{Ep}/G_{Mp}$ decreases smoothly with increasing $Q^2$. In the low $Q^2 (< 1~\mathrm{GeV}^2$) region, the world existing data appear to be less conclusive about where this deviation starts. On the other hand, from the fits performed by Friedrich and Walcher, the data somehow indicate the existence of structure.

\begin{figure}
  \begin{center}
    \includegraphics[angle=0, width=0.90\textwidth]{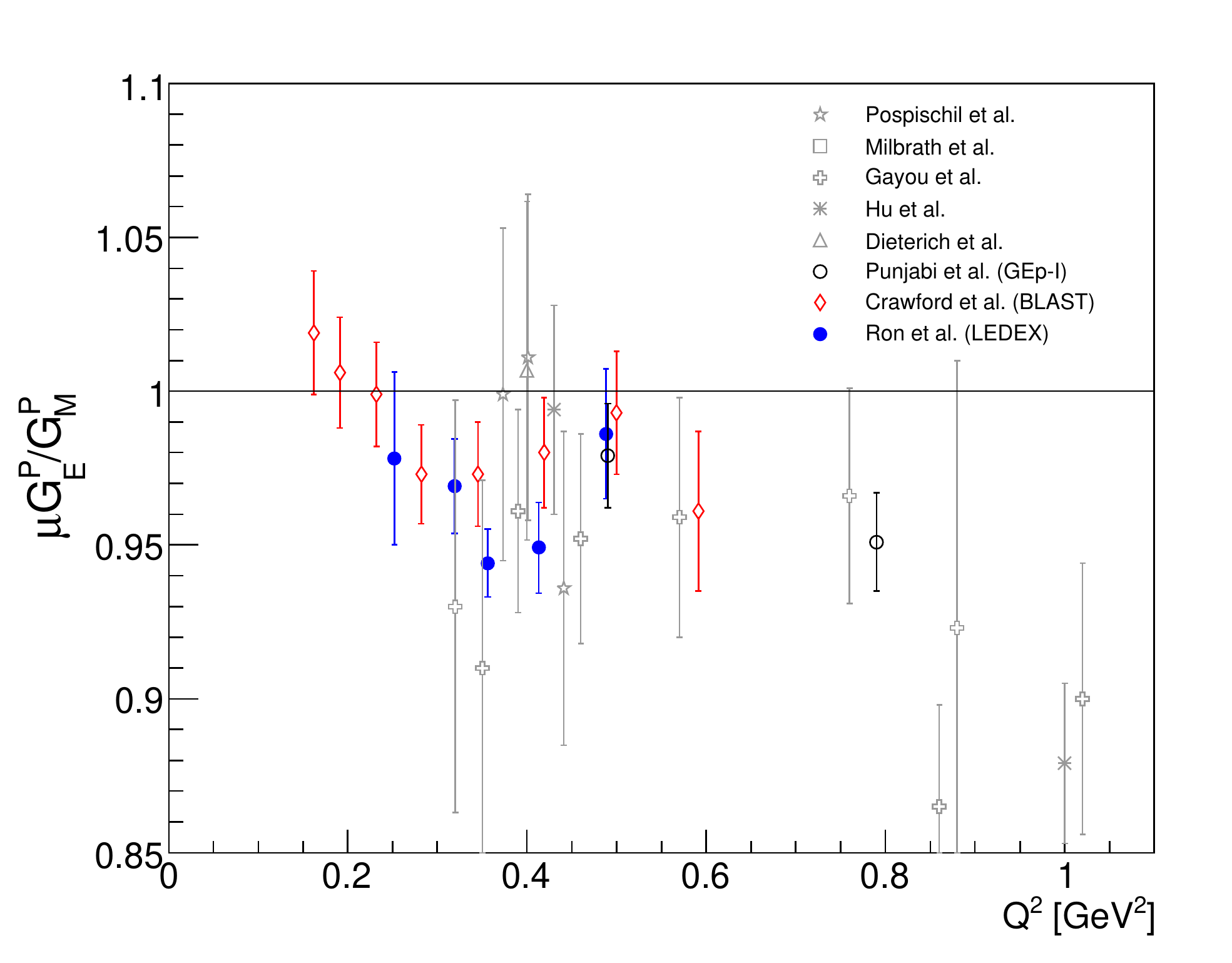}
    \caption{The world data from polarization measurements. Data plotted are from~\cite{posp, Mil, gou_01, hu, dieterich, punj, BLAST, LEDEX}}
    \label{fig:alldata}
  \end{center}
\end{figure}

\begin{figure}
  \begin{center}
    \includegraphics[angle=0, width=0.90\textwidth]{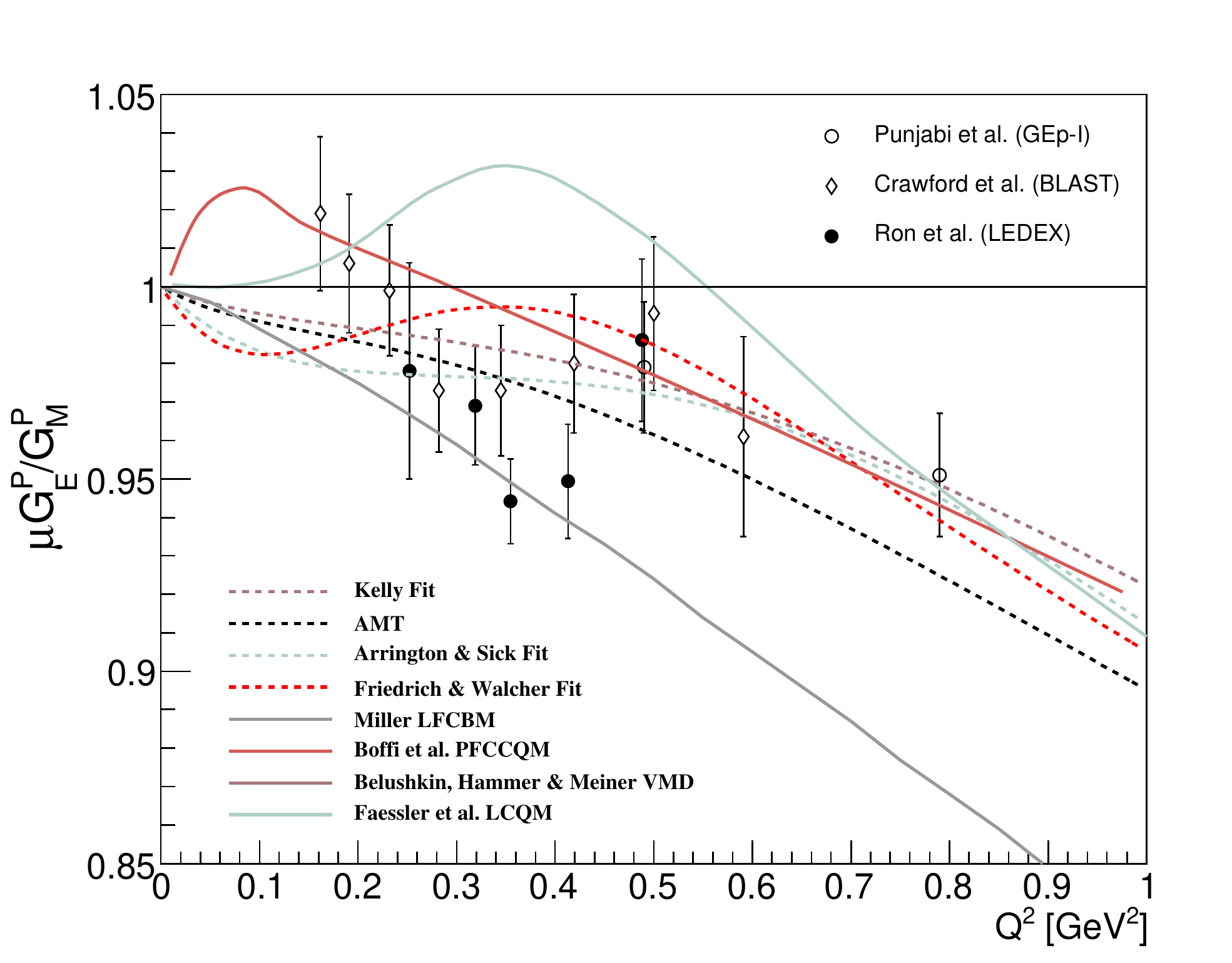}
    \caption{Recent world high precision polarization data~\cite{punj, BLAST, LEDEX} compared to several fits~\cite{kelly_par, john:TPE, arrington_sick, fried} and parameterizations~\cite{Miller, bof01, belu, faessler}.}
    \label{fig:wdata}
  \end{center}
\end{figure}

Fig.~\ref{fig:alldata} presents all world polarization data for $Q^2<1~\mathrm{GeV}^2$ and Fig.~\ref{fig:wdata} presents only the high precision ones ($\sigma_{tot} < 3\%$). The earliest recoil polarization measurement at Jefferson Lab~\cite{punj} has two points below 1 GeV$^2$.   Later on, BLAST (MIT-Bates) performed the first measurement of $\mu_pG_{Ep}/G_{Mp}$ from $^1\vec H(\vec e, e'p)$ in the $Q^2$ region between 0.15 and 0.65 $\mathrm{GeV}^2$~\cite{BLAST}. The extracted ratio from these data is consistent with unity. In 2006, Jefferson Lab~\cite{LEDEX} performed another recoil polarization measurement focusing at low $Q^2$, which overlaps the region covered by BLAST. While both results gave similar behavior over the whole range, a strong deviation from unity is observed at $Q^2\sim 0.35\mathrm{GeV}^2$ in LEDEX. However, due to limited statistics during the experiment and the background issue~\cite{guy_thesis}, such a deviation is not conclusive at that moment. Interestingly, both data sets are inconsistent with Friedrich and Walcher fit.

The experiment reported in this thesis aimed to provide a high precision survey of the proton form factor ratio ( $\sigma_{stat.}< 1\%$) in the region of $Q^2=0.3-0.7~\mathrm{GeV}^2$. With the proposed accuracy, we will be able to either confirm or refute the existence of any deviation from unity and local ``structure'' in this low momentum transfer region. In addition, the range that we cover is particularly important for tests of effective field theory predictions, future precision results from lattice QCD and also helps to quantify the pion cloud effect in nucleon structure. Besides, improved form factor measurements also have implications in the extraction of other physics quantities, such as the ultra-high precision test of QED from the hydrogen hyperfine splitting and the strange quark content of the nucleon through parity violation experiments.

%% file: Exp.tex
\chapter{Experimental Setup}
Experiment E08-007 was completed in the summer of 2008 at Thomas
Jefferson National Accelerator Facility in Hall A. The polarized electron beam was produced and
accelerated by the Continuous Electron Beam Accelerator Facility
(CEBAF). With a 1.19 GeV beam on a liquid hydrogen target, the elasticly scattered electrons were detected by the
BigBite spectrometer in coincidence with the recoil proton detected by
the left High Resolution Spectrometer (HRS). The transferred proton
polarization was measured in the focal plane polarimeter (FPP). The
proton form factor ratio were measured at 8 kinematics, which are listed in Table~\ref{tab:ksummary}. This chapter will describe the experimental setup and instrumentation used for this experiment.
\begin{table}
  \begin{center}
    \begin{tabular}{|c|c|c|c|c|}
      \hline
      Kine. & $Q^2~[\mathrm{GeV}^2]$ & $\theta_e$ [deg] & $\theta_p$
      [deg] & $\varepsilon$ \\
      \hline
      K1 & 0.35 & 31.3 & 57.5 & 0.85 \\
      K2 & 0.30 & 28.5 & 60.0 & 0.88 \\
      K3 & 0.45 & 36.7 & 53.0 & 0.80 \\
      K4 & 0.40 & 34.2 & 55.0 & 0.82 \\
      K5 & 0.55 & 41.9 & 49.0 & 0.75 \\
      K6 & 0.50 & 39.2 & 51.0 & 0.78 \\
      K7 & 0.60 & 44.6 & 47.0 & 0.72 \\
      K8 & 0.70 & 49.8 & 50.0 & 0.66 \\
      \hline
    \end{tabular}
    \caption{E08-007 kinematics.}
    \label{tab:ksummary}
  \end{center}
\end{table}
\section{The Accelerator and the Polarized Electron Source}
\begin{figure}
  \begin{center}
    \includegraphics[angle=0, width=0.8\textwidth]{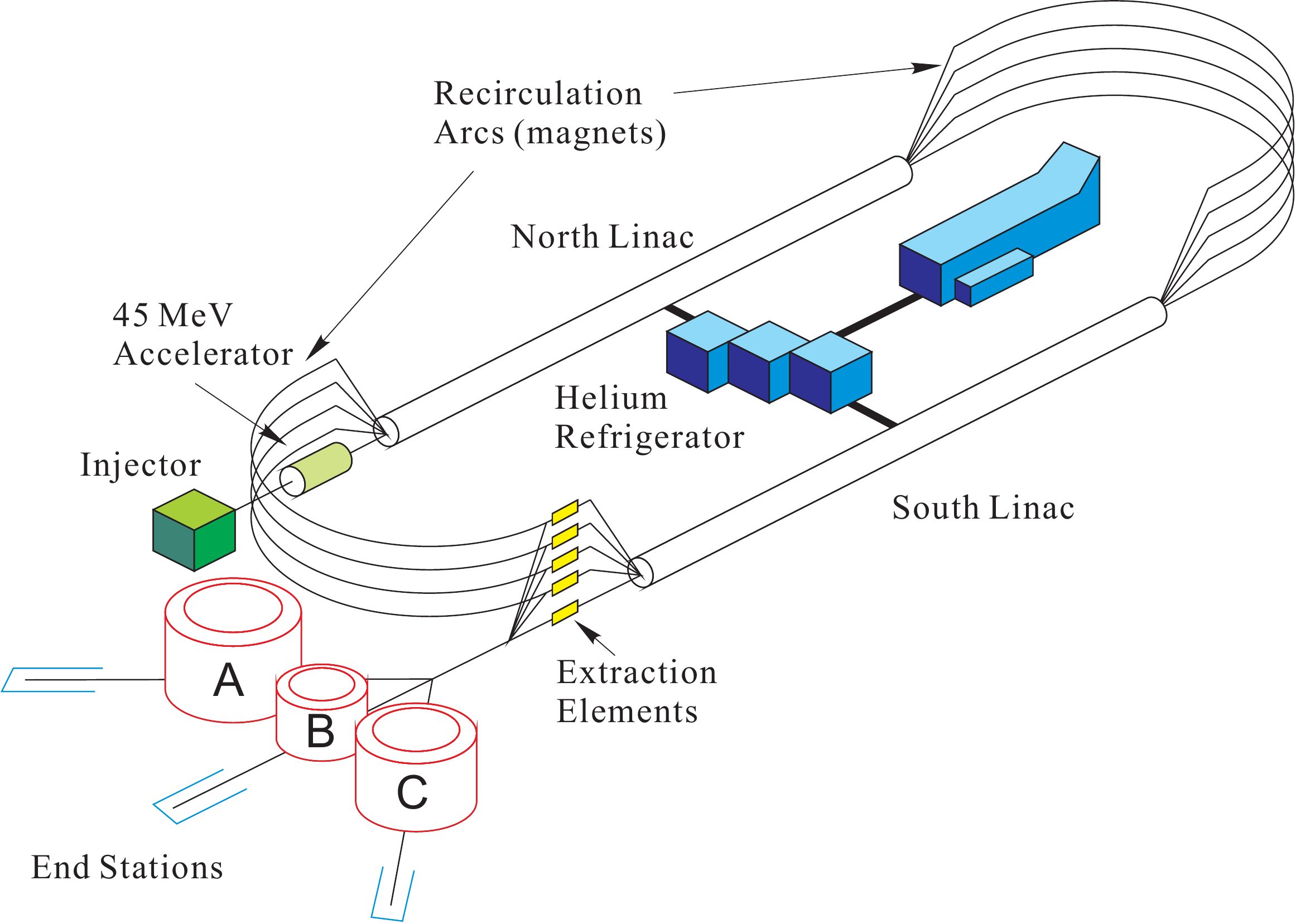}
    \caption{Layout of the CEBAF facility. The electron beam is produced at the injector and further accelerated in each of two superconduction linacs. The beam can be extracted simultaneously to each of the three experimental halls.}
    \label{fig:cebaf}
  \end{center}
\end{figure}

CEBAF (see Fig.~\ref{fig:cebaf}) accelerates electrons up to 5.7 GeV by recirculating the beam up
to five times through two superconducting linacs. Each linac contains
20 cryo-modules with a design accelerating gradient of 5 MeV/m,
producing a nominal energy gain of 400 MeV per pass, and this gain can be tuned up to about 500 MeV per pass if required by the experimental halls. Ongoing insitu processing has already resulted in an average gradient in excess of 7
MeV/m, which has made it possible to accelerate up to about 5.7 GeV.

 Electrons can be injected into the
accelerator from either a thermionic or a polarized gun. With the
polarized gun a strained GaAs cathode is illuminated by a 1497 MHz
gain-switched diode laser, operated at 780 nm. The absorption of a
right or left circularly polarized laser light preferentially produces
electrons with a spin down or up respectively in the conduction band,
thus longitudinally polarizing the beam, up to 85$\%$. The laser light
is circularly polarized using a Pockels cell. The electron beam polarization is
measured at the injector with a 5 MeV Mott polarimeter~\cite{mott} and the
polarization vector can be oriented with a Wien filter~\cite{wien_filter}. The sign of
the beam helicity is flipped pseudo-randomly at a rate of 30 Hz by
switching the circular polarization of the laser, which is achieved by
changing the voltage of the Pockels cell. The current
sent to the three Halls A, B and C can be controlled
independently. The design maximum current is 200 $\mu$A in CW (continuous wave) mode, which can
be split arbitrarily between three interleaved 499 MHz bunch
trains. One such bunch train can be peeled off after each linac pass
to any one of the Halls using RF separators and septa. CEBAF can
deliver 100 $\mu$A beam to one or both of the Hall A and Hall C, while
maintaining high polarization low current (1 nA) to Hall B. Hall C has
been operational since November 1995, Hall A since May 1997 and Hall B
since December 1997.

For this experiment (E08-007), a 1.19 GeV CW beam was delivered into Hall A,
with current $4-20\mu~\mathrm{A}$ for production data taking for various
kinematics. The average beam polarization during the experiment was $\sim83\%$.
\section{Hall A}
\begin{figure}
  \begin{center}
    \includegraphics[angle=0, width=0.85\textwidth]{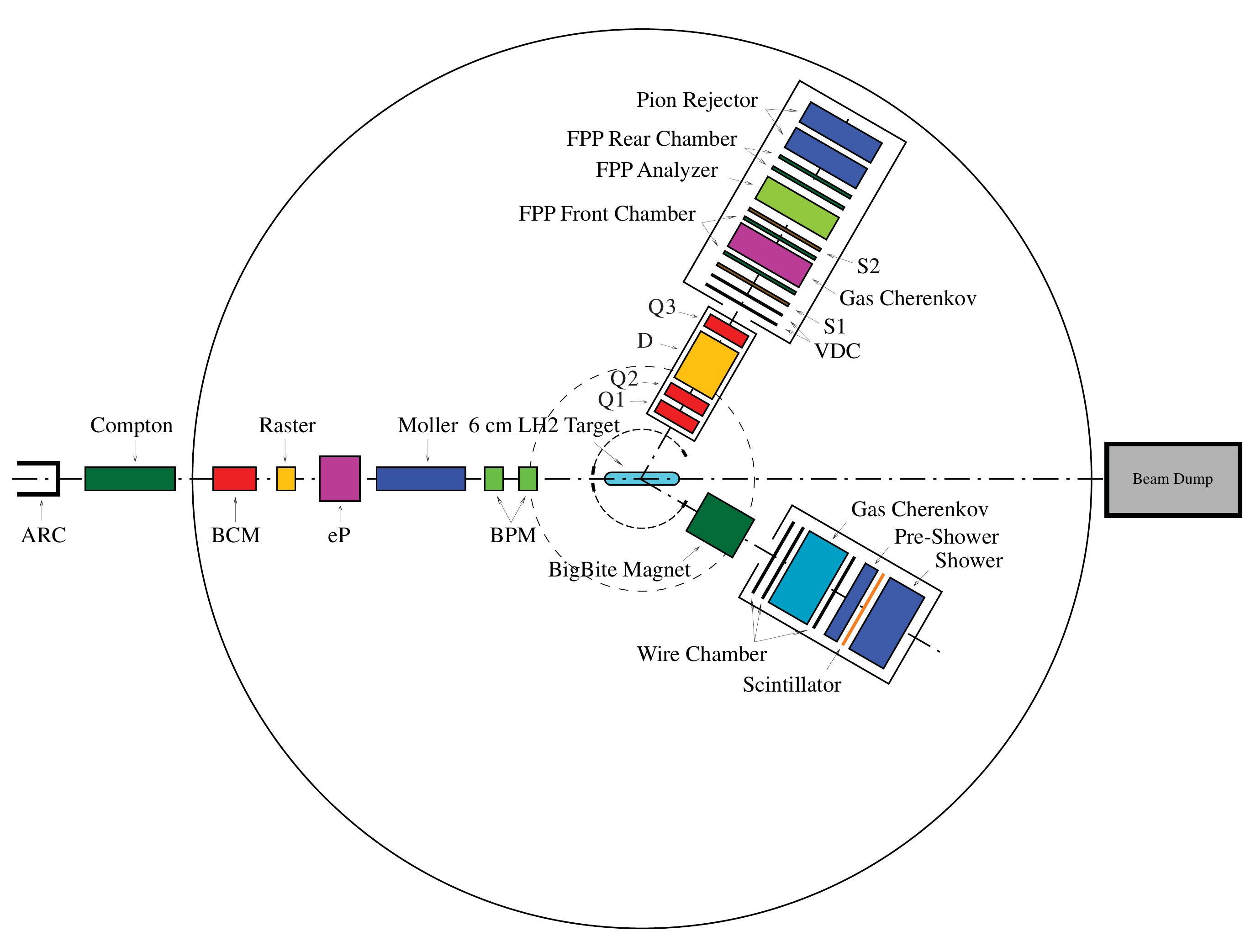}
    \caption{Hall A floor plan during E08-007.}
    \label{fig:hallover}
  \end{center}
\end{figure}

All three experimental halls have their bulk volumes underground with a
shield of concrete and a thick layer of earth. Hall A is the largest
one with a diameter of 53 m. The layout of Hall A during E08-007 is shown in Fig.~\ref{fig:hallover}. The key
elements include the beamline, cryogenic target in the scattering
chamber, the left High Resolution Spectrometers (LHRS) and the BigBite
spectrometer.
\section{Beam Line}
\subsection{Beam Energy Measurement}
The beam energy during the experiment was monitored by ``Tiefenbach''
energy~\cite{hallanim}. The value is calculated by the current values of Hall A arc
B$\cdot dl$ and Hall A arc beam position monitors (BPM). This number is
continuously recorded in the data stream and is calibrated against the
Arc energy of the 9th dipole regularly. The accuracy from this
measurement is about 0.5 MeV. For this experiment, the results are not
sensitive to the absolute beam energy; therefore, there were no
invasive measurements performed during the experiment.
\subsection{Beam Current Monitor}
The beam current is measured by the beam current monitors (BCMs)~\cite{hallanim} in
Hall A, which provides a stable, low-noise, no-invasive
measurement. It consists of an Unser monitor, two RF cavities,
associated electronics and a data-acquisition system. The cavities and
the Unser monitor are enclosed in a temperature-stabilized magnetic
shielding box which is located 25 m upstream of the target.

\begin{figure}
  \begin{center}
    \includegraphics[angle=0, width=0.7\textwidth]{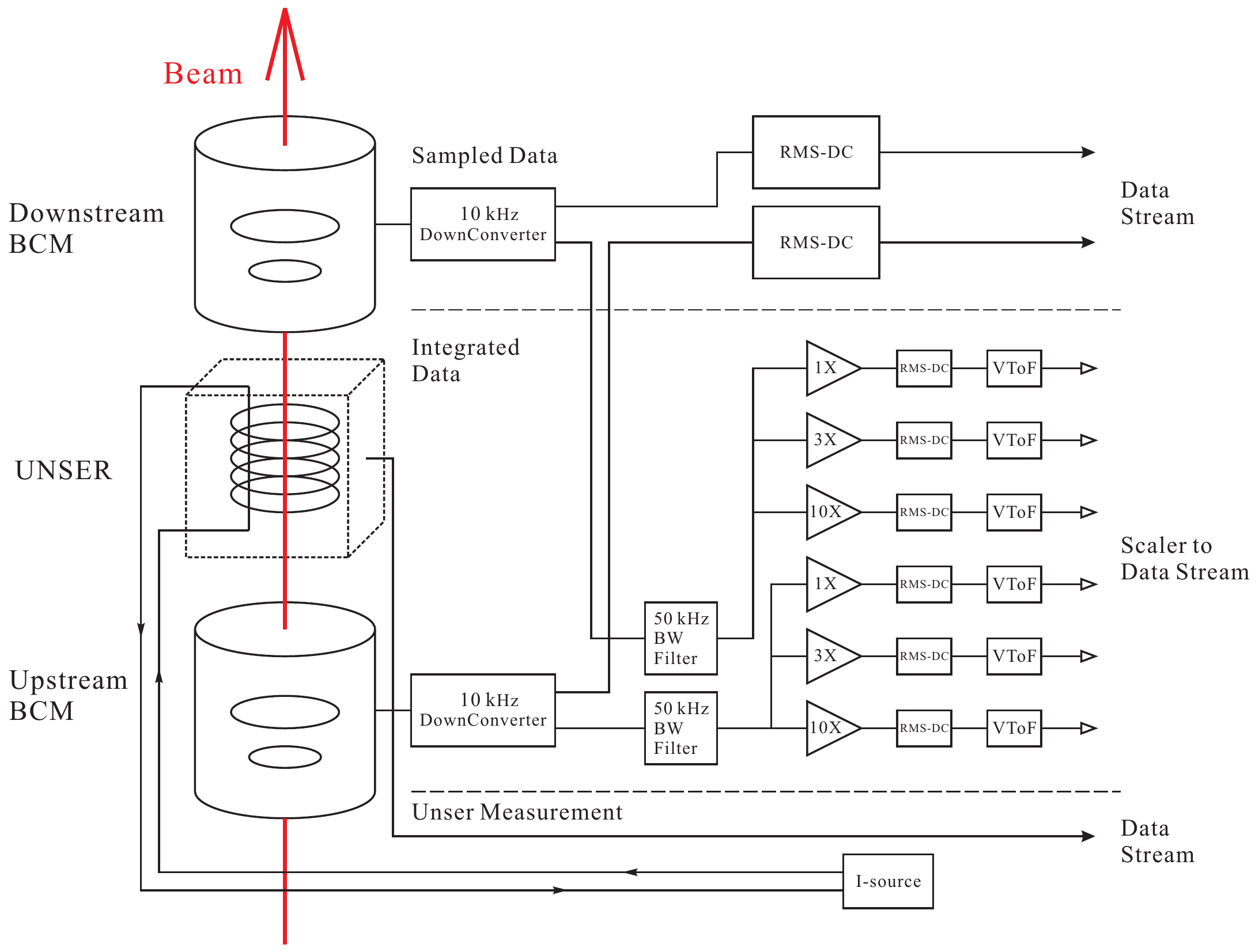}
    \caption{Schematic of beam current monitors.}
    \label{fig:bcm}
  \end{center}
\end{figure}

Fig.~\ref{fig:bcm} shows the schematics of BCMs. The Unser monitor is a Parametric
Current Transformer which provides an absolute measurement~\cite{bcm_1}. The
monitor is calibrated by passing a known current through a wire inside
the beam pipe and has a nominal output of 4 mV/$\mu$A. As the Unser
monitor's output signal drifts significantly on a time scale of several
minutes, it is not suitable for continuous monitoring. However, the
drift can be measured during the calibration runs and the net measured
value is used to calibrate the two RF BCMs. The two resonant RF cavity
monitors on either side of the Unser monitor are stainless steel
cylindrical high-Q ($\sim$3000) waveguides which are tuned to the frequency
of the beam (1497 MHz) resulting in voltage levels at their outputs
which are proportional to the beam current. Each of the RF output
signals from the two cavities in split into two parts, to be sampled
or integrated.

The signals to be sampled are processed by a high-precision digital
multi-meter (DMM), HP3458A. Each second this device gives a
digital output proportional to the RMS of beam current.
 Signals from both cavities' and Unser's multimeters are
transported through GPIB ports and are recorded by the data logging
process every $1-2$ s. The signals to be integrated are sent to an
RMS-to-DC converter to produce an analog DC voltage level, and this
level drives a Voltage-To-Frequency (VTOF) converter. These frequency
signals are then fed to 200 MHz VME scalers, and the outputs are
injected into the data stream along with other scaler
information. These logged scalers accumulate during the run and
provide a number proportional to the time-integrated voltage level
which accurately represents the total delivered charge. The regular
RMS to DC output is linear for currents from about 5 $\mu$A to 200
$\mu$A. A set of amplifiers has been introduced with gain factors
of 1, 3, and 10 in order to allow for lower currents at the expense of saturation at high
currents. Hence, there is a set of three signals coming from each RF
BCM. These six signals are fed to scaler inputs of each
spectrometer, providing redundant beam charge information.

The beam charge can be derived from BCM scaler reading by
\begin{equation}
Q_{\mathrm{BCM}\times A,H}=\frac{\frac{N_{\mathrm{BCM}\times
      A,H}}{\mathrm{clock}_H}-\mathrm{offset}_{\times A,H}}{\mathrm{constant}_{\times A}}\mathrm{clock}_H,
\end{equation}
where $A$ = 1, 3 or 10 is the gain factor,$H$=plus, minus or 0 (ungated) is
the beam helicity state, and $\mathrm{clock}_H$ is the total clock time
of corresponding helicity gate. The BCM calibration is typically
performed every $2 - 3$ months and the results are fairly stable within
$\pm 0.5\%$ down to a current of 1 $\mu$A.
\subsection{Raster and Beam Position Monitor}
The position and direction of the beam at the target
location is determined by two Beam Position Monitors (BPMA and BMPB)
which are located at 7.345 m and 2.214 m upstream of the Hall A center
respectively.

The standard difference-over-sum technique is used to determine the
relative position of the beam to within 100 $\mu$m for currents above 1
$\mu$A~\cite{hallanim,bpm1}. The absolute position of the beam can be determined
from the BPMs by calibrating them with respect to wire scanners
(superharps) which are located adjacent to each BPM. The wire scanners
are regularly surveyed with respect to the Hall A coordinates; the
results are reproducible at the level of 200 $\mu$m. The position
information from the BPMs are recorded in the raw data stream by two
ways: average value and event-by-event. The real beam position and
direction at the target can be reconstructed using the BPM positions
calculated from 8 BPM antennas' readout ($2\times 4$):
\begin{eqnarray}
x,y_{\mathrm{target}}&=&\frac{x,y_{\mathrm{BPMa}}\cdot
  \Delta z_{\mathrm{BPMb}}-x,y_{\mathrm{BPMb}}\cdot \Delta
  z_{\mathrm{BPMa}}}{z_{\mathrm{BPMb}}-z_{\mathrm{BPMa}}}\\
\vec {x}_{\mathrm{beam}} &=& \frac{\vec x_{\mathrm{BPMb}}-\vec
  {x}_{\mathrm{BPMa}}}{|\vec {x}_{\mathrm{BPMb}}-\vec
  {x}_{\mathrm{BPMa}}|},
\end{eqnarray}
where $\Delta z = z_{\mathrm{BPM}}-z_{\mathrm{target}}$.

For liquid or gas targets, high current beam ($>5~\mu$A) may damage
the target cell by overheating it. To prevent this, the beam is
rastered by two pairs of horizontal (X) and vertical (Y) air-core
dipoles located 23 m upstream of the target, and the size of rastered
beam is typically several millimeters. The raster can be used in two
modes, sinusoidal or amplitude modulated. In the sinusoidal mode both
the X and Y magnet pairs are driven by pure sine waves with relative
$90^{\circ}$ phase and frequencies $\sim$18.3 kHz, which do not produce a
closed Lissajous pattern. In the amplitude modulated mode both X and Y
magnets are drive at 18 kHz with a $90^{\circ}$ phase between X and Y,
producing a circular pattern. The radius of this pattern is changed by
amplitude modulation at 1 kHz.

During the experiment, a new triangular raster was used, which copied the
Hall C design~\cite{hallc_raster}. The new raster provides a major improvement
over the sinusoidal raster by reducing dwell time at the peaks. A uniform density
distribution of beam on the target is achieved by moving the beam position with
a time-varying dipole magnetic field with a triangular waveform. The raster contains two dipole magnets,
one vertical and one horizontal, which are located 23 m upstream from the target.

\begin{figure}
  \begin{center}
    \includegraphics[angle=0, width=0.6\textwidth]{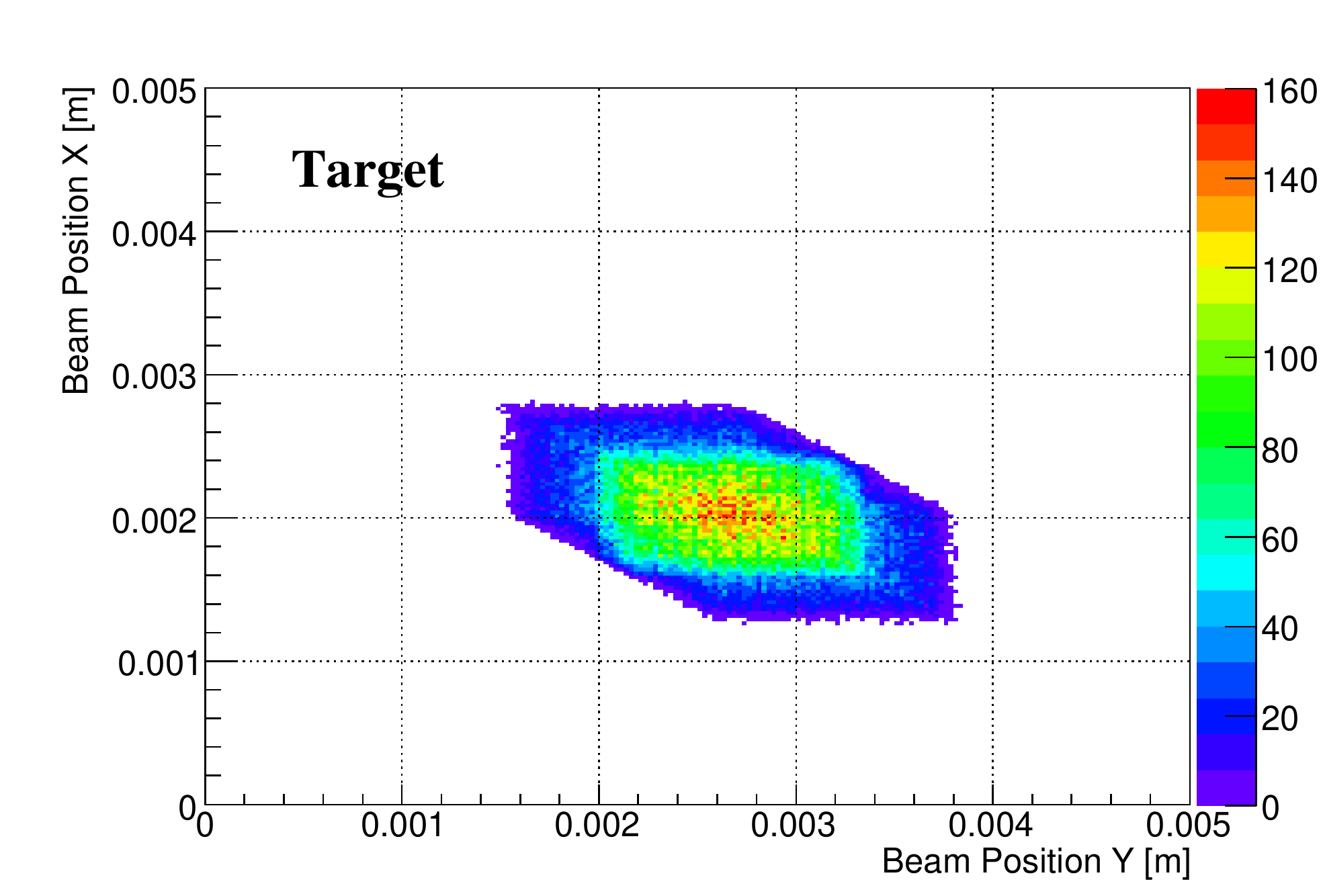}
    \caption{Beam spot at target.}
    \label{fig:raster}
  \end{center}
\end{figure}

In the electronics design, an ``H-bridge'' is used that allows one pair of
switches to open and another pair to close simultaneously and rapidly at
25 kHz. the current is drawn from HV supplies and rises according to
\begin{equation}
I(t)=\frac{\epsilon}{R}(1-e^{-t/\tau})
\end{equation}
where $\tau=L/R$ is the time constant with resistance $R$ and inductance $L$ of the controlling electronics.
The time and applied voltage are $t$ and $\epsilon$, respectively.
Fig.~\ref{fig:raster} is a sample beam spot at target with raster on. In this experiment,
a 1.5 mm $\times$ 1.5 mm raster was used.
\subsection{Beam Polarization Measurement}
There are three methods to measure the electron beam polarization:
\begin{itemize}
\item Mott method.
\item M\o ller polarimetry.
\item Compton polarimetry.
\end{itemize}
The Mott measurement~\cite{mott} is performed at the polarized electron source, and the other two polarimetries are performed
in the experimental Hall. During this experiment, since the beam polarization is canceled in the result, continuous monitoring of the polarization is not required, so only the M\o ller measurement was performed during this experiment.
\subsubsection{M\o ller Polarimetry}
The M\o ller polarimetry~\cite{moller} measures the process of M\o ller scattering
of the polarized beam electrons off polarized atomic electrons in a magnetized
foil $\vec e^-+\vec e^-\to e^-+e^+$. The cross section of the M\o ller scattering
depends on the beam and target polarization $P_b$ and $P_t$ as
\begin{equation}
\sigma\propto (1+\sum_{i=X,Y,Z}(A_{ii}P_{b,i}P_{t,i})),
\end{equation}
where $i$ = $X,Y,Z$ defines the projections of the polarizations. $A_{ii}$ is the
analyzing power, which depends on the scattering angle in the center of mass
(CM) frame $\theta_{CM}$. Assuming that the beam direction is along the
 $Z$-axis and that the scattering happens in the $ZX$ plane, we have
 \begin{eqnarray}
 A_{ZZ} &=& -\frac{\sin^2\theta_{CM}\cdot (7+\cos^2\theta_{CM})}{(3+\cos^2\theta_{CM})^2} {}
 \nonumber\\
 {}A_{XX}& =& -A_{YY} = -\frac{\sin^4\theta_{CM}}{(3+\cos^2\theta_{CM})^2}.
 \end{eqnarray}
 The analyzing power does not depend on the beam energy. At $\theta_{CM}$ = $90^\circ$,
 the analyzing power has its maximum $A_{ZZ,max}$ = 7/9. The M\o ller polarimeter of
 Hall A detects pairs of scattered electrons in a range of $75^\circ<\theta_{CM}<105^\circ$.
 The average analyzing power is about$<A_{ZZ}>=0.76$. A transverse polarization also
 produces an asymmetry, though the analyzing power is lower: $A_{XX,max}$ = $A_{ZZ,max}/7$.
 The main purpose of the polarimeter is to measure the longitudinal component of the
 beam polarization.

 The polarized electron target consists of a thin magnetically saturated ferromagnetic
 foil. An average electron polarization of about $8\%$~\cite{moller} can be obtained. The foil is
 magnetized along its plane and can be tilted at angles from $20^\circ$ to $160^\circ$ to the beam.

 \begin{figure}
  \begin{center}
    \includegraphics[angle=0, width=0.75\textwidth]{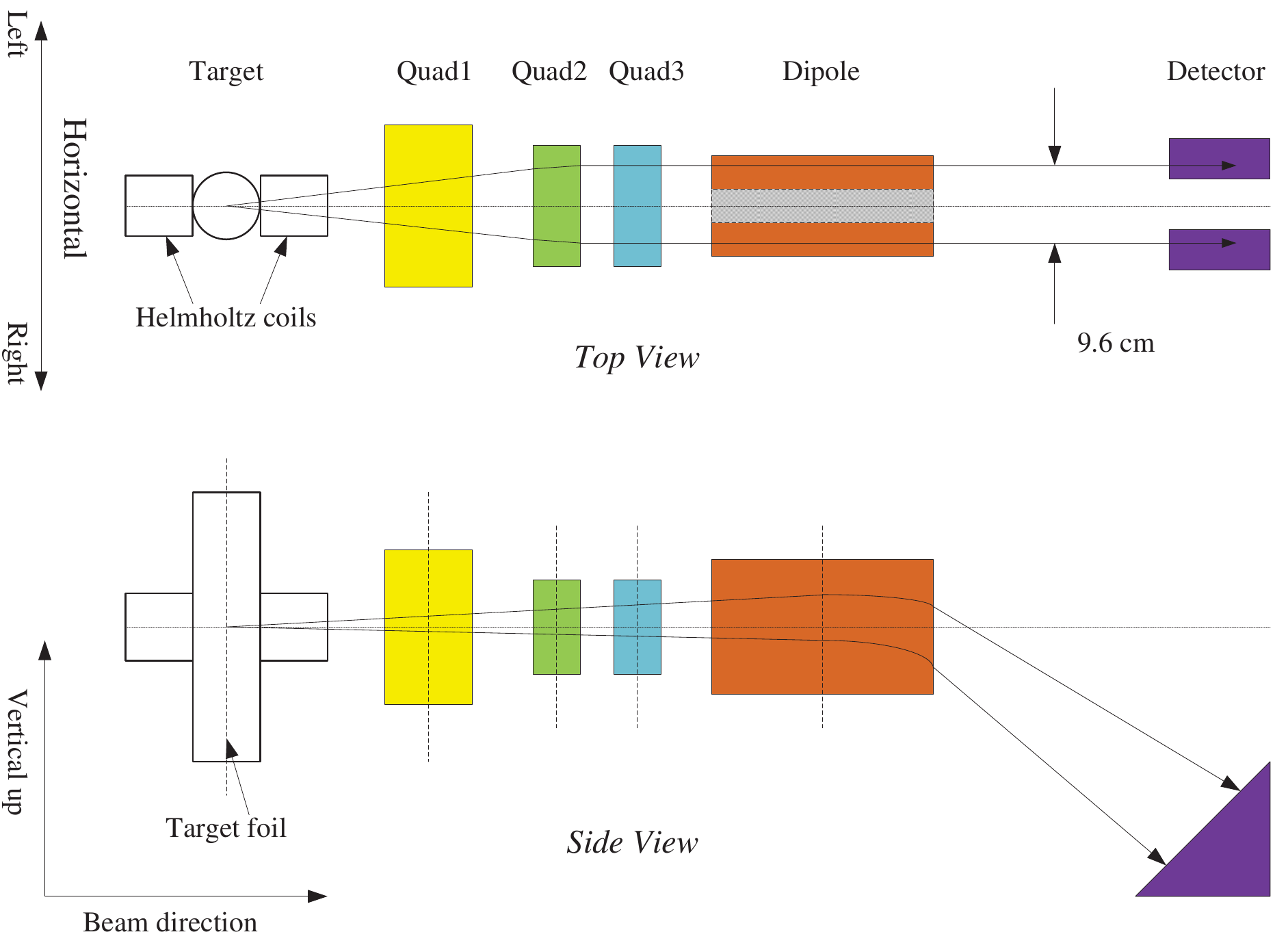}
    \caption{Layout of the M\o ller polarimeter.}
    \label{fig:moller}
  \end{center}
\end{figure}
 The scattered electrons are detected by a magnetic spectrometer (see Fig.~\ref{fig:moller}).
 The spectrometer consists of a sequence of three quadrupole magnets and a dipole magnet.
 The detector consists of scintillators and lead-glass calorimeter modules, and split
 into two arms in order to detect the two scattered electrons in coincidence. The beam
 longitudinal polarization is measured as:
 \begin{equation}
 P_{b,Z} = \frac{N_+-N_-}{N_++N_-}\cdot\frac{1}{P_t\cdot\cos\theta_t\cdot<A_{ZZ}>},
 \end{equation}
 where $N_+$ and $N_-$ are the measured counting rates with two opposite mutual
 orientation of the beam and target polarization. While $<A_{ZZ}>$ is obtained
 using Monte-Carlo calculation of the M\o ller spectrometer acceptance, $P_t$ is
 derived from special magnetization measurements of the foil samples, $\theta_t$
 is measured using a scale which is engraved on the target holder and seen with an TV
 camera, and also using the counting rates measured at different target angles.

 The M\o ller polarimeter can be used at beam energies from 0.8 to 6 GeV. The measurement
 is invasive, since the beam needs to be tuned through the M\o ller chicane, and
 the measurement is performed with low current ($\sim 0.5~\mu$A). One measurement
 typically takes an hour, providing a statistical accuracy of about $0.2\%$. The
 total relative systematic error is about $3\%$~\cite{hallanim}. During this
 experiment, two M\o ller measurements were performed at Wien angle $-33.7^\circ$
 and $-13.5^\circ$ respectively. The results are reported in Table~\ref{tab:moller}.
 The experiment were mostly running with the latter setting.
 \begin{table}
  \begin{center}
    \begin{tabular}{|c|c|c|}
      \hline
      Date & Wien & $P_b\pm\Delta P_b(stat.)$\\
      \hline
      2008/05/16 & $-33.7^\circ$ & $-67.9\pm0.2\%$\\
      \hline
      2008/05/22 & $-13.5^\circ$ & $-83.3\pm0.2\%$\\
      \hline
    \end{tabular}
    \caption{Results of the M\o ller measurements during E08-007.}
    \label{tab:moller}
  \end{center}
\end{table}
\subsection{Beam Helicity}
\begin{figure}
  \begin{center}
    \includegraphics[angle=0, width=0.75\textwidth]{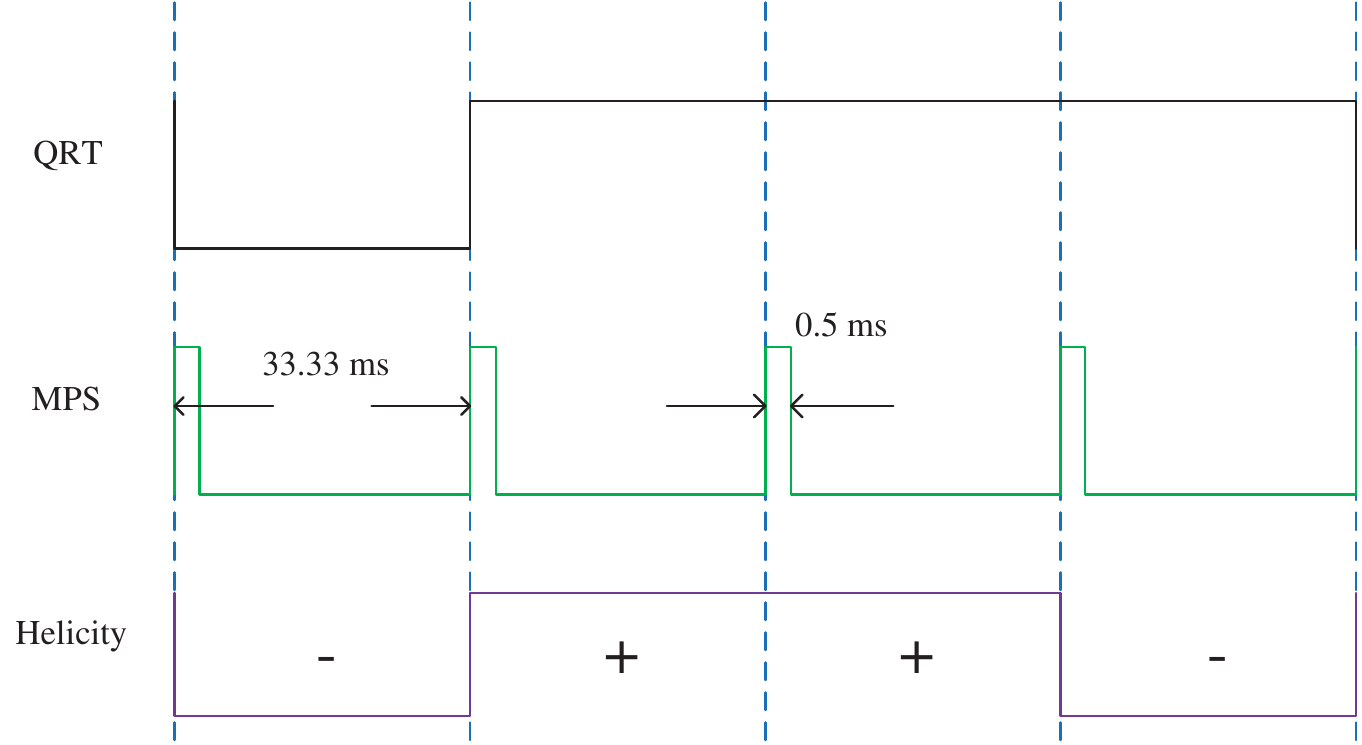}
    \caption{Beam helicity sequence used during experiment E08-007.}
    \label{fig:g0}
  \end{center}
\end{figure}

For experiment E08-007, the ``G0 helicity scheme''~\cite{g0hel} was used.
The schematics is shown in Fig.~\ref{fig:g0}. There are three relevant signals:
macro-pulse trigger (MPS), quartet trigger (QRT), and Helicity.
The characteristics of this scheme are:
\begin{itemize}
\item MPS is the master pulse at 30 Hz which is used as a gate to define
periods when the helicity is valid.
\item The helicity sequence has a quartet structure ($+--+$ or $-++-$). The
helicity of the first MPS gate is chosen pseudorandomly.
\item Quartet trigger (QRT) denotes when a new random sequence of four helicity states has begun.
\end{itemize}
There is a blank-off period of about 0.5 $\mu$s for each 33.3 ms gate period.
This blank-off is the time during which the Pockel cell at the source is changing
and settling. The quartet sequence provides for exact cancelation of linear drifts
over the sequence's timescale.  All three bits (helicity, QRT, gate) are read in
the datastream for each event, and the copies are sent to scalers which have input
registers. The delay of the helicity reporting breaks any correlations with the
helicity of the event by suppressing crosstalk. For this experiment, we used
the configuration with no delay.

\section{Target}
\subsection{Scattering Chamber}
The scattering vacuum chamber~\cite{scat} consists of several rings, and is
supported on a 607 mm diameter central pivot post. The stainless steel
base ring has one vacuum pump-out port and other ports for viewing and
electrical feed-throughs. The middle ring is made out of aluminum and
located at beam height with 152 mm vertical cutouts on each side of
the beam over the full angular range ($12.5^{\circ} \le \theta \le
167.5^{\circ}$). The cutouts are covered with a pair of flanges with
thin aluminum foils. It also has entrance and exit beam ports. The
upper ring is used to house the cryotarget. The chamber vacuum is
maintained at $10^{-6}$ Torr to insulate the target and to reduce the
effect of multiple scattering.

\subsection{Cryogenic Target}
\begin{figure}
  \begin{center}
    \includegraphics[angle=0, width=0.5\textwidth]{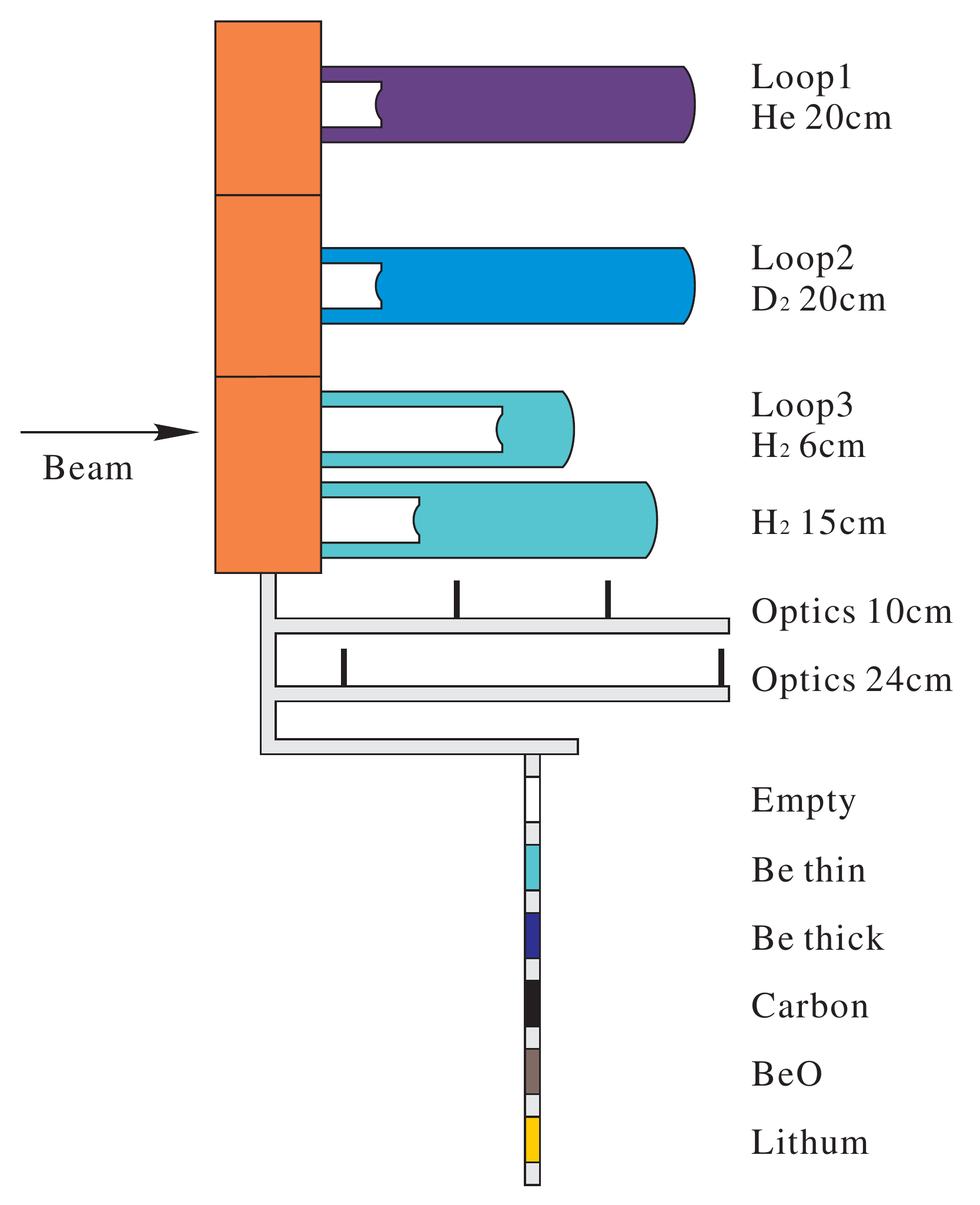}
    \caption{Target ladder.}
    \label{fig:target}
  \end{center}
\end{figure}

A 6 cm liquid hydrogen cryogenic target was used for this experiment.
The target system was mounted inside the
scattering chamber along with sub-systems for cooling, gas handling,
temperature and pressure monitoring, target control and motion, and an
attached calibration and solid target ladder (see Fig.~\ref{fig:target}).

The target system had three independent target loops: a liquid
hydrogen (LH$_2$) loop, a liquid deuterium (LD$_2$) loop and a gaseous
helium loop. The LH$_2$ loop had two aluminum cylindrical target
cells, 15 cm and 6 cm length, mounted on the vertical stack which could
be moved from one position to another by remote control. Both the
LD$_2$ and gaseous helium loops had only single 20 cm aluminum
cell. All the liquid target cells had diameter $\phi = 63.5$ mm, and
the side walls were $178\mu$m thick, width entrance and exit windows
approximately 71 and 102 $\mu$m thick, respectively. The upstream
window consisted of a thick ring holder with an inner diameter of 19
mm, large enough for the beam to pass through.

Below the cryogenic targets were two sets of carbon foil optics targets
constructed of two thin pieces of carbon foils spaced by 10 or 24
cm. A solid target, attached at the bottom, had six target positions:
an empty target, two Be targets with different thickness, a single
carbon foil (can also be used for optics data taking), a BeO foil
(typically used for direct beam observation), and a lithium target.

The LH$_2$ (LD$_2$) target were cooled at 19 K (22 K) with
pressure of 0.17 MPa (0.15 MPa), about 3 K below their boiling
temperature. Under these conditions, they have a density of 0.0723
g/$\mathrm{cm}^3$ and 0.167 g/$\mathrm{cm}^3$. The nominal operating
condition for $^4\mathrm{He}(^3\mathrm{He})$ was 6.3 K at 1.4 MPa (1.1
MPa). The coolant (helium) was supplied by the End Station
Refrigerator (ESR). The helium from ESR is available at 15 K with a
maximum cooling power of 1 kW, and at 4.5 K with a lower maximum
cooling capacity near 600 W. Typically 15 K coolant is used for liquid
cells while 4.5 K for gaseous cells. At the full 1 kW load of 15 K
coolant, up to 130 $\mu$A beam current may be incident on the liquid
target with temperature slightly over 20 K. In this configuration the
beam heating alone deposits 700 W in the target where the rest of
power arises from circuiting fans and small heaters required to
stabilize the target's temperature. The coolant supply is controlled
with Joule-Thompson (JT) valves, which can be adjusted either remotely
or locally.

\section{\label{sec:hrs}High Resolution Spectrometers}
\begin{figure}
  \begin{center}
    \includegraphics[angle=0, width=0.8\textwidth]{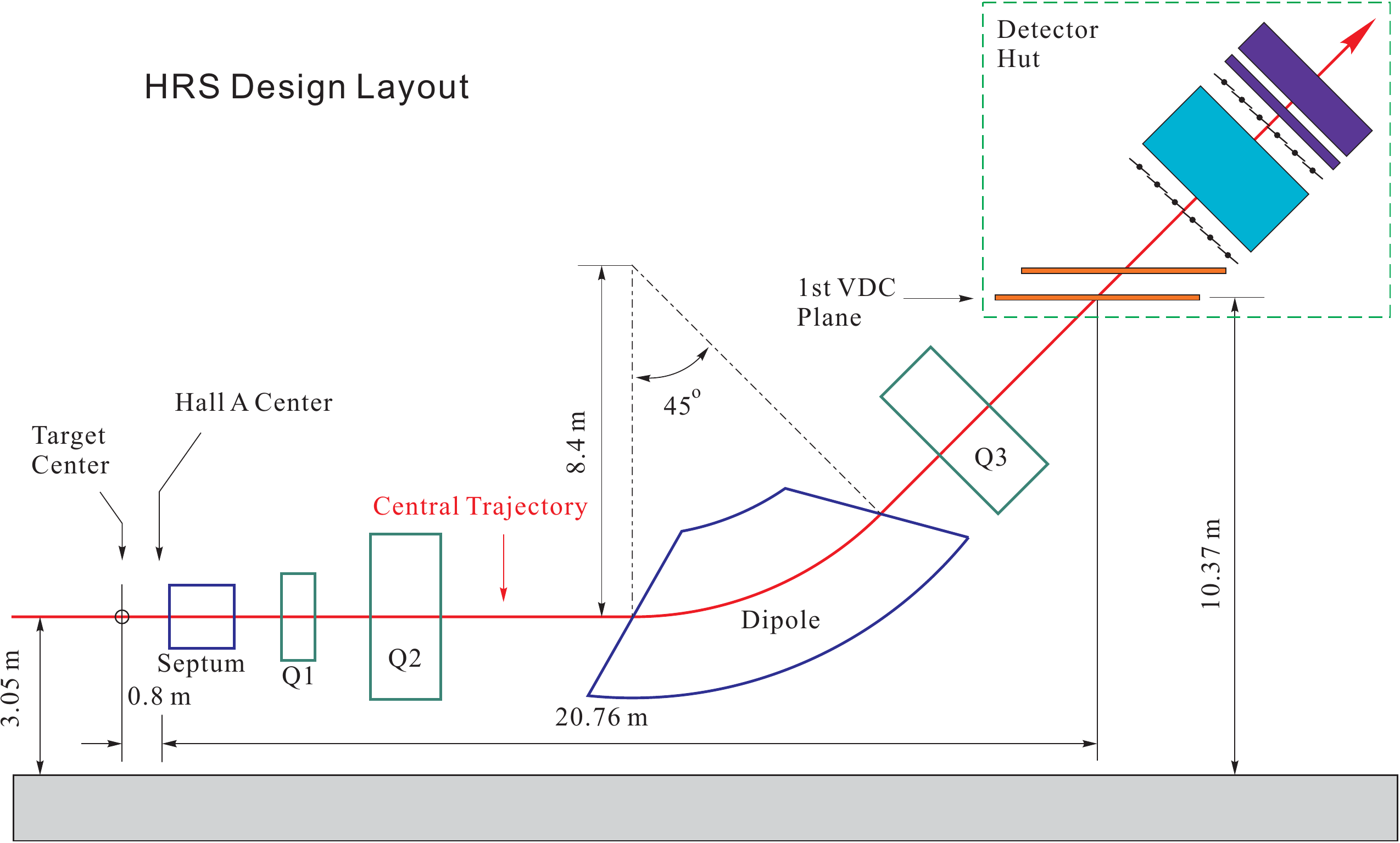}
    \caption{Schematic of Hall A High Resolution Spectrometer and the detector hut.}
    \label{fig:hrs}
  \end{center}
\end{figure}

One of the key pieces of equipments for this experiment is the left High Resolution
Spectrometers (HRS), which was used to detect the recoil proton. A
schematic view of the HRS is shown in Fig.~\ref{fig:hrs}, and the main design
characteristics are provided in Table~\ref{tab:hrs}. The vertically bending
design includes a pair of superconducting $\cos(2\theta$) quadrupoles
followed by a 6.6 m long dipole magnet with focusing entrance and
exit polefaces, including additional focusing from a field
gradient, n, in the dipole. Following the dipole is a third
superconducting $\cos(2\theta$) quadrupole. The first quadrupole Q1 is
convergent in the dispersive (vertical) plane. Q2 and Q3 are identical
and both provide transverse focusing. In this configuration, the spectrometer
can provide a momentum resolution better than $2\times 10^{-4}$ with a
$9\%$ momentum acceptance.

\begin{table}
  \begin{center}
    \begin{tabular}{l l}
      \hline
      Configuration & QQDQ vertical bend \\
      Bending angle & $45^{\circ}$\\
      Optical lengh & 24.2 m\\
      Momentum range & 0.3-4.0 $\mathrm{GeV}/c$\\
      Momentum acceptance & $\pm 4.5\% (\delta p/p)$\\
      Momentum resolution & $2\times 10^{-4}$\\
      Dispersion at the focus (D) & 12.4 m\\
      Radial linear magnification (M) & -2.5 \\
      $D/M$ & 5.0\\
      Horizontal angular acceptance & $\pm 30$ mrad\\
      Vertical angular acceptance & $\pm$ 60 mrad\\
      Horizontal resolution & 1.5 mrad \\
      Vertical resolution & 4.0 mrad \\
      Solid angle at $\delta p/p=0, y_0=0$ & 6 msr\\
      Transverse length acceptance & $\pm 5$ cm\\
      Transverse position resolution & 2.5 mm\\
      \hline
    \end{tabular}
    \caption{Main characteristics of Hall A High Resolution
    Spectrometers; the resolution values are for the FWHM.}
    \label{tab:hrs}
  \end{center}
\end{table}
\subsection{Detector Packages}
The detector packages of the spectrometer were designed to provide
various information in the characterization of charged particles passing
through the spectrometer. These include: a trigger to
readout the data-acquisition electronics, tracking
information (position and direction), coincidence determination, and particle identification.
\begin{figure}
  \begin{center}
    \includegraphics[angle=0, width=0.7\textwidth]{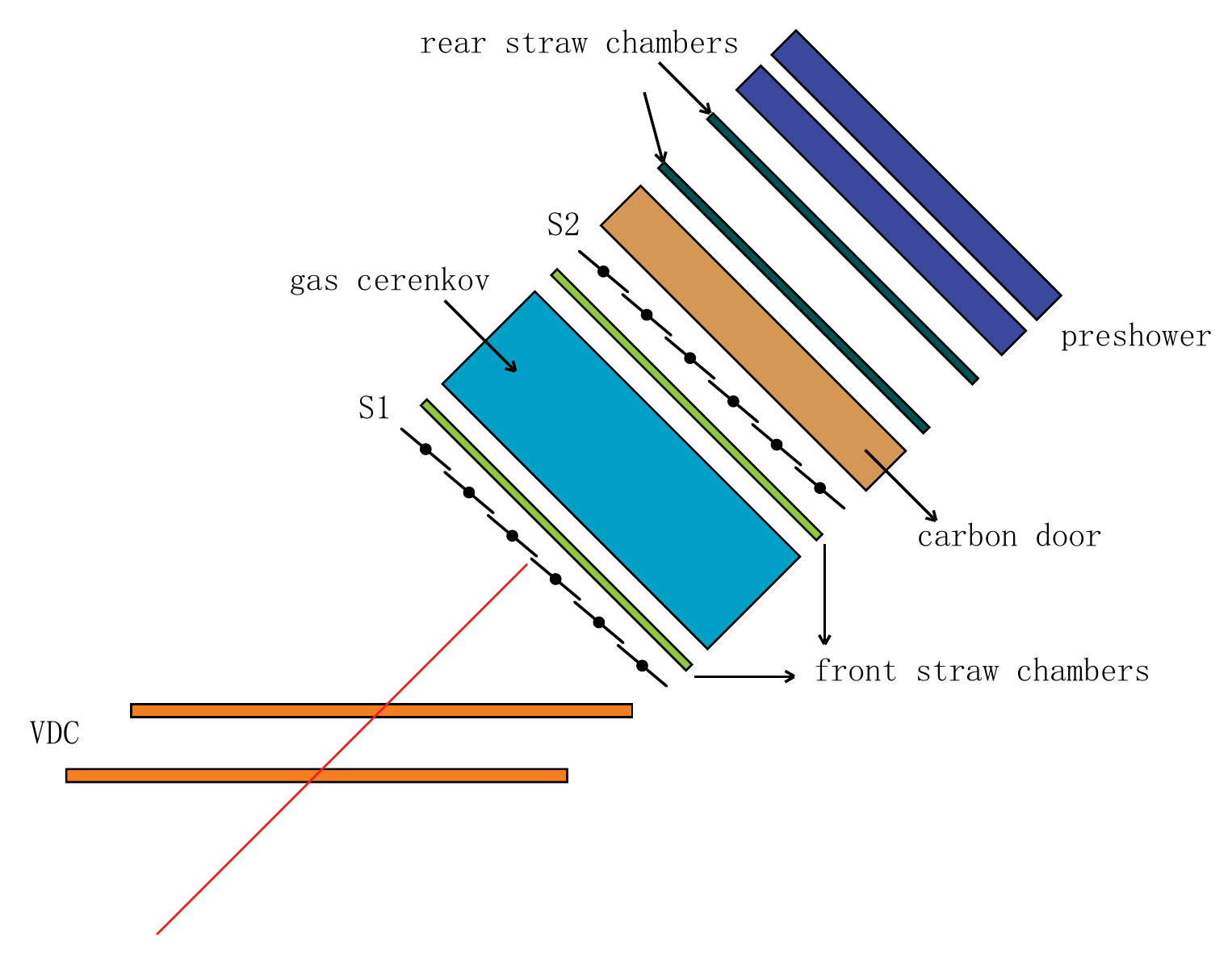}
    \caption{Left HRS detector stack during E08007.}
    \label{fig:detector}
  \end{center}
\end{figure}

The configuration of the detectors on the left spectrometer for this experiment
 is shown in Fig.~\ref{fig:detector}. The detector package includes:
\begin{itemize}
\item a set of two vertical drift chambers (VDCs) which provide tracking
  information.
\item two scintillator planes which provide basic triggers.
\item a CO$_2$ gas Cerenkov detector for particle identification.
\item the focal plane polarimeter (FPP) measure the recoil proton
  polarization.
\item a pair of lead glass pion rejectors for PID.
\end{itemize}
For this experiment, the key instruments are the scintillator planes, VDCs and the FPP.

\subsection{Vertical Drift Chambers}
The vertical drift chamber (VDCs)~\cite{vdc, vdc_1}, provides a precise
measurement of the incident position and angle of the charged
particles at the spectrometer focal plane\footnote{The focal plane is a plane
associated with the lower VDC of each spectrometer. A detailed description
and the definition of related coordinate systems can be found in~\cite{optics}}.
The tracking information from the VDC measurement is combined with the knowledge of the
spectrometer optics to reconstruct the position, angle and momentum of
the particles in the target coordinate system.

\begin{figure}
  \begin{center}
    \includegraphics[angle=0, width=0.75\textwidth]{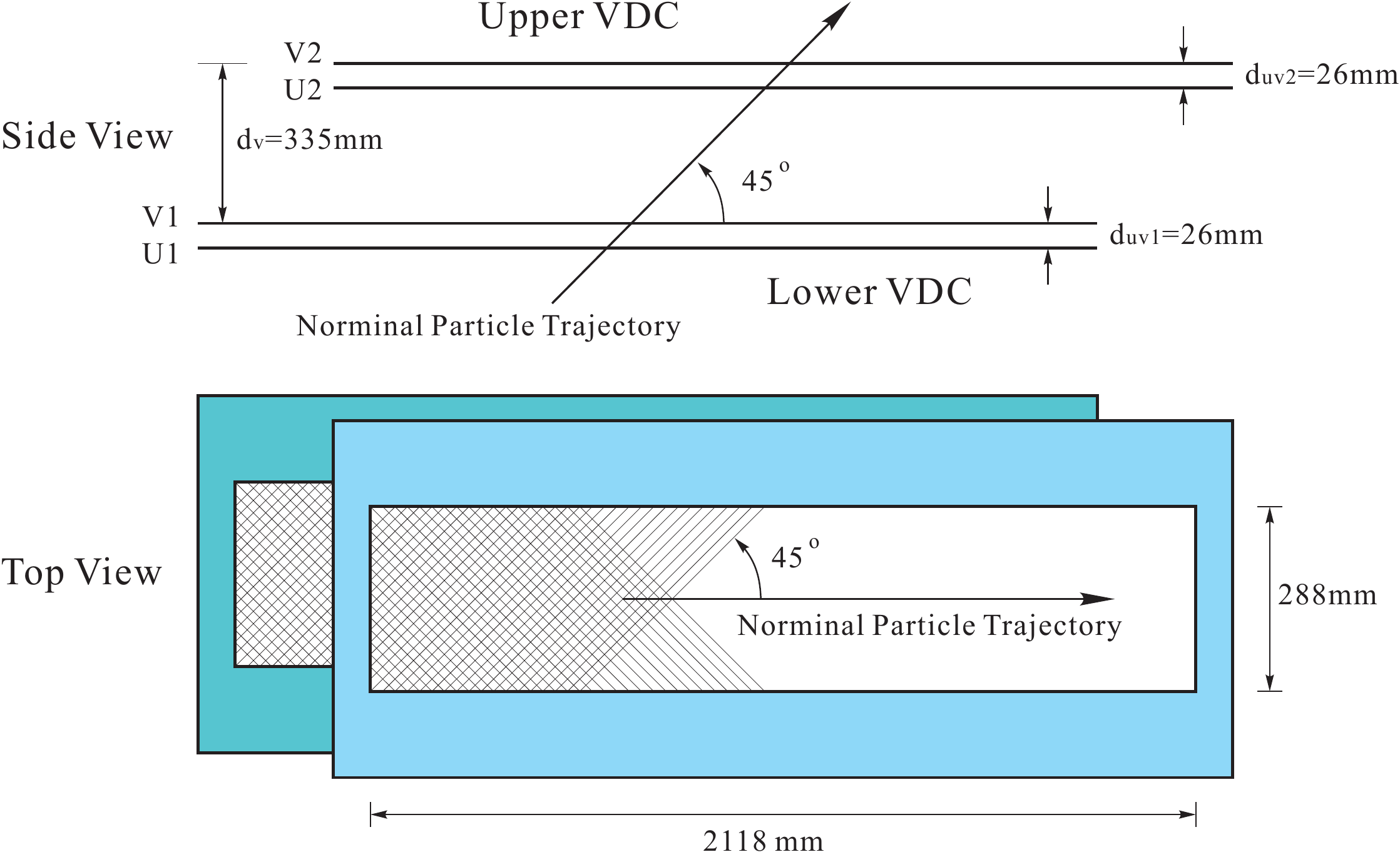}
    \caption{Schematic diagram and side view of VDCs.}
    \label{fig:vdcs}
  \end{center}
\end{figure}

The pair of VDC chambers are laid horizontally. The top VDC is placed 33.5 cm
above the bottom VDC and shifted by another 33.5 cm in the dispersive
direction to account for the $45^{\circ}$ central trajectory (see
Fig.~\ref{fig:vdcs}). Each VDC consists of two planes of wires in a standard UV
configuration: the wires of each successive plane are oriented at
$90^{\circ}$ to one another. There are a total of 368 sense wires in
each plane, spaced 4.24 mm apart.

\begin{figure}
  \begin{center}
    \includegraphics[angle=0, width=0.75\textwidth]{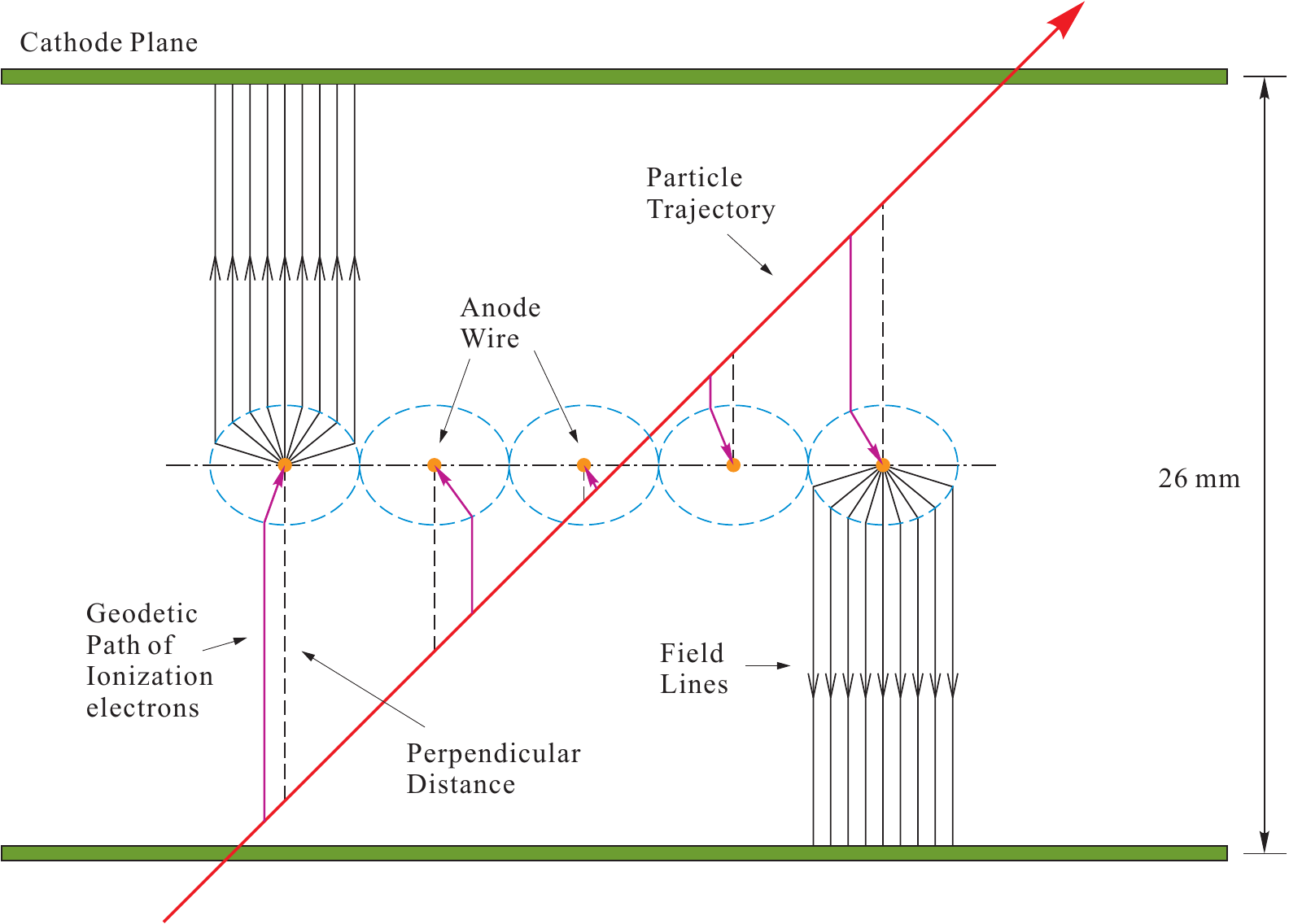}
    \caption{Configuration of wire chambers.}
    \label{fig:track}
  \end{center}
\end{figure}

During operation, the VDC chambers have their cathode plane at about $-4$ kV
and the wires at ground. The gas supplied to the VDCs is a $62\%/38\%$
argon-ethane ($\mathrm{C}_2\mathrm{H}_6$) mixture, with a flow rate of 10
liter/hour~\cite{hallanim}. When a charged particle travels through the
chamber, it ionizes the gas inside the chamber and leaves a track of
electrons and ions along its trajectory behind. The ionized
electrons accelerate toward the wires along the path of least time
(geodetic path). The Hall A VDCs feature a five cell design, i.e a
typical $45^{\circ}$ track will fire five wires as shown in
Fig.~\ref{fig:track}.
The fired wires are read out with time-to-digital
converters (TDCs) operating in common stop mode. In this
configuration, a smaller TDC signal corresponds to a larger drift
time. With a 50 $\mu$m/ns drift velocity and time shift constants, the
distances of the track to each fired wires are precisely
reconstructed. The position and direction of the track is then
determined. In the focal plane, the position resolution
$\sigma_{x(y)}\sim 100~\mu$m, and the angular resolution
$\sigma_{\theta(\phi)}\sim 0.5$ mrad.

\subsection{Scintillator Trigger Plane}
\begin{figure}
  \begin{center}
    \includegraphics[angle=0, width=0.6\textwidth]{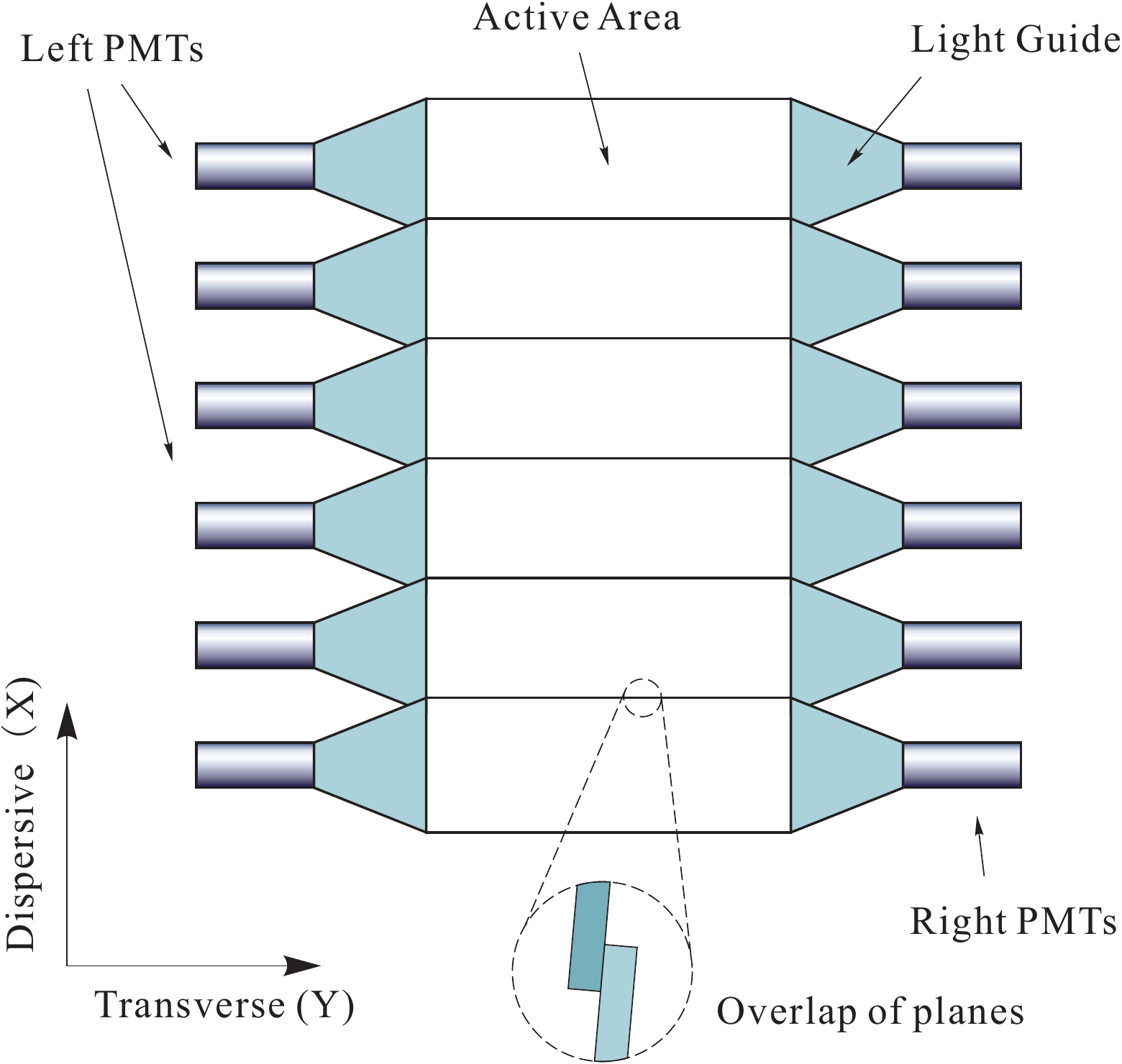}
    \caption{Layout of scintillator counters.}
    \label{fig:scint}
  \end{center}
\end{figure}

There are two planes of trigger scintillators S1 and S2 in the left
HRS, separated by a distance of about 2 m. Each plane is composed of
six overlapping paddles made of thin plastic scintillator (5 mm BC408)
to minimize hadron absorption (see Fig.~\ref{fig:scint}). The active area for the
scintillator paddles are $29.5\times 35.5~\mathrm{cm}^2$ (S1) and
$37.0\times 54.0~\mathrm{cm}^2$ (S2), and are viewed by two photomultiplier
tubes (PMTs) (Burle 8575).
The scintillators were used to generate triggers for the data
acquisition system. The time resolution of each plane is about 0.30
ns. The scintillators can also be used for particle identification by
measuring the Time-of-flight (TOF) between the S1 and S2 planes.

Additionally, the S0 scintillator counter is usually used for
trigger efficiency analysis. It was removed for this experiment to reduce the
energy loss of the low momentum protons ($\sim 550$ MeV/$c$).

\subsection{\label{sec:fpp}Focal Plane Polarimeter}

The Focal Plane Polarimeter (FPP) measured the polarization of protons in
the hadron spectrometer~\cite{fpp_bimbot}. It was developed by the
College of William \& Mary, Rutgers University, Norfolk State
University and the University of Georgia.

\begin{figure}
  \begin{center}
    \includegraphics[angle=0, width=0.75\textwidth]{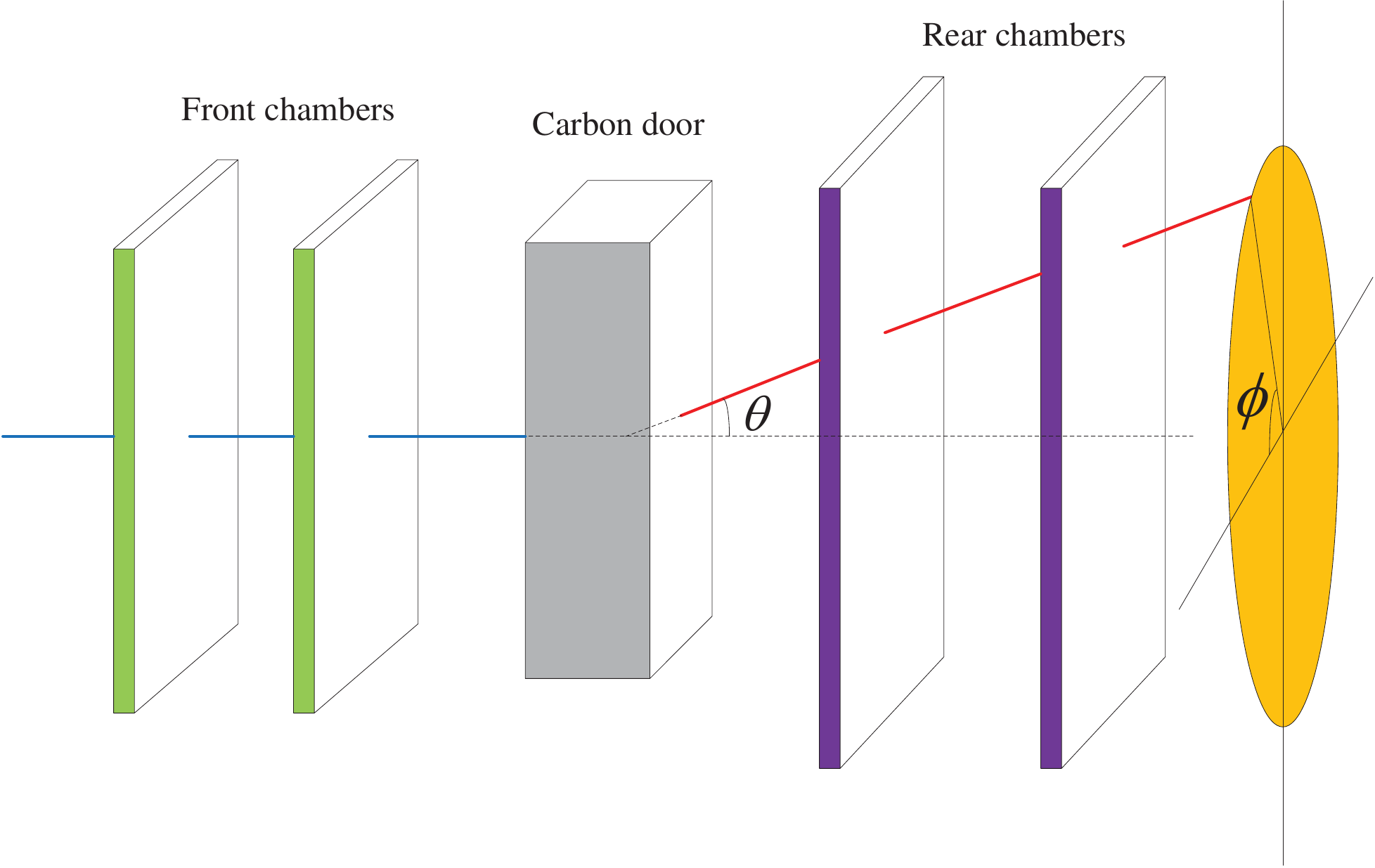}
    \caption{Layout of the Focal Plane Polarimeter.}
    \label{fig:fpp}
  \end{center}
\end{figure}

The FPP is located between the VDCs and the lead glass counter, it consists
of 4 straw chambers and a carbon analyzer (see Fig.~\ref{fig:fpp}).
When the polarized protons pass through the carbon analyzer, the nuclear
spin-orbit force leads to an azimuthal asymmetry due to the scattering from
carbon nuclei. The particle trajectories, in particular the scattering
angles in the carbon analyzer, are determined by the front and rear chambers.

The front straw chambers are separated by about 114 cm, and are
located before and after the gas Cerenkov detector. The second chamber
is followed by S2, which is in turn followed by the FPP carbon
analyzer. The rear chambers, chamber 3 and 4 are separated by 38 cm
and are immediately behind the carbon analyzer.

\begin{figure}
  \begin{center}
    \includegraphics[angle=0, width=0.75\textwidth]{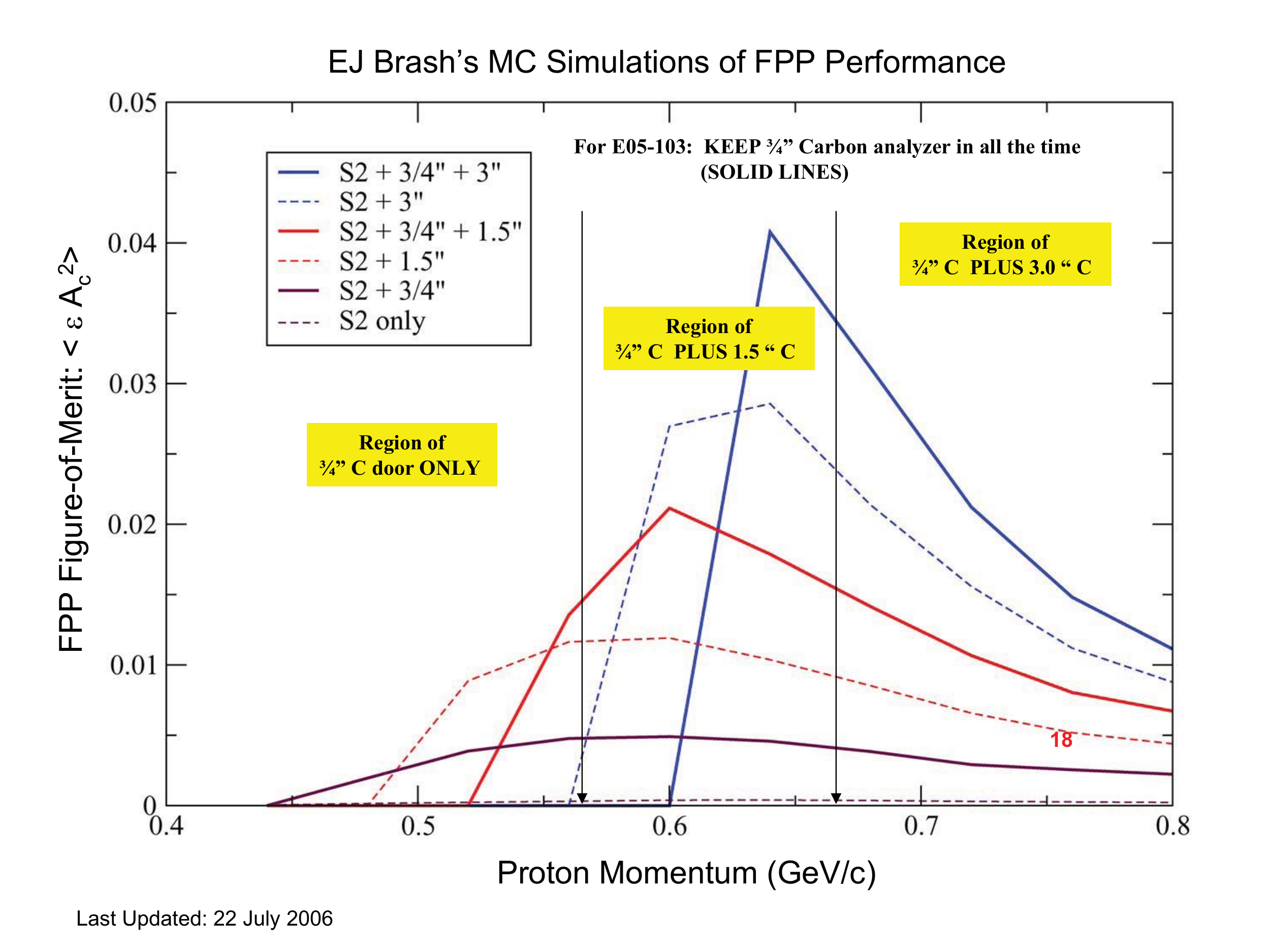}
    \caption{The simulated FPP figure of merit with different carbon door thicknesses~\cite{ed_fpp}.}
    \label{fig:fpp_sim}
  \end{center}
\end{figure}
The carbon analyzer consists of 5 carbon blocks. Each block is split
in the middle so that it can be moved in or out of the proton paths.
The total thickness of the carbon analyzer can be adjusted accounting
for different proton momentum. The block thicknesses, from
front to rear are 9'', 6'', 3'', 1.5'' and 0.75''. The block positions
are controlled through EPICS~\cite{epics}. For this experiment, the proton momentum
was between 550 $\mathrm{MeV}/c$ and 930 $\mathrm{MeV}/c$. We adjusted
the carbon door thicknesses based on a Monte Carlo simulation (see Fig.~\ref{fig:fpp_sim}). The
thicknesses of the carbon door used for different kinematics are
listed in Table~\ref{tab:thick}.
\begin{table}
  \begin{center}
    \begin{tabular}{|c|c|c|c|}
      \hline
      Kine. & $Q^2~[(\mathrm{GeV}/c)^2]$ & $P_p~[\mathrm{GeV}/c]$ & Carbon thickness
      [inch]\\
      \hline
      K1 & 0.35 & 0.616 & 2.25\\
      K2 & 0.30 & 0.565 & 2.25\\
      K3 & 0.45 & 0.710 & 3.75\\
      K4 & 0.40 & 0.668 & 3.75\\
      K5 & 0.55 & 0.794 & 3.75\\
      K6 & 0.50 & 0.752 & 3.75\\
      K7 & 0.60 & 0.836 & 3.75\\
      K8 & 0.70 & 0.913 & 3.75\\
      \hline
    \end{tabular}
    \caption{Carbon thickness along the proton momentum at each kinematics.}
    \label{tab:thick}
  \end{center}
\end{table}

\begin{figure}
  \begin{center}
    \includegraphics[angle=0, width=0.5\textwidth]{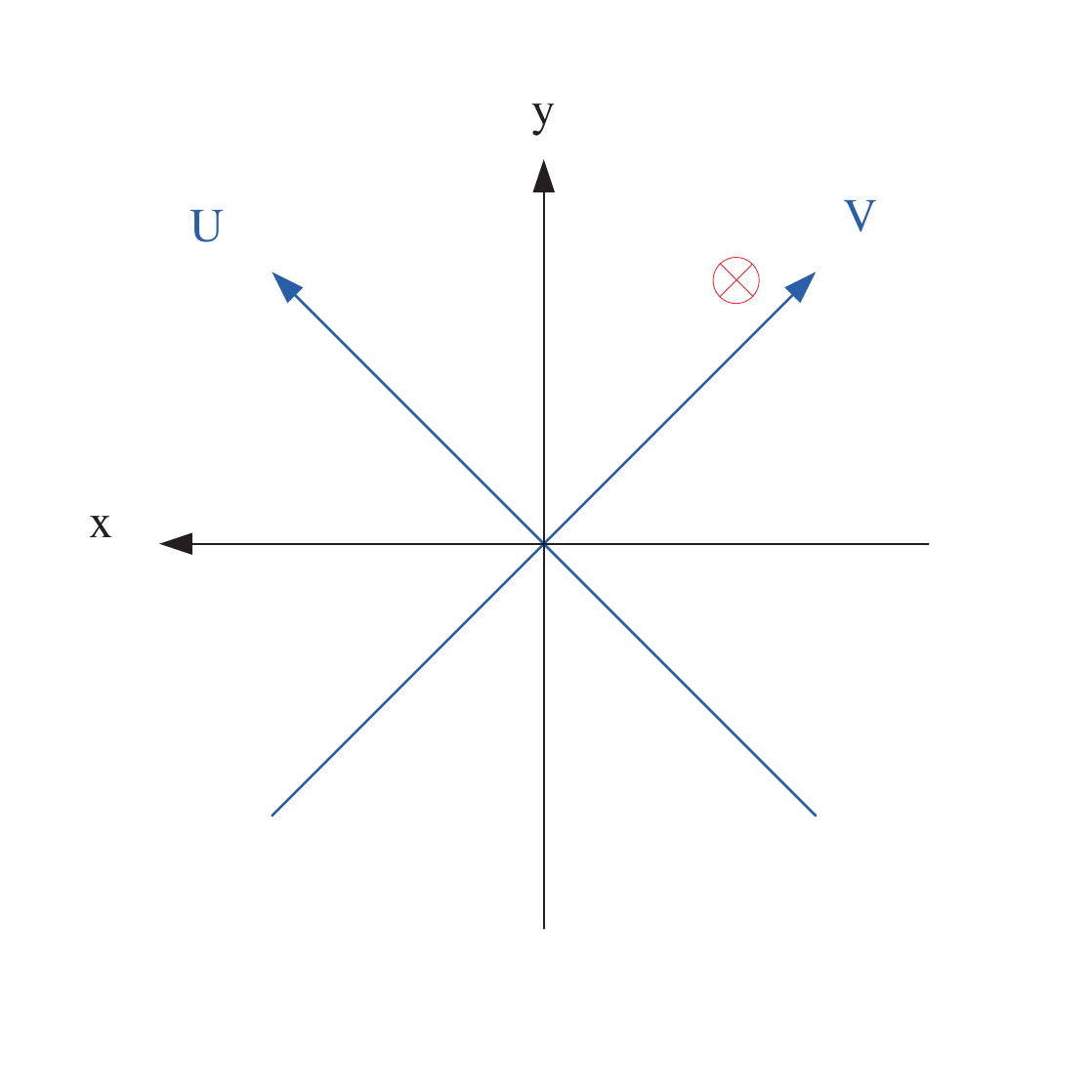}
    \caption{FPP coordinate system.}
    \label{fig:fpp_orient}
  \end{center}
\end{figure}
The straw chambers include X, U, and V planes. The
central ray defines the $z$-axis. X wires are along the horizontal
direction and measure position along the dispersive direction. As illustrated
in Fig.~\ref{fig:fpp_orient}, the UV planes are oriented at $45^{\circ}$ with respect to
the transverse plane of the XY coordinate system, with +U between the +X and +Y axes,
and +V between the +Y and -X axes. The configurations for each chamber are listed
in Table~\ref{tab:fpp_dim}. The FPP has angular resolution better than 1 mrad and
accepts second scattering angles of at least $20^{\circ}$.

\begin{figure}
  \begin{center}
    \includegraphics[angle=0, width=0.80\textwidth]{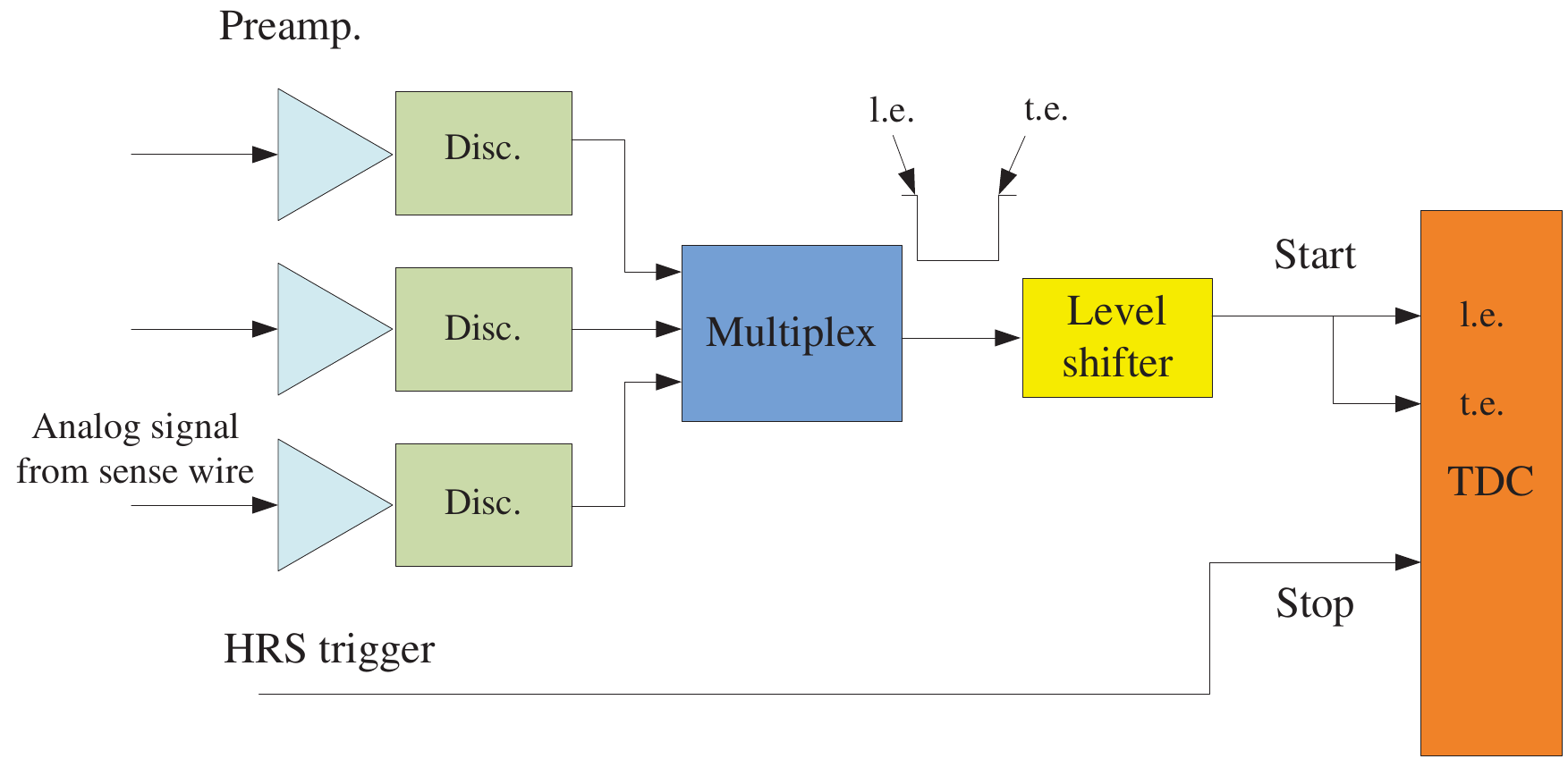}
    \caption{Block diagram for the logic of the FPP signal. (l.e. = leading edge, t.e. = trailing edge).}
    \label{fig:fpp_logic}
  \end{center}
\end{figure}

\begin{table}
  \begin{center}
    \begin{tabular}{|c|c|c|c|c|}
      \hline
      Chamber & Ch.1 & Ch.2 & Ch.3 & Ch.4\\
      \hline
      Active legnth (cm) & 209.0 & 209.0 & 267.5 & 292.2\\
      Active width (cm) & 60.0 & 60.0 & 122.5 & 140.6 \\
      Wire spacing (cm) & 1.095 & 1.095 & 10.795 & 1.0795 \\
      \hline
      Configuration & 3U + 3V & 3U + 3V & 2U+ 2V + 2X & 3U + 3V\\
      Straws per plane & 170 & 170 & 249 & 292 \\
      \hline
    \end{tabular}
    \label{tab:fpp_dim}
    \caption{Dimensions of the FPP straw chambers.}
  \end{center}
\end{table}
\begin{figure}
  \begin{center}
    \includegraphics[angle=0, width=0.75\textwidth]{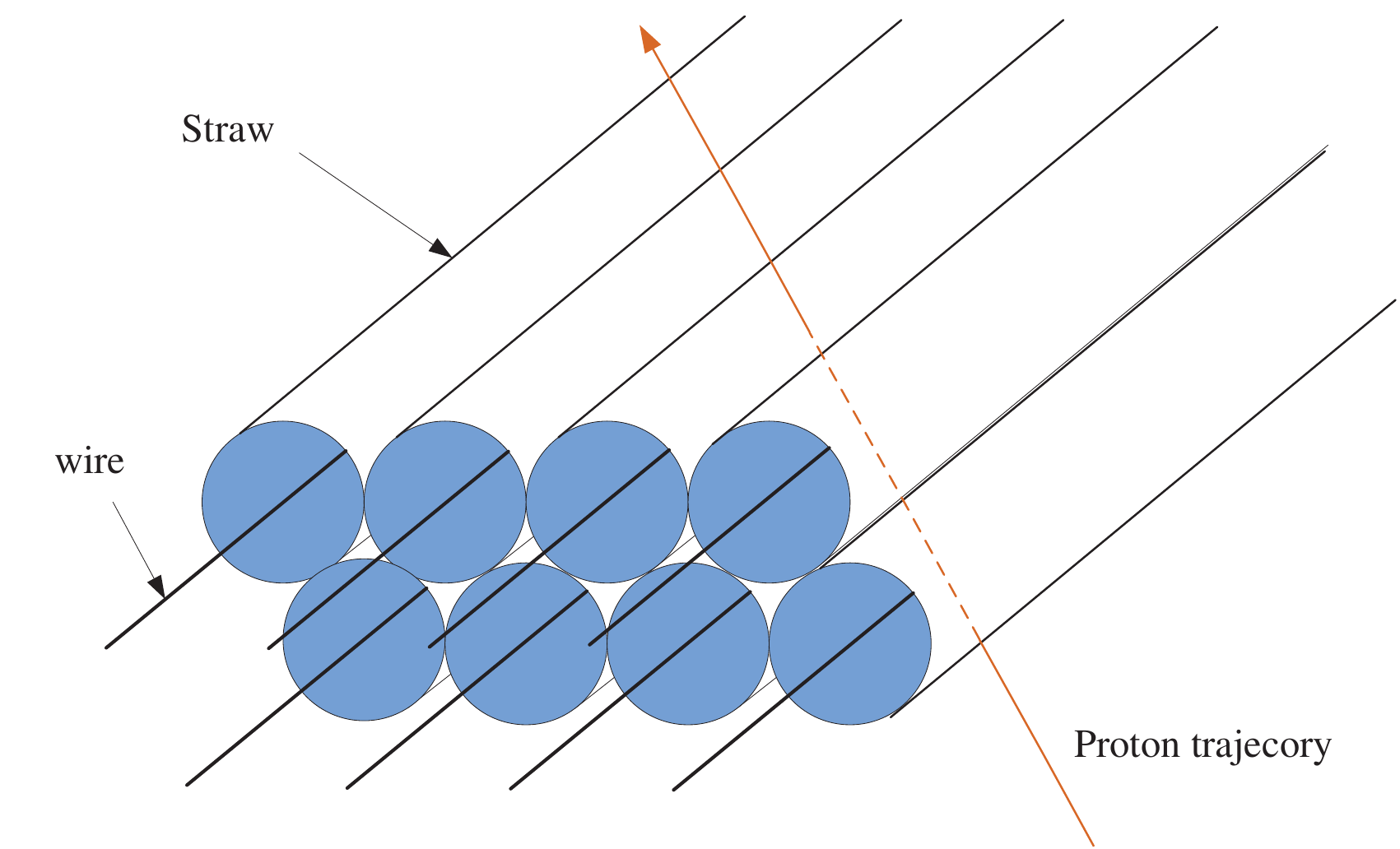}
    \caption{Straws in two different planes of a FPP straw chamber.}
    \label{fig:fpp_straw}
  \end{center}
\end{figure}
The straw chambers are a set of cylindrical tubes of radius 0.5 cm,
with a thin wire running along a central axis of each tube (straw), as
shown in Fig.~\ref{fig:fpp_straw}. The wire is at positive high voltage ($\sim$ 1.8
kV) relative to the straw. Each tube is individually supplied with a
gas mixture of Argon (62$\%$) and Ethane (38$\%$). When a charged
particle passes through the straw, it ionizes the Argon gas atoms,
leaving behind a track of electrons. These electrons drift
towards the anode wire, at a constant velocity of about 50 $\mu$
m/s. When the electrons get within about 100 $\mu$m of the wire, the
increase in electric field strength is larger enough so that addition atoms ionize;
this leads to an avalanche effect and produces a gain of about $10^5$ per primary ionization.
The movement of the positive and negative ions leads to a voltage drop on the wire
and produces a negative electrical signal. The analog signal is then sent to the read
out board, where it is pre-amplified and discriminated to give a logic
pulse (see Fig.~\ref{fig:fpp_logic}).

Because of the straw around each wire forms a physical ground, a
proton track leaves a signal only in one wire of a plane. Multiplexing
the signal in groups of eight neighboring wires, and reading out the entire
group by the same multiplexing chip, it significantly reduces the amount of
electronics required for the FPP. This multiplexing chip is setup to give a
logic pulse whose width depends on which wire fired. This 45 mV signal is converted to
a 800 mV signal in the level shifter and is sent to the FastBus TDC modules,
whose output is readout to the data stream. The multi-hit TDCs records the
arrival of the leading edge and the trailing edge of the logic signal.
The time difference between the leading edge and the common stop given by
the trigger gives the drift time.

\section{BigBite Spectrometer}
Due to the constraints from the preceding experiments, the BigBite
spectrometer was used to detect the electrons instead of the originally proposed right HRS.
Compared with the standard HRS, the BigBite spectrometer has larger angular
and momentum acceptance. Recently, the spectrometer has been
upgraded to detect electrons with adequate momentum and angular
resolution for a series of experiments~\cite{gen,transversity}.

\begin{figure}
  \begin{center}
    \includegraphics[angle=0, width=0.80\textwidth]{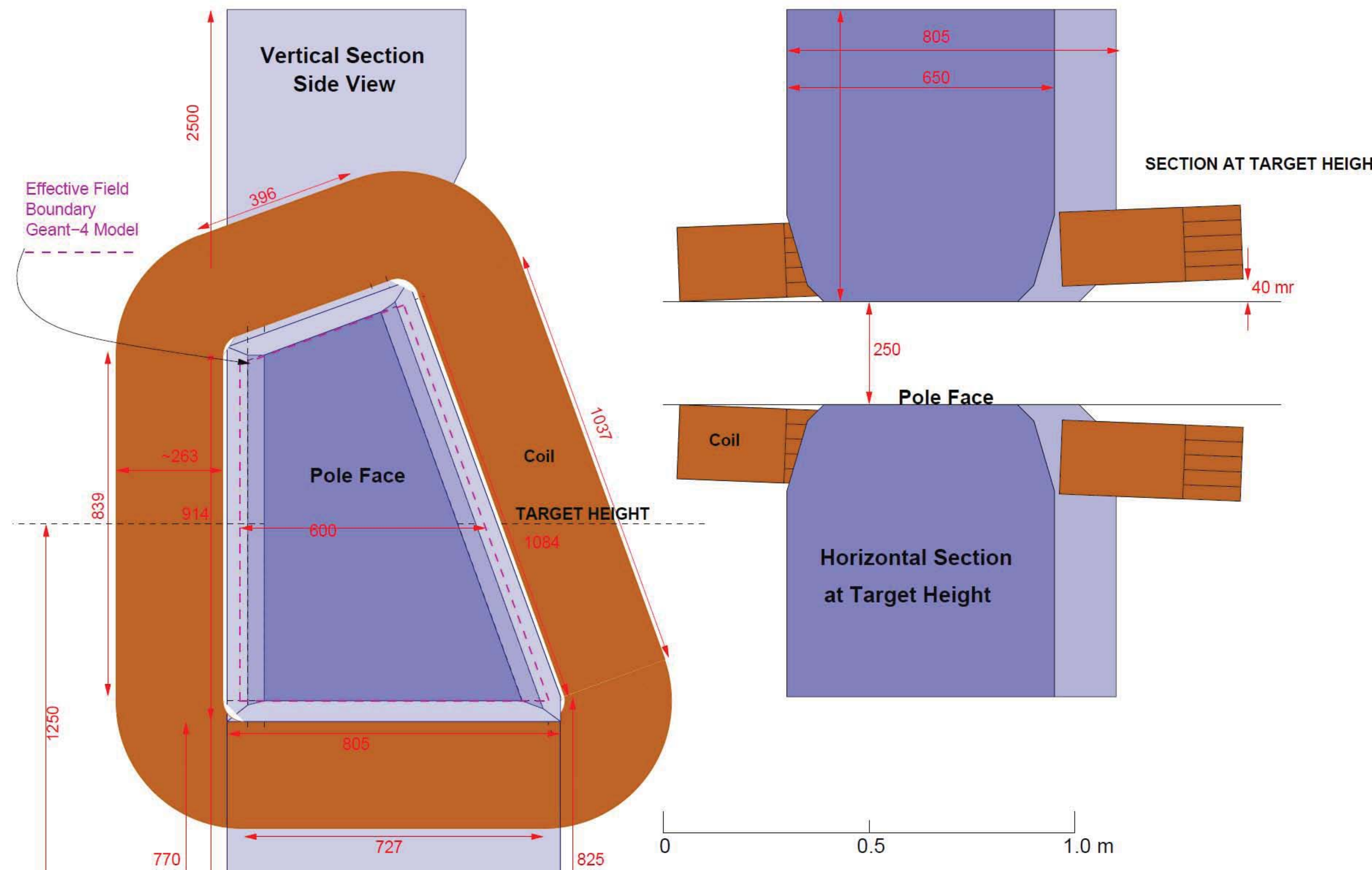}
    \caption{A side view (left) and top view (right) of the BigBite magnet showing the magnetic field boundary and the large pole face gap.}
    \label{fig:bb_mag}
  \end{center}
\end{figure}
The central component of the spectrometer is a large acceptance,
non-focusing dipole magnet. The magnet was originally designed and built for
use at NiKHEF in the Netherlands~\cite{bb_0,bb_1}. The large pole-face gap
(25 cm in the horizontal and 84 cm in the vertical directions) allows
for a larger bite of scattered particles in the angular
acceptance (see Fig.~\ref{fig:bb_mag}).

\begin{figure}
  \begin{center}
    \includegraphics[angle=0, width=0.45\textwidth]{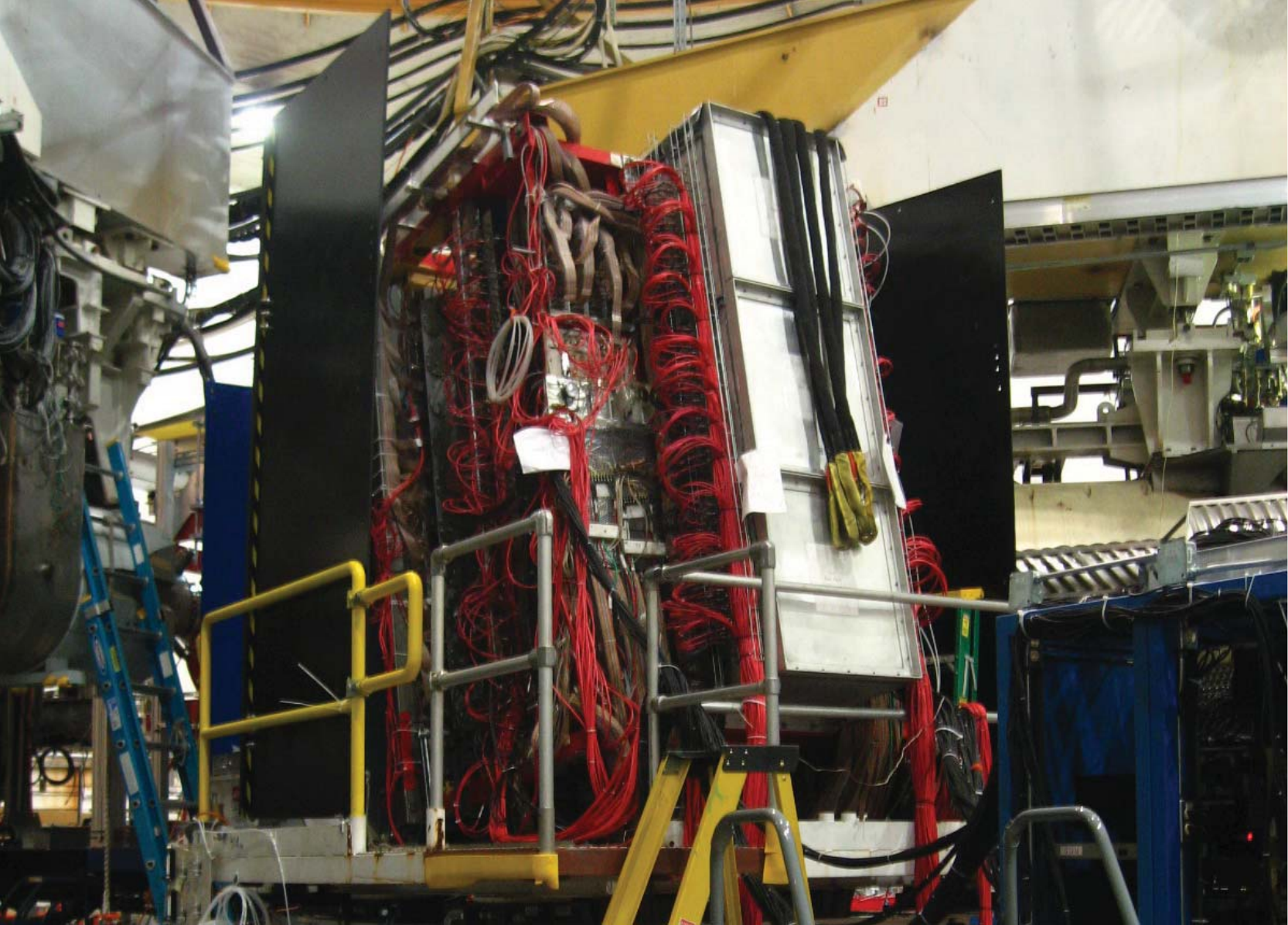}
    \includegraphics[angle=0, width=0.45\textwidth]{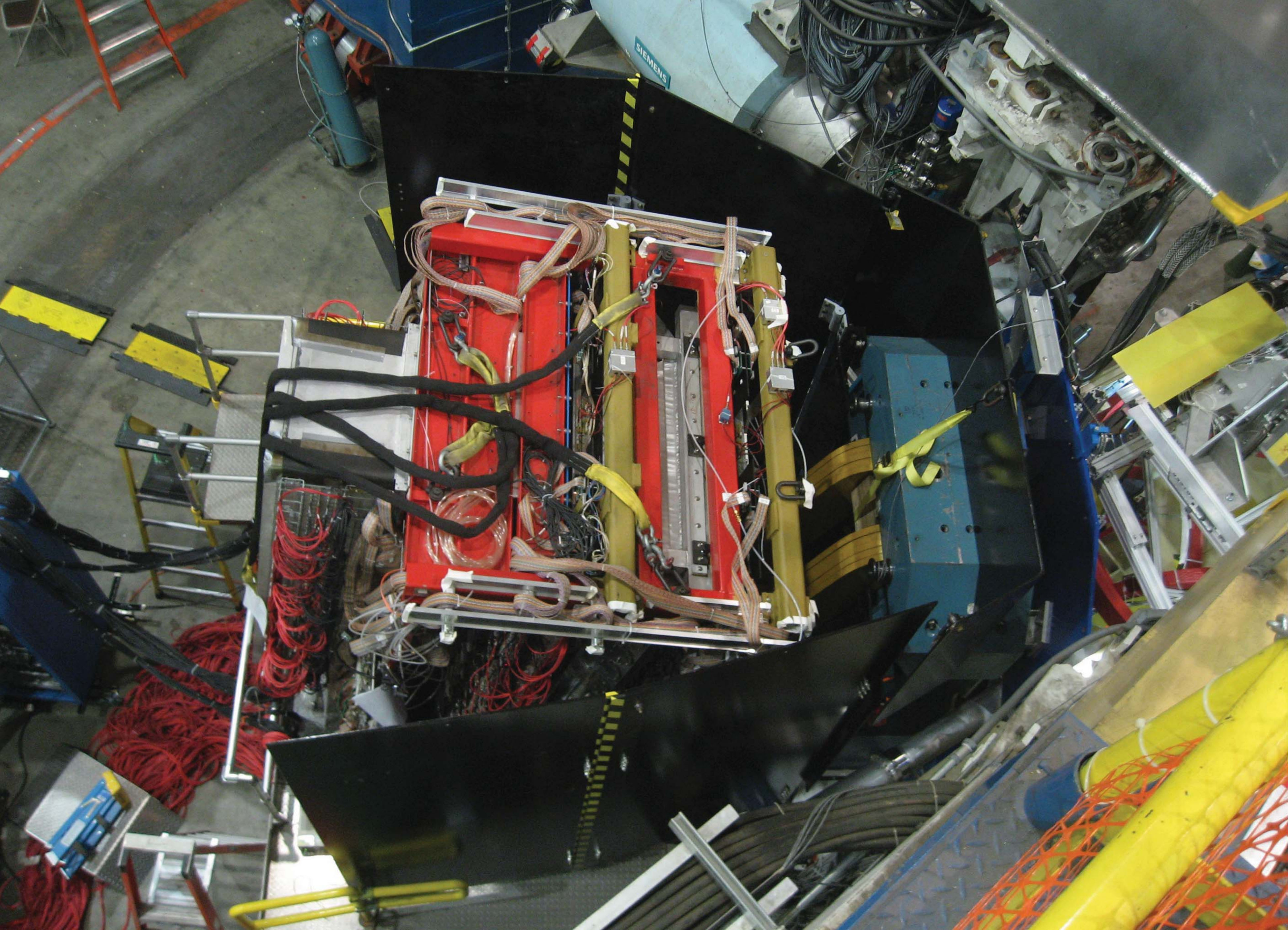}
    \caption{A side view (left) and top view (right) of the BigBite spectrometer during this experiment.}
    \label{fig:bb_pic}
  \end{center}
\end{figure}

In this experiment, the magnet was located $\sim 1.1$ m from the target (see Fig.~\ref{fig:bb_pic})
and can provide a field strength of up to 1.2 T.
The nominal momentum acceptance is $200\sim 900 \mathrm{MeV}/c$,
and the solid angle acceptance is $\sim 96$ msr, roughly sixteen
times larger than the nominal HRS acceptance.

\begin{figure}
  \begin{center}
    \includegraphics[angle=0, width=0.7\textwidth]{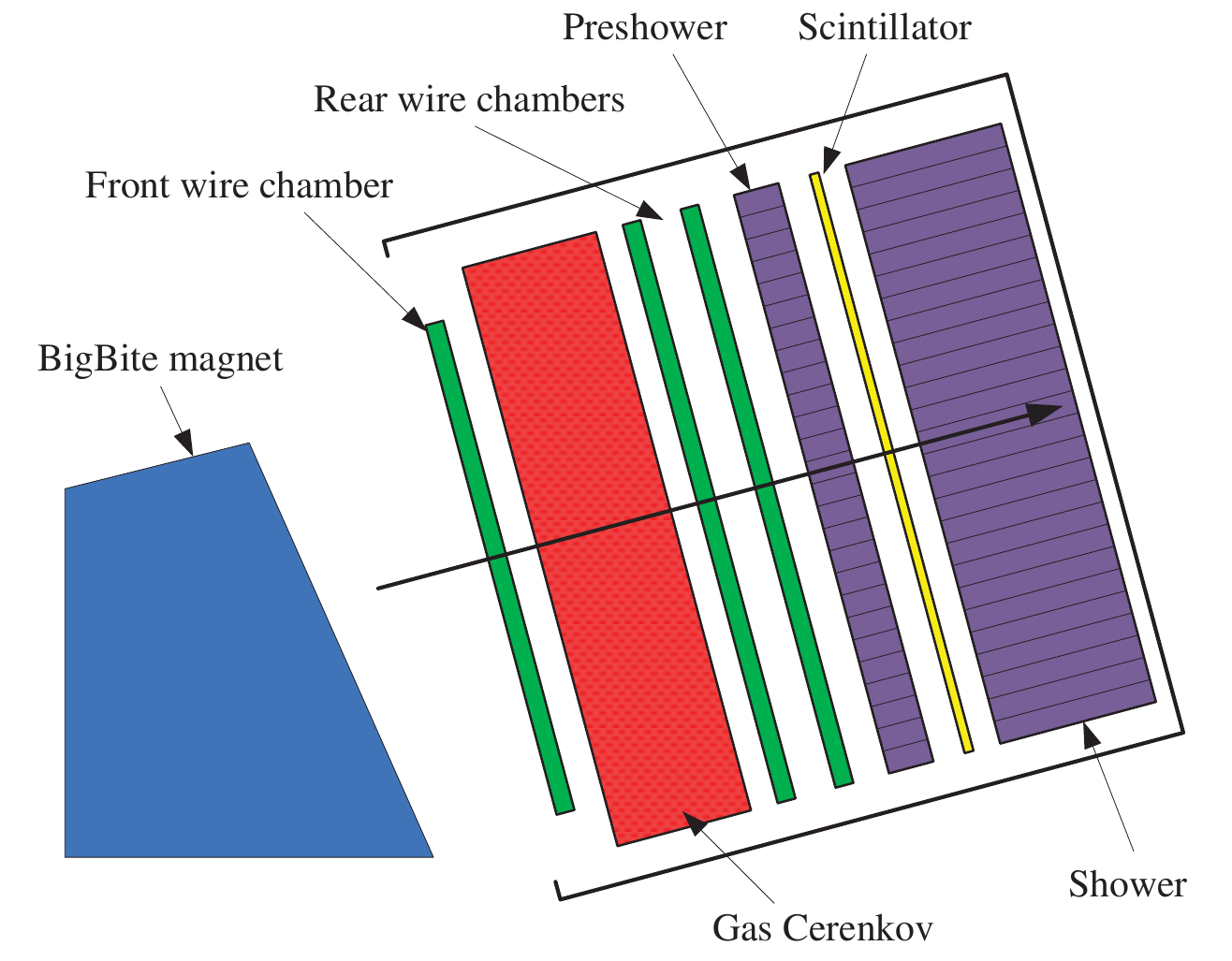}
    \caption{A side view of the BigBite detector package during this experiment.}
    \label{fig:bb_det}
  \end{center}
\end{figure}
As shown in Fig.~\ref{fig:bb_det}, the BigBite electron package consists of:
\begin{itemize}
\item 3 sets of multiple wire drift chambers.
\item a gas Cerenkov counter.
\item a pre-shower counter.
\item a scintillator plane.
\item a shower counter.
\end{itemize}
The scintillator plane consists of 13 scintillator paddles with PMTs on both sides,
and each paddle has a size of $17 \times 64\times 4$ cm.
The pre-shower counter has $2\times 27$ lead glass blocks ($8.5\times
8.5\times 37$ cm), and each block is oriented
perpendicular to the particle tracks. The shower counter has $7\times 27$ lead
glass blocks and are aligned parallel to the tracks. The signal detected by
lead glass blocks is linearly proportional to the energy deposited by the
incoming particle~\cite{grupen}. Electromagnetic showers develop
in the counter, whereas hadronic showers do not due to the longer
hadronic mean free path. Therefore, the longitudinal distribution
of the energy deposited in the counter can be used to identify the incident particles.

\begin{figure}
  \begin{center}
    \includegraphics[angle=0, width=0.7\textwidth]{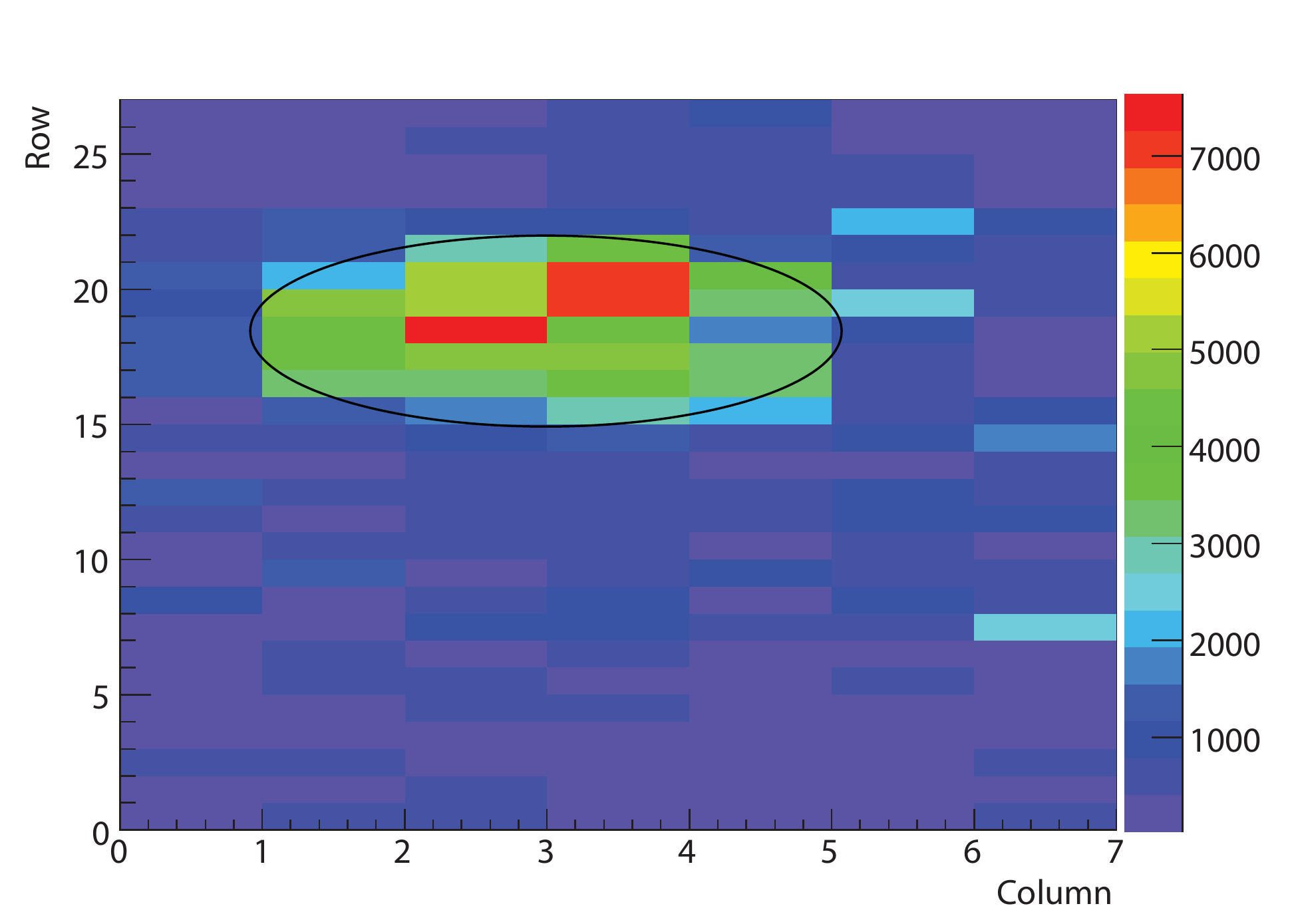}
    \caption{The BigBite shower counter hit pattern for kinematics K8, $\delta_p$ = -2$\%$. The hot region corresponds to the elastic electrons. For production data taking, only the shower blocks inside the ellipse were on.}
    \label{fig:bb_rate}
  \end{center}
\end{figure}

The HV for both the pre-shower and shower counters were
calibrated by cosmics before the experiment. Since the kinematics can be well determined
from the hadron arm for the elastic events, trajectory
information is not required on the BigBite side. Therefore, only the
shower counters were turned on during the production data taking to tag
the electrons and form the coincidence trigger. The pre-shower counter
was turned off to further reduce the background. Fig.~\ref{fig:bb_rate} is an
example of the shower rate pattern for one of the kinematics settings. The hot
region corresponds to the elastic peak on the left HRS.
\section{Hall A Data Acquisition System}
The Hall A data acquisition (DAQ) system used CODA (CEBAF On-line Data
Acquisition)~\cite{coda} developed by the Jefferson Lab Data
Acquisition Group.

CODA is a tool kit composed of a set of software and hardware packages from which
a data acquisition system can be constructed which will manage the
acquisition, monitoring and storage of data of nuclear physics
experiments. The DAQ includes front-end Fastbus and VME digitization devices
(ADCs, TDCs and scalers), the VME interface to Fastbus, single-board VME computers
running VxWorks operating system, Ethernet networks, Unix or Linux workstations, and a
mass storage tape silo (MSS) for long-term data storage. The custom
software components of CODA are:
\begin{itemize}
\item a readout controller (ROC) which runs on the front-end crates to
  facilitate the communication between CODA and the detectors.
\item an event builder (EB) which caches incoming buffers of events
  from the different controllers then merges the data streams in such
  a way that data which was taken concurrently in time appears
  together.
\item an event recorder (ER) to write the data built by EB to the
  disk.
\item an event transfer (ET) system which allows distributed access to
  the data stream from user processes and inserts additional data into
  the data stream every a few seconds from the control system.
\item a graphical user interface (Run Control) to set experimental
  configuration, control runs, and monitor CODA components.
\end{itemize}

A recorded CODA file consists the following major components:
\begin{itemize}
\item Header file including a time stamp and other run information
  like run number, pre-scale factors and event number.
\item CODA physics events from the detectors.
\item CODA scaler events: the DAQ reads the scaler values every $1 - 4$
  seconds and feeds them into the main data stream. Since counted by
  stand-alone units, the scaler values are not effected by the DAQ
  dead time; therefore, they can be used to correct DAQ dead time.
\item EPICS~\cite{epics} data from the slow control software used at JLab,
  e.g., the spectrometer magnet settings and angles, target temperature
  and pressure, etc.
\end{itemize}

\section{Trigger Setup}
In this experiment, six different types of triggers were generated and
used in the data acquisition. T1 and T3 are singles triggers from the
electron arm (BigBite) and the hadron arm (L-HRS) respectively. T4 is the left
HRS scintillator trigger used for trigger efficiency. T5 is the
coincidence trigger of T1 and T3. T7 is the BigBite cosmic trigger for testing. T8 is the EDTM
pulser trigger used to measure the trigger efficiency. The trigger system
was built from commercial CAMAC and NIM discriminators,
delay units, logic units and memory lookup units (MLU).

\subsection{Signal Arm Trigger}
\begin{figure}
  \begin{center}
    \includegraphics[angle=0, width=1.0\textwidth]{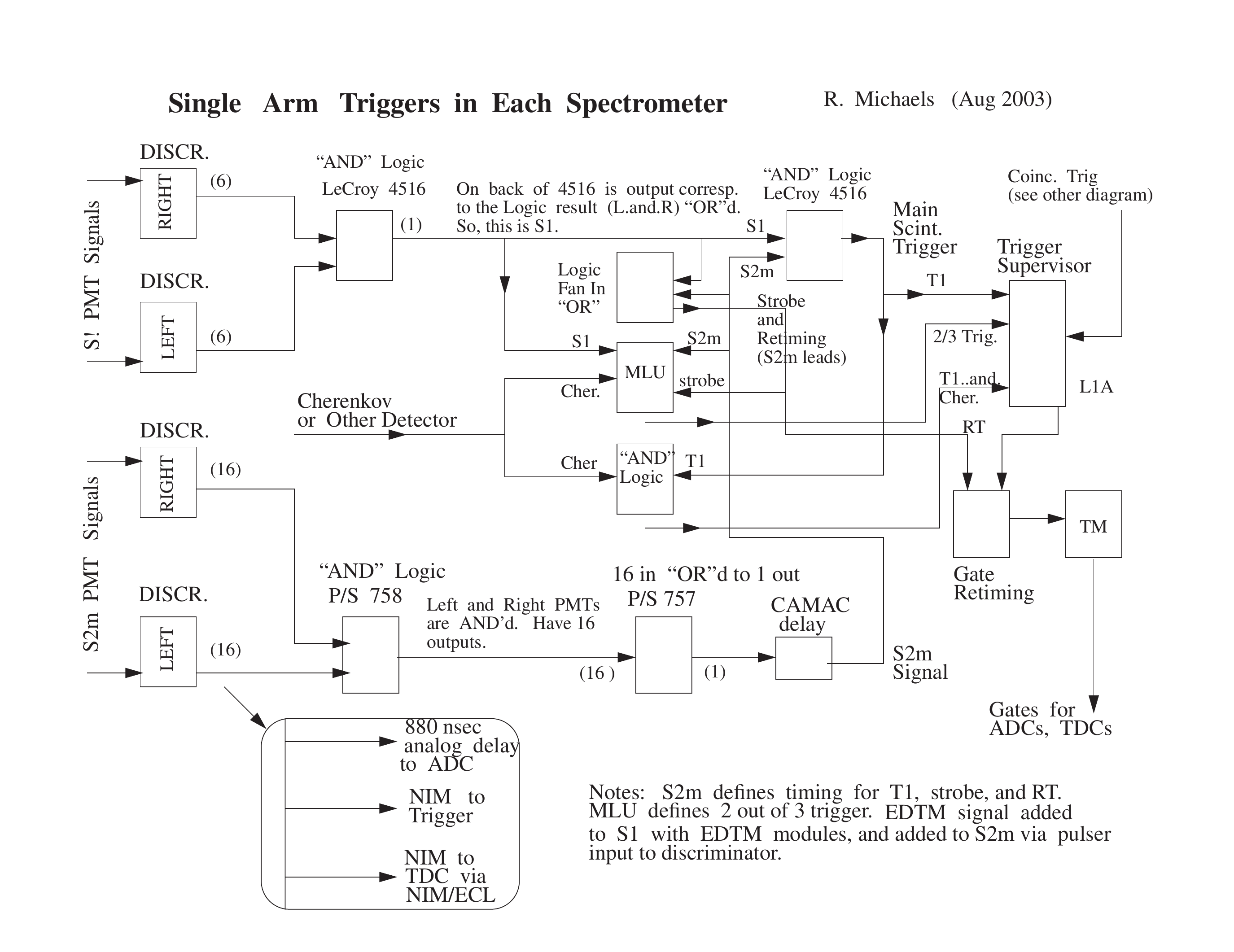}
    \caption{Left HRS single arm triggers diagram during E08-007.}
    \label{fig:strig}
  \end{center}
\end{figure}

T3 was formed by requiring that both scintillator planes S1 and
S2 have at least one fired scintillator bars (both phototubes
fired) and they are close enough to form a valid track. Thus, this main trigger requires
four fired PMTs. The T3 trigger diagram is illustrated in Fig.~\ref{fig:strig}.

\begin{figure}
  \begin{center}
    \includegraphics[angle=0,
    width=1.0\textwidth]{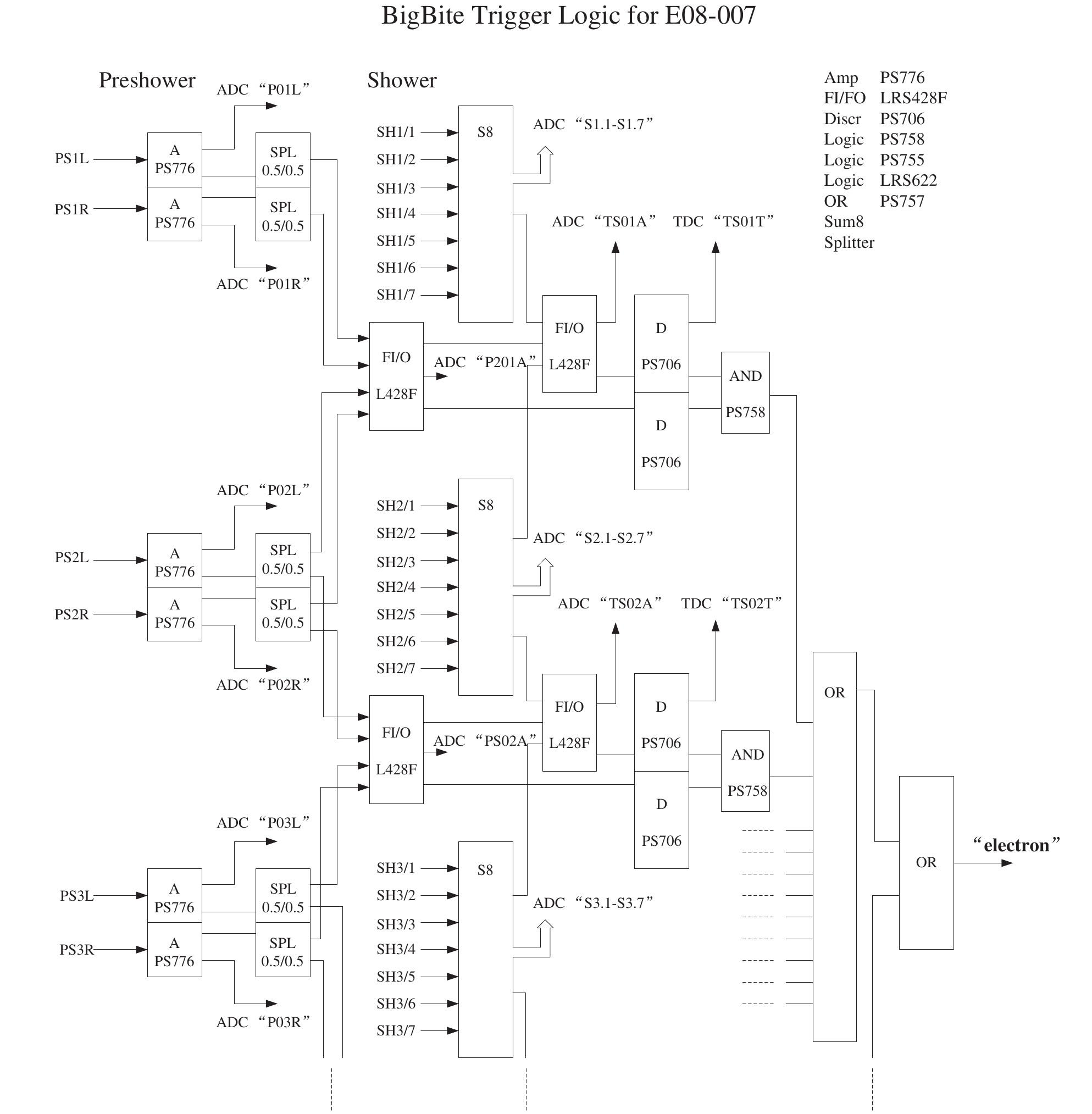}
    \caption{The BigBite trigger diagram during E08-007.}
    \label{fig:bbtrig}
  \end{center}
\end{figure}

T1 was formed by the BigBite shower total sum as illustrated in
Fig.~\ref{fig:bbtrig}. The total sum (TS) was defined as the sum of all the
pre-shower (PS) and shower (SH) ADCs of the two adjacent rows, e.g.,
\begin{eqnarray}
PS1_{sum}&=&PS1L+PS1R+PS2L+PS2R,\\
SH1_{sum}&=&SH1_1+SH1_2+\cdots +SH1_7+SH2_1+SH2_2+\cdots + SH2_7,\\
TS1& = &PS1_{sum}+SH1_{sum}.
\end{eqnarray}
The electron trigger was given by the ``OR'' of the total sum signals.

\subsection{Coincidence Trigger}
\begin{figure}
  \begin{center}
    \includegraphics[angle=0, width=1.0\textwidth]{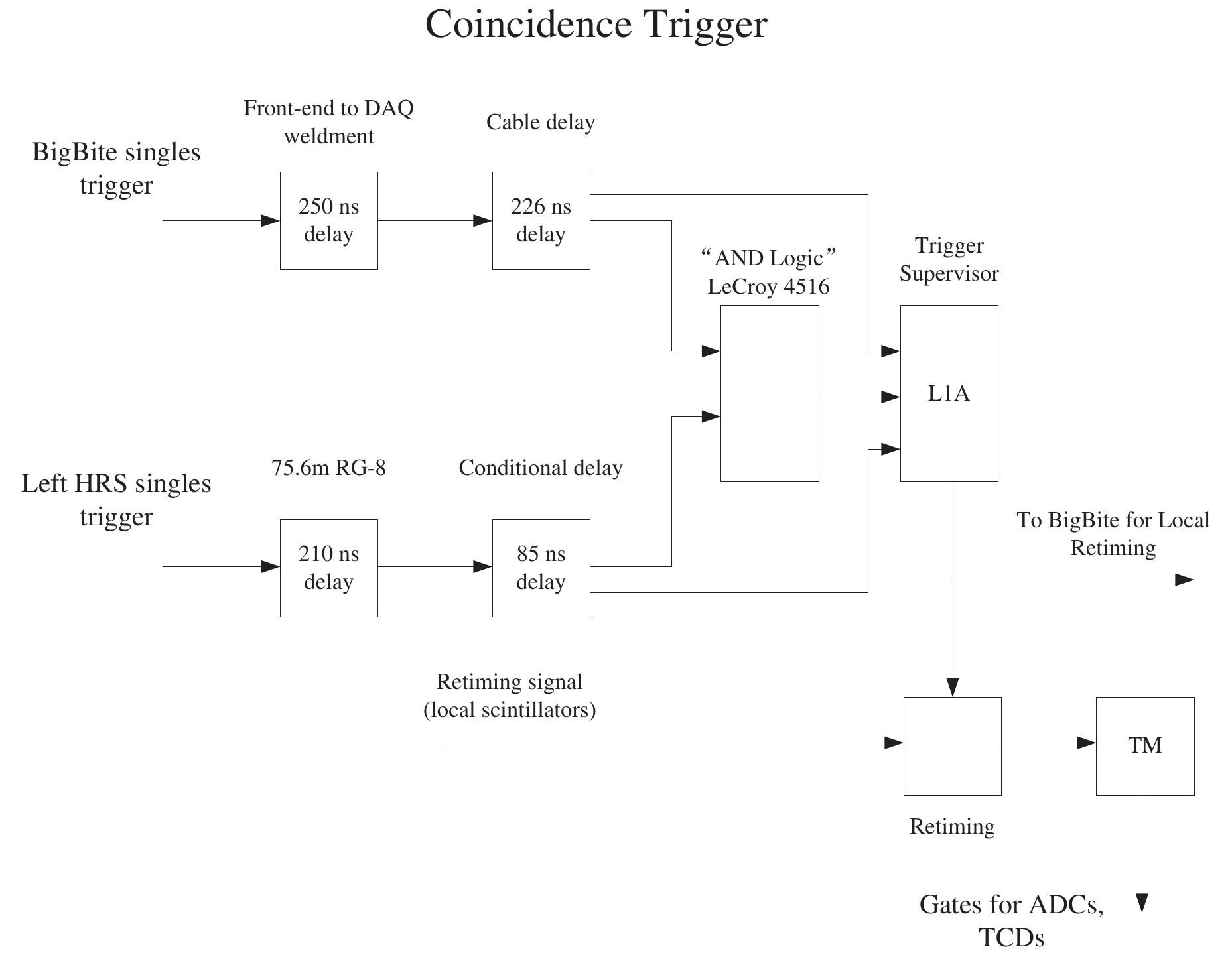}
    \caption{Coincidence trigger diagram during E08-007.}
   \label{fig:coin_trig}
  \end{center}
\end{figure}

The diagram of coincidence triggers is shown in Fig.~\ref{fig:coin_trig}.
Coincidence trigger T5 is simply an ``AND'' of T1 and T3
triggers.
\subsection{Trigger Selection}
A summary of triggers used in E08-007 is listed in Table~\ref{tab:trig_sum}.
After generated, all types of triggers have their copies sent to a scaler unit
for counting and a trigger supervisor (TS) unit to trigger data acquisition.
The TS unit has a pre-scale function. If the pre-scale factor for a specific
trigger type is $N$, then only 1 out of $N$ triggers of that type is recorded
in the data stream. This function is very useful to decrease the computer dead
time caused by frequent data recording while keeping all the events with useful
physics information. Therefore, during the production data taking, all the single
arm triggers were highly pre-scaled, and all the T5 (coincidence) trigger events
were kept in the data stream. The rates of each trigger after the pre-scale
factors are also listed in Table~\ref{tab:trig_sum}.
\begin{table}
  \begin{center}
    \begin{tabular}{|c|c|c|}
     \hline
    Trigger & Definition & Rate after pre-scale\\
      \hline
      T1 & electron arm singles (total shower sum) & $\sim$ 20 Hz\\
      \hline
      T3 & hadron arm singles (S1 AND S2)& $\sim$ 20 Hz \\
      \hline
      T4 & hadron arm efficiency (S1 OR S2) & $\sim$ 10 Hz\\
      \hline
      T5 & coincidence (T1 AND T3) & $\sim$ 2200 Hz\\
      \hline
      T8 & EDTM pulser (1024 Hz)& 10 Hz\\
      \hline
    \end{tabular}
    \caption{Trigger summary for E08-007.}
    \label{tab:trig_sum}
  \end{center}
\end{table}

%% file: Analysis.tex
\chapter{Data Analysis I}
\section{Analysis Overview}
The Hall A C++ Analyzer~\cite{ana} was used to replay the raw data and
generate the processed data files for this experiment. The Analyzer was developed
by Hall A software group and is based on ROOT~\cite{root}, a powerful
object-oriented framework that has been developed at CERN by and for
the nuclear and particle physics community. From the replayed data files, the proton form factor ratio was extracted by the weighted sums technique~\cite{Besset}.

\begin{figure}
  \begin{center}
    \includegraphics[angle=0, width=0.85\textwidth]{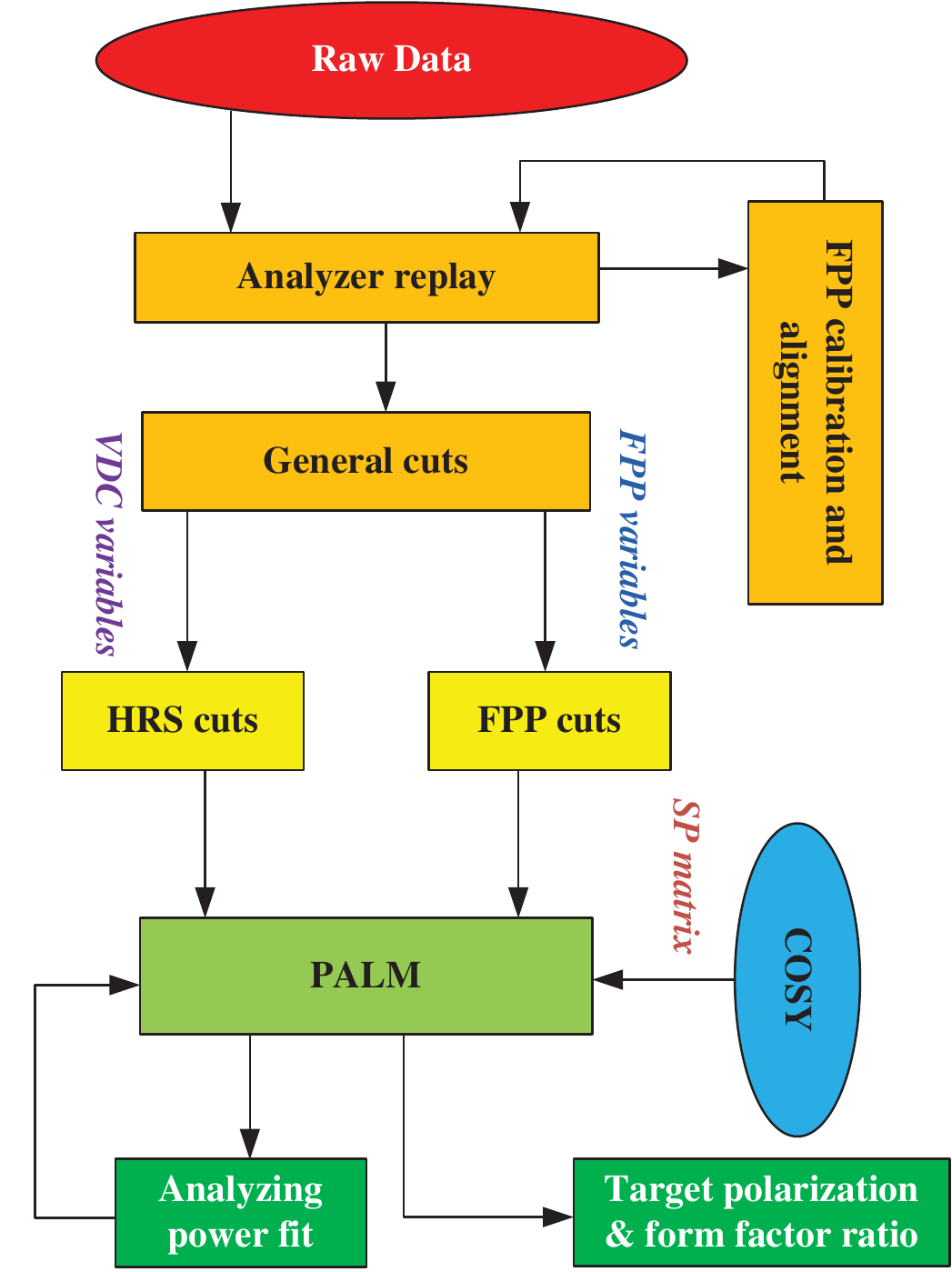}
    \caption{The flow-chart of the E08-007 analysis procedure.}
    \label{fig:flow}
  \end{center}
\end{figure}

The flow-chart of the E08-007 analysis procedure is illustrated in
Fig.~\ref{fig:flow}. The raw data recorded from the detectors were first transformed
into ntuples by the Analyzer after calibration. The recoil proton's second scattering angle was extracted from the FPP reconstruction. The spin transport matrix were
generated by COSY (a model simulating the spectrometer transport system). With
these inputs, the recoil proton polarization and hence the form factor
ratios were extracted by the main analysis code PALM~\cite{palm}.
\section{HRS analysis}
The particle trajectory at the focal plane of the left HRS is
determined by raw wire hits and drift times in the VDCs. These
trajectories are transported from the focal plane to the target using a calibrated
``optics'' matrix of the spectrometer. The reconstructed target quantities (momentum and angles)
allow for the determination of the kinematics of each event. For this experiment, these target quantities are important in another way as the inputs for the spin transport matrix calculation, which determines the recoil proton polarization at the target.
\subsection{Definition of Hall A coordinate systems}
In this section, a short overview of Hall A coordinate conventions is
presented. More details can be found in reference~\cite{optics}.
\subsubsection{Hall Coordinate System (HCS)}
\begin{figure}
  \begin{center}
    \includegraphics[angle=0, width=0.65\textwidth]{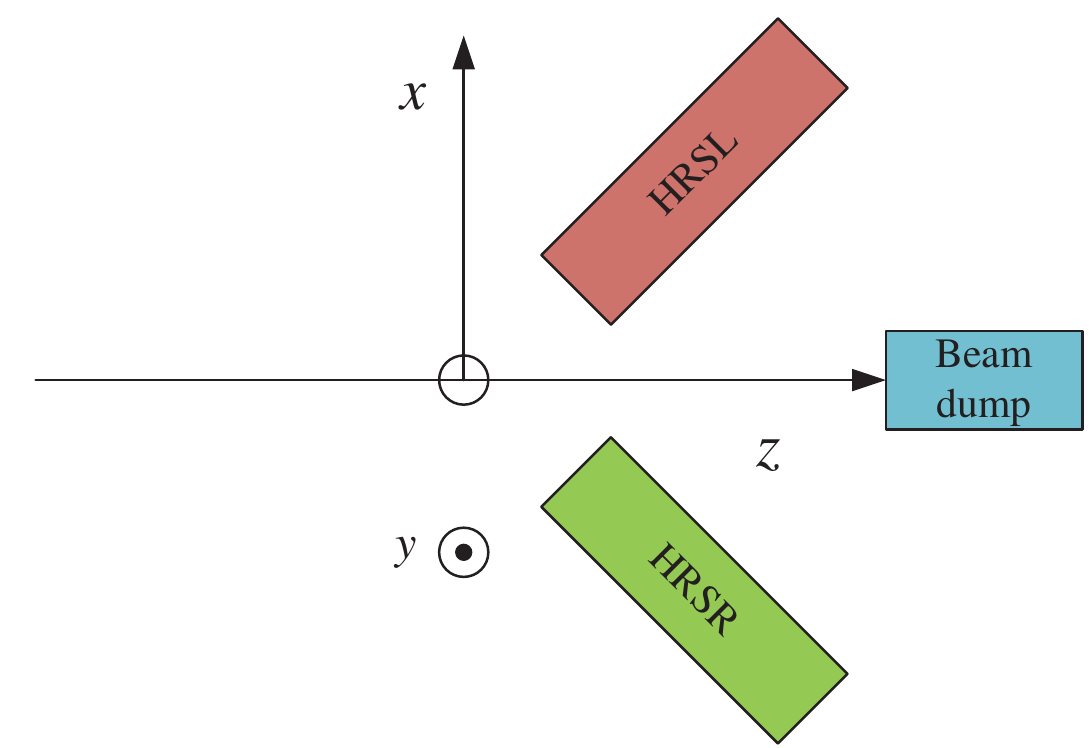}
    \caption{Hall coordinate System (top view).}
    \label{fig:hall_cor}
  \end{center}
\end{figure}

The origin of the HCS is defined by the intersection of the electron
beam and the vertical symmetry axis of the target system. $\vec z$ is
along the beam line and points in the direction of the beam dump, and
$\vec y$ is vertically up, see Fig.~\ref{fig:hall_cor}.
\subsubsection{\label{sec:tcs}Target Coordinate System (TCS)}
\begin{figure}
  \begin{center}
    \includegraphics[angle=0, width=0.75\textwidth]{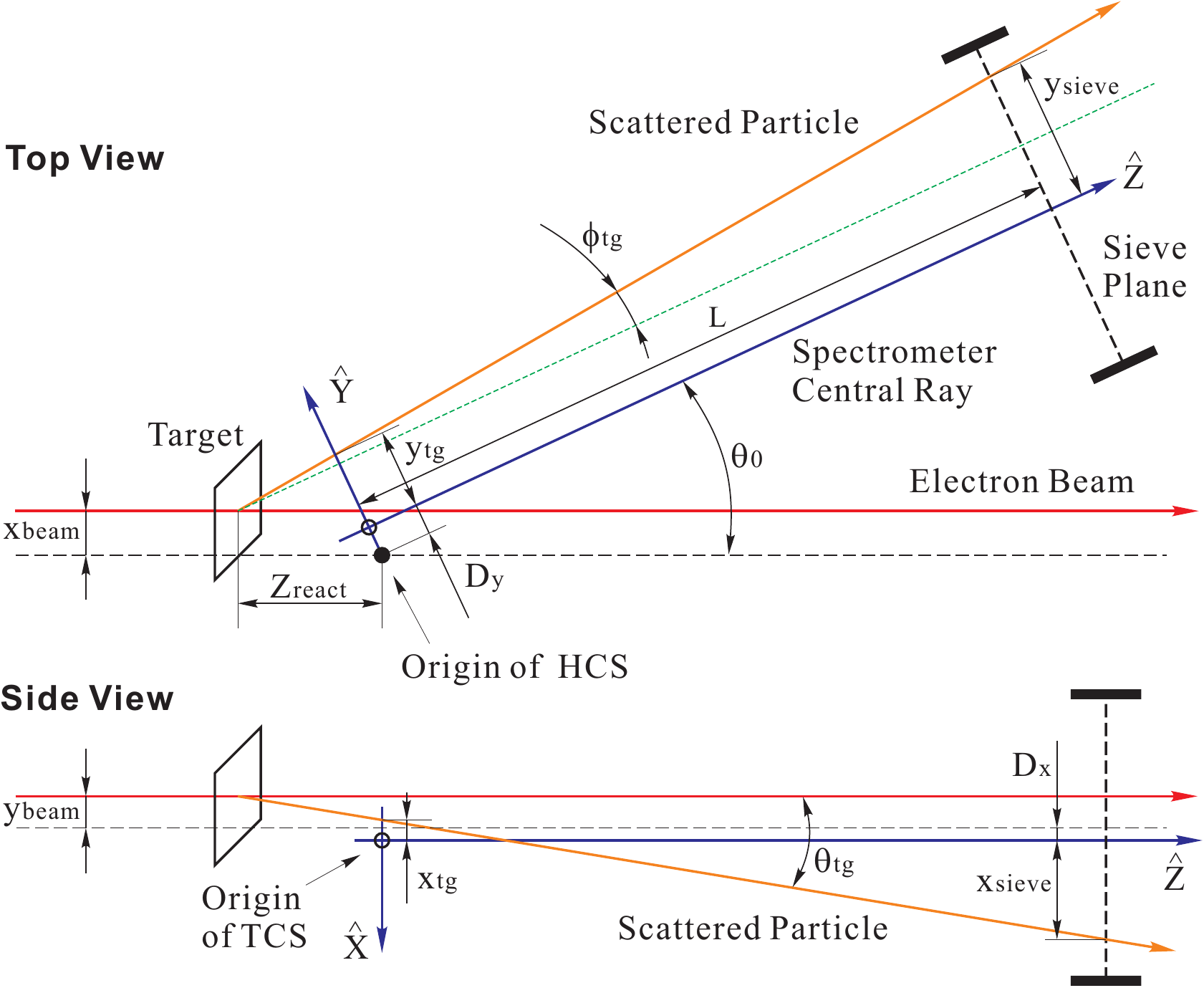}
    \caption{Target coordinate system (top and side views).}
    \label{fig:tcs}
  \end{center}
\end{figure}

The TCS is defined with respect to the central axis of the
spectrometer. A line perpendicular to the sieve slit surface of the
spectrometer and going through the midpoint of the central sieve slit
hole define the $\vec z_{\mathrm{tg}}$-axis. The $\vec y_{\mathrm{tg}}$-axis points to the
right facing the spectrometer, and $\vec x_{\mathrm{tg}}$-axis is vertically down as
illustrated in Fig.~\ref{fig:tcs}.
In the ideal case where the spectrometer is
pointing directly at the hall center and the sieve slit is perfectly
centered on the spectrometer, the TCS has the same origin as
HCS. However, it typically deviates from HCS center by $D_x$ and $D_y$
in the vertical and horizontal directions in TCS, respectively, and the offsets are given by surveys. The distance of the midpoint of the collimator from the TCS
origin is defined to be the length $L$ for the spectrometer. The
out-of-plane angel $\theta_{\mathrm{tg}}$ and the in-plane angle $\phi_{\mathrm{tg}}$
are given by the tangent of the real angle, $dx_{\mathrm{sieve}}/L$ and $dy_{\mathrm{sieve}}/L$.

The TCS variables are used to calculate the scattering angle and the
reaction point along the beam line for each event. Combined with the beam positions (measured
in the Hall coordinate system), the scattering angle and reaction
point are given by:
\begin{eqnarray}
\theta_{\mathrm{scat}}&=&arc\cos\left(\frac{\cos (\theta_0)-\phi_{tg}\sin
  (\theta_0)}{\sqrt{1+\theta_{tg}^2+\phi_{tg}^2}}\right)\\
z_{\mathrm{react}}&=&\frac{-(y_{tg}+D_y)+x_{\mathrm{beam}}(\cos (\theta_0)-\sin
  (\theta_0))}{\cos (\theta_0)\phi_{tg}+\sin (\theta_0)},
\end{eqnarray}
where $\theta_0$ denotes the spectrometer central angle. The in-plane
and out-of-plane angles can be determined using sieve hole positions:
\begin{eqnarray}
\phi_{\mathrm{tg}}&=&\frac{y_{\mathrm{sieve}}+D_y-x_{\mathrm{beam}}\cos
  (\theta_0)+z_{\mathrm{react}}\sin (\theta_0)}{L-z_{\mathrm{react}}\cos
  (\theta_0)-x_{\mathrm{beam}}\sin (\theta_0)}\\
\theta_{\mathrm{tg}}&=&\frac{x_{\mathrm{sieve}}+D_x+y_{\mathrm{beam}}}{L-z_{\mathrm{react}}\cos
(\theta_0)-x_{\mathrm{beam}}\sin (\theta_0)}
\end{eqnarray}
and the position at the target is given by:
\begin{eqnarray}
y_{\mathrm{tg}}&=&y_{\mathrm{sieve}}-L\phi_{\mathrm{tg}}\\
x_{\mathrm{tg}}&=&x_{\mathrm{sieve}}-L\theta_{\mathrm{tg}}.
\end{eqnarray}

\subsubsection{Detector Coordinate System (DCS)}
\begin{figure}
  \begin{center}
    \includegraphics[angle=0, width=0.75\textwidth]{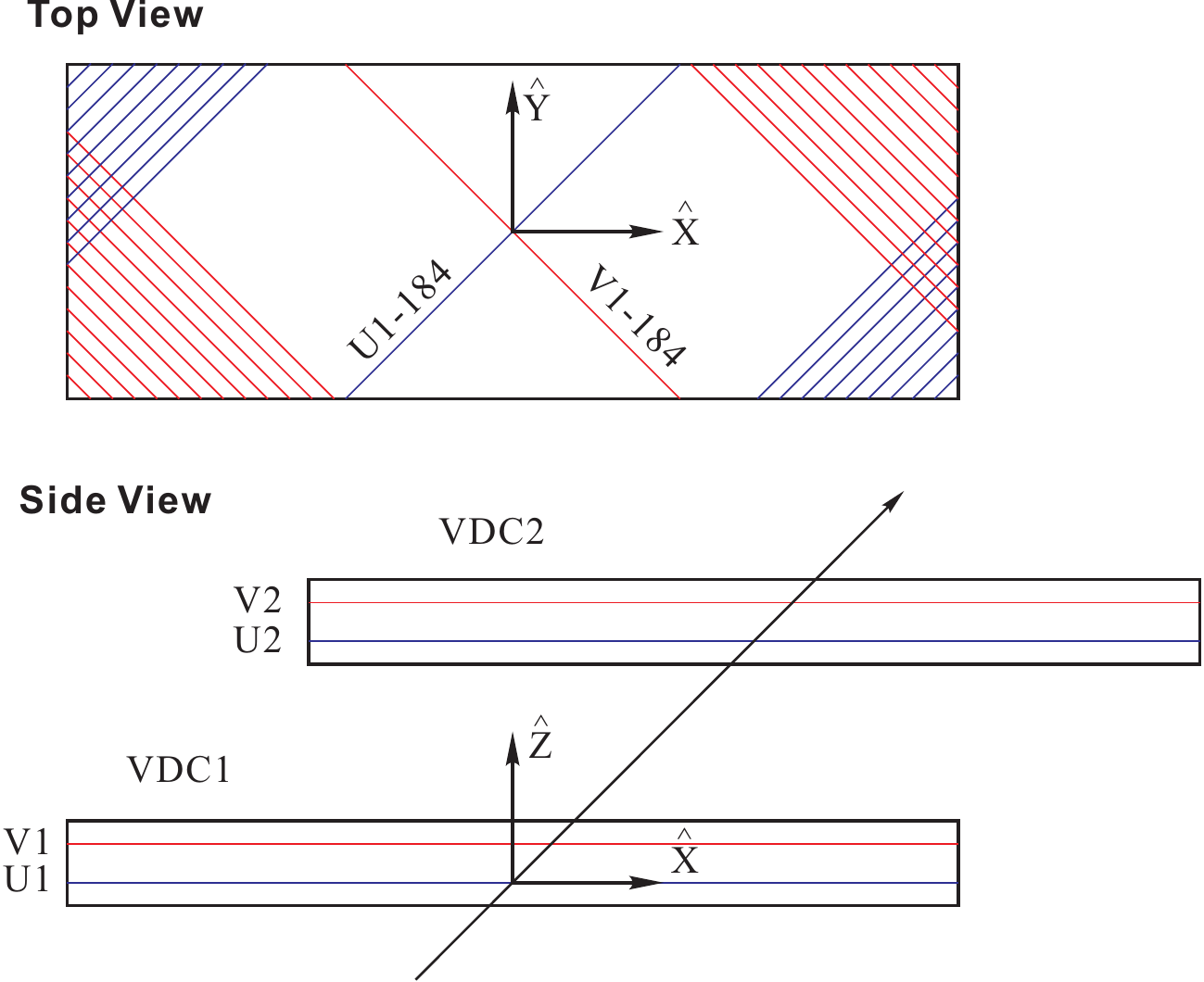}
    \caption{Detector coordinate system (top and side views).}
    \label{fig:dcs}
  \end{center}
\end{figure}

The Detector Coordinate System (DCS) is defined by the positions of the
VDC planes. The intersection of wire 184 of the VDC1 U1 plane and the
perpendicular projection of wire 184 in the VDC1 V1 plane onto the VDC
U1 plane defines the origin of the DCS. $\vec z$ is perpendicular to
the VDC planes pointing vertically up, $\vec x$ is along the long
symmetry axis of the lower VDC pointing away from the hall center
(see Fig.~\ref{fig:dcs}).
Using the trajectory intersection points $p_n$ (where $n=$ U1, V1, U2,
V2) with the four VDC planes, the coordinates of the detector vertex
can be calculated from the following expressions:
\begin{eqnarray}
\tan (\eta_1)& = & \frac{p_{\mathrm{U2}}-p_{\mathrm{U1}}}{d_2}\\
\tan (\eta_2)&= & \frac{p_{\mathrm{V2}}-p_{\mathrm{V1}}}{d_2}\\
\theta_{\mathrm{det}}&=& \frac{1}{\sqrt{2}}(\tan (\eta_1)+\tan
(\eta_2))\\
\phi_{\mathrm{det}}&=&\frac{1}{\sqrt{2}}(-\tan (\eta_1)+\tan
(\eta_2))\\
x_{\mathrm{det}}&=&\frac{1}{\sqrt{2}}(p_{\mathrm{U1}}+p_{\mathrm{V1}}-d_1\tan
(\eta_2))\\
y_{\mathrm{det}}&=&\frac{1}{\sqrt{2}}(p_{\mathrm{U1}}+p_{\mathrm{V1}}-d_1\tan
(\eta_2))
\end{eqnarray}
where $d_1=0.115$ m is the distance between the U and V planes in both
chambers, and $d_2=0.335$ m is the distance between the two planes.

\subsubsection{\label{sec:trcs}Transport Coordinate System (TRCS)}
\begin{figure}
  \begin{center}
    \includegraphics[angle=0, width=0.75\textwidth]{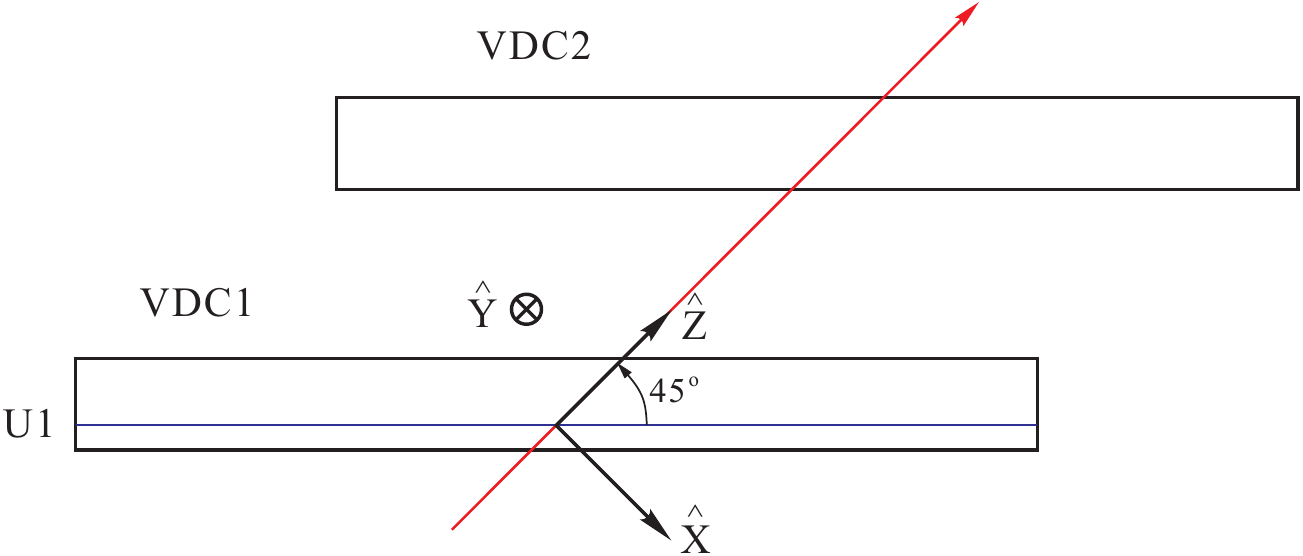}
    \caption{Transport coordinate system.}
    \label{fig:trcs}
  \end{center}
\end{figure}

The TRCS at the focal plane is generated by rotating the DCS clockwise
around its $y$-axis by $45^{\circ}$. It's typically used as a intermediate position
state from DCS to the FCS (focal plane coordinate system), which will
be described in the next section; the bending angle related to
the spin transport can also be calculated from the difference of the out-of-plane angles $(\theta_{\mathrm{tg}}-\theta_{\mathrm{tr}})$ between the
TCS and TRCS. The transport coordinates can be expressed in terms of
the detector coordinates as follows:
\begin{eqnarray}
\theta_{\mathrm{tr}}&=&\frac{\theta_{\mathrm{det}}+\tan
  (\rho_0)}{1-\theta_{\mathrm{det}}\tan (\rho_0)}\\
\phi_{\mathrm{tr}}&=& \frac{\phi_{\mathrm{det}}}{\cos
  (\rho_0)-\theta_{\mathrm{det}}\sin (\rho_0)}\\
x_{\mathrm{tr}}&=& x_{\mathrm{det}}\cos
(\rho_0)(1+\theta_{\mathrm{tr}}\tan (\rho_0))\\
y_{\mathrm{tr}}&=& y_{\mathrm{det}}+\sin
(\rho_0)\phi_{\mathrm{tr}}x_{\mathrm{det}},
\end{eqnarray}
where $\rho_0=-45^{\circ}$ is the rotation angle, see Fig.~\ref{fig:trcs}.

\subsubsection{Focal Plane Coordinate System (FCS)}
\begin{figure}
  \begin{center}
    \includegraphics[angle=0, width=0.75\textwidth]{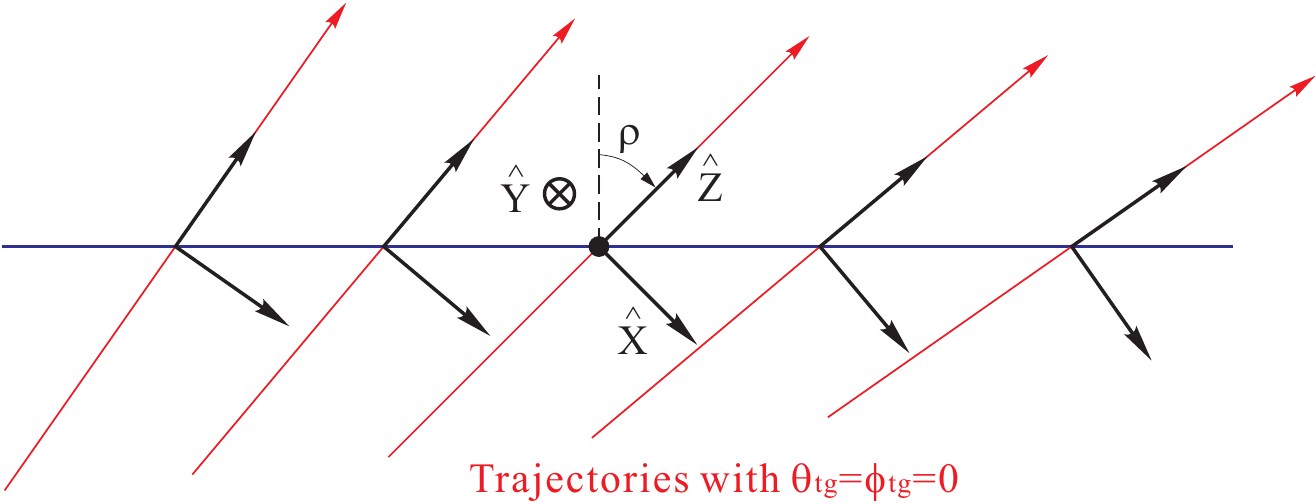}
    \caption{Rotated focal plane coordinate system.}
    \label{fig:fcs}
  \end{center}
\end{figure}

The focal plane coordinate system (FCS) chosen for the HRS analysis is
a rotated coordinate system. Because of the focusing of the HRS magnet
system, particles from different scattering angles with the same
momentum will be focused at the focal plane. Therefore, the relative
momentum from the central momentum of the spectrometer, which is selected by the
HRS dipole magnet field setting,
\begin{equation}
\delta=\frac{\Delta p}{p_0}=\frac{p-p_0}{p_0},
\end{equation}
is approximately only a function of $x_{\mathrm{tr}}$, and $p_0$ in
the formula stands for the central momentum setting of the HRS.
The FCS is obtained by rotating the DCS around its $y$-axis by an
varying angle $\rho~(x_{\mathrm{tr}})$ to have the new $z$-axis
parallel to the local central ray, which has the scattering angle
$\theta_{\mathrm{tg}}=\phi_{\mathrm{tg}}=0$ for the corresponding
$\delta$ at position $x_{\mathrm{tr}}$ (see Fig.~\ref{fig:fcs}). In this rotated coordinate
system, the dispersive angle $\theta_{\mathrm{fp}}$ is small for all the points
across the focal plane, and the distribute is approximately symmetric with respect to
$\theta_{\mathrm{fp}}=0$. This symmetry greatly simplifies further optics optimization.

With proper systematic offsets added, the coordinates of focal plane
vertex can be written as follows:
\begin{eqnarray}
x_{\mathrm{fp}}&=&x_{\mathrm{tr}}\\
\tan (\rho)&=&\sum t_{i000}x_{\mathrm{fp}}^i\\
y_{\mathrm{fp}}&=& y_{\mathrm{tra}}-\sum y_{i000}x_{\mathrm{fp}}^i\\
\theta_{\mathrm{fp}}&=&\frac{x_{\mathrm{det}}+\tan
  (\rho)}{1-\theta_{\mathrm{det}}\tan (\rho)}\\
\phi_{\mathrm{fp}}&=&\frac{\phi_{\mathrm{det}}-\sum
  p_{i000}x_{\mathrm{fp}}^i}{\cos (\rho_0)-\theta_{\mathrm{det}}\sin (\rho_0)}.
\end{eqnarray}
The coordinate transformation is not unitary and we have $x_{\mathrm{fp}}$ equal to
$x_{\mathrm{tr}}$ for simplicity.

\subsection{Target Variables Reconstruction}
For each event, two angular coordinates ($\theta_{\mathrm{det}}$ and
$\phi_{\mathrm{det}}$) and two spatial coordinates ($x_{\mathrm{det}}$
and $y_{\mathrm{det}}$) are measured at the focal plane detectors. The
position of the particle and the tangent of the angle made by its
trajectory along the dispersive direction are given by
$x_{\mathrm{det}}$ and $\theta_{\mathrm{det}}$, while
$y_{\mathrm{det}}$ and $\phi_{\mathrm{det}}$ give the position and
tangent of the angle perpendicular to the dispersive direction. These
variables are corrected for any detector offsets from the ideal
central ray of the spectrometer to obtain the focal plane coordinates
$x_{\mathrm{fp}},~\theta_{\mathrm{fp}},~y_{\mathrm{fp}}$ and
$\phi_{\mathrm{fp}}$. The focal plane observables are used to
reconstruct the variables in the target system by matrix inversion.

The first order  optics matrix can be expressed as,
\begin{equation}
\left( \begin{array}{c}
\delta \\
\theta \\
y\\
\phi
\end{array} \right)_{\mathrm{tg}}
=\left( \begin{array}{cccc}
<\delta|x> & <\delta|\theta> & 0 & 0 \\
<\theta|x> & <\delta|\theta> & 0 & 0 \\
0 & 0 & <y|y> & <y|\phi>\\
0 & 0 & <\phi|y> & <\phi|\phi>
\end{array} \right)\cdot
\left( \begin{array}{c}
x \\
\theta \\
y \\
\phi
\end{array} \right)_{\mathrm{fp}}.
\end{equation}
The null tensor elements result from the mid-plane symmetry of the
spectrometer. In practice, the expansion of the focal plane
coordinates is performed up to the fifth order. A set of tensors
$D_{jkl},T_{jkl},Y_{jkl}$ and $P_{jkl}$ relates the focal plane
coordinates to the target coordinates according to~\cite{berz}
\begin{eqnarray}
\delta &=& \sum_{jkl}
D_{jkl}\theta_{\mathrm{fp}}^jy_{\mathrm{fp}}^k\phi_{\mathrm{fp}}^l\\
\theta_{\mathrm{tg}} &=& \sum_{jkl}
T_{jkl}\theta_{\mathrm{fp}}^jy_{\mathrm{fp}}^k\phi_{\mathrm{fp}}^l\\
y_{\mathrm{tg}} &=& \sum_{jkl}
Y_{jkl}\theta_{\mathrm{fp}}^jy_{\mathrm{fp}}^k\phi_{\mathrm{fp}}^l\\
\phi_{\mathrm{tg}} &=& \sum_{jkl}
P_{jkl}\theta_{\mathrm{fp}}^jy_{\mathrm{fp}}^k\phi_{\mathrm{fp}}^l,
\end{eqnarray}
where the tensors $D_{jkl},T_{jkl}$ and $P_{jkl}$ are polynomials in
$x_{\mathrm{fp}}$. For example,
\begin{equation}
D_{jkl}=\sum_{i=0}^mC_{ijkl}^Dx_{\mathrm{fp}}^i.
\end{equation}
The optics matrix used in this experiment was optimized for the Transversity~\cite{transversity} experiment. The core of the optimization program is the TMinuit package of
ROOT~\cite{root}. This package varies the optics matrix parameters to
minimize the variance $\sigma^2$ of the reconstructed data from their
actual values.

\subsection{Focal Plane Polarimeter Reconstruction}
As the key instrument to measure the recoil proton polarization, the FPP
reconstructs the second scattering angles of the proton in the
analyzer. There are basically four steps: identifying the wires that
have fired, calculating the drift distances, reconstructing the tracks
in the front and rear chambers, and determining the second scattering
angles. All the steps are done in the Analyzer program by
incorporating the FPP tracking library.

\subsubsection{Demultiplexing}
\begin{figure}
  \begin{center}
    \includegraphics[angle=0, width=0.6\textwidth]{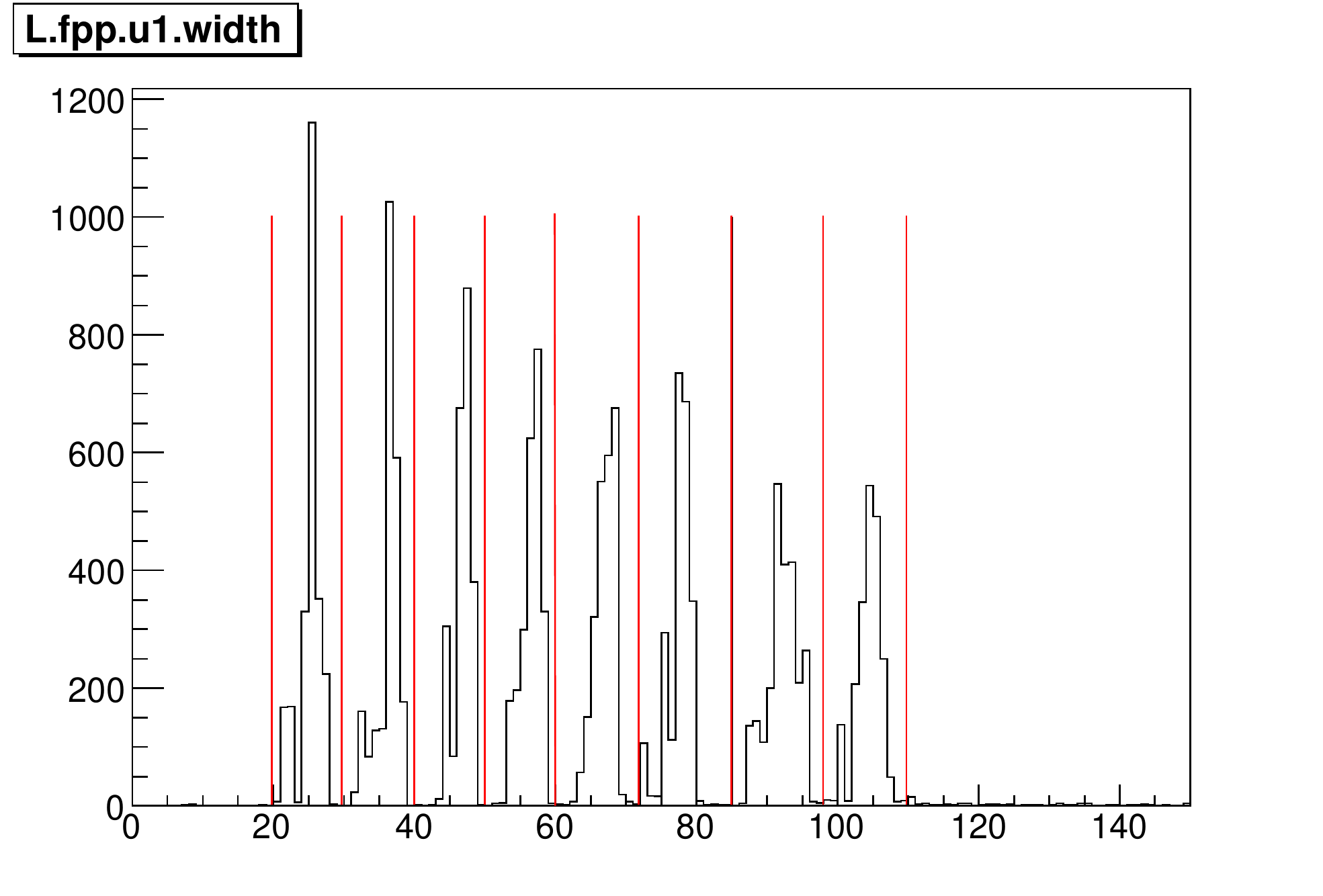}
    \caption{The TDC width of the u1 wire group and the demultiplexing cut.}
    \label{fig:demulx}
  \end{center}
\end{figure}

As noted in Section~\ref{sec:fpp}, the signals from the sense wires are
multiplexed in groups of eight to decrease the number of TDCs. By
assigning a different pulse width to each straw of the group, one can
make a cut to identify which wire fired. Fig.~\ref{fig:demulx} is an example of
the raw pulse width spectrum from one wire group. Then the signal has to be
demultiplexed in the analysis. The straw group, the leading edge and
the trailing edge of the TDC signal are fed into Analyzer, which
calculates two time differences: the difference between the trigger
signal (common stop) and the leading edge gives the drift time, while
the difference between the leading edge and the trailing edge
identifies which straw fired in the group.

Once the drift time for each wire that fired has been determined, one
can convert it into the drift distance; and hence, the tracks can be
reconstructed in the chambers. First, an offset is applied to the drift
time spectrum, to correct for various delays in the electronics. Except
when the event passes very close to the anode wire, the drift distance
is proportional to the drift time.

When the particle approaches the anode wire, the electric field becomes strong
enough for secondary ionization, which starts an
avalanche. In this region, the drift velocity increases near the sense
wire, and the drift distance $d$ is obtained from a fifth-order
polynomial in drift time $t$:
\begin{equation}
d=\sum_{n=0}^5T(j,n)t^n,
\end{equation}
where $T(j,n)$ are obtained from fitting the integrated drift time
spectra for a plane $j$. These coefficients were all
re-calibrated for this experiment. More details of the
FPP calibration can be found in~\cite{fpp_cal}.

\subsubsection{Track Reconstruction}
Using the FPP library in the Analyzer, the raw data were replayed and
the tracks were reconstructed in the straw chambers. The front and rear chambers were
analyzed separately to produce both a rear and front track. For each set
of chambers, the $u$ and $v$ directions are also analyzed
separately. The $x$ planes in chamber 3 were not used\footnote{The original design of the $x$ plane is to provide additional information of the out-of-plane position, but it was found later that the $u$ and $v$ planes were sufficient.}.

\begin{figure}
  \begin{center}
    \includegraphics[angle=0, width=0.8\textwidth]{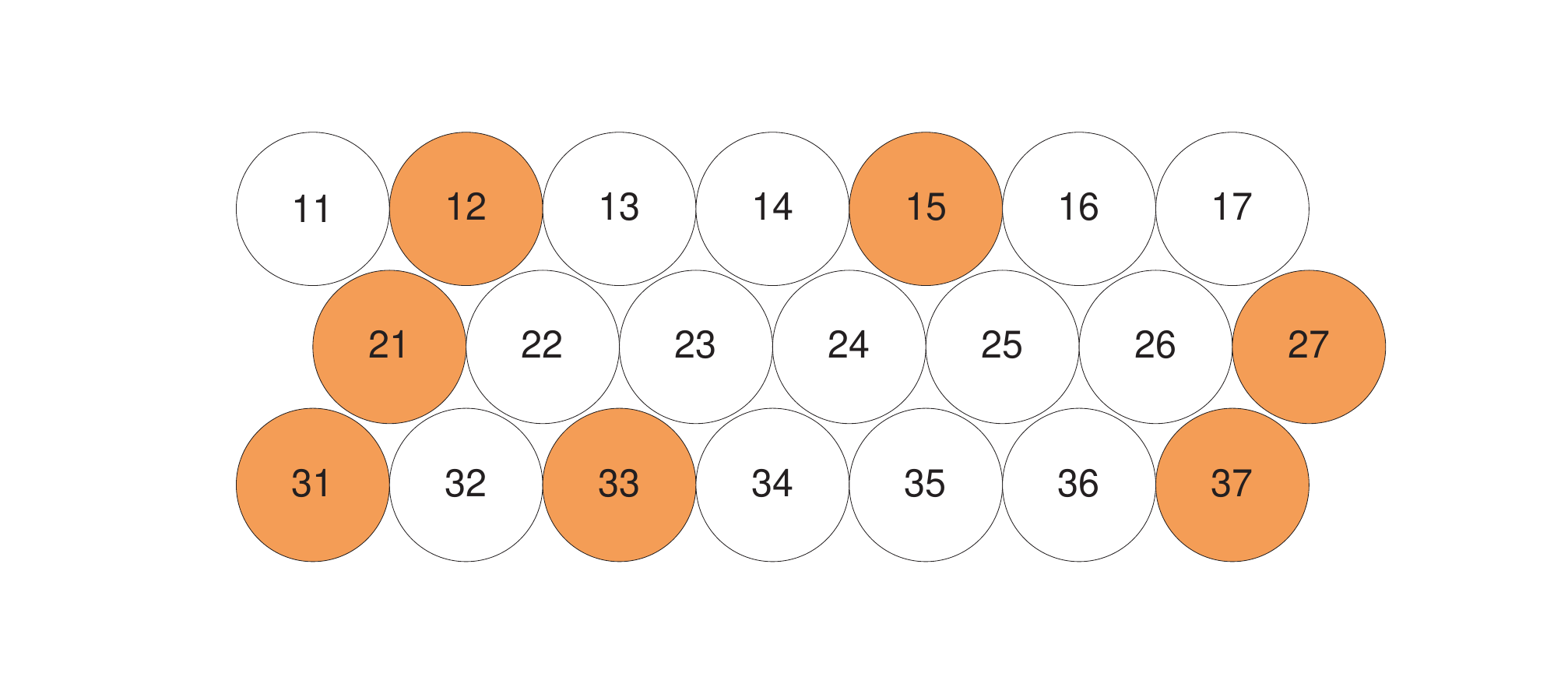}
    \caption{Illustration of the procedure to find clusters in a FPP
      chamber. The three layers represent the three planes, and the
      circles are cross-sectional cuts of the straws. The filled circles represent the fired straws.}
    \label{fig:fired}
  \end{center}
\end{figure}

The first step is to identify hit clusters in the sets of $u$ planes of each
chamber. In this set, a cluster can have at most one hit per
plane. The code searches for a track by looking at the adjacent straws
first. In Fig.~\ref{fig:fired}, the colored circles stands for the fired
straws. The code looks at the top plane and finds a hit in $S_{12}$,
then it looks in the second plane at the straws adjacent to
$S_{12}$. It finds that $S_{21}$ fired, then $S_{12}$ and $S_{21}$
start to form a cluster. When it looks further to the third plane, at
straws that are adjacent to $S_{21}$ or $S_{22}$ which are both
adjacent to $S_{12}$ even though $S_{22}$ didn't fire. It finds
$S_{31}$ and forms the first cluster ($S_{12}\to S_{21} \to S_{31}$),
$S_{33}$ also fired and forms another cluster ($S_{12}\to S_{22} \to
S_{33}$). The area around $S_{12}$ is now all scanned, so the code
starts looking at the rest of the first plane. It finds $S_{15}$, and
finds nothing else in this cluster on the next planes. When the entire
first plane has been scanned, it goes to the second plane. A hit is
found at $S_{27}$, which forms a cluster with $S_{37}$. When looking
at the third plane, no hit is found that is not already included in a
cluster so the procedure is complete. As a results, the code has found a total of
four clusters: ($S_{12}\to S_{21} \to S_{31}$), ($S_{12}\to S_{22} \to
S_{33}$), ($S_{15}$), ($S_{27}\to S_{37}$).

\begin{figure}
  \begin{center}
    \includegraphics[angle=0, width=0.65\textwidth]{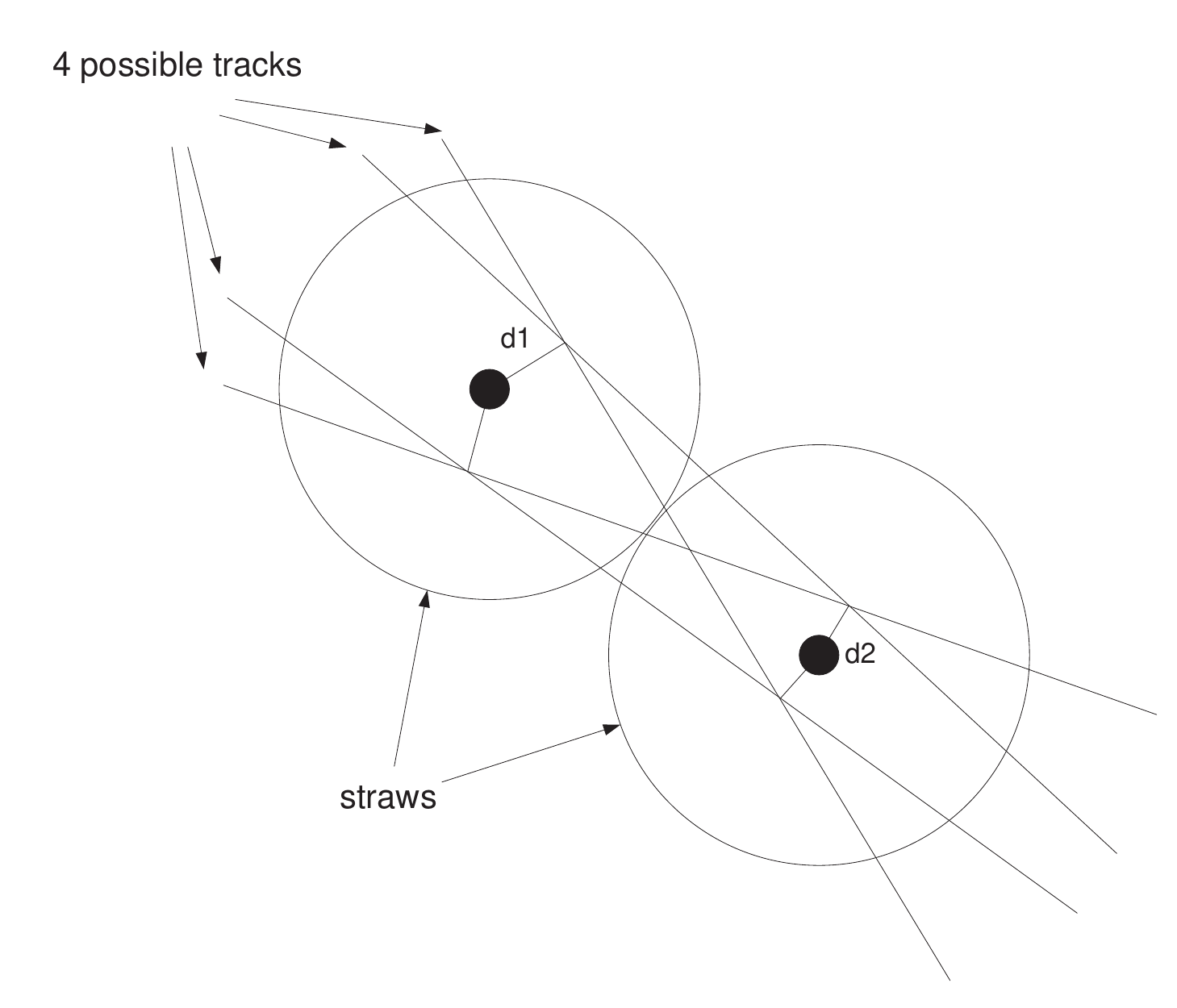}
    \caption{4 possible tracks for two given fired straws with given
      drift distances $d_1$ and $d_2$. The good track is the one
      with the lowest $\chi^2$ when taking into account all planes
    of all chambers.}
    \label{fig:tracks}
  \end{center}
\end{figure}

The same procedures are applied to the second chamber. All
combinations of pairs of clusters in both chambers are considered. For
each combination, several tracks are reconstructed. From the
drift distance, the track can be passing left or
right of the sense wire of every fired straw, therefore, there are 4
track possibilities with two given drift distance, as illustrated in
Fig.~\ref{fig:tracks}. Straight lines are then fitted, and a $\chi^2$ for each
possible trajectory is calculated. Since it is easier for a cluster
with very few hits to give a very good $\chi^2$, a weight is give to
the $\chi^2$ corresponding to the number of hits for the track. The
track with the lowest $\chi^2$ is then considered as the good
track. The procedure is repeated for the $v$ direction.

\subsubsection{\label{sec:align}Chamber Alignment}
In order to determine the proton scattering angle in the carbon analyzer, the
positions of the chambers have to be well known so that
the second scattering angles $\phi_{fpp}$ and $\theta_{fpp} $ are correctly reconstructed.
To achieve the precision of $\Delta\phi_{fpp}\sim\Delta\theta_{fpp}\sim1$ mrad, a software alignment was applied.
This procedure is crucial for two reasons. First, what we measured is the phase shift of the
azimuthal angle $\phi_{fpp}$, therefore, any rotation between the front chambers and the rear chambers
will directly shift $\phi_{fpp}$. Second, what we really care about is the proton polarization at the target;
therefore, the FPP front and rear chambers have to be aligned with respect to a
well known coordinate system so that the second scattering angle is calibrated referring to that
coordinates system and can easily be related to the target frame. As described
in Section~\ref{sec:trcs}, the transport coordinate system (TRCS) defined by the VDCs is
a convenient choice. By taking ``straight-through'' data
with the carbon door open, the trajectory determined by the FPP should
coincide with the trajectory reconstructed by the VDCs after alignment has been completed.

There are two different methods to do the alignment. In the first approach, a
software procedure is applied to fit the alignment offset parameters
$u_0$, $v_0$, $z_0$, and the rotation angles $\theta_{zu}$,
$\theta_{zv}$, $\theta_{uv}$, by minimizing the trajectory difference
between the VDCs and the FPP. The advantage of this method is the
direct link between the alignment parameters and the physical
offsets of the chambers.

For this experiment, the alignment
procedure was done by the second approach, which was developed for experiment
E93-049~\cite{rikkie}. Compared to the chamber alignment approach mentioned above,
this method directly applies a correction to the reconstructed track instead of the individual chambers.
It first aligns the front chamber track with respect to the VDC track,
and then the rear chamber track is aligned with respect to the well aligned front chamber track.
The alignment parameters obtained with this method are not easily
related to the physical offsets or rotations, but the extension to higher
order corrections is straight forward. The detailed alignment algorithm is presented in
Appendix B. For high precision measurements, the
previous experiment analysis~\cite{rikkie} showed that using the second
method by extending the corrections with higher order terms can achieve better results.

\begin{figure}
  \begin{center}
    \includegraphics[angle=0, width=0.85\textwidth]{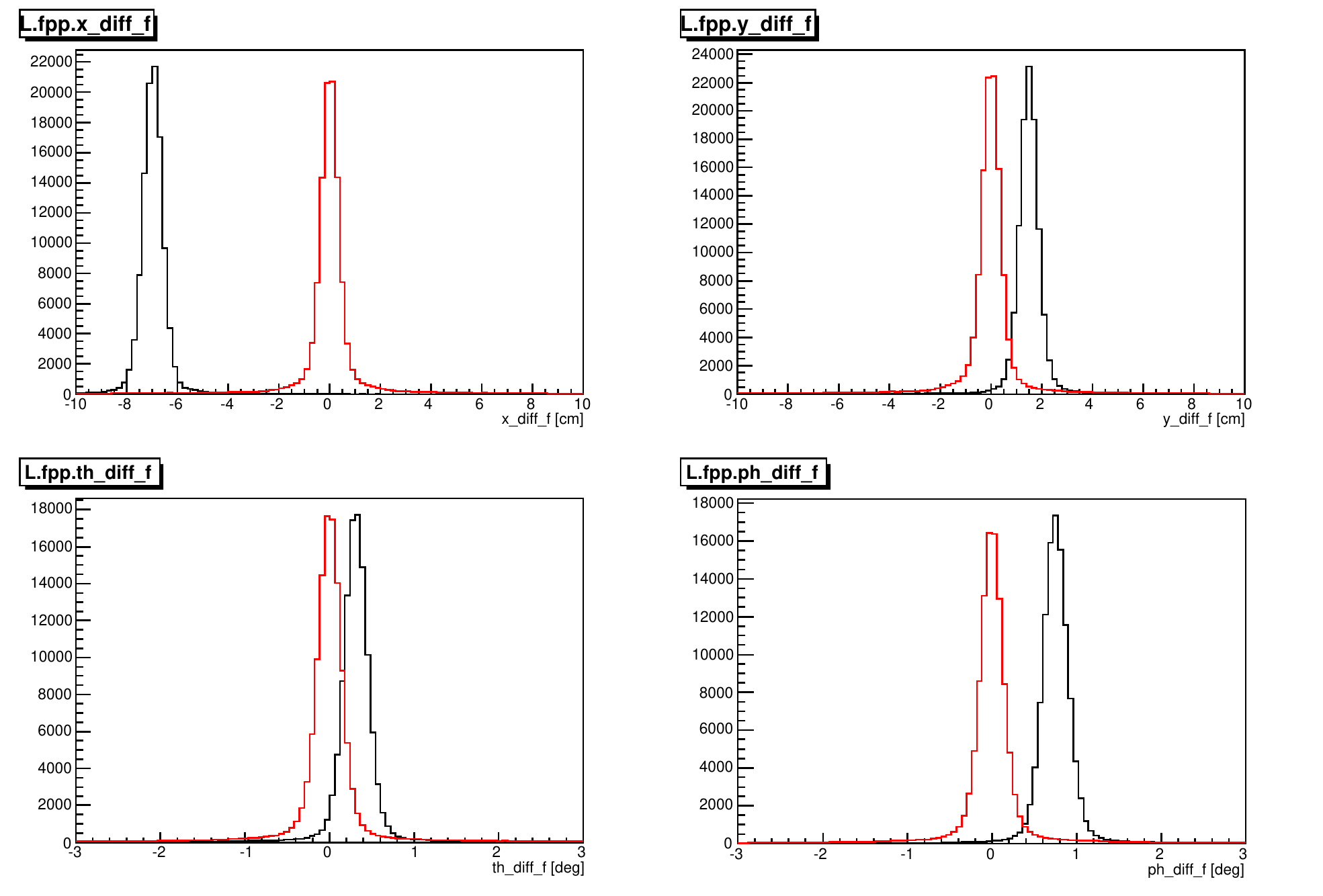}
    \caption{The difference between the VDC track and the FPP front
      track before (in black) and after (in red) the chamber alignment.
      The difference is centered at 0 after the alignment.}
    \label{fig:align_0}
  \end{center}
\end{figure}

The ``straight-through'' data
(electron) was taken during experiment E04-007~\cite{pi0}, which ran just before
this experiment\footnote{The FPP chambers were installed before experiment E04-007 took data
and were not touched until this experiment was finished.}. The histograms of the track
difference ($x_{diff}$, $y_{diff}$, $\theta_{diff}$, $\phi_{diff}$) between
the VDCs and the FPP front chambers before (black) and after (red) the software
alignment are shown in Fig.~\ref{fig:align_0}.
As one can see, the differences are well centered at 0 after the alignment.

\begin{figure}
  \begin{center}
    \includegraphics[angle=0, width=0.45\textwidth]{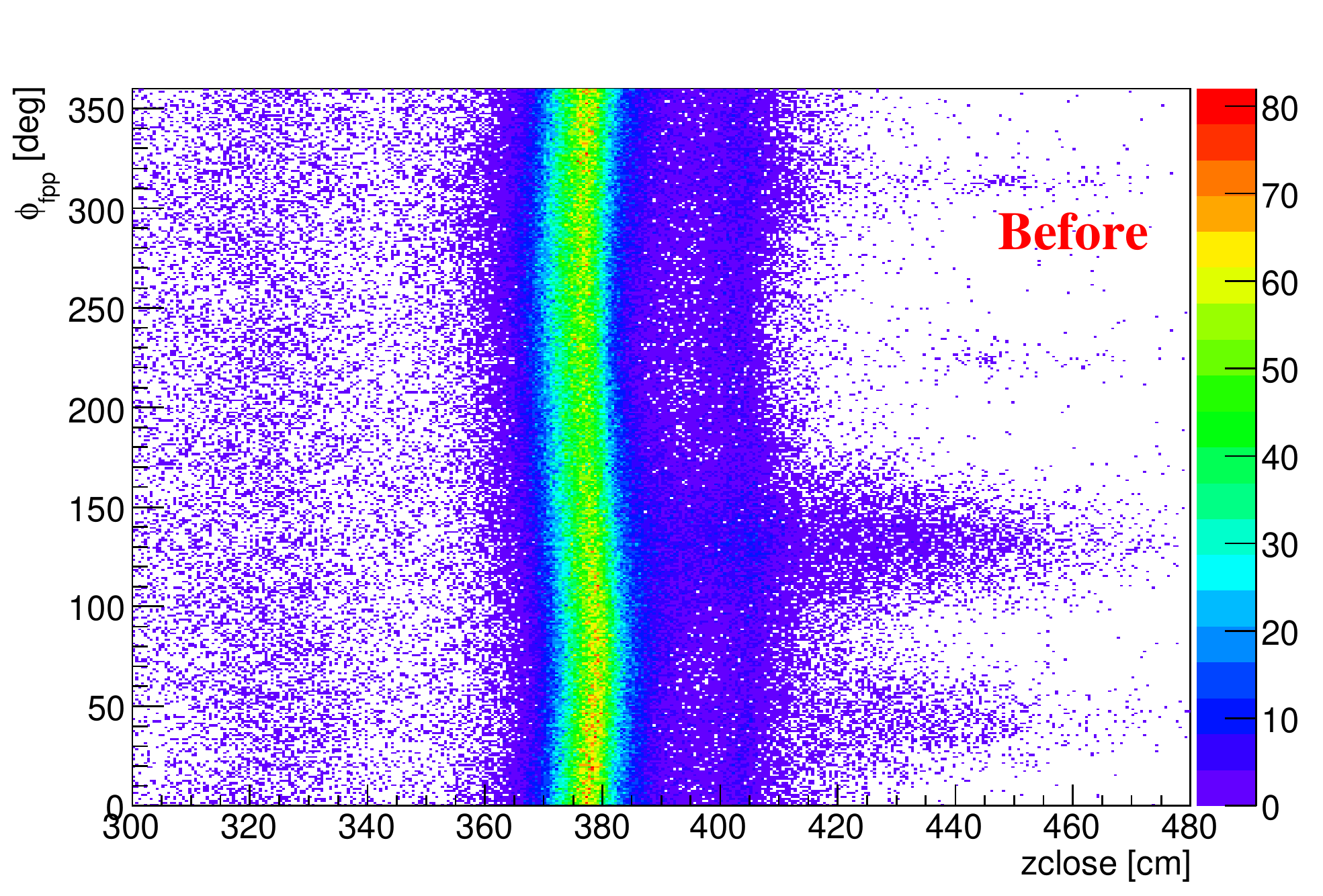}
    \includegraphics[angle=0, width=0.45\textwidth]{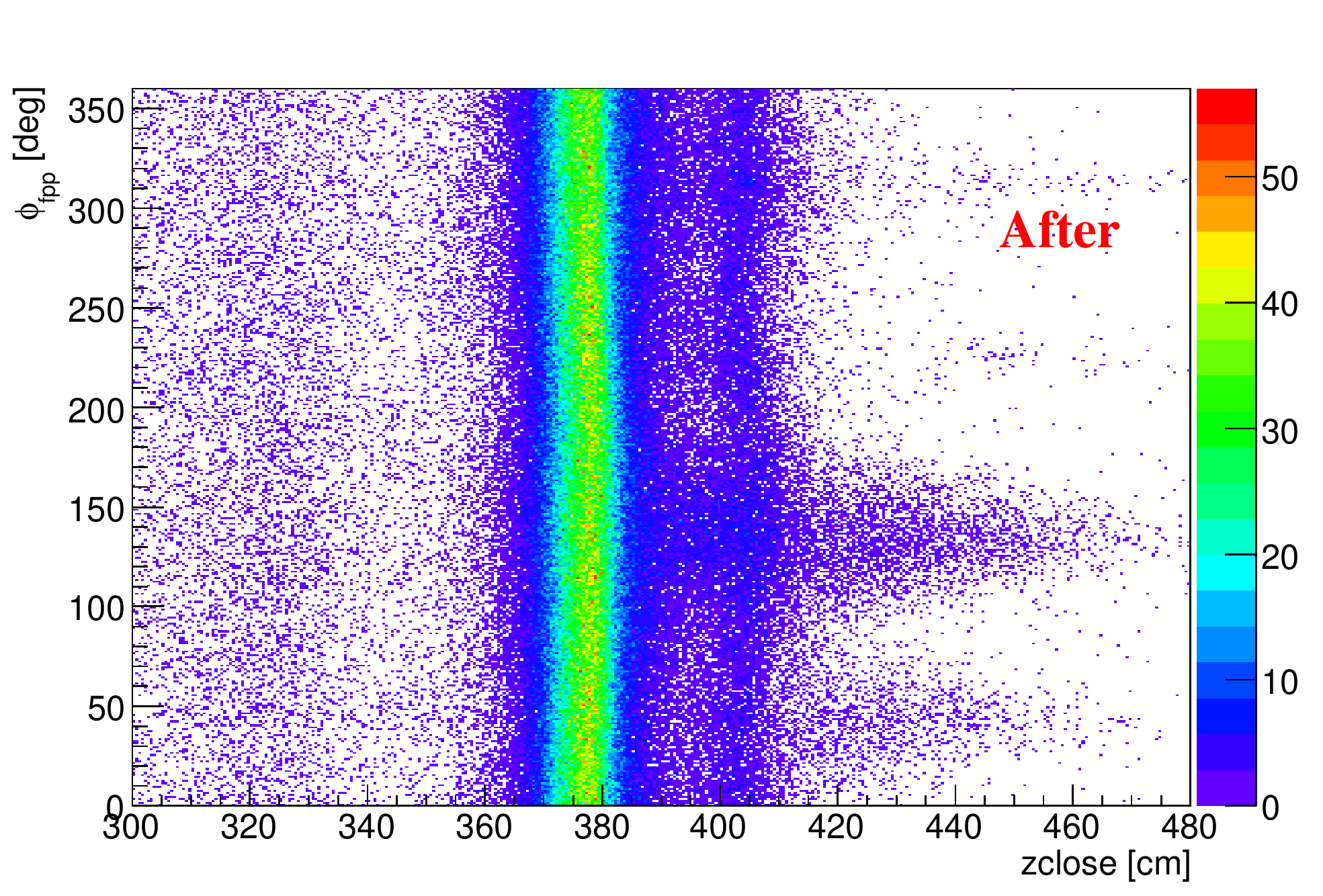}
    \caption{$\phi_{fpp}$ versus $zclose$ before and after the FPP chamber alignment.}
    \label{fig:align_1}
  \end{center}
\end{figure}

Another way to see the alignment quality
is by looking at $zclose$, which is the location along the spectrometer axis of closest
approach between the front and rear FPP tracks and stands for the second scattering vertex
in the carbon analyzer. For the ideal alignment, the reaction vertex should not depends on the azimuthal angle $\phi_{fpp}$, so the plot of $zclose$ versus $\phi_{fpp}$ should be ``straight'' in the $zclose$ dimension, with sharply defined edges centered at the physical position of the carbon analyzer. Fig.~\ref{fig:align_1} shows a plot before and after the alignment. One can obviously see the ``snake'' shape is gone after the alignment.

\subsubsection{Scattering Angle Calculation}
For the determination of the polar and azimuthal angles of the second
scattering, one first needs to rotate the coordinates system so that its
$z$-axis is along the momentum of the incident track, and then express
the scattered track in this new coordinate system.

\begin{figure}
  \begin{center}
    \includegraphics[angle=0, width=0.6\textwidth]{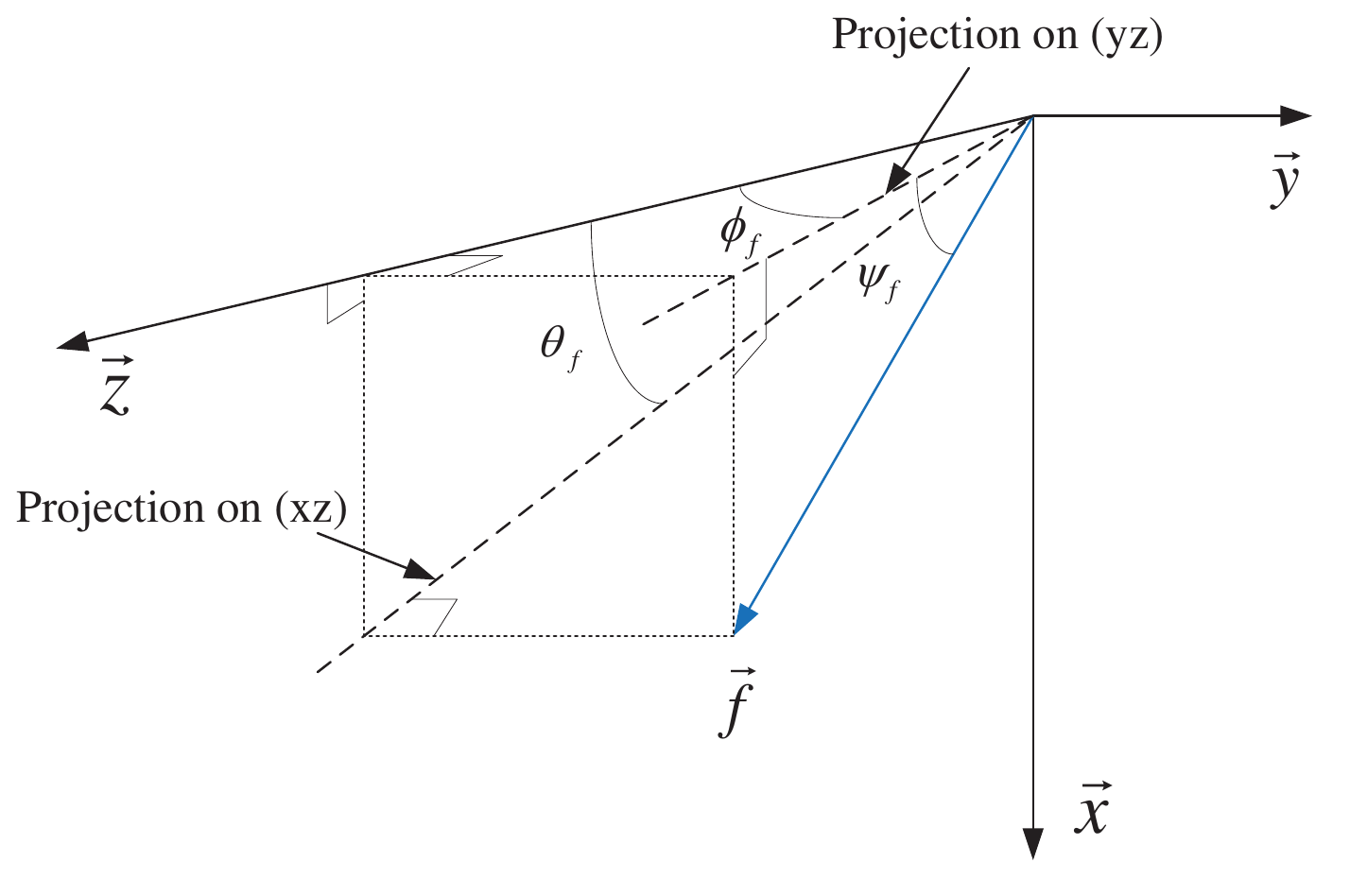}
    \caption{Cartesian angles for tracks in the transport coordinates system.}
    \label{fig:carte}
  \end{center}
\end{figure}
\begin{figure}
  \begin{center}
    \includegraphics[angle=0, width=0.6\textwidth]{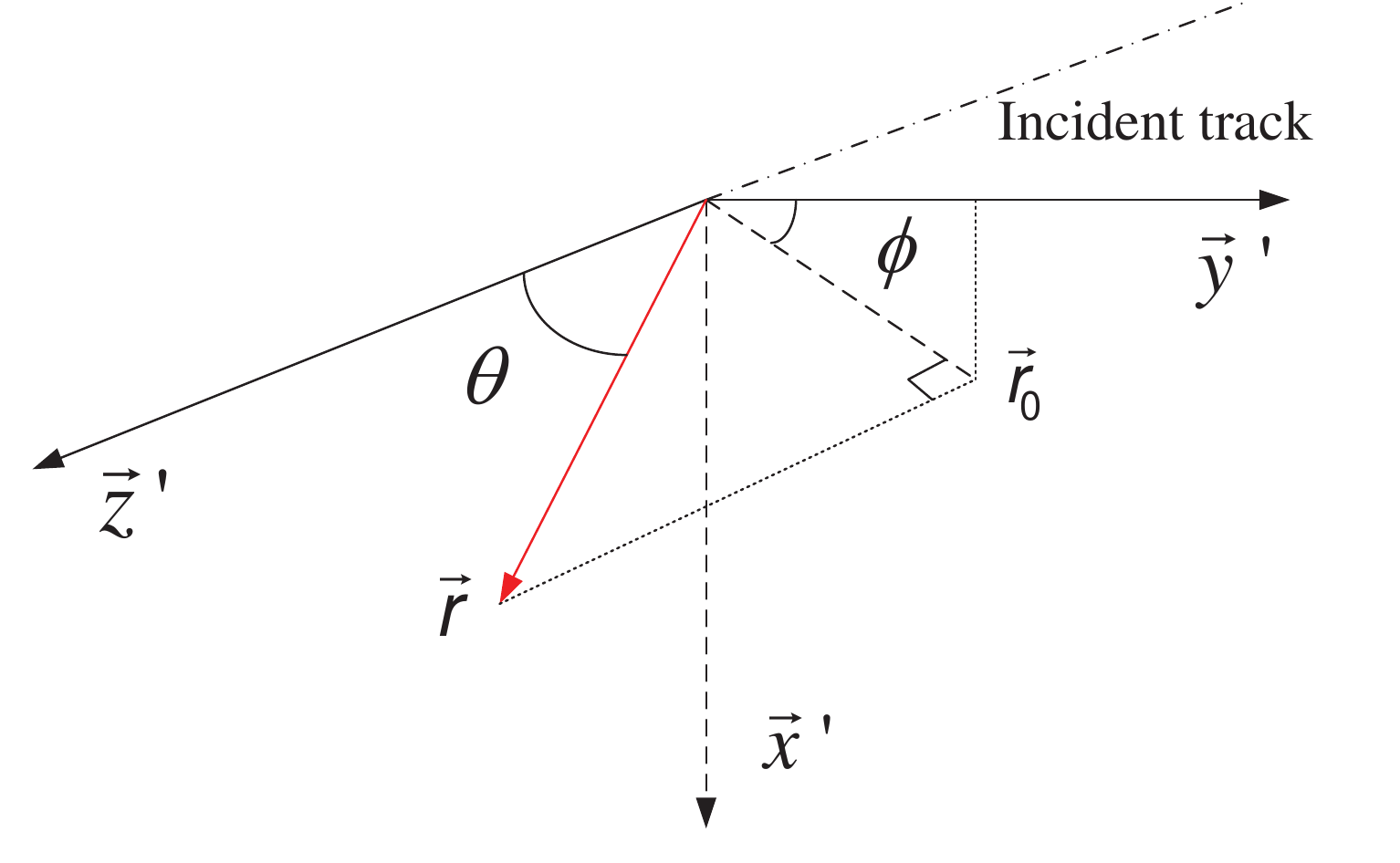}
    \caption{Spherical angles of the scattering in the FPP.}
    \label{fig:rotate}
  \end{center}
\end{figure}

As shown in Fig.~\ref{fig:carte}, for the incident track $\vec f$ in the
transport coordinates system, $\vec z$ is along the spectrometer axis
at the focal plane, $\vec x$ is perpendicular to $\vec z$ and
vertically down, and $\vec y=\vec z \times \vec x$. $\theta_{f}$ and
$\phi_f$ are the Cartesian angles: $\theta_f$ is the angle between the
projection of the track on the $x-z$ plane and the $z$-axis, and
$\phi_f$ is the angle between the projection on the $y-z$ plane and
the $z$-axis. For convenience, we define $\psi_f$ as the angle between
the track and its projection on the $y-z$ plane, and the relation
between the angles is:
\begin{equation}
\tan\psi_f=\tan\theta_f\cos\phi_f,
\end{equation}
Therefore, the rotation can be decomposed into two
rotations: first, a rotation of the $y-z$ plane around the $x$-axis by
an angle $\phi_f$, and followed by a second rotation by angle $\psi_f$
so that the new $z'$-axis lies along the incident track. The new
projection of the incident track $\vec f$ is given by:
\begin{equation}
\left( \begin{array}{c}
f'_x \\
f'_y\\
f'_x
\end{array}\right)
=\left( \begin{array}{c}
0\\
0\\
1
\end{array}\right)
\left( \begin{array}{ccc}
\cos\psi_f & 0 & -\sin\psi_f\\
0 & 1 & 0\\
\sin\psi_f & 0 & \cos\psi_f
\end{array}\right)
\left( \begin{array}{ccc}
1 & 0 & 0\\
0 & \cos\phi_f & -\sin\phi_f \\
0 & \sin\phi_f & \cos\phi_f
\end{array}\right)
\left( \begin{array}{c}
f_x\\
f_y\\
f_z
\end{array}\right).
\end{equation}
Similarly, the new projection of the scattered track $\vec r$ is
now:
\begin{equation}
\left( \begin{array}{c}
r'_x \\
r'_y\\
r'_x
\end{array}\right)=
\left( \begin{array}{ccc}
\cos\psi_f & 0 & -\sin\psi_f\\
0 & 1 & 0\\
\sin\psi_f & 0 & \cos\psi_f
\end{array}\right)
\left( \begin{array}{ccc}
1 & 0 & 0\\
0 & \cos\phi_f & -\sin\phi_f \\
0 & \sin\phi_f & \cos\phi_f
\end{array}\right)
\left( \begin{array}{c}
r_x\\
r_y\\
r_z
\end{array}\right).
\end{equation}
We can now define the scattering angles $(\theta_{fpp},\phi_{fpp})$ as
the spherical angles of the scattered track in this new coordinate
system as illustrated in Fig.~\ref{fig:rotate}.
If $\vec {r}_0$ is the projection of $\vec r$ on the $x'-y'$ plane, we have:
\begin{eqnarray}
r_0^2=r_x'^2+r_y'^2,\\
\theta_{fpp}=\tan^{-1}(\frac{r_0}{r_z'}),\\
\phi_{fpp}=\tan^{-1}(\frac{r_x'}{r_y'}).
\end{eqnarray}
\section{Events Selection}
Before we extracts the physics asymmetries, a series of cuts were applied to select the elastic events and to minimize the experimental systematic uncertainties.
\subsection{HRS Cuts}
\subsubsection{One-Track-Only Cut}
\begin{figure}
  \begin{center}
    \includegraphics[angle=0, width=0.65\textwidth]{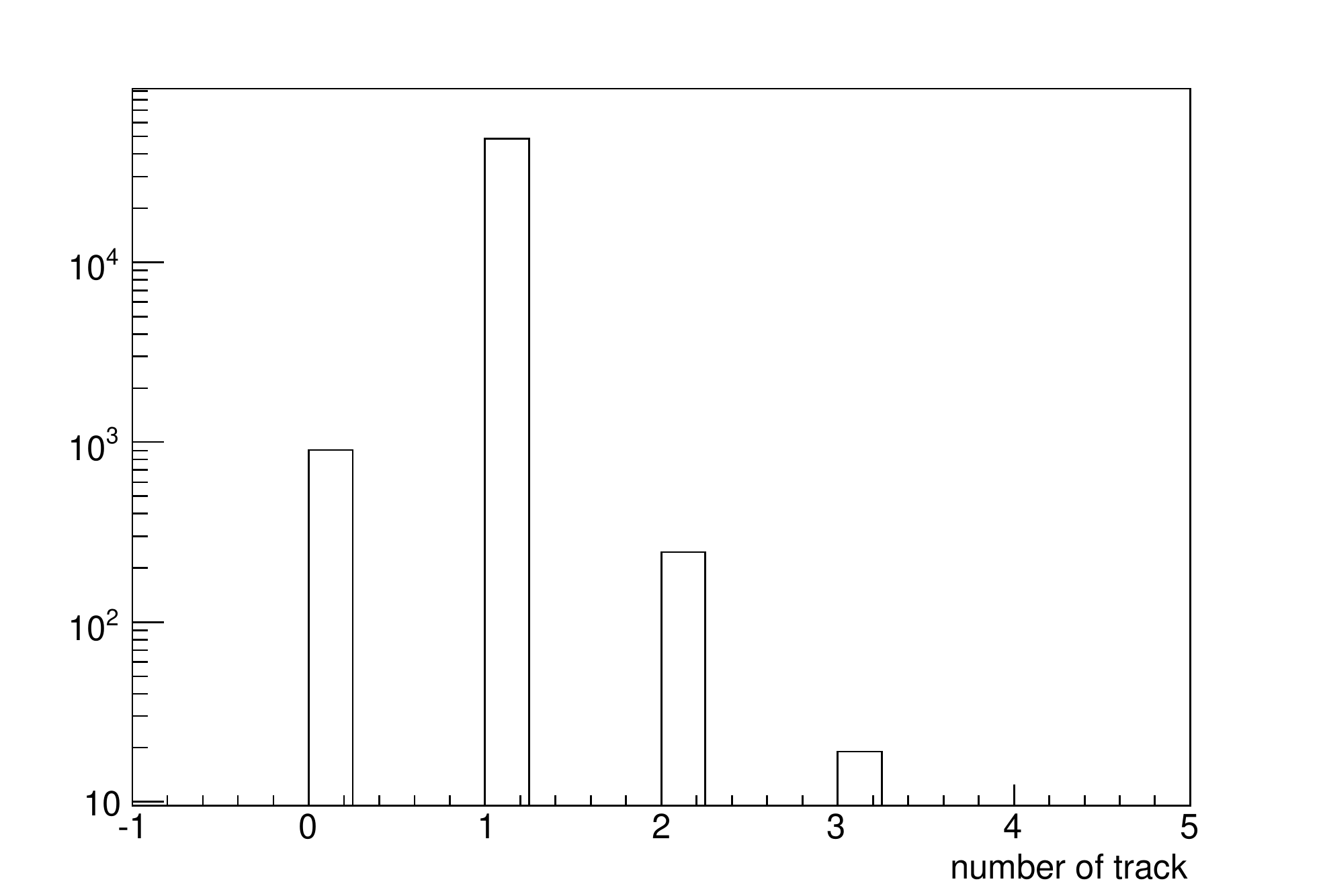}
    \caption{Left HRS VDC track number distribution.}
    \label{fig:ntrack}
  \end{center}
\end{figure}

First, the one-track-only
cut was applied to the events reconstructed from VDC clusters. The
drift times range from 0 to 360 ns. In the Analyzer, a software cut
of 400 ns is applied after the first wire fires to ensure the
completeness of the track searching. If only one track is observed in
an event, the track reconstruction will be accurate. If multiple tracks
for an event are found in the analysis, the first track reconstruction
may be distorted due to the interference of a nearby second track. This
cut removes $\sim 1.5\%$ of the total events (see Fig.~\ref{fig:ntrack}).
\begin{equation}
\eta (ONE)=\frac{N(n_{track}=1)}{N(n_{track})>0}\sim 98.5\%
\end{equation}
\subsubsection{HRS Acceptance Cut}
\begin{figure}
  \begin{center}
    \includegraphics[angle=0, width=0.45\textwidth]{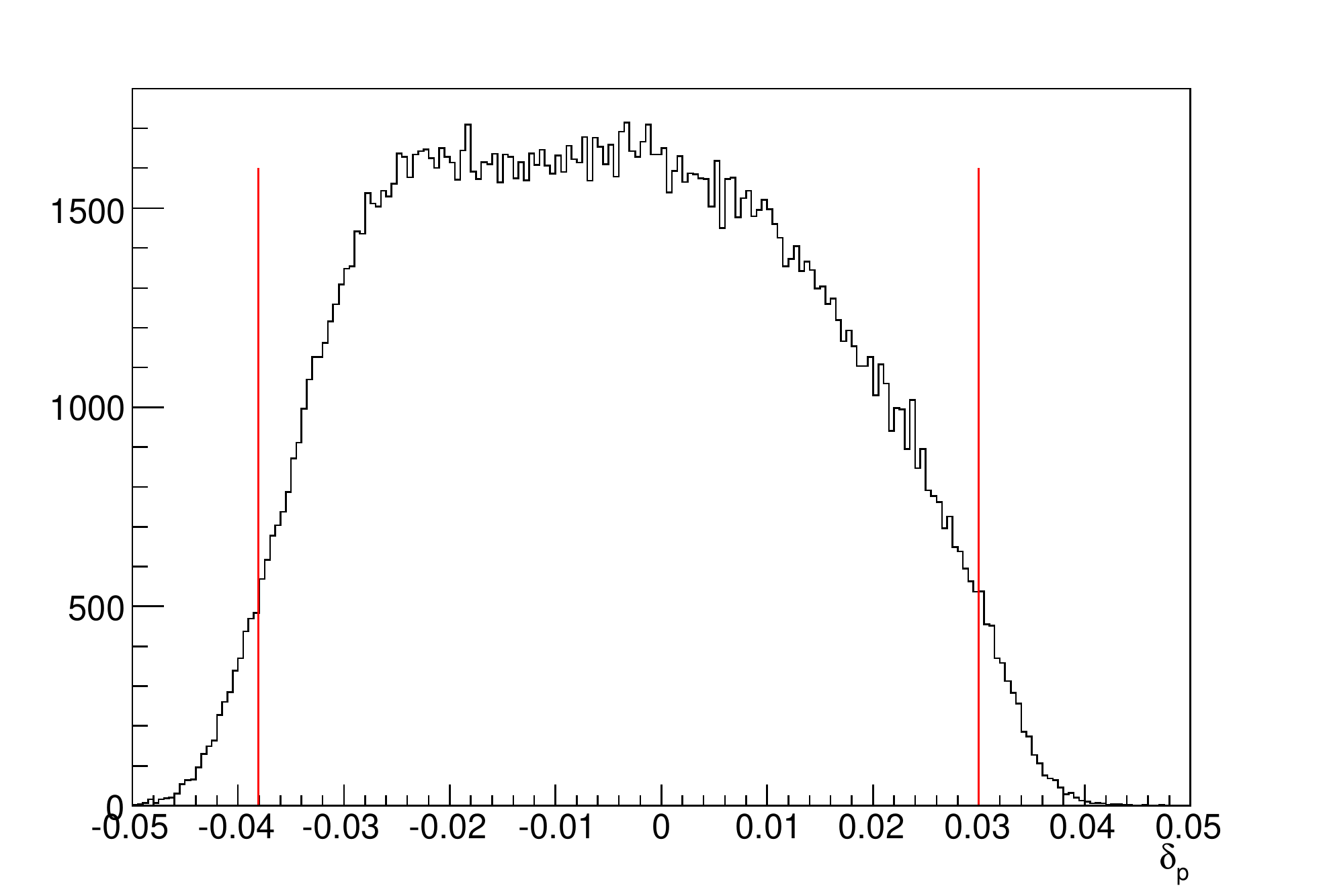}
    \includegraphics[angle=0, width=0.45\textwidth]{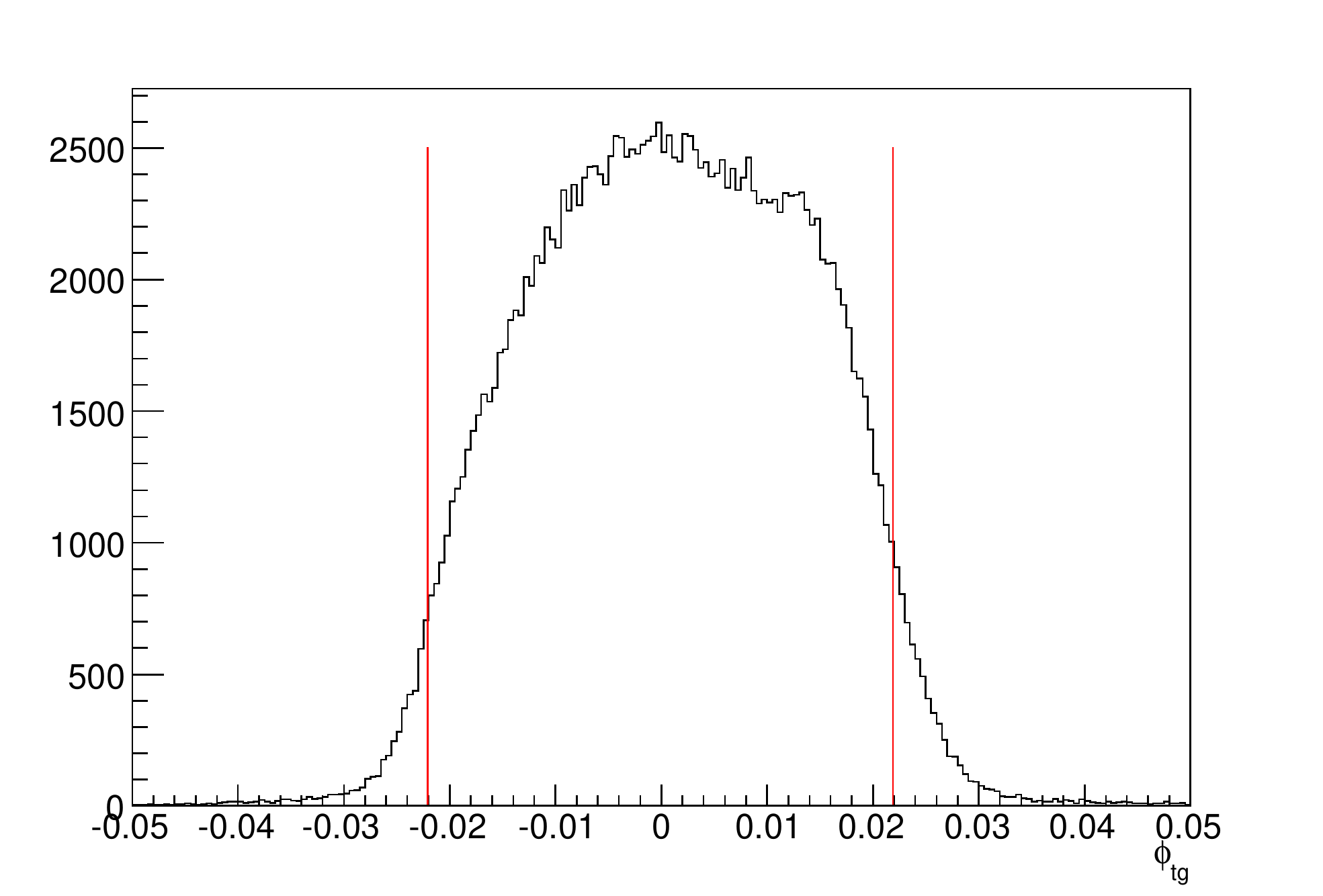}
    \includegraphics[angle=0, width=0.45\textwidth]{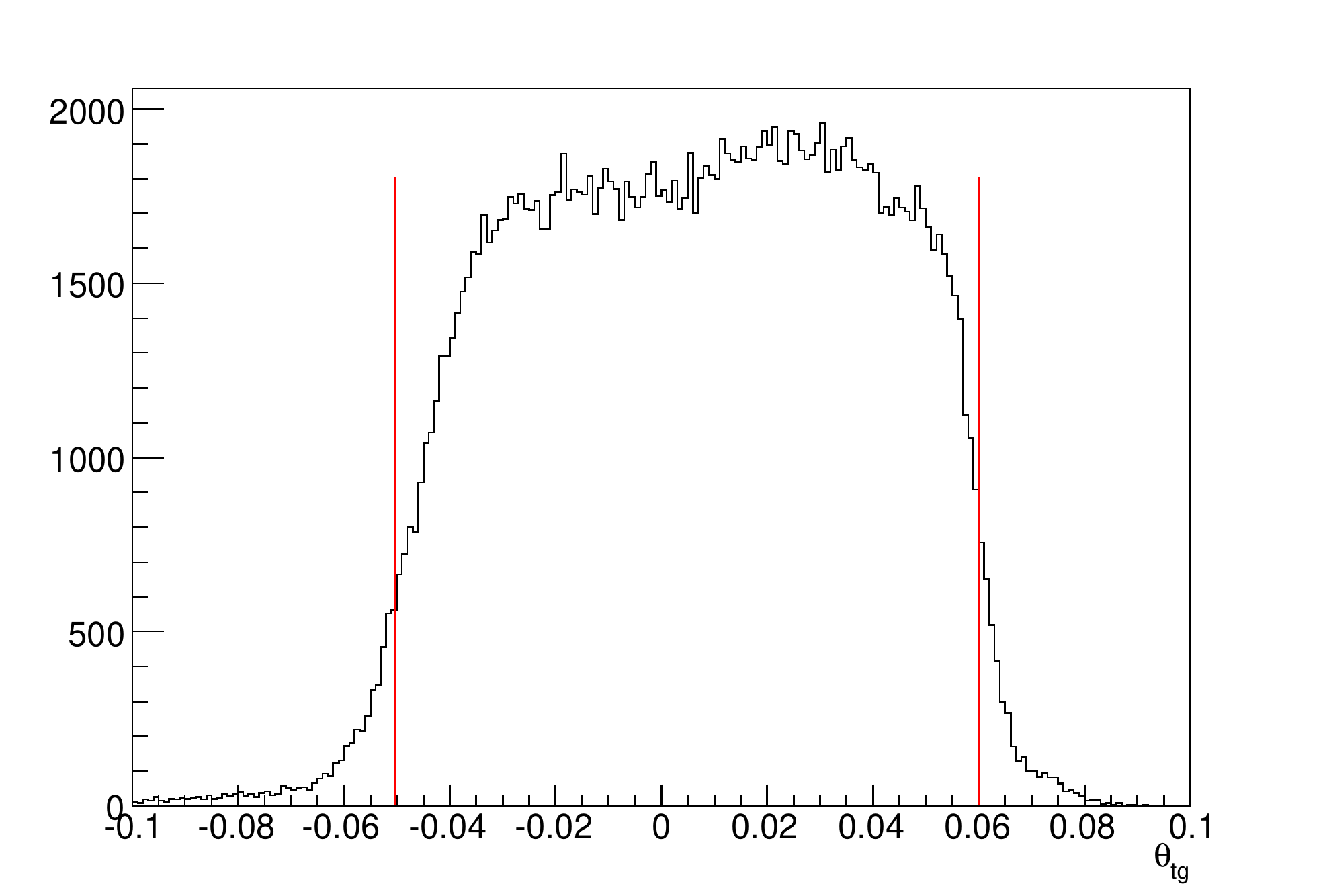}
    \includegraphics[angle=0, width=0.45\textwidth]{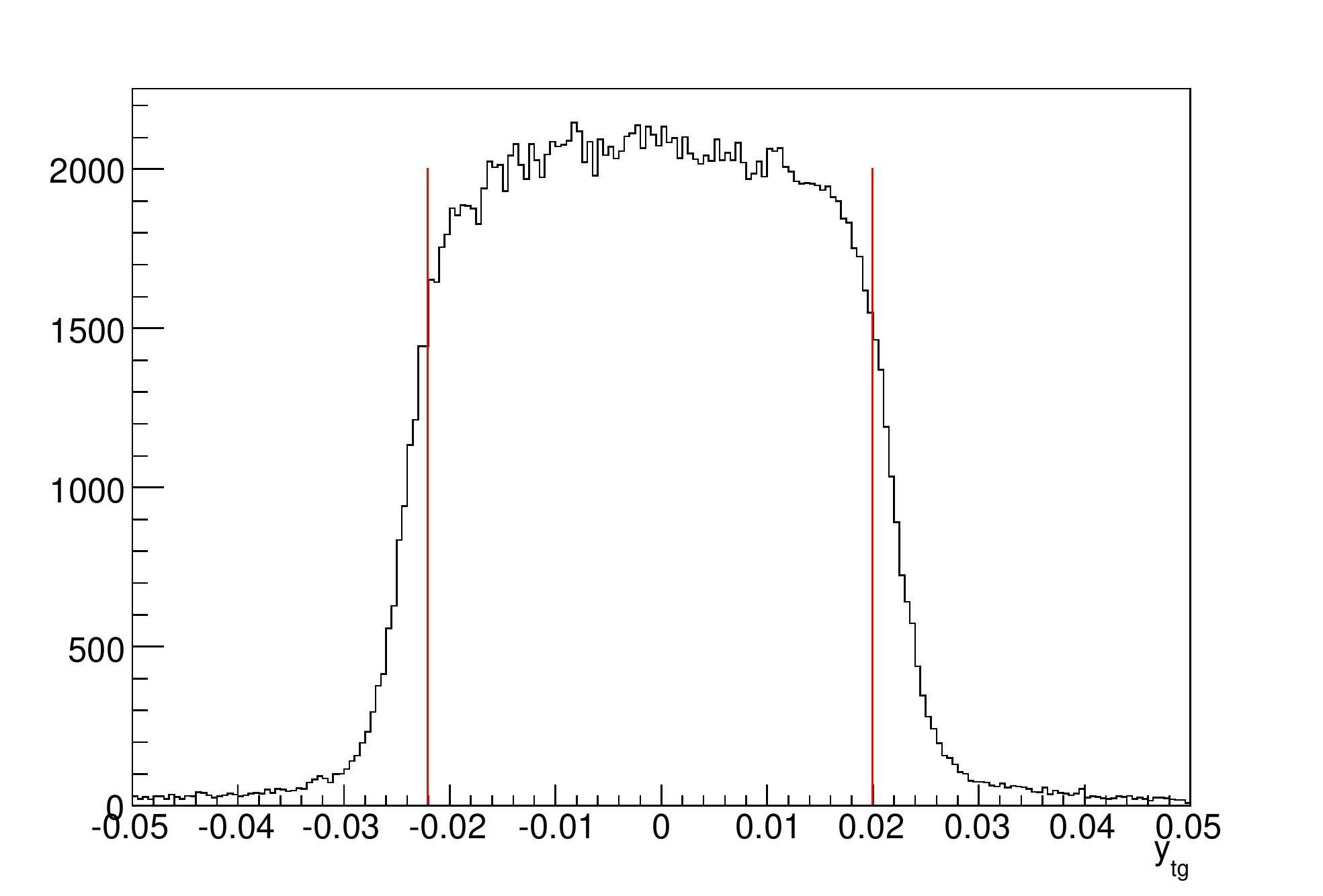}
    \caption{HRS acceptance cuts for kinematic setting K5 $\delta_p=0\%$.}
    \label{fig:accpcut}
  \end{center}
\end{figure}

As mentioned in Section~\ref{sec:hrs}, the HRS has a finite momentum and angular
acceptance. Events with the target coordinates reconstructed outside the
physical acceptance need to be cut out. On the other hand, since this
experiment measures the helicity dependent asymmetry difference at the
focal plane, precise knowledge of the acceptance is not required
compared to an absolute cross section measurement. In order to avoid
potential problems arising from the spin transport at the edge of the
acceptance, relatively tight cuts were applied compared to the HRS nominal acceptance.
The reaction vertex cut was also applied ($y_{\mathrm{tg}}$) to
reduce the number of events from the quasi-elastic scattering off the aluminum end cap.
Typical cuts on different target variables are shown in Fig.~\ref{fig:accpcut}.
\subsubsection{Elastic Cut for the Hadron Arm}
Since during the production of this experiment, only part of the
BigBite shower blocks were turned on, we cannot reconstruct the electron kinematics.
However, due to the small acceptance and high resolution of the HRS, the
elastic kinematics are well determined by the hadron arm.
The beam energy ($\sim1.2$ GeV) is low  enough so that the inelastic background
channel is highly suppressed. By placing constraints on the proton elastic
kinematics, the background is further suppressed within the acceptance.

\begin{figure}
  \begin{center}
    \includegraphics[angle=0, width=0.49\textwidth]{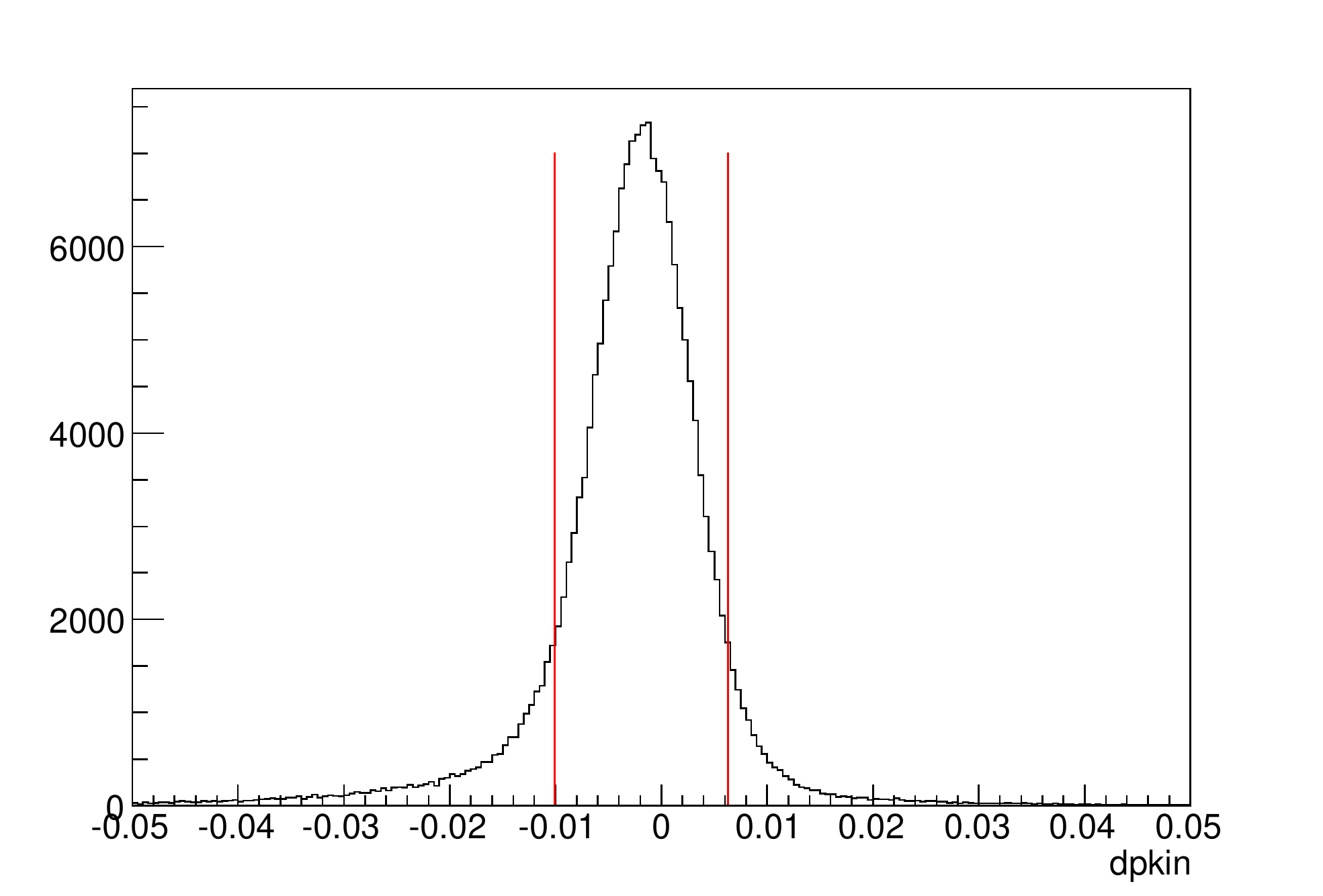}
    \includegraphics[angle=0, width=0.49\textwidth]{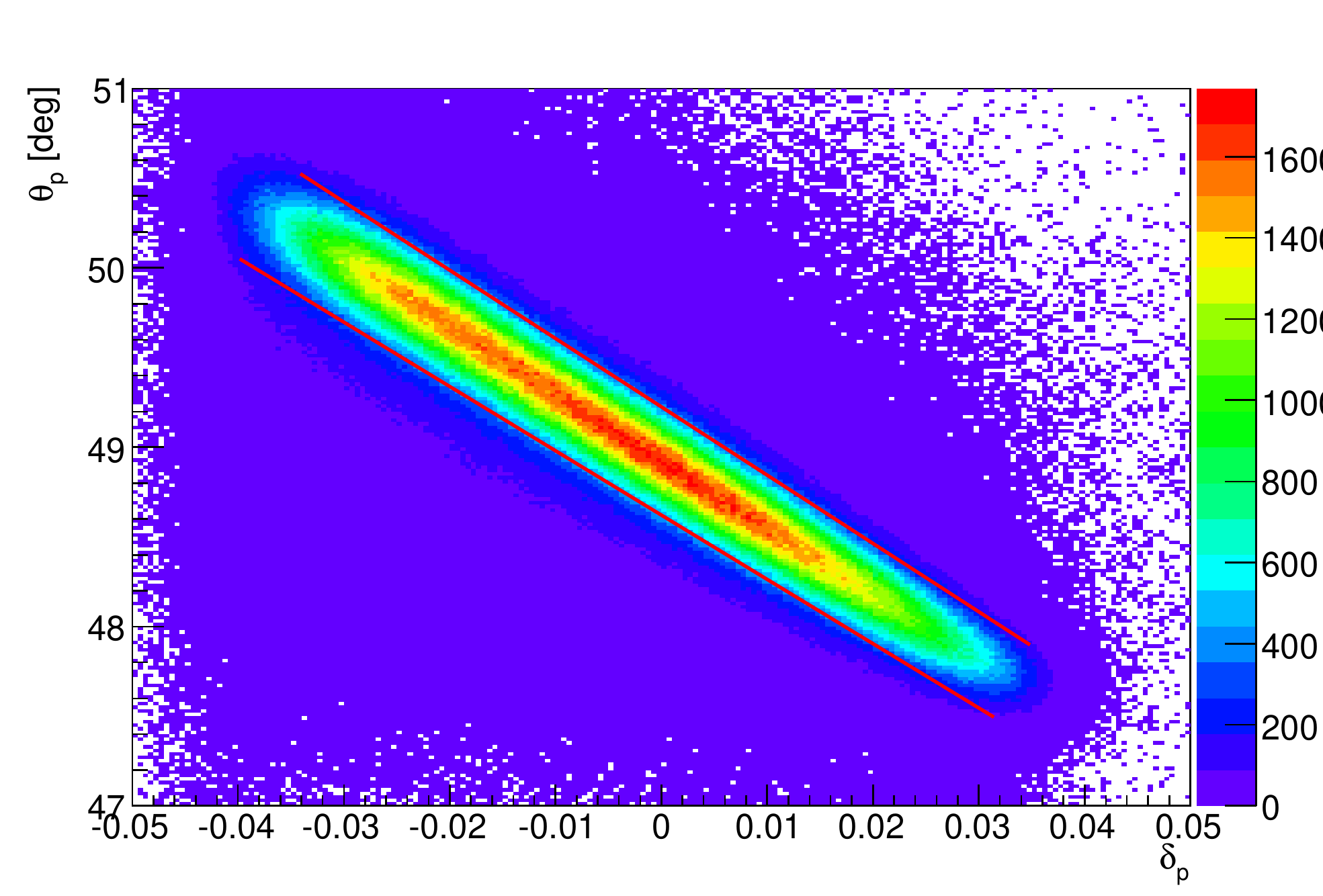}
    \caption{Elastic cut on dpkin (left), and the corresponding 2D cut on the proton angle $\theta_p$ versus momentum $\delta_p$.}
    \label{fig:dpkin_cut}
  \end{center}
\end{figure}

We applied an elastic cut on the proton ``dpkin'', which is the angle-corrected $\delta_p$.
The resolution of ``dpkin'' represents the momentum and angular resolution of the hadron spectrometer.
With sufficient statistics, we applied a tight cut on the proton elastic peak to keep $\sim 80\%$
of the elastic events. This cut corresponds to a 2 dimensional
cut on the proton angle versus momentum (see Fig.~\ref{fig:dpkin_cut}).

This elastic cut minimizes the contributions from the radiative tail,
inelastic events and the background due to the proton re-scattering inside the
spectrometer\footnote{This becomes crucial for low momentum protons,
since the re-scattering can change the momentum and direction of the
proton at the focal plane while the reconstruction is still within the
acceptance. The spin transport is totally different for this type
of events.} without sacrificing too many of the elastic events.
\subsection{Other Cuts}
Although most the singles triggers were pre-scaled away during the
experiment, there were still $\sim 20$ Hz T1 and T3 events left in the
data stream. An event-type cut was applied to select the coincidence
trigger T5.

A coincidence timing cut was applied to the TDC spectrum of T3.
The accidental background under the elastic cut is $\sim
0.3\%$. Since the accidental background is still dominated by the elastic singles, the
reconstructed proton polarization outside the cut is similar to the
events inside; hence, we do not expect background from these events to produce any noticeable effect.

\subsection{BigBite Replay}
\begin{figure}
  \begin{center}
    \includegraphics[angle=0, width=0.75\textwidth]{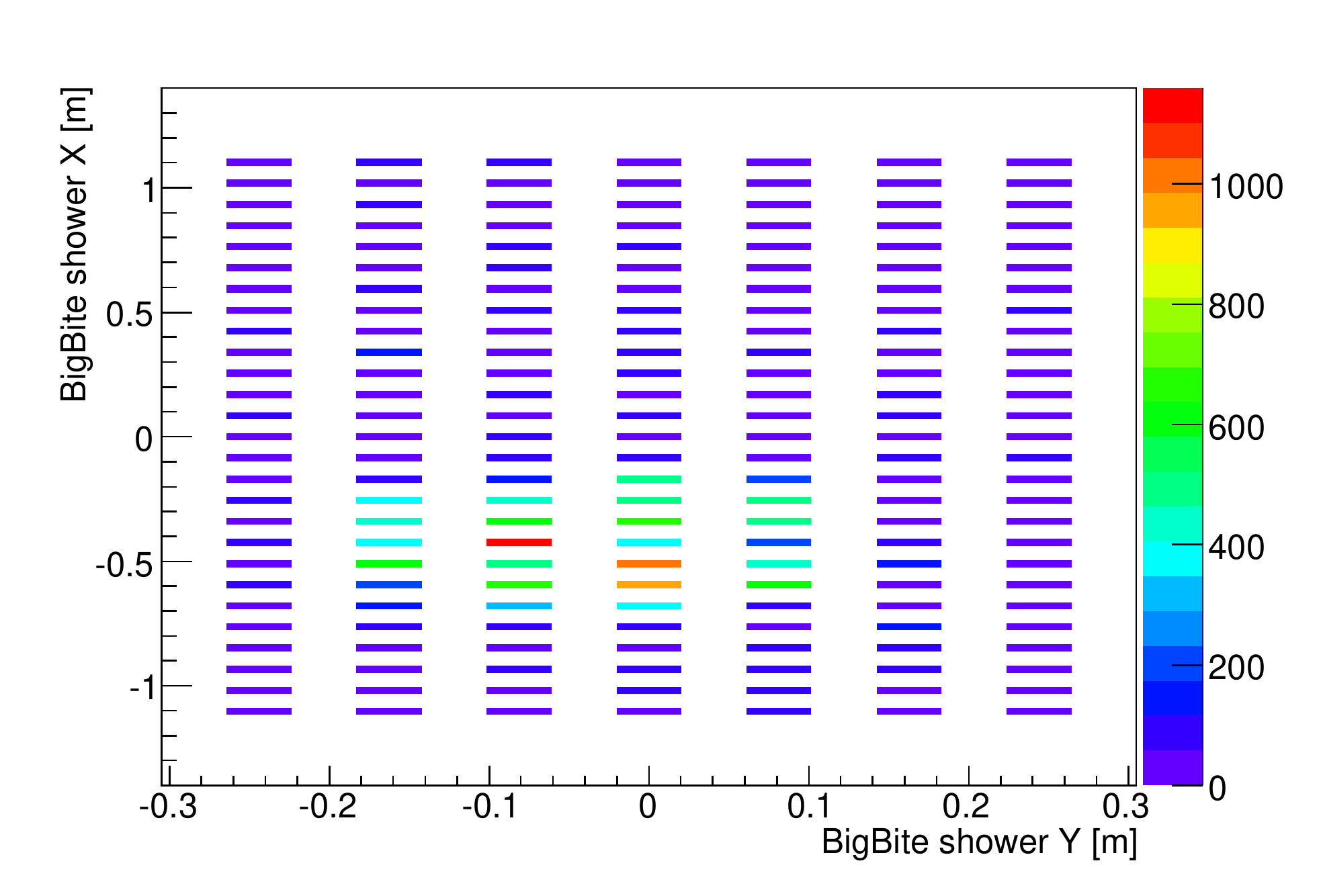}
    \includegraphics[angle=0, width=0.45\textwidth]{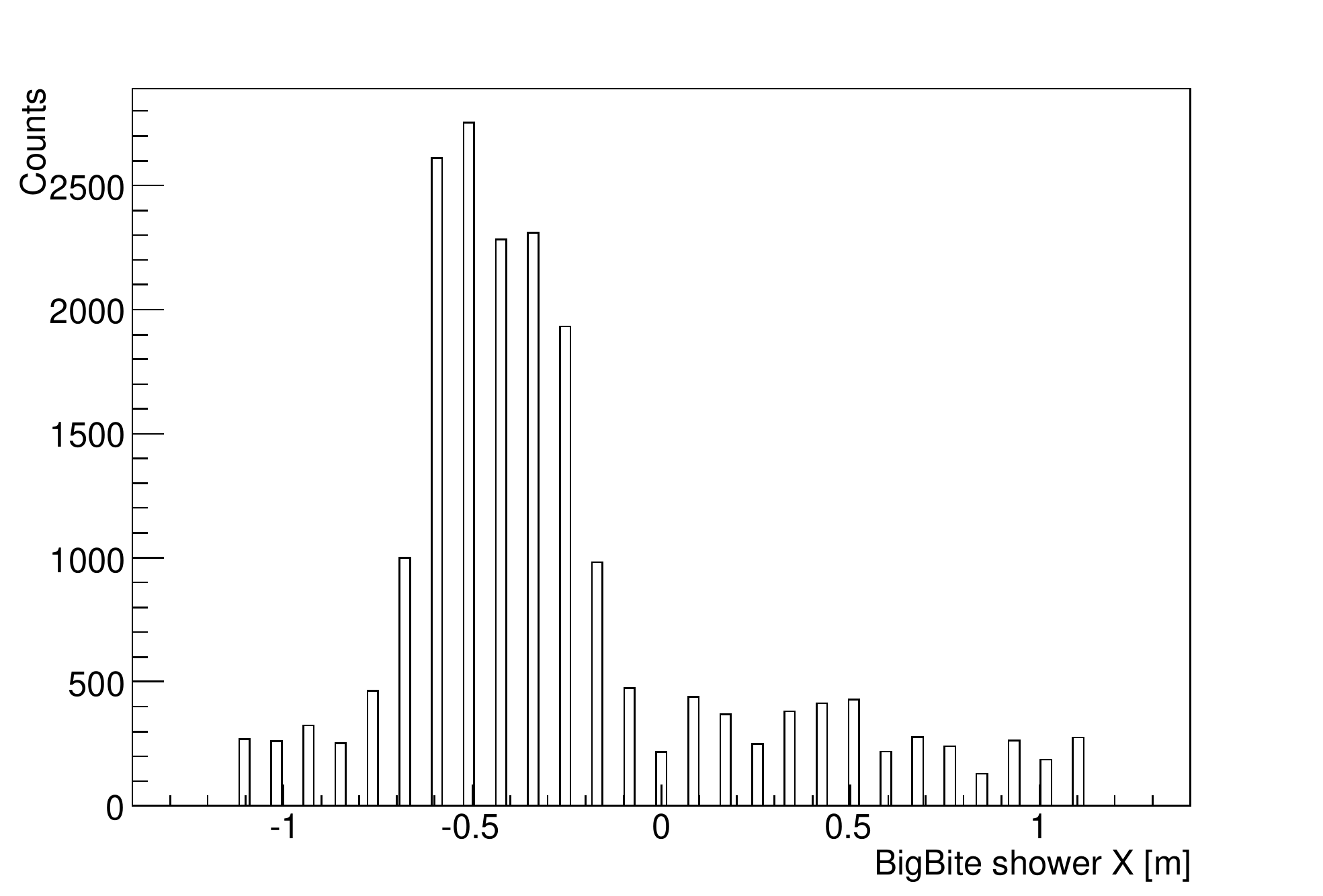}
    \includegraphics[angle=0, width=0.45\textwidth]{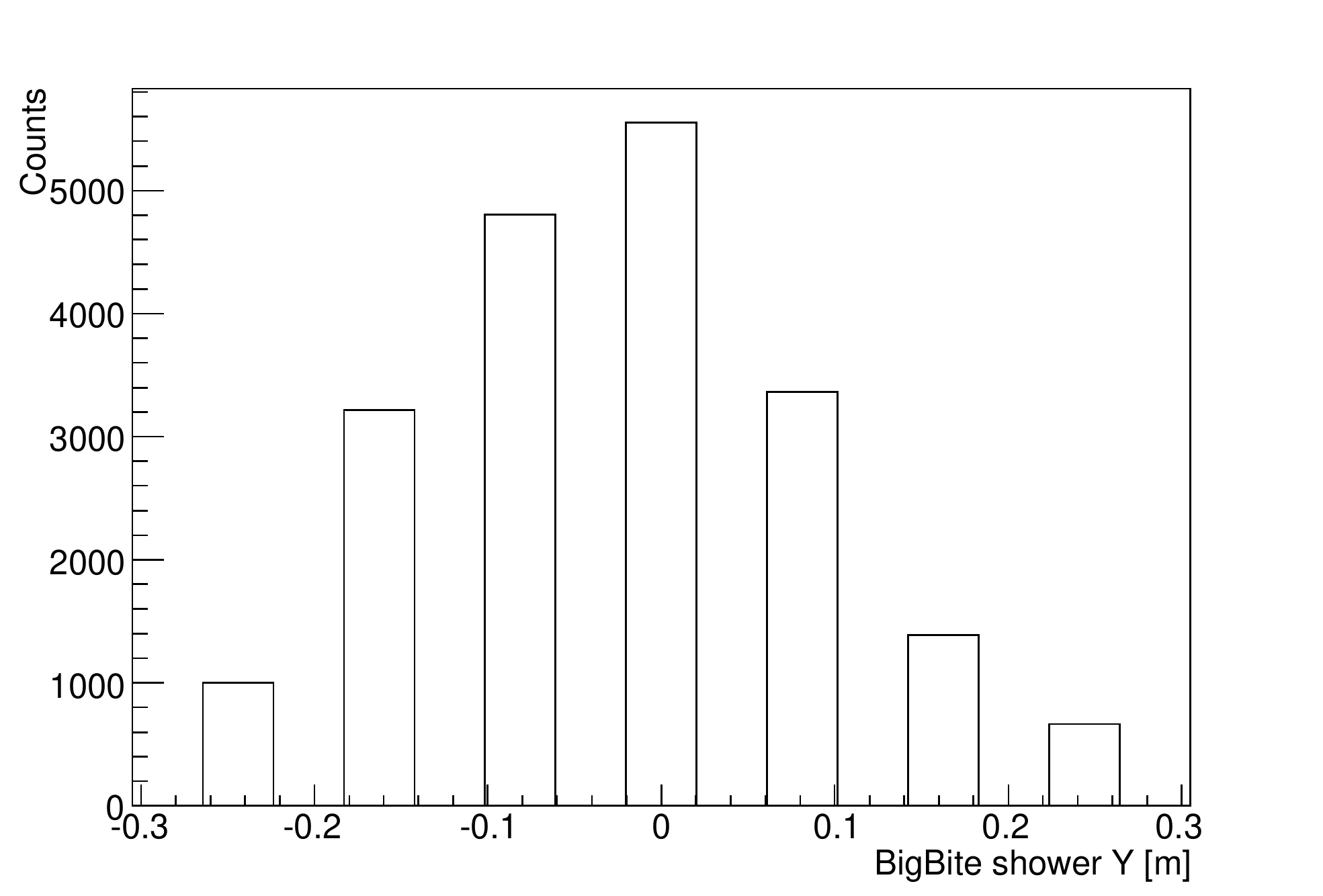}
    \caption{BigBite shower counter hit pattern in the upper panel and the profiles on $x$ (vertical) and $y$ (horizontal) in the left and right panels, respectively.}
    \label{fig:peak}
  \end{center}
\end{figure}

During this experiment, the BigBite shower counter was used to tag the
scattered electrons and form the coincidence trigger. The entire pre-shower counter and part of the shower blocks outside the elastic peak
were turned off during the production data taking to reduce the background. To ensure we turned on the right shower
counter region, we did test runs after every kinematic setting change in which both the
pre-shower and shower counters were turned on to locate the electron elastic peak
(see Fig.~\ref{fig:peak}).

\begin{figure}
  \begin{center}
    \includegraphics[angle=0, width=0.45\textwidth]{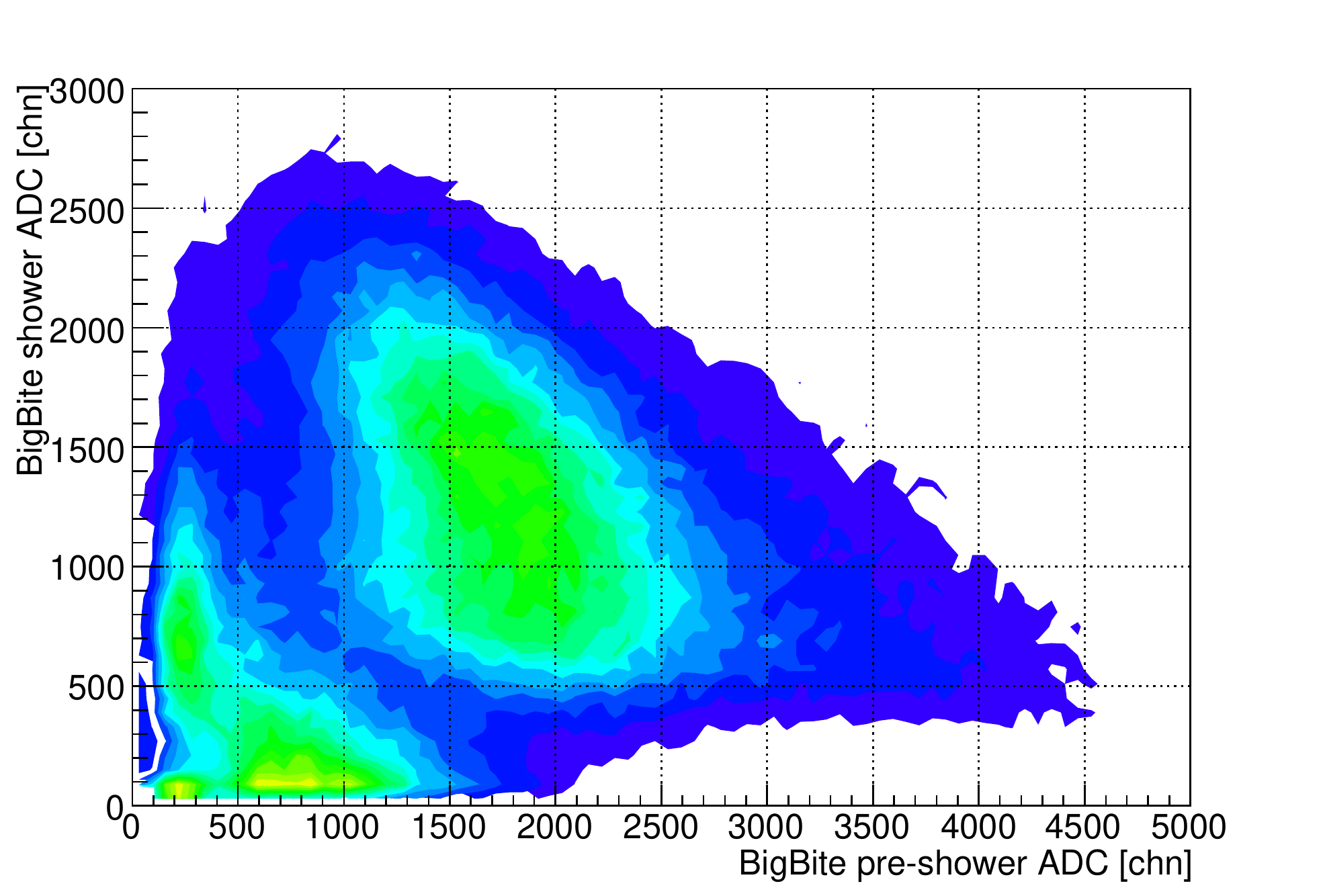}
    \includegraphics[angle=0, width=0.45\textwidth]{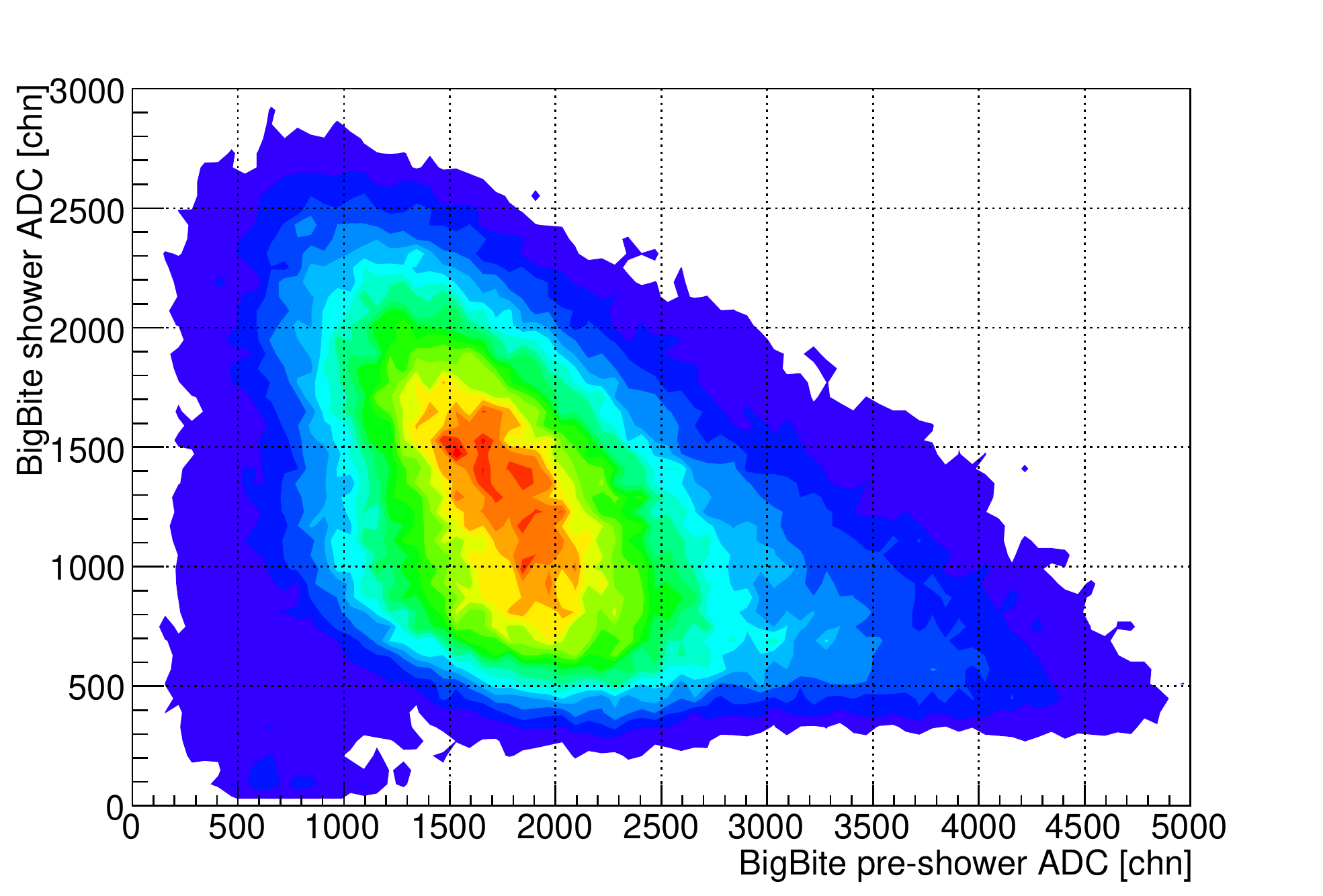}
    \caption{The BigBite pre-shower ADC sum versus shower ADC sum with (right panel) and without (left panel) the coincidence trigger cut (T5). The low energy background were highly suppressed with the coincidence configuration.}
    \label{fig:shower_all}
  \end{center}
\end{figure}

\begin{figure}
  \begin{center}
    \includegraphics[angle=0, width=0.46\textwidth]{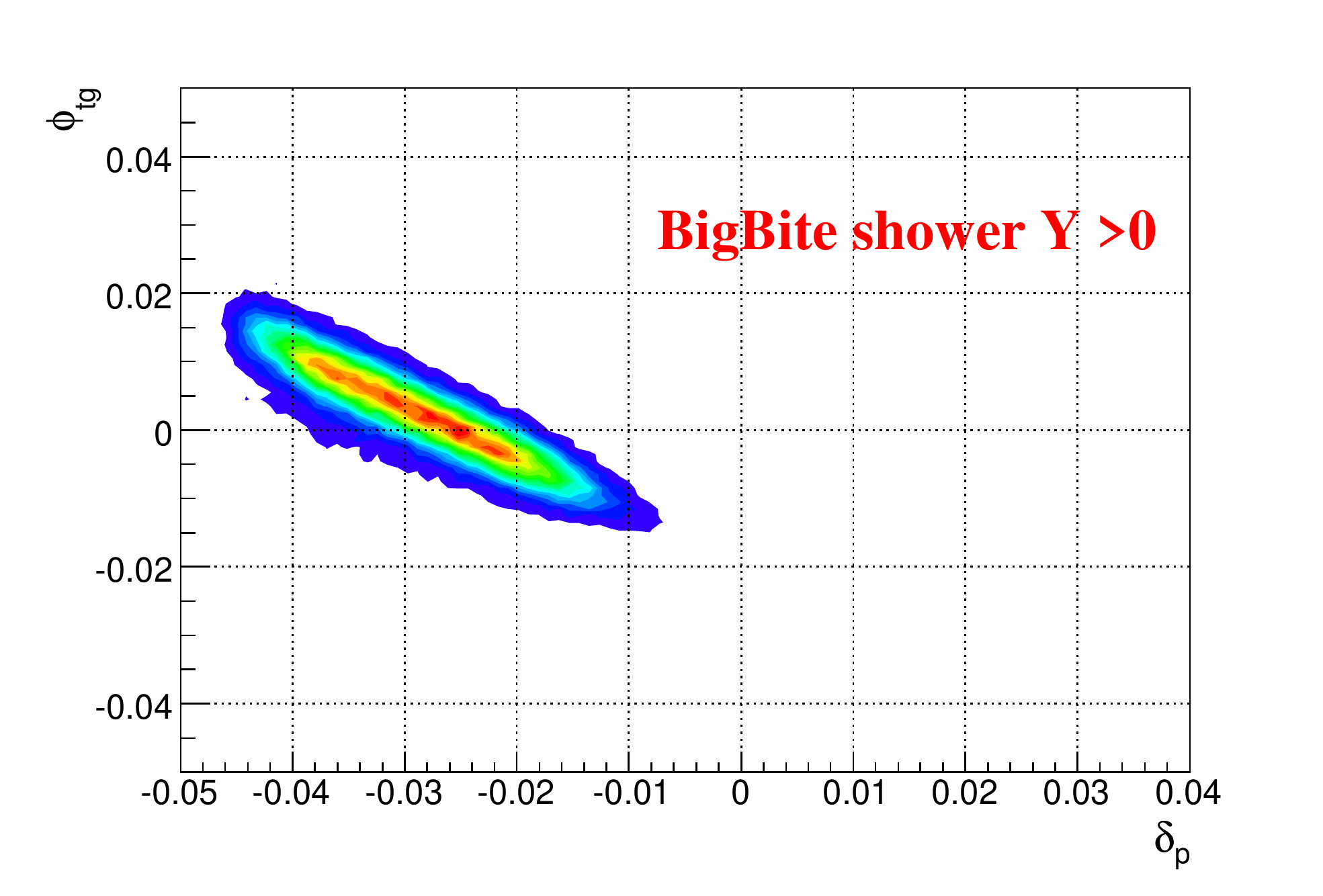}
    \includegraphics[angle=0, width=0.46\textwidth]{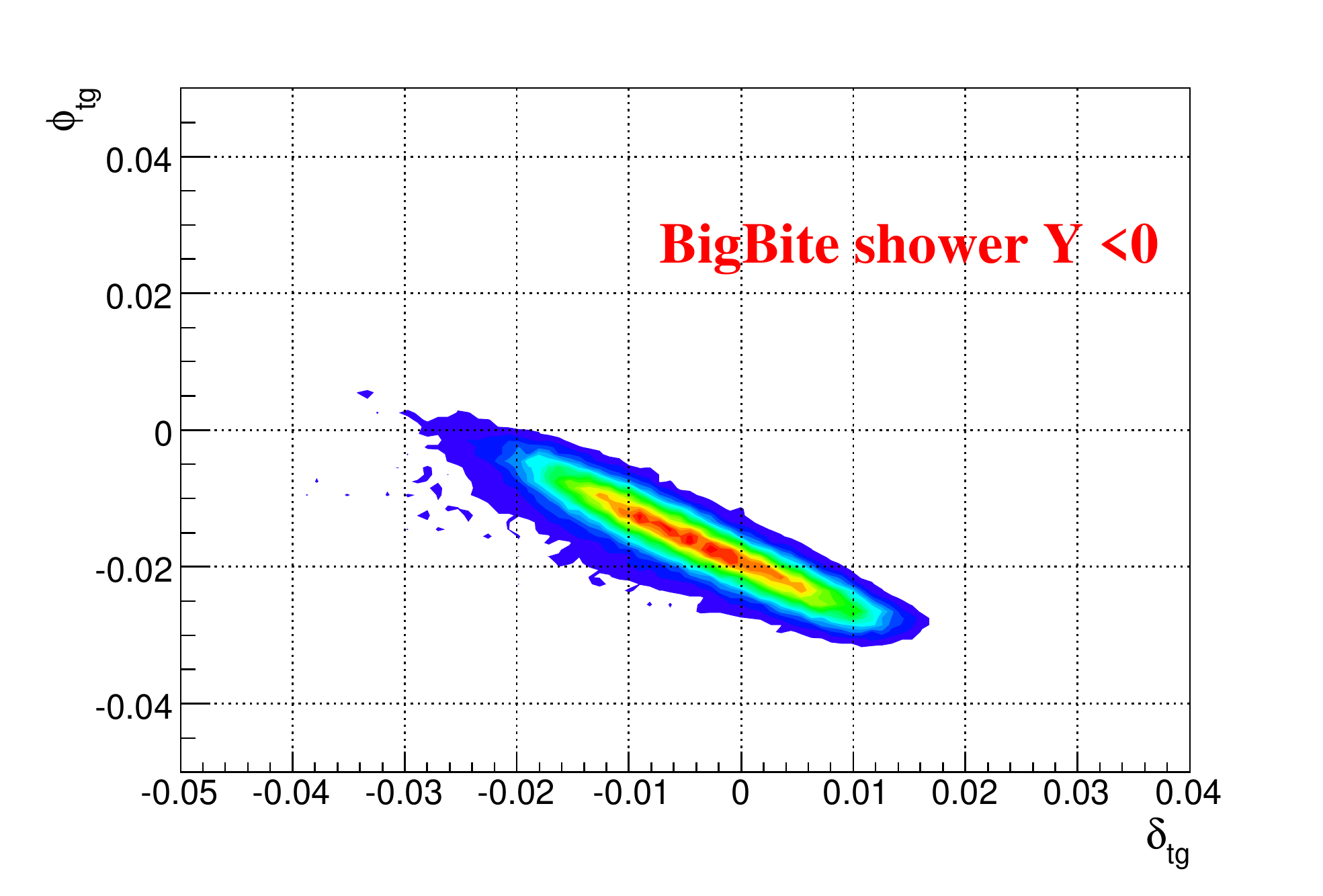}
    \caption{Proton acceptance (angle versus momentum) with BigBite shower $y>0$ and $y<0$.}
    \label{fig:correl}
  \end{center}
\end{figure}

From these test runs, the electron energy deposited in the pre-shower and shower counters were reconstructed. Fig.~\ref{fig:shower_all} shows the BigBite shower ADC sum versus the pre-shower ADC sum when both of them were
turned on, with and without the coincidence cut (T5). Clearly, the
coincidence trigger can effectively suppress the pions and low energy electrons.
Additionally, the plots of the proton acceptance with BigBite
shower $y>0$~($y<0$) (see Fig.~\ref{fig:correl}), directly demonstrate the correlation
between the electrons and the protons.

\subsection{FPP Cuts}
\subsubsection{Scattering Angle Cut}
\begin{figure}
  \begin{center}
    \includegraphics[angle=0,
    width=0.6\textwidth]{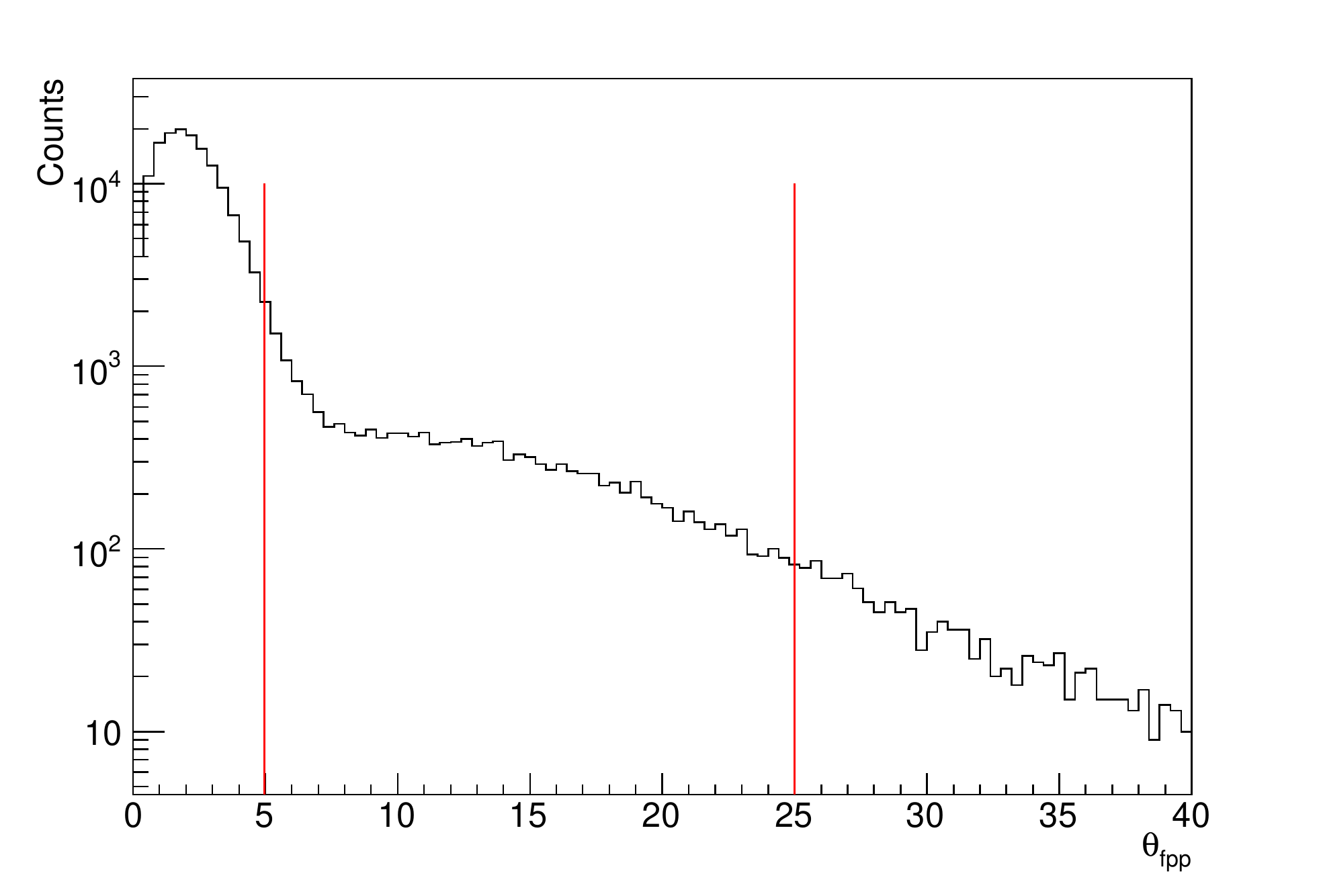}
    \caption{The distribution of the FPP polar scattering angle $\theta_{fpp}$ and the applied cut.}
    \label{fig:thfppcut}
  \end{center}
\end{figure}

In order to select the correct reconstructions of the second scattering
in the FPP, several cuts were applied on the FPP variables. First, a
cut was applied on the polar scattering angle:
$5^{\circ}<\theta_{fpp}<25^{\circ}$. This cut removes the small
scattering angle events, which are dominated by Coulomb
scattering with little analyzing power, and the larger scattering angle events,
which have large instrumental asymmetry and smaller analyzing power.
Fig.~\ref{fig:thfppcut} shows an example
of the $\theta_{fpp}$ distribution and the applied cut.

\subsubsection{Scattering Vertex Cut}
\begin{figure}
  \begin{center}
    \includegraphics[angle=0,
    width=0.70\textwidth]{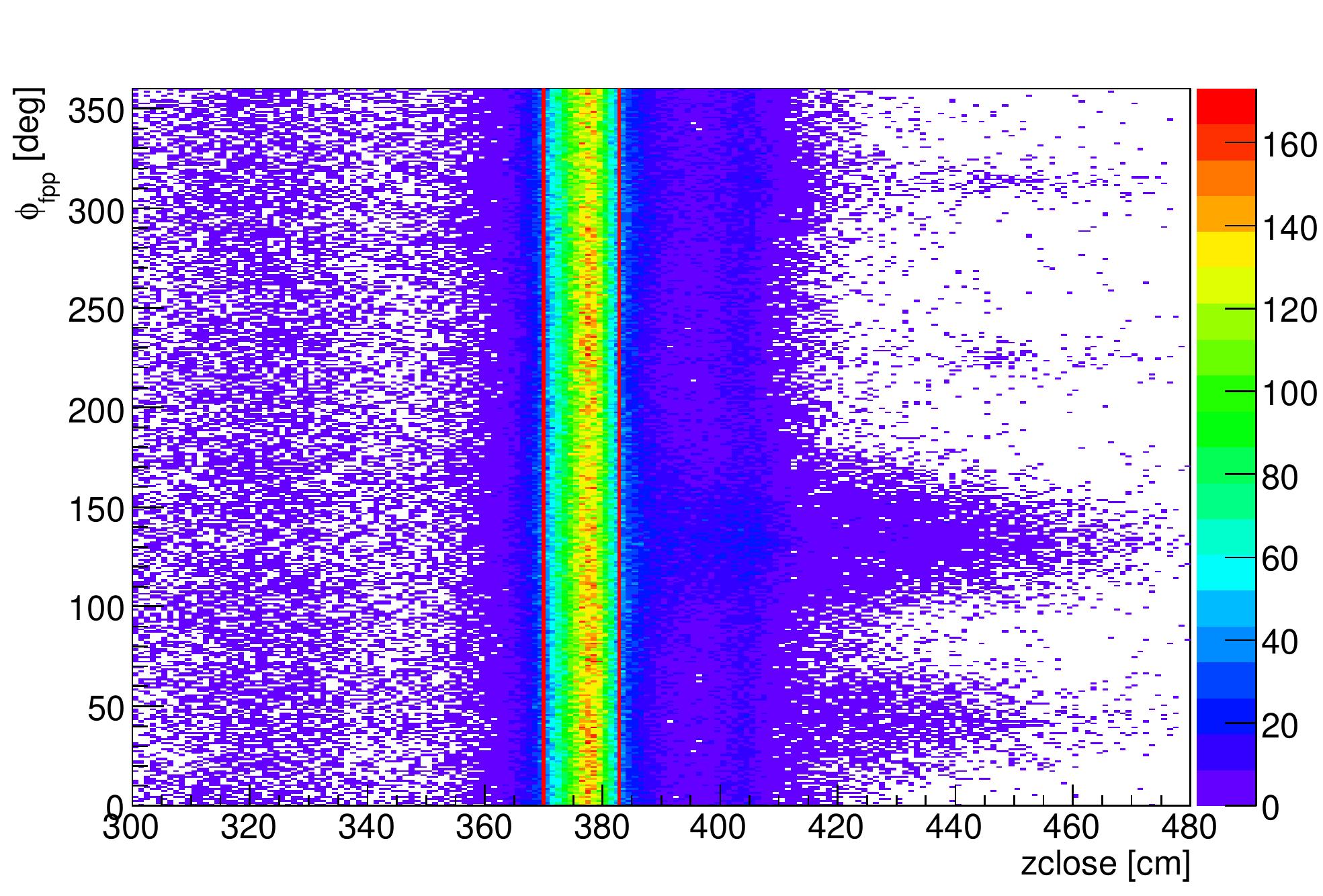}
    \caption{Cut applied to $zclose$ after the manual correction for setting K2 $\delta_p=0\%$.}
    \label{fig:zcut}
  \end{center}
\end{figure}

In order to ensure that the scattering originated from within the carbon block,
a tight cut on the reaction vertex was applied. Due to the
imperfect alignment of the FPP chambers, a manual correction was applied to
$zclose$ along the azimuthal scattering angle $\phi_{fpp}$. In this
procedure, a set of coefficients were generated along $\phi_{fpp}$ by
the profile of the 2D plot of $\phi_{fpp}$ versus $zclose$. After this
correction, a straight line cut was applied to the corrected $zclose$.
Fig.~\ref{fig:zcut} shows the plot of $\phi_{fpp}$ versus $zclose$ after the
correction, and the applied cut\footnote{The purpose of this
correction is to make the cut simpler, and it doesn't change the FPP alignment.}.
\begin{figure}
  \begin{center}
    \includegraphics[angle=0,
    width=0.70\textwidth]{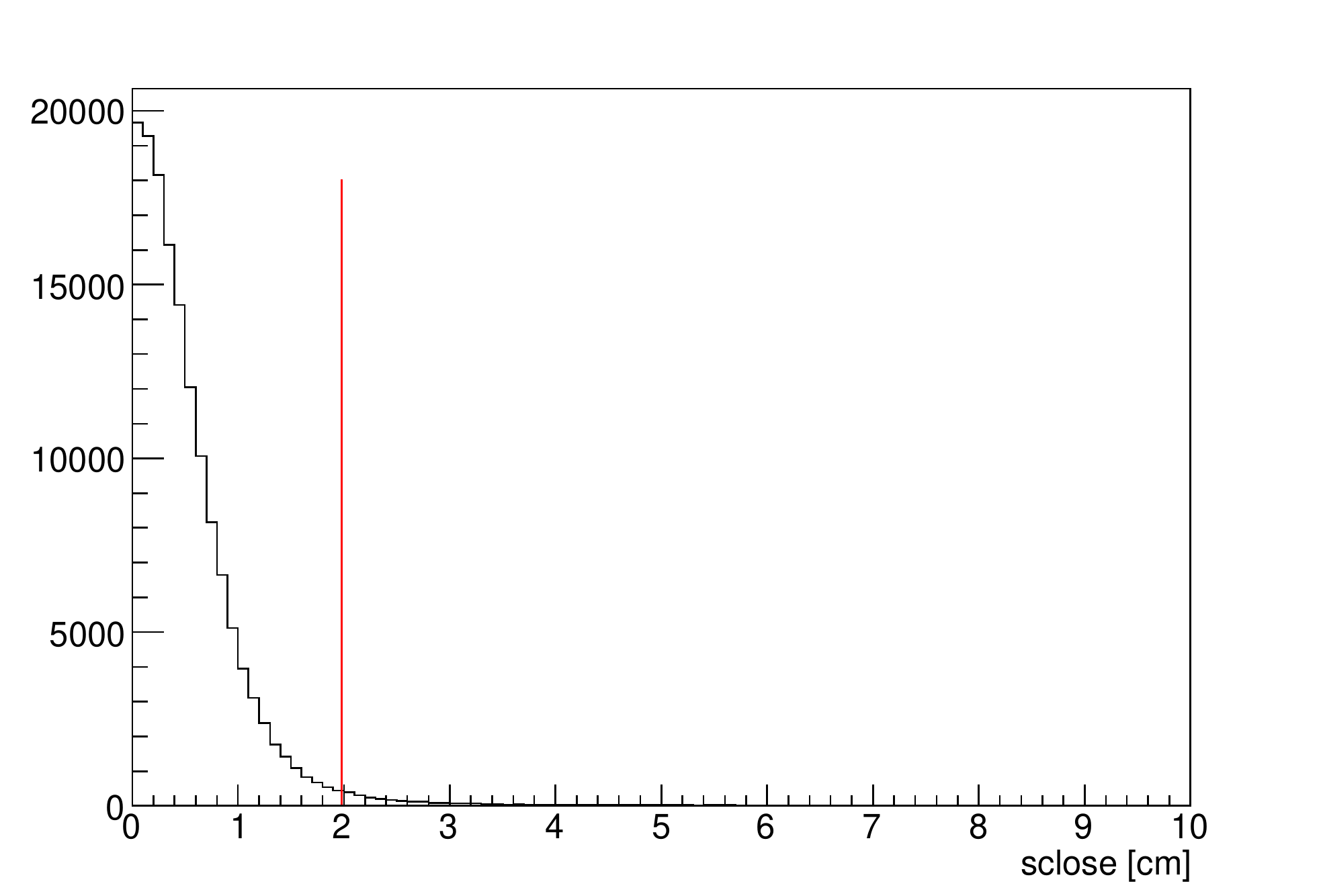}
    \caption{$sclose$ distribution and cut applied to it for setting K2 $\delta_p=0\%$.}
    \label{fig:scut}
  \end{center}
\end{figure}

 The correlation between the FPP front track and rear track is represented by $sclose$, which is the distance of the closest approach between these two tracks. To ensure the quality of the FPP tracking, a cut on $sclose$ of 2 cm or less was applied (see Fig.~\ref{fig:scut}).

\subsubsection{Cone-test Cut}
\begin{figure}
  \begin{center}
    \includegraphics[angle=0,width=0.60\textwidth]{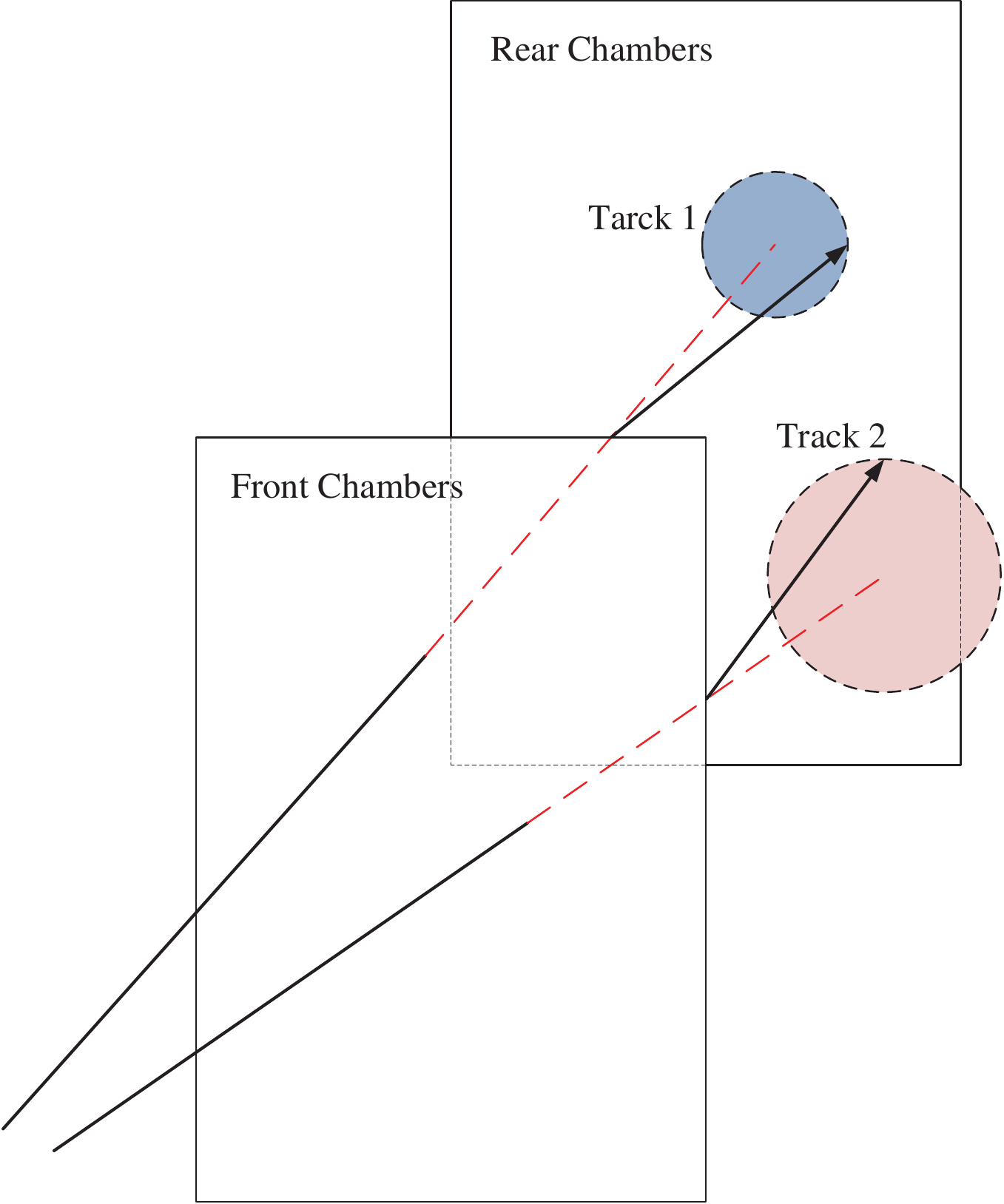}
    \caption{The cone-test in the FPP. The cone of angle
      $\theta_{fpp}$ around track 1 is entirely within the rear
      chambers acceptance, while the one around track 2 is not.
      Track 2 fails the cone-test and is rejected.}
    \label{fig:conetest}
  \end{center}
\end{figure}

To avoid large non-physical asymmetries arising at the edges of the rear
chambers due to the limited size, a cone-test was applied. For a
scattering angle $\theta_{fpp}$, if the entire cone of angle
$\theta_{fpp}$ around the incoming track is within the acceptance of
the rear chamber, this event passes the cone-test. As illustrated
in Fig.~\ref{fig:conetest}, track 1 passes the cone test, while track 2 fails and
is rejected. This test eliminated $\sim 15\%$ of the events. Most of
the rejected events have a scattering angle larger than $20^{\circ}$.
\section{Recoil Polarization Extraction}
For the events passing all the cuts described in the previous section, the recoil
polarization and the form factor ratio was extracted. In this section, the
distribution of the scattering angle of the recoil proton in the
FPP is analyzed. With the reconstruction of the spin precession through the spectrometer,
the proton polarization is extracted. In addition, the discussion of the carbon analyzing power is presented.

\subsection{Angular Distribution}
\begin{figure}
  \begin{center}
    \includegraphics[angle=0,width=0.70\textwidth]{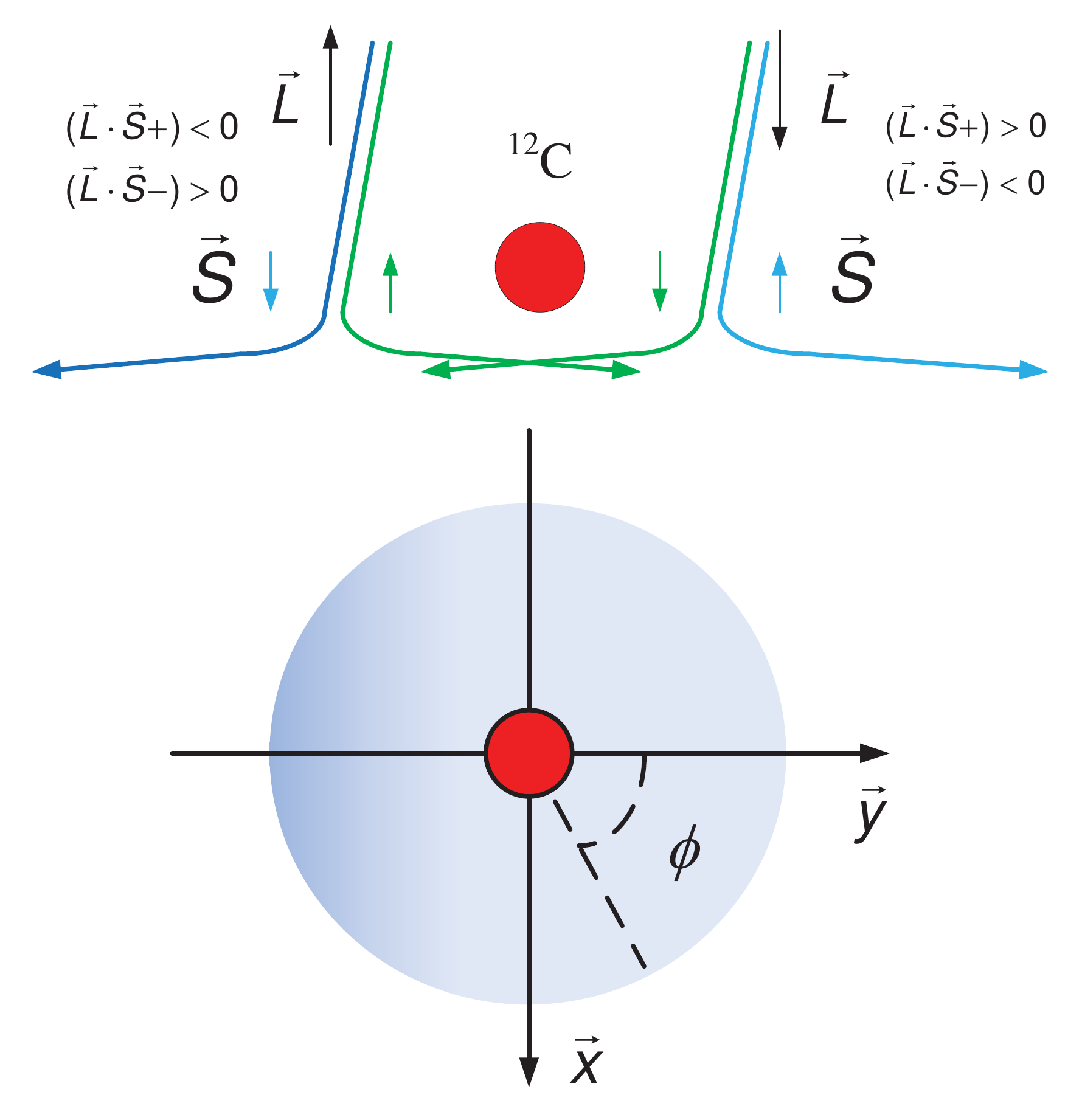}
    \caption{Polarimetry principle: via a spin-orbit coupling, a
      left-right asymmetry is observed if the proton is vertically polarized.}
    \label{fig:analyzer}
  \end{center}
\end{figure}

For the polarization measurement of the recoil proton, the events of interest are those that
have scattered in the carbon analyzer via the strong interaction with a
carbon nucleus. As illustrated in Fig.~\ref{fig:analyzer}, the interaction between the polarized proton and an analyzer
nucleus is sensitive to the direction of the incident proton's spin through a
spin-orbit coupling. A left-right asymmetry in the scattering will be
occurred if the proton spin is preferentially up or down. The sign of the force is
determined by the sign of $\vec L\cdot \vec S$ scalar product, where
$\vec L$ is the orbital angular momentum of the proton with respect to
the analyzer nucleus, and $\vec S$ is the proton spin. Protons are
scattered to the left with spins up and to the right with spin down
(corresponding to the polarization of the incident proton). Hence, an asymmetry in the horizontal direction will be
observed. Similarly, an vertical asymmetry will be observed when
the polarization is along the horizontal direction. However, the longitudinal
component does not result in an asymmetry.

In general, the angular distribution for a
large sample of incident polarized protons is expressed by a sinusoidal function of the vertical $P_x^{fpp}$ and horizontal $P_y^{fpp}$ polarization components:
\begin{equation}
  f^{\pm}(\theta,\phi)=\frac{1}{2\pi}\epsilon(\theta,\phi)(1\pm
  A_y(\theta,T_p)(P_x^{fpp}\cos \phi-P_y^{fpp}\sin \phi)),
  \label{eq:pro0}
\end{equation}
where $\pm$ refers to the sign of the beam helicity. In this expression, $\epsilon(\theta,\phi)$
is the normalized efficiency, which describes
the non-uniformities in the acceptance due to chamber
misalignment and detector inefficiency. $A_y(\theta,T_p)$ is the analyzing power of the reaction
$A(p,N)X$, which represents the strength of the spin-orbit coupling of
the nuclear scattering, thus the sensitivity to the incident particle
polarization. The analyzing power depends on the scattering polar
angle $\theta$ and the proton kinetic energy $T_p$.\footnote{The details of the analyzing
power analysis are presented in Section~\ref{sec:ay}.} For Coulomb
scattering, there is no analyzing power, since there is no spin-orbit coupling.
\subsection{From Focal Plane to the Target Frame}
The FPP measures the proton polarization at the focal plane, however, the
form factor ratio $G_{Ep}/G_{Mp}$ is obtained from the polarization
in the target frame; hence, the measured polarization
at the FPP has to be transported to the one at the target. The relation between the
polarization components in these two frames is complicated due to the
proton spin precession through the spectrometer magnets.
\subsubsection{Dipole Approximation}
\begin{figure}
  \begin{center}
    \includegraphics[angle=0,width=0.70\textwidth]{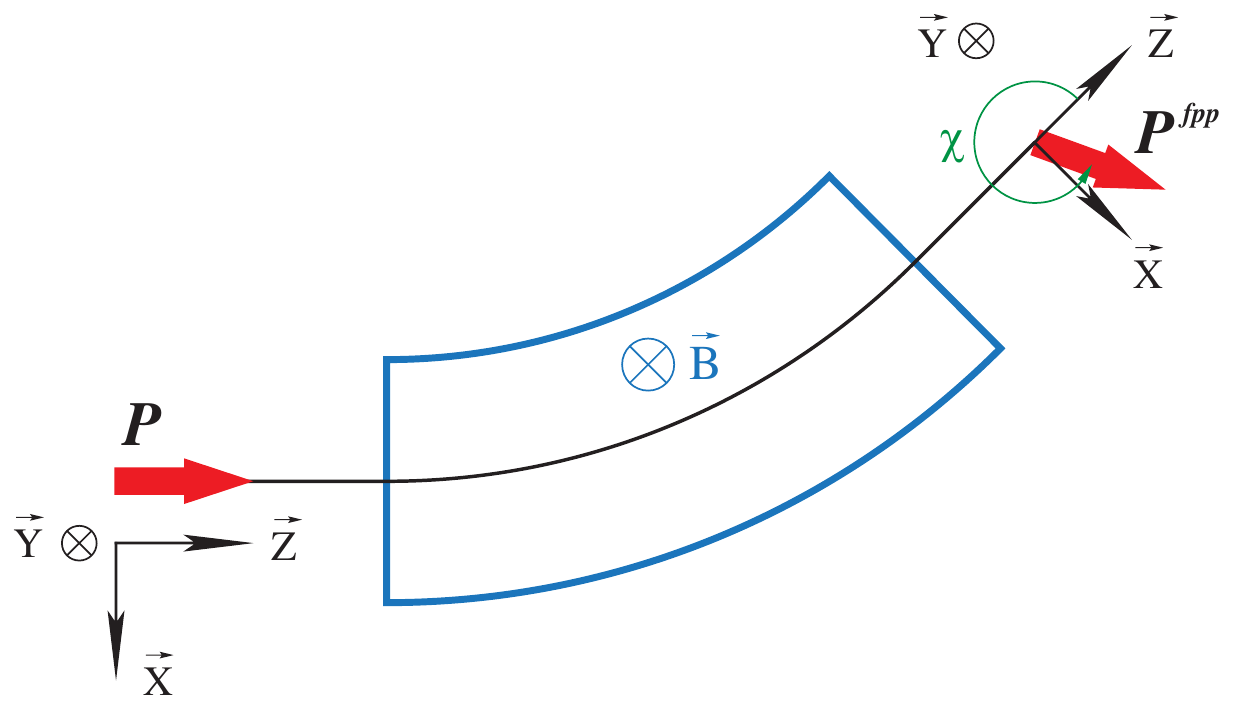}
    \caption{The dipole approximation of the spin transport in the
      spectrometer: only a perfect dipole with sharp edges and a
      uniform field. The proton spin only processes along the out-of-plane direction.}
    \label{fig:dipole}
  \end{center}
\end{figure}

Before we try to fully describe the spin transport through the
spectrometer, a simple approximation can be used by considering a single
perfect dipole, as illustrated in Fig.~\ref{fig:dipole}.
With only a transverse field with respect to the particle momentum, the spin rotates
along the $y$-axis. In this case, the spin precession angle is a simple function of the trajectory bending
angle $\Theta_{bend}$:
\begin{eqnarray}
\chi=\gamma(\mu_p-1)\Theta_{bend},
\end{eqnarray}
where $\gamma = 1/\sqrt{1-\beta^2}$. The HRS dipole central bending angle is $\sim 45^{\circ}$;
in this approximation, the relation between the polarization components at the target and at the focal plane is:
\begin{equation}
\left(\begin{array}{c}
P_x^{fpp}\\
P_y^{fpp}\\
P_z^{fpp}
\end{array}\right)=
\left(\begin{array}{ccc}
\cos\chi &0& \sin\chi\\
0 &1 & 0 \\
-\sin\chi & 0& \cos\chi
\end{array}\right)
\left(\begin{array}{c}
P_x\\
P_y\\
P_z
\end{array}\right).
\label{eq:dipole}
\end{equation}
Note that the transverse component $P_y$ does not precess, since
it is parallel to the magnetic field. As mentioned earlier, in the
one-photon-exchange approximation, the $ep$ elastic scattering process has no
induced polarization, which means that the normal part of the
polarization is:
\begin{equation}
P_x=0.
\end{equation}
Since the FPP can measure only the two perpendicular components to the
momentum at the focal plane, the relation in Eq.~\ref{eq:dipole} is further simplified:
\begin{equation}
\left(\begin{array}{c}
P_x^{fpp}\\
P_y^{fpp}
\end{array}\right)=
\left(\begin{array}{cc}
0 & \sin\chi\\
1 & 0
\end{array}\right)
\left(\begin{array}{c}
P_y\\
P_z
\end{array}\right).
\label{eq:fptotg}
\end{equation}
Using the angular distribution function from Eq.~\ref{eq:pro0}, the
polarization components at the focal plane can be extracted. By taking
the difference of the distributions with respect to two beam helicities,
the efficiency term cancels in the first order. Assuming
the efficiency is fairly uniform over the FPP so that the higher order terms
can be ignored, the asymmetry difference distribution has the simple form:
\begin{equation}
f^{diff}=f^+-f^-\approx\frac{1}{\pi}[A_y(P^{fpp}_x\cos\phi-P^{fpp}_y\sin\phi)].
\label{eq:diff}
\end{equation}
This expression can be written equivalently as:
\begin{equation}
f^{diff}=C\cos(\phi+\delta),
\end{equation}
where:
\begin{eqnarray}
C&=&\frac{1}{\pi}\sqrt{(P^{fpp}_x)^2+(P_y^{fpp})^2}{}
\nonumber\\
{}\tan\delta &=&\frac{P_y^{fpp}}{P_x^{fpp}}.
\end{eqnarray}
In the simple dipole approximation (Eq.~\ref{eq:fptotg}), $P_y^{fpp}$
is equal to the transverse component at the target frame $P_y$,
which is proportional to the product $G_{Ep}G_{Mp}$, and $P_x^{fpp}$
is related to the longitudinal component which is proportional to
$G_{Mp}^2$, via $P_x^{fpp}=\sin\chi P_z$. Therefore, the phase shift
of the helicity difference distribution is a direct measure of $G_{Ep}/ G_{Mp}$:
\begin{eqnarray}
\frac{G_{Ep}}{G_{Mp}}=K\frac{P_y}{P_z}\approx
K\sin\chi \left(\frac{P_y^{fpp}}{P_x^{fpp}}\right),
\end{eqnarray}
where $K=\frac{E+E'}{m}\tan^2(\theta_e/2)$.

\begin{figure}
  \begin{center}
    \includegraphics[angle=0,width=0.85\textwidth]{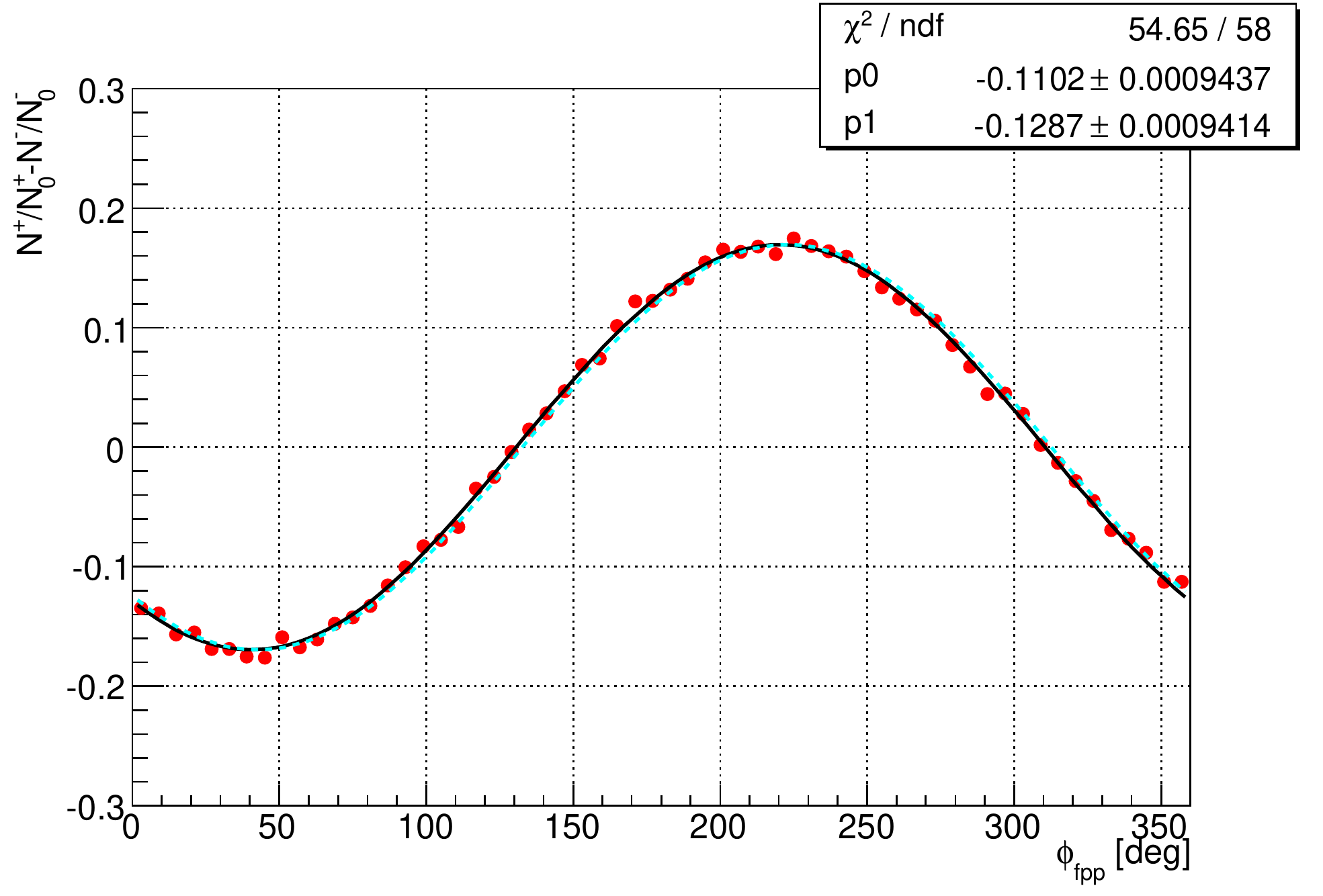}
    \caption{Asymmetry difference distribution along the azimuthal scattering
      angle $\phi_{fpp}$ at kinematics K6 ($Q^2=0.5~\mathrm{GeV}^2$). The black solid curve represents the
      sinusoidal fit to the data ($\chi^2/ndf=0.94$). The dashed light blue
    curve corresponds to a hypothetical distribution assuming $\mu_pG_{Ep}/G_{Mp}=1$ in dipole approximation.}
    \label{fig:hdiff}
  \end{center}
\end{figure}

\begin{figure}
  \begin{center}
    \includegraphics[angle=0,width=0.75\textwidth]{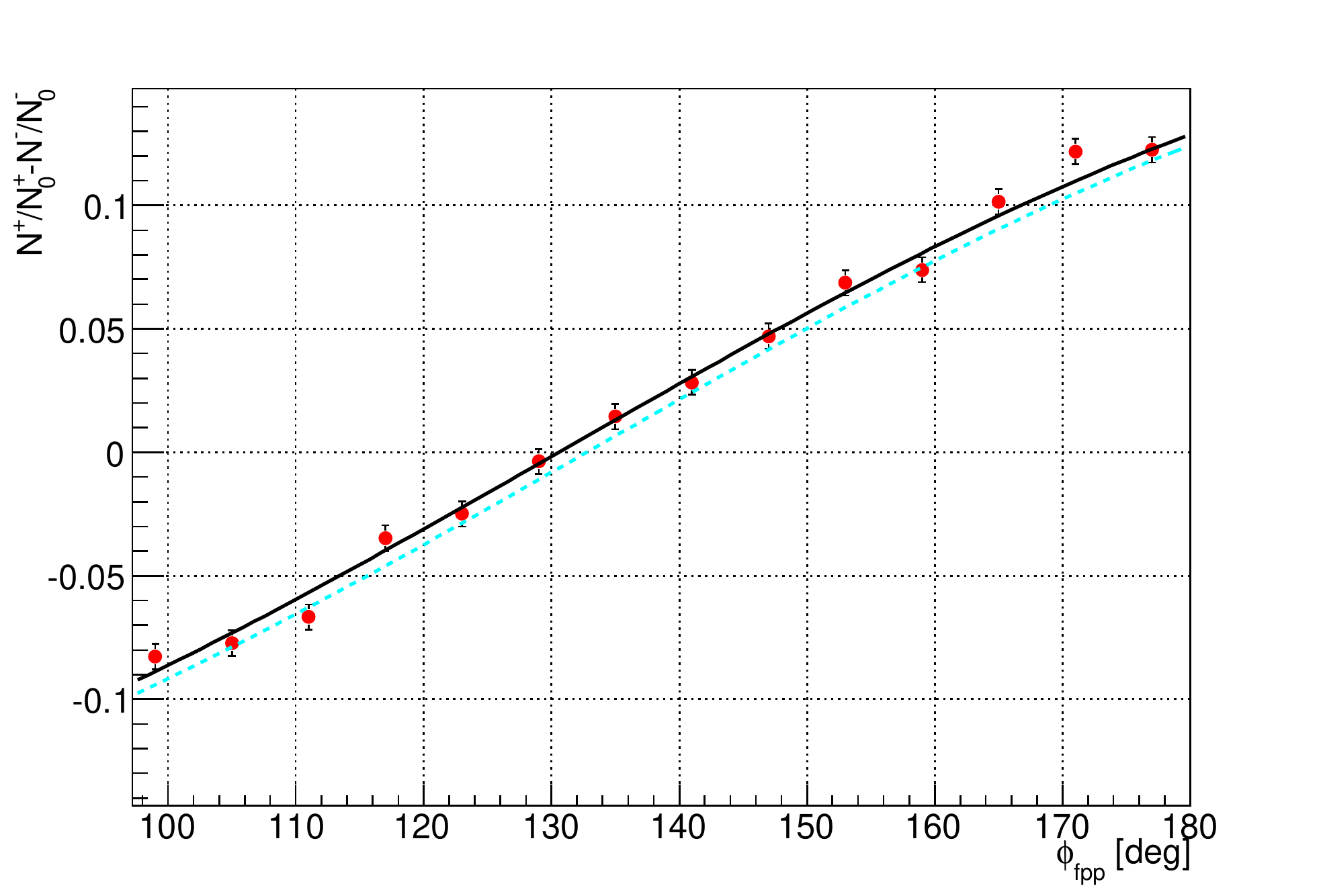}
    \caption{Close up view of Fig.~\ref{fig:hdiff}. The black solid curve represents the
      sinusoidal fit to the data, while the dashed light blue curve corresponds to a hypothetical
      distribution assuming $\mu_pG_{Ep}/G_{Mp}=1$ in dipole approximation.
    There is $\sim 2^{\circ}$ shift between these two curves at the zero crossing.}
    \label{fig:hdiff1}
  \end{center}
\end{figure}

Fig.~\ref{fig:hdiff} presents the helicity difference $f^{diff}$ and a fit to the data.
The black solid curve is a sinusoidal fit to the data (K6,
$\delta_p=0\%$), with a $\chi^2$ of 0.94 per degree of freedom. The
dashed light blue curve is a hypothetical distribution assuming
$\mu_pG_{Ep}/G_{Mp}=1$, as predicted by the dipole model. By zooming
in this figure, one can see a small but clear deviation between these two curves in Fig.~\ref{fig:hdiff1},
which is a direct indication that the form factor ratio deviates from unity in dipole approximation.
\subsubsection{Full Spin Precession Matrix and COSY}
In reality, the spectrometer magnets are more complicated than just a
simple perfect dipole. First, the field is not uniform inside the
dipole, it is distorted by the fringe fields at the entrance and exit
apertures. In addition, there are three quadrupoles that have field components in both
$x$ and $y$ directions; hence, the matrix that relates the two
polarizations measured in the FPP and in the target frame takes the general form:
\begin{equation}
\left(\begin{array}{c}
P_x^{fpp}\\
P_y^{fpp}\\
P_z^{fpp}
\end{array}\right)=
\left(\begin{array}{ccc}
S_{xx} & S_{xy} & S_{xz}\\
S_{yx} & S_{yy} & S_{yz}\\
S_{zx} & S_{zy} & S_{zz}
\end{array}\right)
\left(\begin{array}{c}
P_x\\
P_y\\
P_z
\end{array}\right).
\label{eq:sp_matrix}
\end{equation}
The coefficients $S_{ij}$ depend on the trajectory of the proton as it passes through
the spectrometer. Within the HRS acceptance, the protons recoiling with
different angles and momenta at the target frame have different trajectories
inside the spectrometer, they experience different magnetic
fields along their trajectories, and hence, their spin precession is
different. Therefore, the coefficients are calculated event by
event to account for this difference in path length.

The COSY model was used to calculate the spin precession matrices. It is a
differential algebra-based code written by M. Berz of the Michigan
State University~\cite{cosy}. This model is originally developed for the simulation, analysis
and design of particle optics systems. COSY takes the dimensions and
positions of the magnetic elements, such as the diameter and the
path length of the magnet, the central momentum of the particle,
etc. as the inputs. The fringe fields are also taking into account by a set of
coefficients that were determined from measurements when Hall A was commissioned. With all these ingredients, COSY calculates a table of the expansion
coefficients $C_{ij}^{klmnp}$ of the rotation matrix. This matrix
is calculated event by event based on the particle trajectory
variables located at the target coordinate system (TCS), which is defined in Section~\ref{sec:tcs}:
\begin{equation}
S_{ij}=\sum_{k,l,m,n,p}C_{ij}^{klmnp}r_1^kr_2^lr_3^mr_4^nr_5^p
\label{eq:coef}
\end{equation}
where:
\begin{eqnarray}
r_1&=&x\\
r_2&=&p_x/p_0\\
r_3&=&y\\
r_4&=&p_y/p_0\\
r_5&=&\delta_K=(K-K_0)/K_0.
\end{eqnarray}
$x$ and $y$ are the positions, $p_0$ and $K$ are the particle momentum
and kinetic energy\footnote{The particle mass is assigned in the code so that the matrix is calculated according to the correct momentum.} respectively. From Eq.~\ref{eq:sp_matrix},
the transverse polarization component at the focal
plane is $P_{y}^{fpp}=S_{yy}P_y+S_{yz}P_z$. Compared to the dipole
approximation, the non-zero term $S_{yz}$ brings the contribution from the
longitudinal target component $P_z$; this term is mainly due to the
precession of the spin in the non-dispersive direction from the
quadrupoles, which is neglected in the dipole approximation.

The spin rotation matrix given by COSY only relates
the polarization at the target coordinate system (TCS) to the
transport coordinate system (TRCS). Therefore, two addition rotations, from the target
scattering frame to the target coordinate system (TCS) and from the transport
coordinate system (TRCS) to the focal plane frame at the FPP, are needed.

\begin{figure}
  \begin{center}
    \includegraphics[angle=0,width=0.65\textwidth]{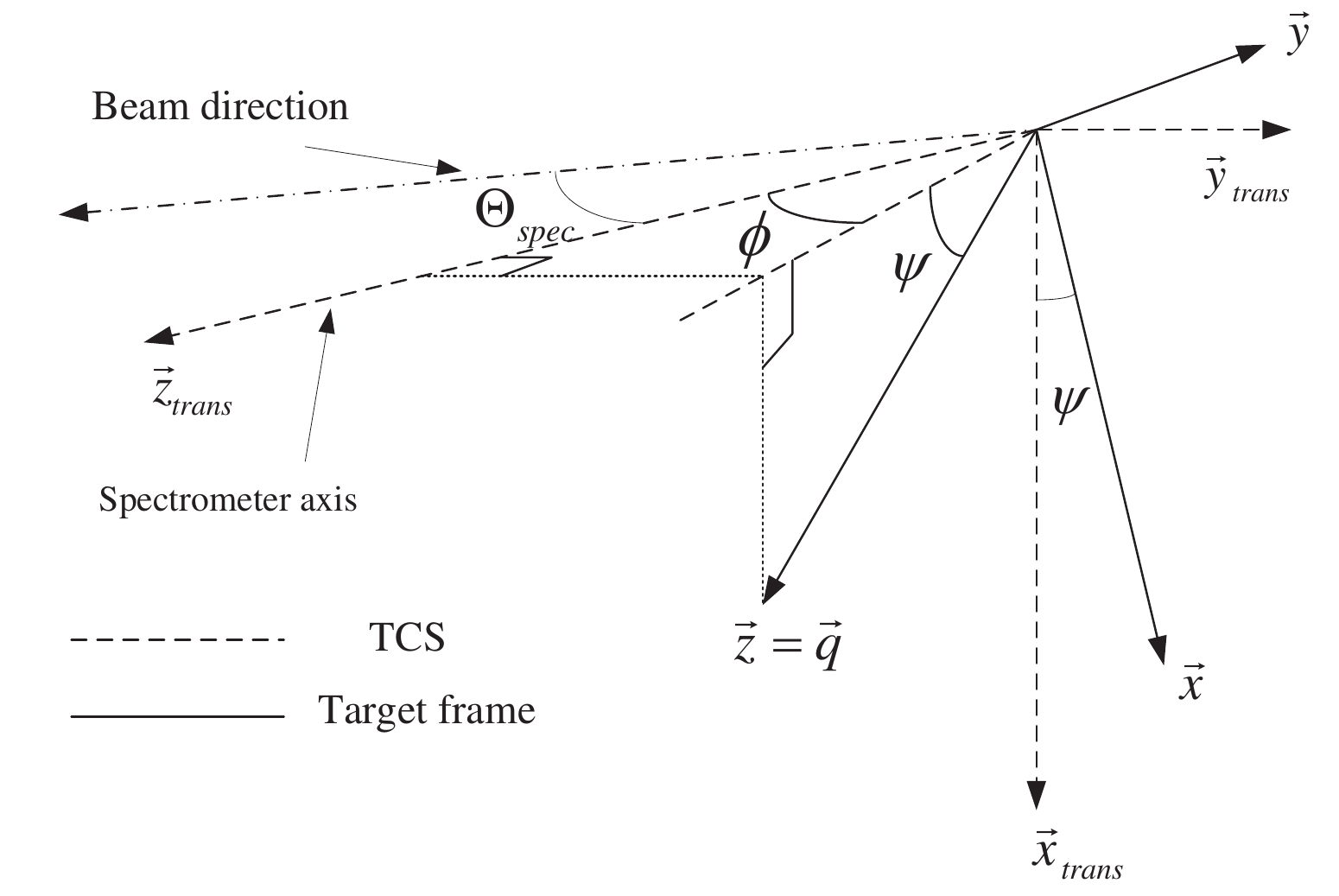}
    \caption{The target scattering coordinate system (solid lines) is the
      frame where the polarization is expressed while the
   TCS (dashed lines) is the one in which COSY does the calculation.}
    \label{fig:coord1}
  \end{center}
\end{figure}

First, we need to express the proton track in the TCS. As illustrated in
Fig.~\ref{fig:coord1},
the target scattering frame is defined as:
\begin{eqnarray}
\vec x &=& \frac{\vec k_i\times \vec k_f}{|\vec k_i\times \vec k_f|}{}
\nonumber\\
{}\vec y &=& \vec z\times \vec x{}
\nonumber\\
{}\vec z &=& \frac{\vec k_i-\vec k_f}{|\vec k_i- \vec k_f|},
\label{eq:scaframe}
\end{eqnarray}
where $\vec k_i$ and $\vec k_f$ are vectors along the incident and
scattered electron momenta, respectively. In the elastic case, $\vec
q$ is the vector along the momentum of the recoil proton:
\begin{eqnarray}
\vec q & = & \vec k_i-\vec k_f,
\end{eqnarray}
so that:
\begin{equation}
\vec k_i\times\vec k_f=\vec k_i\times\vec k_i-\vec k_i\times\vec
q=\vec q\times\vec k_i.
\end{equation}
Eq.\ref{eq:scaframe} becomes:
\begin{eqnarray}
  \vec x &=& \frac{\vec q\times \vec k_i}{|\vec q\times \vec k_i|}{}
  \nonumber\\
  {}\vec y &=& \vec z\times \vec x{}
  \nonumber\\
  {}\vec z &=& \frac{\vec q}{|\vec q|}.
  \label{eq:scaframe1}
\end{eqnarray}
In the lab frame, $\vec k_i$ is the beam direction, which is along the
$z$-axis. The momentum transfer $\vec q$ is in the direction of the outgoing proton, they both
can be expressed in the TCS:
\begin{equation}
\vec k_i=\left(\begin{array}{c}
0\\
-\sin\Theta_{spec}\\
\cos\Theta_{spec}
\end{array}\right),
~\vec q = \left(\begin{array}{c}
\sin\psi\\
\cos\psi\sin\phi\\
\cos\psi\cos\phi
\end{array}\right).
\label{eq:trans}
\end{equation}
Finally, the matrix of the transformation from the target frame to the TCS,
$\mathbf{T_0}$ can be obtained by Eq.~\ref{eq:scaframe} and
Eq.~\ref{eq:trans}.

\begin{figure}
  \begin{center}
    \includegraphics[angle=0,width=0.45\textwidth]{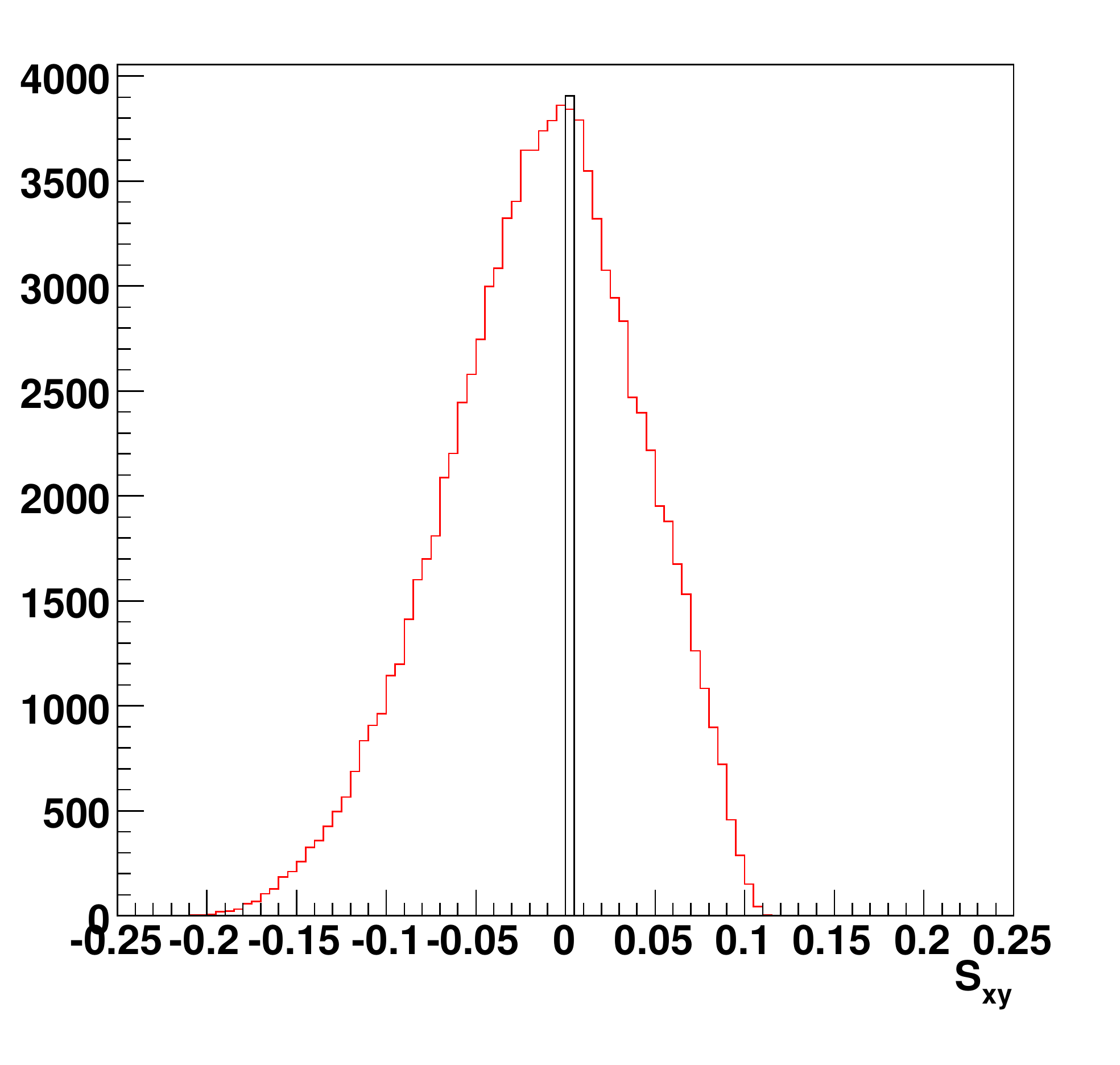}
    \includegraphics[angle=0,width=0.45\textwidth]{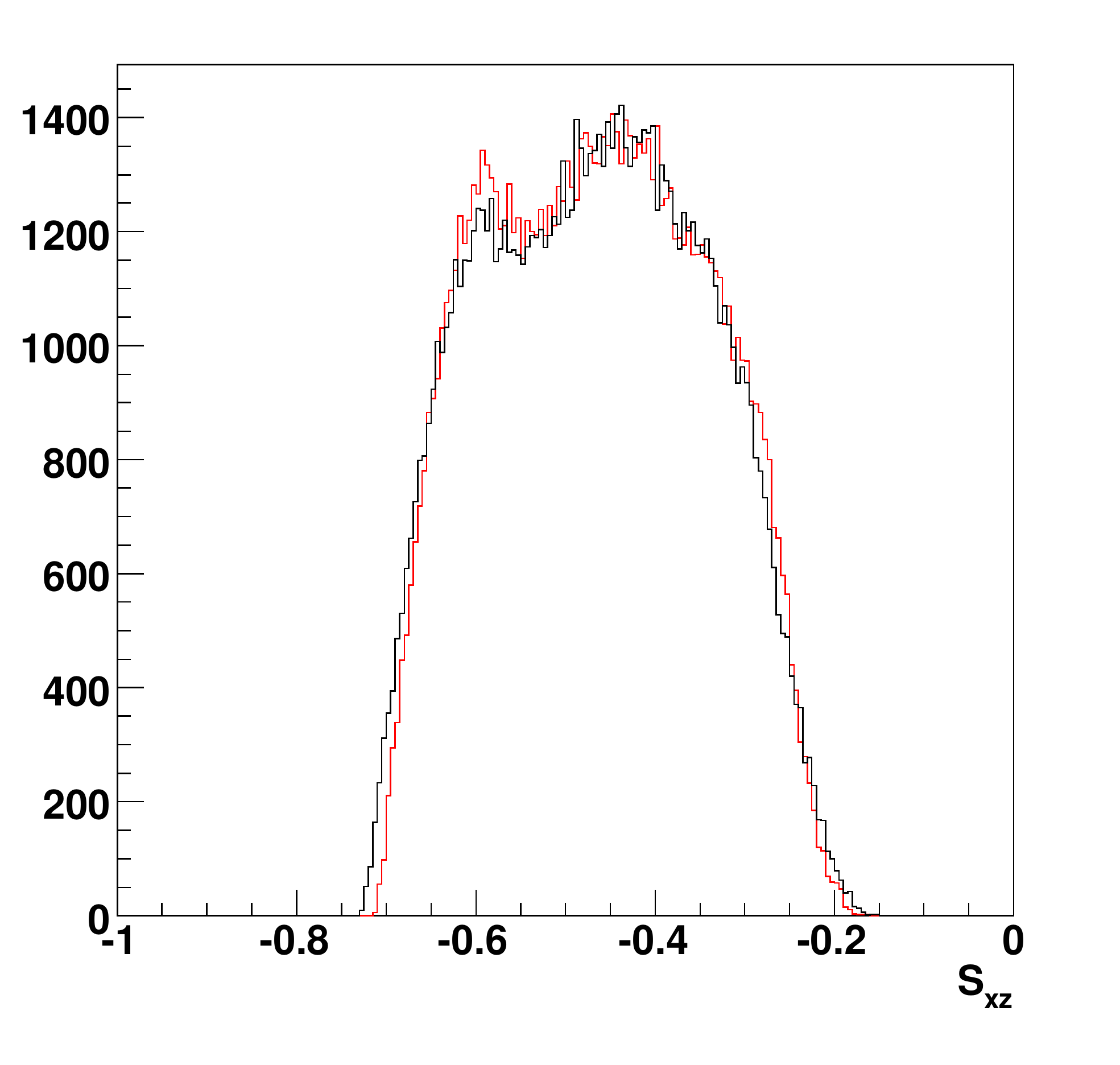}
    \includegraphics[angle=0,width=0.45\textwidth]{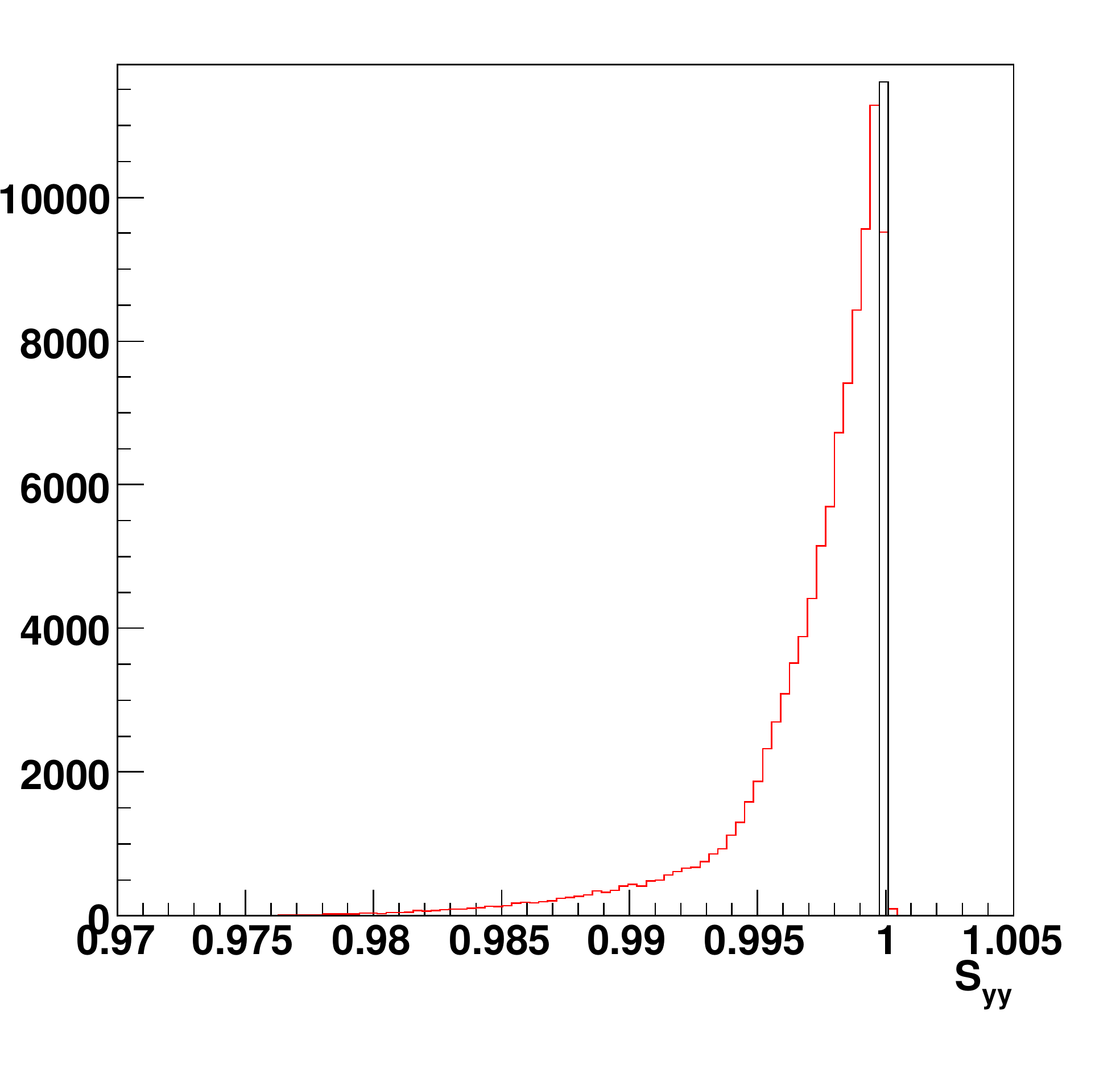}
    \includegraphics[angle=0,width=0.45\textwidth]{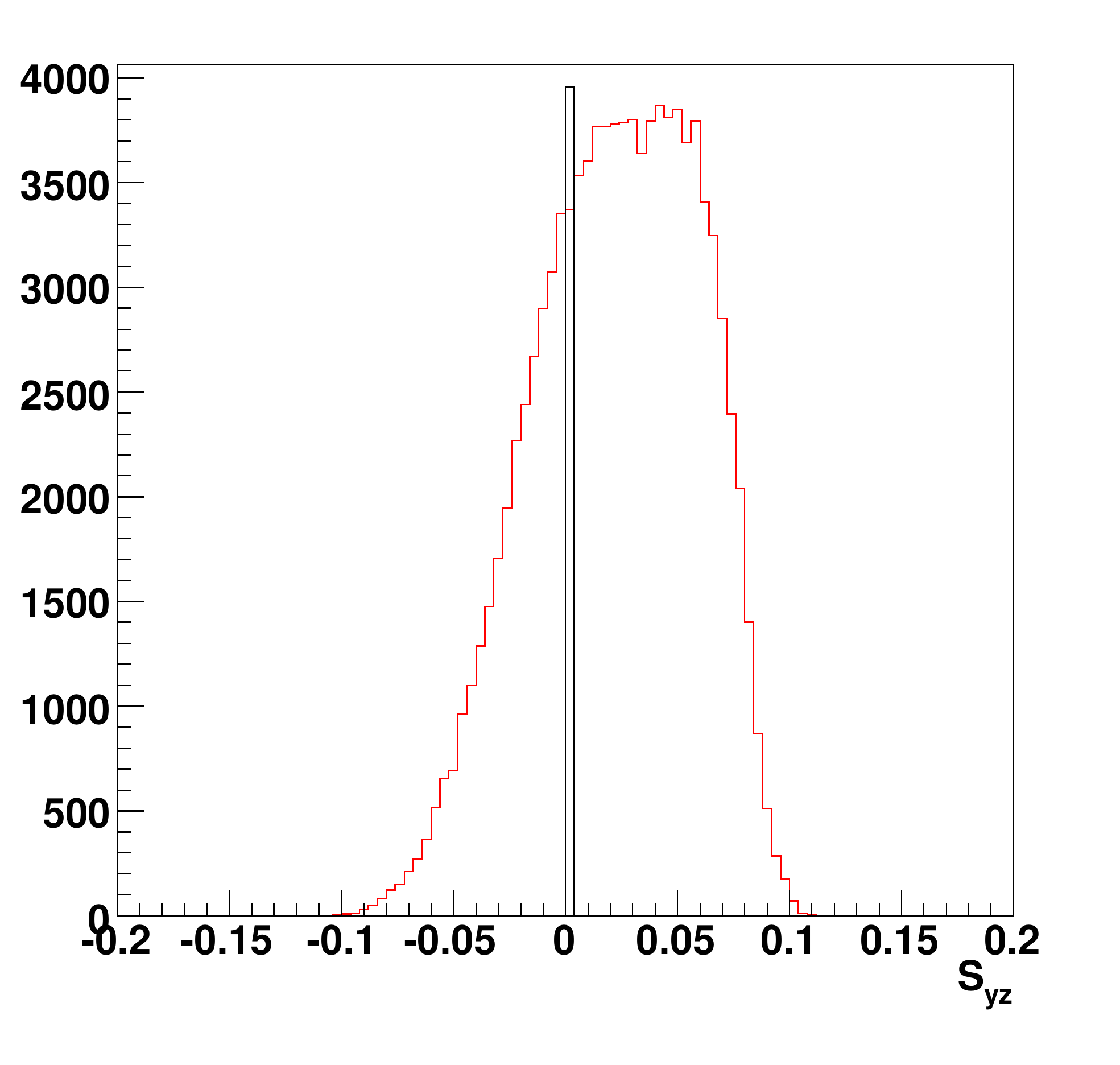}
    \caption{Histograms of the four spin transport matrix elements, $S_{xy}$ (upper left), $S_{xz}$ (upper right), $S_{yy}$ (lower left) and $S_{yz}$ (lower right) at $Q^2 = 0.7$ GeV$^2$ for the elastic events. The ones plotted in black are from dipole approximation, and the ones in red are from the full spin transport matrix generated by COSY. For the dipole approximation, $S_{xy}$ and $S_{yz}$ are exactly zero, and $S_{xy}=1$ by ignoring the transverse components of the field. The full spin precession matrix gives broad distributions for these elements which represent the effect from the quadrupoles and the dipole fringe field.}
    \label{fig:full_sp}
  \end{center}
\end{figure}

Second, we need to perform a rotation from the TRCS to the FPP local
frame, whose $z$-axis is along the proton momentum. In a similar way
as defined in Fig.~\ref{fig:carte}, the transformation can be done by a rotation
around the $x$-axis by an angle $\phi_{f}$, which is then followed by a rotation by
an angle $\psi_{f}$ around the new $y$-axis. For this transformation, the
coordinates are related by the matrix $\mathbf{T_1}$:
\begin{equation}
\left(\begin{array}{c}
P_x^{fpp}\\
P_y^{fpp}\\
P_z^{fpp}
\end{array}\right)
=
\underbrace{
\left(\begin{array}{ccc}
\cos\phi_f & -\sin\psi_f\sin\phi_f & -\sin\psi_f\cos\phi_f\\
0 & \cos\phi_f & -\sin\phi_f\\
\sin\psi_f & \cos\psi_f\sin\phi_f & \cos\psi_f\cos\psi_f
\end{array}\right)}_{\mathbf{T_1}}
\left(\begin{array}{c}
P_x^{tr}\\
P_y^{tr}\\
P_z^{tr}
\end{array}\right).
\end{equation}
Therefore, the total rotation matrix $\mathbf{S}$ consists of
$\mathbf{T_0},~\mathbf{T_1}$ and the spin rotation matrix
$\mathbf{S_{sp}}$ given by COSY. The measured polarization at the focal plane can be
expressed as
\begin{equation}
\mathbf{P^{fpp}}=\underbrace{\mathbf{T_1}\mathbf{S_{sp}}\mathbf{T_0}}_{\mathbf{\large{S}}}\mathbf{P^{tg}}.
\label{eq:total_rot}
\end{equation}
As an example, Fig.~\ref{fig:full_sp} shows the four major elements of the full spin transport matrix for one of the kinematic settings.

\subsection{Extraction of Polarization Observables}
With the scattering angles reconstructed by the FPP and the rotation
matrix calculated by COSY, we are able
to extract the polarization components at the target. There are 3 different methods to extract the polarization
observables, as discussed in ~\cite{Besset}. For the transferred polarization analysis, the weighted-sum method is used. The advantage with this technique is that by using different beam helicities, we can ignore the efficiency of the acceptance in extracting the transferred polarization. The detailed formalism of the weighted-sum method will be presented in this section.
\subsubsection{Weighted-sum}
As noted earlier, the probability that a proton scatters in
the analyzer with angles $(\theta,\phi)$ with a polarization
$(P_x^{fpp}, P_y^{fpp})$ is given by Eq.~\ref{eq:pro0}:
\begin{equation}
f^{\pm}(\phi)=\frac{1}{2\pi}\epsilon(1\pm A_y(P_y^{fpp}\sin \phi - P_x^{fpp}\cos \phi)),
\end{equation}
where $\epsilon$ is the normalized instrumental efficiency (acceptance):
\begin{equation}
\epsilon(\phi_i)=\frac{f^+ + f^-}{\pi}.
\end{equation}
By considering the spin transport, the probability function can be written in terms of the polarization components at the target frame:
\begin{equation}
f(\phi)=\frac{1}{2\pi}\epsilon(1+\lambda_{x}P_x^{tg}+\lambda_{y}hP_y^{tg}+\lambda_{z}hP_z^{tg}),
\label{eq:diff_full}
\end{equation}
where
\begin{eqnarray}
\lambda_{x}&=&A_y(S_{yx}\sin\phi-S_{xx}\cos\phi){}
\nonumber\\
{}\lambda_{y}&=&\eta hA_y(S_{yy}\sin\phi-S_{xy}\cos\phi){}
\nonumber\\
{}\lambda_{z}&=&\eta hA_y(S_{yz}\sin\phi-S_{xz}\cos\phi),
\end{eqnarray}
where $\eta$ is the sign for the beam helicity, and $h$ is the beam
polarization. Note that the contribution from the induced (normal) polarization
$P_x^{tg}$ is beam helicity independent. In the Born approximation, $P_x^{tg}=0$; hence, Eq.~\ref{eq:diff_full} reduces to:
\begin{equation}
f(\phi)=\frac{1}{2\pi}\epsilon(\phi)(1+\lambda_{y}hP_y^{tg}+\lambda_{z}hP_z^{tg}).
\label{eq:diff_red}
\end{equation}

As derived in~\cite{Besset}, for Eq.~\ref{eq:diff_red}, with different beam helicities we can always construct an effective acceptance that has a symmetry period of $\pi$ in $\phi$ so that the acceptance $\epsilon$ cancels in the integral. We can obtain the equations:
\begin{eqnarray}
\int_0^{2\pi}f(\phi)\lambda_y d\phi &=& hP_y^{tg}\int_0^{2\pi}f(\phi)\lambda^2_yd\phi+{}
\nonumber\\
&&{}hP_z^{tg}\int_0^{2\pi}f(\phi)\lambda_y\lambda_z d\phi\\
\int_0^{2\pi}f(\phi)\lambda_z d\phi &=& hP_y^{tg}\int_0^{2\pi}f(\phi)\lambda_y\lambda_zd\phi+{}
\nonumber\\
&&{}hP_z^{tg}\int_0^{2\pi}f(\phi)\lambda^2_z d\phi,
\label{eq:integral}
\end{eqnarray}
since for $n+m$ odd,
\begin{equation}
\int_0^{2\pi}\epsilon(\phi)\sin^m\phi\cos^n\phi d\phi =0.
\end{equation}
By replacing the integrals in Eqs.~\ref{eq:integral} with corresponding sums over the observed events, we have
\begin{equation}
\left(\begin{array}{c}
\sum_i\lambda_{y,i}\\
\sum_i\lambda_{z,i}
\end{array}\right)
=\left(\begin{array}{cc}
\sum_i\lambda_{y,i}\lambda_{y,i}&\sum_i\lambda_{z,i}\lambda_{y,i}\\
\sum_i\lambda_{y,i}\lambda_{z,i}&\sum_i\lambda_{z,i}\lambda_{z,i}
\end{array}\right)
\left(\begin{array}{c}
P_y^{tg}\\
P_z^{tg}
\end{array}\right).
\label{eq:sums}
\end{equation}

With the accumulation of a large event sample, Eq.~\ref{eq:sums}
can be solved to obtain $P_y^{tg}$ and $P_z^{tg}$. Eq.~\ref{eq:sums} is rewritten as:
\begin{eqnarray}
\mathbf{B}&=&\mathbf{M}\cdot\mathbf{P}{}
\nonumber\\
{}\mathbf{P}&=&\mathbf{M}^{-1}\cdot \mathbf{B}.
\end{eqnarray}
The statistical error is given by:
\begin{eqnarray}
\Delta(P_i)=\sqrt{(\mathbf{M}^{-1})_{ii}}
\end{eqnarray}
with $i=y,z$, and the correlation factor between the two is
\begin{equation}
\rho_{ij}=\frac{(\mathbf{M}^{-1})_{ij}}{\sqrt{(\mathbf{M}^{-1})_{ii}(\mathbf{M}^{-1})_{jj}}}.
\label{eq:rho}
\end{equation}
Then the form factor ratio is given by:
\begin{equation}
\mu_p\frac{G_{Ep}}{G_{Mp}}=Kr
\label{eq:ratio}
\end{equation}
where $a=P_y$, and $b=P_z$, and $r=a/b$. $K$ is the kinematic
factor:
\begin{equation}
K=-\mu_p\frac{E_e+E_{e'}}{2M_p}\tan\frac{\theta_e}{2}.
\end{equation}
The statistical errors are calculated by:
\begin{equation}
\Delta (\frac{G_{Ep}}{G_{Mp}})=\sqrt{(\frac{dr}{da})^2(\Delta
  a)^2+(\frac{dr}{db})^2(\Delta b)^2+2\rho\frac{dr}{da}\Delta
  a\frac{dr}{db}\Delta b}
\end{equation}
where:
\begin{eqnarray}
\frac{dr}{da}&=&K\frac{1}{b}{}
\nonumber\\
{} \frac{dr}{db}&=&-K\frac{a}{b^2}.
\end{eqnarray}
The weighted-sum technique is valid under the condition that
there is no induced polarization in the physics asymmetry,
since this helicity independent term breaks the symmetry period
of $\epsilon$. In reality, non-zero $P_x^{tg}$ may arise from the $2\gamma$ exchange process.
From a detailed study which considered the non-zero induced polarization in Appendix C, we have concluded that
the weighted-sum method is valid given the required precision.
\subsection{\label{sec:ay}Analyzing Power}
From Eq.~\ref{eq:ratio}, one can see that since the polarization components are measured simultaneously, for the ratio of $P_y$ and $P_z$, the knowledge of the beam polarization $h$ and the analyzing power $A_y$, which cancel out in the ratio is not necessary. However, certain properties of the analyzing power are useful in giving the correct statistical uncertainty. As noted earlier, the analyzing power $A_y$ depends only on the scattering angle $\theta_{fpp}$ and the proton kinematic energy $T_p$. For example, by taking the $\theta_{fpp}$ dependence of the analyzing power into account provides more weight to events scattered at angles corresponding to high analyzing power and less weight to events scattered at smaller angles with low analyzing power, which is dominated by Coulomb scattering.

Although the absolute value of the analyzing power is irrelevant in the extraction of the form factor ratio, it is a byproduct of this measurement. In the first pass of the data analysis, the analyzing power $A_y$ was ignored by setting it to be 1. Since the beam polarization is well know from the M\o ller measurements and is included in the analysis, the solutions of Eq.~\ref{eq:sums} become $A_yP_y$ and $A_yP_z$. We can rewrite the proton polarization components as a function of the form factor ratio only, independent of the beam polarization and the the analyzing power:
\begin{eqnarray}
P_y&=&\frac{-2\sqrt{\tau(1+\tau)}\tan\frac{\theta_e}{2}G_EG_M}{G^2_E+(\tau/\epsilon)G^2_M}=
\frac{-2\sqrt{\tau(1+\tau)}\tan\frac{\theta_e}{2}\frac{G_E}{G_M}}{(\frac{G_E}{G_M})^2+(\tau/\epsilon)}\\
P_z&=&\frac{\frac{E+E'}{m}\sqrt{\tau(1+\tau)}\tan^2\frac{\theta_e}{2}G^2_M}{G^2_E+(\tau/\epsilon)G^2_M}
=\frac{\frac{E+E'}{m}\sqrt{\tau(1+\tau)}\tan^2\frac{\theta_e}{2}}{(\frac{G_E}{G_M})^2(\tau/\epsilon)}.
\end{eqnarray}
With the measured ratio $G_E/G_M$, we can calculate $P_y$ and $P_z$. By comparing them with the measured $A_yP_y$ and $A_yP_z$, we can extract the analyzing power:
\begin{eqnarray}
A_y &=& \alpha\frac{a^2}{b}+\beta b\\
\Delta A_y&=&\sqrt{\left(\frac{dA_y}{da}\right)^2(\Delta a)^2+\left(\frac{dA_y}{db}\right)^2(\Delta b)^2+2\rho\frac{dA_y}{da}\Delta a\frac{dA_y}{db}\Delta b},
\end{eqnarray}
with:
\begin{eqnarray}
\frac{dA_y}{da} &=& 2\alpha\frac{a}{b}\\
\frac{dA_y}{db} &=& -\alpha\left(\frac{a}{b}\right)^2+\beta,
\end{eqnarray}
where
\begin{eqnarray}
\alpha &=&\frac{E_e+E_{e'}}{4m\sqrt{\tau(1+\tau)}}\\
\beta &=&\frac{\tau[1+2(1+\tau)\tan^2(\frac{\theta_e}{2})]}{\frac{E_eE_{e'}}{m}\sqrt{\tau(1+\tau)}\tan^2(\frac{\theta_e}{2})}.
\end{eqnarray}
In the above expression, $\rho$ is the correlation factor as defined in Eq.~\ref{eq:rho}, and $a=A_y P_y$ and $b=A_yP_z$ are the output of the first pass analysis with $A_y=1$.

A parameterization of the analyzing power for large solid angle spectrometers
was first suggested by Ransome {\it et al.}~\cite{ransome} and was later
expanded by McNaughton {\it et al.}~\cite{mcnau} for inclusive $p^{12}C$
experiments at Los Alamos. The parameterization is divided at $T_p = 450$
MeV into a ``low energy region'' and a ``high energy region'', where $T_p$
is the proton kinetic energy at the center of the carbon analyzer.

For the low energy fit, the suggested fitting function in~\cite{mcnau,ransome} is:
\begin{equation}
A_y = \frac{ar}{1+br^2+cr^4},
\label{eq:low_ay}
\end{equation}
where $r = p_p\sin(\theta_{fpp})$ and $p_p$ is the proton momentum in GeV/$c$
at the center of the carbon analyzer. The coefficients $a$, $b$, $c$ are
polynomials of the momentum. In 2006, the LEDEX~\cite{jg_ay} experiment extracted the carbon
analyzing power for proton energies from 82 to 217 MeV. A similar functional
form as shown in Eq.~\ref{eq:low_ay} was used:
\begin{equation}
A_y = \frac{ar}{1+br^2+cr^4+dr^6},
\label{eq:low_jg}
\end{equation}
where the $dr^6$ term was added in order to improve the quality of fit. The coefficients
are expanded as follows:
\begin{eqnarray}
a = \sum_{i=0}^4a_i(p_p-p_0)^i\\
b = \sum_{i=0}^4b_i(p_p-p_0)^i\\
c = \sum_{i=0}^4c_i(p_p-p_0)^i\\
d = \sum_{i=0}^4d_i(p_p-p_0)^i,
\end{eqnarray}
where $p_0$, $a_i$, $b_i$, $c_i$ and $d_i$ are the parameters of the fit.

For this experiment, we have much better statistics and much larger proton energy
coverage (90 to 360 MeV). We used the same functional form
in Eq.~\ref{eq:low_jg}. The analyzing power was extracted by binning the data with respect to
$\theta_{fpp}$ and $T_p$, and the average values of
each bin were used to fit the parameters. The parameterization based on the new data is provided in Appendix D.

\begin{figure}
  \begin{center}
    \includegraphics[angle=0,width=0.7\textwidth]{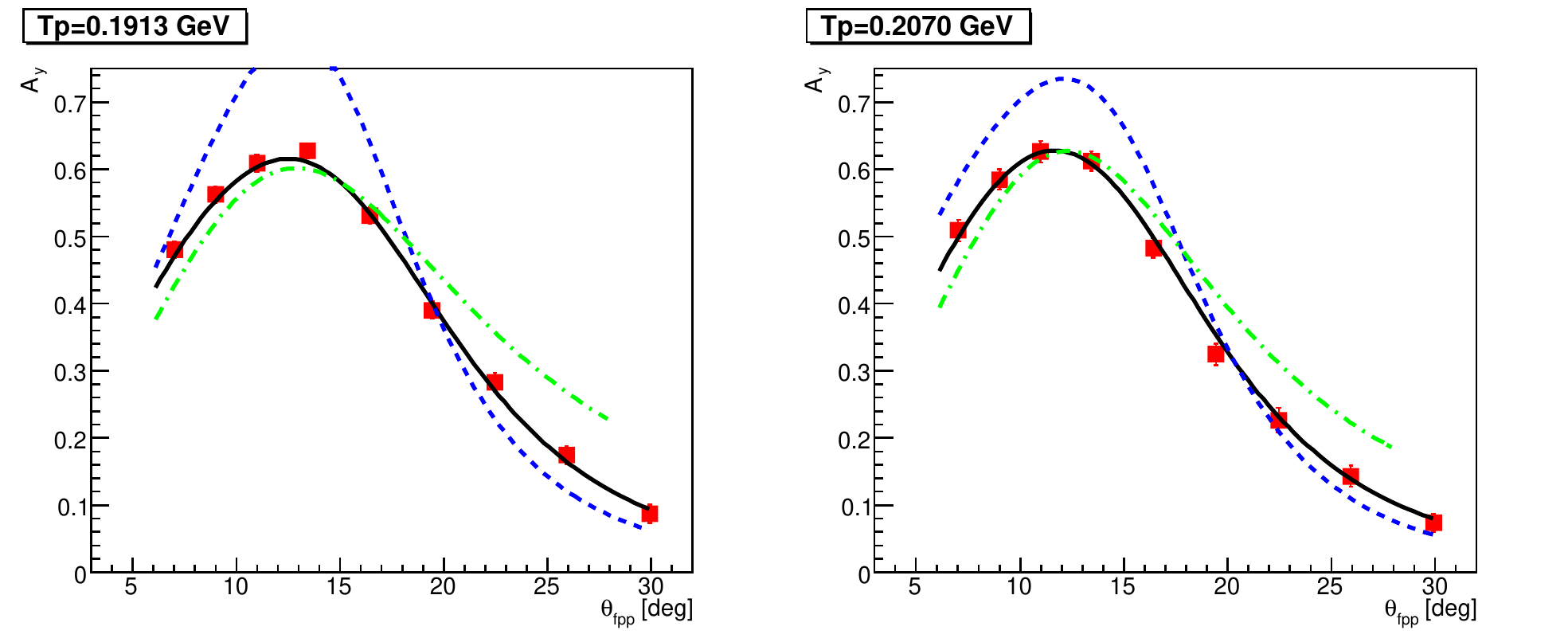}
     \includegraphics[angle=0,width=0.7\textwidth]{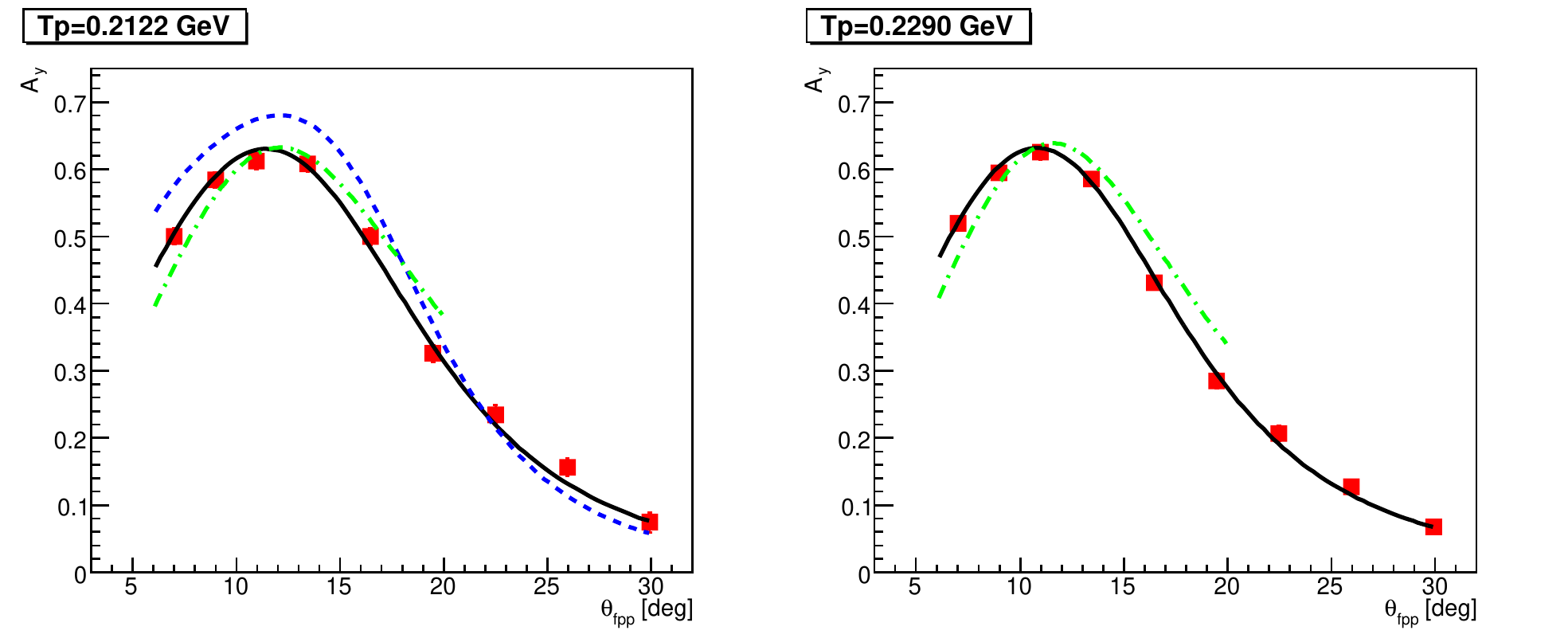}
     \includegraphics[angle=0,width=0.7\textwidth]{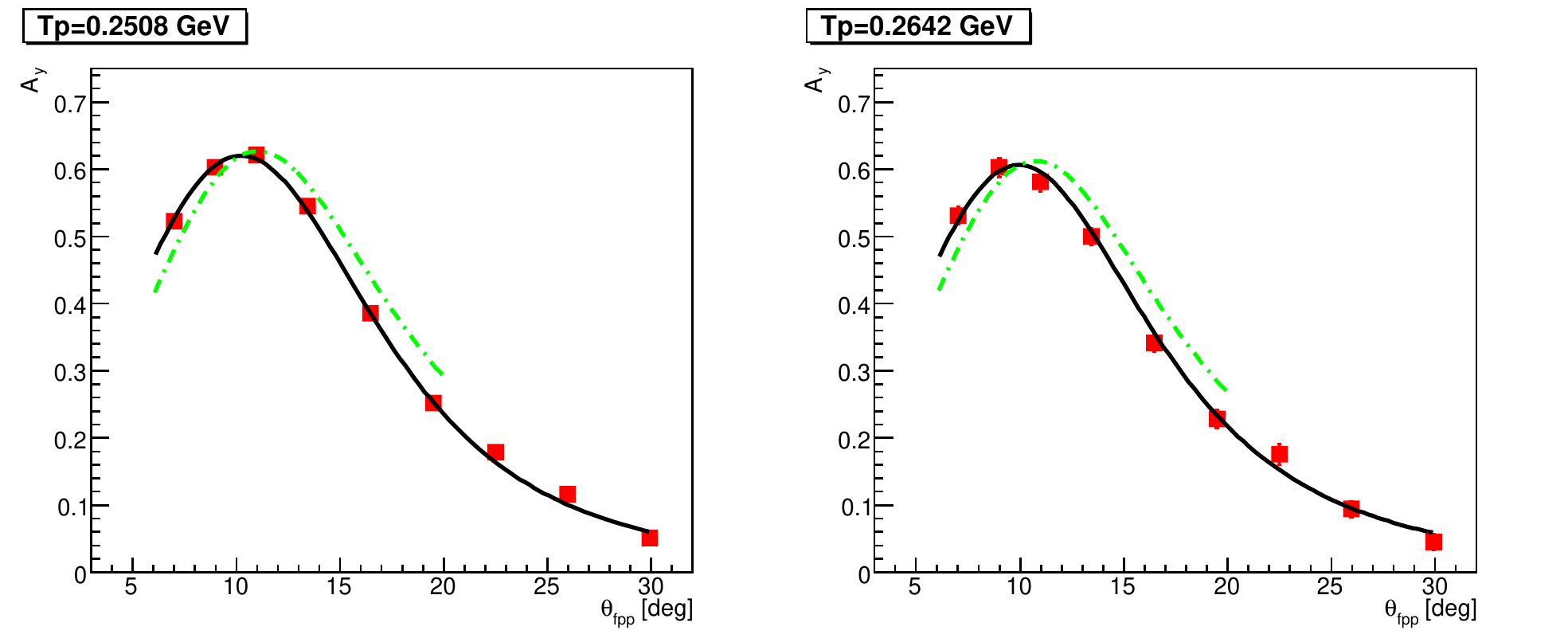}
     \includegraphics[angle=0,width=0.7\textwidth]{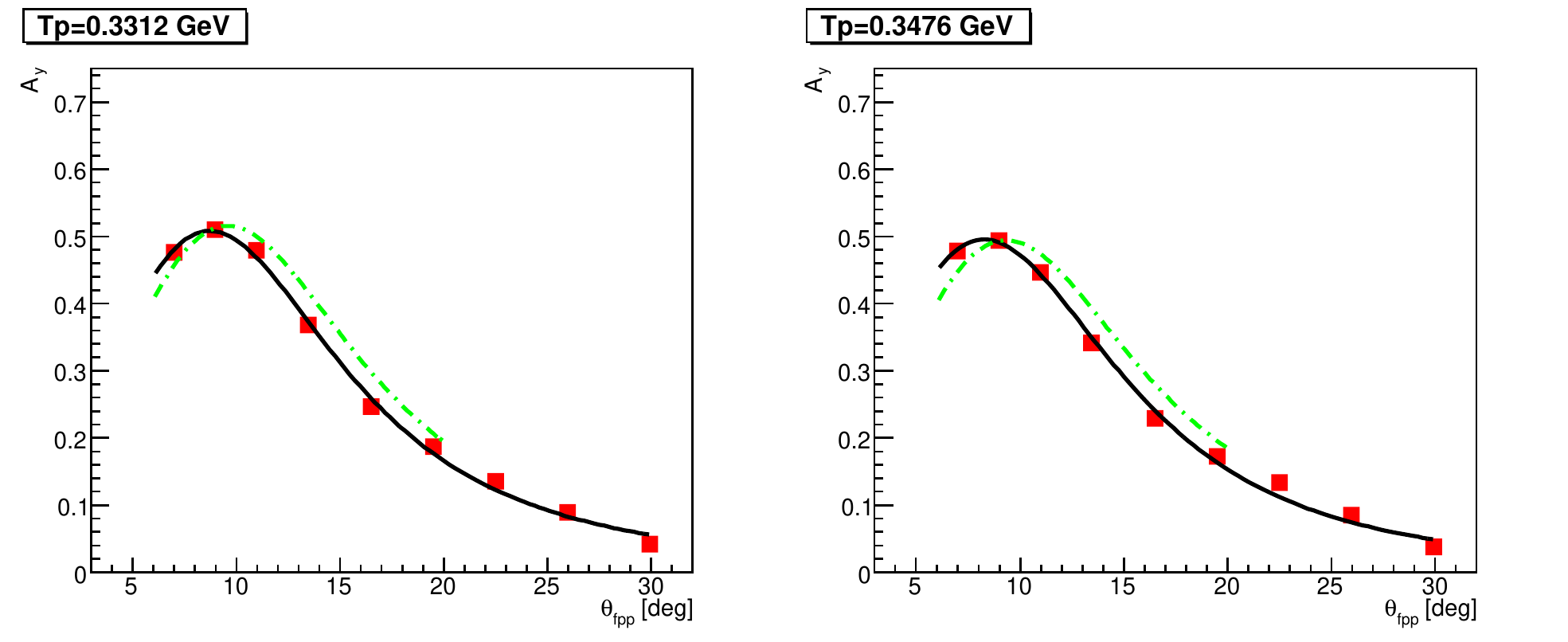}
    \caption{Analyzing power fit part 2: $A_y$ plotted with different parameterization in the high energy region ($T_p>170$ MeV). The error bars shown are statistical only. The dashed lines are from the LEDEX~\cite{jg_ay} parameterization, the dashed dotted lines are from the ``low energy'' McNaughton parameterization~\cite{mcnau}, and the solid lines are from the new parameterization for experiment E08-007.}
    \label{fig:ay_fit1}
  \end{center}
\end{figure}

As illustrated in Figs.~\ref{fig:ay_fit} and \ref{fig:ay_fit1}, the analyzing power
in the low energy region ($T_p<130$ MeV) rises slowly with respect to the scattering angle $\theta_{fpp}$.
For $T_p>150$ MeV, the analyzing power peaks around 10 to 12$^\circ$ and decreases
rapidly at very small angles and angles larger than 25$^\circ$.
In the final analysis, events with angle below 5$^\circ$ and above 25$^\circ$ were rejected.

The new parameterization based on this experiment is in good agreement with
both the McNaughton~\cite{mcnau} and LEDEX~\cite{jg_ay} parameterizations
in the energy/angle regimes for which they were intended, considering all
fits were done for different polarimeters and for different carbon block
thicknesses. Compared to the older fits, the new parameterization extends
the kinematics coverage and provides a smooth transition from the low energy
to the high energy region.
\begin{figure}
  \begin{center}
    \includegraphics[angle=0,width=0.7\textwidth]{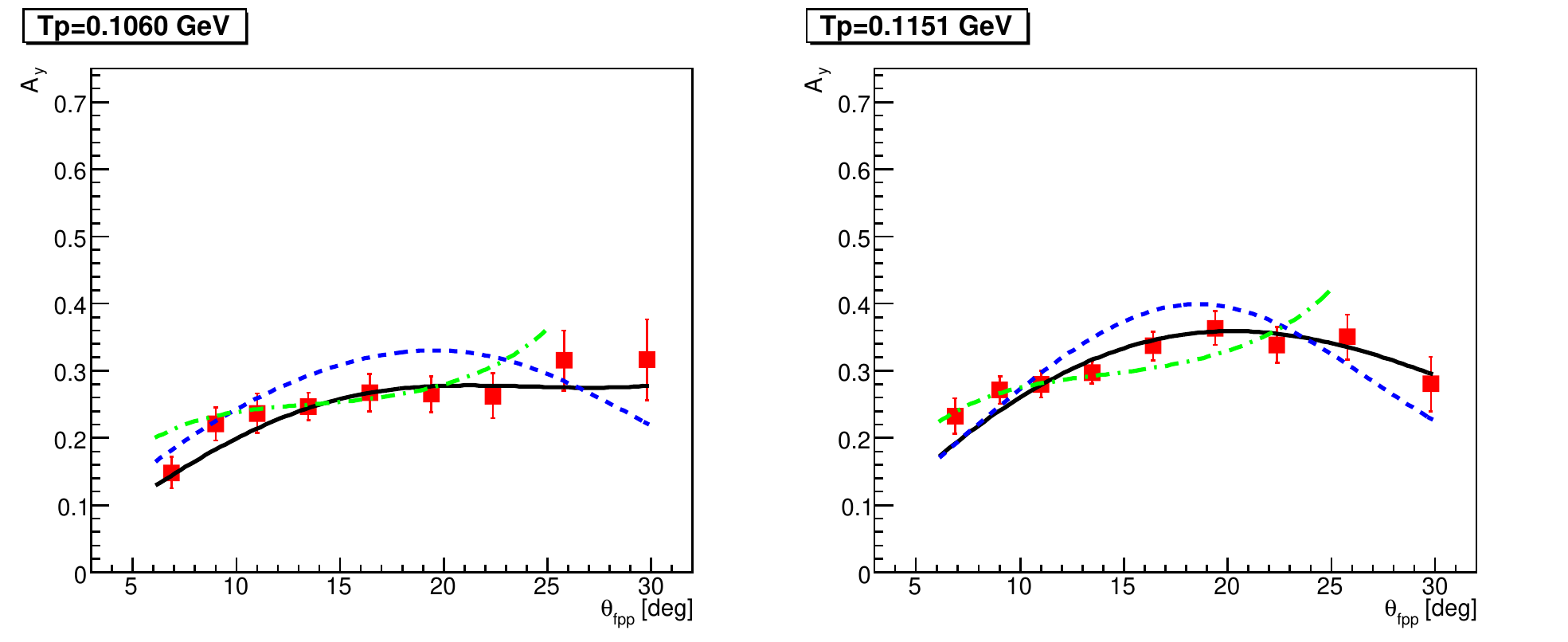}
    \includegraphics[angle=0,width=0.7\textwidth]{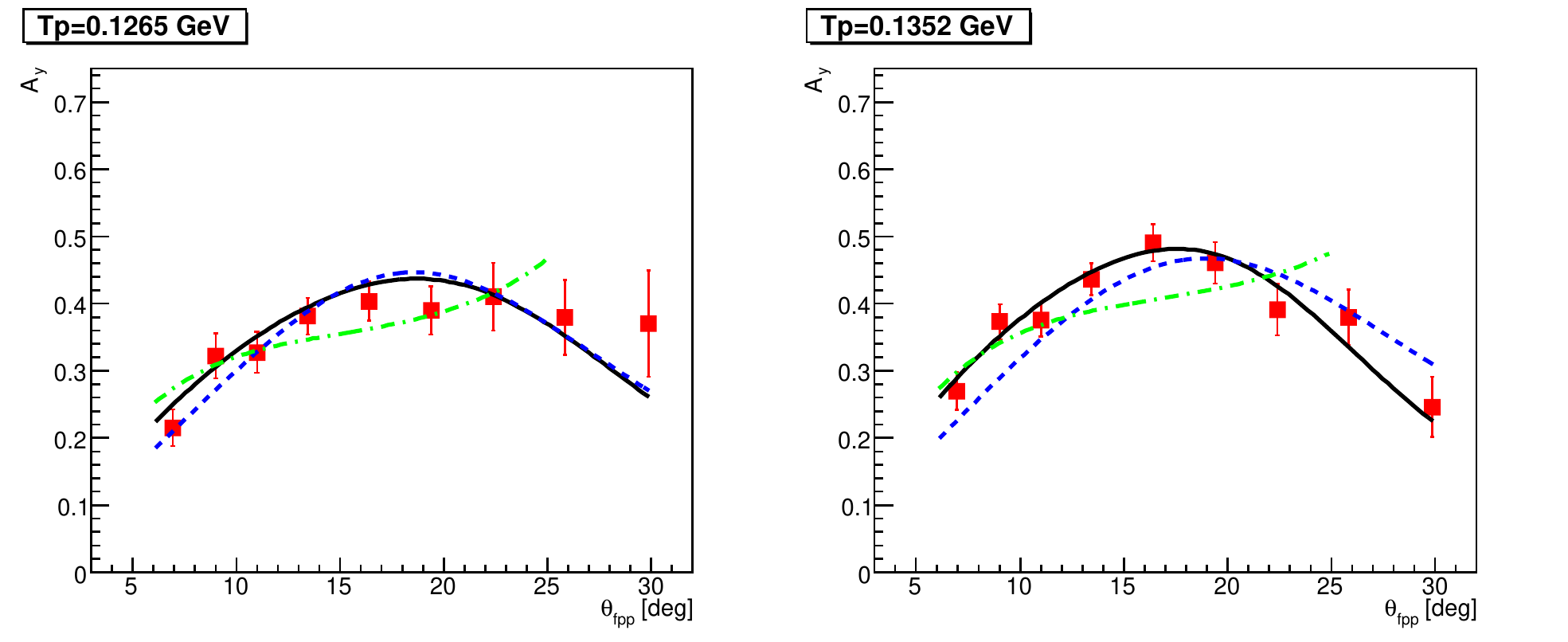}
    \includegraphics[angle=0,width=0.7\textwidth]{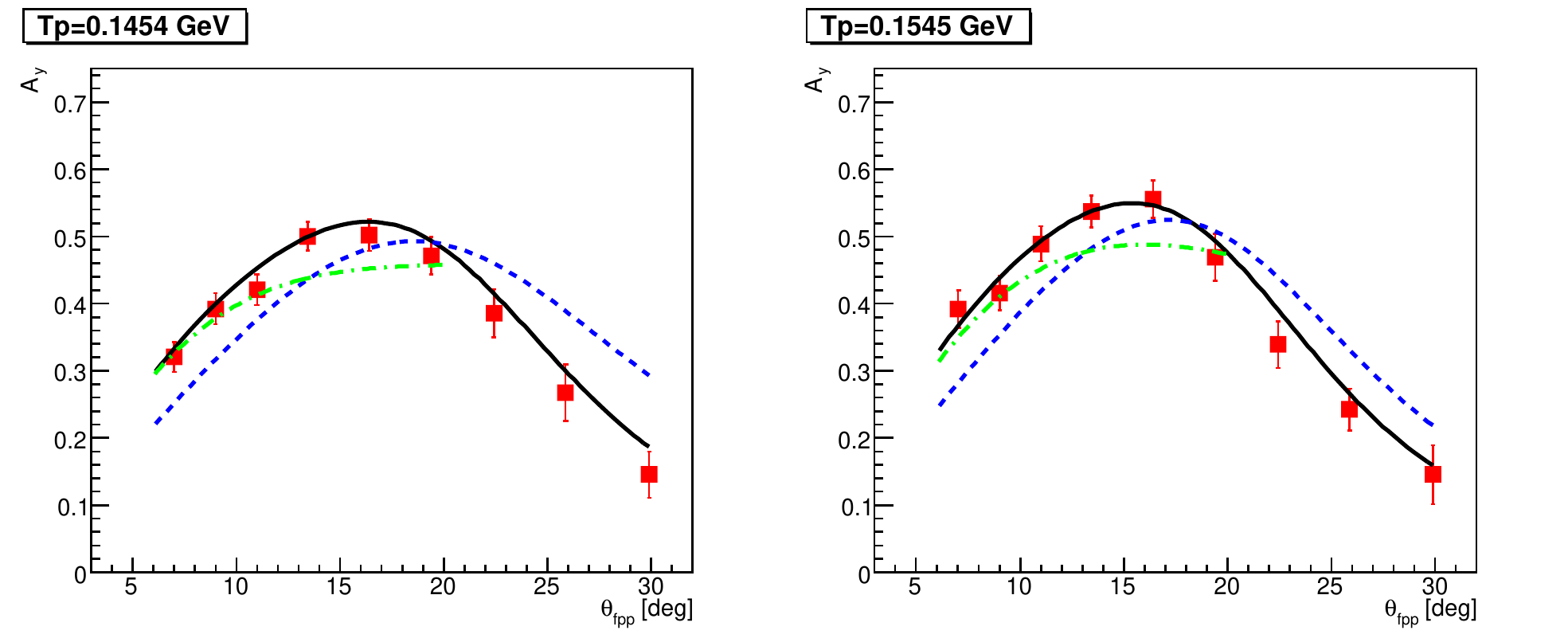}
    \includegraphics[angle=0,width=0.7\textwidth]{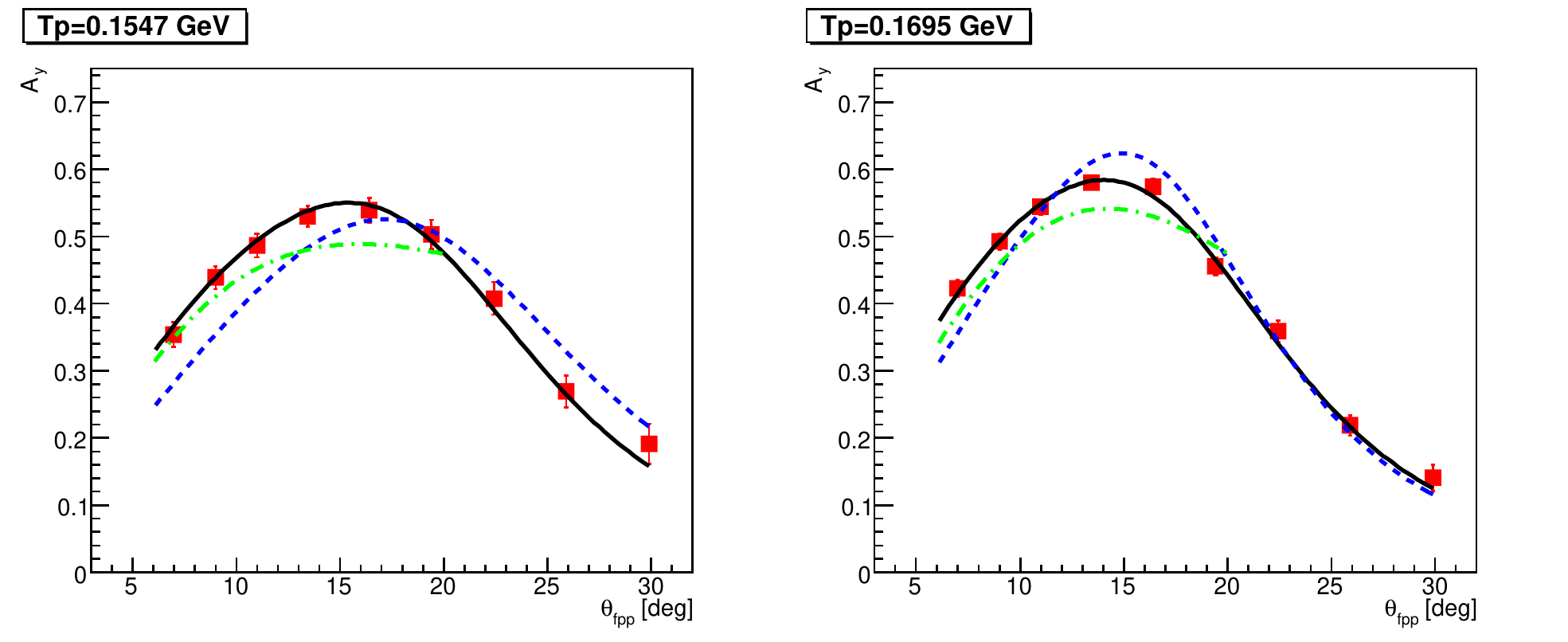}
    \caption{Analyzing power fit part 1: $A_y$ plotted with different parameterization in the low energy region ($T_p<170$ MeV). The error bars shown are statistical only. The dashed lines are from the LEDEX~\cite{jg_ay} parameterization, the dashed dotted lines are from the ``low energy'' McNaughton parameterization~\cite{mcnau}, and the solid lines are from the new parameterization for experiment E08-007.}
    \label{fig:ay_fit}
  \end{center}
\end{figure}

The statistical uncertainty of the ratio $\mu_pG_{Ep}/G_{Mp}$ depends on the uncertainty of the
asymmetries' amplitudes at the focal plane $hA_yP_x^{fpp},hA_yP_y^{fpp}$, which is
proportional to the number of events $N$ that contribute to the amplitude via the strong
interaction in the analyzer:
\begin{equation}
\Delta(hA_yP_{x(y)}^{fpp})\propto\sqrt{\frac{1}{N}}.
\label{eq:err}
\end{equation}
First we define the efficiency of the polarimeter
\begin{equation}
\epsilon(\theta) = \frac{N_{eff}(\theta)}{N_0}.
\end{equation}
$N_0$ is the number of incoming protons, and $N_{eff}$ is the number of valid outgoing tracks that
passed a series of the FPP cuts (cone-test, $zclose$, $sclose$, etc.) and scattered with a
polar angle $\theta$. In other words, $N_{eff}(\theta)$ is the effective number of events
which participated in the measurement of the asymmetry.

\begin{figure}
  \begin{center}
     \includegraphics[angle=0,width=0.7\textwidth]{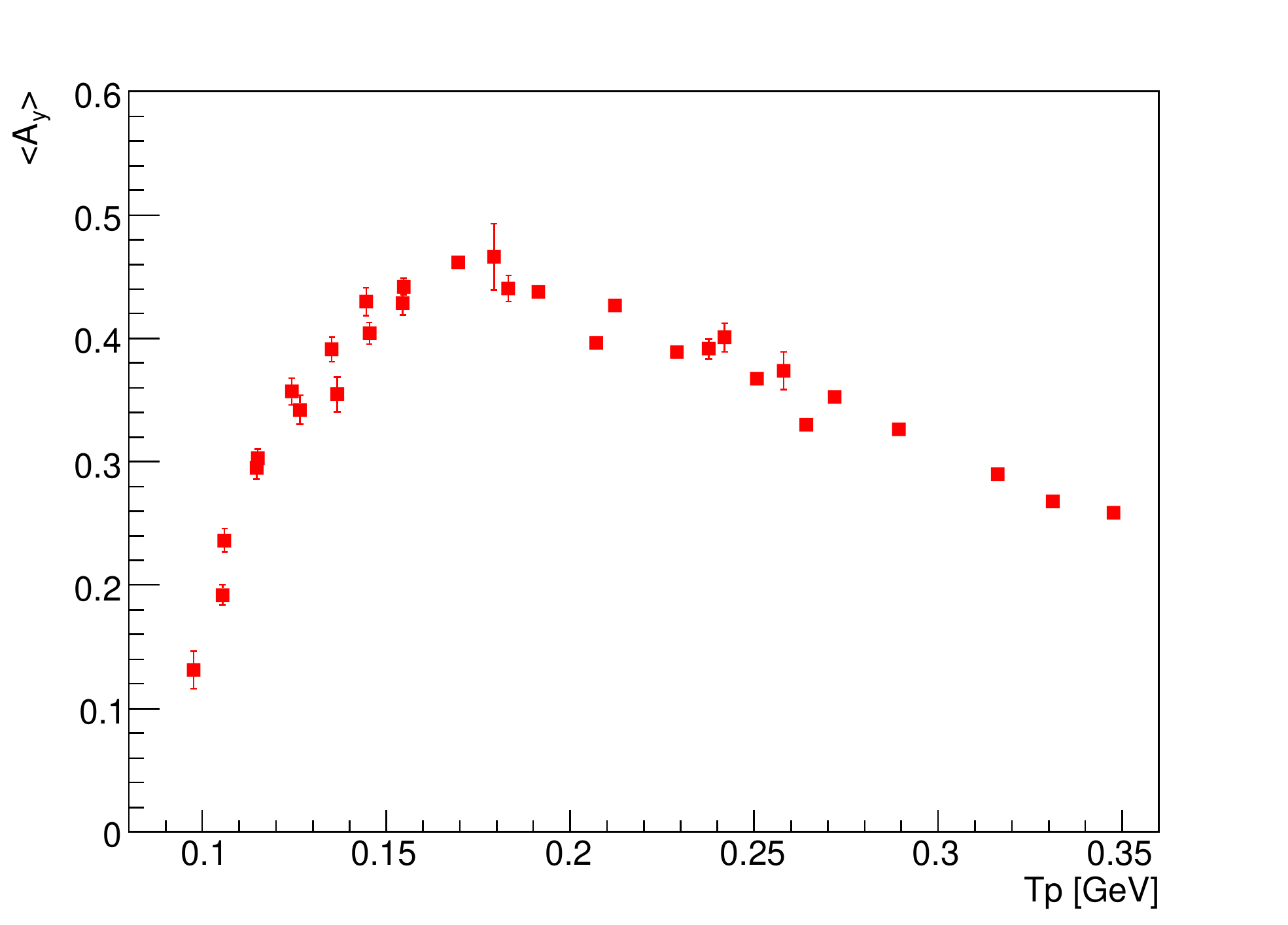}
    \caption{Weighted average analyzing power $\langle A_y\rangle$ with respect to $T_p$ for scattering angles $5^\circ\le\theta_{fpp}\le30^\circ$.}
    \label{fig:aver_ay}
  \end{center}
\end{figure}

Since the analyzing power $A_y$ has a dependence on the scattering angle $\theta$, from Eq.~\ref{eq:err}
the effective number of events has to be multiplied by a weight, $A_y^2({\theta})$; hence, the weighted
effective number of events $N(\theta)$ is
\begin{equation}
N(\theta) = N_0\epsilon(\theta)A_y^2(\theta).
\end{equation}
The total effective number of events $N$ is obtained by integrating over the scattering
angle $\theta$:
\begin{equation}
N=\int N(\theta)d\theta = N_0\int_{\theta_{min}}^{\theta_{max}}\epsilon(\theta)A_y^2(\theta)d\theta=N_0\cdot FOM,
\end{equation}
where
\begin{equation}
FOM = \int_{\theta_{min}}^{\theta_{max}}\epsilon(\theta)A_y^2(\theta)d\theta
\end{equation}
is the Figure of Merit (FOM) and is an intrinsic characteristic of the polarimeter. Then, Eq.~\ref{eq:err} can be
expressed as:
\begin{equation}
\Delta(hA_yP_{x(y)}^{fpp})\propto\sqrt{\frac{1}{N}}=\sqrt{\frac{1}{N_0\cdot FOM}}
\end{equation}
The weighted average analyzing power $\langle A_y\rangle$ for $T_p=90$ to 360 MeV is shown in Fig.~\ref{fig:aver_ay}, and
the FOM for each kinematics is summarized in Table~\ref{tab:fom}

\begin{table}[t]
\caption{FPP performance for E08-007 with $5^\circ<\theta_{fpp}<25^\circ$. $T_p$ is the proton average kinetic energy at the center of the carbon door.}
\begin{center}
\begin{tabular}{c c c c c c}
\hline
Kinematics & $Q^2$ [$(\mathrm{GeV}/c)^2$] &  $T_p$ [MeV] & $\langle A_y\rangle$ &  $\epsilon_{fpp}~[\%]$ & FOM [$\%$]\\
\hline
K1 & 0.35 & 141.2 & 0.3938 & 3.67 & 0.57 \\
K2 & 0.30 & 109.8 & 0.2191 & 5.30 & 0.25 \\
K3 & 0.45 & 195.4 & 0.4876 & 4.09 & 0.97 \\
K4 & 0.40 & 165.3 & 0.4662 & 4.36 & 0.95 \\
K5 & 0.55 & 252.5 & 0.4305 & 4.34 & 0.81 \\
K6 & 0.50 & 221.4 & 0.4659 & 3.81 & 0.83 \\
K7 & 0.60 & 282.2 & 0.3923 & 4.41 & 0.68 \\
K8 & 0.70 & 335.6 & 0.3343 & 4.74 & 0.53 \\
\hline
\label{tab:fom}
\end{tabular}
\end{center}
\end{table} 

%% file: Analysis_2.tex
\chapter{Data Analysis II}
In this chapter, the inelastic background, systematic errors,
and the radiative effects will be discussed in detail.
\section{Background Study}
In addition to the $ep$ elastic events, there are three major types of background that can
potentially contaminate the measurement. The first background is the scattering off the
aluminum (Al) end cap of the liquid hydrogen (LH$_2$) cell through the reaction $^{27}$Al($\vec e,e'\vec p$);
the second is the accidental background under the coincidence timing peak, and the final
one is from the photoproduction of pions. In this section, the
background analysis and the impact to the final results are discussed.
\subsection{Aluminum Background}
\begin{figure}
  \begin{center}
    \includegraphics[angle=0, width=0.65\textwidth]{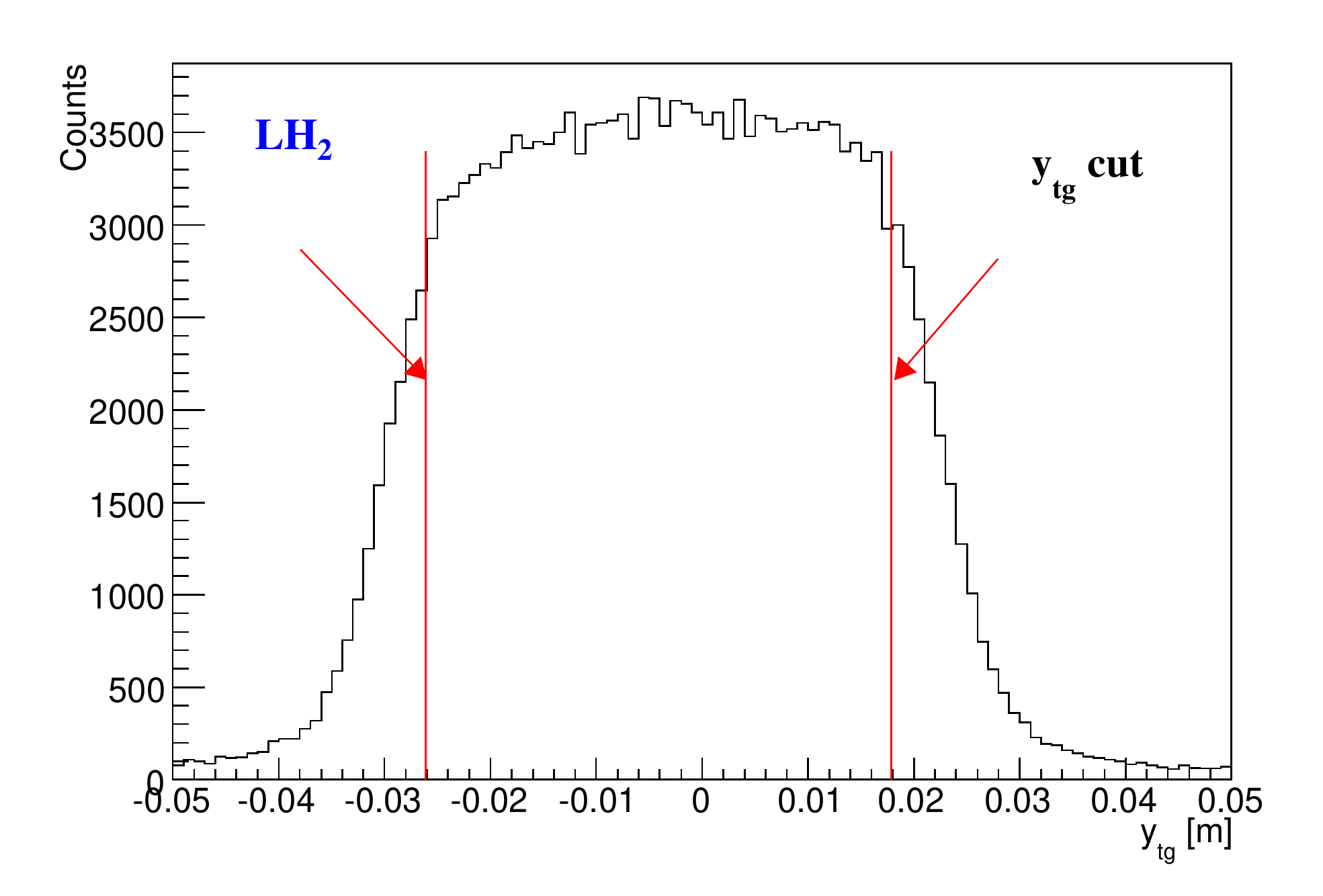}
    \includegraphics[angle=0, width=0.65\textwidth]{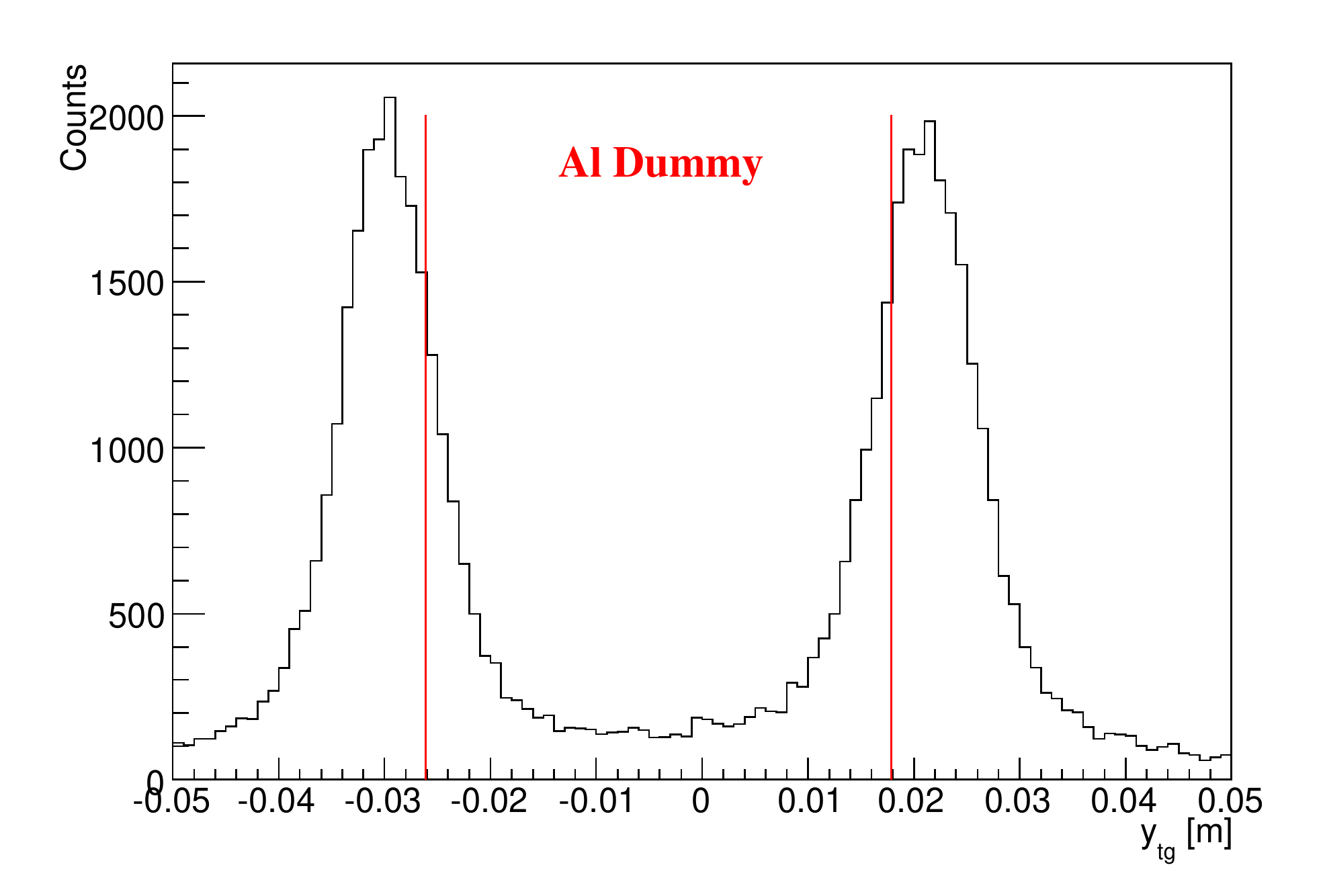}
    \caption{The $y_{tg}$ spectrum for LH$_2$ and Al dummy data with the cut shown by the vertical solid lines.}
    \label{fig:al_y}
  \end{center}
\end{figure}

\begin{figure}
  \begin{center}
    \includegraphics[angle=0, width=0.70\textwidth]{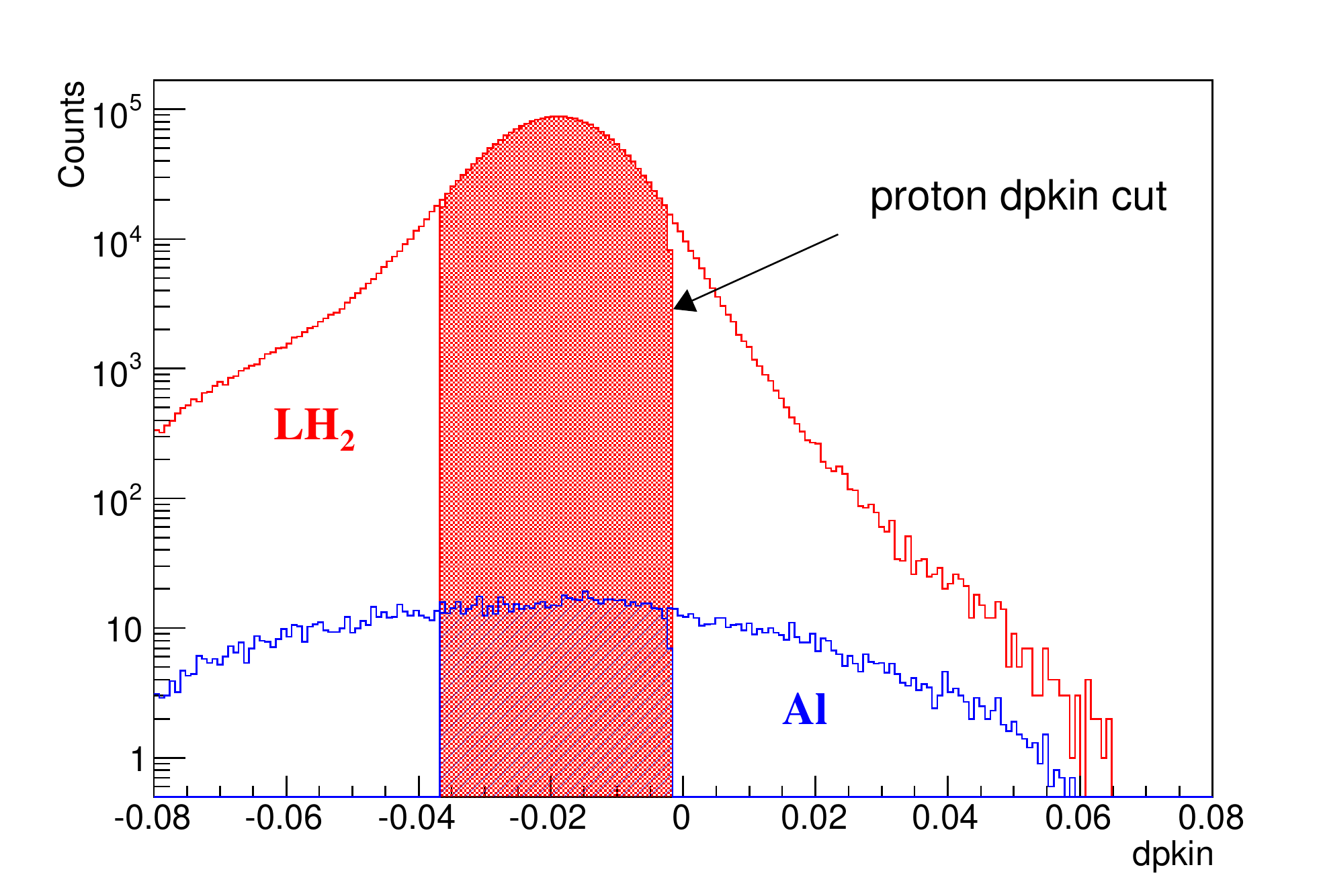}
    \caption{The normalized dpkin spectrum for LH$_2$ and Al dummy at setting K2 $\delta_p = -2\%$. The unfilled and filled spectra are with and without the proton dpkin cut respectively.}
    \label{fig:LH_AL_spec}
  \end{center}
\end{figure}

To estimate the Al background from the target end cap, we took Al dummy
runs for every kinematic setting. The elastic polarization results need to be corrected if
there is a significant amount of Al events passing the cuts, which can have a different proton polarization.
The corrected target polarization $P_{y(z)}$ is calculated by using:
\begin{eqnarray}
 Y_{el.} &=& Y_{H}-Y_{Al},\\
 Y_{el.}P_{y,el.} &=& Y_{H}P_{y,H}-Y_{Al}P_{y,Al},\\
 Y_{el.}P_{z,el.}&=&Y_{H}P_{z,H}-Y_{Al}P_{z,Al},
\end{eqnarray}
where $Y$ is the normalized yield. First, we first need
to estimate the fraction of Al events in the elastic data to obtain the corrected proton polarization. In order to be consistent with
the elastic proton polarization extraction, the same relevant cuts were applied:
\begin{itemize}
\item HRS acceptance cut ($\phi_{tg}$, $\theta_{tg}$, $\delta_p$).
\item Coincidence event type cut (T5).
\item Coincidence timing cut.
\item Elastic proton peak on dpkin;
\end{itemize}
The fraction of Al in LH$_2$ data was estimated by using the charge normalization method\footnote{Due to the small
acceptance of the HRS, it's difficult to select a pure Al sample spectra in LH$_2$ data; hence, the normalization factors obtained
from comparing the Al and LH$_2$ spectra could highly overestimate the Al contamination.}. By assuming
the running conditions (beam energy, position and size, trigger setup, etc.) were the same between the
LH$_2$ and the Al dummy run and the polarization of the background polarization is
independent of the reaction location ($y_{tg}$), the fraction of Al in LH$_2$ can be extracted by:
\begin{equation}
R=Y_{Al}/Y_{H} = f\cdot \frac{N_{Al}\times C_H\times (1-DT_H)}{N_H \times C_{Al} \times (1-DT_{Al})},
\end{equation}
where $N_{H(Al)}$ is the number of events in the LH$_2$ (Al) run after applying the same cuts\footnote{These include the HRS acceptance cut, coincidence trigger and timing cut, but no target vertex cut was applied to avoid the inconsistency due to the position shift between the LH$_2$ target cell and the Al dummy target.}, $C_{H(Al)}$ is the charge, and $DT_{H(Al)}$ is the DAQ dead time. In the expression, $f$ is the ratio of the Al foil thickness for the LH$_2$ and the Al dummy target. From the Al foil thicknesses reported in Table.~\ref{tab:al_thick}, $f = 0.113$. The fraction of Al background in LH$_2$, $R$, for each kinematic setting is summarized in Table~\ref{tab:al_r}\footnote{The first two $\delta_p$ settings of kinematics K1 were with the entire BigBite shower counter on; hence, more Al background was included for these data compared to the other kinematic settings, which had only a limited set of shower blocks turned on.}. These are the upper limits of $R$, since in the elastic analysis a cut on the target reaction vertex was applied ($y_{tg}$), and the events from the target end caps were further suppressed. Fig.~\ref{fig:al_y} illustrates the spectrum of $y_{tg}$ for the LH$_2$ and Al dummy runs respectively with the location of the target vertex cut indicated by the vertical lines. Fig.~\ref{fig:LH_AL_spec} gives an example of the normalized LH$_2$ and Al spectra after applying all the cuts (including the target vertex cut).
 \begin{table}[t]
 \caption{Aluminum foil thickness.}
 \begin{center}
 \begin{tabular}{| c | c |}
 \hline
 Target & Thickness [cm]\\
 \hline
 LH$_2$ (6cm) &  0.0113 \\
 Al dummy (6cm) & 0.100 \\
 \hline
 \end{tabular}
 \label{tab:al_thick}
 \end{center}
 \end{table}
 \begin{table}[t]
  \caption{The upper limit of the Al background fraction $R_{max}$ for each kinematics. The numbers listed
 are the average over all $\delta_p$ settings.}
 \begin{center}
 \begin{tabular}{|c|c|c|}
 \hline
 Kinematics & $Q^2$ [$(\mathrm{GeV}/c)^2$] & $R_{max}$\\
 \hline
 K1 & 0.35 &0.0021\\
 K2 & 0.30 &0.0001\\
 K3 & 0.45 &0.0001\\
 K4 & 0.40 &0.0001\\
 K5 & 0.55 &0.0001\\
 K6 & 0.50 &0.0001\\
 K7 & 0.60 &0.0001\\
 K8 & 0.70 &0.0001\\
 \hline
 \end{tabular}
 \label{tab:al_r}
 \end{center}
 \end{table}

The recoil proton polarization of the LH$_2$ and Al dummy targets
for each kinematics were extracted. As an example, the results of kinematics K1 ($Q^2 = 0.35$ GeV$^2$) are reported in Table~\ref{tab:pol_corr}. As can be seen that the correction to the elastic form factor ratio $\mu_pG_E/G_M$
is less than 0.001, which is negligible compared to the statistical error. The corrections for the other kinematic settings are at the same level.
 \begin{table}[t]
 \caption{Polarization $P_{y(z)}$ of LH$_2$, Al dummy and corrected values
 for kinematics K1 ($Q^2 = 0.35$ GeV$^2$).}
 \begin{center}
 \begin{tabular}{|c| c| c| c| c|}
 \hline
 Pol. & LH$_2$ & Al & LH$_2$ corrected\\
 \hline
 $P_y$ & -0.2624$\pm 0.0017$ & -0.0562$\pm$0.0995 & -0.2628$\pm0.0017$\\
 $P_z$ & 0.2536$\pm 0.0017$ & 0.2709$\pm$0.1016 &0.2536$\pm0.0017$\\
 \hline
 \end{tabular}
 \label{tab:pol_corr}
 \end{center}
 \end{table}
\subsection{Accidental Background}
In this experiment, the coincidence trigger helped to significantly reduce
the inelastic background. The accidental background can be estimated by using the
same method used as for the Al case. With the coincidence timing cut, the accidental
background was estimated by interpolating the timing spectrum under the elastic peak region;
the typical background to signal ratio was found to be $\le 0.003$.

The polarization of the accidental background outside the timing cut was extracted. Unlike the Al
background, the proton polarization of the accidental background is very close to
the polarization of the elastic events within the timing cut. This behavior is expected since the accidental events are dominated by the elastic events. Using a similar procedure as described in the Al case,
the correction to the elastic results from the accidental background is $\le 0.001$,
which is also negligible. As an example, the polarization of the accidental background
outside the timing cut was extracted for kinematics K8 ($Q^2 = 0.7$ GeV$^2$), the results
are reported in Table~\ref{tab:acc_corr}. The correction for the other kinematics settings are at the same level.
 \begin{table}[t]
  \caption{Polarization $P_{y(z)}$ of LH$_2$ inside, outside the coincidence
 timing cut and the corrected values for kinematics K8 ($Q^2 = 0.7$ GeV$^2$).}
 \begin{center}
 \begin{tabular}{|c| c| c| c| c|}
 \hline
 Pol. & LH$_2$ & Accidental & LH$_2$ corrected\\
 \hline
 $P_y$ & -0.3636$\pm 0.0015$ & -0.3295$\pm$0.0188 & -0.3637$\pm0.0015$\\
 \hline
 $P_z$ & 0.5552$\pm 0.0016$ & 0.5320$\pm$0.0208 &0.5553$\pm0.0016$\\
 \hline
 \end{tabular}
 \label{tab:acc_corr}
 \end{center}
 \end{table}
\subsection{Pion Photoproduction}
Due to the reduced detector configuration of the BigBite spectrometer, events cannot be easily distinguished between an electron or a photon that decayed from a $\pi^0$, which fired the shower counter,
since the coincidence trigger could be formed by pion photoproduction via $\gamma + p \to p + \pi^0$.
A study was made to estimate the pion contamination which is elaborated in Appendix E.
Due to the small acceptance and high resolution of the HRS combined with the
tight elastic cut applied on the proton kinematics, we have concluded that
the contribution from pion photoproduction is less than $10^{-4}$, and
the correction to the proton polarization is also at $<10^{-4}$ level, which is negligible.

\begin{figure}
  \begin{center}
    \includegraphics[angle=0, width=0.65\textwidth]{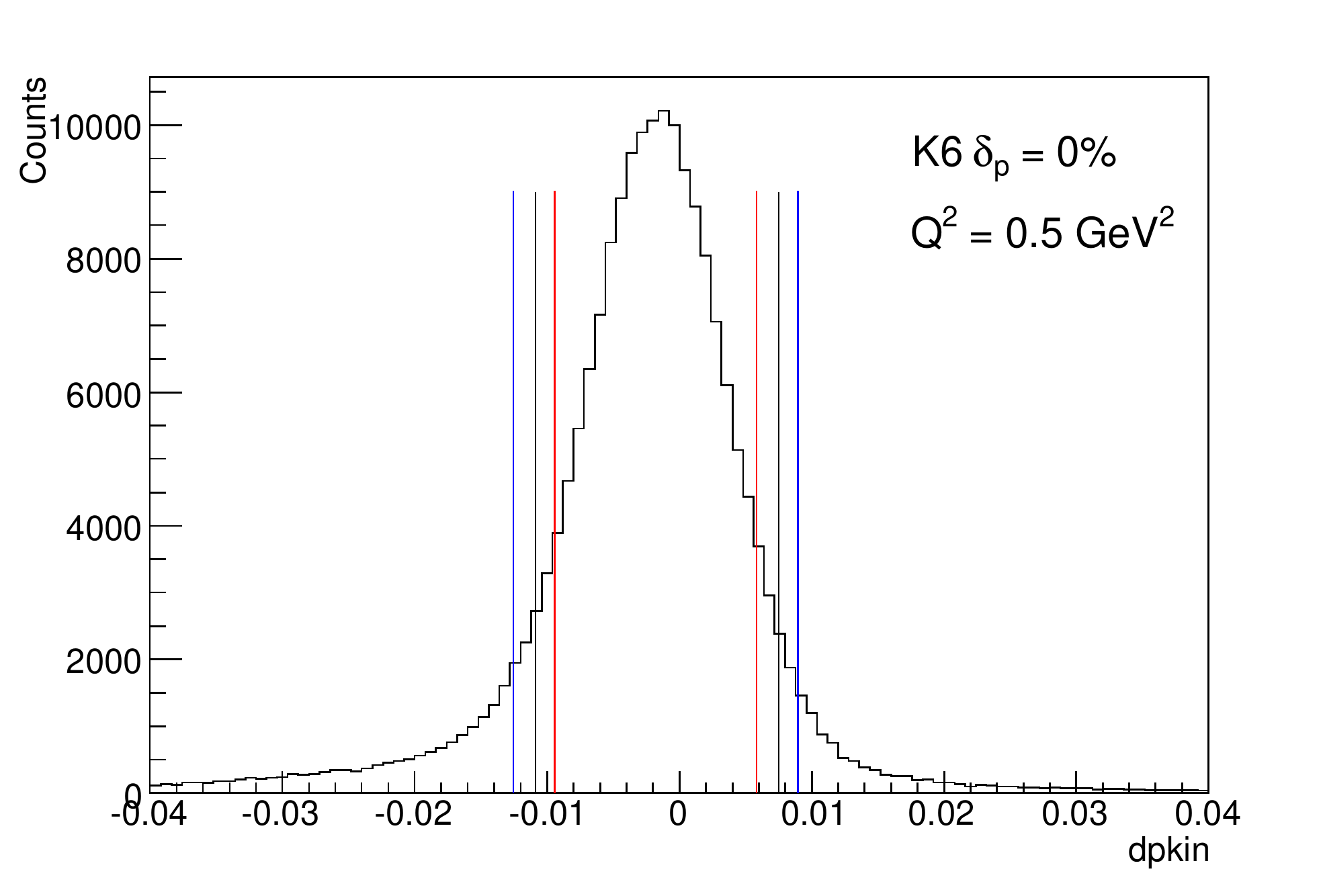}
    \caption{Different elastic cuts on proton dpkin.}
    \label{fig:diffcut}
  \end{center}
\end{figure}

\begin{figure}
  \begin{center}
    \includegraphics[angle=0, width=0.65\textwidth]{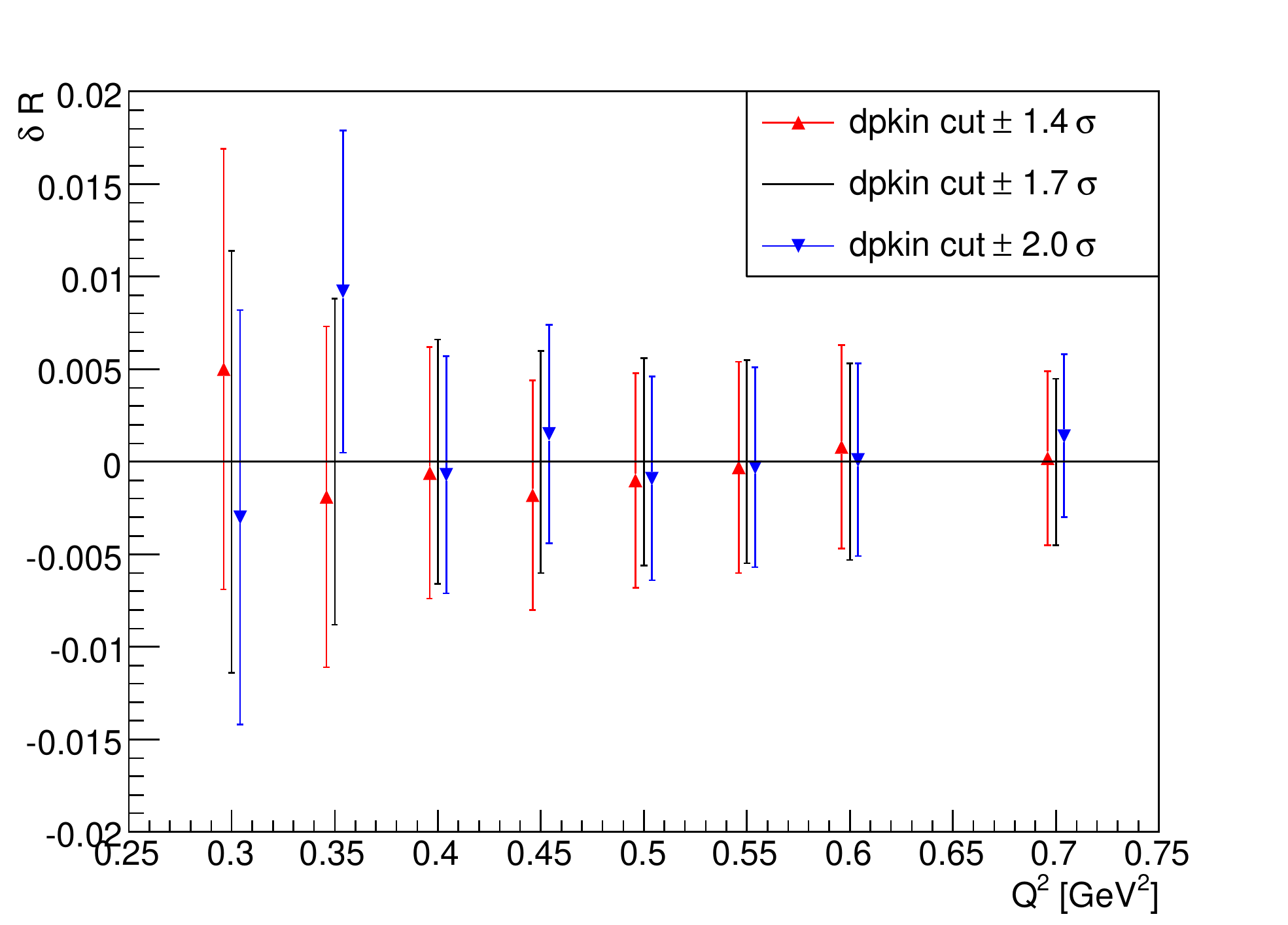}
    \caption{The ratio difference with different elastic cuts. The $y$-axis is the difference between the results, the $x$-axis which was manually shifted for different cuts for a better view.}
    \label{fig:cuts}
  \end{center}
\end{figure}

As a simple demonstration to test whether the results are sensitive to the elastic cut applied in this work, 3 different cuts were applied on the peak of the proton dpkin as shown in Fig.~\ref{fig:diffcut}: $\pm 1.4\sigma$, $\pm 1.7\sigma$ and $\pm 2.0\sigma$. As shown in Fig.~\ref{fig:cuts}, the results with different elastic cuts are consistent within the statistical uncertainty. In the final analysis, a $\pm 1.7\sigma$ cut was applied.
\section{Systematic Analysis}
For this experiment, the proposed statistical uncertainties were achieved ($\le1\%$),
and hence, the systematic uncertainties will dominate the total errors.
The details of the systematic analysis are presented in the following sections, which includes a discussion of:
  \begin{itemize}
  \item Spin precession: HRS optics and the COSY model.
  \item Scattering angle reconstruction in the FPP.
  \item Beam energy and HRS mis-pointing.
  \item Charge asymmetry.
  \end{itemize}
The helicity independent factors such as the acceptance, beam current, target
density etc., cancel in the polarization ratio. The beam energy and
the spectrometer setting are used to calculate the kinematic factors; however, the form
factor ratio is less sensitive to these parameters with the current experimental precision.
The spin precession and the FPP reconstruction are directly related to the
extraction of the proton polarization at the target, and therefore, they
are the most important components for this type of measurement. In this section,
the analysis for all the significant systematic uncertainties will be discussed.
\subsection{Spin Precession}
What we measured in this experiment is the proton polarization detected at the focal
plane: $P_x^{fpp},P_y^{fpp}$. However, the polarization at the target is directly related to the physics of interests. In reality, the magnetic structure of the spectrometer
is more complicated than just a simple perfect dipole; COSY~\cite{cosy} was
used to calculate the the full precession matrix $S_{ij}$ to relate the
polarization at the target to the one detected at the focal plane by Eq.~\ref{eq:sp_matrix}.

To calculate the matrix $S_{ij}$, two inputs are required. The first input is a table of the expansion coefficients $C_{ij}^{klmnp}$ which is generated by COSY, and the second is the target coordinates of
each event which are reconstructed by the HRS optics matrix. Hence, it is natural to
separate the spin precession systematics error into two parts: HRS optics and COSY.
\subsubsection{HRS Optics}
The optics database used for this experiment was optimized for experiment E06-010~\cite{transversity}\footnote{Experiment E06-010 took the optics data for the left HRS at a similar momentum
setting ($p_0 =$ 1.2 GeV). We also have the
optics acquired in 2000 during experiment E89-044~\cite{doug_opt}, which was also carefully
optimized. Both sets of optics were utilized and produced similar results, which indicates that the spectrometer optics
reconstruction is fairly stable over the past ten years.}. We used two steps to estimate the systematic uncertainty
due to the optics. First, the uncertainties in the central deviation of each target
quantity ($\Delta\delta_p,\Delta\phi_{tg}, \Delta\theta_{tg}, \Delta y_{tg}$) were estimated.
Then they were shifted separately by the amount of the estimated uncertainties to
determine the impact on the form factor ratio $\mu_pG_E/G_M$.
The sensitivities of the ratio $\mu_pG_E/G_M$ to each target quantity are
summarized in Table~\ref{tab:tg_uncer}. Clearly
$\phi_{tg}$ is the most important quantity and hence requires additional attention.
\begin{table}[t]
\caption{Shifts of the form factor ratio associated with shifts of the
individual target quantities for each kinematic setting.}
\begin{center}
\begin{tabular}{|c| c |c |c |c|}
\hline
Kinematics & $\delta_p$ (+0.001) & $\phi_{tg}$ (+1 mrad) & $\theta_{tg}$ (+1 mrad) & $y_{tg}$ (+1 mm)\\
\hline
K1 & 0.0015  & 0.0064 & -0.0004 & 0.0011\\
K2 & 0.0018  & 0.0064 & -0.0002 & 0.0011\\
K3 & 0.0005  & 0.0064 & -0.0006 & 0.0015\\
K4 & 0.0013  & 0.0066 & -0.0005 & 0.0010\\
K5 & 0.0004  & 0.0064 & -0.0009 & 0.0019\\
K6 & 0.0006  & 0.0064 & -0.0008 & 0.0015\\
K7 & 0.0007  & 0.0064 & -0.0010 & 0.0022\\
K8 & 0.0005  & 0.0064 & -0.0014 & 0.0027\\
\hline
\end{tabular}
\label{tab:tg_uncer}
\end{center}
\end{table}

To evaluate the quality of the optics, especially the uncertainty in $\phi_{tg}$,
we take advantage of the proton elastic kinematics, since the angle is well
constrained when the beam energy and the proton momentum are fixed. Beforehand, we need to evaluate all
the parameters which are relevant in determining $\phi_{tg}$ and convert their
uncertainties into $\Delta\phi_{tg}(x)$. Then, the offset between the anticipated
proton elastic peak position and the reconstructed proton spectrum is quoted
as $\Delta\phi_{tg}(off)$. The total error in $\phi_{tg}$ is quoted conservatively as:
\begin{equation}
\Delta\phi_{tg} = \sqrt{\Delta^2\phi_{tg}(x) + \Delta^2\phi_{tg}(off)}.
\end{equation}

The relevant parameters which would affect the anticipated proton elastic peak position are:
\begin{itemize}
\item Spectrometer central angle $\theta_s$.
\item Beam energy $E_e$.
\item Proton central momentum $P_0$.
\item $\delta_p$ reconstruction.
\end{itemize}
Each one of them is discussed in the following subsections.

1. Spectrometer Central Angle

\begin{figure}
\begin{center}
\includegraphics[angle=0,width=.6\textwidth]{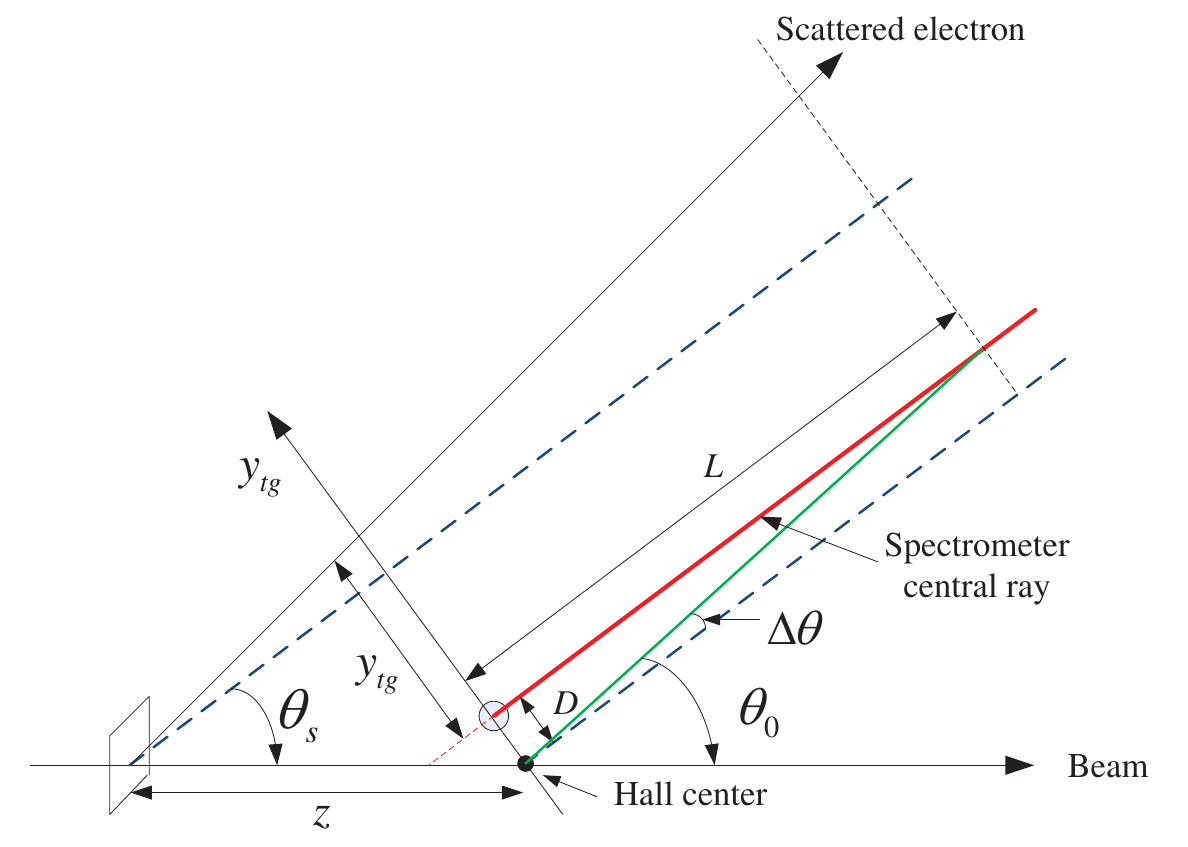}
\caption{Coordinates for electrons scattering from a thin foil target. $L$
is the distance from Hall center to the floor mark, and $D$ is the
horizontal displacement of the spectrometer axis from its ideal position.
The spectrometer set angle is $\theta_0$ and the true angle is denoted
by $\theta_s$ when the spectrometer offset is considered.}
\label{fig:spec_offset}
\end{center}
\end{figure}

\begin{figure}
\begin{center}
\includegraphics[angle=0,width=.65\textwidth]{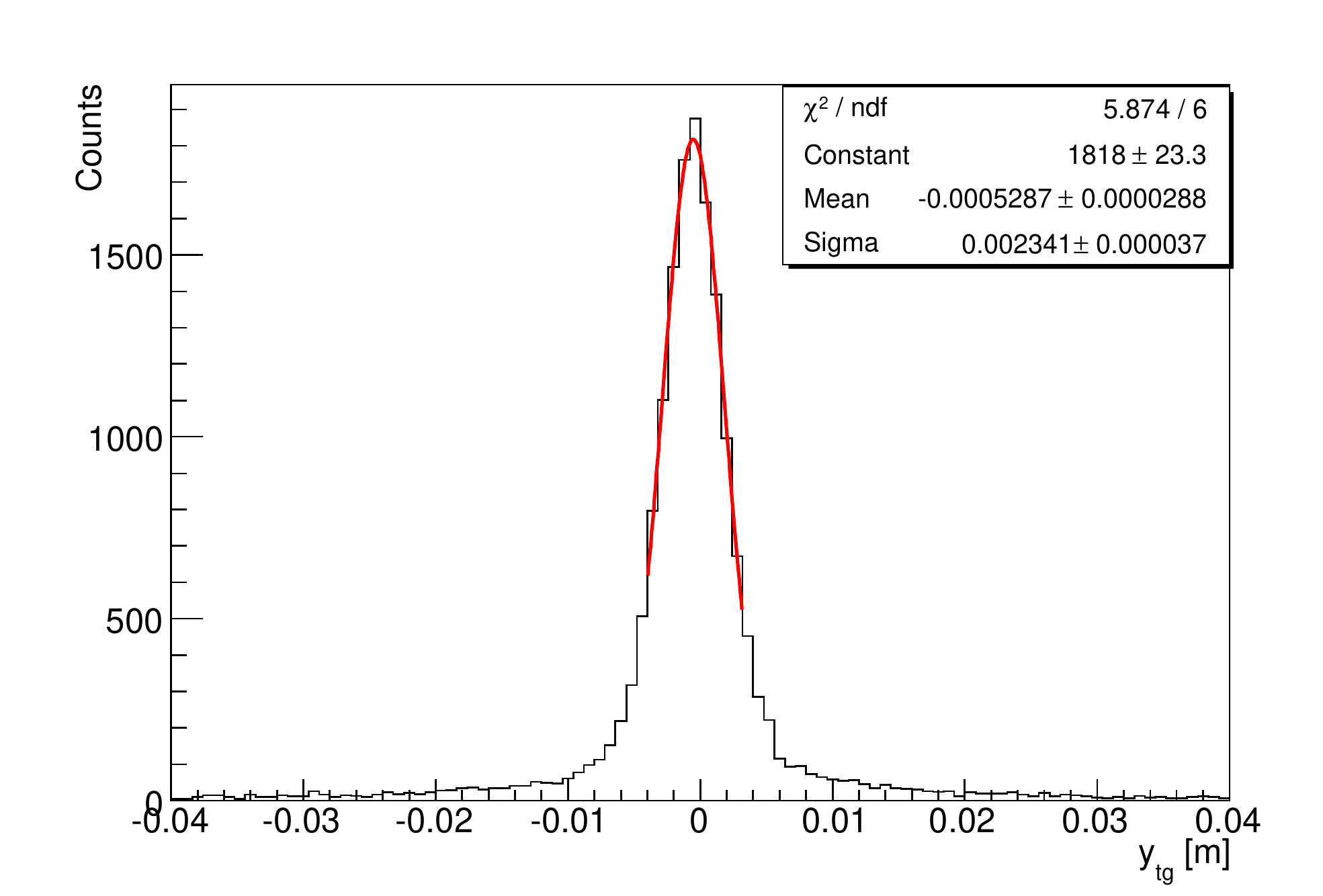}
\caption{Carbon pointing $y_{tg}$ for kinematics K8 ($Q^2=0.7$ GeV$^2$).}
\label{fig:ytg}
\end{center}
\end{figure}

Due to the misplacement between the front and the end of
the spectrometer during movement, the HRS central angle can be off by a small
amount as illustrated in Fig~\ref{fig:spec_offset}. During the experiment, we took
carbon foil data at each kinematics to determine the spectrometer central angle.
With the target position survey and ignoring the higher
order terms introduced by $\phi_{tg}$, we can determine the spectrometer
horizontal offset $D$ from its ideal position by:
  \begin{equation}
  z=-(y_{tg}+D)/\sin\theta_0 + x_{beam}\cot\theta_0,
  \label{eq:ytg}
  \end{equation}
where $x_{beam}$ is the horizontal beam position. The $y_{tg}$ is the peak
value, which is fit as shown in Fig.~\ref{fig:ytg}.
The actual spectrometer angle $\theta_s$ is corrected by $D$ in the first order:
  \begin{equation}
  \theta_s \approx \theta_0 - \frac{D}{L},
  \end{equation}
where $L$ is the distance between the hall center and the floor marks where
the angles are scripted (8.458 m). By considering the uncertainty of the survey
($\pm 1$ mm), and the uncertainty in $y_{tg}$ ($\pm 1$ mm) , the error in $D$
is derived by:
\begin{equation}
\Delta D = \sqrt{\Delta^2 y_{tg}+\sin^2 \theta_0 \Delta^2 z}
\end{equation}
The spectrometer central angle for each kinematics was corrected using the
pointing method, and the results are reported in Table~\ref{tab:angle}.
One can see that the angle mispointing is small which is consistent with
previous records~\cite{fatiha}. This observation was anticipated due to the large value of $L$.
\begin{table}[t]
\caption{Spectrometer nominal ($\theta_0$) and real ($\theta_s$) central angle for each kinematic setting.}
\begin{center}
\begin{tabular}{ |c |c| c| c|}
\hline
Kinematics & $\theta_0$ [deg] & $\theta_s$ [deg] & $\Delta\phi_{tg}(\theta_s)$ [mrad]\\
\hline
K1 & 57.5 & 57.478 $\pm$ 0.008 & 0.14\\
K2 & 60.0 & 59.975 $\pm$ 0.008 & 0.14\\
K3 & 53.0 & 52.991 $\pm$ 0.008 & 0.14\\
K4 & 55.0 & 54.986 $\pm$ 0.008 & 0.14\\
K5 & 49.0 & 48.990 $\pm$ 0.008 & 0.14\\
K6 & 51.0 & 50.974 $\pm$ 0.008 & 0.14\\
K7 & 47.0 & 46.990 $\pm$ 0.008 & 0.14\\
K8 & 43.5 & 43.484 $\pm$ 0.007 & 0.12\\
K1ext & 57.5 & 57.494 $\pm$ 0.008 & 0.14\\
K2ext & 60.0 & 59.977 $\pm$ 0.008 & 0.14\\
\hline
\end{tabular}
\label{tab:angle}
\end{center}
\end{table}

2. Beam Energy

During the experiment, the beam energy was given by the Tiefenbach value.
According to~\cite{hallanim}, the uncertainty for this non-invasive
measurement is $ 0.5$ MeV. The average beam energy loss in the target is also
taken into account. The target material thicknesses are summarized in Table~\ref{tab:material}.
\begin{table}[t]
\caption{Target materials in the beam energy loss calculation.}
\begin{center}
\begin{tabular}{|l| c |}
\hline
Material & Thickness \\
\hline
Al vacuum chamber window & 0.0406 cm\\
\hline
Al entrance window & 0.0113 cm\\
\hline
LH$_2$ & 3 cm\\
\hline
\end{tabular}
\label{tab:material}
\end{center}
\end{table}
The average total energy loss in the target for a 1.19 GeV beam is $1.5$
MeV; hence, the beam energy we used to calculate the elastic kinematics is:
\begin{equation}
E_{e} = (E_{tiefenbach}-1.5)\pm 0.5 \mathrm{MeV}.
\end{equation}
Table~\ref{tab:beam} gives the converted uncertainty of $\phi_{tg}$
for each kinematics due to the uncertainty of $E_{e}$\footnote{There
is also some uncertainty in the value of the beam energy loss due to the possible non-uniformity of the material thicknesses; however, this uncertainty is much less than 0.5 MeV given
the precision of the survey.}.
\begin{table}[t]
\caption{Converted uncertainty in $\phi_{tg}$ with $\Delta(E_e)=0.5$ MeV.}
\begin{center}
\begin{tabular}{|c| c| }
\hline
Kinematics & $\Delta\phi_{tg}(E_e)$  [mrad]\\
\hline
K1 & 0.11\\
K2 & 0.11\\
K3 & 0.14\\
K4 & 0.14\\
K5 & 0.16\\
K6 & 0.14\\
K7 & 0.18\\
K8 & 0.18\\
\hline
\end{tabular}
\label{tab:beam}
\end{center}
\end{table}

3. Proton central momentum $P_0$

The momentum we reconstructed is the relative momentum $\delta_p$, which refers
to the central momentum $P_0$. At the beginning of the experiment, we
switched to NMR probe D instead of probe A, which is typically used. From the calibration
study at 1 GeV/c, the offset between probe A and D is $1.07\times
10^{-4}$~\cite{log_nmr}. The NMR values for each momentum setting are
listed in Table~\ref{tab:nmr}.
\begin{table}[t]
\caption{Recorded magnetic field $B_0$ in kG with probe D for each momentum setting.}
\begin{center}
\begin{tabular}{|c| c| c| c| }
\hline
Kinematics & $\delta_p=-2\%$ & $\delta_p=0\%$ & $\delta_p=2\%$ \\
\hline
K1 & 2.647  & 2.595 & 2.543 \\
K2 & 2.426  & 2.378 & 2.330 \\
K3 & 3.049  & 2.989 & 2.929 \\
K4 & 2.869  & 2.813 & 2.757 \\
K5 & 3.410  & 3.342 & 3.276 \\
K6 & 3.229  & 3.166 & 3.102 \\
K7 & 3.591  & 3.521 & 3.450 \\
K8 & 3.923  & 3.846 & 3.769 \\
\hline
\end{tabular}
\label{tab:nmr}
\end{center}
\end{table}

\begin{figure}
\begin{center}
\includegraphics[angle=0,width=.45\textwidth]{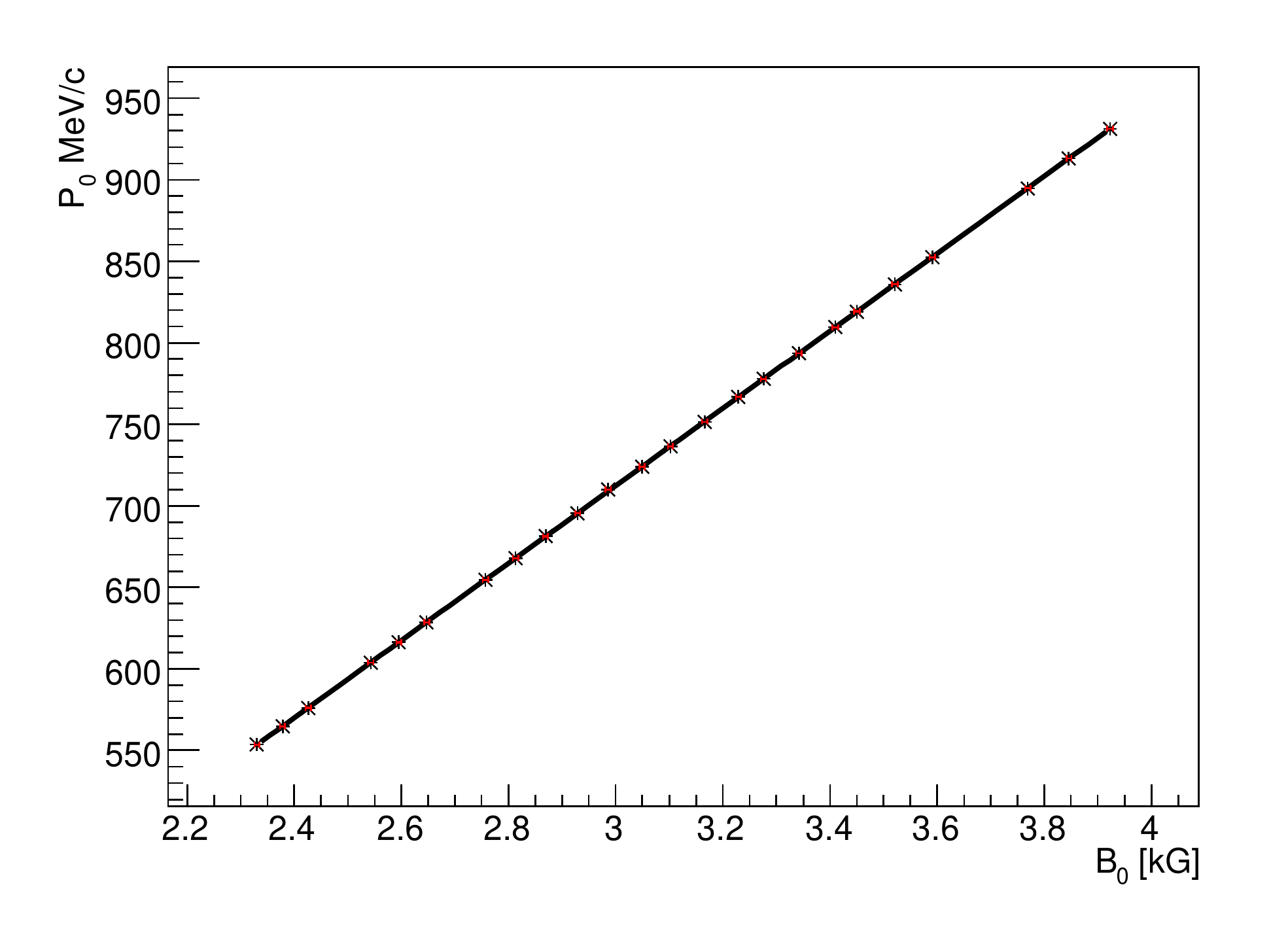}
\includegraphics[angle=0,width=.45\textwidth]{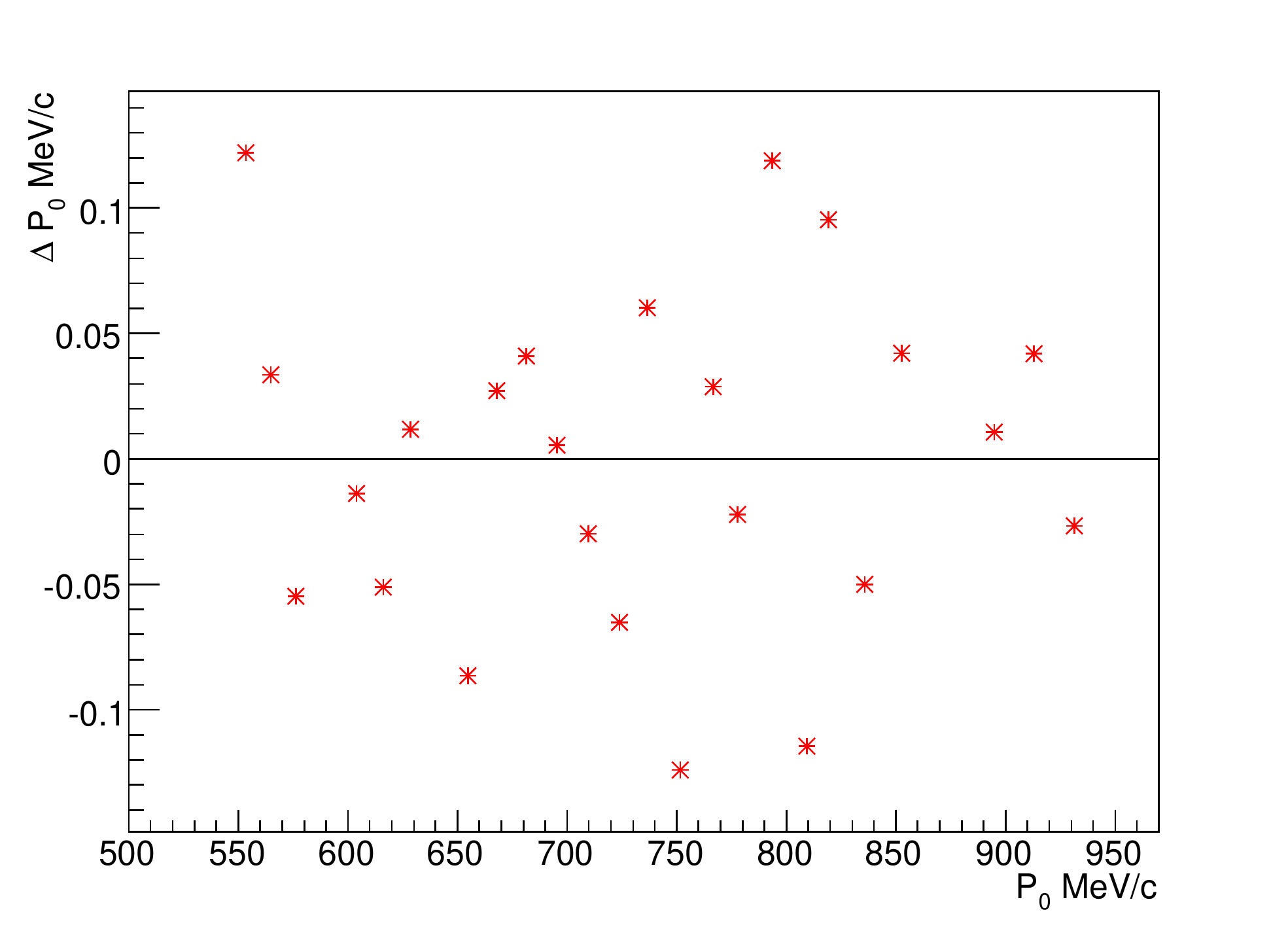}
\caption{NMR reading with probe D versus the central momentum setting (left
panel), and the deviation between the value from the linear fit function
and the set value.}
\label{fig:probe_d}
\end{center}
\end{figure}

From a previous calibration study~\cite{nmr_calib} with NMR probe A, we know
that the central momentum $P_0$ is fairly linear with the central magnetic
field $B_0$. The relation between $P_0$ and $B_0$ is given by:
\begin{equation}
P_0 = \Gamma_1 B_0 + \Gamma_3 B_0^3,
\end{equation}
where $B_0$ is measured in kG. For the left HRS, $\Gamma_1 = 270.2\pm 0.15$, and $\Gamma_3
= -1.6\times 10^{-3}\pm 0.7\times 10^{-3}$, which is much smaller than $\Gamma_1$.
With probe D, a linear fit yields:
\begin{equation}
P_0 = \Gamma_1 B^d_0.
\end{equation}
As shown in Fig.~\ref{fig:probe_d}, the linearity was well preserved when the
probed was switched. Based on the differences between the set values
and the ones derived from the new fit, we conservatively estimate $\pm 0.15$ MeV/c as
the uncertainty on the proton central momentum. The converted uncertainty in
$\phi_{tg}$ for each kinematics is summarized in Table~\ref{tab:p0}.
\begin{table}[t]
\caption{Converted uncertainty in $\phi_{tg}$ from $P_0$.}
\begin{center}
\begin{tabular}{|c| c|}
\hline
Kinematics & $\Delta\phi_{tg}(P_0)$ [mrad]\\
\hline
K1 & 0.12  \\
K2 & 0.13  \\
K3 & 0.13  \\
K4 & 0.13  \\
K5 & 0.12  \\
K6 & 0.12  \\
K7 & 0.13  \\
K8 & 0.12  \\
\hline
\end{tabular}
\label{tab:p0}
\end{center}
\end{table}

4. Proton momentum loss in the target

The recoil protons passed through a few materials before they entered the spectrometer.
The materials are summarized in Table~\ref{tab:ploss_ma}.
\begin{table}[t]
\caption{Target materials that the proton passed through before entering the spectrometer.}
\begin{center}
\begin{tabular}{|l | c|}
\hline
Material & Thickness [cm] \\
\hline
LH$_2$ & 1.27$\pm0.01$ (radius)\\
\hline
Al wall & 0.0127$\pm 0.0005$ \\
\hline
Al vacuum chamber window & 0.0406$\pm0.0005$\\
\hline
Air & 65.1$\pm0.1$\\
\hline
Kapton window & 0.0355$\pm0.0005$\\
\hline
\end{tabular}
\label{tab:ploss_ma}
\end{center}
\end{table}
The proton momentum loss $P_{loss}$ for each momentum setting is summarized in Table~\ref{tab:ploss}.
We conservatively quote $\pm 0.1 \mathrm{MeV}/c$ as the uncertainty in the average proton
momentum loss in the materials by considering the uncertainty in the material thicknesses.
   \begin{table}[t]
   \caption{Proton momentum loss [MeV/c] for each kinematics.}
   \begin{center}
   \begin{tabular}{|l |c|c|c|}
   \hline
   Kinematics &  $\delta_p =-2\%$ & $\delta_p = 0\%$ & $\delta_p = 2\%$\\
   \hline
   K1  & 3.69 & 3.83 & 3.98\\
   K2  &  4.30 & 4.48 & 4.67\\
   K3  &  2.95 & 3.05 & 3.16\\
   K4  &  3.23 & 3.35 & 3.48\\
   K5  &  2.54 & 2.62 & 2.70\\
   K6  &  2.72 & 2.81 & 2.91\\
   K7  & 2.39 & 2.46 & 2.54\\
   K8  &  2.19 & 2.25 & 2.31\\
  \hline
  \end{tabular}
  \label{tab:ploss}
  \end{center}
  \end{table}

5. $\delta_p$ reconstruction

The last parameter we need to consider is the uncertainty of the reconstructed momentum $\delta_{tg}$. From the optimization results~\cite{jin_optics}, we conservatively quote $\pm 5\times 10^{-4}$ as the uncertainty of $\delta_p$, and convert it to an uncertainty in $\phi_{tg}$. The results for each kinematics are listed in Table~\ref{tab:dp_phi}.
  \begin{table}[t]
  \caption{Uncertainty of $\phi_{tg}$ with $\Delta\delta_p = 0.0005$ }
  \begin{center}
  \begin{tabular}{|c| c| }
  \hline
  Kinematics & $\Delta\phi(\delta_p)$ [mrad]\\
  \hline
  K1 & 0.25\\
  K2 & 0.25\\
  K3 & 0.30\\
  K4 & 0.30\\
  K5 & 0.33\\
  K6 & 0.30\\
  K7 & 0.35\\
  K8 & 0.35\\
  \hline
  \end{tabular}
  \label{tab:dp_phi}
  \end{center}
  \end{table}

  In Table~\ref{tab:tot_uncer}, the uncertainty in $\phi_{tg}$ converted from the uncertainties
  of the external parameters as discussed above are given. $\Delta\phi(x)$ is defined as:
  \begin{equation}
   \Delta\phi(x) = \sqrt{\sum_{i=0}^N\Delta^2\phi(x_i)},
   \end{equation}
   where $\Delta\phi(x_i)$ are the converted uncertainties in $\phi_{tg}$ from the related parameters.
   \begin{table}[t]
   \caption{Total uncertainty in $\phi_{tg}$ from the external parameters.}
   \begin{center}
   \begin{tabular}{|c |c |c| c| c |c |c| }
   \hline
   Kinematics & $\Delta\phi(E_e)$ & $\Delta\phi(\theta_s)$ & $\Delta\phi(\delta_p)$ & $\Delta\phi(P_0)$ & $\Delta\phi(P_{loss})$ & $\Delta\phi(x)$ [mrad] \\
   \hline
   K1 & 0.11 & 0.14 & 0.25 & 0.12 & 0.08 & 0.34\\
   K2 & 0.11 & 0.14 & 0.25 & 0.13 & 0.09 & 0.34\\
   K3 & 0.14 & 0.14 & 0.30 & 0.13 & 0.09 & 0.39\\
   K4 & 0.16 & 0.14 & 0.30 & 0.13 & 0.08 & 0.40\\
   K5 & 0.16 & 0.14 & 0.33 & 0.12 & 0.08 & 0.42\\
   K6 & 0.14 & 0.14 & 0.30 & 0.12 & 0.08 & 0.39\\
   K7 & 0.18 & 0.14 & 0.35 & 0.13 & 0.09 & 0.45\\
   K8 & 0.18 & 0.13 & 0.35 & 0.12 & 0.08 & 0.44\\
   \hline
   \end{tabular}
   \label{tab:tot_uncer}
   \end{center}
   \end{table}

\begin{figure}[hbt]
   \begin{center}
    \includegraphics[angle=0,width=.55\textwidth]{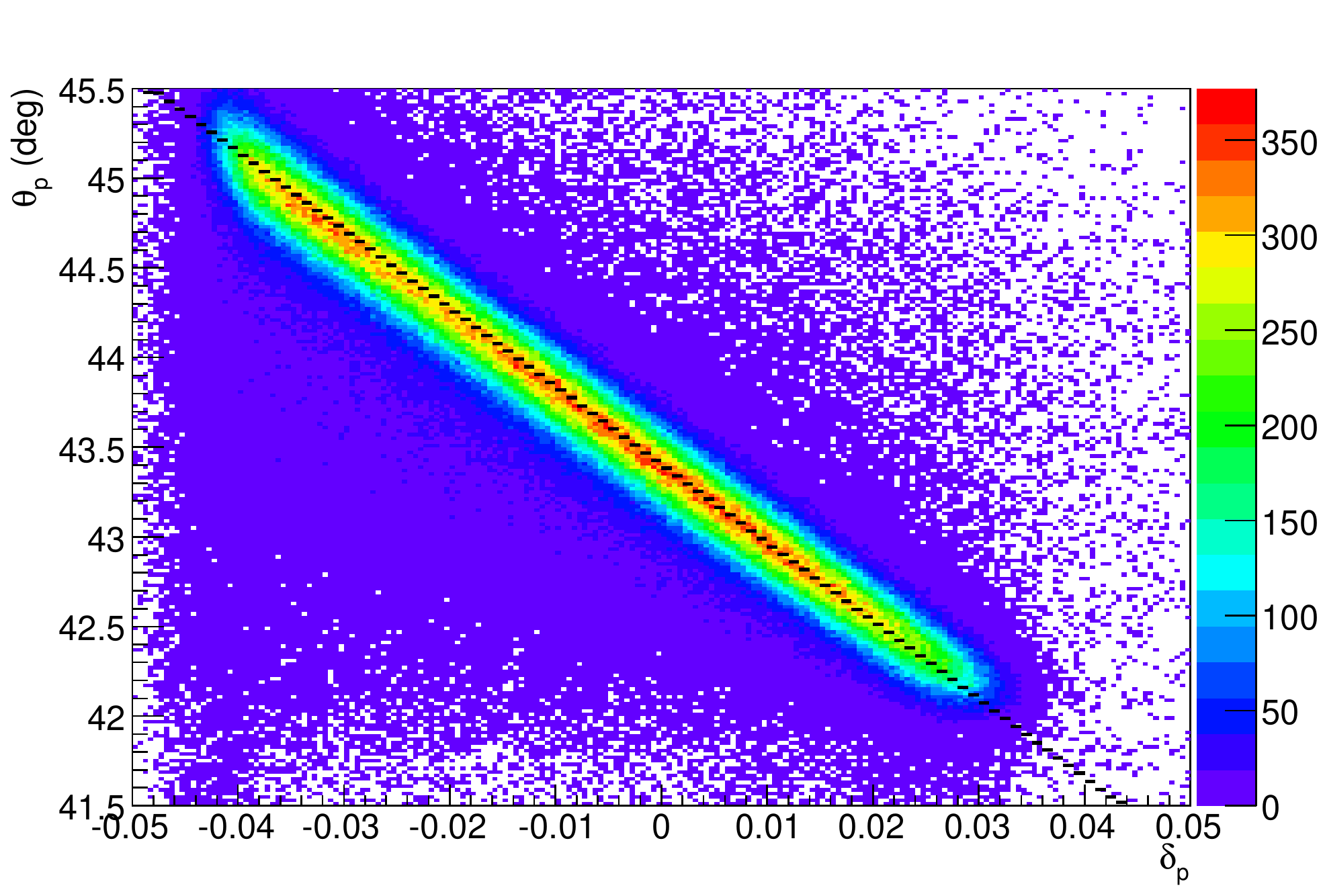}
    \caption{Proton scattering angle $\theta_{p}$ versus the momentum $\delta_p$ for kinematics K8 $\delta_p = 0\%$. The anticipated elastic peak position is plotted as the black dash line.}
    \label{fig:optics1}
   \end{center}
   \end{figure}

   The next step is to quote $\Delta\phi(off)$, which is the average deviation of the replayed proton kienmatics from the anticipated elastic peak position. From Figure~\ref{fig:optics1}, we see that the slope of the elastic strips generally matches the predicted slopes. The average offset across the acceptance between the predicted peak position and the center of the data $\Delta\phi(off)$ is a combined effect of the optics and external parameters (beam energy, proton momentum, etc.). The systematic uncertainty from the optics is given by:
  \begin{eqnarray}
  \Delta\phi_{tg}  = \sqrt{\Delta^2\phi(off) + \Delta^2\phi(x)}.
  \label{eq:dev_phi}
  \end{eqnarray}
   The final uncertainty in $\phi_{tg}$ is summarized in Table~\ref{tab:phi_offset}.
   \begin{table}[t]
   \caption{$\phi_{tg}$ uncertainty for each kinematics.}
   \begin{center}
   \begin{tabular}{|c |c| c| c|}
   \hline
   Kinematics & $\Delta\phi_{off}$ [mrad] & $\Delta\phi(x)$ [mrad] & $\Delta\phi_{tot}$ [mrad]\\
   \hline
   K1 & 1.30 & 0.34 &  1.34\\
   K2 & 0.76 & 0.34 &  0.83\\
   K3 & 0.87 & 0.39 &  0.95\\
   K4 & 0.87 & 0.40 &  0.96\\
   K5 & 0.70 & 0.42 &  0.67\\
   K6 & 0.87 & 0.39 &  0.95\\
   K7 & 1.22 & 0.45 &  1.30\\
   K8 & 1.05 & 0.44 &  1.14\\
   \hline
   \end{tabular}
   \label{tab:phi_offset}
   \end{center}
   \end{table}
  The uncertainties of the other target quantities ($\theta_{tg},\delta_p, y_{tg}$) are quoted according to their difference when the data were replayed by using different HRS optics:
 \begin{equation}
 \Delta\theta_{tg} = 2~\mathrm{mrad},
 \Delta\delta_p = 0.001,
 \Delta y_{tg} = 1~\mathrm{mm}
 \label{eq:uncer}
 \end{equation}

Combining the results in Table~\ref{tab:phi_offset}, Eq.~\ref{eq:uncer} and Table~\ref{tab:tg_uncer}, the total systematic uncertainty from the left HRS optics is summarized in Table~\ref{tab:uncer_optics}.
\begin{table}[t]
\caption{Systematic uncertainty in $R = \mu_pG_E/G_M$ for each kinematics associated with left HRS optics.}
\begin{center}
\begin{tabular}{|c| c| }
\hline
Kinematics &  $\Delta R $ (optics)\\
\hline
K1 & 0.0087\\
K2 & 0.0057\\
K3 & 0.0062\\
K4 & 0.0068\\
K5 & 0.0051\\
K6 & 0.0063\\
K7 & 0.0090\\
K8 & 0.0084\\
\hline
\end{tabular}
\label{tab:uncer_optics}
\end{center}
\end{table}

\subsubsection{COSY}
\begin{figure}[hbt]
  \begin{center}
    \includegraphics[angle=0,width=.49\textwidth]{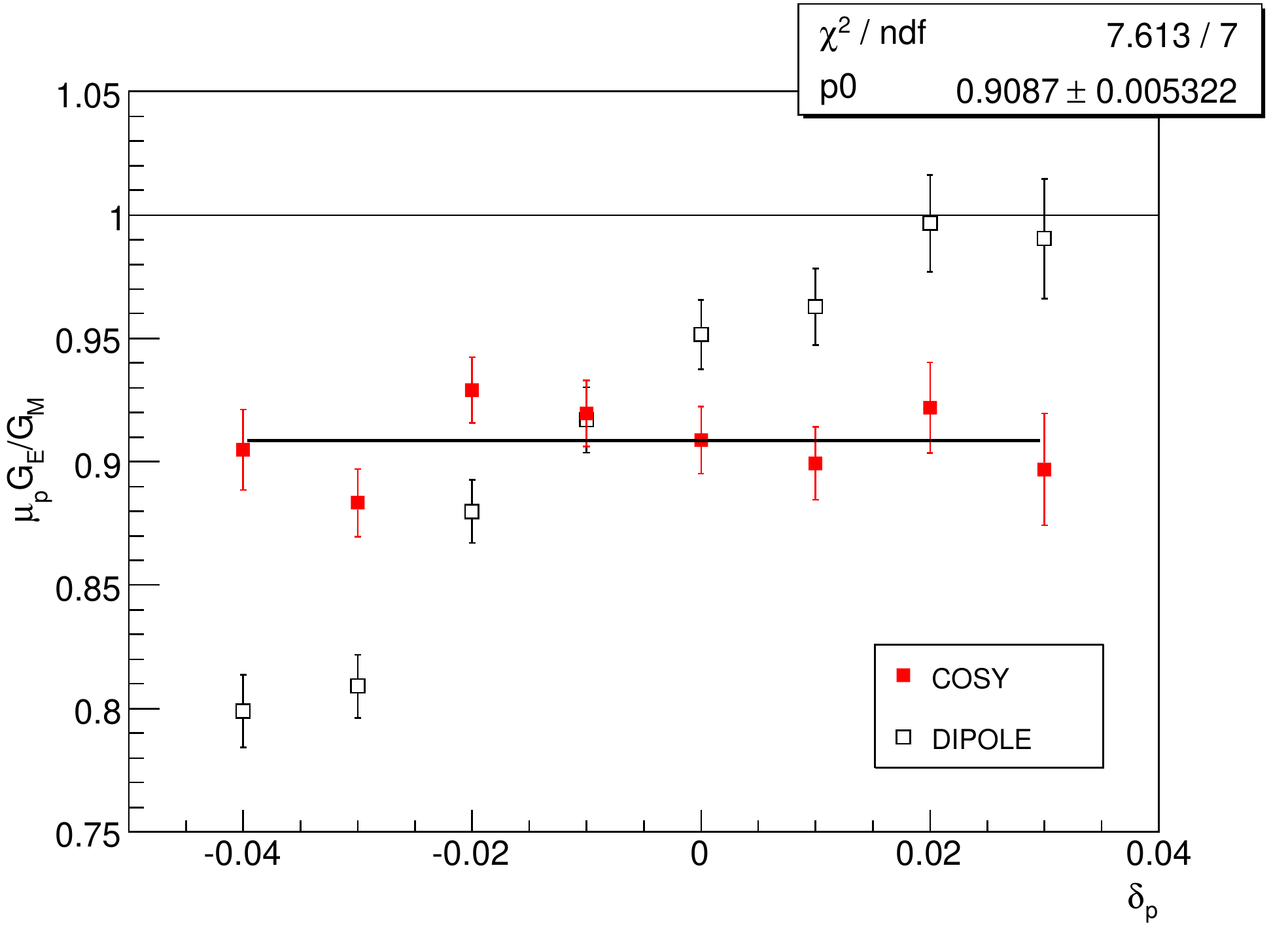}
    \includegraphics[angle=0,width=.49\textwidth]{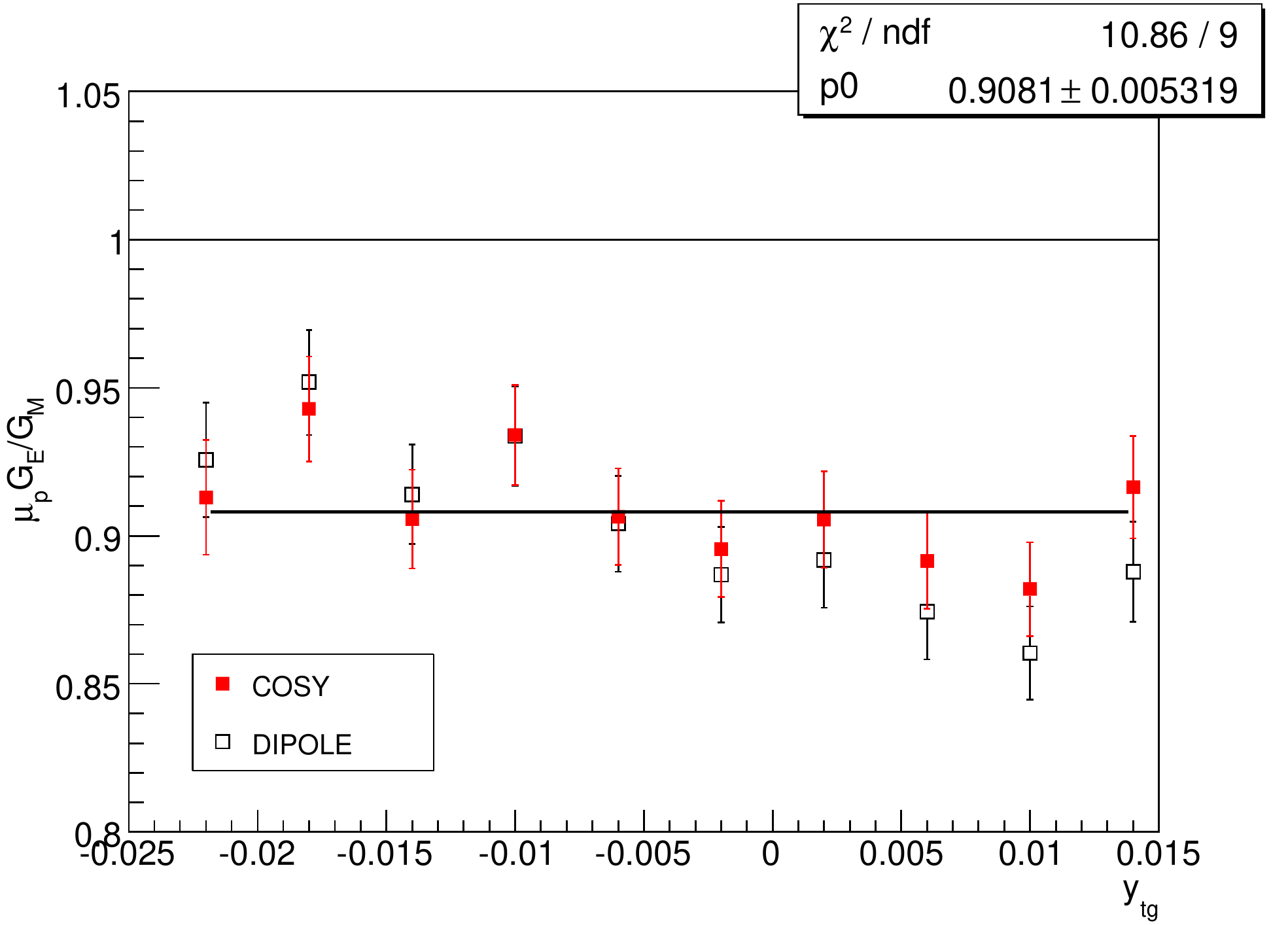}
    \includegraphics[angle=0,width=.49\textwidth]{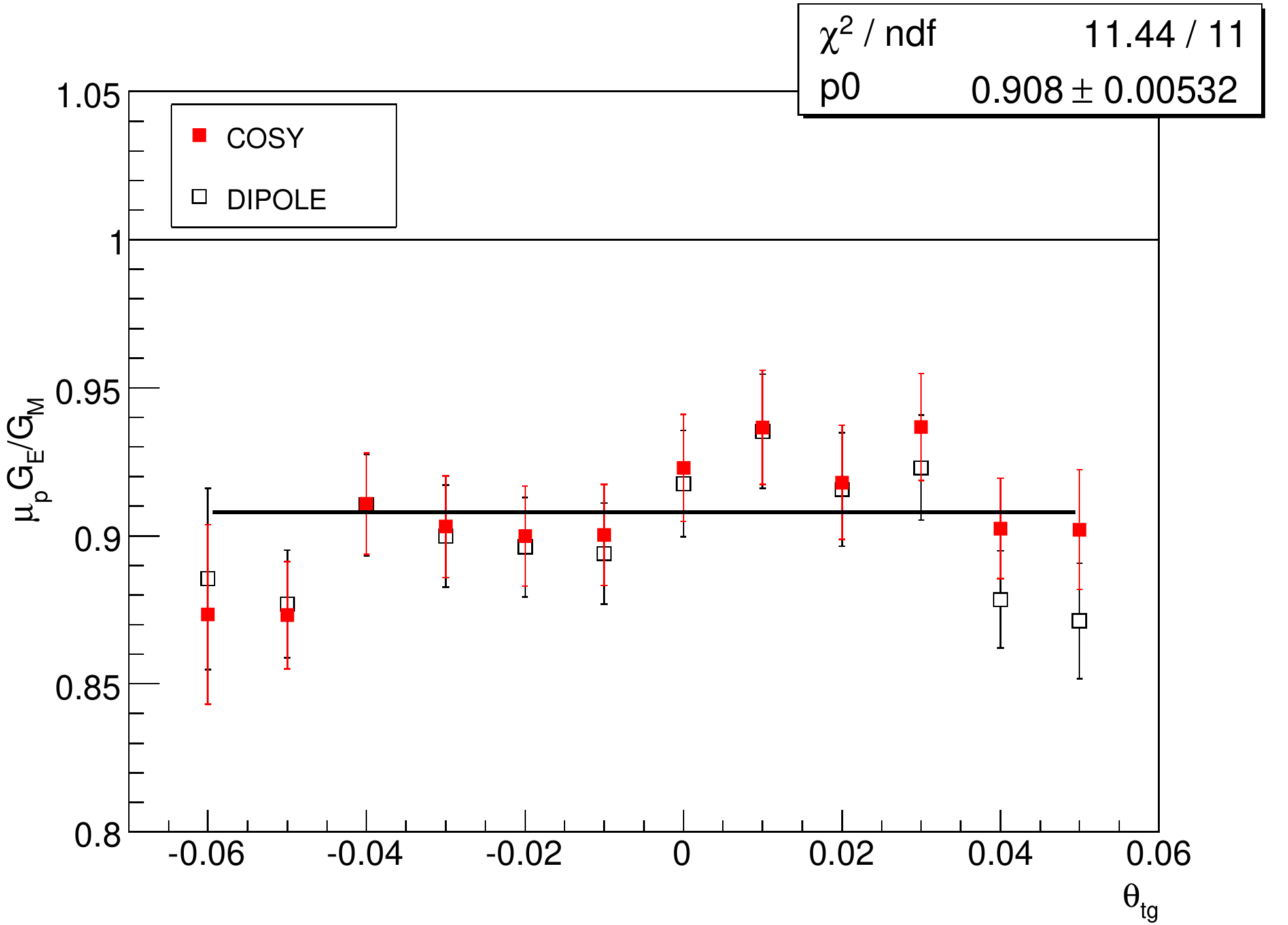}
    \includegraphics[angle=0,width=.49\textwidth]{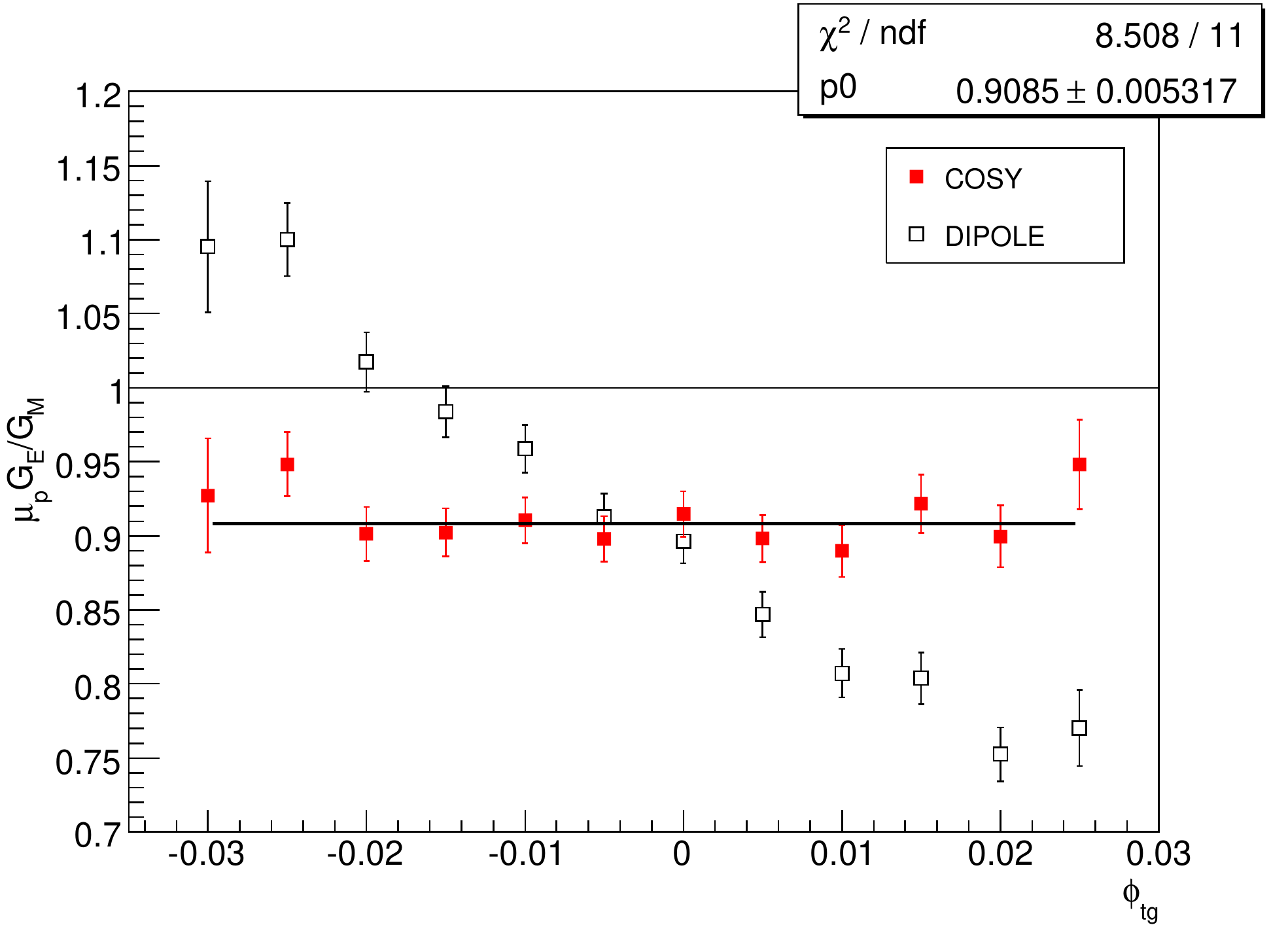}
    \caption{Dependence of $\mu_pG_{Ep}/G_{Mp}$ on the proton target quantities for kinematics K7 ($Q^2 = 0.6$ GeV$^2$). The full precession matrix calculated by COSY (solid quare) is compared to the dipole approximation (open square) and a constant fit. The data points are shown with statistical error bars only. }
    \label{fig:bin}
  \end{center}
 \end{figure}

\begin{figure}[hbt]
  \begin{center}
    \includegraphics[angle=0,width=.65\textwidth]{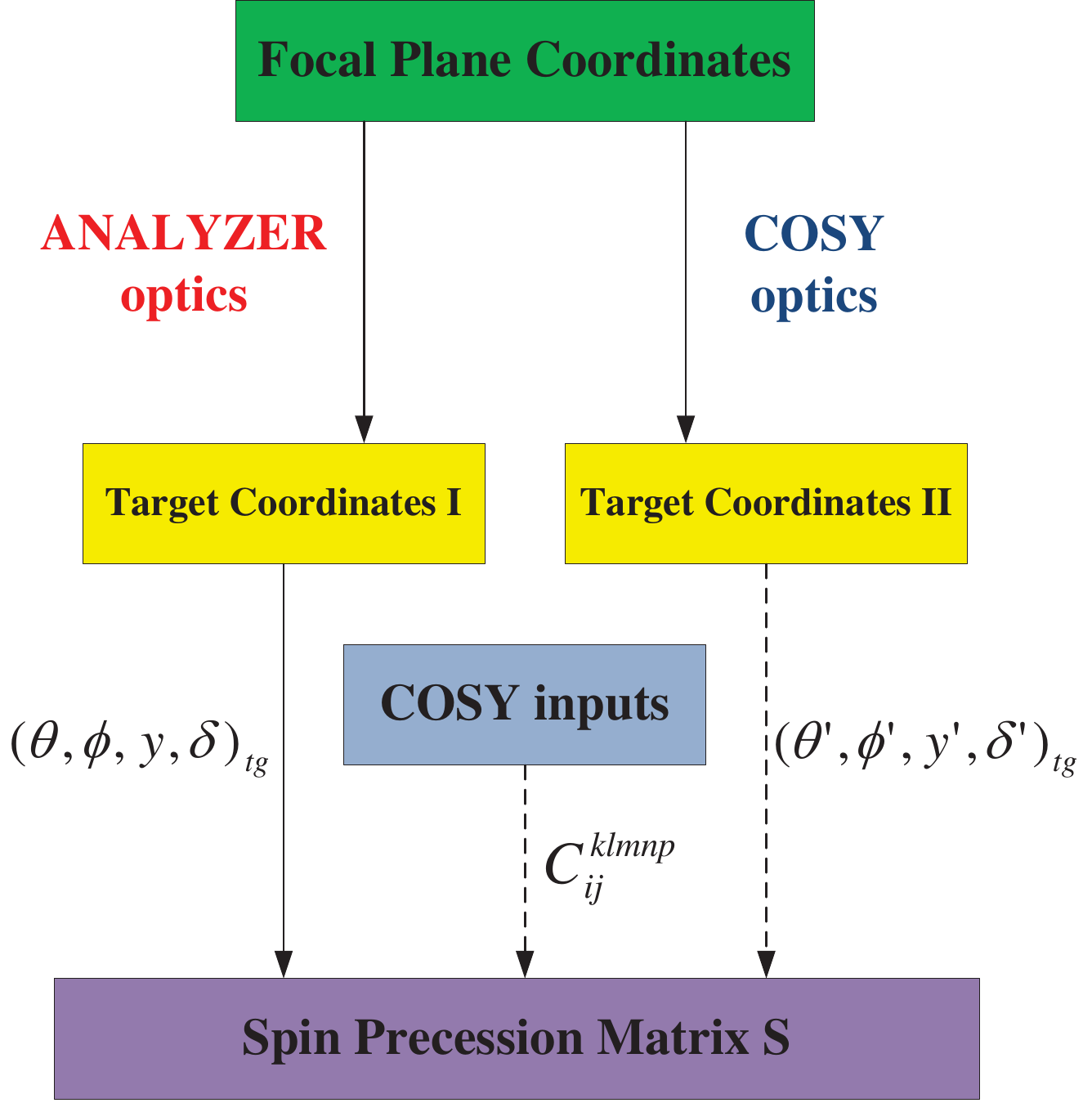}
    \caption{Alternative ways to calculate the spin precession matrix $S_{ij}$.}
    \label{fig:cosytest}
  \end{center}
 \end{figure}

Another source of systematic error, which is related to the spin precession, is COSY. If the precession matrix determined by COSY is correct, the form factor ratio $\mu_pG_{Ep}/G_{Mp}$ should not depend on any target quantities. As illustrated in Fig.~\ref{fig:bin}, while the results with dipole approximation show a strong dependence on $\delta_p$ and $\phi_{tg}$, COSY provides a nice correction to these quantities and gives a reasonable $\chi^2$ with a constant fit.
To estimate the systematic error of COSY, more detailed studies were carried out. The COSY systematic error was separated into two parts:
 \begin{itemize}
 \item The first one is associated with the spectrometer configuration and settings defined in the COSY input file. Through a series of tests, the most sensitive parameters were identified. Then, those parameters were changed and the spin procession was calculated in different ways to see the variation in the form factor ratio.
 \item The other part was determined from the COSY optics map, which also reflects the quality of the model. We used the target quantities reconstructed by COSY instead of the ones from the ANALYZER to calculate the spin precession matrix $S_{ij}$.
 \end{itemize}
Fig.~\ref{fig:cosytest} demonstrates the alternative ways to calculate the spin precession and estimate the model's systematics uncertainty.

  1. Configuration Inputs

\begin{figure}[hbt]
  \begin{center}
  \includegraphics[angle=0,width=.5\textwidth]{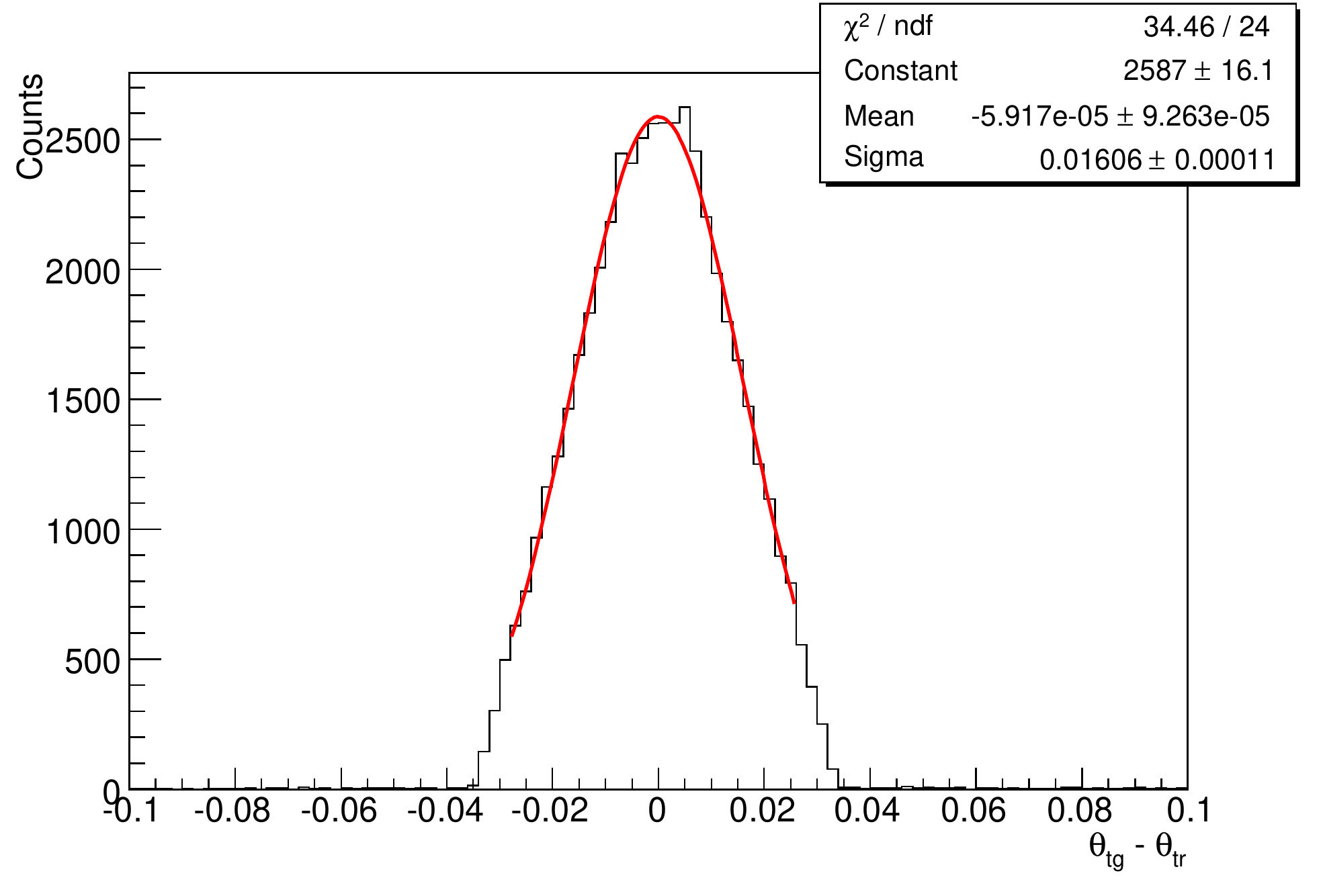}
    \caption{Fit of the out-of-plane angle difference between the target and the focal plane. $\theta_{tr}=\theta_{det}-45^{\circ}$ (K6 $\delta_p = 0\%$). The peak at zero corresponds to a $45^{\circ}$ bending angle in the spectrometer.}
    \label{fig:fit}
  \end{center}
 \end{figure}

  In the COSY input file, geometries and settings of the magnets were defined. Many of the parameters were determined by comparing them with the field maps. We focused on the ones that are either intuitive or examined in the previous study~\cite{alignment}. The tested parameters included:
   \begin{itemize}
   \item Dipole bending angle $\Theta_0$.
   \item Dipole radius.
   \item Drift distances between magnets.
   \item Quadrupole alignment coefficients.
   \end{itemize}
   The dipole bending angle was found to be the most important parameter in this measurement; on the other hand, the impact from the other parameters was negligible. The default setting for the dipole bending angle is $45^{\circ}$. Ideally we should be able to check the central bending angle from the trajectory determined by the VDCs, when combined with the VDC position survey~\cite{vdc_survey}. However, it's very difficult to define the spectrometer central trajectory\footnote{Usually the central sieve hole position was used to define the spectrometer central trajectory.}. To minimize bias, we cut on a very small region of the central part of the HRS acceptance and treated the events in this region as the central trajectories. By fitting the out-of-plane angle difference ($\theta_{tg}-\theta_{tr}$) between the target frame and the focal plane, the dipole central bending angle was verified. The cuts applied to select the central trajectories were:
  \begin{itemize}
  \item $-0.01<\delta_p<0.01$.
  \item $-0.01<y_{tg}<0.01$.
  \item $-0.01<\theta_{tg}<0.01$.
  \end{itemize}
  A fit to the out-of-plane angle difference is illustrated in Figure~\ref{fig:fit}.
 We can see that the mean value of the peak is very close to 0, which corresponds to a $45^{\circ}$ bending angle between the target and the focal plane. In a previous analysis~\cite{oliv}, a $\pm$5.5 mrad uncertainty in the dipole bending angle $\Theta_0$ was quoted from a fit to the $180^{\circ}$ rotation data, and this uncertainty was also used in this analysis. Therefore, the central dipole bending angle $\Theta_0$ was changed by 5.5 mrad in the COSY input file and another set of spin precession matrices were generated to extract the ratio. The difference in the resulting form factor ratios was quoted as the systematic error associated with the central bending angle. The results are provided in Table~\ref{tab:cosy}.

 2. COSY optics map

 COSY not only generates the spin precession matrix but also produces the optics map. With the quantities measured at the VDCs, we could use the COSY optics map to reconstruct the target quantities. The COSY reconstructed target quantities are in general agreement with the target quantities determined via the ANALYZER. The central peak differences are a couple of mrad for the angles ($\phi_{tg},\theta_{tg}$) and a couple of mm for the position ($y_{tg}$). By using the COSY reconstructed target quantities in the spin precession matrix calculation, the difference in the form factor ratio was quoted as another part of the systematic error from COSY. The results from this study are reported in Table~\ref{tab:cosy}.
\begin{table}[t]
\caption{Systematic error in $\mu_pG_E/G_M$ associated with COSY.}
\begin{center}
\begin{tabular}{|c| c| c|}
\hline
Kinematics & Bending angle (+5.5 mrad) & COSY optics\\
\hline
K1 & -0.0018  & 0.0012 \\
K2 & -0.0012  & 0.0018 \\
K3 & -0.0029  & 0.0011 \\
K4 & -0.0022  & 0.0002 \\
K5 & -0.0043  & 0.0005 \\
K6 & -0.0035  & 0.0006 \\
K7 & -0.0048  & 0.0004 \\
K8 & -0.0062  & 0.0002 \\
\hline
\end{tabular}
\label{tab:cosy}
\end{center}
\end{table}

\subsection{FPP Alignment and Reconstruction}
As previously mentioned, the second scattering angles at the FPP are directly related to the proton polarizations measured in the focal plane; to make sure the angles at the FPP were determined correctly, the software alignment was completed, which is elaborated in Section~\ref{sec:align}. Ideally, the straight-through data should uniformly cover the FPP chamber's full acceptance. However, full coverage is difficult to achieve for the rear chambers due to their larger area, which was designed for the second scattering; the lack of uniform coverage inevitably changes the weight of the data over the acceptance and can affect the fits of the alignment coefficients.

\begin{figure}[hbt]
  \begin{center}
    \includegraphics[angle=0,width=.65\textwidth]{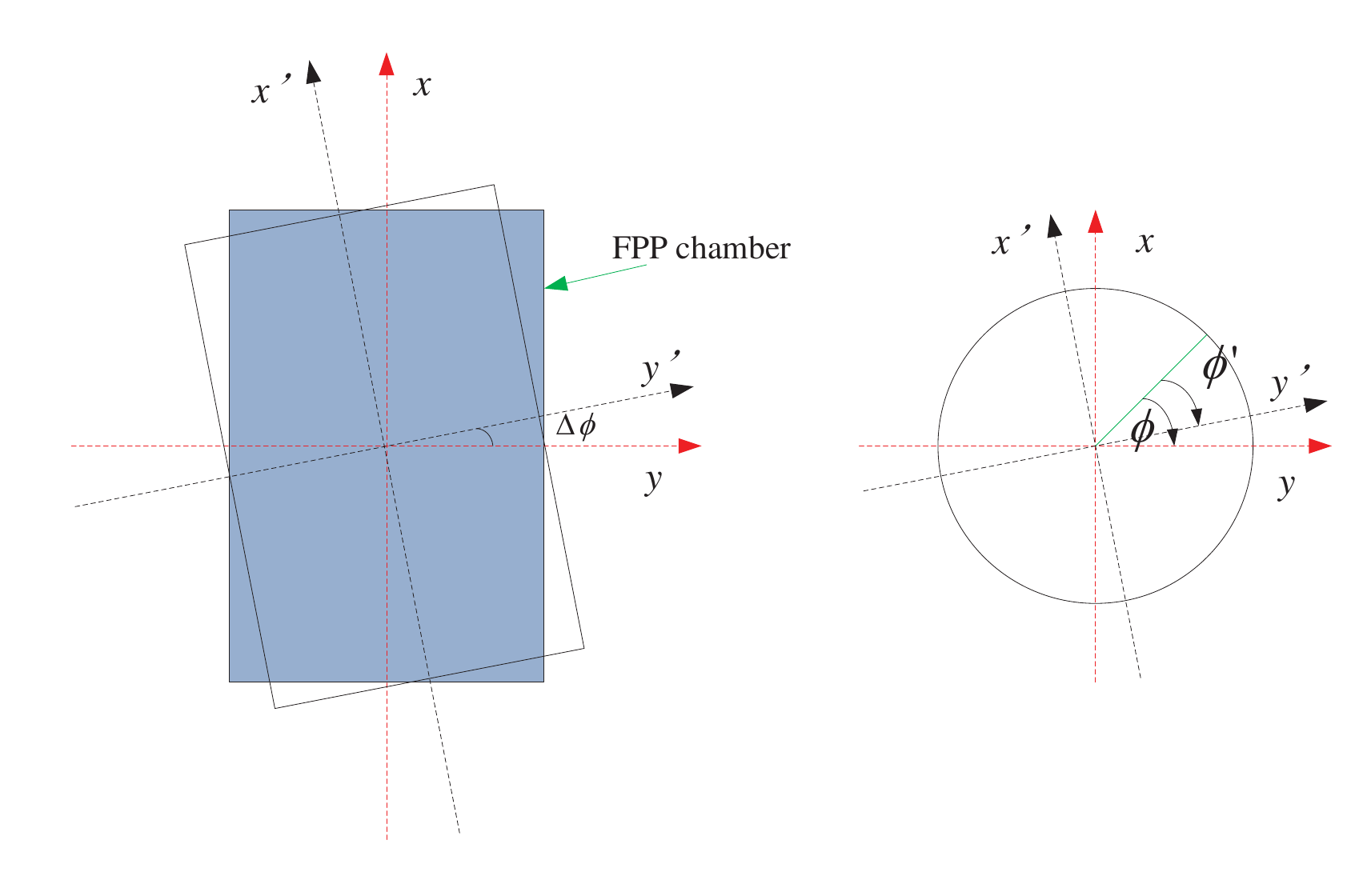}
    \caption{FPP chamber rotation along $z$ and the shift of the azimuthal angle $\phi$.}
    \label{fig:rot}
  \end{center}
 \end{figure}

\begin{figure}[hbt]
  \begin{center}
    \includegraphics[angle=0,width=.65\textwidth]{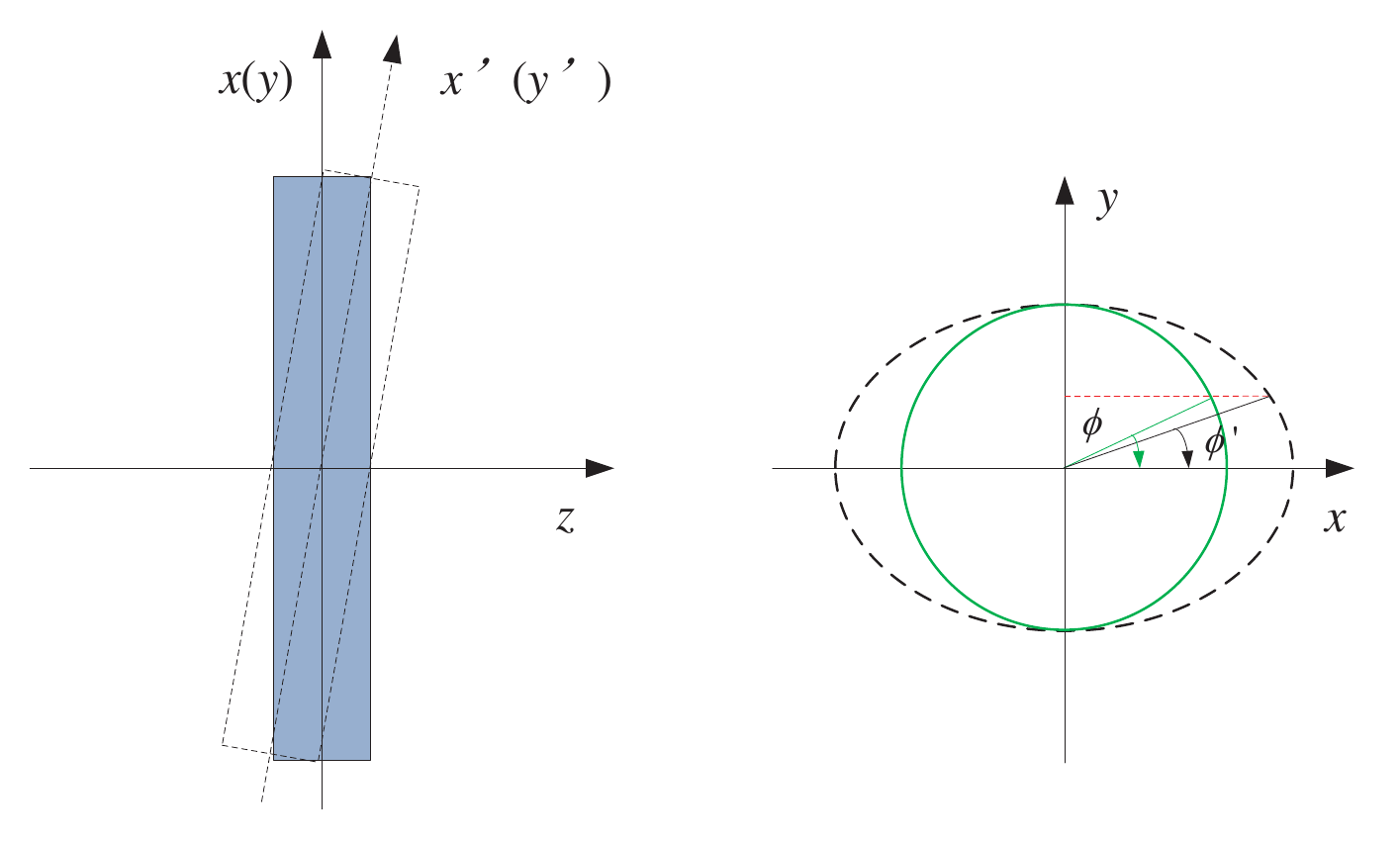}
    \caption{FPP chamber rotation along $x(y)$ and the change of $\phi$ distribution.}
    \label{fig:rot1}
  \end{center}
 \end{figure}

The misalignment of the chambers involves both offsets and rotations. For offsets, the effect is equivalent to a non-uniform acceptance $A(\phi)$, whose effect can be canceled by flipping the beam helicity. For rotations, we can separate them into two types as illustrated in Figs.~\ref{fig:rot} and \ref{fig:rot1}. The chamber rotations along $x$ and $y$-axis induce an elliptical acceptance which can also be absorbed in the non-uniform acceptance; hence, they are not our primary concern. The rotation along the $z$-axis, which is the particle's incident direction, will shift $\phi$ by an additional offset and cannot be canceled. This type of rotation will directly change the result of the ratio $\mu_p G_{Ep}/G_{Mp}$.

\begin{figure}[hbt]
  \begin{center}
    \includegraphics[angle=0,width=.60\textwidth]{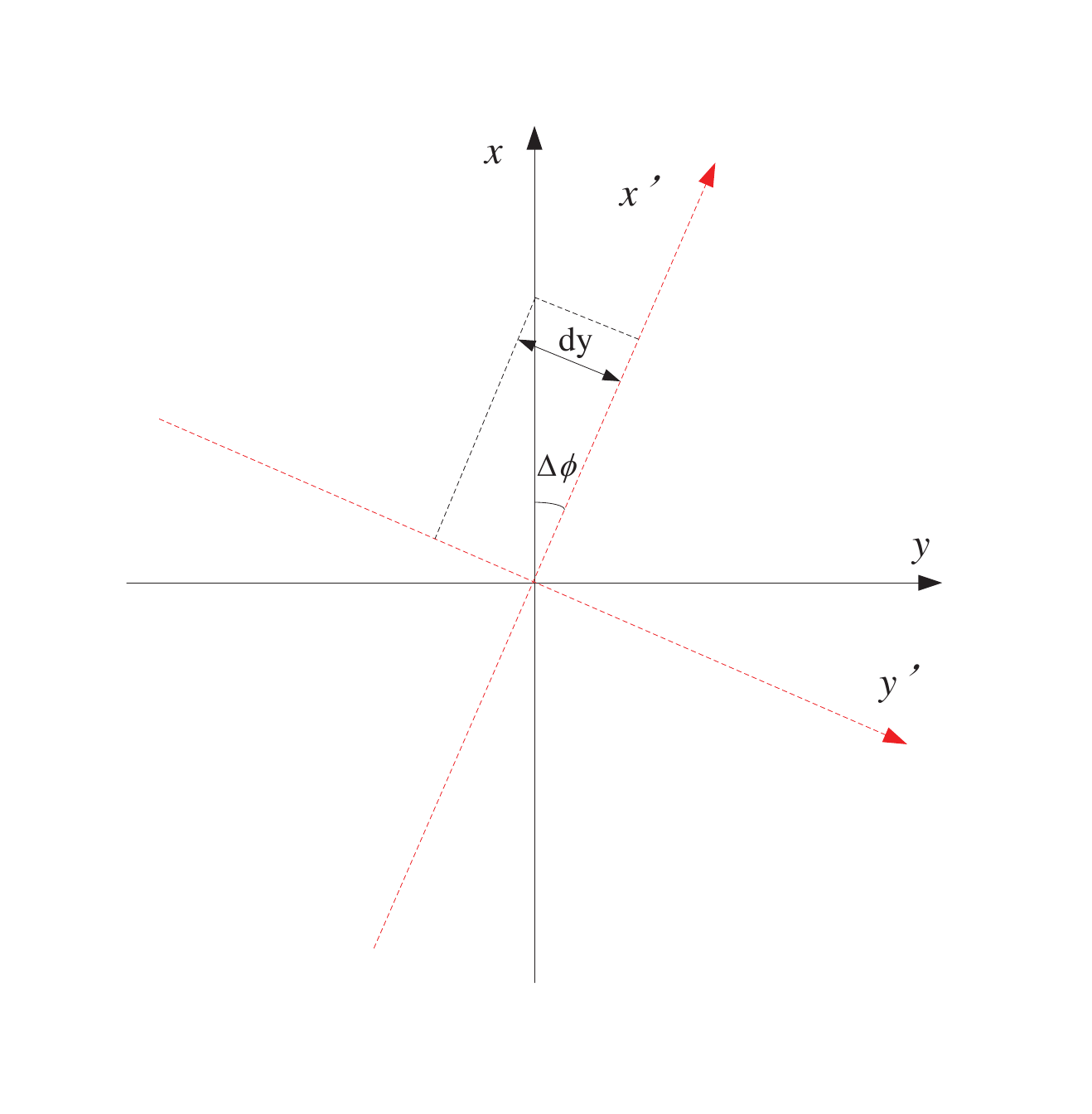}
    \caption{The non-zero $y$ component in the rotated frame.}
    \label{fig:angle}
  \end{center}
 \end{figure}

\begin{figure}[hbt]
  \begin{center}
  \includegraphics[angle=0,width=.48\textwidth]{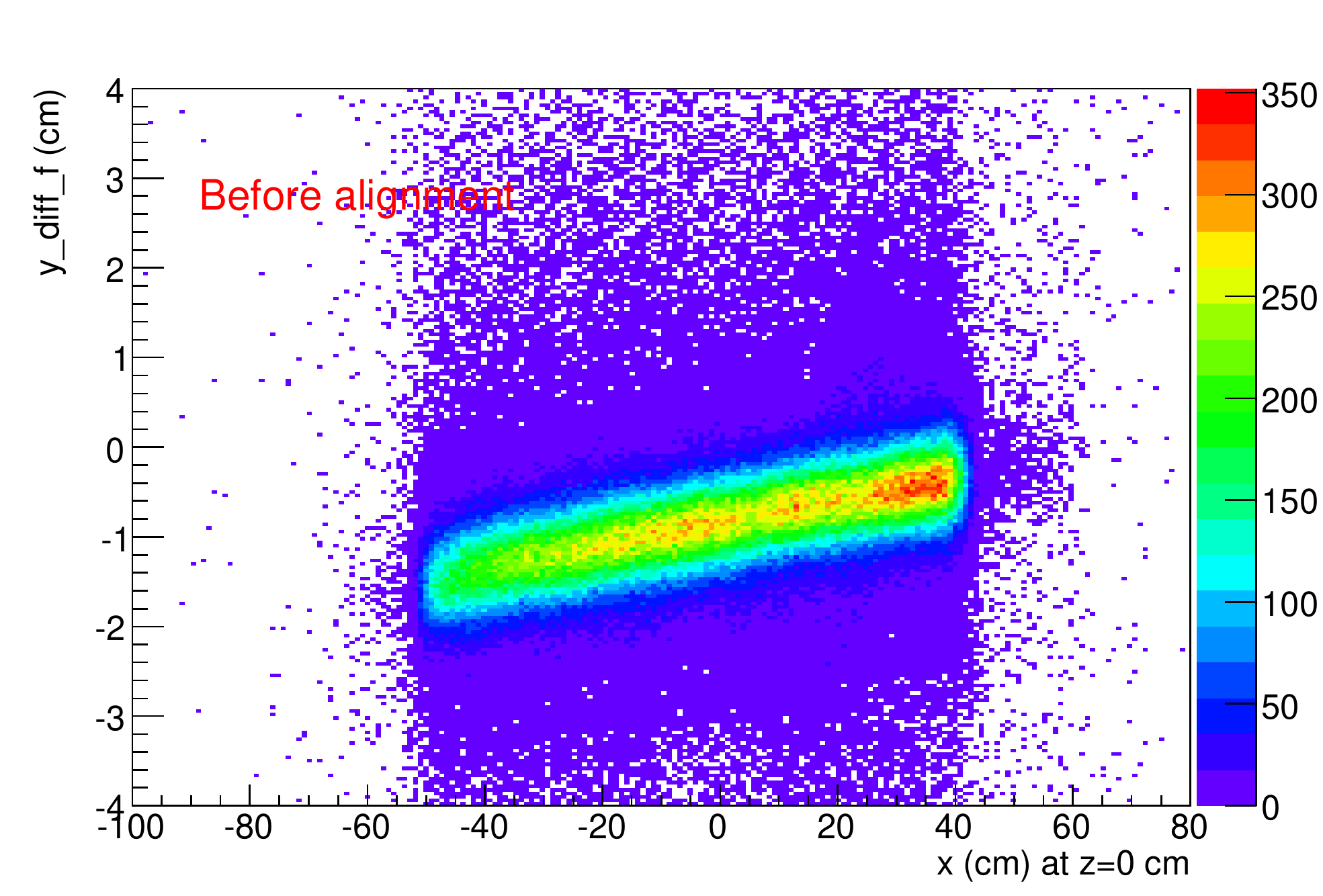}
  \includegraphics[angle=0,width=.48\textwidth]{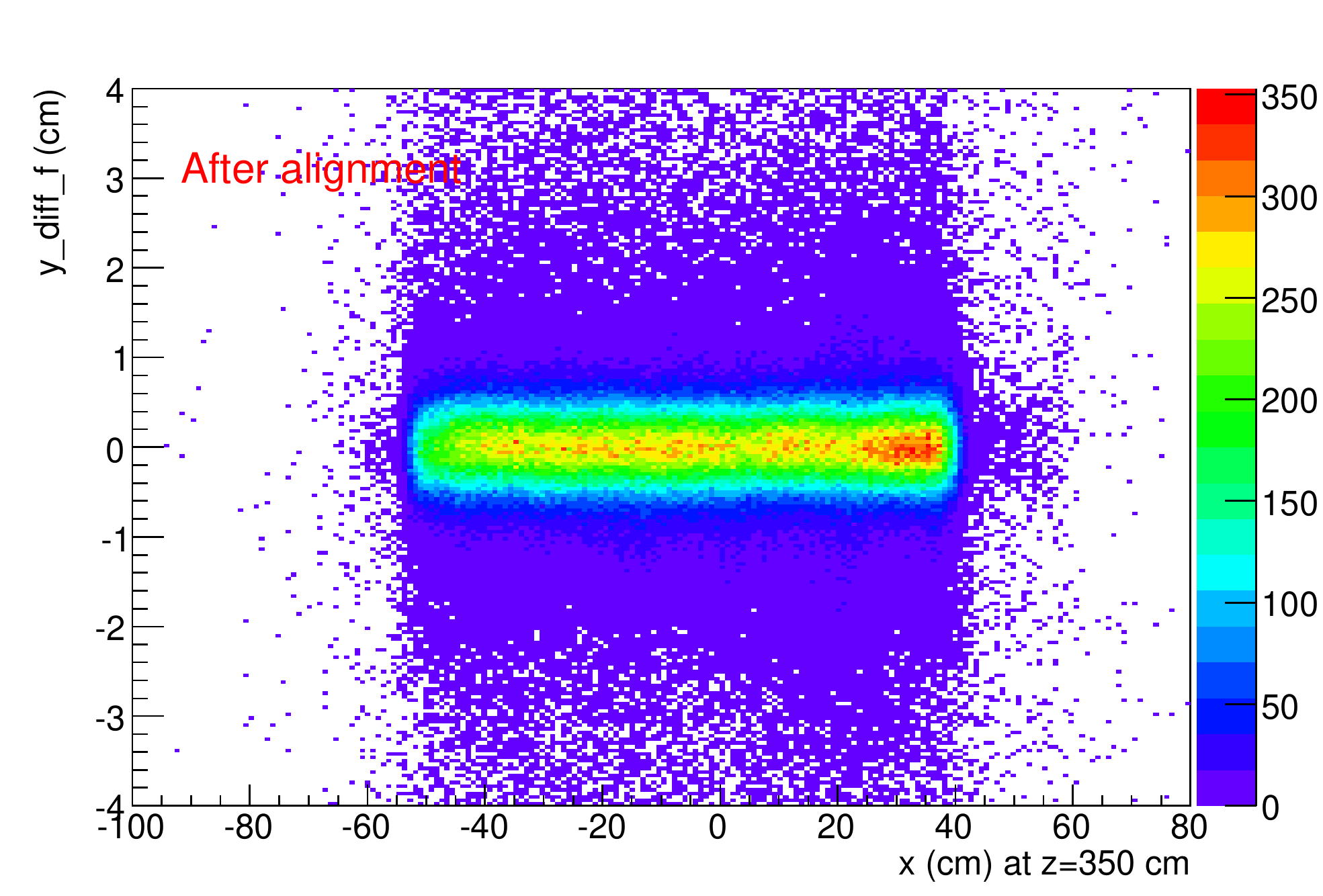}
    \caption{The track difference in $y$ versus $x$ before and after the software alignment.}
    \label{fig:diffy}
  \end{center}
 \end{figure}

\begin{figure}[hbt]
  \begin{center}
  \includegraphics[angle=0,width=.55\textwidth]{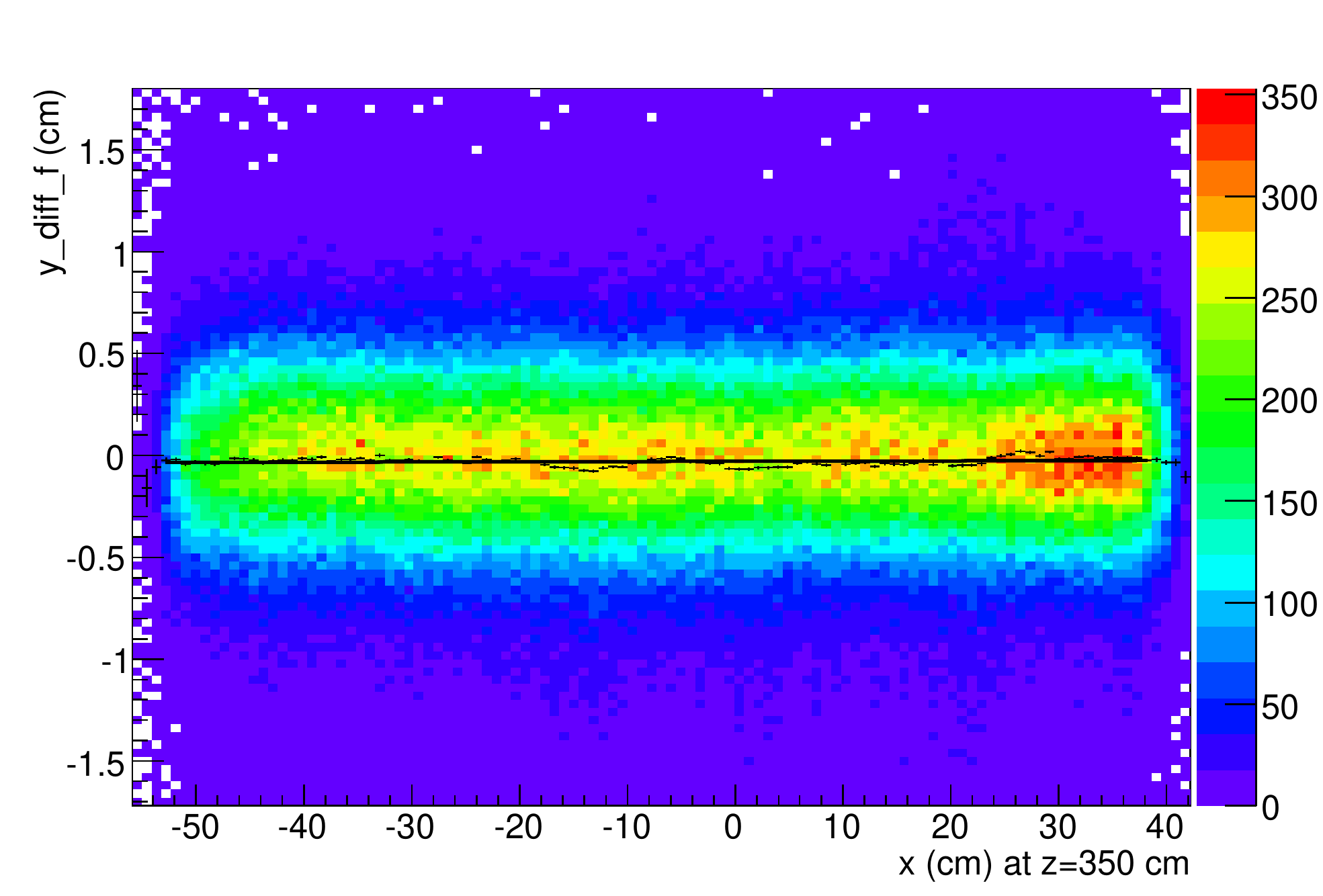}
    \caption{The track difference ($y$) and its profile versus $x$ after the software alignment. The solid line is a linear fit to the profile with a slope of $1\times10^{-4}$.}
    \label{fig:diffy_fit}
  \end{center}
 \end{figure}

The events at the FPP are mostly dispersed in $x$-direction (vertical), whereas they are close to zero in $y$. To make the estimation simpler, the reasonable assumption of $y=0$ is made. As illustrated in Fig.~\ref{fig:angle}, if there is a small rotation along $z$ or $x$, the difference in $y$ between the VDC track and the FPP track will depend on $x$ by:
\begin{equation}
dy\approx\Delta\phi \times x.
\end{equation}
   Before applying the software alignment, there is an obvious slope between $dy$ and $x$ as shown in Fig.~\ref{fig:diffy}, which indicates a rotation around $z$. After the software alignment, the slope is gone. If we zoom in and fit the spectrum after alignment, the residual slope is at the $1\times 10^{-4}$ level as shown in Fig.~\ref{fig:diffy_fit}. The same fit was applied to the rear track, and the slope is at the same order of magnitude ($\sim -3\times 10^{-4}$). Since in the first order this slope can only be caused by a rotation along $z$, we conservatively quote twice the residual slope value to be the uncertainty in the angle of rotation along $z$. The slopes of the front and rear alignment were added together as the final uncertainty of $\phi_{fpp}$, which is $\sim 1$ mrad. From this study, the systematic uncertainty associated with the FPP angle uncertainty is summarized in Table~\ref{tab:scatter_angle} for each kinematic setting.
 \begin{table}[t]
 \caption{Errors in the FPP scattering angles and the associated systematic error in $\mu_pG_E/G_M$.}
 \begin{center}
 \begin{tabular}{|c |c| c|}
 \hline
 Kinematics & $\theta_{fpp}$ (+1 mrad) & $\phi_{fpp}$ (+1 mrad)\\
 \hline
 K1 & -0.0003  & 0.0018 \\
 K2 & 0.0002   & 0.0018 \\
 K3 & 0.0001   & 0.0018 \\
 K4 & -0.0002  & 0.0019 \\
 K5 & -0.0001  & 0.0018 \\
 K6 & -0.0002  & 0.0018 \\
 K7 & -0.0001  & 0.0019 \\
 K8 & -0.0001  & 0.0019 \\
 \hline
 \end{tabular}
 \label{tab:scatter_angle}
 \end{center}
 \end{table}

\begin{figure}[hbt]
  \begin{center}
  \includegraphics[angle=0,width=.49\textwidth]{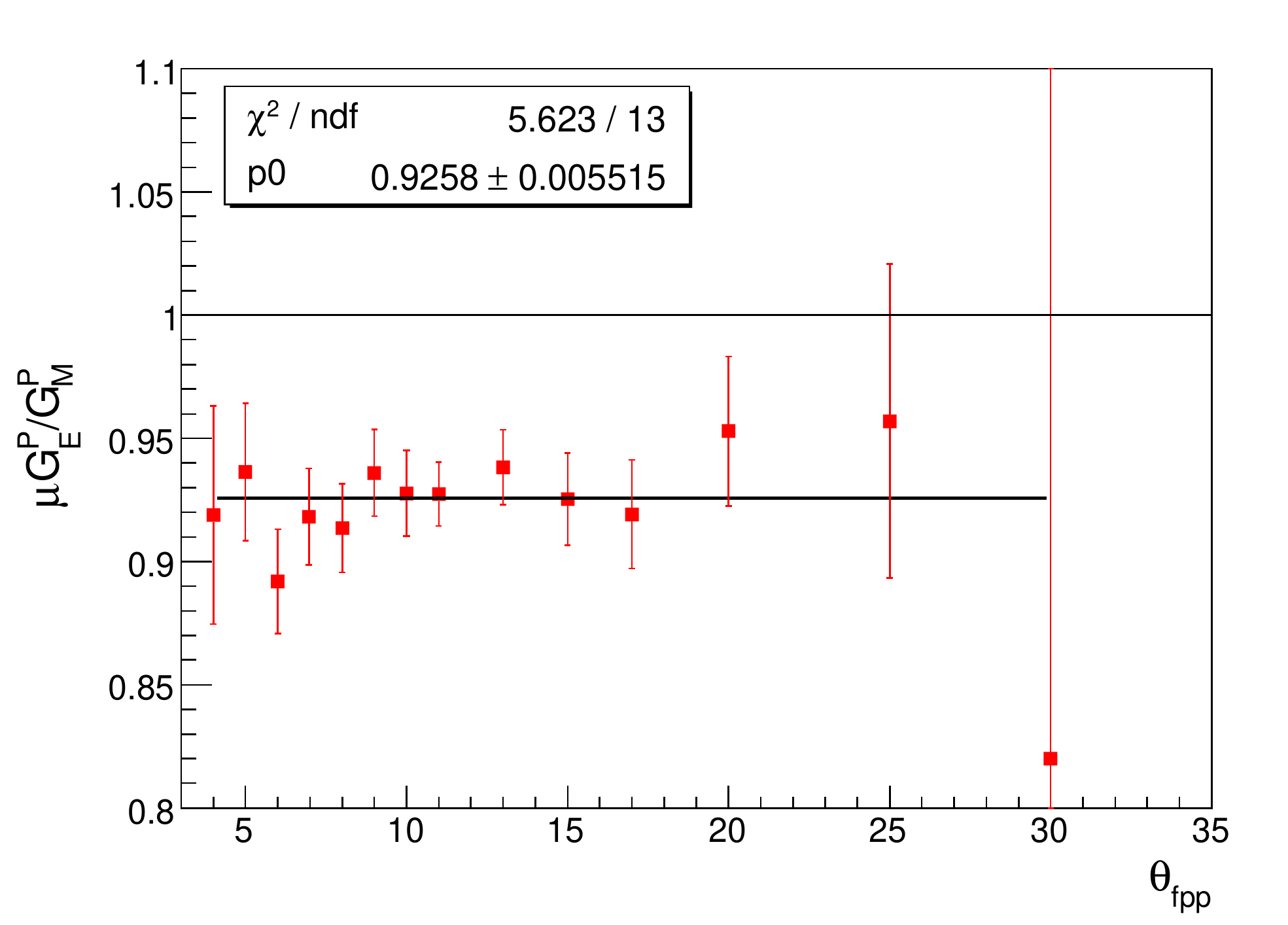}
  \includegraphics[angle=0,width=.49\textwidth]{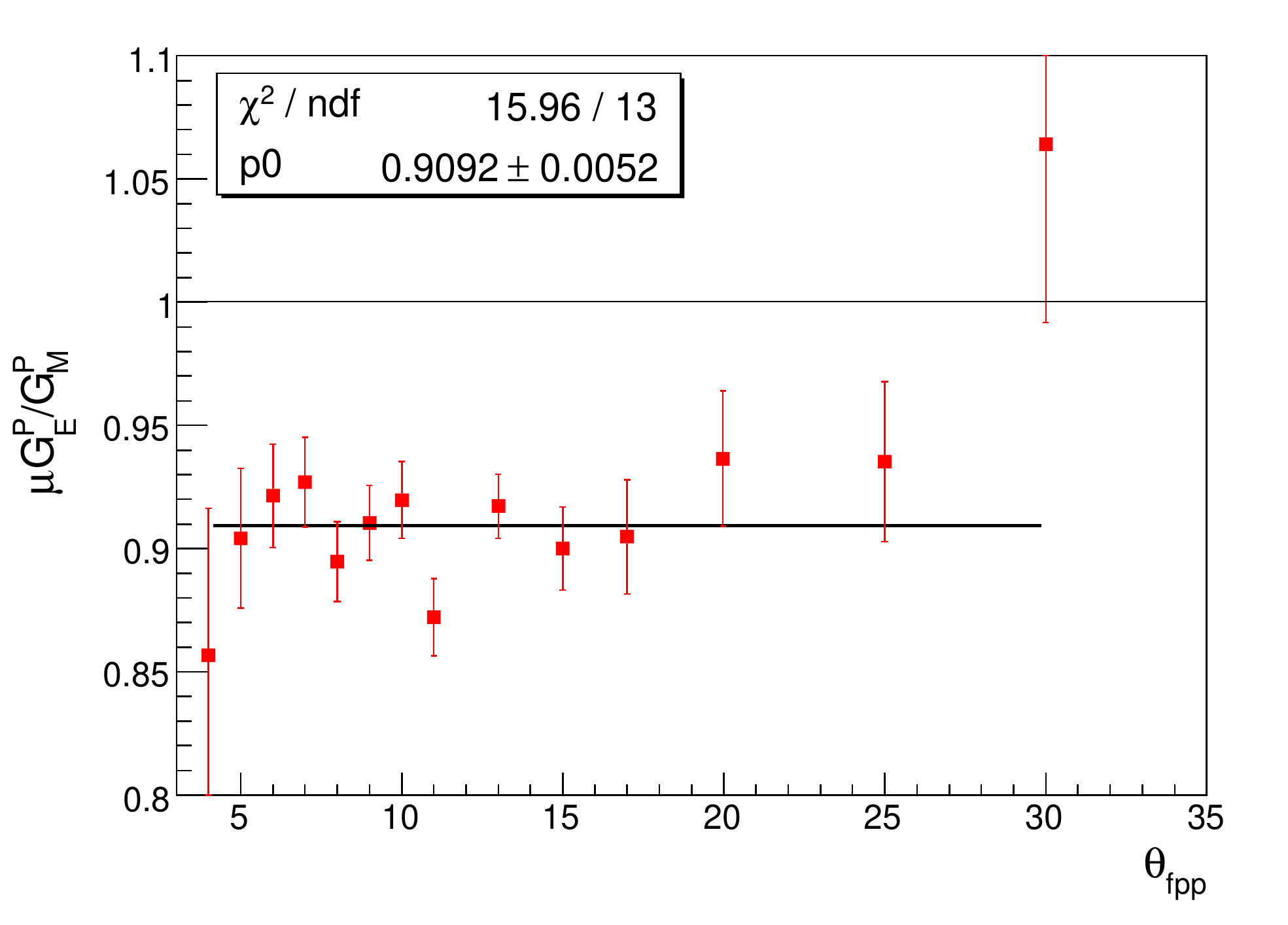}
    \caption{The form factor ratio binning on the FPP polar scattering angle $\theta_{fpp}$ for kinematic setting K6 ($Q^2 = 0.5$ GeV$^2$) and K7 ($Q^2 = 0.6$ GeV$^2$).}
    \label{fig:thfppbin}
  \end{center}
 \end{figure}

 As another demonstration of the FPP alignment quality, Fig.~\ref{fig:thfppbin} shows the form factor ratio binning results on the FPP polar scattering angle $\theta_{fpp}$ with a constant fit, and there is no indication of any systematic dependence on this variable with the current precision.
 \subsection{VDC Resolution}
 The VDC quantities $\theta_{tr}$ and $\phi_{tr}$ were used to calculate the spin rotation matrix between the transport frame and the FPP local frame. By manually shifting these variables, the systematic error on the ratio was obtained and reported in Table.~\ref{tab:vdc}.
 \begin{table}[t]
 \caption{Errors of the VDC angles and associated systematic error in $\mu_pG_E/G_M$.}
 \begin{center}
 \begin{tabular}{|c| c| c|}
 \hline
 Kinematics & $\theta_{tr}$ (+1 mrad) & $\phi_{tr}$ (+1 mrad)\\
 \hline
 K1 & -0.0002  & 0.0002 \\
 K2 & -0.001   & -0.0002 \\
 K3 & 0.0003   & -0.0002 \\
 K4 & -0.0004  & -0.0002 \\
 K5 & -0.0006  & -0.0004 \\
 K6 & -0.0005  & -0.0003 \\
 K7 & -0.0002  & -0.0004 \\
 K8 & -0.0001  & -0.0007 \\
 \hline
 \end{tabular}
 \label{tab:vdc}
 \end{center}
 \end{table}
\subsection{Other Systematics}
\subsubsection{Charge Asymmetry}
 In the analysis code, we randomly throw out a small fraction of events with one beam helicity state to test the sensitivity to the charge asymmetry. With the charge asymmetry ($<1000$ ppm) from this experiment, the change of the form factor ratio is negligible ($\le 0.001$). This result is expected since the charge asymmetry only introduces a high order effect from the instrumental efficiency.
 \subsubsection{Kinematics factors}
 From the form factor ratio formula:
 \begin{equation}
 R = \mu_p G_E/G_M = -\mu_p\frac{P_y}{P_z}\frac{E+E'}{2m_p}\tan(\frac{\theta_e}{2}) = -\mu_pK\frac{P_y}{P_z},
 \end{equation}
 knowledge of the kinematic factor $K$ is also required. In the analysis code, the initial inputs are the beam energy and the proton scattering angle\footnote{The reconstruction of the electron kinematics is not available due to reduced configuration of the BigBite spectrometer.}; therefore, the kinematic factor $K$ is a function of $E$ and $\theta_p$ in this analysis. Based on the systematic studies mentioned earlier, we quote $\pm 0.5$ MeV as the beam energy uncertainty and $0.02^{\circ}$ as the proton scattering angle uncertainty. As an example, Table~\ref{tab:kine} lists the uncertainty of each factor and the resulting uncertainty in the ratio for one of our kinematics (K7). Clearly, the change of the form factor ratio is negligible ($< 0.001$) and is at the same level for the other kinematics.
 \begin{table}[t]
 \caption{Errors of the kinematic factors and the resulting uncertainty in the form factor ratio $R$ for kinematics K7 ($Q^2=0.6$ GeV$^2$).}
 \begin{center}
 \begin{tabular}{|c |c| c|}
 \hline
 $\delta_R(E_0) (\pm 0.5 \mathrm{MeV})$ & $\Delta R(\theta_0) (\pm 0.02^{\circ})$ & $\Delta R$\\
 \hline
 0.0003 & 0.0005 & 0.0006\\
 \hline
 \end{tabular}
 \label{tab:kine}
 \end{center}
 \end{table}

 \section{Summary of Uncertainties}
\begin{figure}[hbt]
  \begin{center}
  \includegraphics[angle=0,width=.7\textwidth]{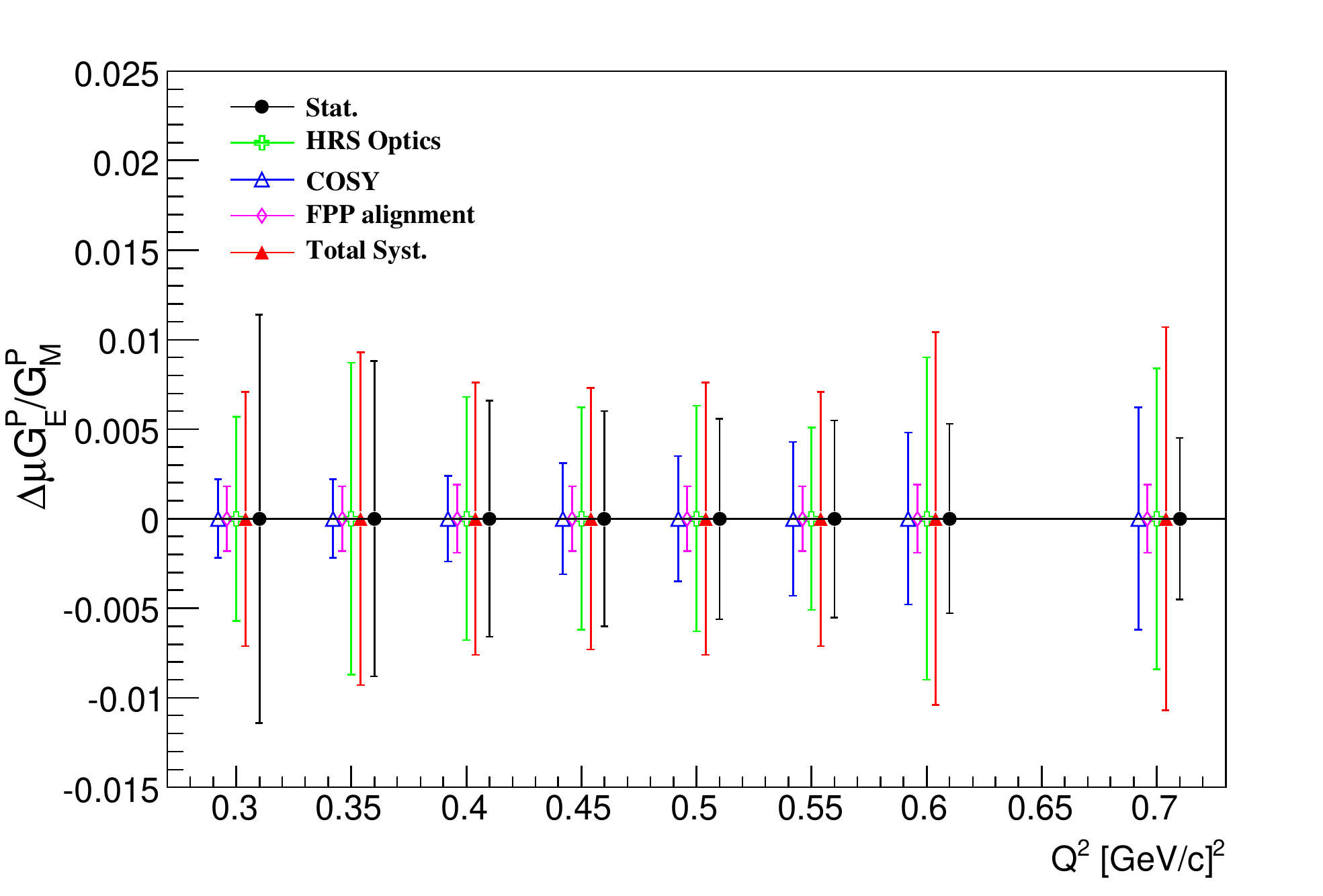}
    \caption{Comparison of the major contributions to the systematic uncertainties and the statistical uncertainty for each kinematics.}
    \label{fig:tot_sys}
  \end{center}
 \end{figure}

 As a summary, Fig~\ref{fig:tot_sys} shows the major uncertainties for each $Q^2$ point. All these contributions are added quadratically to obtain the total systematic error for this experiment. Table.~\ref{tab:sum} presents the final results with both the statistical and systematic errors. As $Q^2$ increases, the systematic error starts to dominate the total uncertainty.

 \begin{table}[t]
 \caption{Final results with statistical and systematic uncertainties for each kinematics.}
 \begin{center}
 \begin{tabular}{|c|c|c|c|c|}
 \hline
 Kinematics & $\langle Q^2\rangle$ [$(\mathrm{GeV}/c)^2$] & $R$ & $\Delta R_{sys.}$ & $\Delta R_{stat.}$\\
 \hline
 K1 & 0.3458 & 0.9433 & 0.0093 & 0.0088 \\
 K2 & 0.2985 & 0.9272 & 0.0071 & 0.0114 \\
 K3 & 0.4487 & 0.9314 & 0.0073 & 0.0060 \\
 K4 & 0.4017 & 0.9318 & 0.0076 & 0.0066 \\
 K5 & 0.5468 & 0.9274 & 0.0071 & 0.0055 \\
 K6 & 0.4937 & 0.9264 & 0.0076 & 0.0056 \\
 K7 & 0.5991 & 0.9084 & 0.0104 & 0.0053 \\
 K8 & 0.6951 & 0.9122 & 0.0107 & 0.0045 \\
 \hline
 \end{tabular}
 \label{tab:sum}
 \end{center}
 \end{table}
 \section{Radiative Correction}
 For electron scattering, the radiative process is inevitably involved. This includes the electron initial and final state Bremsstrahlung, loop correction, as well as 2$\gamma$ exchange effects. The radiative correction to this experiment is discussed by providing the results from recent theoretical calculations.

\begin{figure}[hbt]
  \begin{center}
  \includegraphics[angle=0,width=.6\textwidth]{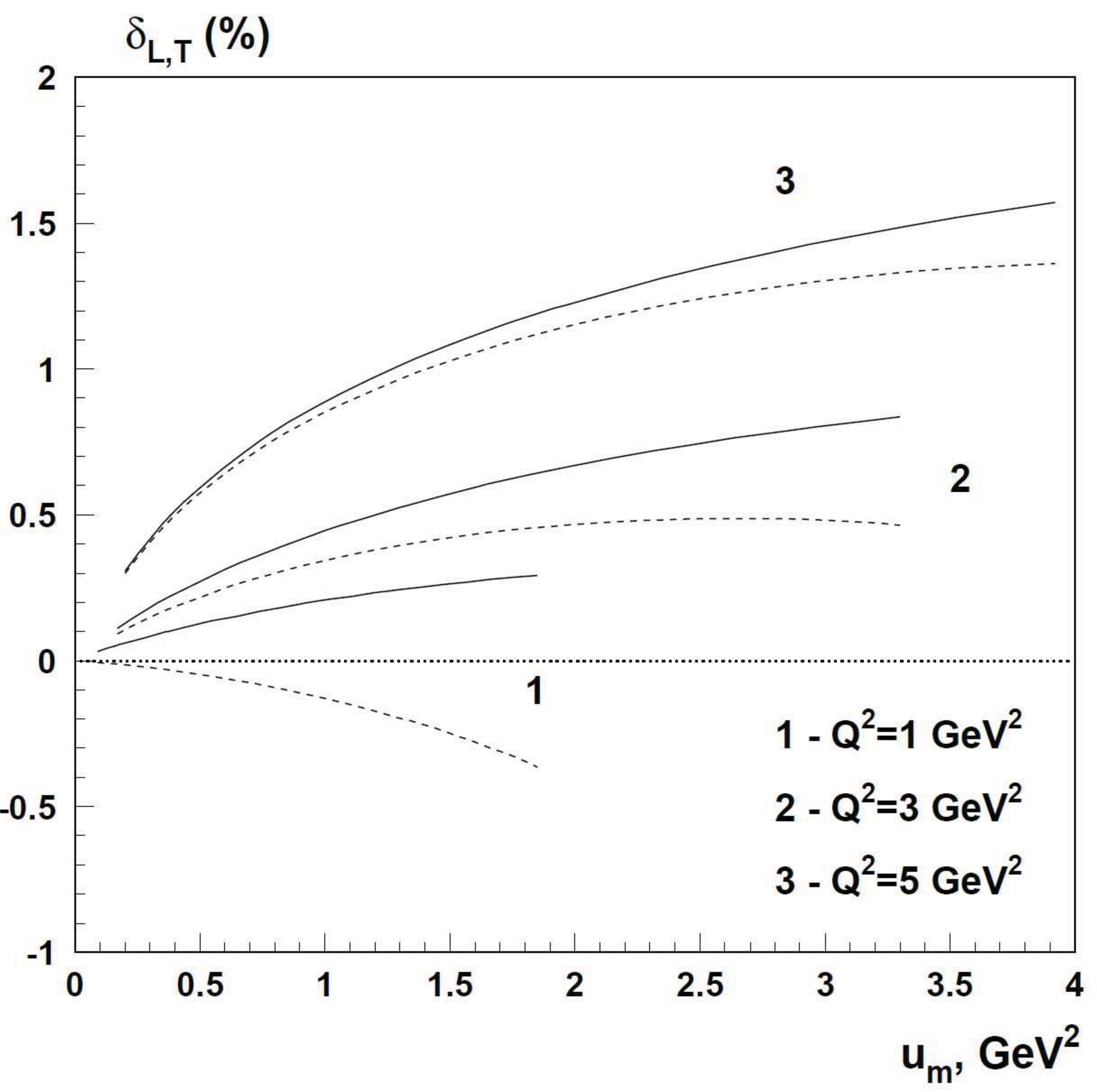}
  \caption{Radiative corrections to the recoil polarization. The solid and dashed lines correspond to the longitudinal and transverse components with $s=8$ GeV$^2$. Figure from~\cite{afana_rad}.}
  \label{fig:afa_rad}
  \end{center}
 \end{figure}

\begin{figure}[hbt]
  \begin{center}
  \includegraphics[angle=0,width=.6\textwidth]{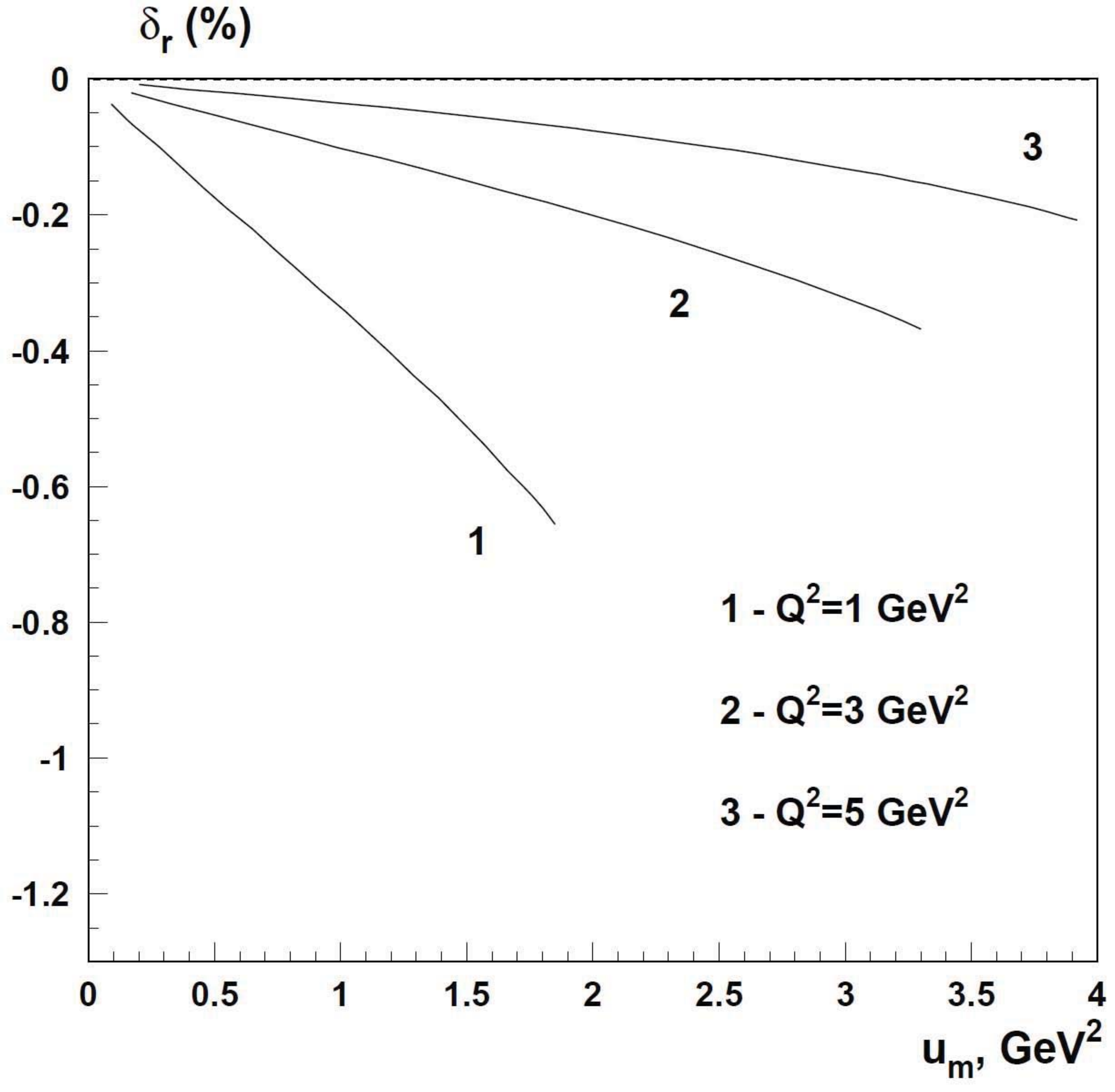}
  \caption{Radiative corrections to the ratio of the recoil proton polarization in the region where the invariant mass of the unobserved state is close to the pion mass and $s=8$ GeV$^2$. Figure from~\cite{afana_rad}.}
  \label{fig:afa_rad1}
  \end{center}
 \end{figure}

 Afanasev {\it et al.}~\cite{afana_rad} performed a numerical analysis for the radiative corrections in elastic $ep$ scattering when the kinematic variables are only reconstructed from the recoil proton. This study calculated the radiative correction to the cross sections and asymmetries differential in $Q^2$. Fig.~\ref{fig:afa_rad} shows the correction to the longitudinal and transverse polarization components as a function of the inelasticity $u_m = (k_1+p_1-p_2)^2-m^2$, where $m$ is the electron mass, $k_1$ is the electron initial momentum, and $p_{1(2)}$ is the initial (final) proton momentum at $s=8$ GeV$^2$. The magnitude of the correction does not exceed $1.5\%$, though it does rise with increasing $Q^2$ and inelasticity cut. Fig.~\ref{fig:afa_rad1} gives the correction to the measured ratio of final proton polarization; the correction is negative and does not exceed $1\%$.

\begin{figure}[hbt]
  \begin{center}
  \includegraphics[angle=0,width=.90\textwidth]{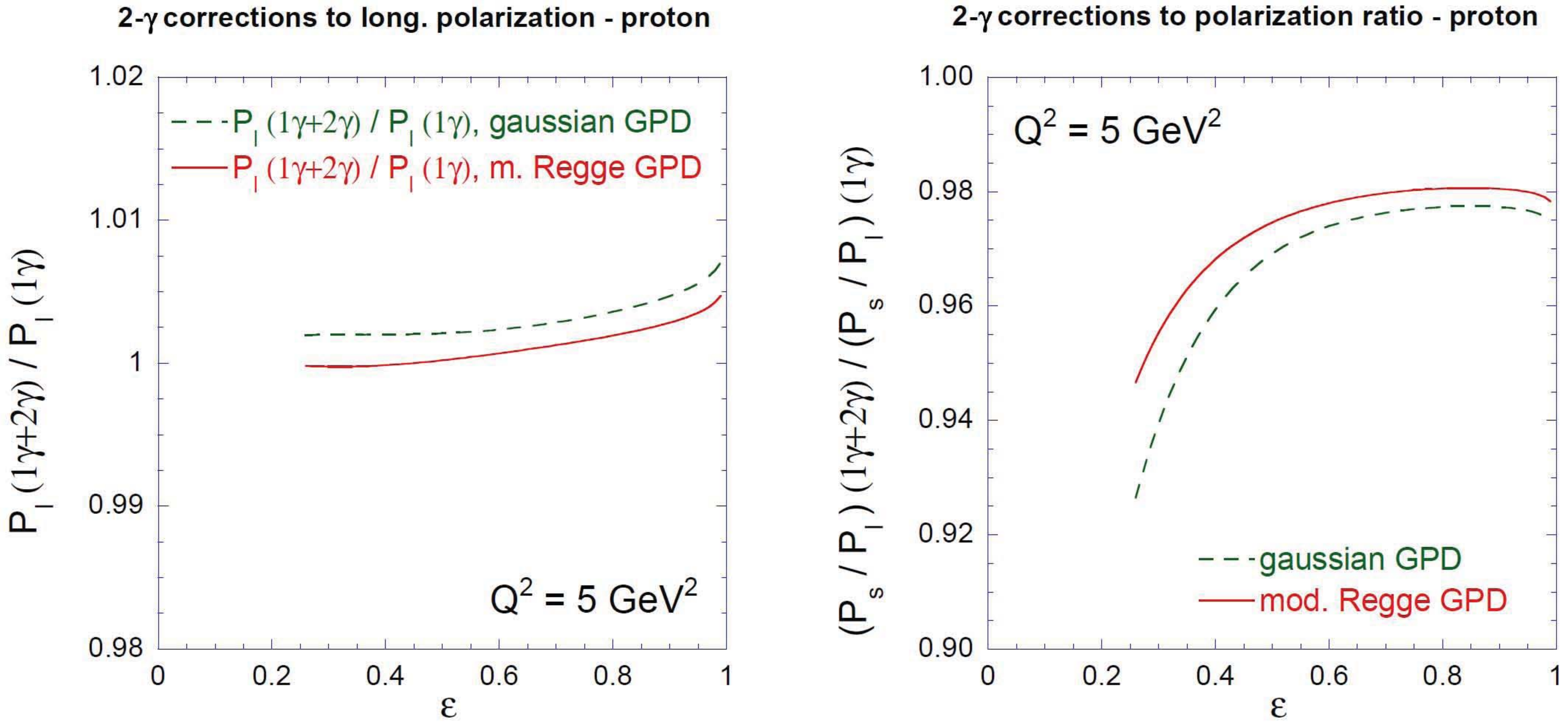}
  \caption{The 2$\gamma$ exchange correction to the recoil proton longitudinal polarization components $P_l$ and the ratio of the transverse to longitudinal component for elastic $ep$ scattering at $Q^2=5$ GeV$^2$. Figure from~\cite{afana}.}
  \label{fig:2gamma_afa}
  \end{center}
 \end{figure}

 Afanasev {\it et al.}~\cite{afana} also estimated the 2$\gamma$ exchange contribution to elastic $ep$ scattering at large momentum transfer by using a quark-parton representation of virtual Compton scattering. While the correction is significant for cross-section measurements, the impact upon the recoil polarization measurement is small. Fig.~\ref{fig:2gamma_afa} shows the calculated transferred proton polarization with and without the 2$\gamma$ exchange terms, for 100$\%$ right-handed electron polarization and with a fixed $Q^2$ of 5 GeV$^2$.

\begin{figure}[hbt]
  \begin{center}
  \includegraphics[angle=0,width=.7\textwidth]{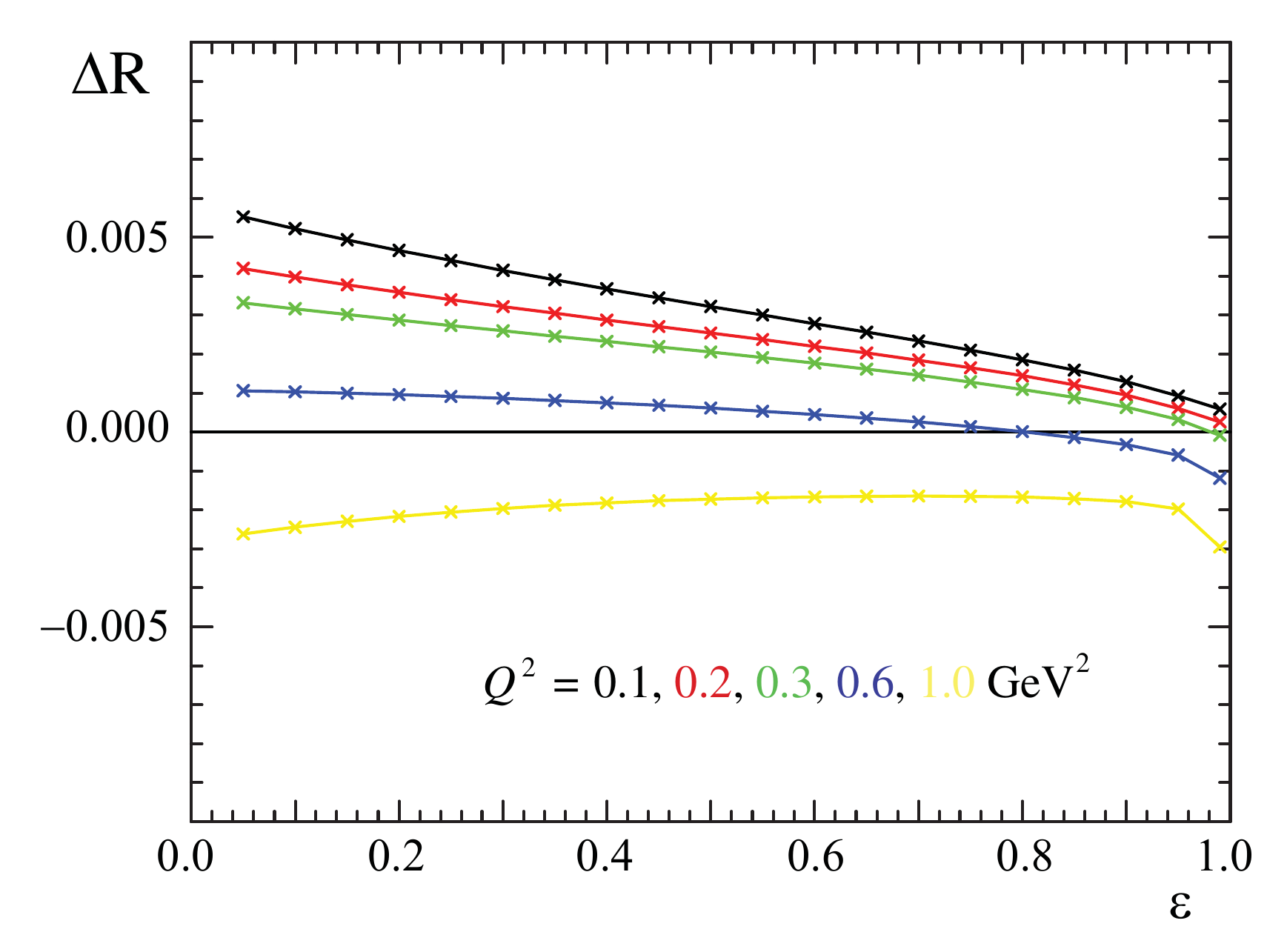}
    \caption{The relative correction to the proton form factor ratio from 2$\gamma$ exchange as a function of $\varepsilon$ for 5 different $Q^2$~\cite{blunden,blunden_melnitch}.}
    \label{fig:2gamma_corr}
  \end{center}
 \end{figure}

 Blunden {\it et al.}~\cite{blunden,blunden_melnitch} performed an explicit calculation of the 2$\gamma$ exchange diagram in which nucleon structure effects were fully incorporated. They also applied it to systematically calculate the effects in a number of electron-nucleon scatterings. Fig.~\ref{fig:2gamma_corr} shows the relative correction of the proton form factors ratio $\mu_pG_E/G_M$ as a function of $\varepsilon$ at different $Q^2$.

 For the kinematic condition of this experiment ($Q^2<1$ GeV$^2$, $0.66<\varepsilon<0.85$, $s=3.12$ GeV$^2$ ), we have concluded that the radiative corrections to the form factor ratio is less than $0.3\%$ based on the current theoretical calculations.

%% file: impacts.tex
\chapter{Discussion and Conclusion}
In this chapter, the experimental data are compared with the world data and various models
and fits. In addition, the impacts of the new results to other physics quantities are discussed, and the future outlook to access lower $Q^2$ is also presented.
\section{Comparison with World Data}
\begin{figure}
  \begin{center}
    \includegraphics[angle=0, width=0.9\textwidth]{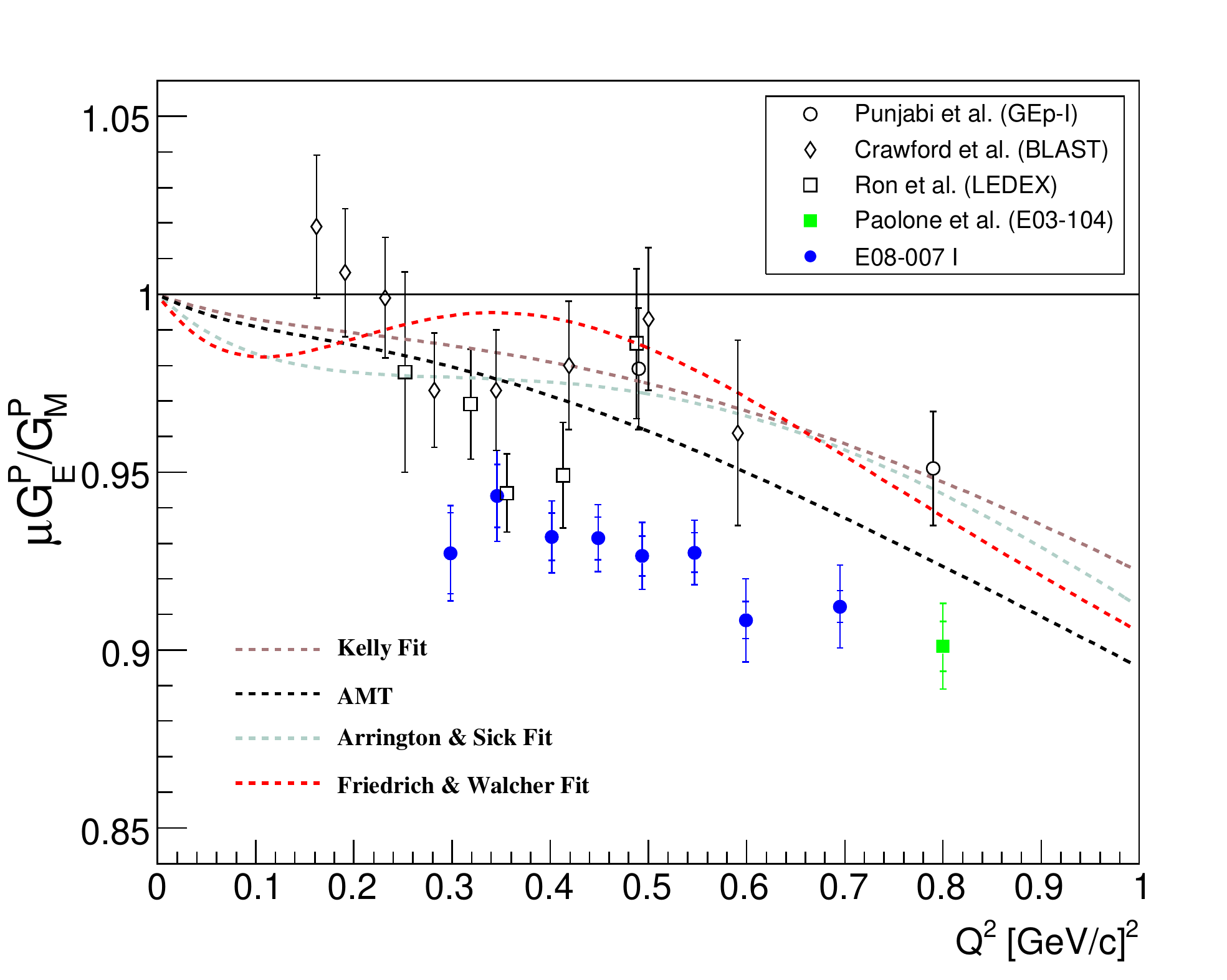}
    \caption{The proton form factor ratio $\mu_pG_E/G_M$ as a function of $Q^2$ with world high precision data~\cite{punj,BLAST,LEDEX} ($\sigma_{tot}<3\%$). For the new data, the inner error bars are statistical, and the outer ones are total errors. For the world data sets, the total errors are plotted. The dashed lines are fits~\cite{kelly,john:TPE,arrington_sick,fried}.}
    \label{fig:final1}
  \end{center}
\end{figure}

Fig.~\ref{fig:final1} and \ref{fig:final2} show the new results of this work, $\mu_pG_E/G_M$ as a function of $Q^2$ together with previous high precision measurements ($\sigma_{tot}<3\%$). The green point at $Q^2 = 0.8$ GeV$^2$ which will also be published soon is from one of the LEDEX experiments E03-104~\cite{e03104}. The new data
have the following features:
\begin{itemize}
\item The new results are in good agreement with the high precision point at $Q^2=0.8$ GeV$^2$, which was taken in 2006 with a different configuration\footnote{E03-104 used right HRS to detect the electron and the electron kinematics was well known.} and analyzed independently.
\item The whole data set slowly decrease along $Q^2$ in the region of $Q^2=0.3\sim0.8$ GeV$^2$; no obvious indication of any ``narrow structure''.
\item The new data strongly deviate from unity by several percent which is unexpected from the previous measurements.
\end{itemize}
\begin{figure}
  \begin{center}
    \includegraphics[angle=0, width=0.9\textwidth]{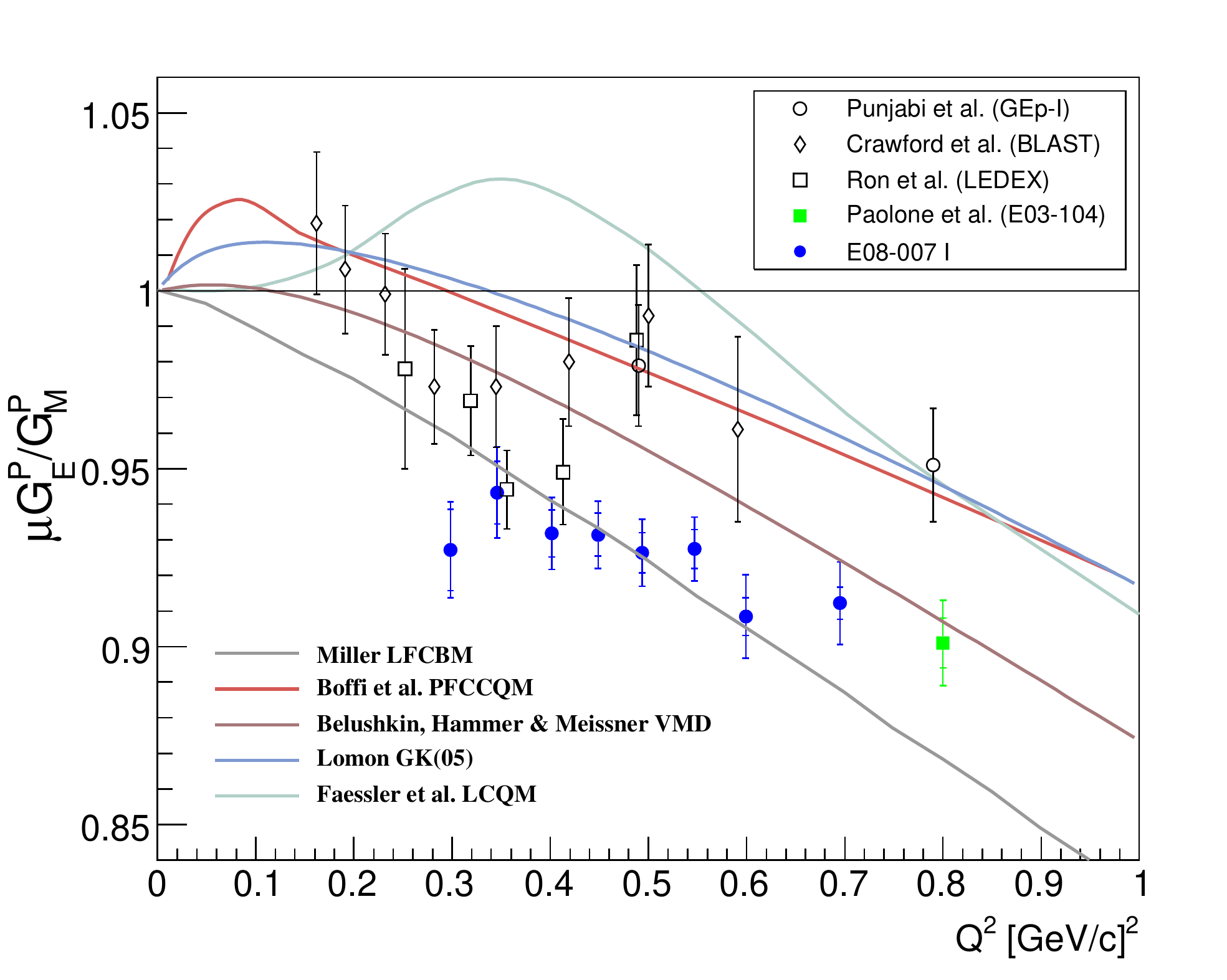}
    \caption{The proton form factor ratio $\mu_pG_E/G_M$ as a function of $Q^2$ shown with world high precision data~\cite{punj,BLAST,LEDEX} ($\sigma_{tot}<3\%$). For the new data, the inner error bars are statistical, and the outer ones are total errors. For the world data sets, the total errors are plotted. The solid lines are from vector-meson dominance calculations~\cite{belu,lomon_3}, a light-front cloudy-bag model calculation~\cite{Miller}, a light-front quark model calculation~\cite{faessler}, and a point-form chiral constituent quark model calculation~\cite{bof01}.}
    \label{fig:final2}
  \end{center}
\end{figure}

Although the LEDEX results (Ron {\it et al.}) overlap with the new data in the vicinity of
$Q^2 = 0.36$ GeV$^2$, the highest $Q^2$ point is $\sim 3~\sigma$ above the new data. To investigate this potential discrepancy, we reanalyzed the LEDEX data and found that the Al background was overestimated in the original analysis~\cite{guy_thesis}; hence, the data were overcorrected for dilution effect from the Al end cap. The preliminary results of the LEDEX reanalysis are in good agreement with the new data, and we expect to publish the erratum soon.

The other two data points contributed by JLab in this region are from GEp-I measurement~\cite{punj}, which was performed in 1998. The point at $Q^2=0.5$ GeV$^2$ is $\sim 3.5~\sigma$ higher than the new results, and the point at $Q^2=0.8$ GeV$^2$ is $\sim 2.5~\sigma$ higher than the E03-104 result (Paolone {\it et al.}). The investigation of the original GEp-I analysis is still underway, which includes the consistency check of different analysis codes, the accuracy of the kinematic parameters\footnote{In the very early days of the Hall A running, the beam energy and spectrometer momentum were not very well known.}, the discussion of the cuts and the systematic error analysis\footnote{GEp-I used the right HRS to detect the recoil proton instead of the left HRS; therefore, the optics and spin transport were different.}.

 Another discrepancy is in the comparison with the BLAST~\cite{BLAST} results. The new data are systematically lower by 2 to 3 $\sigma$, which is hard to explain by statistical fluctuations. Since BLAST used the beam-target asymmetry technique, the origin of the systematic uncertainty is different. While the investigation of this discrepancy is needed, a third measurement by using the beam-target asymmetry technique in this region is strongly recommended to uncover any unknown systematic errors in the recent measurements.

In summary, Fig.~\ref{fig:final2} shows the new results plotted with a different scale together with the world polarization data, which includes the preliminary results of the GEp-III measurement~\cite{gep3}.
\begin{figure}
  \begin{center}
    \includegraphics[angle=0, width=0.85\textwidth]{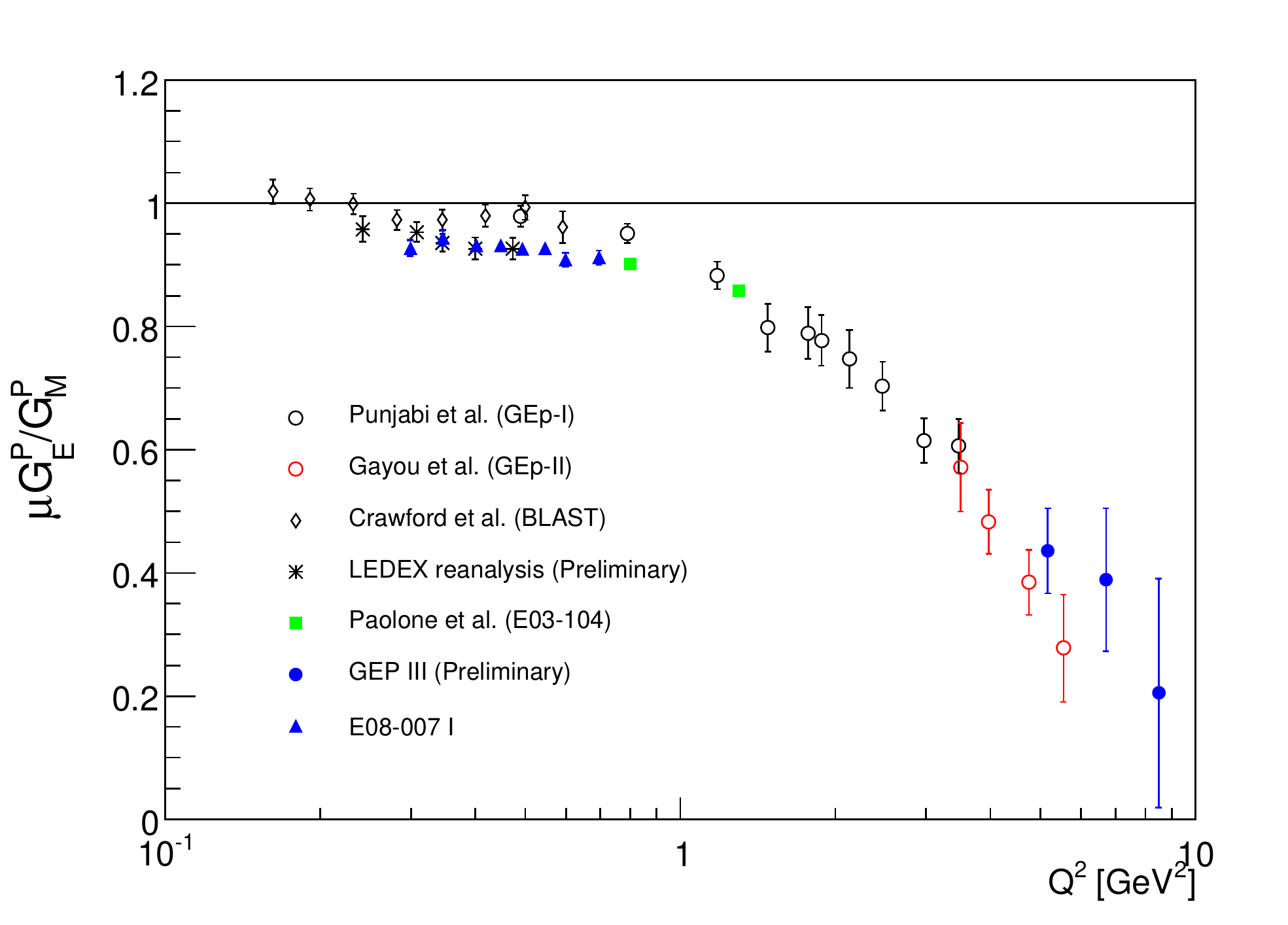}
    \caption{The proton form factor ratio $\mu_pG_E/G_M$ as a function of $Q^2$ shown with world high precision polarization data~\cite{punj,BLAST,LEDEX,Gayou,gep3}.}
    \label{fig:final2}
  \end{center}
\end{figure}

\section{Discussion with Theoretical Models and Fits}
In Fig.~\ref{fig:final1} and Fig.~\ref{fig:final2}, the data are shown together with
a representative set of the existing theoretical models and fits.
Analytical fits from Kelly~\cite{kelly} and AMT~\cite{john:TPE} are based on the
data over all $Q^2$, while the fits from Arrington and Sick~\cite{arrington_sick} and Friedrich and Walcher~\cite{fried} concentrate on the lower $Q^2$ data. Due to the absence of physical
interpretation and the dominance of the old data, it is plausible to expect that
the global fits are substantially above our new data. On the other hand, the new results
cannot completely rule out the existence of the structure given by the phenomenological fit of Friedrich
and Walcher~\cite{fried}; however, the average value of the structure would be much lower then what they predicted
if there is any.

The existing theoretical models also cannot accurately predicts the results.
A chiral constituent quark model by Boffi {\it et al.}~\cite{bof01}, and a
Lorentz covariant chiral quark model by Faessler {\it et al.}~\cite{faessler}
are both above the new data. A Light-front cloudy bag model by Miller~\cite{Miller},
which includes the pion cloud effect, generally reproduces the large
deviation from unity in this region; however, this calculation decreases
too rapidly compared to the data.
The VMD calculations by Belushkin {\it et al.}~\cite{belu} and Lomon~\cite{lomon_3}
are also above the new data. Although this type of calculation is known to be very successful in representing the existing world data, the large number of tunable parameters in these models inevitably weaken the predictive power; therefore, the current disagreement is not surprising.
\section{Individual Form Factors and Global Fits}
\begin{figure}
  \begin{center}
    \includegraphics[angle=0, width=0.90\textwidth]{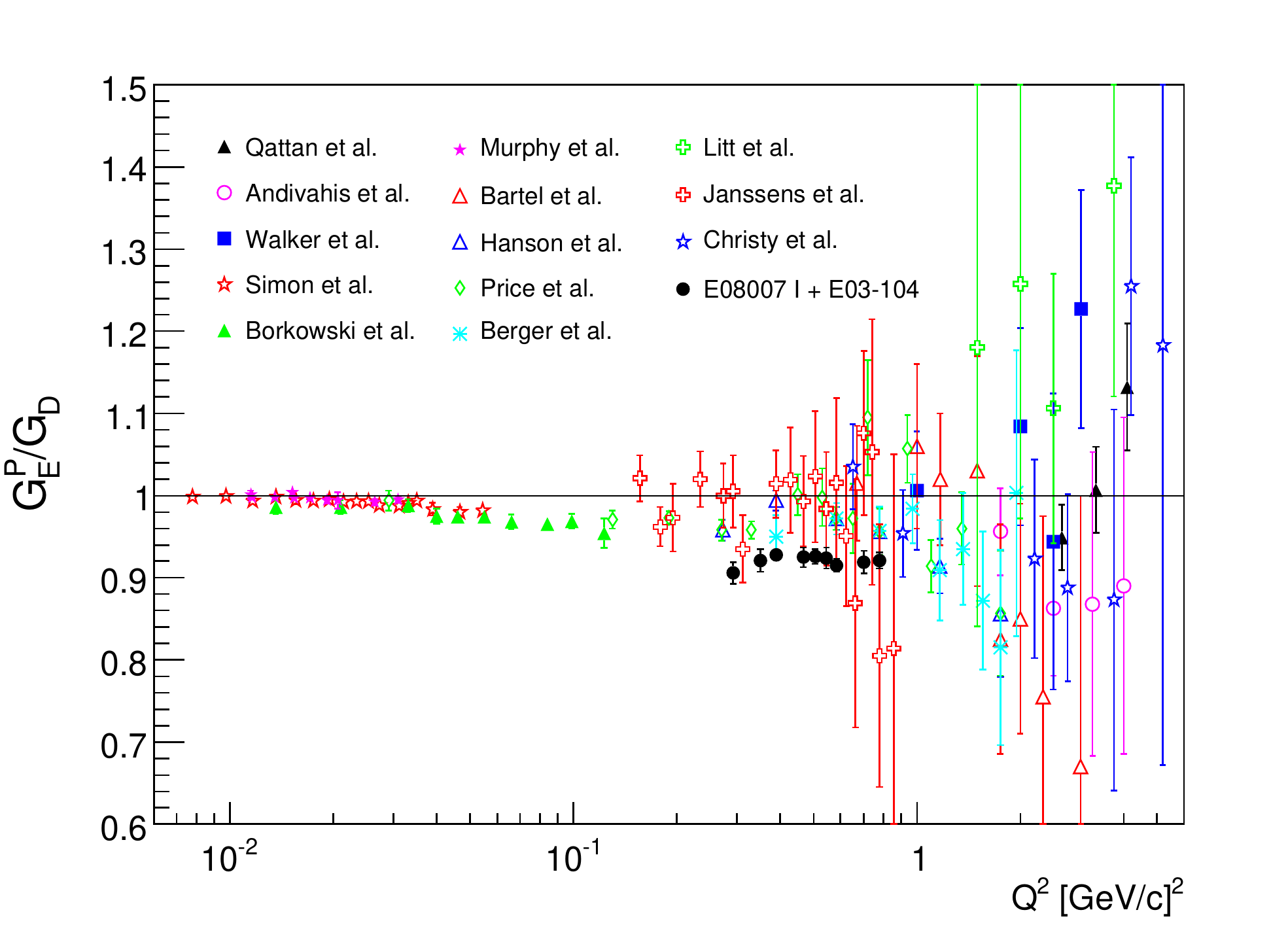}
    \includegraphics[angle=0, width=0.90\textwidth]{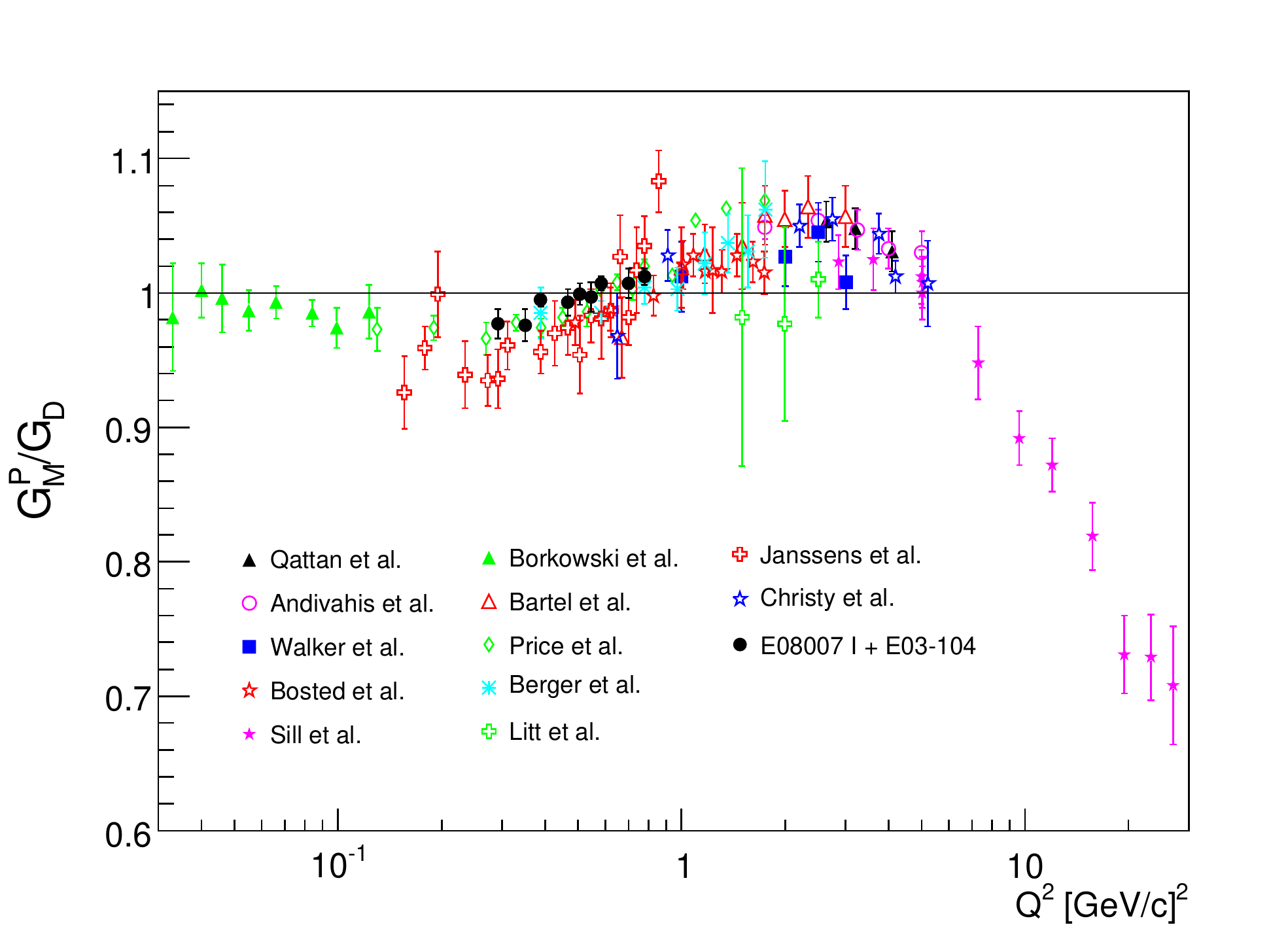}
    \caption{The new extraction of $G_E$ and $G_M$ plotted together with the world unpolarized data.}
    \label{fig:gegm1}
  \end{center}
\end{figure}

To extract the individual form factors, the data must be combined with cross section measurements to
determine the absolute magnitudes of $G_E$ and $G_M$. From Eq.~\ref{eq:red}
\begin{equation}
\sigma_{red} = \varepsilon(1+\tau)\frac{d\sigma/d\Omega}{(d\sigma/d\Omega)_{Mott}} = \varepsilon G_E^2+\tau G_M^2,
\end{equation}
if the ratio $R = \mu_pG_E/G_M$ is completely fixed, there is only degree of freedom
left in the linear fit of the reduced cross section. A new set of $G_E$ and $G_M$ were extracted by forcing the ratio $\mu_pG_E/G_M$ to be the experimental value of the new results (E08-007 I and E03-104). The cross sections used in this extraction are listed in Appendix F. Fig.~\ref{fig:red_fit} shows the fits of the reduced cross sections at 9 different $Q^2$s, which are in the vicinity of the new ratio measurements. Table~\ref{tab:gegm} provides the results of the extractions of $G_E$ and $G_M$ by the standard Rosenbluth separation and the constrained fit. The new extracted $G_E$ and $G_M$ are plotted with the world data in Fig.~\ref{fig:gegm1}.
\begin{figure}
  \begin{center}
    \includegraphics[angle=0, width=0.32\textwidth]{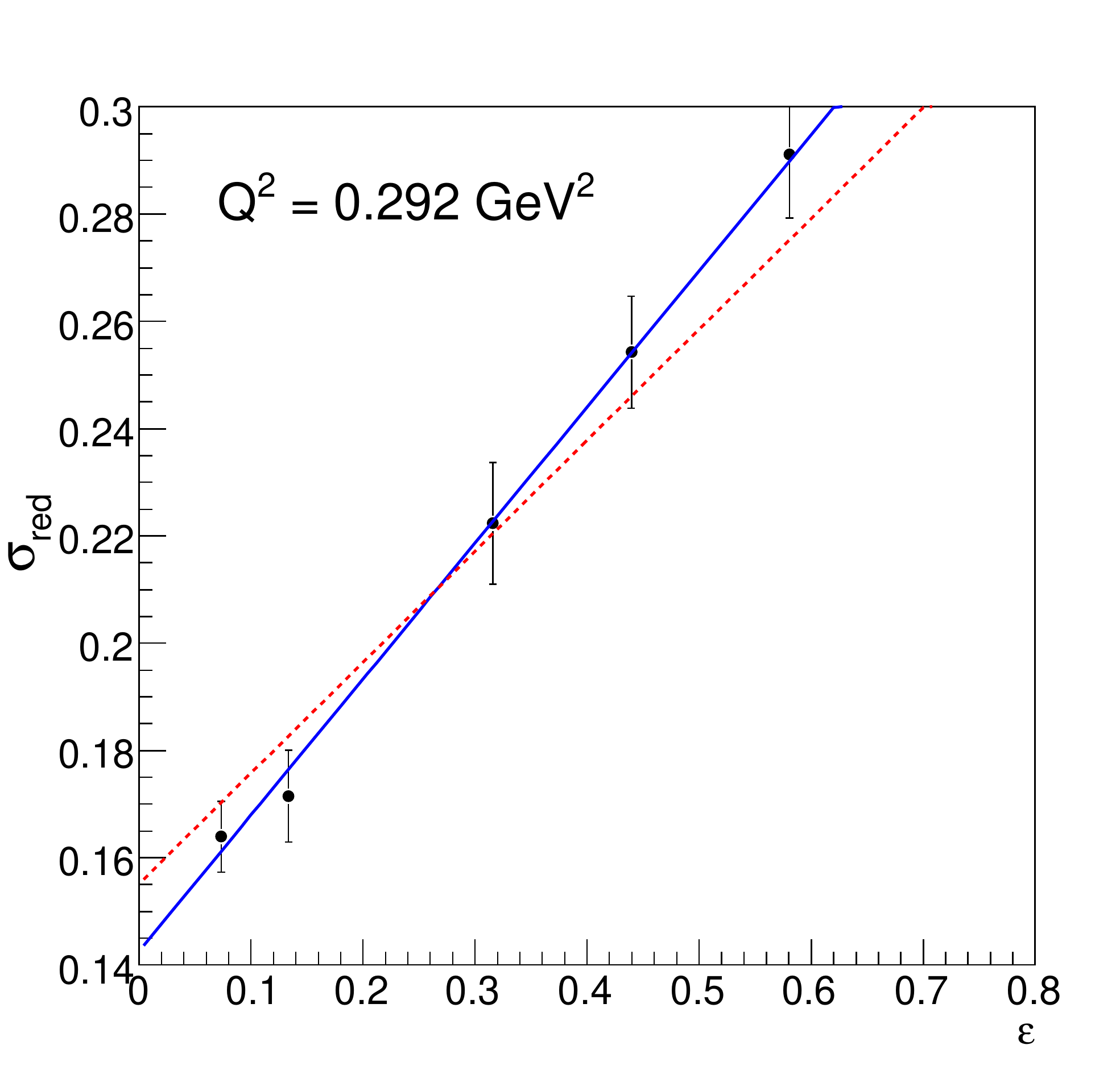}
    \includegraphics[angle=0, width=0.32\textwidth]{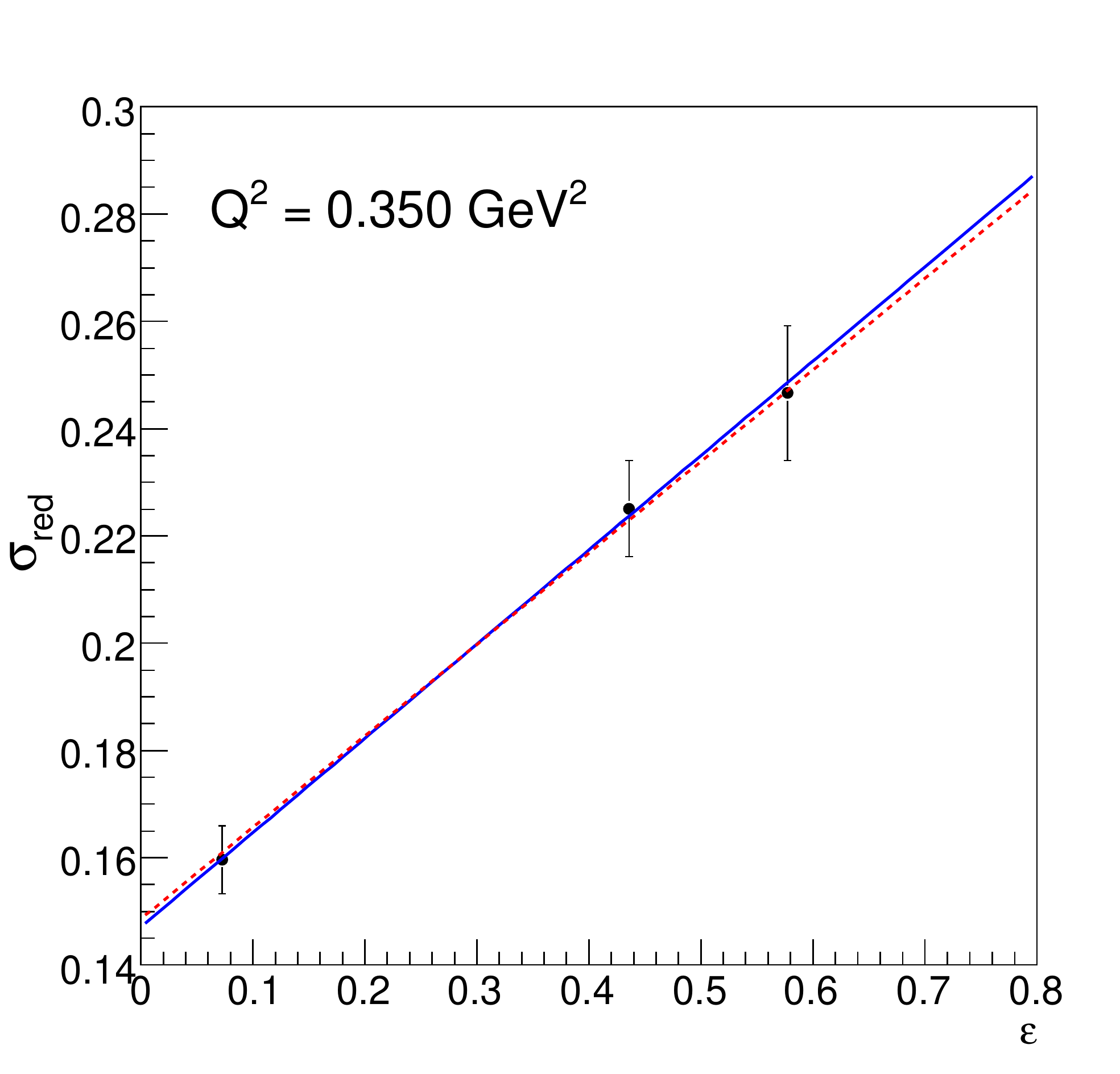}
    \includegraphics[angle=0, width=0.32\textwidth]{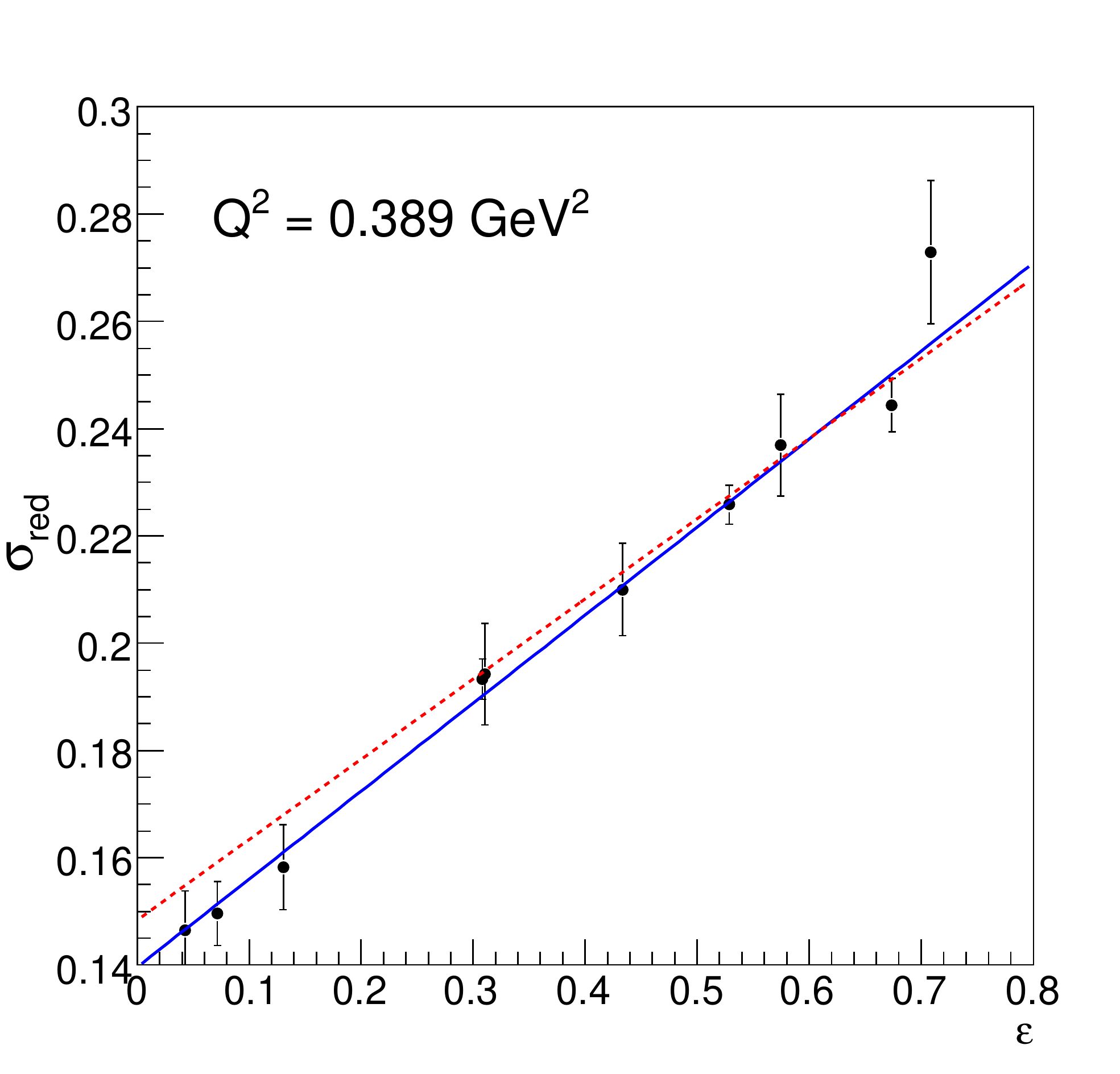}
    \includegraphics[angle=0, width=0.32\textwidth]{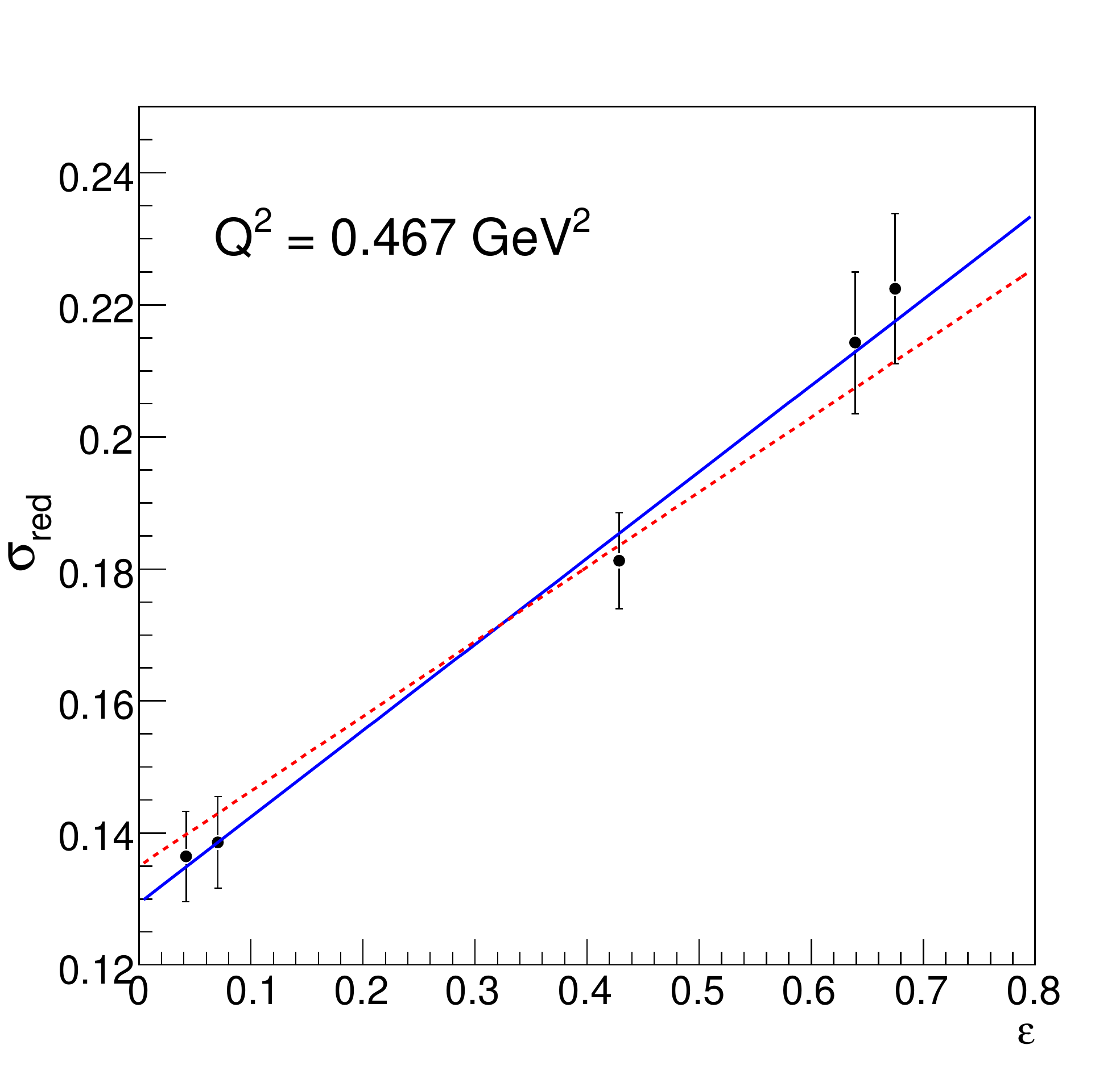}
    \includegraphics[angle=0, width=0.32\textwidth]{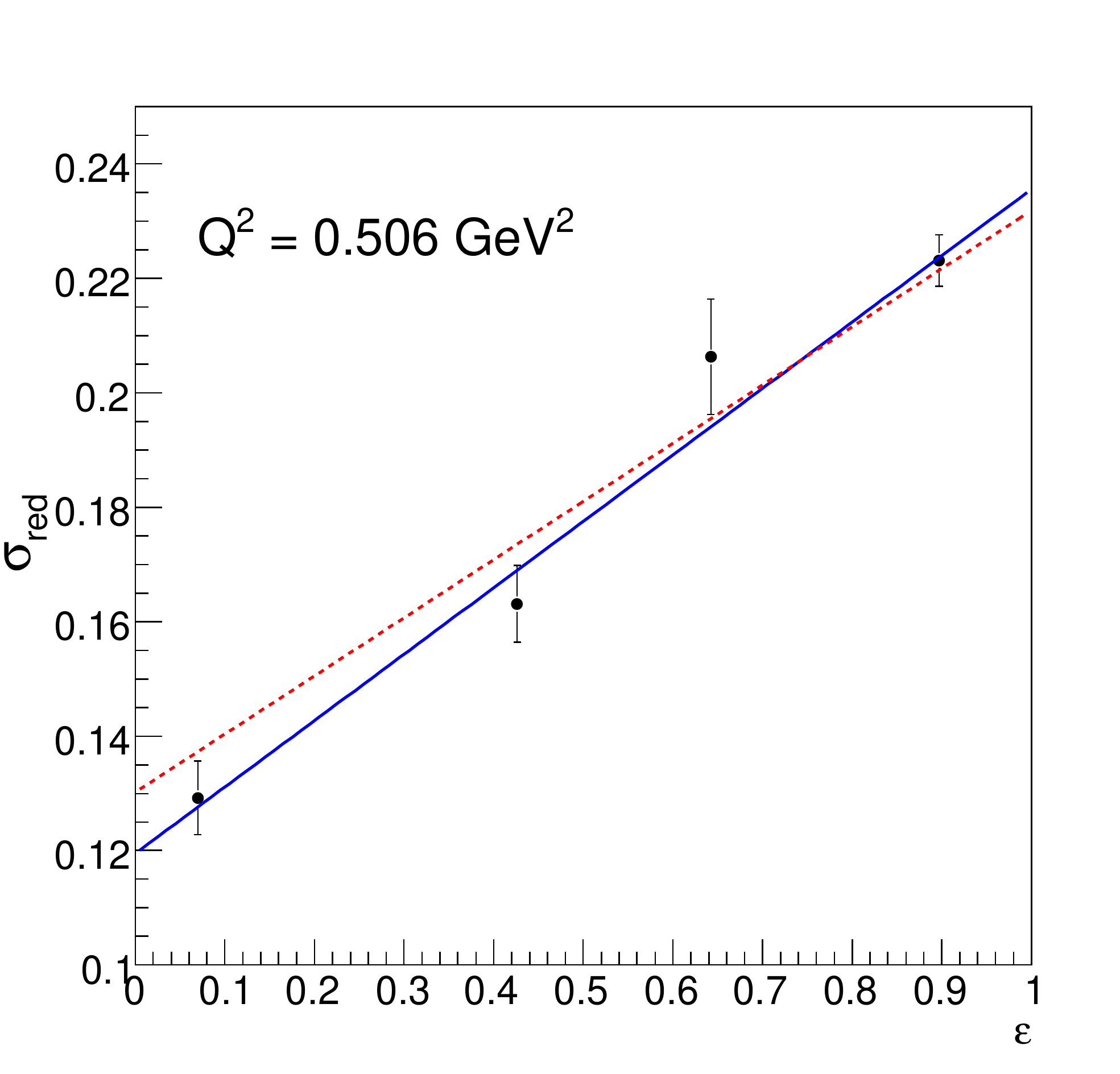}
    \includegraphics[angle=0, width=0.32\textwidth]{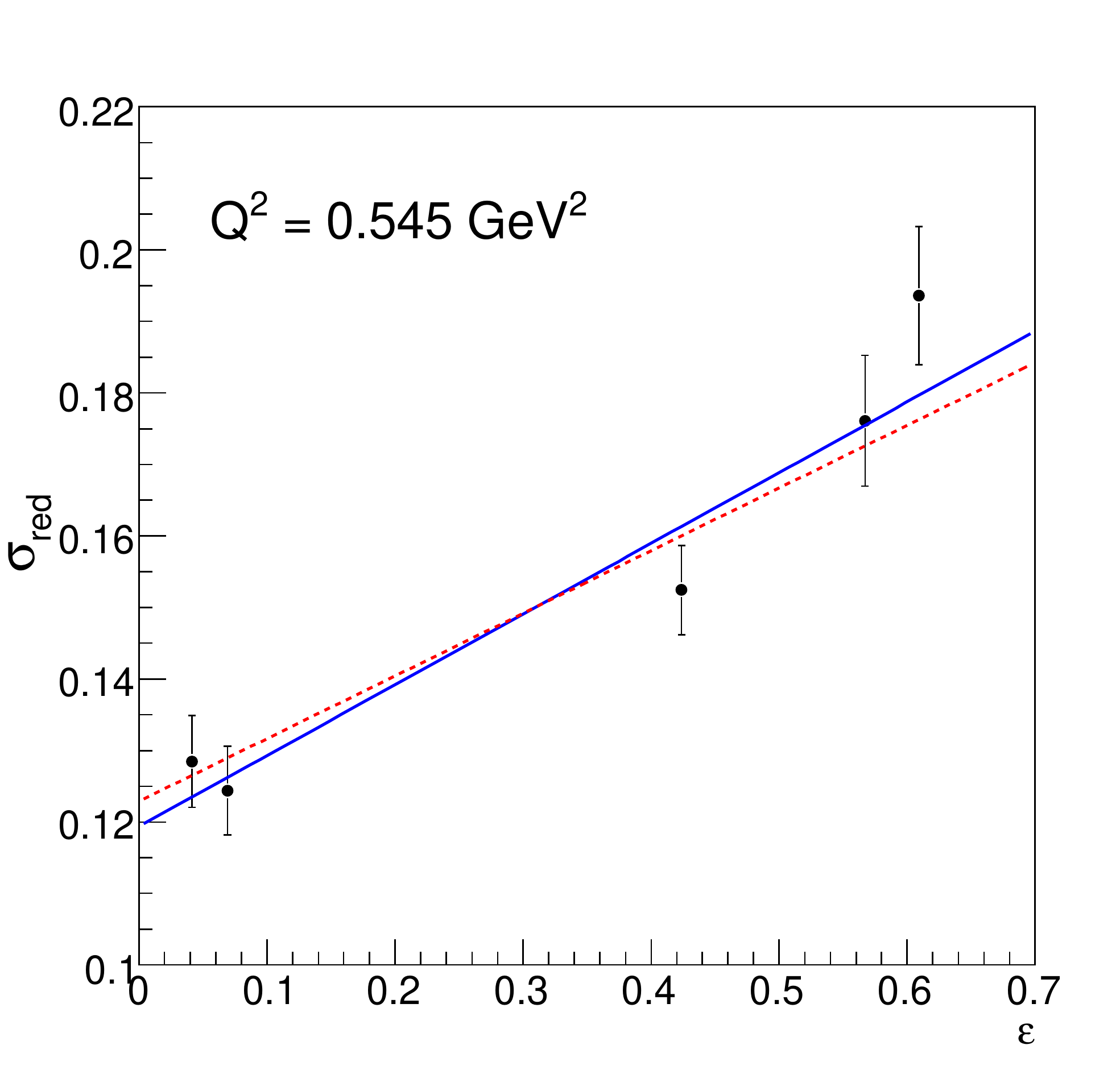}
    \includegraphics[angle=0, width=0.32\textwidth]{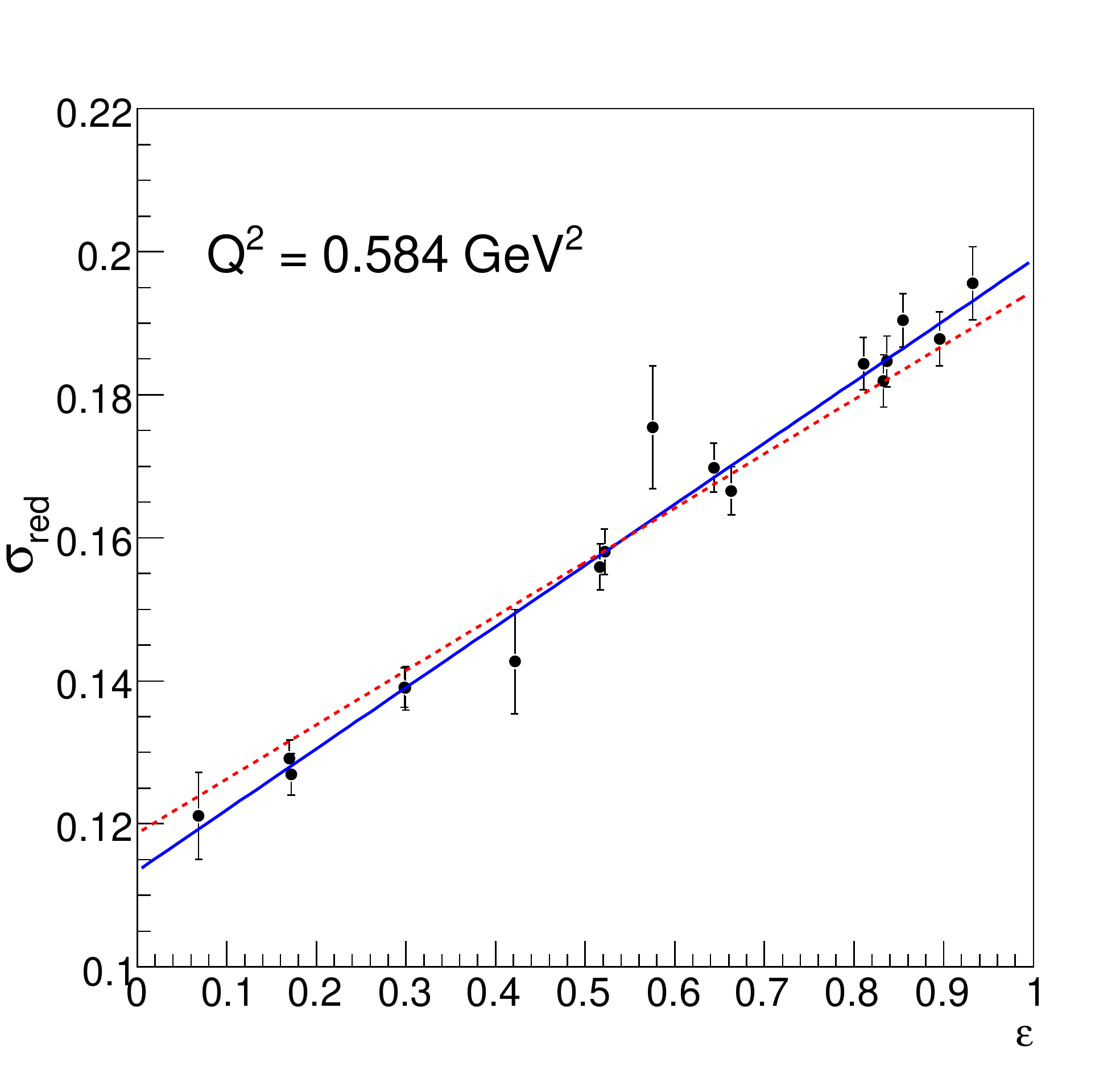}
    \includegraphics[angle=0, width=0.32\textwidth]{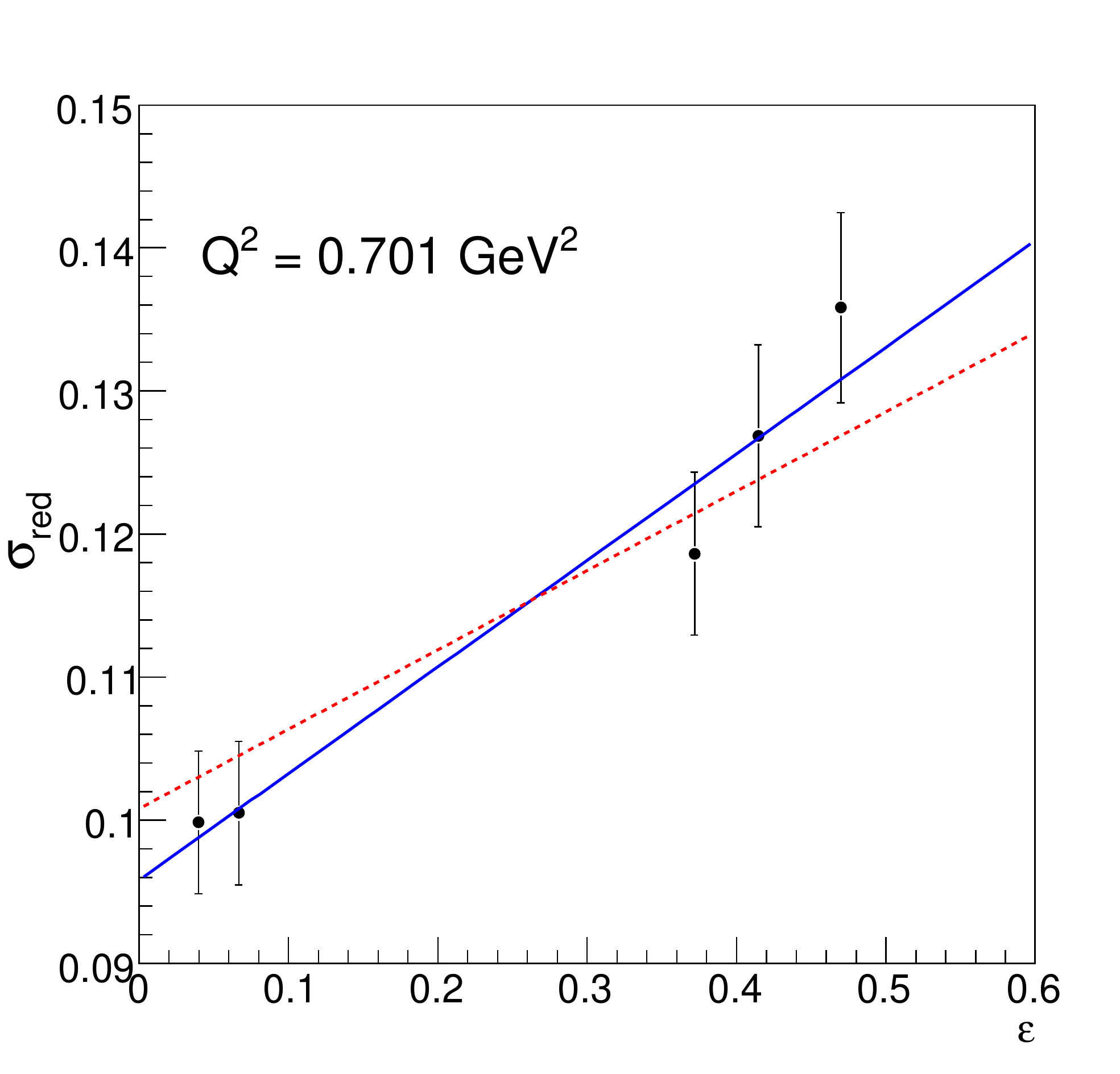}
    \includegraphics[angle=0, width=0.32\textwidth]{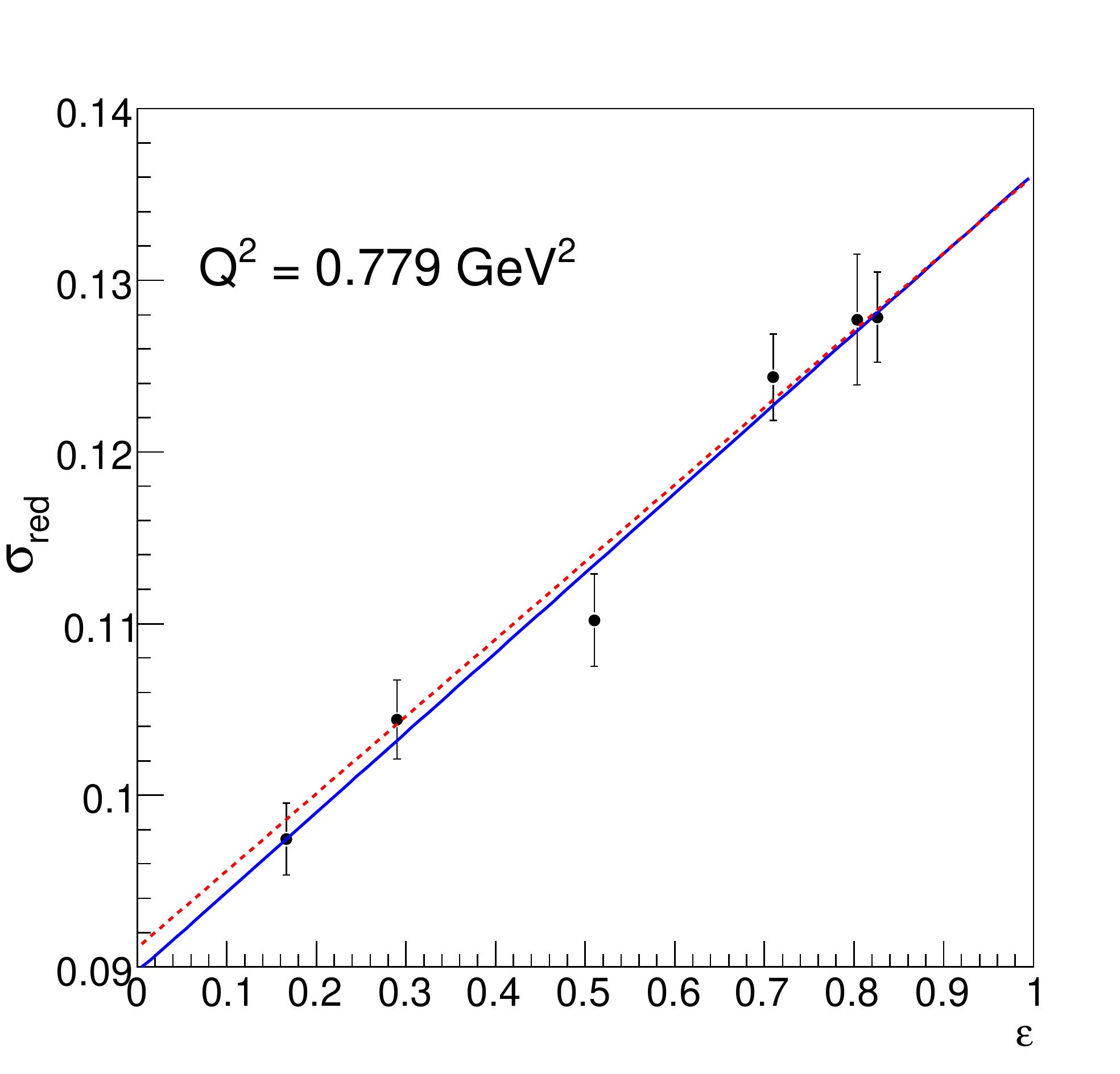}
    \caption{Rosenbluth separation of $G_E$ and $G_M$ constrained by $R=\mu_pG_E/G_M$. For each $Q^2$, the reduced cross section $\sigma_R$ is plotted against $\varepsilon$. The solid blue line is the standard Rosenluth separation fit without any constraint on $R$. The dotted red line is fit with an exact ratio constraint.}
    \label{fig:red_fit}
  \end{center}
\end{figure}
 \begin{table}[t]
 \caption{The extracted values of $G_E$ and $G_M$, with and without the constraint of $\mu_pG_E/G_M$ from the new measurements. The errors are indicated in parentheses.}
 \begin{center}
 \begin{tabular}{|c|ccc|ccc|}
 \hline
 $Q^2$ & \multicolumn{3}{|c|}{Unconstrained LT Separation} & \multicolumn{3}{|c|}{Constrained Fit}\\
 $[(\mathrm{GeV}/c)^2]$& $G_E/G_D$ & $G_M/\mu_pG_D$ & $\chi^2$/ndf & $G_E/G_D$ & $G_M/\mu_pG_D$ & $\chi^2$/ndf\\
 \hline\hline
 0.292 & 1.003(44) & 0.936(22) & 0.17 & 0.906(13) & 0.977(11) & 1.27 \\
 0.350 & 0.935(61) & 0.971(25) & 0.05 & 0.921(14) & 0.976(12) & 0.05 \\
 0.389 & 0.972(16) & 0.965(11) & 1.14 & 0.928(07) & 0.995(05) & 1.60 \\
 0.467 & 0.993(54) & 0.972(20) & 0.19 & 0.925(12) & 0.993(10) & 0.52 \\
 0.506 & 0.999(40) & 0.957(25) & 1.15 & 0.926(09) & 0.999(08) & 1.87 \\
 0.545 & 0.982(69) & 0.983(20) & 1.60 & 0.924(13) & 0.997(11) & 1.35 \\
 0.584 & 0.971(18) & 0.984(08) & 0.50 & 0.915(08) & 1.007(05) & 1.05 \\
 0.701 & 1.078(10) & 0.981(21) & 0.47 & 0.919(14) & 1.007(11) & 0.90 \\
 0.779 & 0.949(41) & 1.004(12) & 0.55 & 0.921(10) & 1.012(06) & 0.52 \\
 \hline
 \end{tabular}
 \label{tab:gegm}
 \end{center}
 \end{table}

With the constraint of the new ratio results, the uncertainty of the individual form factors are significantly improved. While the new $G_E$ obviously deviates from unity by a few percent, $G_M$ is slightly higher than the world unpolarized data, and both of them show a relatively smooth evolution along $Q^2$ in this region.

However, forcing the fit to match the ratio results gives too much weight to the polarization data.
To avoid this issue, a global combined fit~\cite{John_fitnew} was performed by John Arrington. This new fit followed the same procedure as in~\cite{john:TPE,arrington_2,jfit1} with a treatment for the TPE effect in the cross section data. The $\chi^2$ of the combined fit is the contribution from the cross section measurements plus the additional contribution from the polarization ratio measurements:
\begin{equation}
\chi^2 = \chi^2_\sigma +\sum_{i=1}^{N_R}\frac{(R_i-R_{fit})^2}{(dR_{stat})^2}+\sum_{i=1}^{N_{exp}}\frac{(\Delta_j)^2}{(dR_{sys})^2},
\end{equation}
where $R=\mu_pG_E/G_M$, $dR_{stat}$ and $dR_{sys}$ are the statistical and systematics uncertainties in $R$, and $R_{fit}$ is the new ratio parameterization by including the new results. $N_R$ is the total number of polarization measurements of $R$, $\Delta_j$ is the offset for each data set and $N_{exp}$ is the number of the polarization data sets.

\begin{figure}
  \begin{center}
    \includegraphics[angle=0, width=0.75\textwidth]{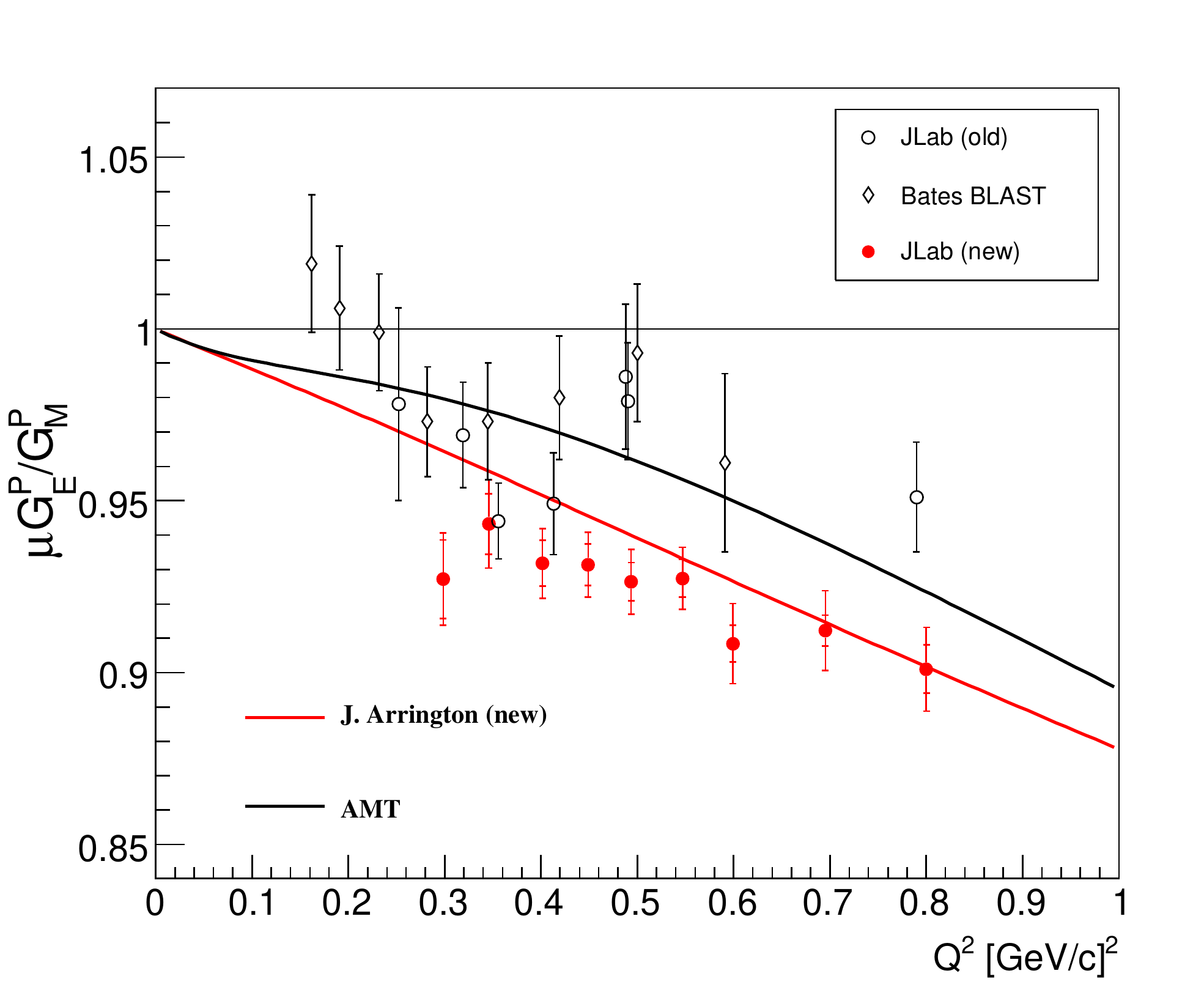}
    \caption{The global fit for the proton form factor ratio with world high precision data. The red points are the new results (E08-007 I and E03-104), the other points are from previous polarization measurements~\cite{punj,BLAST,LEDEX}. The black line is the AMT fit to the world 2$\gamma$ exchange corrected cross section and polarization data. The red line is the new fit by including the new data.}
    \label{fig:fit0}
  \end{center}
\end{figure}

\begin{figure}
  \begin{center}
    \includegraphics[angle=0, width=0.75\textwidth]{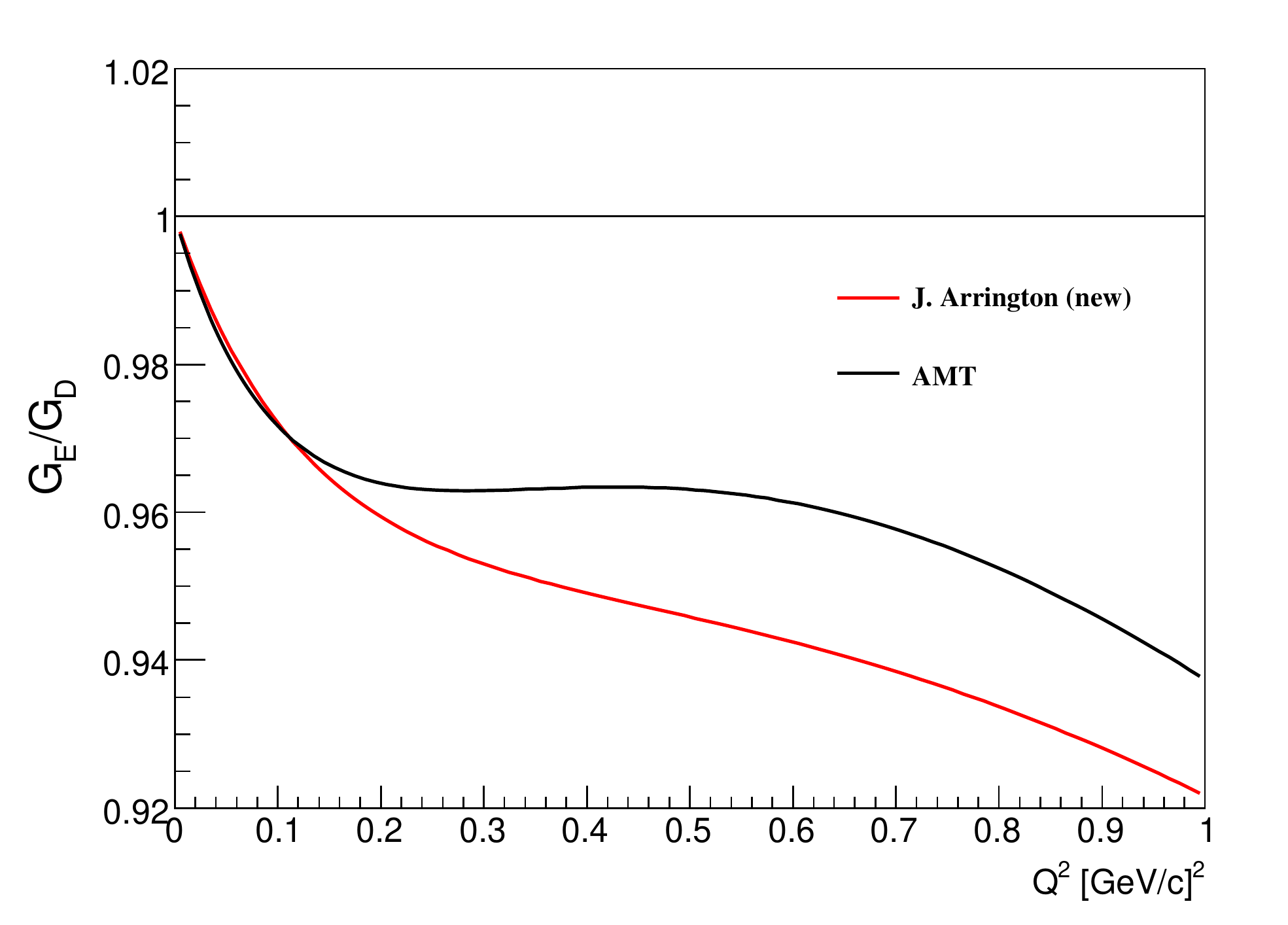}
    \caption{The global fit for the proton electric form factor $G_E$. The black line is the AMT fit to the world 2$\gamma$ exchange corrected cross section and polarization data. The red line is the new fit by including the new data.}
    \label{fig:fit_ge}
  \end{center}
\end{figure}

\begin{figure}
  \begin{center}
    \includegraphics[angle=0, width=0.75\textwidth]{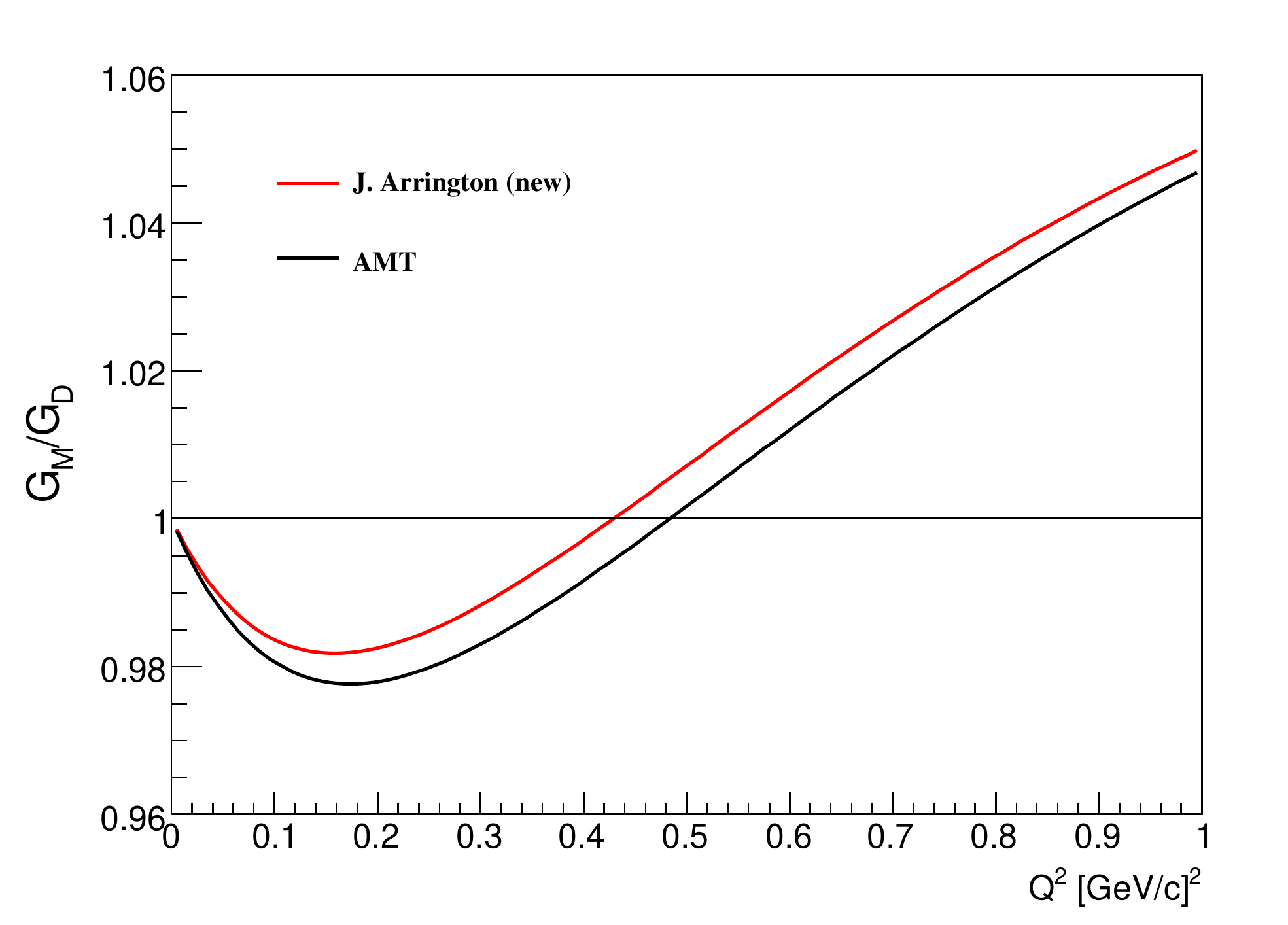}
    \caption{The global fit for the proton magnetic form factor $G_M$. The black line is the AMT fit to the world 2$\gamma$ exchange corrected cross section and polarization data. The red line is the new fit by including the new data.}
    \label{fig:fit_gm}
  \end{center}
\end{figure}

The form factors are fit to the following functional form:
\begin{equation}
G_E(Q^2),G_M(Q^2)/\mu_p = \frac{1+\sum_{i=1}^na_i\tau^i}{1+\sum_{i=1}^{n+2}b_i\tau^i},
\end{equation}
where $\tau = Q^2/4M^2$. The first pass of the new fit~\cite{John_fitnew} was performed by removing the lowest
$Q^2$ Punjabi {\it et al.} (GEp-I) point and highest $Q^2$ Ron {\it et al.} (LEDEX) point,
since the reanalysis is still underway. The other data points from the world data sets kept the same so that we have a conservative estimate of how much the fit changed. The new fit has a slightly increased $\chi^2$ compared to the previous AMT fit~\cite{john:TPE}, which is mainly due to the change in the polarization data set. Fig.~\ref{fig:fit0} shows the high precision world data with the previous AMT fit and the new fit. As one can see, the new
fit is still slightly above the new data set. The fit to the individual form factors are shown in
Fig.~\ref{fig:fit_ge} and Fig.~\ref{fig:fit_gm}, respectively. While $G_M$ stays almost the same, the new
fit indicates a $\sim 2\%$ decrease in $G_E$ in this low $Q^2$ region.
\section{Proton RMS Radius}
 In the past, the proton root-mean-square (rms) radius in general has been determined from the low $Q^2$ form factor measurements. In the non-relativistic limit, the proton charge radius is related to the electric form factor as:
 \begin{equation}
 r_p = \left(-6\frac{dG_E(Q^2)}{dQ^2}\right)_{Q^2\to 0}.
 \end{equation}

 The most cited value is from the analysis of Simon {\it et al.}~\cite{simon}, which gives $r_p= 0.862\pm0.012$ fm by using the unpolarized data up to $Q^2<2$ fm$^{-2}$. Occasionally, fits with 2- or 4-pole expressions~\cite{borkow_1} were performed, and significantly bigger values (0.88$\pm$0.02 fm and 0.92$\pm$0.02 fm) were found. The difference was partially understood as a consequence of different treatments of the $\langle r^4\rangle$ term. In parallel, fits based on dispersion relations and VMD~\cite{hoehler,mergell} models give 0.854$\pm$0.012 fm.

 As mentioned in Section~\ref{sec:den}, Kelly~\cite{kelly_par} defined the intrinsic density $\rho(r)$ as the density in the nucleon rest frame, and the moment is defined by
 \begin{equation}
 M_\alpha = \int_0^\infty dr r^{2+\alpha}\rho (r),
 \end{equation}
 where $\alpha$ is an even integer. For a charge density, these moments are related to the electric form factor by
 \begin{eqnarray}
 M_0 &=& G_E(0),\\
 M_2 &=& \left(-6\frac{dG_E(Q^2)}{dQ^2}\right)_{Q^2\to 0}-\frac{3\lambda}{2m^2}G_E(0).
 \end{eqnarray}
 While the definition for the intrinsic charge radius depends upon the choice of $\lambda_E$ employed to fit the form factor, the radius parameter
\begin{equation}
\xi_p = \left(-6\frac{d\ln G(Q^2)}{dQ^2}\right)^{1/2}_{Q^2\to 0} = \left(\frac{M_2}{M_0}+\frac{3\lambda}{2m^2}\right)^{1/2}
\end{equation}
is a model-independent quantity to be compared with the Lamb shift results and other form factor fits. This approach yields that $\xi_p=0.88\pm0.01$ fm, which represents a model-independent property of the data even if its interpretation as a charge radius depends upon the choice of $\lambda_E$. Kelly~\cite{kelly} also provided a simple fit with a rational function of $Q^2$, which is consistent with dimensional scaling at high $Q^2$. It provides excellent fits to the existing data, and the rms radii are consistent with those in~\cite{kelly_par}.

Recently, Sick~\cite{sick} used the Continued-fraction (CF) expansions to deal properly with the higher moments after accounting for the Coulomb distortion, and this leads to a radius of 0.895$\pm0.018$ fm, which is significantly larger than the radii used in the past. Later on, Blunden and Sick~\cite{blunden_sick} investigated the effect of 2$\gamma$ exchange processes in the analysis; they found that the change in the radius by removing the contribution of 2$\gamma$ exchange is small (+0.0052 fm). With the new fit presented in the previous section, we give an updated proton charge rms-radius and compare it with recent representative parameterizations in Table~\ref{tab:rms_r}.
 \begin{table}[t]
 \caption{Proton charge rms-radius from different parameterizations.}
 \begin{center}
 \begin{tabular}{l c c }
 \hline\hline
 Form factor & $r_p$ [fm] & year\\
 \hline
 Dipole & 0.851 & - \\
 FW~\cite{fried} & 0.808 & 2003\\
 Kelly~\cite{kelly} & 0.878 & 2004\\
 AS~\cite{arrington_sick} & 0.879 & 2007\\
 AMT~\cite{john:TPE} & 0.885 & 2007 \\
 BS~\cite{blunden_sick} & 0.897 & 2008\\
 New (pre.) & 0.868 & 2009\\
 \hline\hline
 \end{tabular}
 \label{tab:rms_r}
 \end{center}
 \end{table}
\section{Proton Zemach Radius}
High-precision measurements and calculations of the hydrogen hyperfine-splitting (hfs) provide very high precision tests of QED~\cite{udem,schwob,beauvoir,pachu}. Experimentally, the hfs of the hydrogen ground state is known to 13 significant figures in frequency units~\cite{karshen},
 \begin{equation}
 E_{hfs}(e^{-1}p) = 1420.4057517667(9) \mathrm{MHz}.
 \end{equation}
 One the theoretical side, the QED corrections have reached a level of a ppm accuracy. The major theoretical uncertainty comes from nuclear structure-dependent contributions, which are determined exclusively by the spatial distribution of the charge and magnetic moment of the proton.

The calculated hfs can be given as~\cite{volotka,dupays}
\begin{equation}
E_{hfs}(e^{-1}p) = (1+\Delta_{\mathrm{QED}}+\Delta_{\mathrm{hvp}}^p+
\Delta_{\mu\mathrm{vp}}^p+\Delta_{\mathrm{weak}}^p+\Delta_{\mathrm{S}})E_F^p,
\end{equation}
where $E_F^p$ is the Fermi energy
\begin{equation}
E_F^p = \frac{8\alpha^3m_r^3}{3\pi}\mu_B\mu_p=\frac{16\alpha^2}{3}\frac{\mu_p}{\mu_B}\frac{R_\infty}{(1+m_l/m_p)}.
\end{equation}
The mass $m_r = m_lm_p/(m_p+m_l)$ is the reduced mass, and $R_\infty$ is the Rydberg constant (in frequency units).

The first four corrections are due to QED, hadronic vacuum polarization, muonic vacuum polarization, and weak interactions ($Z^0$ exchange), which are all well known. The proton structure dependent corrections are
\begin{equation}
\Delta_{\mathrm{S}}=\Delta_Z+\Delta_R^p+\Delta_{pol},
\end{equation}
where the individual terms stand for ``Zemach'', ``recoil'', and ``polarizability''. The Zemach correction is given by~\cite{zemach}
\begin{equation}
\Delta_Z=-2\alpha m_rr_Z(1+\delta_Z^{rad}),
\end{equation}
where $r_Z$ is the Zemach radius
\begin{equation}
r_Z = -\frac{4}{\pi}\int_0^\infty\frac{dQ}{Q^2}\left(\frac{G_E(Q^2)G_M(Q^2)}{1+\kappa_p}-1\right).
\end{equation}
Note that this term depends on the knowledge of the elastic form factors. Due to the $1/Q^2$ term in the integral, the form factors at low $Q^2$ dominate the contribution.

Carlson {\it et al.}~\cite{Carlson} performed an analysis by including the most recent published data on proton spin-dependent structure functions. Table~\ref{tab:hfs} shows the results of this study. Note that the uncertainty of the polarizalibity term is now comparable with the uncertainty of the Zemach term.
 \begin{table}[t]
 \caption{Summary of corrections for electronic hydrogen.}
 \begin{center}
 \begin{tabular}{l r c}
 \hline\hline
 Quantity & value [ppm] & uncertainty [ppm]\\
 \hline
 $(E_{hfs}(e^-p)/E_F^p)-1$ & 1103.48 & 0.01\\
 \hline
 $\Delta_{\mathrm{QED}}$ & 1136.19 & 0.00 \\
 $\Delta_{\mu\mathrm{vp}}^p+\Delta_{\mathrm{hvp}}^p+\Delta_{\mathrm{weak}}^p$ & 0.14 \\
 $\Delta_Z (using~\cite{john:TPE})$ & -41.43 & 0.44\\
 $\Delta_R^p (using~\cite{john:TPE})$ & 5.85 & 0.07 \\
 $\Delta_{\mathrm{pol}} (using~\cite{john:TPE})$ & 1.88 & 0.64\\
 \hline
 Total & 1102.63 & 0.78\\
 Deficit & 0.85 & 0.78\\
 \hline\hline
 \end{tabular}
 \label{tab:hfs}
 \end{center}
 \end{table}

\begin{figure}
  \begin{center}
    \includegraphics[angle=0, width=0.65\textwidth]{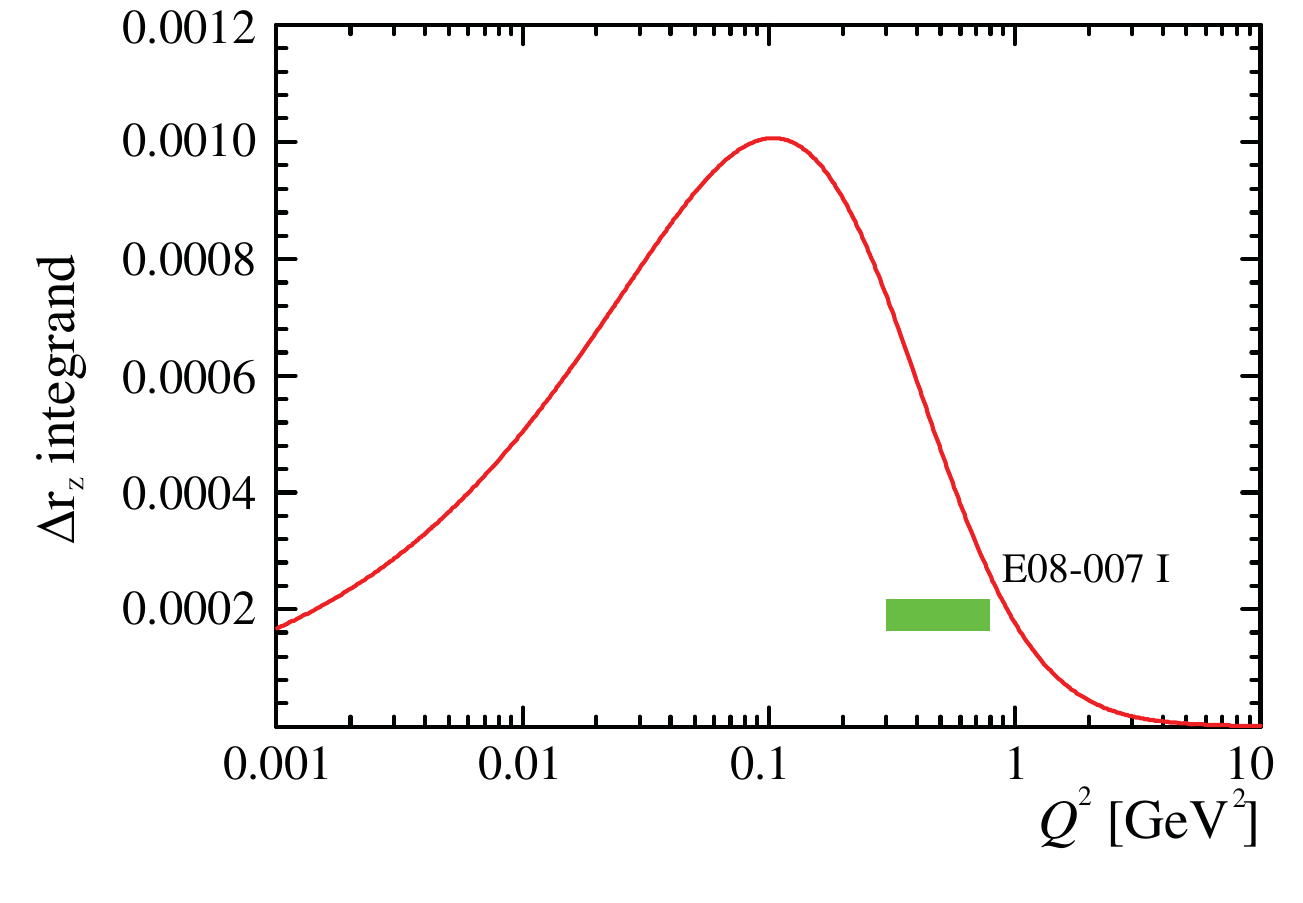}
    \caption{The uncertainty of the Zemach radius as a function of $Q^2$. The green band shows the coverage of the new data.}
    \label{fig:zerror}
  \end{center}
\end{figure}

We calculated the Zemach term with the new fit~\cite{John_fitnew}, and compared it with other parameterizaions in Table~\ref{tab:rz}. The new fit gives a slightly larger Zemach term (+0.22 ppm) which shifts the total calculation in the ``right'' direction. The deficit is now reduced to 0.63 and is within one standard deviation. Fig.~\ref{fig:zerror} shows the uncertainty of the Zemach radius integrand as a function of $Q^2$. The new results ($Q^2 = 0.3\sim 0.8$ GeV$^2$) contributed $\sim 11\%$ of the uncertainty with an optimistic approach as $Q^2$ goes to zero\footnote{This is by assuming a smooth behavior of $G_M$ in the region where it is not well measured ($Q^2<0.3$ GeV$^2$), and the uncertainty goes to zero as $Q^2\to 0$}.
 \begin{table}[t]
 \caption{Zemach radii, $\Delta_Z$ for different parameterizations.}
 \begin{center}
 \begin{tabular}{l r r c }
 \hline\hline
 Form factor & $r_Z$ [fm] & $\Delta_Z$ [ppm] & year\\
 \hline
 Dipole & 1.025 & -39.29 & - \\
 FW~\cite{fried} & 1.049 & -40.22 & 2003 \\
 Kelly~\cite{kelly_par} & 1.069 & -40.99 & 2004 \\
 AS~\cite{arrington_sick} & 1.091 & -41.85 & 2007 \\
 AMT~\cite{john:TPE} & 1.080 & -41.43 & 2007 \\
 New fit (pre.) & 1.075 & -41.21 & 2009 \\
 \hline\hline
 \end{tabular}
 \label{tab:rz}
 \end{center}
 \end{table}
\section{\label{sec:trans}Proton Transverse Densities}
As noted in Section~\ref{sec:den}, unique relativistic relationships between the Sachs form factors measured at finite $Q^2$ and the nucleon densities in the rest frame do not exist. Miller~\cite{miller_ff} showed that the form factor $F_1$ can be interpreted as a two dimensional Fourier transform of charge density in transverse space in the infinite-momentum-frame (IMF)
\begin{equation}
 \rho_{Ch}(\mathrm{\bf b}) \equiv \sum_q e_q\int dx q(x,\mathrm {\bf b}) = \int \frac{d^2q}{(2\pi)^2}F_1(Q^2=q^2)e^{iq\cdot . \mathrm {\bf b}}
\end{equation}

Recently, Miller {\it et al.}~\cite{miller_guy} extended the analysis and showed that the form factor $F_2$ may be interpreted as the two dimensional Fourier transform of the magnetization density by
\begin{equation}
\rho_M(b) = \int\frac{d^2q}{(2\pi)^2}F_2(Q^2)e^{i\bf q\cdot\bf b}.
\end{equation}
For small values of $Q^2$ it is possible to make the following expansion:
\begin{eqnarray}
F_1(Q^2) &\approx & 1-\frac{Q^2}{4}\langle b^2\rangle _{Ch},\\
F_2(Q^2) &\approx & \kappa\left(1-\frac{Q^2}{4}\langle b^2\rangle _M\right),
\end{eqnarray}
where $\langle b^2\rangle _{Ch(M)}$ is the second moment of $\rho_{Ch(M)}(b)$. The effective ( $^*$ ) square radii via the small $Q^2$ expansion of the Sachs form factors are defined as
\begin{eqnarray}
G_E(Q^2)&\approx & 1-\frac{Q^2}{6}R_E^{*2},\\
G_M(Q^2)&\approx & 1-\frac{Q^2}{6}R_M^{*2}.
\end{eqnarray}
Then the form factor ratio can be expanded as
\begin{equation}
R = \mu_pG_E/G_M \approx 1+\frac{Q^2}{6}(R_M^{*2}-R_E^{*2}),
\end{equation}
and the charge and magnetization transverse densities can be related to the ratio $R$ by:
\begin{equation}
\langle b^2\rangle _M-\langle b^2\rangle _{Ch} = \frac{\mu_p}{\kappa}\frac{2}{3}(R_M^{*2}-R_E^{*2})+\frac{\mu_p}{M_p^2},
\end{equation}
where $\frac{u_p}{M_p^2}\approx 0.1235$ fm$^2$ represents the relativistic correction. It is a consequence of the Foldy term~\cite{foldy}, which arises from the interaction of the anomalous magnetic moment of the nucleon with the external magnetic field of the electron.

\begin{figure}
  \begin{center}
    \includegraphics[angle=0, width=0.58\textwidth]{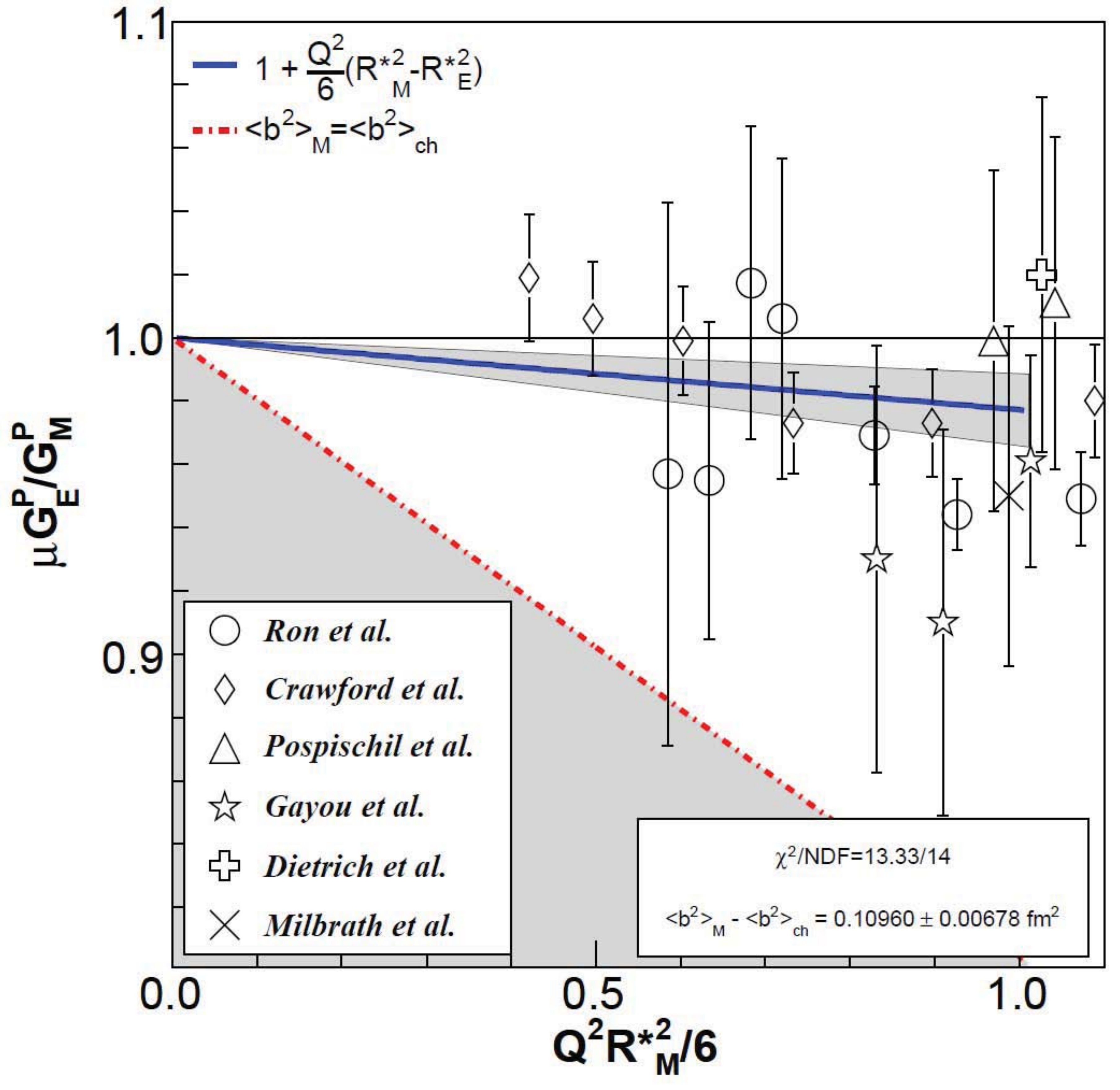}
    \caption{A linear fit to previous world polarization data, shown by the solid (blue) line and error band. The fit was done up to the region of $Q^2=0.35$ GeV$^2$ where the linear expansion is valid for the transverse radii difference. The shaded area indicates $\langle b^2\rangle _{Ch}> \langle b^2\rangle_M$. The dashed (red) line shows the critical slope when $\langle b^2\rangle _M = \langle b^2\rangle _{Ch}$. Figure from~\cite{miller_guy}}
    \label{fig:trans0}
  \end{center}
\end{figure}
\begin{figure}
  \begin{center}
    \includegraphics[angle=0, width=0.6\textwidth]{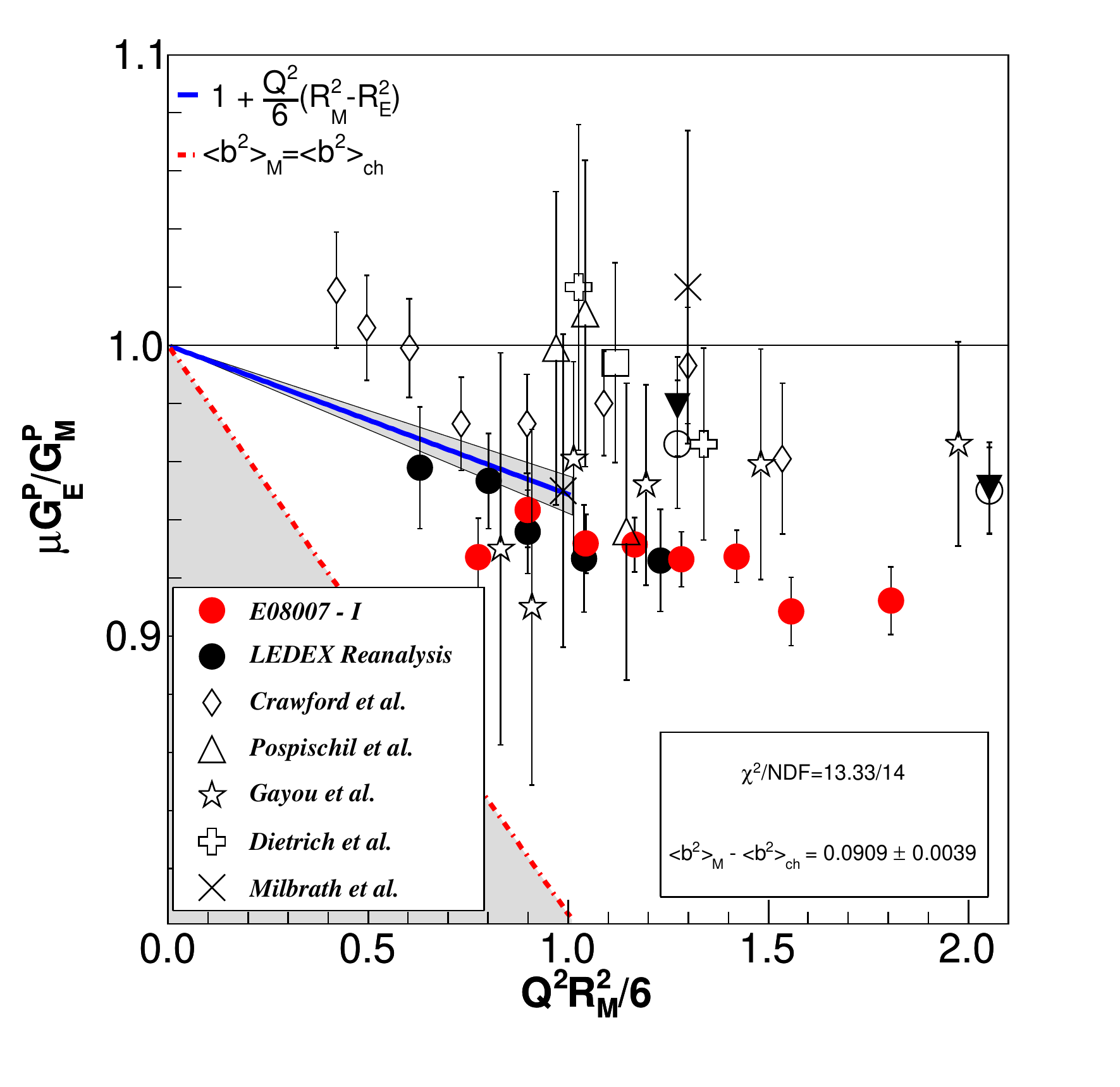}
    \caption{New fit with the E08-007 data, shown by the solid (blue) line and error band. The shaded area indicates $\langle b^2\rangle _{Ch}>\langle b^2\rangle _M$. The dashed (red) line shows the critical slope when $\langle b^2\rangle _M = \langle b^2\rangle _{Ch}$.}
    \label{fig:trans1}
  \end{center}
\end{figure}

Fig.~\ref{fig:trans0} shows the results of a linear fit to the previous world data, and Fig.~\ref{fig:trans1} shows the fit with the new results from experiment E08-007 I and the preliminary results of LEDEX reanalysis~\cite{guy_reana}. The charge and magnetization second moments difference changed from
\begin{equation}
\langle b^2\rangle _M-\langle b^2\rangle _{Ch} = 0.10960\pm 0.00687 \mathrm{fm}^2
\end{equation}
to
\begin{equation}
\langle b^2\rangle _M-\langle b^2\rangle _{Ch} = 0.09093\pm 0.00395 \mathrm{fm}^2
\end{equation}
Note that the new fit improves the uncertainty by a factor of $\sim 2$, and the magnetic density still extends further than the electric density in the transverse space. This result can be related to the failure of quarks spin to account for the total angular momentum of the proton and the expected importance of quark orbital angular momentum~\cite{proton_spin}.

\section{Strangeness Form Factors}
The parity-violating (PV) asymmetry in elastic $ep$ scattering can be used to extract the strangeness form factors~\cite{strange0,strange1,strange2}. The PV asymmetry arises due to interference between photon exchange and $Z$-boson exchange. The asymmetry in the Born approximation is given by~\cite{afan_carlson}:
\begin{equation}
A_{PV} = -\frac{G_FQ^2}{4\pi\alpha\sqrt{2}}\frac{A_E+A_M+A_A}{\tau G_{Mp}^2+\varepsilon G_{Ep}^2},
\end{equation}
where $G_F$ is the Fermi constant, and $\alpha$ is the fine structure constant. The individual asymmetry terms can be written in terms of the proton form factors $G_{Ep}$ and $G_{Mp}$ and the proton neutral weak vector and axial form factors $G_{Ep}^Z$, $G_{Mp}^Z$ and $G_A^Z$:
\begin{eqnarray}
A_E &=& \varepsilon G_{Ep} G_{Ep}^Z,\\
A_M &=& \tau G_{Mp} G_{Mp}^Z,\\
A_A &=& (1-4\sin^2\theta_W)\varepsilon 'G_{Mp}G_A^Z,
\end{eqnarray}
where $\theta_W$ is the weak mixing angle, and $\varepsilon '=\sqrt{\tau(1+\tau)(1-\varepsilon^2)}$. With the assumption of isospin symmetry, the weak vector form factors can be expressed in terms of the proton and neutron form factors together with the strangeness form factors: $G_{Es}$ and $G_{Ms}$. Neglecting the contributions from heavier quarks~\cite{strange2}, $A_{PV}$ is given by:
\begin{eqnarray}
A_{PV} &=& -\frac{G_FQ^2}{4\pi\alpha\sqrt{2}} \biggl[(1-4\sin^2\theta_W)-\frac{\varepsilon G_{Ep}(G_{En}+G_{Es})+\tau G_{Mp}(G_{Mn}+G_{Ms})}{\varepsilon (G_{Ep})^2+\tau(G_{Mp})^2}{}
\nonumber\\
& &{}-\frac{(1-4\sin^2\theta_W)\varepsilon ' G_{Mp}G_A^Z}{\varepsilon(G_{Ep})^2+\tau(G_{Mp})^2}\biggr].
\end{eqnarray}
Clearly, the measurements of the strangeness form factors require the knowledge of the nucleon form factors.

From our data, we estimated the impact of the new fit to the existing strangeness form factor measurements by comparing them with the AMT parameterization~\cite{john:TPE}. The difference in the extracted physics asymmetry is summarized in Table~\ref{tab:strange}.
 \begin{table}[t]
 \caption{The absolute asymmetry difference ($\Delta A_{PV}$), the normalized difference by the experimental uncertainty ($\Delta A_{PV}/\sigma$) and the relative asymmetry difference ($\Delta A_{PV}/A_{PV}$) between using the AMT~\cite{john:TPE} parameterization and the new one.}
 \begin{center}
 \begin{tabular}{c c c c l }
 \hline\hline
 $Q^2$ [GeV$^2$] & $\Delta A_{PV}$ [ppm] & $\Delta A_{PV}/\sigma$ & $\Delta A_{PV}/A_{PV}$ & Experiment\\
 \hline
 0.38 & -0.178 & 0.42 & $1.6\%$ & G0 FWD~\cite{g0for}\\
 0.56 & -0.347 & 0.50 & $1.6\%$ & G0 FWD\\
 1.00 & -0.414 & 0.30 & $0.8\%$ & G0 FWD\\
 \hline
 0.23 & +0.038 & 0.12 & $0.2\%$ & G0 BCK~\cite{g0bak}\\
 0.65 & +0.014 & 0.14 & $0.3\%$ & G0 BCK\\
 \hline
 0.50 & -0.299 & 0.50 & $1.7\%$ & HAPPEX III~\cite{happex3}\\
 \hline\hline
 \end{tabular}
 \label{tab:strange}
 \end{center}
 \end{table}
\section{Future Results and Experiment}
\subsection{The Mainz Cross Section Measurement}
\begin{figure}
  \begin{center}
    \includegraphics[angle=0, width=0.65\textwidth]{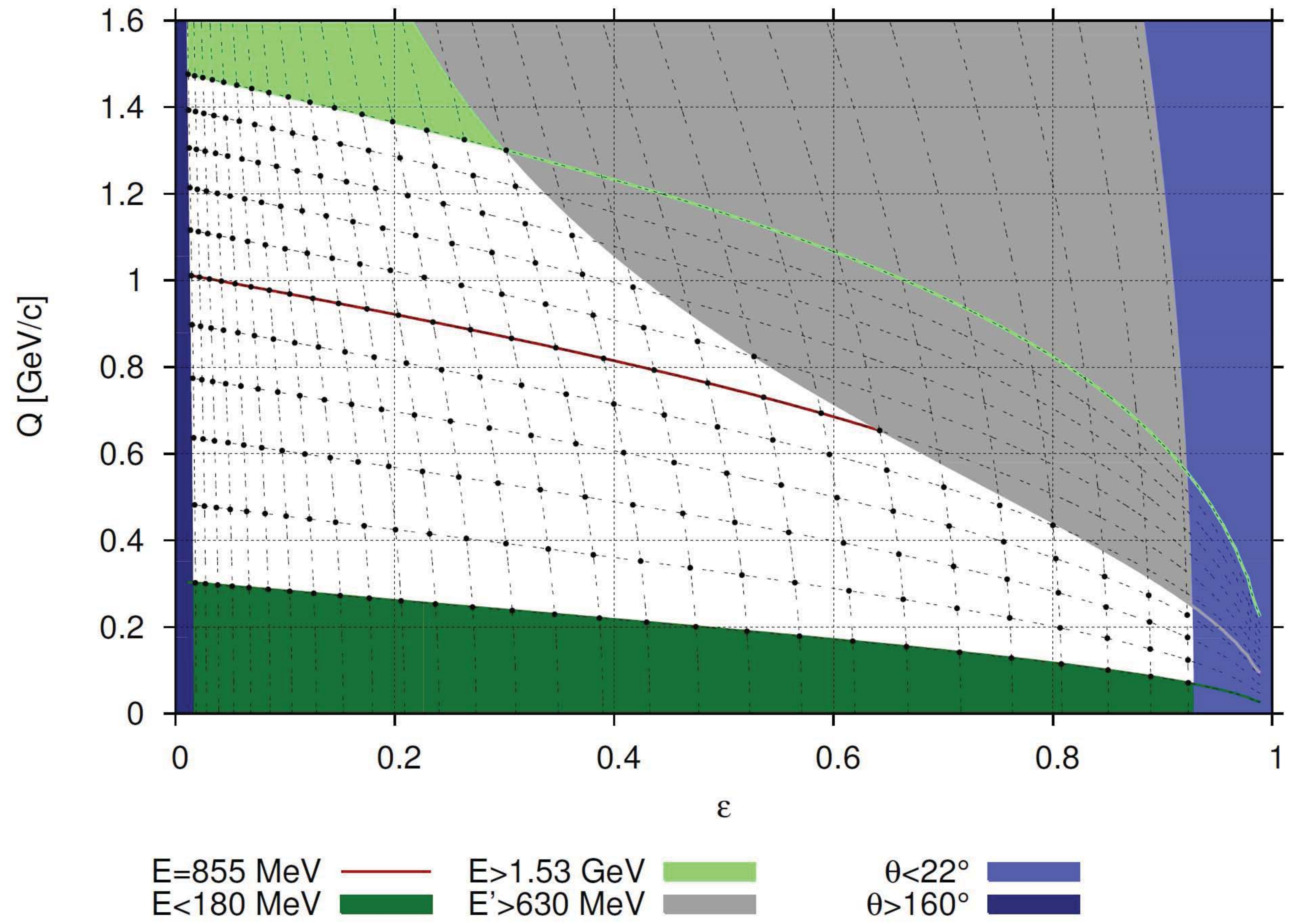}
    \caption{The accessible kinematic region in $\varepsilon/Q$ space. The black dots represent the chosen settings (centers of the respective acceptance). The dotted curves correspond to constant incident beam energies in steps of 135 MeV ("horizontal" curves) and to constant scattering angles in $5^\circ$ steps ("vertical" curves). Also shown are the limits of the facility: the red line represents the current accelerator limit of 855 MeV, with the upgrade, it will be possible to measure up to the light green curve. The dark green area is excluded by the minimal beam energy of 180 MeV. The maximum (minimum) spectrometer angle excludes the dark (light) blue area. The gray shaded region is excluded by the upper momentum of spectrometer A (630 MeV/$c$). Figure from~\cite{mami_205}.}
    \label{fig:mainz}
  \end{center}
\end{figure}

The A1 collaboration~\cite{a1} at Mainz Microtron (MAMI) completed a very high precision elastic $ep$ cross section measurement in the range of $Q^2 = 0.01 - 2$ GeV$^2$~\cite{mami_205}. The experiment aimed to measure the cross section at a fixed $Q^2$ for several settings of $\varepsilon$ to perform the Rosenbluth separation of the individual form factors. The accessible region is determined by the accelerator and the properties of the detector system. Fig.~\ref{fig:mainz} shows the accessible kinematic region for the experiment.

Due to the large cross section in the low $Q^2$ region, a very small statistical uncertainty can be achieved. The collaboration estimated a $<0.5\%$ statistical error plus a $0.5\%$ systematics uncertainty, leading to a total error of $\sim 1\%$ or less for every cross section measurement, which is an unprecedentedly small for cross section measurements.

The Mainz experiment plans to extract the individual form factors using two methods. The first way is by using the standard Rosenbluth separation which utilizes a linear fit to the cross section at constant $Q^2$ but different $\varepsilon$. This works in a completely model independent way except for the larger $Q^2$ where the two photon exchange contribution becomes larger. A second approach is to fit the global ansatz for the form factors directly to the cross sections. With a flexible ansatz, this is quasi model-independent and is an even more powerful method to directly test available models.

\subsection{E08-007 Part II}
The second part of experiment E08-007 is tentatively scheduled in 2012. This part will measure the proton form factor ratio in the range of $Q^2 = 0.015 - 0.4$ GeV$^2$ using the beam target asymmetry technique.

\begin{figure}
  \begin{center}
    \includegraphics[angle=0, width=0.6\textwidth]{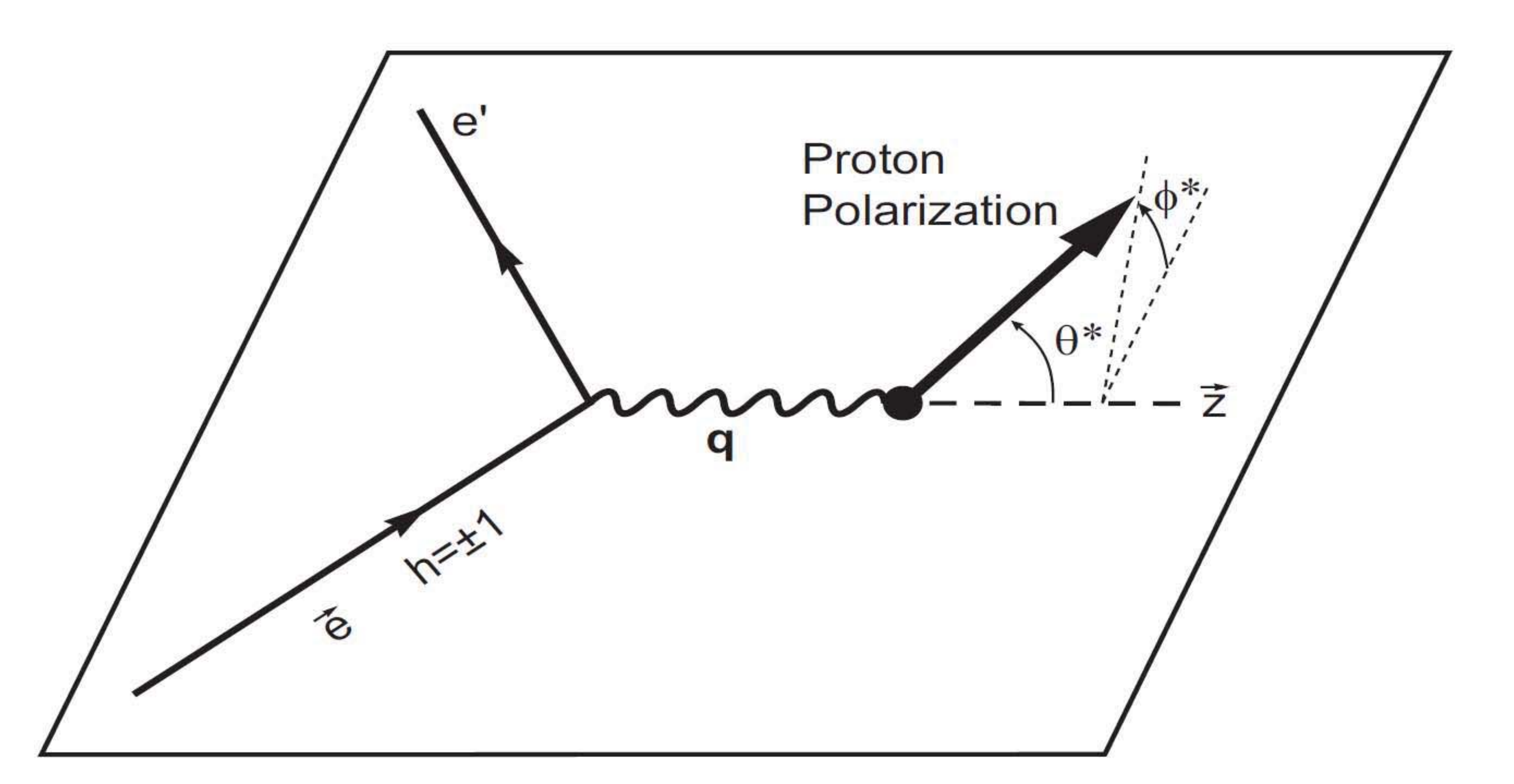}
    \caption{Spin-dependent $ep$ elastic scattering in Born appromixation.}
    \label{fig:dsa}
  \end{center}
\end{figure}

For longitudinally polarized electrons scattering from a polarized proton target, the differential cross section can be written as~\cite{donnelly}:
\begin{equation}
\frac{d\sigma}{d\Omega} = \Sigma +h\Delta,
\end{equation}
where $\Sigma$ is the unpolarized differential cross section, $h$ is the electron helicity and $\Delta$ is the spin-dependent differential cross section given by:
\begin{equation}
\Delta = \left(\frac{d\sigma}{d\Omega}\right)_{Mott}f^{-1}_{recoil}\left[2\tau v_{T'}\cos\theta^* G_M^2-2\sqrt{2\tau(1+\tau)}v_{TL'}\sin\theta^*\cos\phi^*G_MG_E\right],
\end{equation}
where $\theta^*$ and $\phi^*$ are the polar and azimuthal proton spin angles defined with respect to the three-momentum transfer vector $\vec q$ and the scattering plane (see Fig.~\ref{fig:dsa}), and $v_{T'}$ and $v_{TL'}$ are kinematic factors~\cite{donnelly}.

\begin{figure}
  \begin{center}
    \includegraphics[angle=0, width=0.5\textwidth]{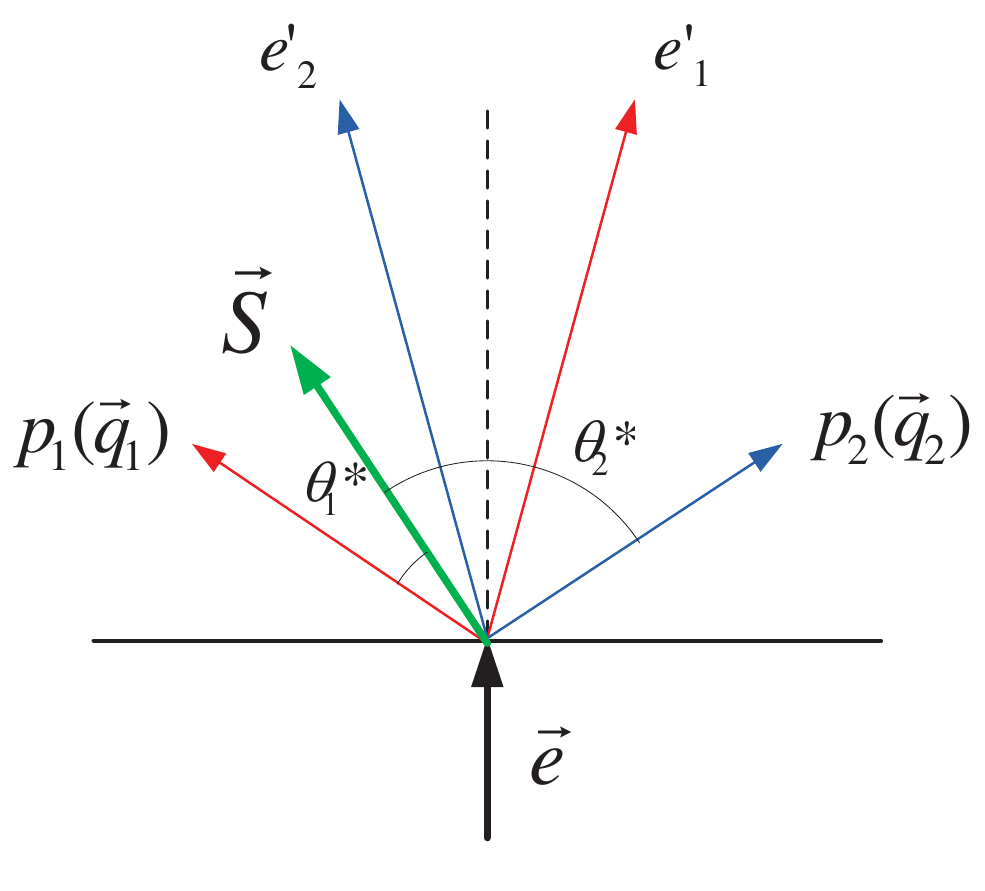}
    \caption{The kinematics for the two simultaneous measurements. The scattered electrons $e_1'$ and $e_2'$ are detected in left and right HRS, respectively. The recoil protons $p_1$ and $p_2$ point in the direction of the q-vector $\vec q_1$ and $\vec q_2$, respectively. $\vec S$ denotes the target spin direction.}
    \label{fig:dsa1}
  \end{center}
\end{figure}

The spin-dependent asymmetry $A$ is defined as:
\begin{equation}
A=\frac{\sigma^+-\sigma^-}{\sigma^++\sigma^-},
\end{equation}
where $\sigma^{+(-)}$ is the differential cross section for the two different helicities of the polarized electron beam. The spin-dependent asymmetry $A$ can be written in terms of the polarized and unpolarized differential cross-sections as:
\begin{equation}
A = \frac{\Delta}{\Sigma} = -\frac{2\tau v_{T'}\cos\theta^*G_M^2-2\sqrt{2\tau(1+\tau)}v_{TL'}\sin\theta^*\cos\phi^*G_MG_E}{(1+\tau)v_{L}G_E^2+2\tau v_{T}G_M^2}.
\end{equation}
The experimental asymmetry $A_{exp}$ is related to the spin-dependent asymmetry by the relation:
\begin{equation}
A_{exp} = P_bP_tA,
\end{equation}
where $P_b$ and $P_t$ are the beam and target polarizations, respectively. By measuring the asymmetry simultaneously in two spectrometers with different angles between the momentum transfer and the target spin as illustrated in Fig.~\ref{fig:dsa1}, the following super ratio is directly related to the ratio $G_E/G_M$:
\begin{equation}
R = \frac{A_1}{A_2} = \frac{\tau v_{T'}\cos\theta_1^*-\sqrt{2\tau(1+\tau)}v_{TL'}\sin\theta_1^*\cos\phi_1^* \frac{G_E}{G_M}}{\tau v_{T'}\cos\theta_2^* - \sqrt{2\tau(1+\tau)v_{TL'}\sin\theta_2^*\cos\phi_2^* \frac{G_E}{G_M}}},
\end{equation}
which is independent of the knowledge of the beam and target polarization.

\begin{figure}
  \begin{center}
    \includegraphics[angle=0, width=0.85\textwidth]{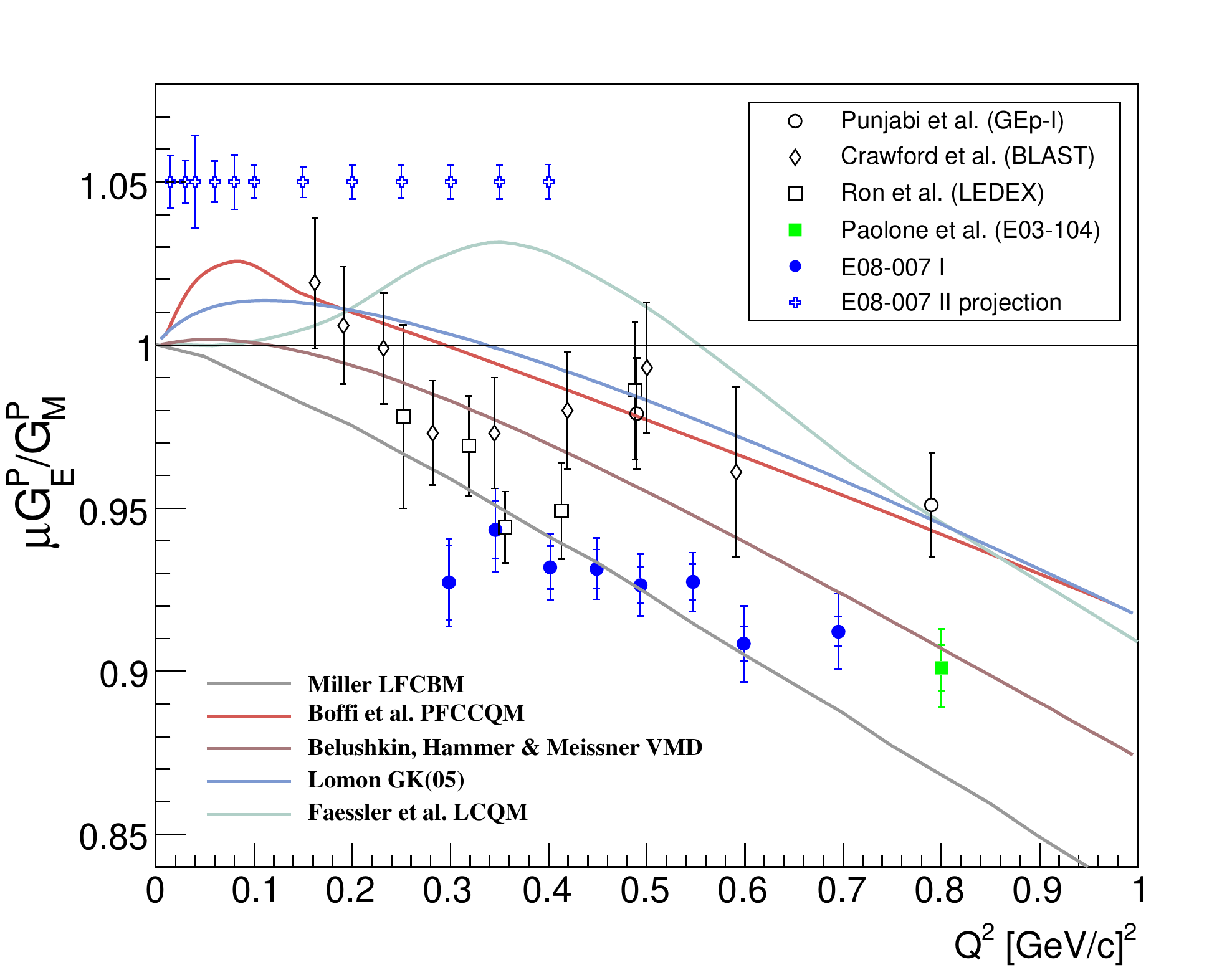}
    \caption{The proposed $Q^2$ points and projected total uncertainties for the second part of E08-007.}
    \label{fig:project}
  \end{center}
\end{figure}

The solid polarized proton target developed by UVa will be used. In this target, $^{15}$NH$_3$ is polarized by Dynamic Nuclear Polarization (DNP)~\cite{dnp} in a strong magnetic field (5 T) at very low temperature ($\sim 1$ K). The left and right HRS together with two septum magnets~\cite{septum} will be used to detect the scattered electrons simultaneously. The proposed $Q^2$ points and projected total errors are shown in Fig.~\ref{fig:project}. This future measurement will overlap with the part I points and the lower range of the BLAST~\cite{BLAST} measurement. This will provide a direct comparison with the BLAST results by using the same technique and allow an examination of any unknown systematic uncertainties of the recent measurements.
\begin{figure}
  \begin{center}
    \includegraphics[angle=0, width=0.65\textwidth]{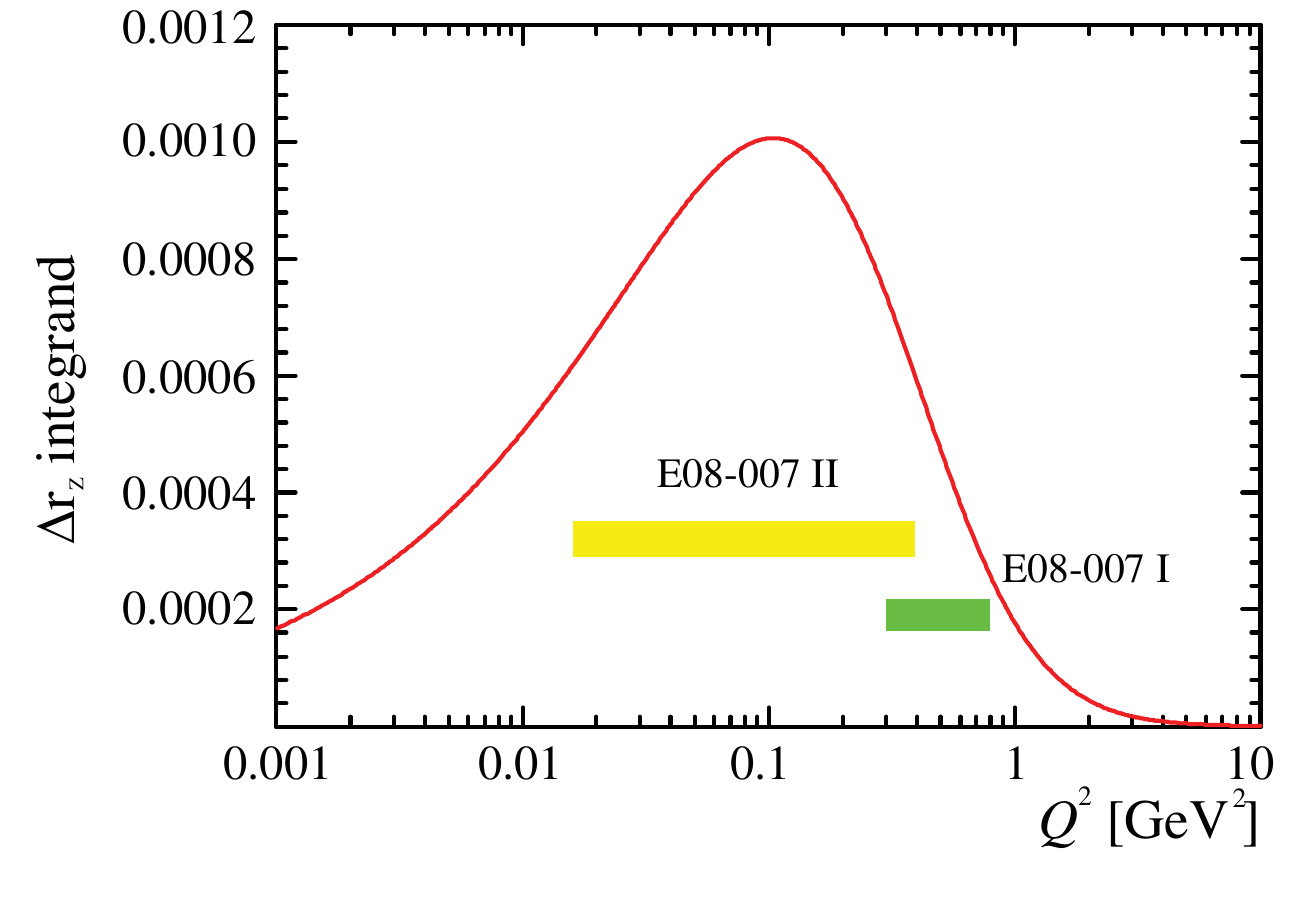}
    \caption{The uncertainty of the Zemach radius as a function of $Q^2$. The green band shows the coverage of the new data from this work, and the yellow band shows the proposed coverage of the second part of E08-007.}
    \label{fig:zemach_2}
  \end{center}
\end{figure}

 Beyond the curiosity in the form factor behavior in the extremely low momentum transfer region, the motivation for the running of the experiment E08-007 part II also comes from the determination of the proton Zemach radius. As illustrated in Fig.~\ref{fig:zemach_2}, the second part of this measurement will cover the peak region where the existing data contribute $\sim 60\%$ of the total uncertainty in $r_Z$. By assuming the form factor ratio follows in a similar trend as the part I data, a conservative change of $\sim 0.4$ ppm in the Zemach term $\Delta_Z$ is expected.

\begin{figure}
  \begin{center}
    \includegraphics[angle=0, width=0.63\textwidth]{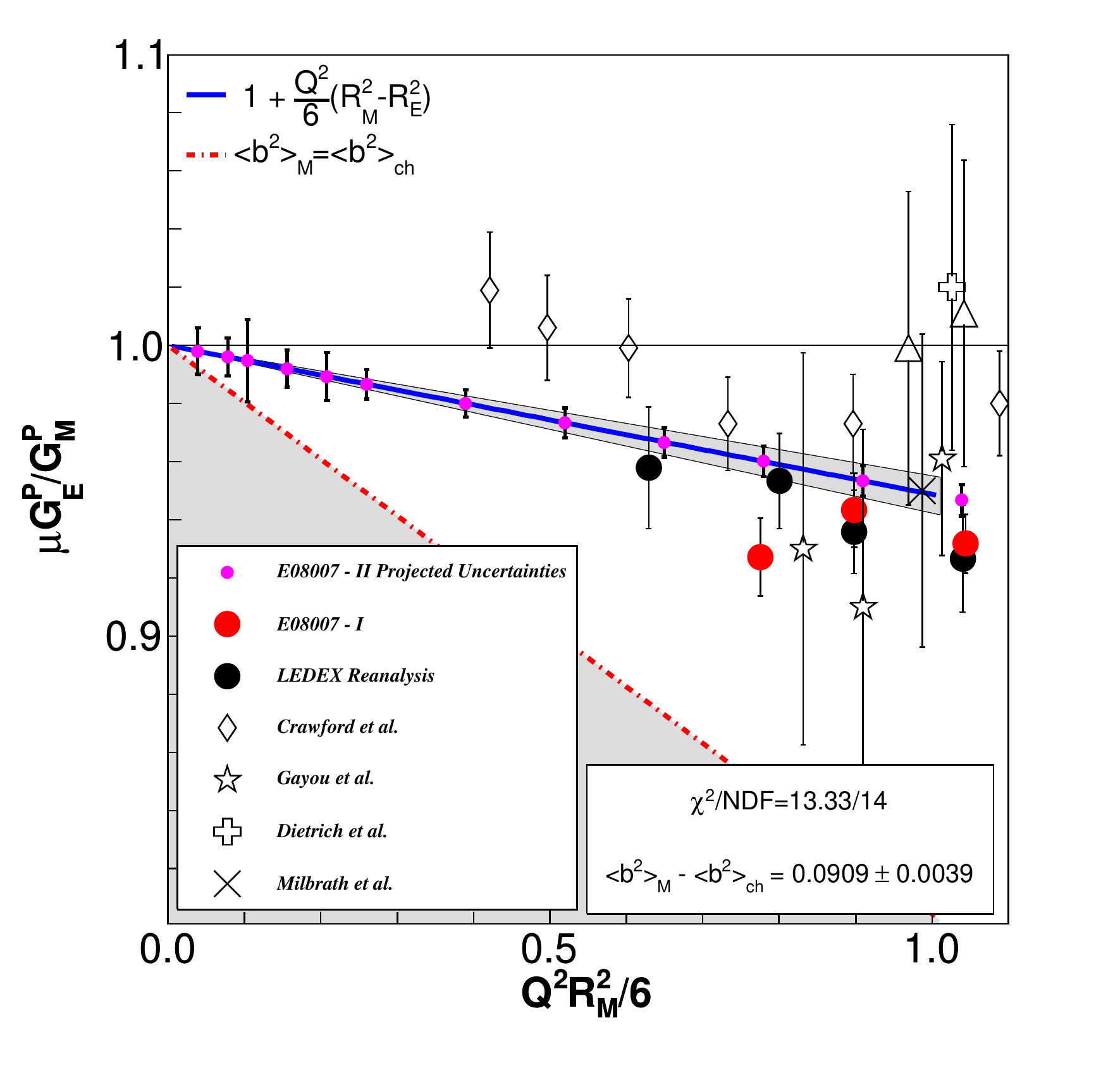}
    \caption{Projection of E08-007 part II measurements on the new fit by assuming the same slope as $Q^2$ decreases.}
    \label{fig:tran_pro}
  \end{center}
\end{figure}

 As mentioned in Section~\ref{sec:trans}, the low $Q^2$ data will also greatly improve our knowledge of the proton transverse densities in the impact parameter space. The expected results from the part II measurement are shown in Fig.~\ref{fig:tran_pro} and will allow us to make a definitive fit to this quantity.
\section{Conclusion}
In conclusion, this thesis presents the details of the proton electric to magnetic form factor ratios measurements at $Q^2 = 0.3- 0.7$ GeV$^2$. This experiment used the standard Hall A experimental system with one of the two high resolution spectrometers. To reduce the inelastic background, the BigBite calorimeter was used to tag the electrons and form the coincidence trigger. The central experimental equipment was the Focal Plane Polarimeter (FPP), which measured the polarization of the recoil proton in the elastic scattering of polarized electrons from an unpolarized liquid hydrogen target. The statistical uncertainty in this experiment is determined by the polarization of the electron beam and the figure of merit of the FPP. The main source of the systematic uncertainty in this measurement comes from the spin precession of the proton in the magnetic field of the spectrometer. With an $85\%$ beam polarization and 21 days of running, we have achieved the best statistics to date. For the most of the $Q^2$ kinematic points, the systematic error dominates the total uncertainty.

The results of this measurement together with a high precision point at $Q^2 = 0.8$ GeV$^2$ (from experiment E03-104) strongly deviate from unity, and are systematically below the world polarization data. The preliminary reanalysis of the LEDEX data is in agreement with the new data, but the discrepancy between the BLAST results and the new data still needs to be investigated. The new results do not favor any narrow structure in this region as suggested by the phenomenological fit~\cite{fried}. At the $Q^2=0$ limit, the ratio is forced to unity by definition, and the current slope of the data in this region appears to be too smooth to meet this condition, which might indicate a change in the slope as $Q^2$ approaches 0.

The low $Q^2$ range measured in this experiment does not allow for a pQCD calculation, which necessitates the development of low energy effective field theories and the use of fits to the data in order to describe the form factors. None of the current theories accurately predicts the entire data set, which is mainly due to the ``free'' parameters that had been tuned to the older data in those calculations. On the other hand, fast developments of computational capabilities may allow theories such as Lattice QCD to offer a complete and model-independent description in the near future.

In the mean time, the new results from this experiment have been used in global fit to extract the individual form factors. The preliminary fit suggests a smaller $G_{Ep}$ in this region, while the change in $G_{Mp}$ is relatively small. The new data also changed the results of the proton transverse density as proposed in~\cite{miller_guy}; the difference between the transverse RMS magnetic and electric radius is smaller with improved precision. The improved knowledge of the individual form factors also has a significant impact in the ultra-high precision test of QED in the hydrogen hyperfine splitting calculations and in the extraction of the strangeness form factors from parity-violation experiments.

The second part of this experiment, which will access the region of $Q^2 = 0.015-0.4$ GeV$^2$ is tentatively scheduled in 2012. In addition to resolving the potential data discrepancy, this part will be the first polarization measurement in the extremely low $Q^2$ region and will offer a great opportunity to vastly improve our knowledge of the nucleon structure.

%% file: app0.tex
\chapter{Kinematics in the Breit Frame}
The Breit frame, also called the ``brickwall frame'', is the frame where the momenta of the initial and final nucleon are equal and opposite:
\begin{equation}
\vec p_B = -\vec {p'}_B = -\frac{\vec q_B}{2},
\end{equation}
so there is no energy transfer in the elastic scattering in this frame:
\begin{eqnarray}
E_{pB} &=& E'_{pB}\\
\omega_B &=& E_{pB}-E'_{pB} = 0.
\label{eq:0trans}
\end{eqnarray}
The four-momentum transfer in the Breit frame is:
\begin{equation}
Q^2 = -q^2_B = \vec q^2_B
\end{equation}

For the electron kinematics, Eq.~\ref{eq:0trans} imposes
\begin{eqnarray}
E_{B} &=& E'_{B}\\
\vec k^2_{B} &=& \vec k'^2_{B}\\
\vec k_{B} &=& \vec q_B + \vec k'_{B}.
\end{eqnarray}
\begin{figure}[hbt]
  \begin{center}
    \includegraphics[angle=0,width=.65\textwidth]{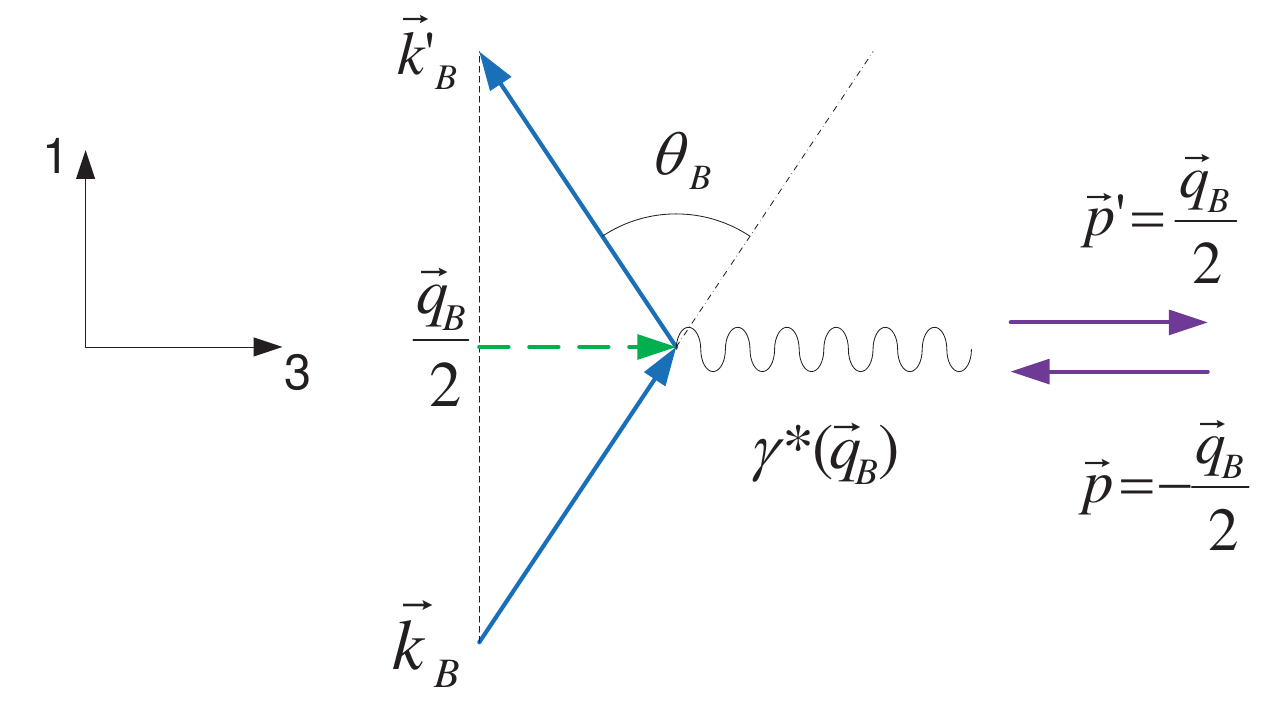}
    \caption{Elastic scattering in the Breit frame.}
    \label{fig:breit}
  \end{center}
\end{figure}
As illustrated in Fig.~\ref{fig:breit}, the three-momentum of the electron have:
\begin{eqnarray}
k_{B1} &=& k'_{B1} = \frac{|\vec q_B|}{2}\cot(\frac{\theta_B}{2}) = \frac{\sqrt{Q^2}}{2}\cot(\frac{\theta_B}{2})\\
k_{B2}&=& k'_{B2} = 0\\
k_{B3}&=&-k'_{B3} = \frac{|\vec q_B|}{2} = \frac{\sqrt{Q^2}}{2}
\end{eqnarray}

Now we can express the scattering angle $\theta_B$ in the Lab frame. The Breit frame is moving along the 3-axis, so that the 1 and 2 components of the electron momentum are left unchanged by the Lorentz transformation:
\begin{eqnarray}
k_1 &=& k_{1B} = k'_1 = k'_{1B} = \frac{\sqrt{Q^2}}{2}\cot\frac{\theta_B}{2}\\
k_2 &=& k_{2B} = k'_2 = k'_{2B} = 0.
\end{eqnarray}
The $\vec q$ is along the 3-axis, so we can write
\begin{equation}
k^3_2 = \frac{(\vec k \cdot \vec q)}{\vec q^2}=
\frac{\vec k\cdot \vec k -\vec k\cdot \vec k'}{\vec q^2} = \frac{(E^2)^2+(EE'\cos\theta_e)^2-2E^2EE'\cos\theta_e}{\vec q^2}.
\end{equation}
By using the relation
\begin{equation}
Q^2 = 4EE'\sin^2\frac{\theta_e}{2}
\end{equation}
we can get
\begin{eqnarray}
k^2_1 &=& \vec k^2 - k^2_3 = \frac{\vec k^2\vec q^2-(\vec k\cdot\vec q)^2}{\vec q^2}{}
\nonumber\\
&=& \frac{[(E^2)^2+E^2E'^2-2E^2EE'\cos\theta_e]-[(E^2)^2+(EE'\cos\theta_e)^2-2E^2EE'\cos\theta_e]}{\vec q^2}{}
\nonumber\\
&=& \frac{E^2E'^2\sin^2\theta_e}{\vec q^2} = \frac{Q^4}{4\vec q^2}\cot^2\frac{\theta_e}{2},
\label{eq:k1}
\end{eqnarray}
where the electron mass is neglected. Since
\begin{eqnarray}
q &=& p'-p\\
p^2 &=& p'^2 = m_p^2,
\end{eqnarray}
we can write
\begin{eqnarray}
p'^2 &=& (q+p)^2 = q^2+2p\cdot q + p^2\\
q^2 &=& -2q\cdot p = -2\omega m_p\\
\omega &=& -\frac{q^2}{2m_p} = \frac{Q^2}{2m_p}.
\end{eqnarray}
Using $Q^2 = -(\omega^2-\vec q^2)$, we can express $\vec q^2$ as
\begin{equation}
\vec q^2 = Q^2(1+\frac{Q^2}{4m_p^2})= Q^2(1+\tau).
\end{equation}
Now Eq.~\ref{eq:k1} can be replaced by
\begin{equation}
k_1^2 = \frac{Q^2}{4(1+\tau)}\cot^2\frac{\theta_e}{2},
\end{equation}
and the angle $\theta_B$ can be expressed as
\begin{equation}
\cot^2\frac{\theta_B}{2} = \frac{\cot^2\frac{\theta_e}{2}}{1+\tau}.
\end{equation}
\clearpage
\newpage

%% file: app1.tex
\chapter{Algorithm for Chamber Alignment}

The FPP chamber alignment by matrix expansion used in this analysis was developed in experiment
E93-049. Instead of doing the physical alignment for each FPP chamber, the
correction is directly applied to the track reconstructed by the FPP
referring to the VDC track. The alignment algorithm is described as
the following.

For the ``straight-through'' events, the VDCs have the reconstructed
track T0 ($x_0,y_0,\theta_0,\phi_0$), and the FPP chambers have the
reconstructed track T1 before the alignment. The difference between the two tracks
is $\Delta\mathrm{T} = \mathrm{T}1-\mathrm{T}0$. The goal is to apply the correction terms $\Delta \mathrm{T}$
for each track of the FPP. Intuitively, this
correction depends on where the track is hitting at, so it's
convenient to expend the correction in terms of the polynomial of the
track position $x_0, y_0$ at the focal plane, which are $1,~x_0,~y_0,~x_0^2,~y_0^2,~x_0\cdot y_0$.

The vector $\mathbf{V}$ is defined as:
\begin{displaymath}
\mathbf{V}=
\left( \begin{array}{c}
1\\
x_0\\
y_0\\
x_0^2\\
y_0^2\\
x_0\cdot y_0
\end{array} \right)
\end{displaymath}
and $\Delta \mathrm{T}$ is:
\begin{displaymath}
\mathbf{\Delta T}=
\left( \begin{array}{c}
x_1-x_0\\
y_1-y_0\\
x_1^2-x_0^2\\
y_1^2-y_0^2
\end{array} \right).
\end{displaymath}
The matrix $\mathbf{A}$ is constructed by:
\begin{equation}
\mathbf{A}=\mathbf{\mathrm{V} \cdot \mathrm{V'}}.
\end{equation}
The matrix $\mathbf{B}$ is constructed by:
\begin{equation}
\mathbf{B}=\mathbf{\Delta \mathrm{T} \cdot \mathrm{V'}}.
\end{equation}
The correction matrix $\mathbf{M}$ is defined by:
\begin{eqnarray}
\mathbf{M}&=&\mathbf{A}^{-1}\cdot \mathbf{B}'
\nonumber\\
&=& (\mathbf{V\cdot V'})^{-1}\cdot (\mathbf{\Delta \mathrm{T}}\cdot \mathrm{V}')'\\
&=& (\mathbf{V}')^{-1}\cdot \mathbf{\Delta \mathrm{T}}.
\end{eqnarray}
The simple matrix calculation leads to:
\begin{equation}
\mathbf{T1}=\mathbf{T0}+\mathbf{V'\cdot \mathbf{M}}
\label{eq:corr_ap}
\end{equation}
From Eq.~\ref{eq:corr_ap}, the FPP track is corrected by the matrix
$\mathbf{M}$. In the real procedure, the front chamber track is first
aligned with respect to the VDC tracks, and the rear chamber
track is aligned by the same way with respect to the aligned front track.

\clearpage
\newpage

%% file: app2.tex
\chapter{Extraction of Polarization Observables}

  \section{Introduction}
  For experiment E08-007 we measured the recoil proton polarization in the elastic reaction $^1H(\vec e,e'\vec p$). With the scattering angles reconstructed by the FPP and the spin rotation matrix generated by COSY, we are able to extract the polarization components at the target. Three different methods to extract the polarization observables are presented in~\cite{Besset}. In this work, the weighted-sum and maximum likelihood method were discussed. Since we are dealing with the $\le 1\% $ statistical uncertainty in this measurement, the validity of the approximations used in the formalism was carefully examined.

  \section{Azimuthal asymmetry at the focal plane}
    The detection probability for a proton scattered by the analyzer with polar angle $\theta$ and azimuthal angle $\phi$ is given by~\cite{punj}:
  \begin{equation}
    f^{\pm}(\theta,\phi)=\frac{1}{2\pi}\epsilon(\theta,\phi)(1\pm A_y(P_y^{fpp}\sin \phi-P_x^{fpp}\cos \phi)),
    \label{eq:pro_ap}
  \end{equation}
  where $\pm$ refers to the sign of the beam helicity, $P_x^{fpp}$ and $P_y^{fpp}$ are the transverse and normal polarization components at the analyzer with plus beam helicity, respectively; $P_z^{fpp}$ is not measured because it does not result in an asymmetry. $\epsilon(\theta,\phi)$ is the normalized efficiency (acceptance) which describes the non-uniformities in the detector response that results from misalignments and inhomogeneities in detector efficiency. $A_y$ is the analyzing power. Based on Eq.~\ref{eq:pro_ap} the efficiency can be extracted by:
  \begin{equation}
    \epsilon(\theta,\phi)=\frac{f^+(\theta,\phi)+f^-(\theta,\phi)}{\pi}.
    \label{eq:sum}
  \end{equation}

 \section{Weighted-sum}
 The spin transport matrix is defined by:
  \begin{equation}
  \left(\begin{array}{c}
  P_x^{fpp}\\
  P_y^{fpp}
  \end{array}\right)=
  \left(\begin{array}{ccc}
  S_{xx} & S_{xy} &S_{xz}\\
  S_{yx} & S_{yy} &S_{yz}
  \end{array}\right)
  \left(\begin{array}{c}
  P_x^{tg}\\
  \eta hP_y^{tg}\\
  \eta hP_z^{tg}
  \end{array}\right),
  \end{equation}
 where $P_x^{tg}, P_y^{tg}, P_z^{tg}$ is the polarization component at the target. By writing Eq.~\ref{eq:pro_ap} in terms of the polarization components at the target, we now have:
\begin{equation}
f(\phi)=\frac{1}{2\pi}\epsilon(1+\lambda_{x}P_x^{tg}+\lambda_{y}hP_y^{tg}+\lambda_{z}hP_z^{tg}),
\end{equation}
where
\begin{eqnarray}
\lambda_{x}&=&A_y(S_{yx}\sin\phi-S_{xx}\cos\phi){}
\nonumber\\
{}\lambda_{y}&=&\eta A_y(S_{yy}\sin\phi-S_{xy}\cos\phi){}
\nonumber\\
{}\lambda_{z}&=&\eta A_y(S_{yz}\sin\phi-S_{xz}\cos\phi).
\label{eq:lambda}
\end{eqnarray}
$\eta$ is the sign for the beam helicity, and $h$ is the beam
polarization. Note that the contribution from the induced (normal) polarization
$P_x^{tg}$ is independent of the beam helicity. In the Born approximation, the induced polarization $P_x^{tg}=0$. As noted in~\cite{Besset}, with different beam helicities, we can always construct an effective acceptance which has a symmetry period of $\pi$ in $\phi$.
The integrals can be expressed as:
\begin{eqnarray}
\int_0^{2\pi}f(\phi)\lambda_y d\phi &=&hP_y^{tg}\int_0^{2\pi}f(\phi)\lambda^2_yd\phi+{}
\nonumber\\
&&{}hP_z^{tg}\int_0^{2\pi}f(\phi)\lambda_y\lambda_z d\phi+{}
\nonumber\\
{}\int_0^{2\pi}f(\phi)\lambda_z d\phi &=&hP_y^{tg}\int_0^{2\pi}f(\phi)\lambda_y\lambda_zd\phi+{}
\nonumber\\
&&{}hP_z^{tg}\int_0^{2\pi}f(\phi)\lambda^2_z d\phi.
\label{eq:int}
\end{eqnarray}
By replacing the integrals in Eqs.~\ref{eq:int} with corresponding sums over the observed events, we have
\begin{equation}
\left(\begin{array}{c}
\sum_i\lambda_{y,i}\\
\sum_i\lambda_{z,i}
\end{array}\right)
=\left(\begin{array}{ccc}
\sum_i\lambda_{y,i}\lambda_{y,i}&\sum_i\lambda_{z,i}\lambda_{y,i}\\
\sum_i\lambda_{y,i}\lambda_{z,i}&\sum_i\lambda_{z,i}\lambda_{z,i}
\end{array}\right)
\left(\begin{array}{c}
hP_y^{tg}\\
hP_z^{tg}
\end{array}\right).
\label{eq:sums_ap}
\end{equation}
So $P_y^{tg}$ and $P_z^{tg}$ can be solved from the equation above. Problems may arise if $P_x^{tg}$ is non-zero from the 2$\gamma$ exchange, since an acceptance with symmetry period of $\pi$ in $\phi$ cannot be constructed.
\section{Maximum likelihood}
The individual polarization components can also be extracted by the maximum-likelihood (ML) technique. Based on Eq.~\ref{eq:pro_ap}, we can express the probability for the experimental angular distribution as the product of all the individual probabilities:
\begin{equation}
      F=\prod_{i=1}^N{\frac{1}{2\pi}\epsilon[1+A_y(P_y^{fpp}\sin\phi_i-P_x^{fpp}\cos\phi_i)]}.
\end{equation}
The likelihood function is given by:
\begin{equation}
L(P_x^{tg},P_y^{tg},P_z^{tg})=\prod_{i=1}^N\frac{1}{2\pi}\epsilon(1+\lambda_y hP_y^{tg}+\lambda_z hP_z^{tg}),
\end{equation}
where $\lambda_y,\lambda_z$ are the same as defined in Eq.~\ref{eq:lambda}.
By maximizing the probability function:
\begin{eqnarray}
\frac{\partial \mathrm{ln}L}{\partial P_y^{tg}}&=&0{}
\nonumber\\
{}\frac{\partial \mathrm{ln}L}{\partial P_z^{tg}}&=&0,
\label{eq:max}
\end{eqnarray}
we can extract $P_y^{tg}$ and $P_z^{tg}$. The normalized efficiency term $\epsilon$ is eliminated in the derivative since it dose not depend on $P$ ($\frac{\partial \mathrm{ln \epsilon}}{{\partial P}}=0$). To linearize the equations, an approximation is applied:
\begin{equation}
\ln (1+x)\approx x-\frac{x^2}{2}+o(x^3),
\label{eq:ln}
\end{equation}
where $x=\lambda_yhP_y^{tg}+\lambda_zhP_z^{tg}$. By omitting the $o(x^3)$ term, the equations is simplified as:
\begin{equation}
\left(\begin{array}{c}
\sum_i\lambda_{y,i}\\
\sum_i\lambda_{z,i}
\end{array}\right)=
\left(\begin{array}{cc}
\sum_i\lambda_{y,i}\lambda_{y,i}&\sum_i\lambda_{y,i}\lambda_{z,i}\\
\sum_i\lambda_{z,i}\lambda_{y,i}&\sum_i\lambda_{z,i}\lambda_{z,i}
\end{array}\right)
\left(\begin{array}{c}
hP_y^{tg}\\
hP_z^{tg}
\end{array}\right),
\label{eq:simp}
\end{equation}
which is the same as Eq.~\ref{eq:sums_ap}, and the weighted-sum and ML methods converge at this point. Here we still assume the induced polarization $P_x^{tg}=0$ to simplify the context, since determining the induced polarization $P_x^{tg}$ which is sensitive to the false asymmetry is not the intent of this experiment.

\section{Simulation}
Although it is clearly derived from the above sections and also in~\cite{Besset} that false asymmetry can be canceled by flipping the beam helicity, it is straight forward to confirm the results within a certain precision and test the statistical sensitivity of the weighted-sum method by simulation.

\begin{figure}[hbt]
  \begin{center}
    \includegraphics[angle=0,width=.95\textwidth]{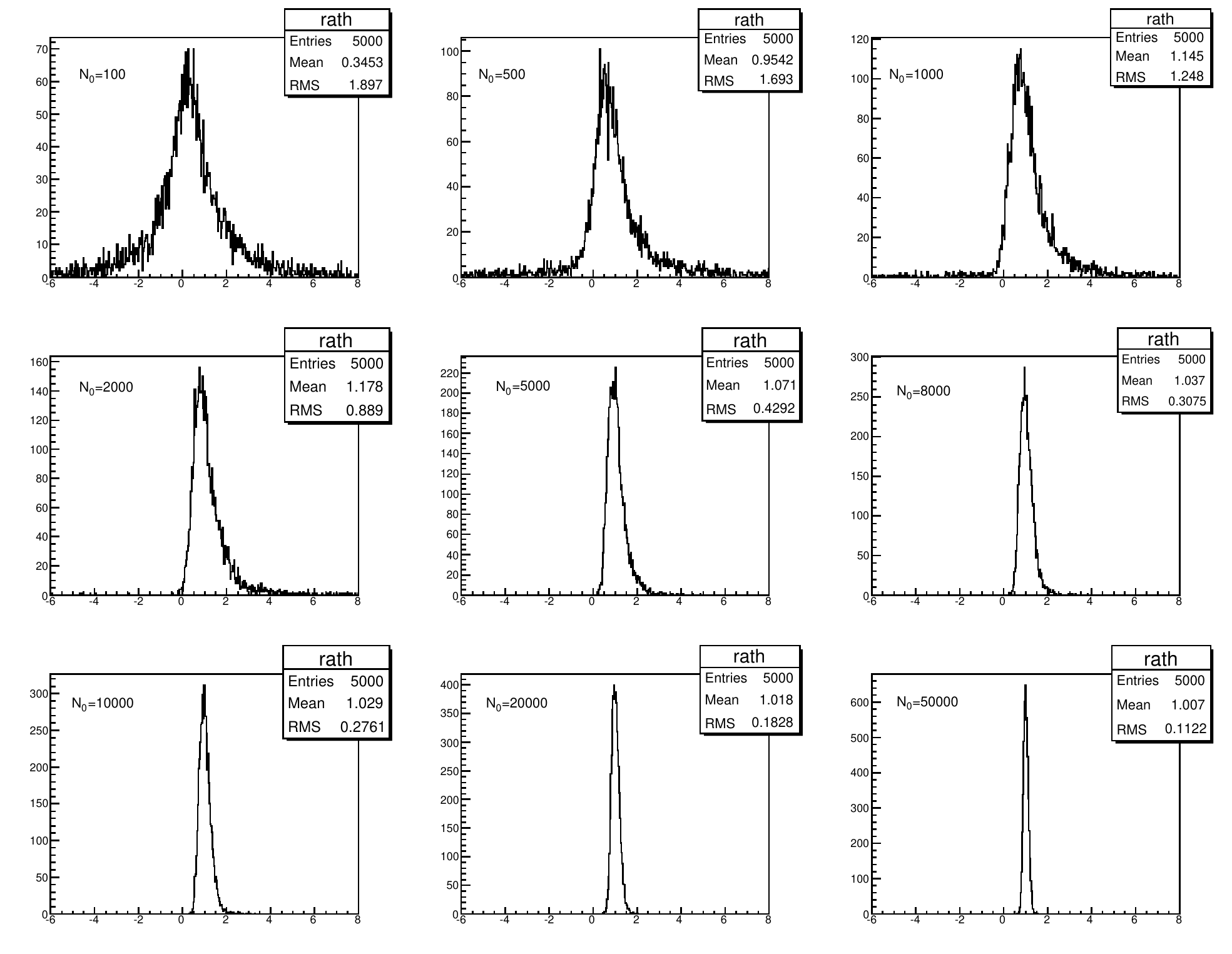}
    \caption{Histograms of the extracted ratio $P_y/P_z$ by weighted-sum method with no false asymmetry ($s_1=s_2=0$) in the simulation. $N_0$ is the sample size of each trial in the simulation. At large statistics, the extracted ratio is in good agreement with the set ratio in the simulation.}
    \label{fig:nofa_dis}
  \end{center}
\end{figure}

\begin{figure}[hbt]
  \begin{center}
    \includegraphics[angle=0,width=.40\textwidth]{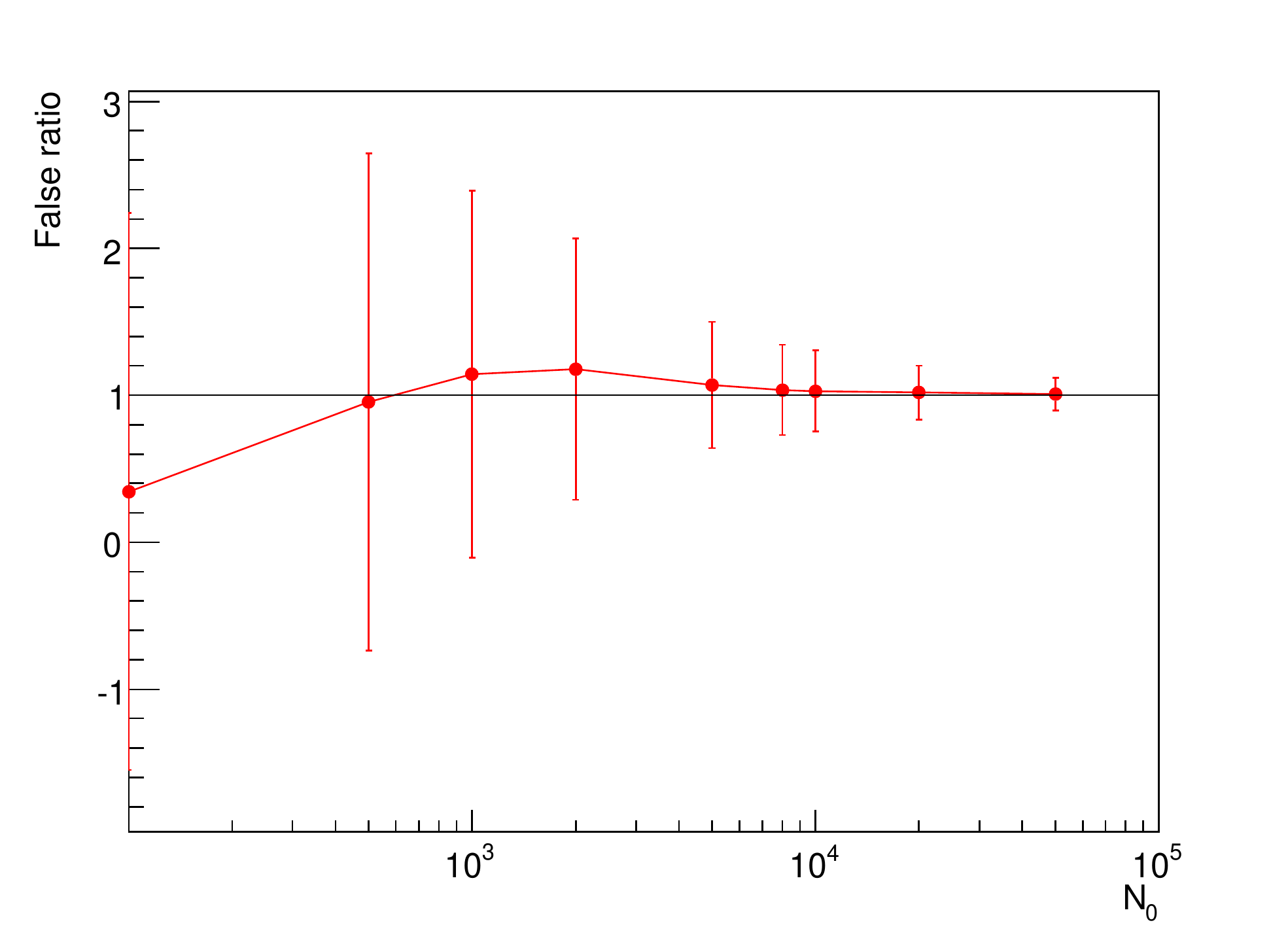}
    \includegraphics[angle=0,width=.40\textwidth]{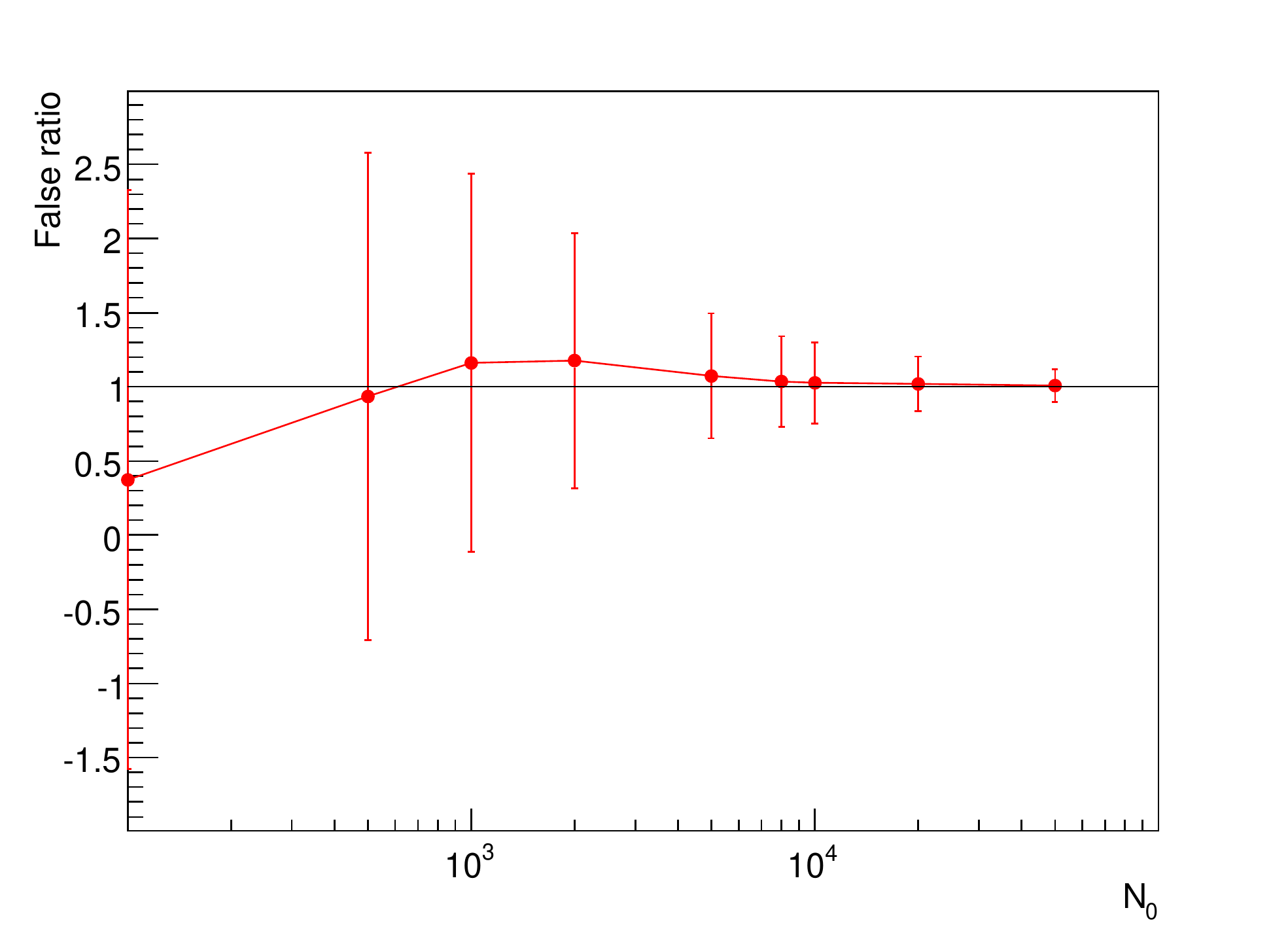}
    \includegraphics[angle=0,width=.40\textwidth]{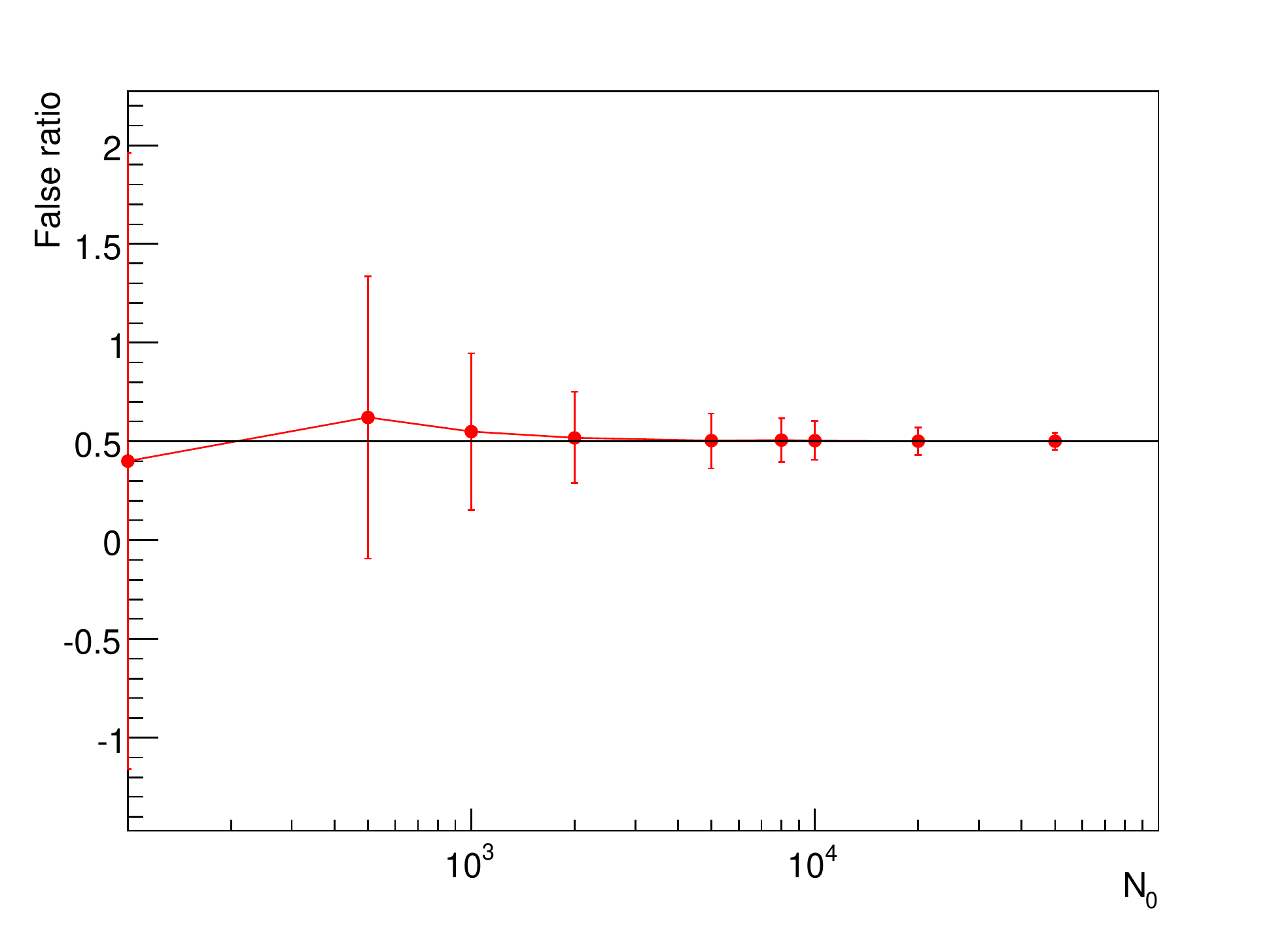}
    \includegraphics[angle=0,width=.40\textwidth]{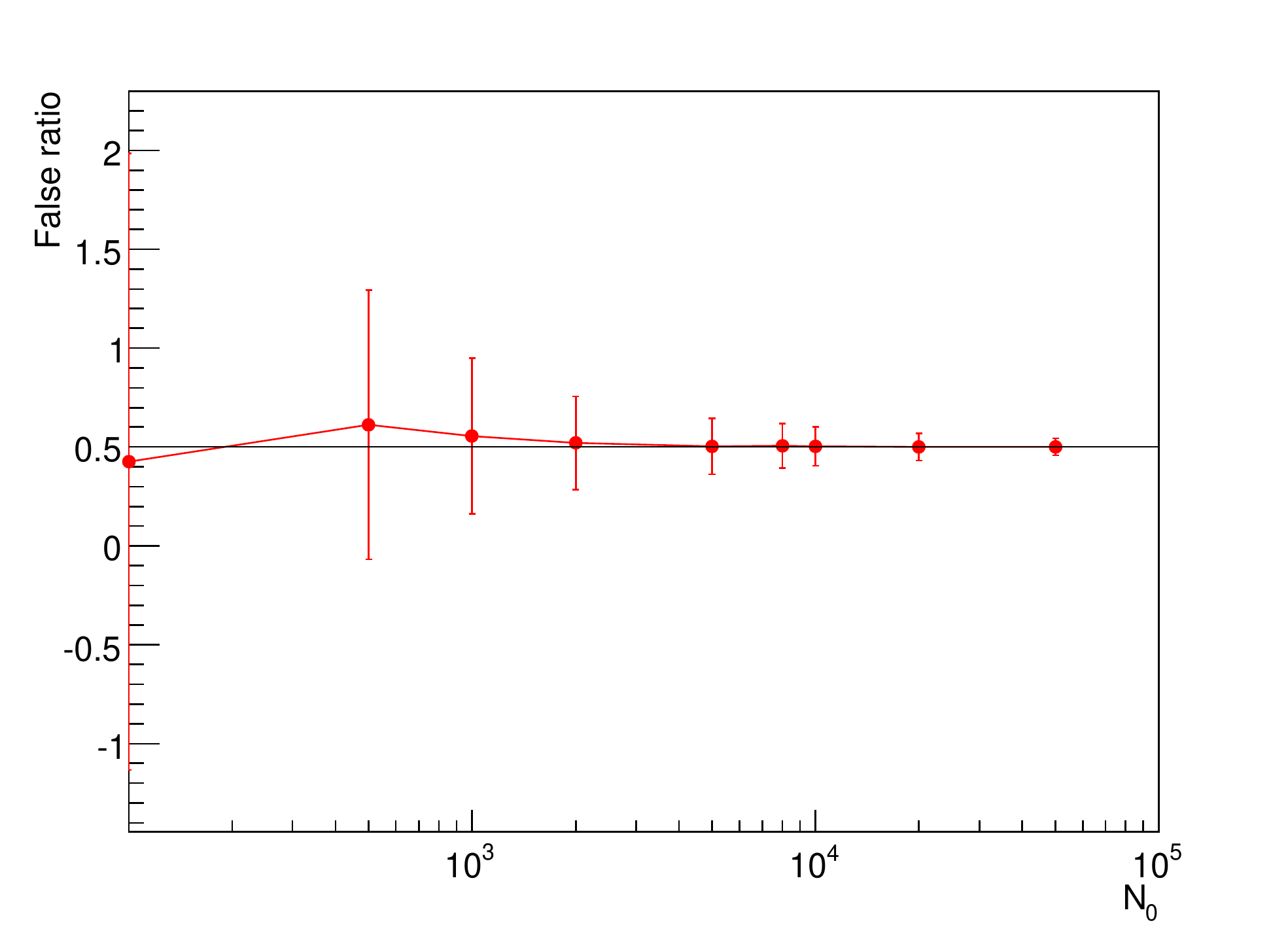}
    \caption{Extracted ratio mean value by weighted-sum method vs. different sample size $N_0$ with false asymmetry $s_1=0$ (left) and $s_1=0.1$ (right). There is no noticeable difference between the two. Upper panel with set polarization $P_y=0.1,P_z=0.1$, lower panel with set polarization $P_y=0.1,P_z=0.2$, showing that the results of the tests do not depend on the value of the set ratio $P_y/P_z$.}
    \label{fig:mean}
  \end{center}
\end{figure}

\begin{figure}[hbt]
  \begin{center}
    \includegraphics[angle=0,width=.45\textwidth]{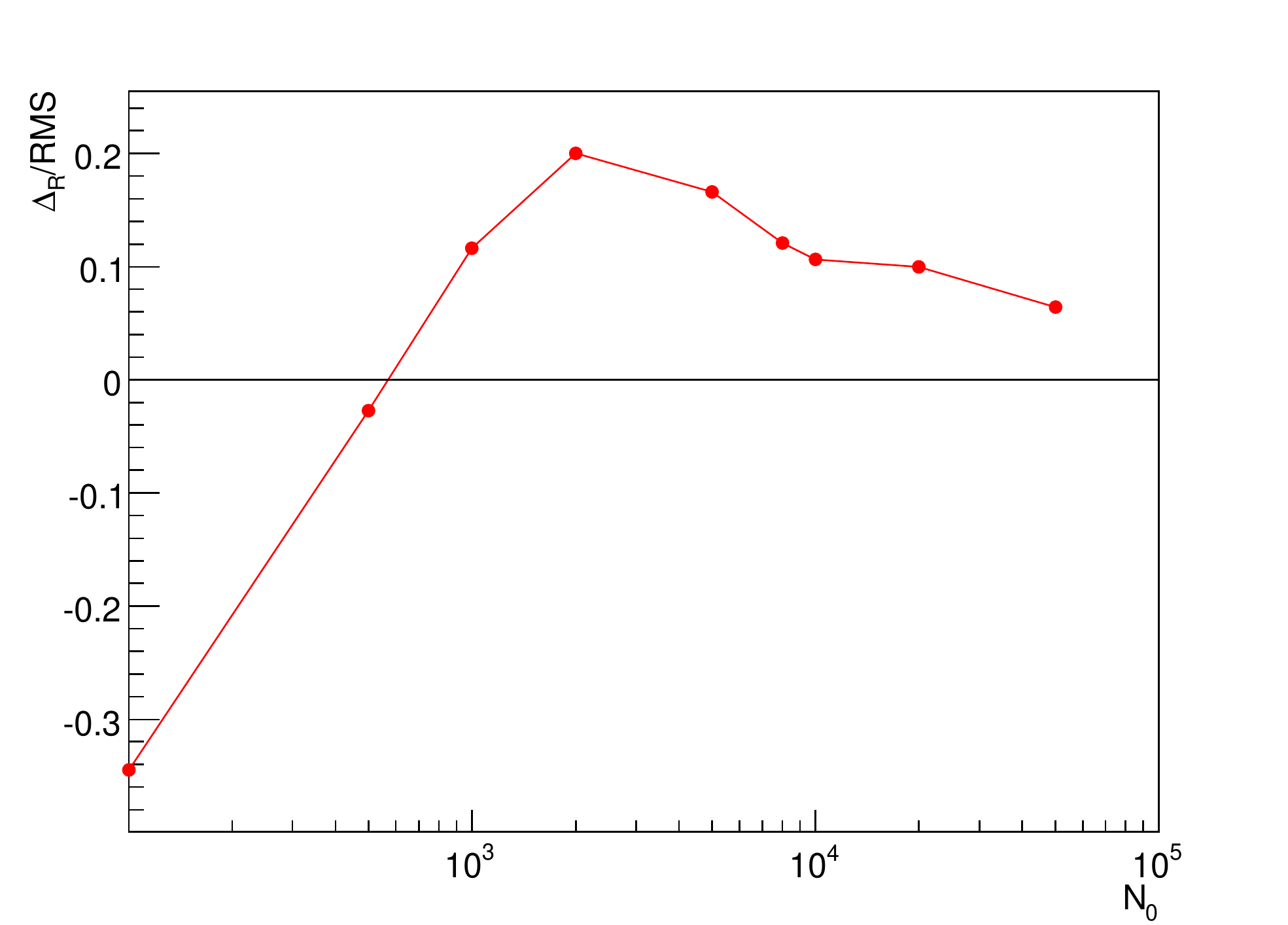}
    \includegraphics[angle=0,width=.45\textwidth]{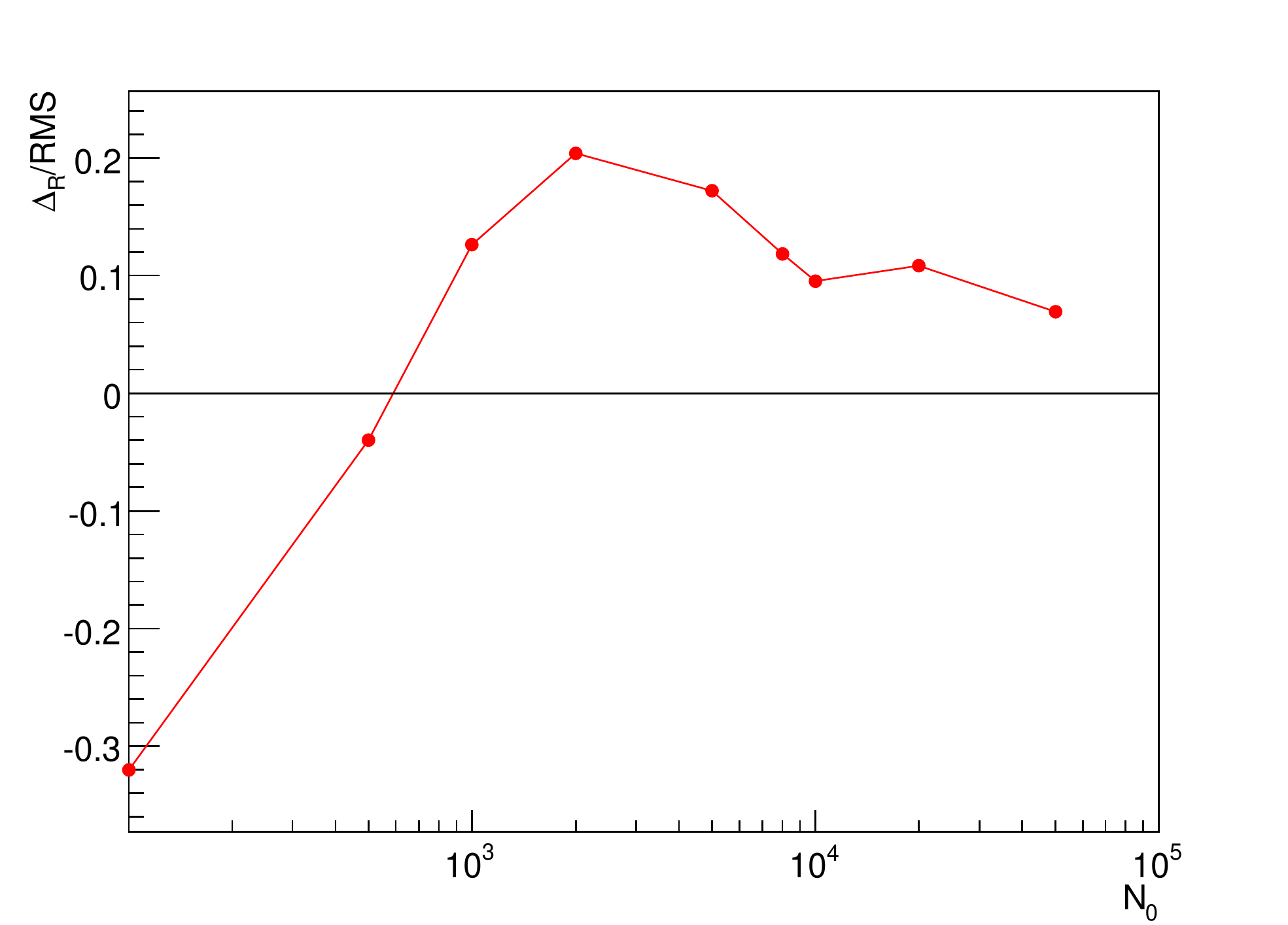}
    \includegraphics[angle=0,width=.45\textwidth]{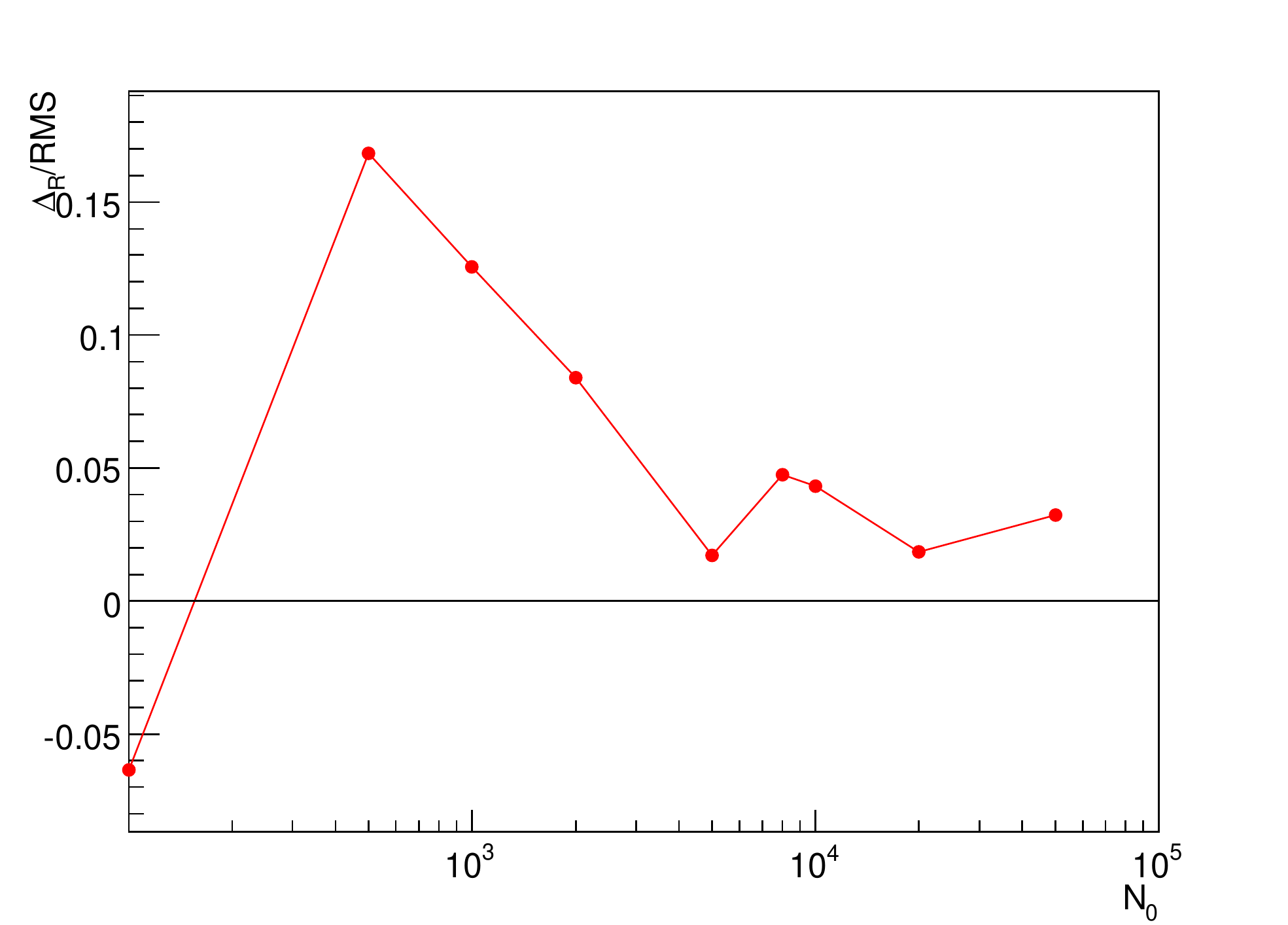}
    \includegraphics[angle=0,width=.45\textwidth]{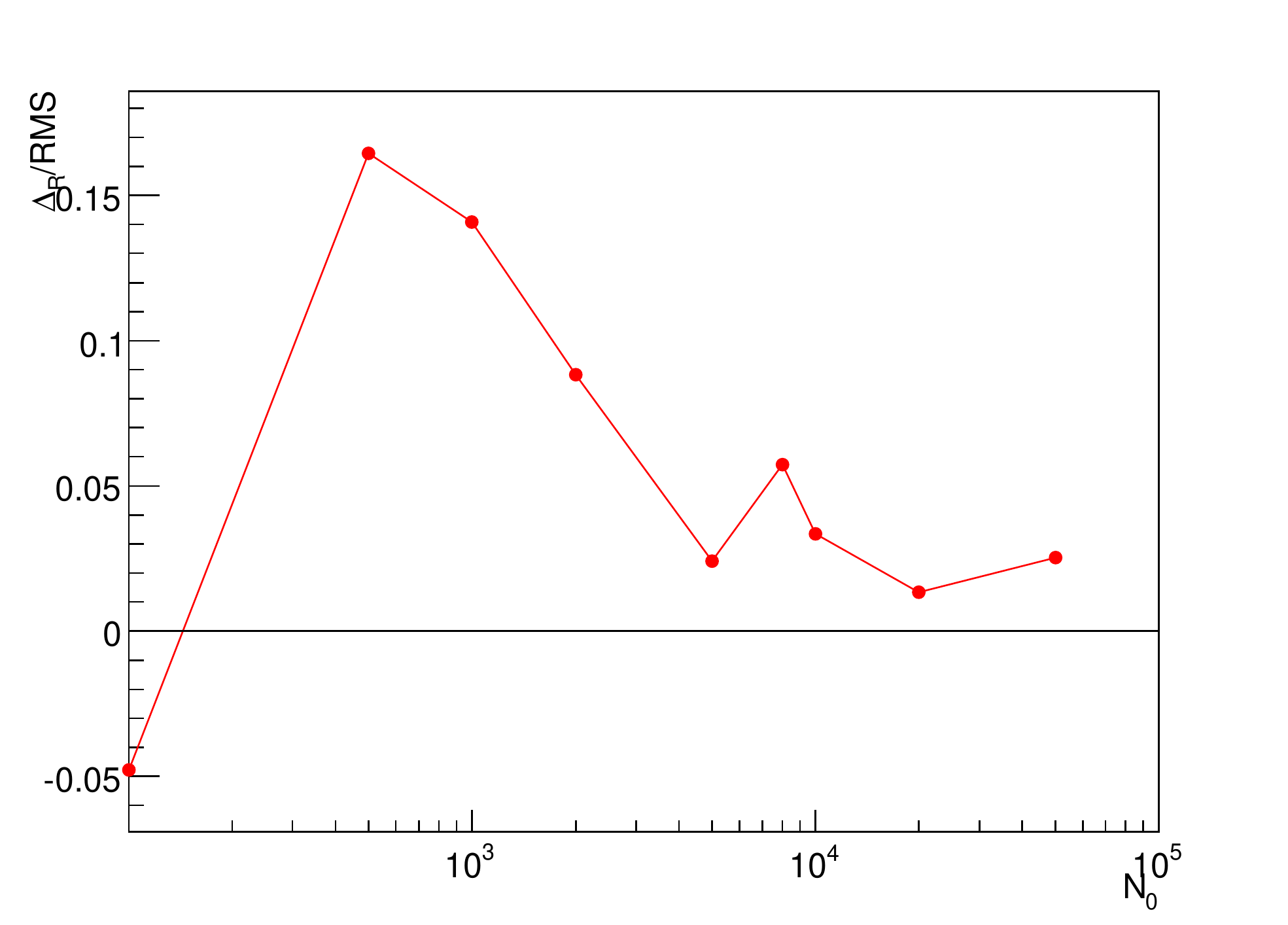}
    \caption{Extracted ratio mean value deviation from the set value divided by the sample standard deviation (RMS) vs. different sample size $N_0$ with false asymmetry $s_1=0$ (left) and $s_1=0.1$ (right). There is no noticeable difference between the two. Upper panel is with set polarization $P_y=0.1,P_z=0.1$, lower panel is with set polarization $P_y=0.1,P_z=0.2$.}
    \label{fig:rms}
  \end{center}
\end{figure}

For simplicity, we use the dipole approximation for the spin transport and assume that there is no induced polarization. Then, for each trial the simulation generates a sample of events by the probability:
\begin{equation}
f^{\pm}(\phi)=\epsilon(\phi)(1\pm P_y\sin\chi\cos\phi \pm P_z\sin\phi)
\label{eq:set}
\end{equation}
Here $P_y,P_z$ represent the transferred polarization components at the target. To be similar to the real case, we choose the spin rotation angle $\chi=90^{\circ}\sim 100^{\circ}$. We set the pseudo efficiency $\epsilon$:
\begin{equation}
\epsilon=1+s_1\sin\phi.
\end{equation}
 False asymmetries with higher order terms ($c_1,s_2,c_2$) were also tested, and the results are similar. With the simulated sample events, we extracted the pseudo ratio $P_y/P_z$ by Eq.~\ref{eq:sums_ap}. We also varied the event sample size $N_0$ used for each trial to test the statistical sensitivity. The extracted ratio distributions are shown in Fig.~\ref{fig:nofa_dis} with 5000 trials per plot, and the sample sizes for each trial $N_0$ is from 100 to 50000. The mean value for the extracted ratio versus the sample size is plotted in Fig.~\ref{fig:mean}, and the deviation from the set value $\Delta_{R}$ divided by the standard deviation of the simulated distribution versus the sample size is plotted in Fig.~\ref{fig:rms}.

\begin{figure}[hbt]
  \begin{center}
    \includegraphics[angle=0,width=.85\textwidth]{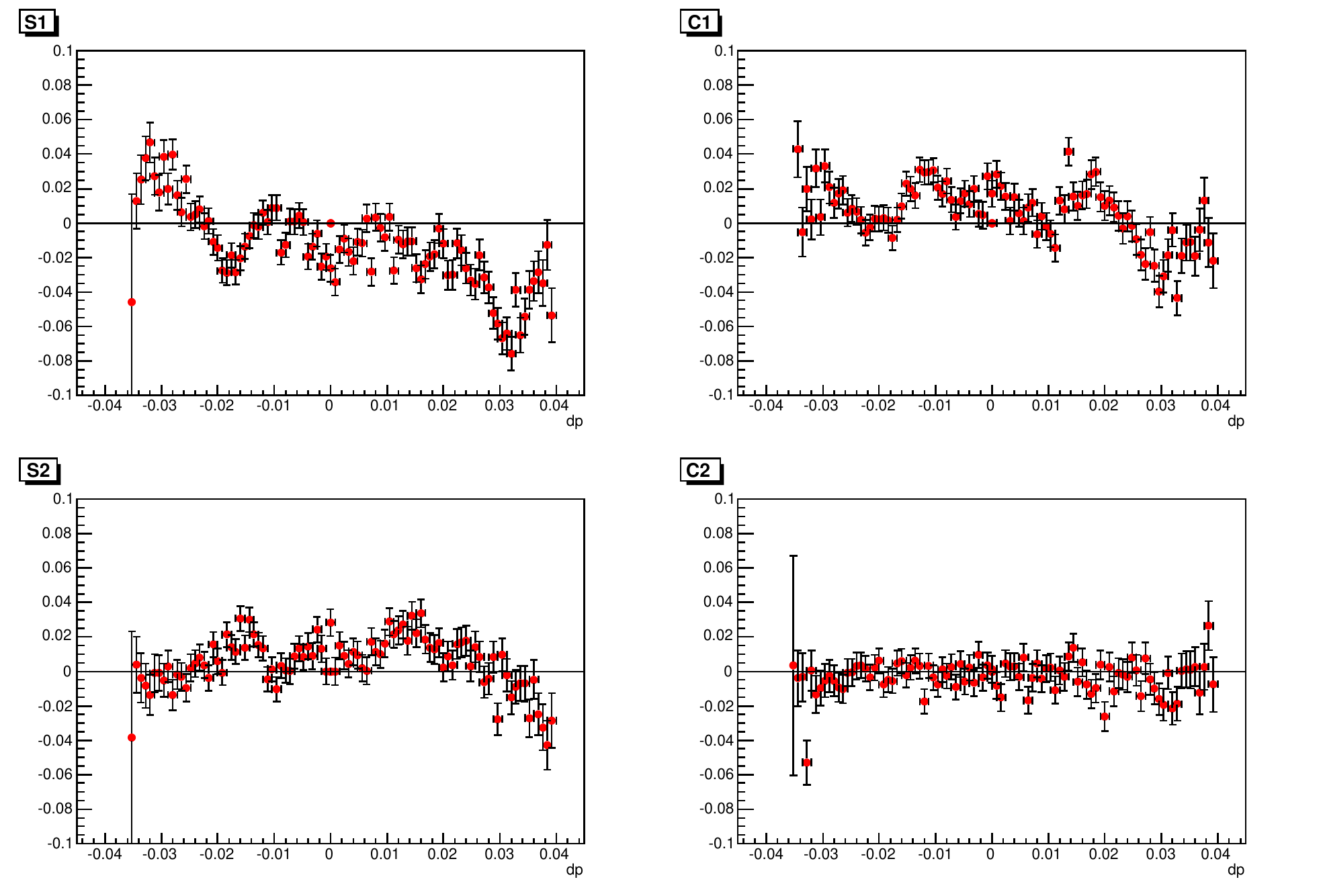}
    \caption{False asymmetry Fourier series coefficients vs. $\delta_p$ for kinematics K6 $\delta_p=2\%$.}
    \label{fig:fa}
  \end{center}
\end{figure}

  Results from the simulation with two different set ratios $P_y/P_z=0.5$ and $P_y/P_z=1$ are shown in Fig.~\ref{fig:mean} and Fig.~\ref{fig:rms}. From these results we can see that the weighted-sum method can extract the ratio without any problem even with a significant size of the false asymmetry. The comparison shown here is between $s_1=0$ and $s_1=0.1$. The real false asymmetry is about a few percent level as shown in Fig.~\ref{fig:fa}. We also notice that when the statistics are low, the distribution is not symmetric as shown in Fig.~\ref{fig:nofa_dis}, and the deviation of the mean value is due to the cutoff of the histogram. The deviation decreases rapidly as the statistics increases and becomes unnoticeable when $N_0>50000$, and there is no noticeable difference with and without the false asymmetry.

\begin{figure}[hbt]
  \begin{center}
    \includegraphics[angle=0,width=.50\textwidth]{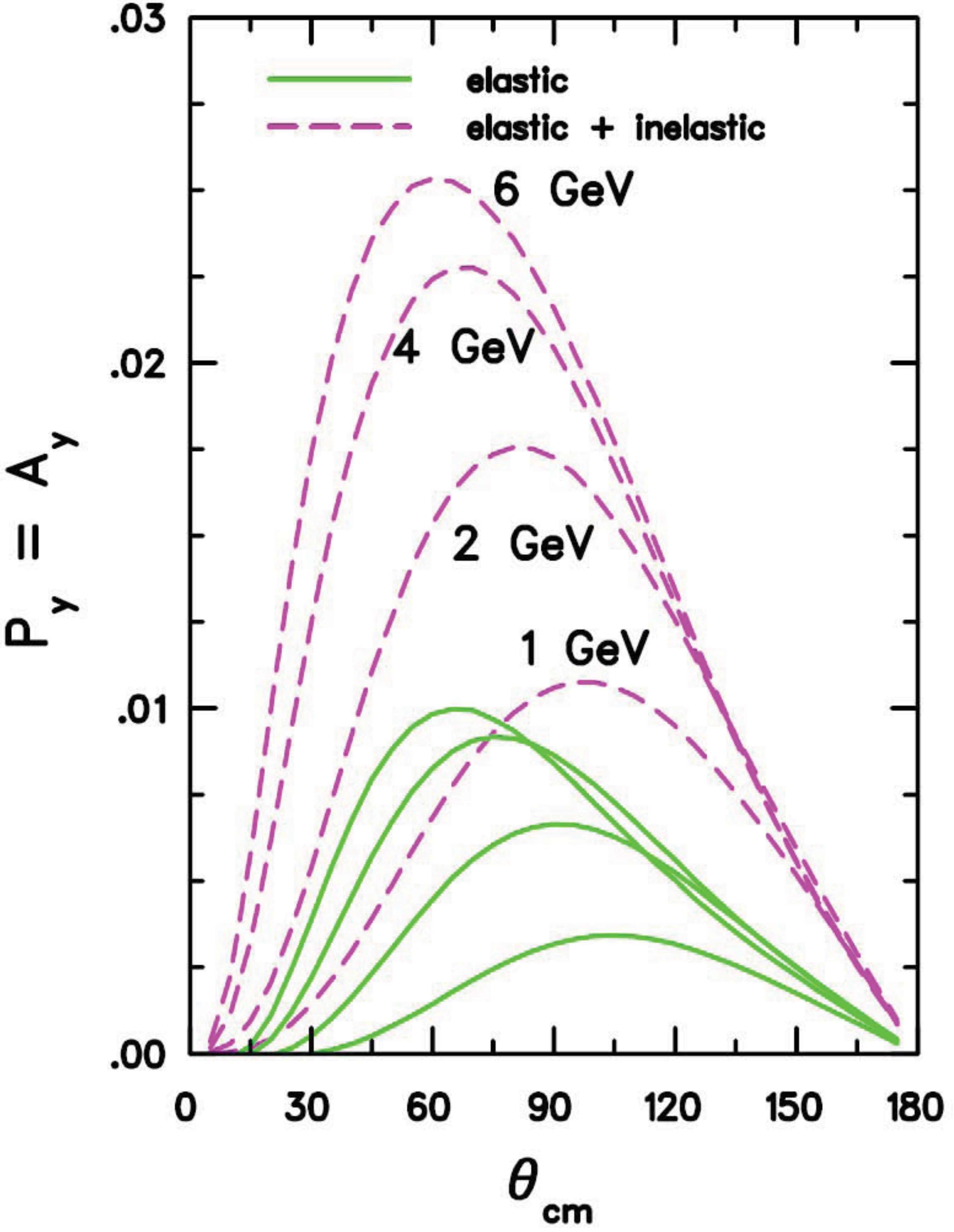}
    \caption{Proton induced polarization component, as a function of the electron $\theta_{cm}$ scattering angle for different beam energies. The dash (solid) line shows the total (elastic only) 2$\gamma$ exchange effect. The y-axis $P_y$ is actually $P_x^{tg}$ for the convention used here. }
    \label{fig:py}
  \end{center}
\end{figure}
From the results presented above, the simulation confirmed the results in~\cite{Besset} and the formalism of weighted-sum works well with our statistics. The only tiny flaw is that we assumed there is no induced polarization ($P_x^{tg}=0$). The problem with non-zero induced polarization is that we cannot exactly construct an acceptance with symmetry period of $\pi$ in $\phi$ as mentioned earlier, so Eq.~\ref{eq:int} is not exactly true. However, through simulation, we can give an estimate with a small amount of induced polarization $P_x^{tg}$ as predicted by~\cite{py_afan}. The predicted induced polarization is shown in Fig.~\ref{fig:py}. The electron scattering angle $\theta_{cm}$ for each kinematic setting is listed in Table~\ref{tab:angle_cm}.
\begin{table}[t]
\caption{Electron scattering angle $\theta_{cm}$ for each kinematics ($\delta_p =0 \% $).}
\begin{center}
\begin{tabular}{c c c}
\hline
Kinematics & $Q^2$ [$(\mathrm{GeV}/c)^2$]&  $\theta_{cm}$ [deg]\\
\hline
K1 & 0.35 & 55.5 \\
K2 & 0.30 & 50.9 \\
K3 & 0.45 & 63.9 \\
K4 & 0.40 & 60.2 \\
K5 & 0.55 & 71.5 \\
K6 & 0.50 & 67.7 \\
K7 & 0.60 & 75.4 \\
K8 & 0.70 & 82.5 \\
\hline
\label{tab:angle_cm}
\end{tabular}
\end{center}
\end{table}
To simulate the case with non-zero induced polarization, the set probability at the focal plane Eq.~\ref{eq:set} becomes:
\begin{equation}
f^{\pm}(\phi)=\epsilon(\phi)(1\pm P_y\sin\chi\cos\phi \pm P_z\sin\phi+P_0\cos\phi+P_1\sin\phi),
\label{eq:set1}
\end{equation}
where $P_0,P_1$ represent the polarization components raised from the induced polarization at the focal plane. To test the extreme case, we set them to be comparable to the physics asymmetry $P_0,P_1=0.2$ which is much larger than predicted, and with false asymmetry $s_1=0.1$ which is the same level as the real case. Different combinations of $P_0,P_1$ were tested and corresponding results are listed in Table~\ref{tab:combo}.
\begin{table}[t]
\caption{Deviation from the set value $\Delta_R$ with different combinations of $P_0$ and $P_1$. The set transferred polarization is $P_y=P_z=0.1$. Simulation with sample size $N_0=10^5$ and number of trial $N_{trial}=10^4$. The standard deviation for extracted values is $\sim 0.075$.}
\begin{center}
\begin{tabular}{c |c c}
\hline
   & $ P_0=0$ & $ P_0=0.2 $ \\
\hline
$P_1=0$   & 0.0015 & 0.002   \\
$P_1=0.2$ & 0.012 & 0.012    \\
\hline
\end{tabular}
\end{center}
\label{tab:combo}
\end{table}

\begin{figure}[hbt]
  \begin{center}
    \includegraphics[angle=0,width=.40\textwidth]{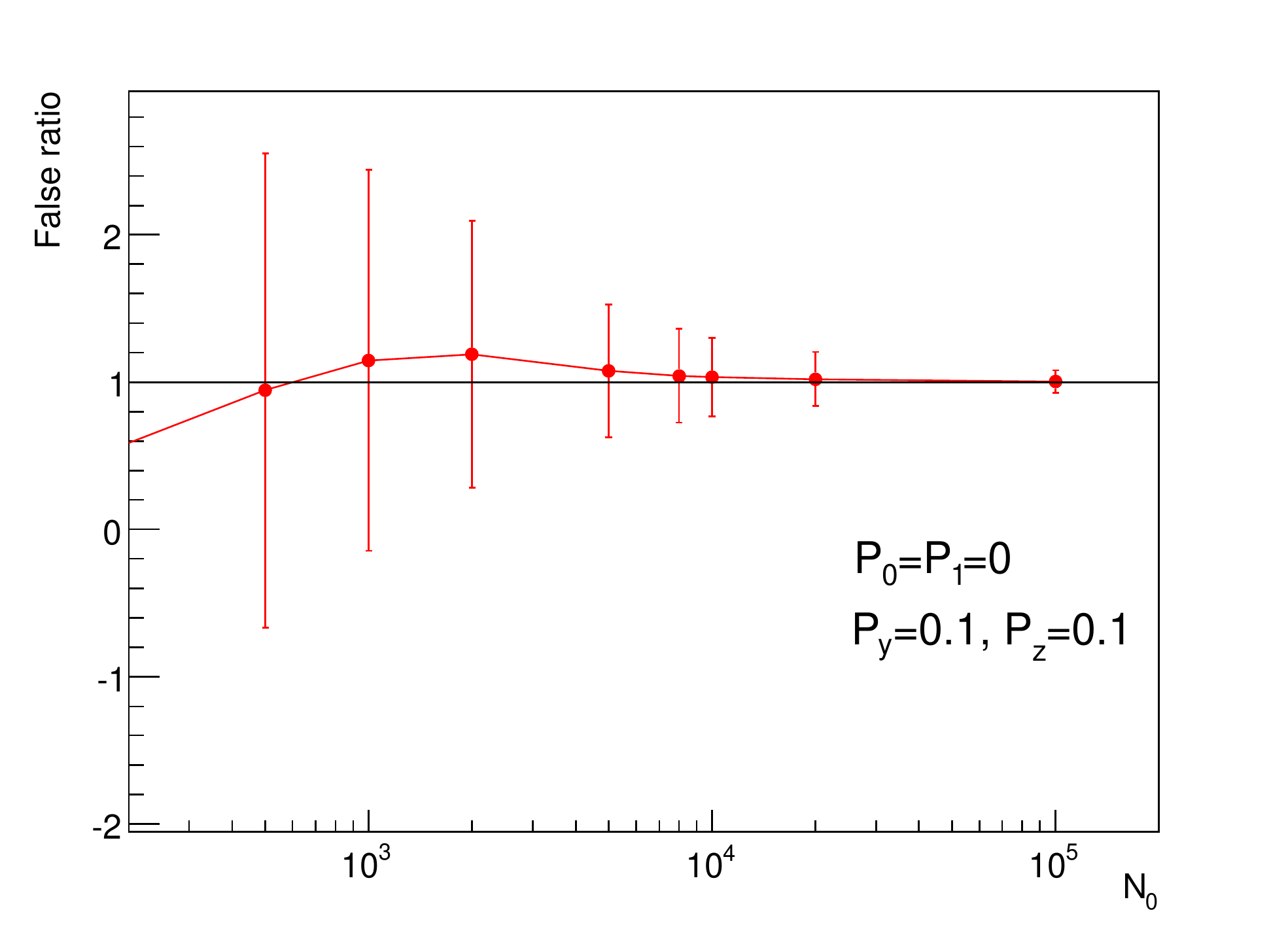}
    \includegraphics[angle=0,width=.40\textwidth]{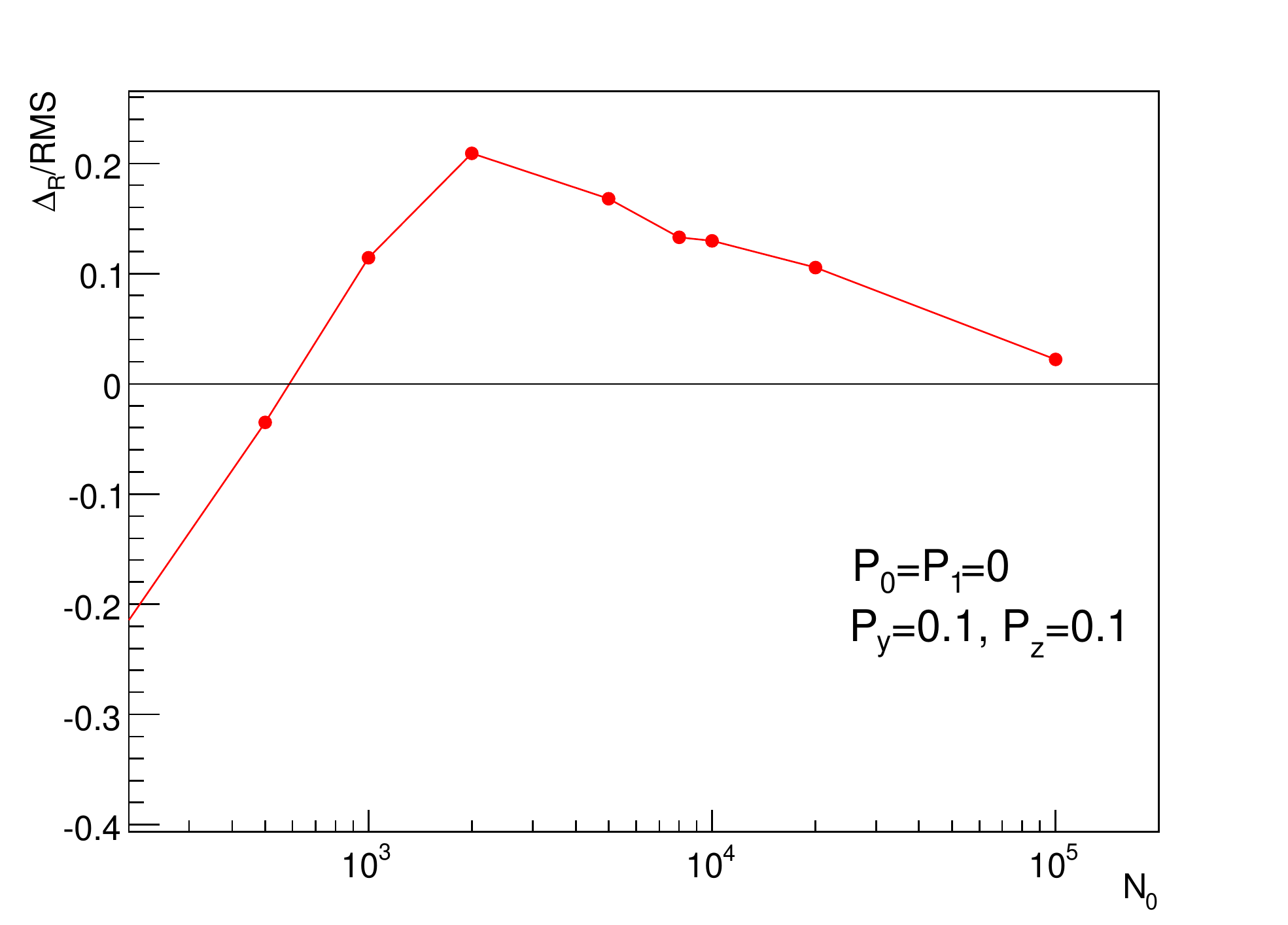}
    \includegraphics[angle=0,width=.40\textwidth]{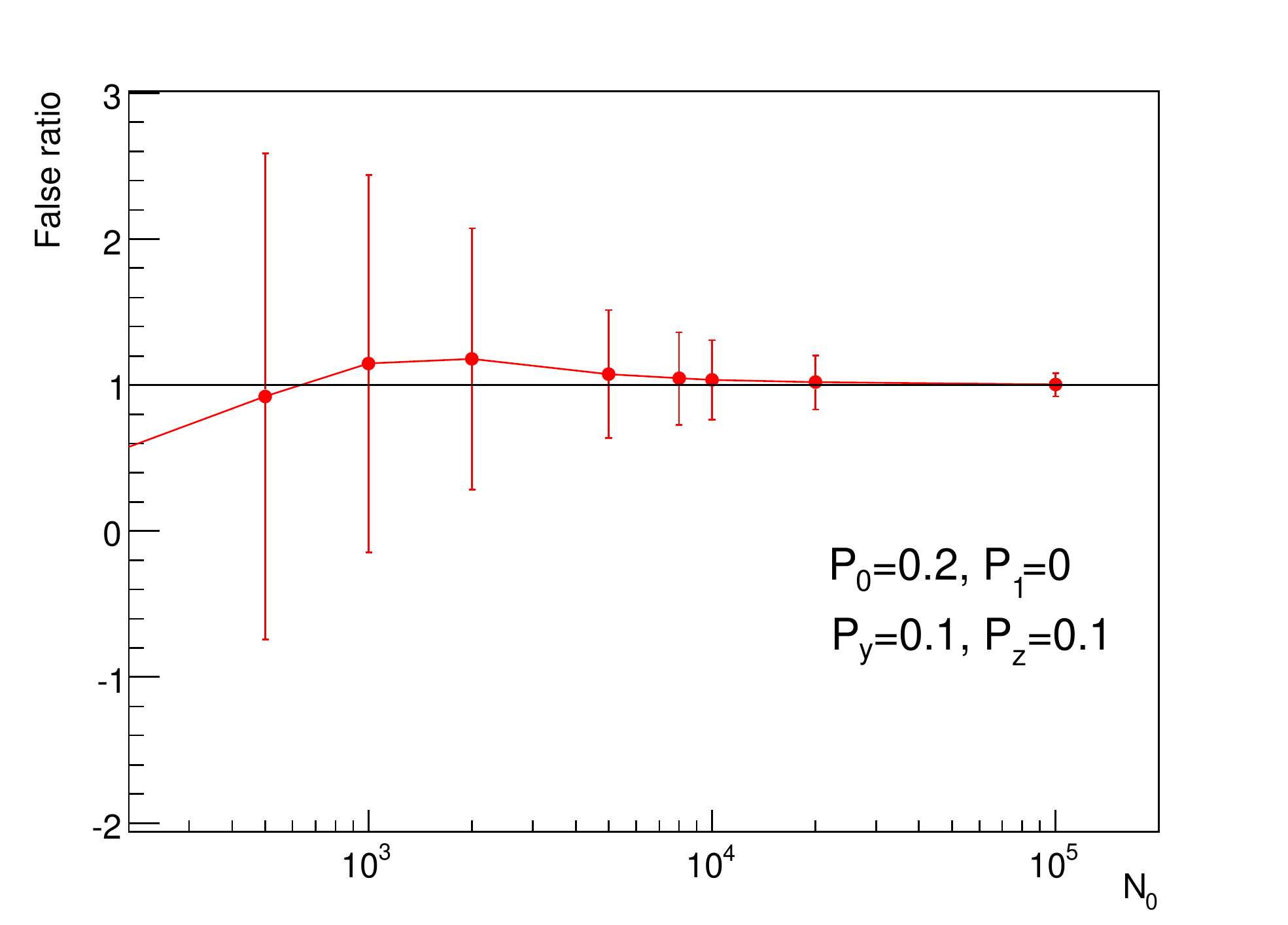}
    \includegraphics[angle=0,width=.40\textwidth]{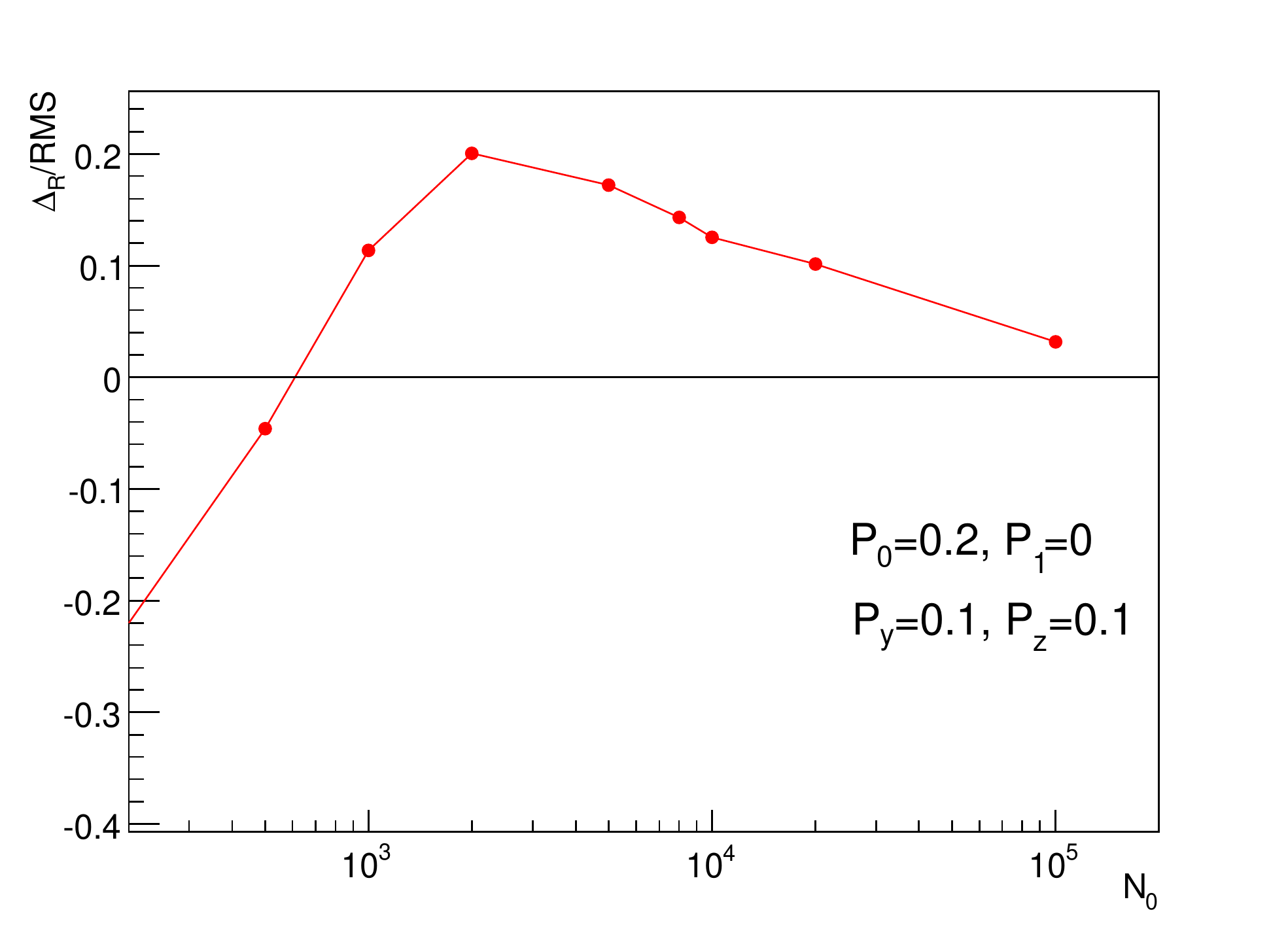}
    \includegraphics[angle=0,width=.40\textwidth]{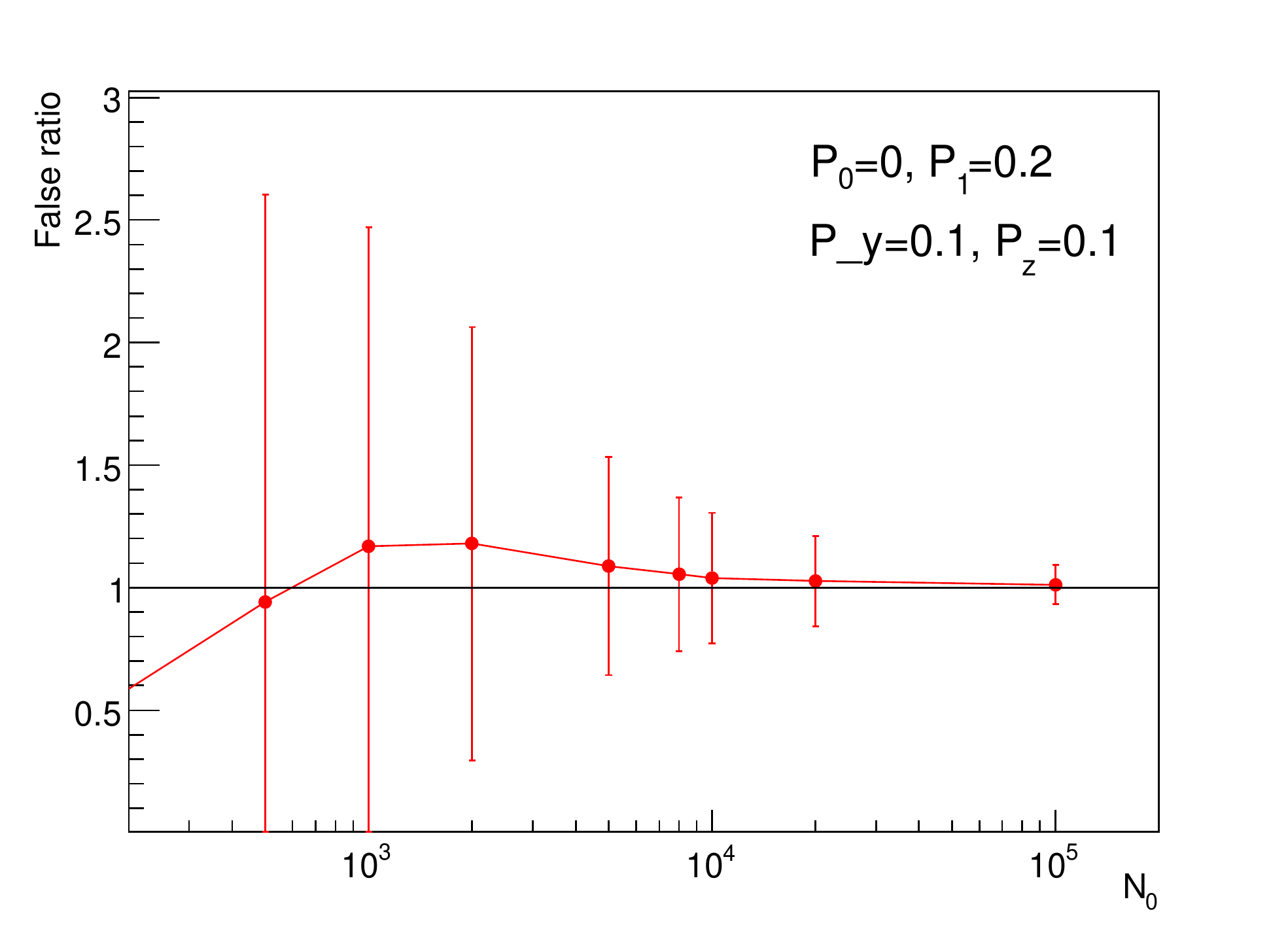}
    \includegraphics[angle=0,width=.40\textwidth]{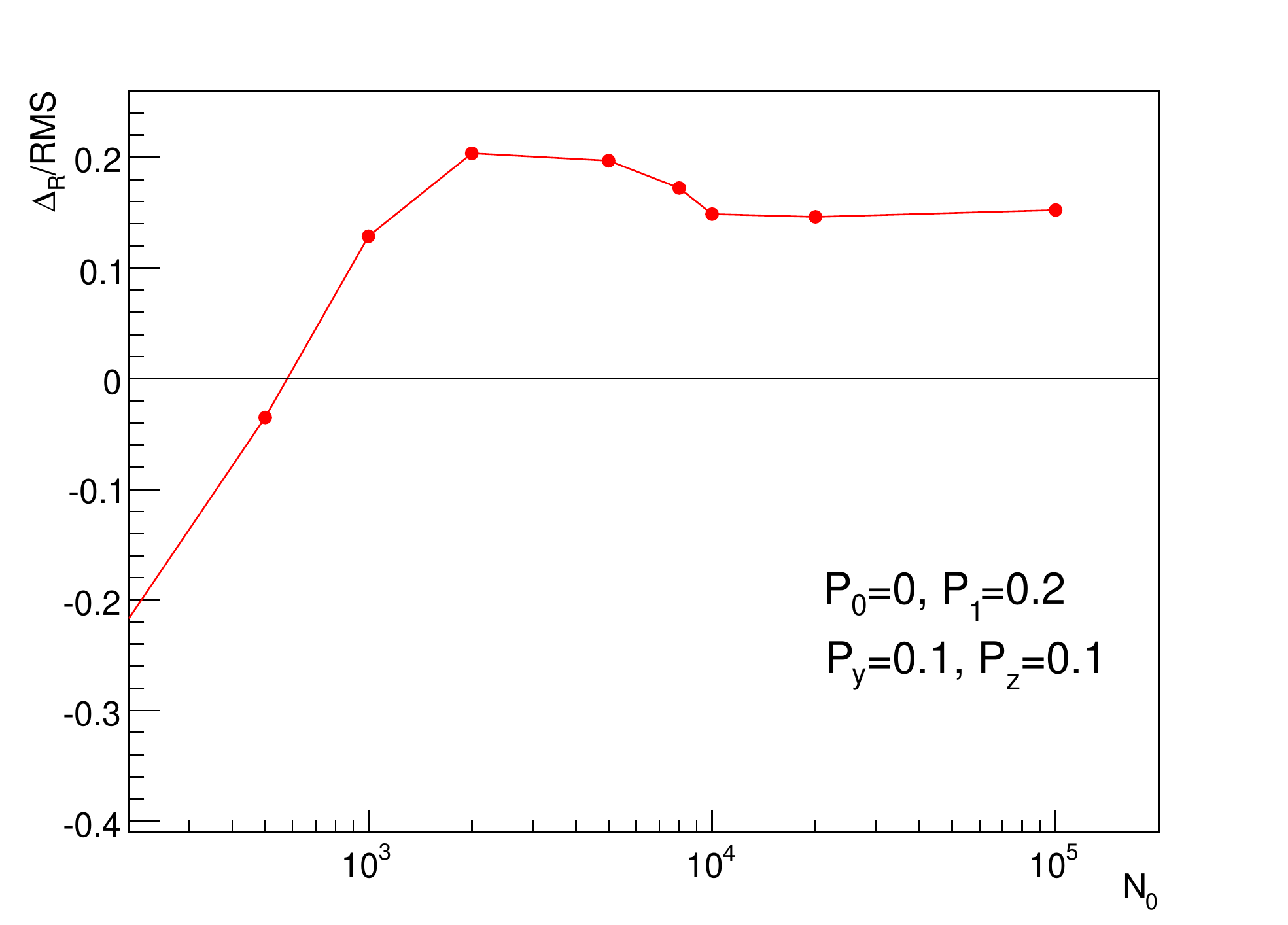}
    \includegraphics[angle=0,width=.40\textwidth]{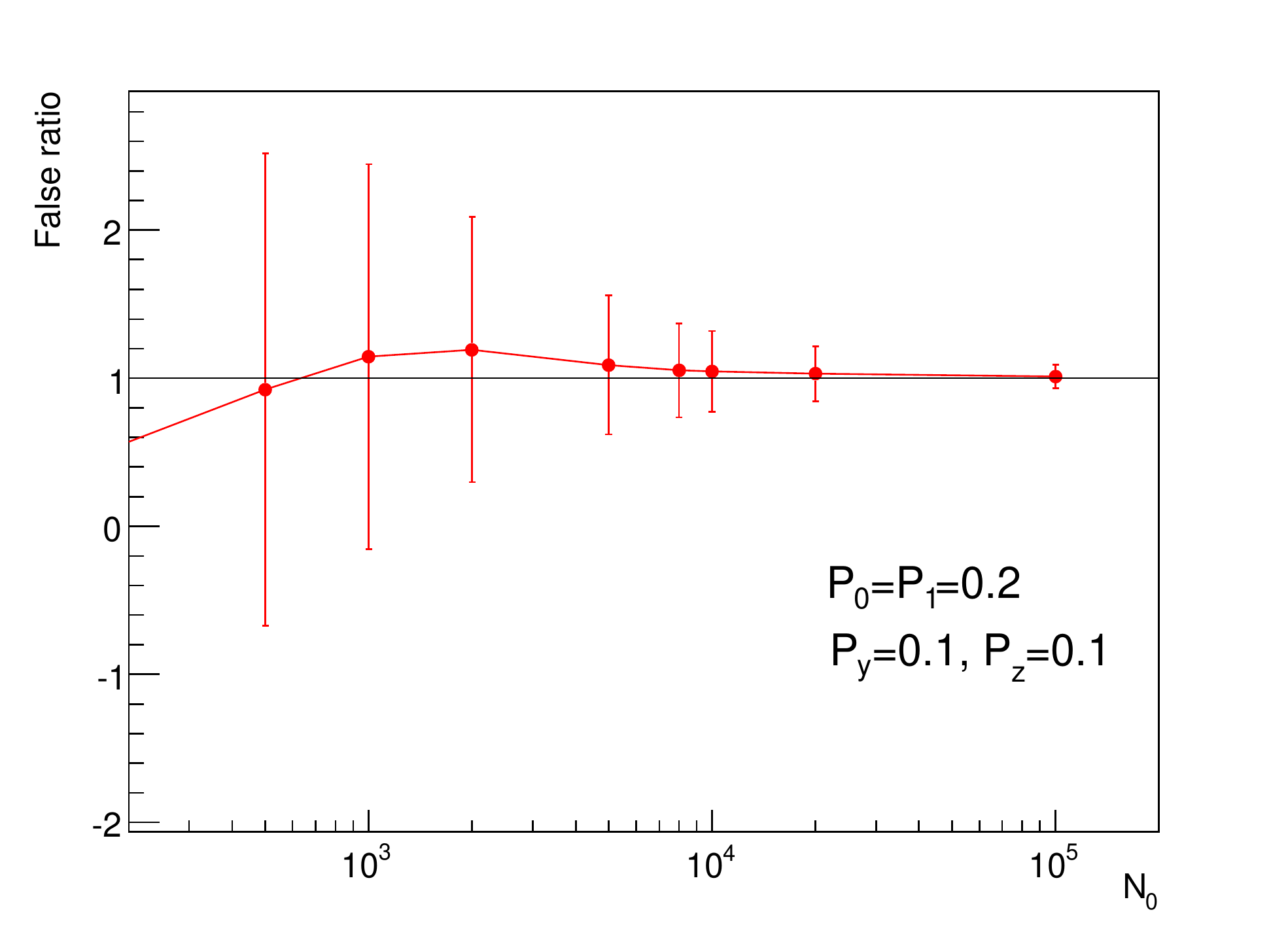}
    \includegraphics[angle=0,width=.40\textwidth]{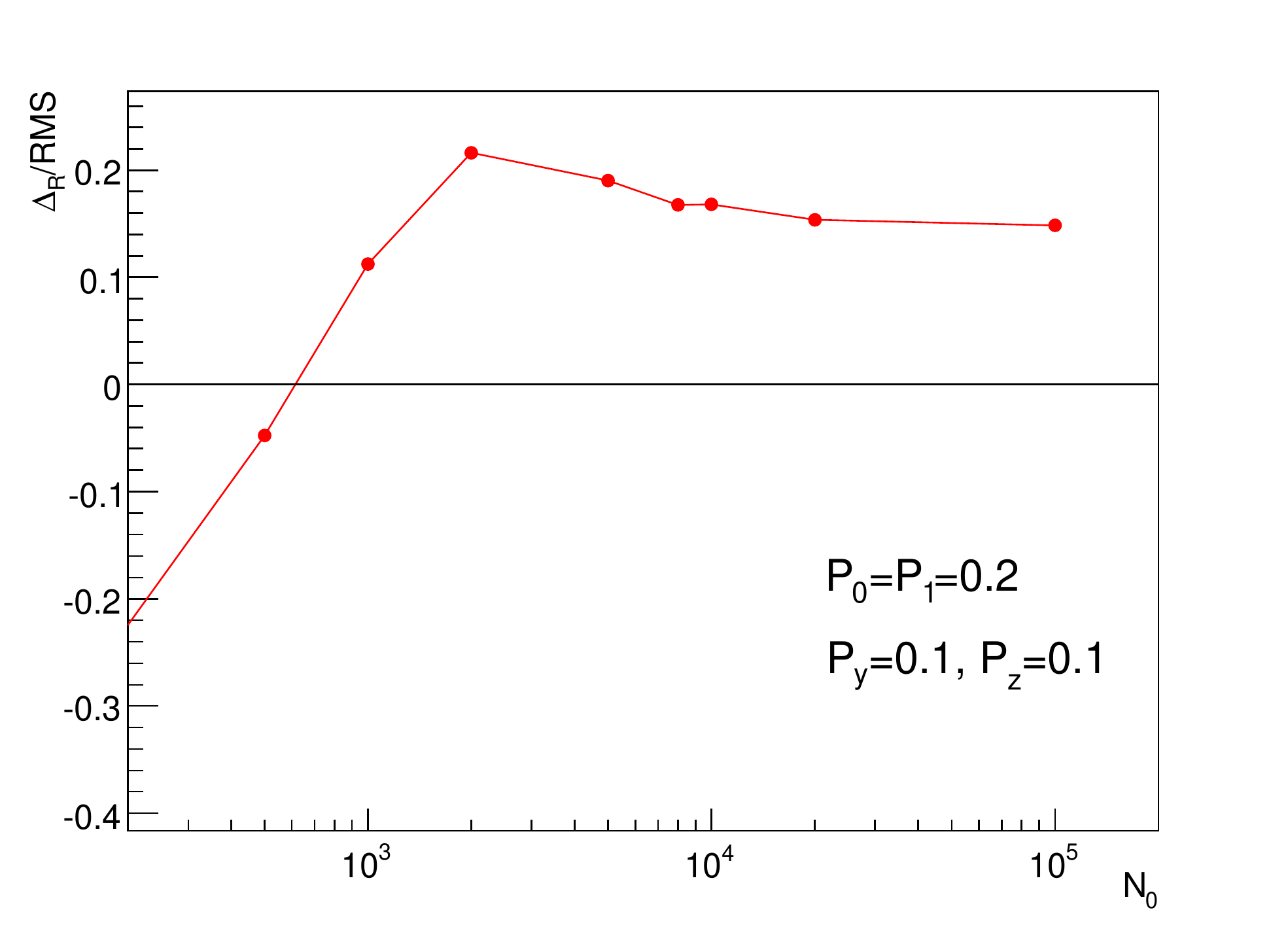}
    \caption{Extracted ratio mean value and relative deviation vs. different sample size $N_0$ with false asymmetry $s_1=0.1$, and different combinations of set polarization $P_0,P_1$.}
    \label{fig:py1}
  \end{center}
\end{figure}
\begin{figure}[hbt]
  \begin{center}
    \includegraphics[angle=0,width=.45\textwidth]{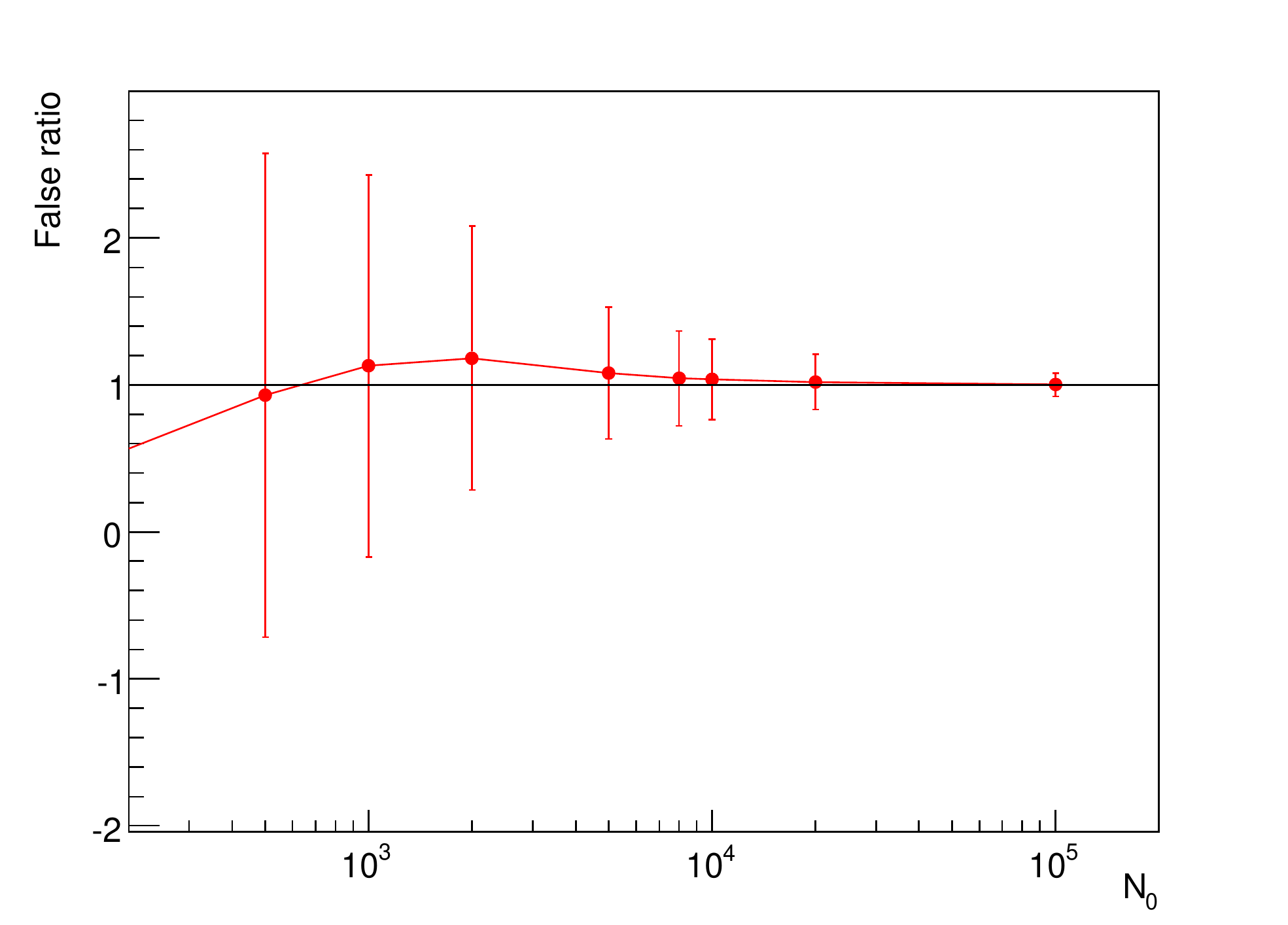}
    \includegraphics[angle=0,width=.45\textwidth]{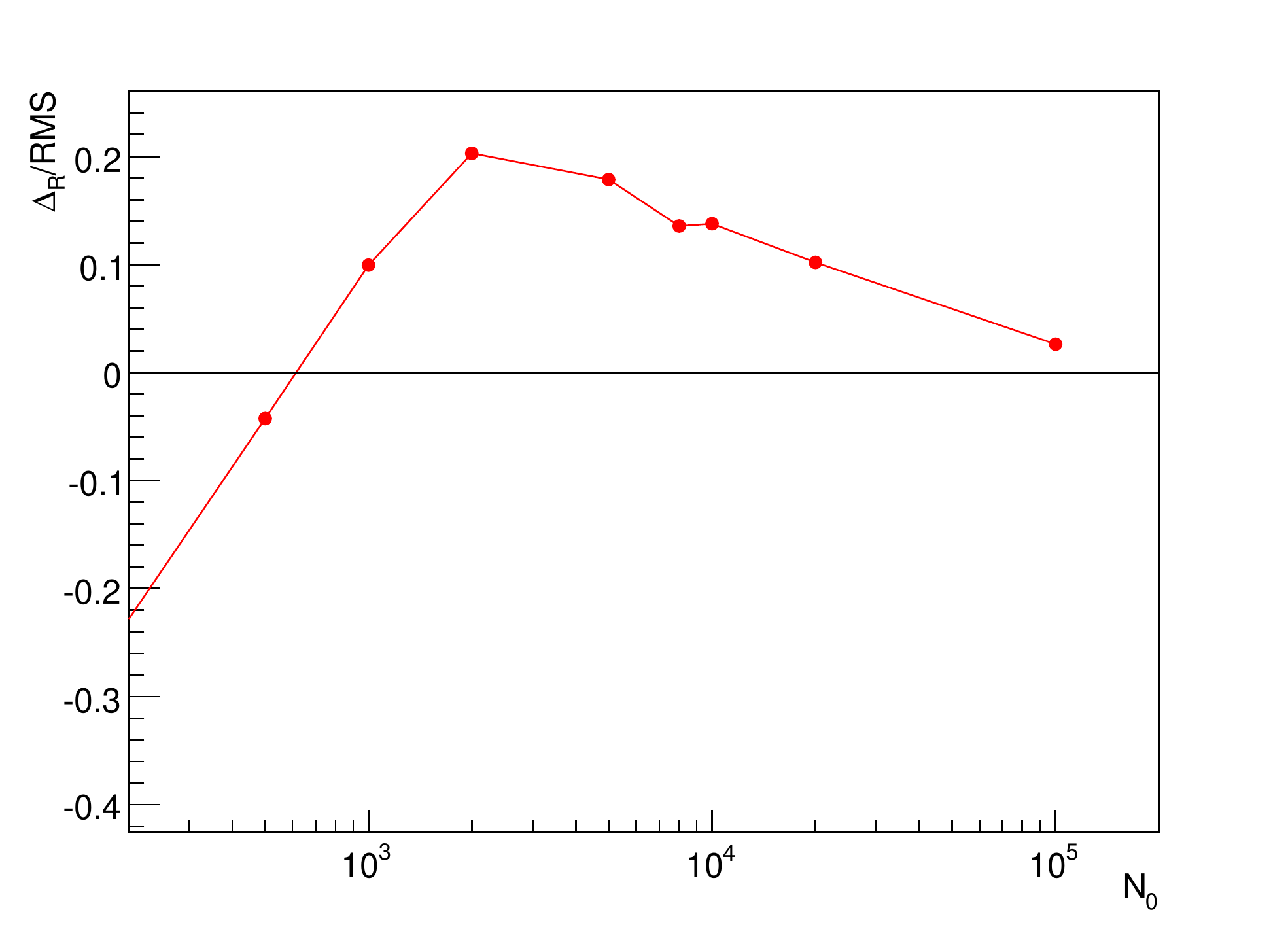}
    \caption{Extracted ratio mean value and relative offset from the set value vs. different sample size $N_0$ with difference false asymmetries: $s_1=0.08,c_1=0.05,s_2=0.05,c_2=0.01$, and set polarizations: $P_0=P_1=0.01,P_y=P_z=0.1$, respectively.}
    \label{fig:real}
  \end{center}
\end{figure}

The simulations for set polarization $P_y=0.1, P_z=0.1$ are shown in Fig.~\ref{fig:py1}. The results show that the deviation is much less within one standard deviation. For the real case, the induced polarization at the target $P_x^{tg}\sim 10^{-3}$\footnote{By considering the spin rotation matrix elements $S_{yx},S_{xx}\ll 1$, the contribution from $P_0$ at the focal plane is actually $\ll 10^{-3}$}. To estimate the effect close to the real case, we used the similar size of $P_0$, and $P_1$ as predicted, and also with the ``full'' false asymmetry:
\begin{equation}
\epsilon=1+s_1\sin\phi+c_1\cos\phi+s_2\sin2\phi+c_2\cos2\phi
\end{equation}
where $s_1=0.08,c_1=0.05,s_2=0.05,c_2=0.01$ which is assigned according to the maximum of their real sizes in this experiment. We assume that $P_0=P_1=0.01$ which is very conservative compared to the real case ($10^{-3}$) after considering the spin rotation. The simulation results are shown in Fig.~\ref{fig:real}.

From the simulation (Fig.~\ref{fig:real}) we can see that the deviation of the ratio $\Delta_R$ is $\sim 0.002$, and since the asymmetries $P_0,P_1$ are even smaller and the statistics are much better in the real case, we do not expect any noticeable effect from the induced polarization.

\section{Summary}
Through this study, we have confirmed the results in~\cite{Besset}. The approximations and assumptions used were carefully examined. From the simulation, we have confirmed that the weighted-sum method is valid and false asymmetry plays a negligible role in extracting the transferred polarization and thus the form factor ratio.

\clearpage
\newpage

%% file: app3.tex
\chapter{$pC$ Analyzing Power Parameterizations}
The carbon analyzing power $A_y$ was extracted for this measurement. We applied two-dimensional binning in the analysis to extract the dependence on $\theta_{fpp}$ and $T_p$. The mean values of the two variables of each bin were used for the fit. The $\theta_{fpp}$ binning is the same for all the kinematics settings as listed in Table~\ref{tab:bin_theta}, and the $T_p$ binning is summarized in Table~\ref{tab:bin_tp}.
\begin{table}[t]
\caption{Binning on $\theta_{fpp}$.}
\begin{center}
\begin{tabular}{c c c}
\hline
Bin & $\theta_{low}$ [deg] &  $\theta_{high}$ [deg]\\
\hline
1 & 4 & 5  \\
2 & 5 & 6  \\
3 & 6 & 8  \\
4 & 8 & 10  \\
5 & 10 & 12 \\
6 & 12 & 15 \\
7 & 15 & 18 \\
8 & 18 & 21 \\
9 & 21 & 24 \\
10& 24 & 28 \\
11& 28 & 36 \\
\hline
\label{tab:bin_theta}
\end{tabular}
\end{center}
\end{table}
\begin{table}[t]
\caption{Binning on $T_p$.}
\begin{center}
\begin{tabular}{c c c c c}
\hline
Kinematics & carbon thickness [inch] &$T_{low}$ [MeV] &  $T_{high}$ [MeV] & Bin size [MeV] \\
\hline
K1 & 3 & 120 & 150 & 10 \\
K2 & 3 & 90 & 120 & 10\\
K3 & 3.75 & 160 & 220 & 20\\
K4 & 3.75 & 140 & 200 & 20\\
K5 & 3.75 & 220 & 280 & 20\\
K6 & 3.75 & 200 & 260 & 20\\
K7 & 3.75 & 240 & 300 & 20\\
K8 & 3.75 & 300 & 360 & 20\\
\hline
\label{tab:bin_tp}
\end{tabular}
\end{center}
\end{table}
The ``low energy'' McNaughton, the LEDEX and the new parameterizations are summarized in Table~\ref{tab:aypara}.
\begin{table}[t]
\caption{Coefficients of different parameterizations for the $pC$ analyzing power $A_y$. The reduced $\chi^2$ of the new fit is 0.74 with a $\chi^2$ of 272.5 and 368 degrees of freedom.}
\begin{center}
\begin{tabular}{c  c  c  c }
\hline
 & LEDEX & McNaughton (low) &  New \\
\hline\hline
Energy Range & 82 $\sim$ 127 MeV & 95 $\sim$ 483 MeV & 90 $\sim$ 360 MeV\\
\hline
$p_0$ & 0.55 & 0.70 & 0.55\\
$a_0$ & 4.0441 & 5.3346 & 5.92823\\
$a_1$ & 19.313 & -5.561 & 24.8291\\
$a_2$ & 119.27 & 2.8353 & -130.046\\
$a_3$ & 439.75 & 61.915 & -111.329\\
$a_4$ & 9644.7 & -145.54 & 834.988\\
$b_0$ & 6.4212 & -12.774 & 34.8843\\
$b_1$ & 111.99 & -68.339 & 28.6809\\
$b_2$ & -5847.9 & 1333.5 & -2207.81\\
$b_3$ & -21750 & -3713.5 & 6089.94\\
$b_4$ & 973130 & 3738.3 & -595.011\\
$c_0$ & 42.741 & 1095.3 & -776.587\\
$c_1$ & -8639.4 & 949.50 & 102.862\\
$c_2$ & 87129 & -28012.0 & 125407\\
$c_3$ & 8.1359$\times 10^5$ & 96833.0 & -530126\\
$c_4$ & -2.1720$\times 10^7$& -118830.0 & 595619\\
$d_0$ & 5826.0 & & 23845.1\\
$d_1$ & 2.4701$\times 10^5$& & 1.16981$\times 10^5$ \\
$d_2$ & 3.3768$\times 10^6$& & -1.99475$\times 10^6$\\
$d_3$ & -1.1201$\times 10^7$& & 5.8203$\times 10^6$\\
$d_4$ & -1.9356$\times 10^7$&  & -4.41281$\times 10^6$\\
\hline
\label{tab:aypara}
\end{tabular}
\end{center}
\end{table}

\clearpage
\newpage

%% file: app4.tex
\chapter{Neutral Pion Photoproduction Estimation}
\section{Introduction}
 For experiment E08007, we measured the recoil proton polarization in the elastic
 reaction $^1$H$(\vec e,e'\vec p$). For the production data taking we required
 a coincidence between a proton detected in the left HRS and a signal in a limited
 set of BigBite shower blocks which a coincident elastic $ep$ electron would be
 expected to hit. Since the particle identification is limited in the BigBite shower counter
 due to the configuration, it may allow
 contamination by background reactions. Therefore, in the data analysis, an elastic cut was
 applied to the proton kinematics (angle vs. momentum).

 This study is to investigate whether there is a significant contribution from pion
 photoproduction $\gamma+p\to p+\pi^0$ with the current event selection. In this work, we do not consider possible backgrounds from virtual Compton scattering which are expected to be much smaller than the backgrounds from pion production. Since the goal is to give an estimate for the order of magnitude, some approximations were applied to simplify the simulation. Based on this study we will see if a full simulation is needed.

For the pion photoproduction estimation, it includes the following inputs:
\begin{itemize}
\item phase space simulation for both $e+p\to e+p$ and $\gamma + p\to p +\pi^0$.
\item real photon flux estimation.
\item BigBite calorimeter acceptance.
\item elastic cross section, pion photoproduction cross section and polarization observable from the world database and calculations.
\end{itemize}

 We took the lowest momentum kinematics setting K2\footnote{The spectrometer resolution becomes worse with lower proton momentum, hence, more difficult to separate the pion background via the elastic cut.} as an example. The procedure described below was applied to every kinematics, and the results are reported in the end.
\section{Phase Space Simulation}
\begin{figure}[hbt]
  \begin{center}
    \includegraphics[angle=0,width=.48\textwidth]{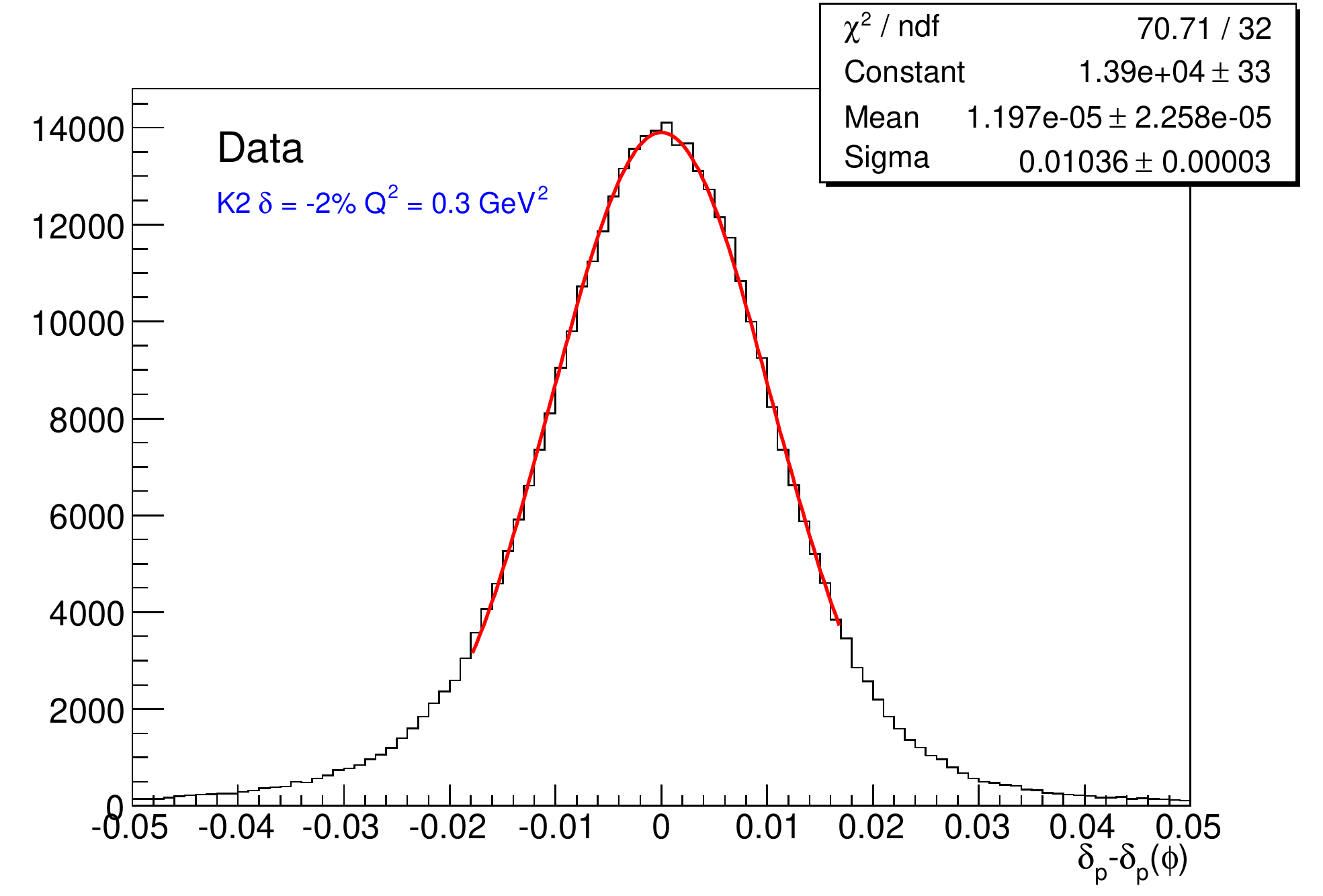}
    \includegraphics[angle=0,width=.48\textwidth]{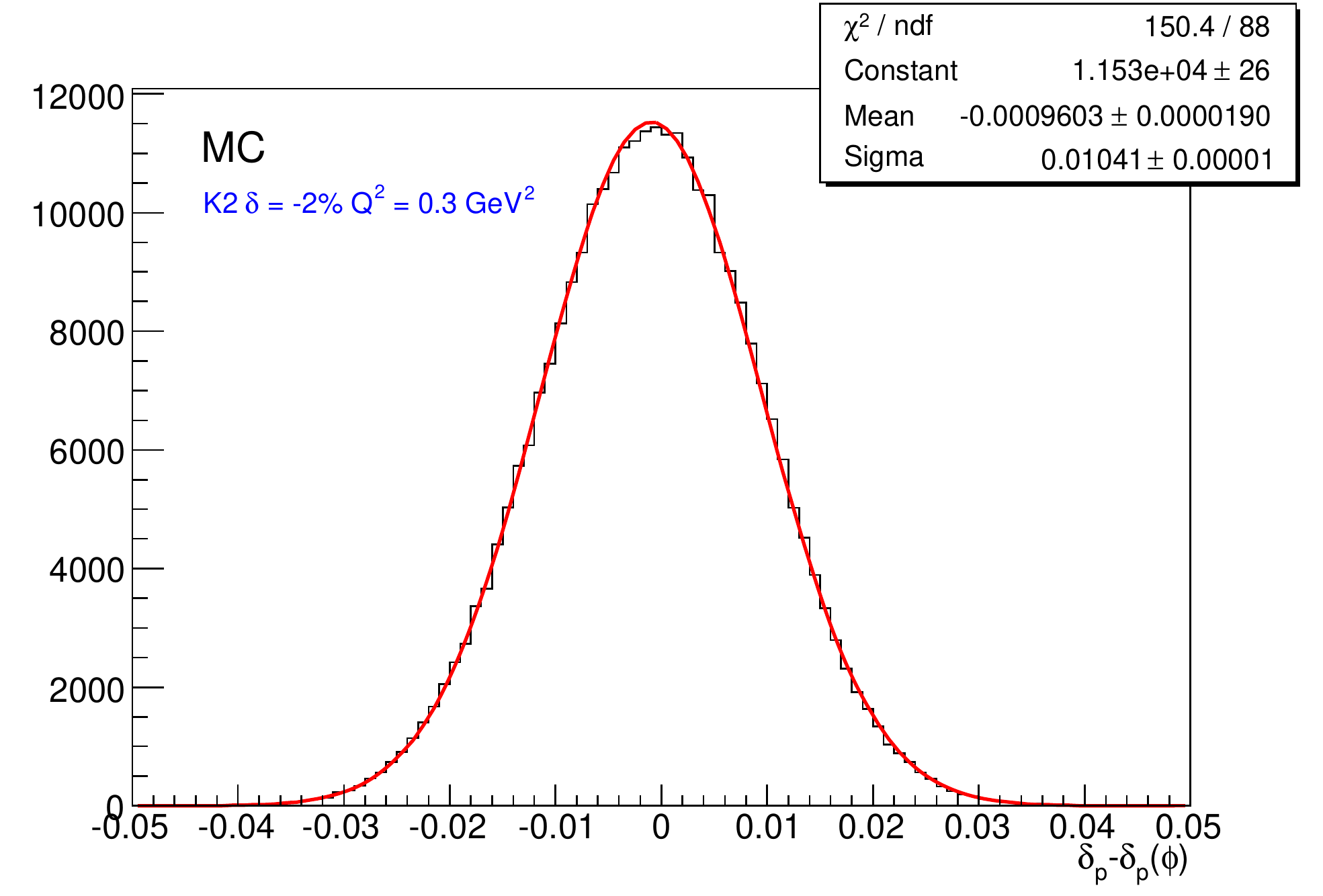}
    \caption{Data and simulated spectrum on $\delta_p-\delta_p(\phi)$.}
    \label{fig:dpmiss}
  \end{center}
\end{figure}

The proton was detected in the left HRS which has a small acceptance and high resolution. To simplify the phase space simulation, we first put constraints on protons only, assuming that all the pions (decayed photons) could be detected in the BigBite shower counter. By applying the same elastic cut on the simulated proton spectrum, we can get the $\pi^0 p$ to $ep$ phase space ratio. The momentum resolution was manually adjusted in the simulation to match the resolution of the data.
\begin{table}[t]
\begin{center}
\begin{tabular}{|c|c|}
\hline
$\phi_{tg}$ & $\pm 30$ mrad\\
\hline
$\theta_{tg}$ & $\pm 60$ mrad\\
\hline
$\delta_p$ & $\pm 0.04$\\
\hline
\end{tabular}
\caption{HRS acceptance.}
\label{tab:accp}
\end{center}
\end{table}
The resolution comparison was made on $\delta_p-\delta_p(\phi)$, which is the difference between the measured momentum and the one reconstructed from the scattering angle via the elastic kinematics. The acceptance cuts were applied according to the HRS default acceptance as reported in Table~\ref{tab:accp}. The comparison between the real data and the simulated spectrum are shown in Fig.~\ref{fig:dpmiss}.

\begin{figure}[hbt]
  \begin{center}
    \includegraphics[angle=0,width=.6\textwidth]{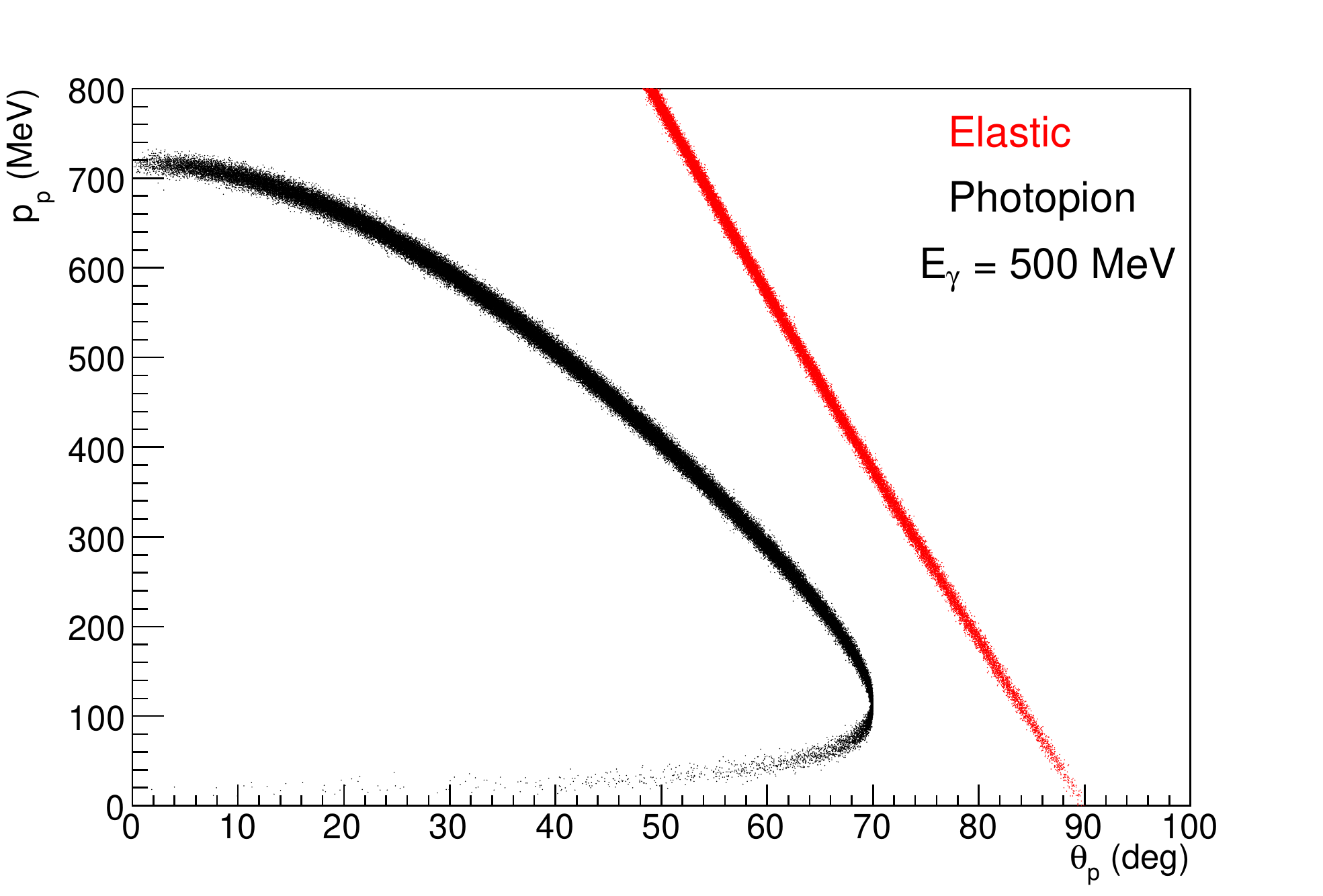}
    \caption{Simulated proton kinematics for $\pi^0 p$ at $E_{\gamma}= 500$ MeV and elastic. $P_p$ is the proton momentum and $\theta_p$ is the scattering angle.}
    \label{pion}
  \end{center}
\end{figure}

\begin{figure}[hbt]
  \begin{center}
    \includegraphics[angle=0,width=.60\textwidth]{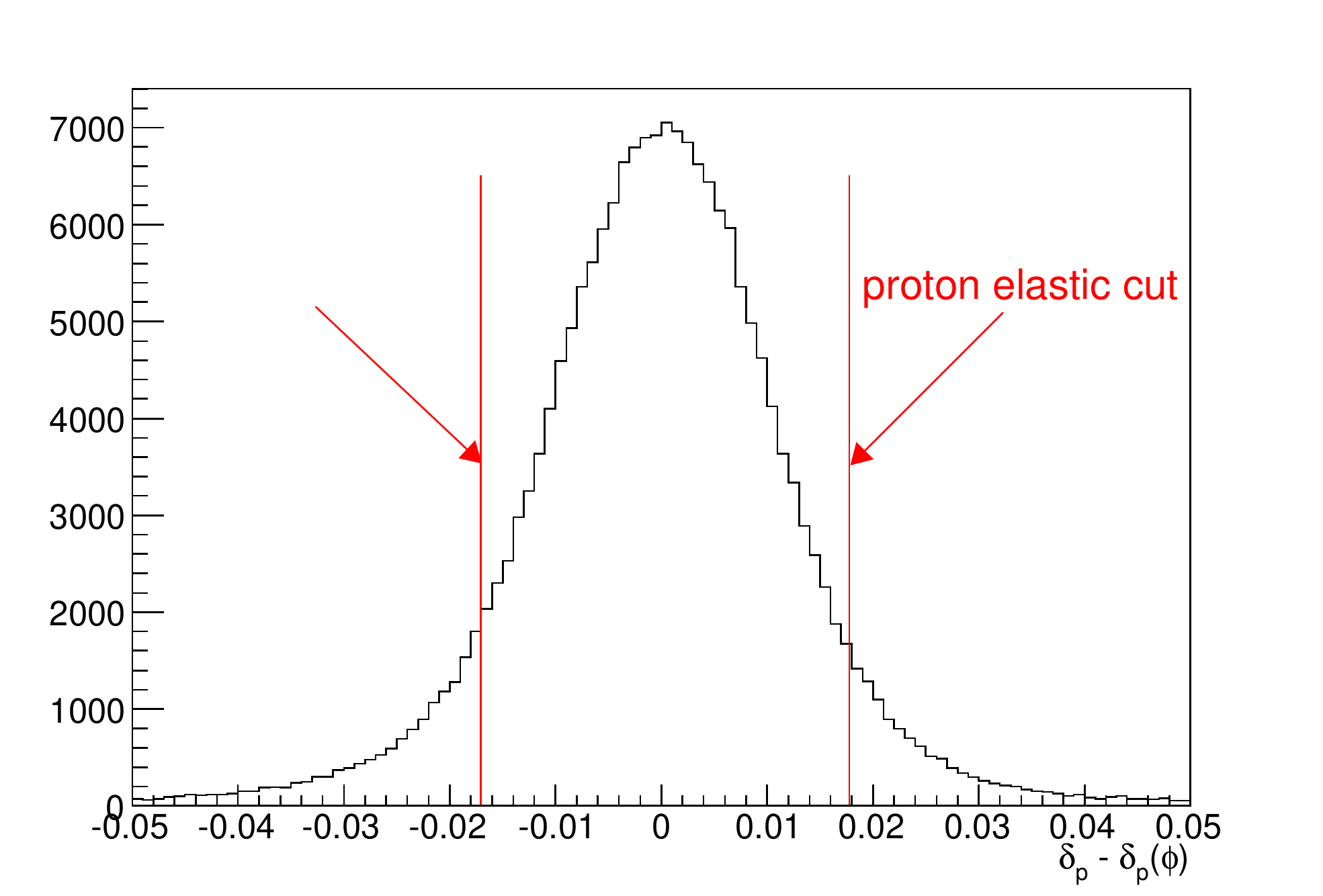}
    \caption{Proton elastic cut on $\delta_p-\delta_p(\phi)$ spectrum for kinematics K2.}
    \label{fig:2D}
  \end{center}
\end{figure}

The simulation also generated the phase space for $\gamma + p\to p +\pi^0$. The range of the photon energy is from 0 to 1192 MeV (beam energy). However, the proton from pion production will only be detected in the HRS acceptance when the photon carries almost all the beam energy. As an example, for kinematics K2, the HRS central angle is $60^\circ$, for the $\delta_p= 0\%$ setting, the central momentum is 565 MeV. As demonstrated in Fig.~\ref{pion}, the proton kinematics for $\pi^0 p$ at $E_{\gamma}=500$ MeV is far away from the HRS setting. To estimate the $\pi^0 p$ to $ep$ phase space ratio, a cut was applied to the simulated spectrum on $\delta_p-\delta_p(\phi)$ according to the elastic cut applied to the data as illustrated in Fig.~\ref{fig:2D}.

It is not surprising to find that only when $E_{\gamma}> 1150$ MeV $\pi^0 p$ phase space becomes noticeable. The procedure is repeated at several photon energy intervals from 1150 MeV to 1192 MeV. Table~\ref{tab:r_phase} gives the $\pi^0 p$ to $ep$ phase space ratio.
\begin{table}[t]
\begin{center}
\begin{tabular}{|c|c|}
\hline
 $E_{\gamma}$ [MeV] & $R_{phase}$\\
\hline
1150 & $1.3\times 10^{-4}$\\
\hline
1160 & $6\times 10^{-4}$\\
\hline
1170 & $2.5\times 10^{-3}$\\
\hline
1180 & $9.2\times 10^{-3}$\\
\hline
1185 & $1.7\times 10^{-2}$\\
\hline
1190 & $2.8\times 10^{-2}$\\
\hline
\end{tabular}
\caption{Simulated $\pi^0 p$ to $ep$ phase space ratio at kinematics K2.}
\label{tab:r_phase}
\end{center}
\end{table}

\begin{figure}[hbt]
  \begin{center}
    \includegraphics[angle=0,width=.65\textwidth]{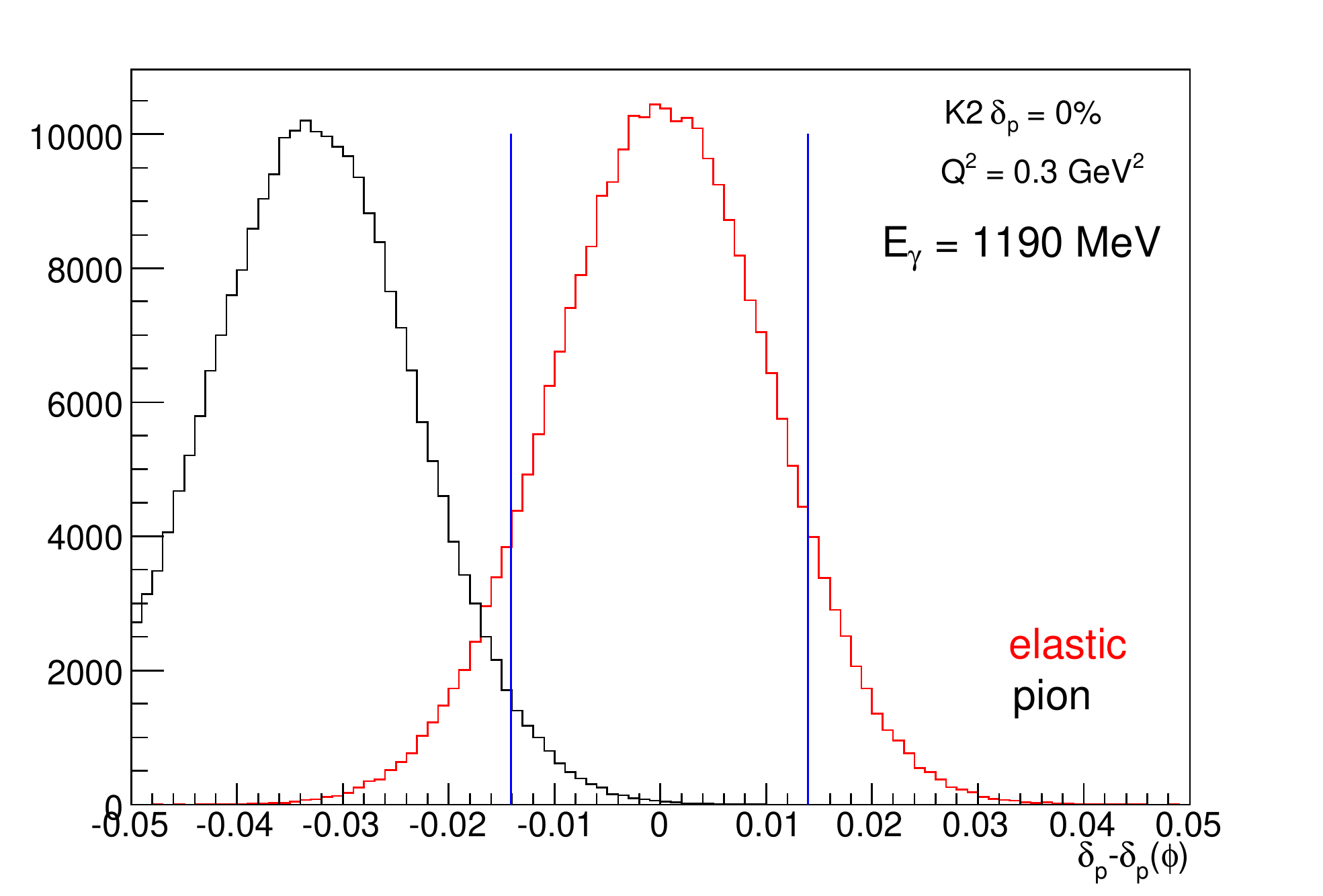}
    \includegraphics[angle=0,width=.65\textwidth]{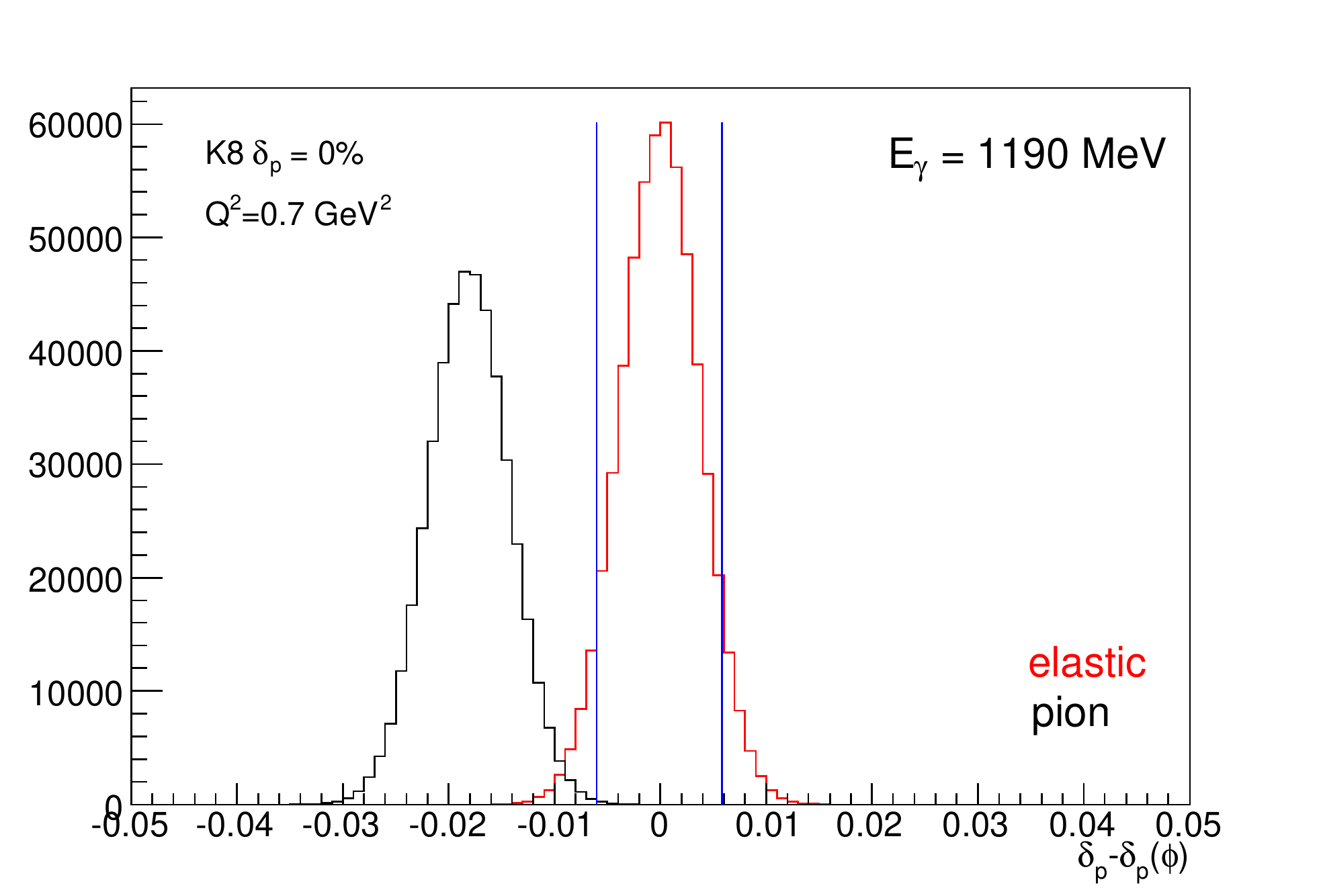}
    \caption{Simulated $ep$ and $\pi^0 p$ spectrum for kinematics K2 and K8. The blue lines are the corresponding elastic cut applied to the data.}
    \label{fig:sim}
  \end{center}
\end{figure}

For higher $Q^2$ settings, although the $\pi^0 p$ kinematics is getting closer to the $ep$ kinematics, the proton momentum resolution improves and $\pi^0 p$ can be more clearly separated by the elastic cut. Fig.~\ref{fig:sim} shows the phase space simulation for kinematics K2 ($Q^2$ = 0.3 GeV$^2$) and K8 ($Q^2$ =0.7 GeV$^2$). It is clear to see that at K8, $\pi^0 p$ are mostly cut away. The same procedure was applied to every kinematics with different photon energies, the $\pi^0 p$ to $ep$ phase space ratios for different kinematics for the simulated $\pi^0 p$ are listed in Table.~\ref{tab:r_kall}.

\begin{table}[hbt]
\begin{center}
\begin{tabular}{|c|c|c|c|c|}
\hline
Kine. & $Q^2$ [GeV]$^2$ & $R_{phase}~[1180$ MeV] & $R_{phase}~[1185$ MeV]& $R_{phase}~[1190$ MeV]\\
\hline
K1 & 0.35& $1.0\times 10^{-2}$ & $2.0\times 10^{-2}$ & $3.4\times 10^{-2}$\\
K2 & 0.3 & $9.2\times 10^{-3}$ & $1.7\times 10^{-2}$ &$2.8\times 10^{-2}$\\
K3 & 0.45 &$8.6\times 10^{-4}$ & $3.0\times 10^{-3}$ & $1.1\times 10^{-2}$\\
K4 & 0.40 & $1.1\times 10^{-3}$ & $3.5\times 10^{-3}$ & $6.8\times 10^{-3}$\\
K5 & 0.55 & $8.0\times 10^{-5}$ & $5.0\times 10^{-4}$ & $2.8\times 10^{-3}$ \\
K6 & 0.50 & $4.4\times 10^{-4} $ & $2.0\times 10^{-3}$ &$7.2\times 10^{-3}$\\
K7 & 0.6  & $3.8\times 10^{-5} $ & $ 4.8\times 10^{-4} $ & $3.5\times 10^{-3}$\\
K8 & 0.7  & $0.0 $ & $3.6\times 10^{-5}$ &$0.7\times 10^{-3}$\\
\hline
\end{tabular}
\caption{$\pi^0 p$ to $ep$ phase space ratio for different kinematics with $E_{\gamma}=$ 1180, 1185, and 1190 MeV.}
\label{tab:r_kall}
\end{center}
\end{table}
\section{Photon flux}
The real photon flux from bremsstrahlung were calculated using~\cite{brem}, with 3 cm liquid hydrogen target. The results are listed in Table.~\ref{tab:flux}
\begin{table}[t]
\begin{center}
\begin{tabular}{|c| c|}
\hline
 $E_{\gamma}$ range [MeV] & $\Gamma_{\gamma}$ \\
\hline
1150-1160 & $5\times 10^{-5}$\\
1160-1170 & $5\times 10^{-5}$\\
1170-1180 & $5\times 10^{-5}$\\
1180-1190 & $3\times 10^{-5}$\\
1185-1190 & $1.5\times 10^{-5}$\\
1190-1192 & $0.5\times 10^{-5}$\\
\hline
\end{tabular}
\caption{Real photon flux at different energies with 1.192 GeV electron beam.}
\label{tab:flux}
\end{center}
\end{table}
\section{Cross Sections}
In order to compare the rate, the cross sections for $ep$ and $\pi^0 p$ are required. The elastic cross section in the lab can be directly estimated from the rate during the experiment. For K2, with 4$\mu$A beam, 6 cm target, the coincidence rate is around 3 kHz. The maximum HRS acceptance (6msr) is used for $d\Omega$. The elastic differential cross section in the lab frame can be estimated by:
\begin{eqnarray}
L&=& 6 cm \cdot 0.07g/cm^3 \cdot 6.02\times 10^{23}/g\cdot 4\times 10^{-6} A \cdot 1.6 \times 10 ^{19}/C{}\\
\nonumber\\
{}&=&16\times 10^{36}/cm^2\cdot s\\
\frac{d\sigma_{el}}{d\Omega}&=&\frac{3\times 10^3/s}{(16\times 10^{36}/cm^2\cdot s)\cdot 6\times 10^{-3}sr}{}\\
\nonumber\\
{}&=& 3.1 \times 10^{-2}\mu b/sr.
\end{eqnarray}

\begin{figure}[hbt]
  \begin{center}
    \includegraphics[angle=0,width=.65\textwidth]{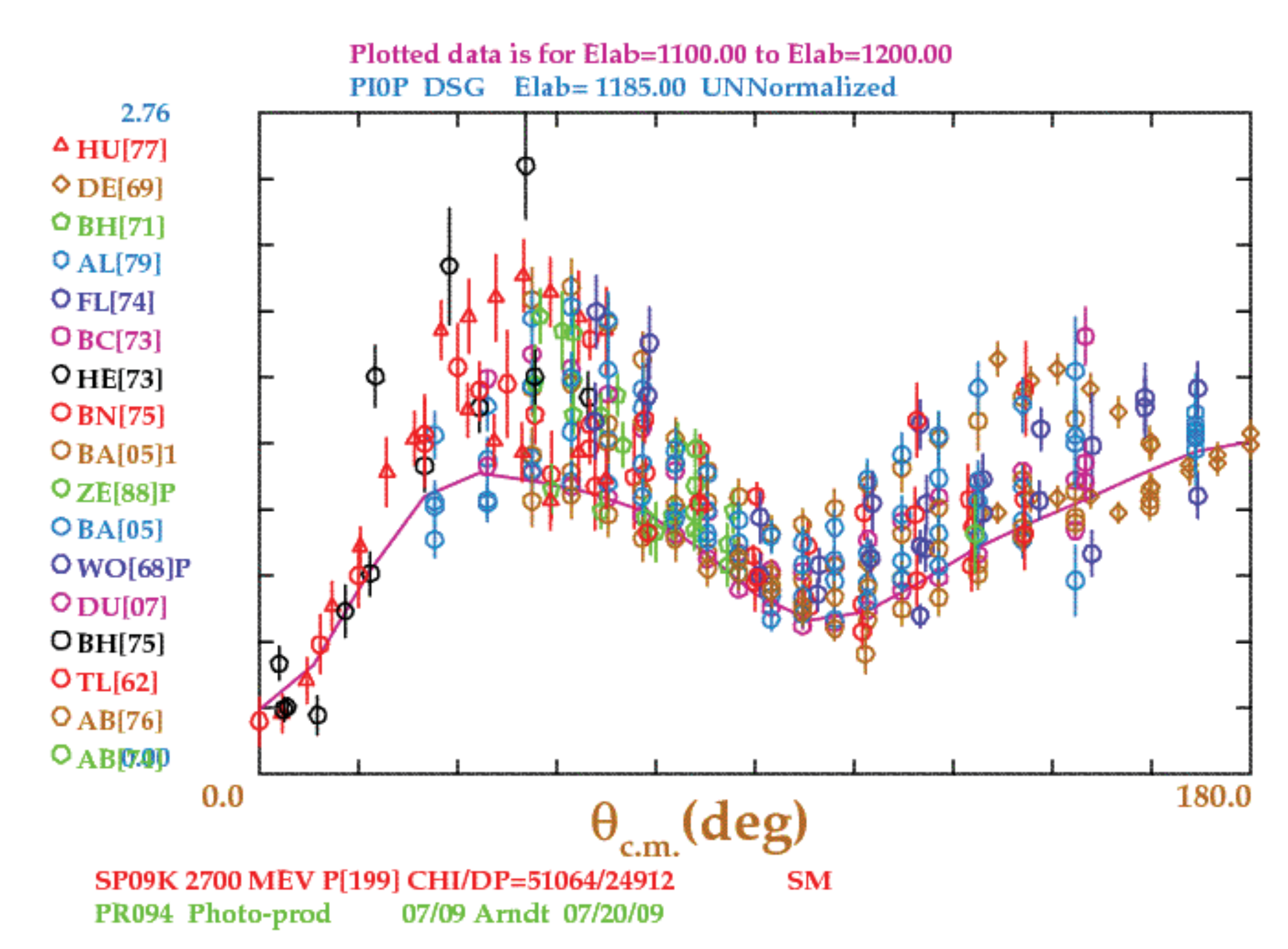}
    \caption{World data and calculations for $\pi^0 p$ differential cross section at E$_{\gamma}$ = 1185 MeV.}
    \label{fig:pionxs}
  \end{center}
\end{figure}
The $\pi^0 p$ differential cross section for $E_{\gamma}\sim 1185$ MeV at the same setting was looked up in the world database~\cite{said} (see Fig.~\ref{fig:pionxs}), which is $\sim 1.2\times\mu$b/sr in the C.M. frame, Jacobian $J=1.6$ for K2, so the cross section in the lab frame is $\sim 1.6\times 1.2 = 2~\mu$b/sr. The cross section ratio can be obtained:
\begin{equation}
R_{XS}=\sigma_{\pi^0 p}/\sigma_{ep}=2/3.1 \times 10^{-2}\sim 60.
\end{equation}
The $ep$ and $\pi^0 p$ differential cross sections in the lab frame for different kinematics are listed in Table~\ref{tab:xs}.
\begin{table}[hbt]
\begin{center}
\begin{tabular}{|c| c |c| c|}
\hline
Kine. & $\frac{d\sigma_{ep}}{d\Omega}$($\mu b/sr$)  & $\frac{d\sigma_{\pi^0 p}}{d\Omega}$ ($\mu b/sr$) & $R_{XS}$ \\
\hline
K1 & $6.7\times 10^{-2}$ & 2.2&  33\\
K2 & $3.3\times 10^{-2}$ & 2.0&  64\\
K3 & $1.8\times 10^{-2}$ & 2.4&  133 \\
K4 & $1.9\times 10^{-2}$ & 2.4&   126\\
K5 & $1.1\times 10^{-2}$ & 2.7&  245 \\
K6 & $1.3\times 10^{-2}$ & 2.6&  200 \\
K7 & $1.0\times 10^{-2}$ & 2.6&  260\\
K8 & $0.6\times 10^{-2}$ & 2.6&  433\\
\hline
\end{tabular}
\caption{$ep$ and $\pi^0 p$ differential cross sections in the lab frame and the ratio $R_{XS}$ for different kinematics.}
\label{tab:xs}
\end{center}
\end{table}
\section{Pion Electroproduction}
\begin{figure}[hbt]
  \begin{center}
    \includegraphics[angle=0,width=.65\textwidth]{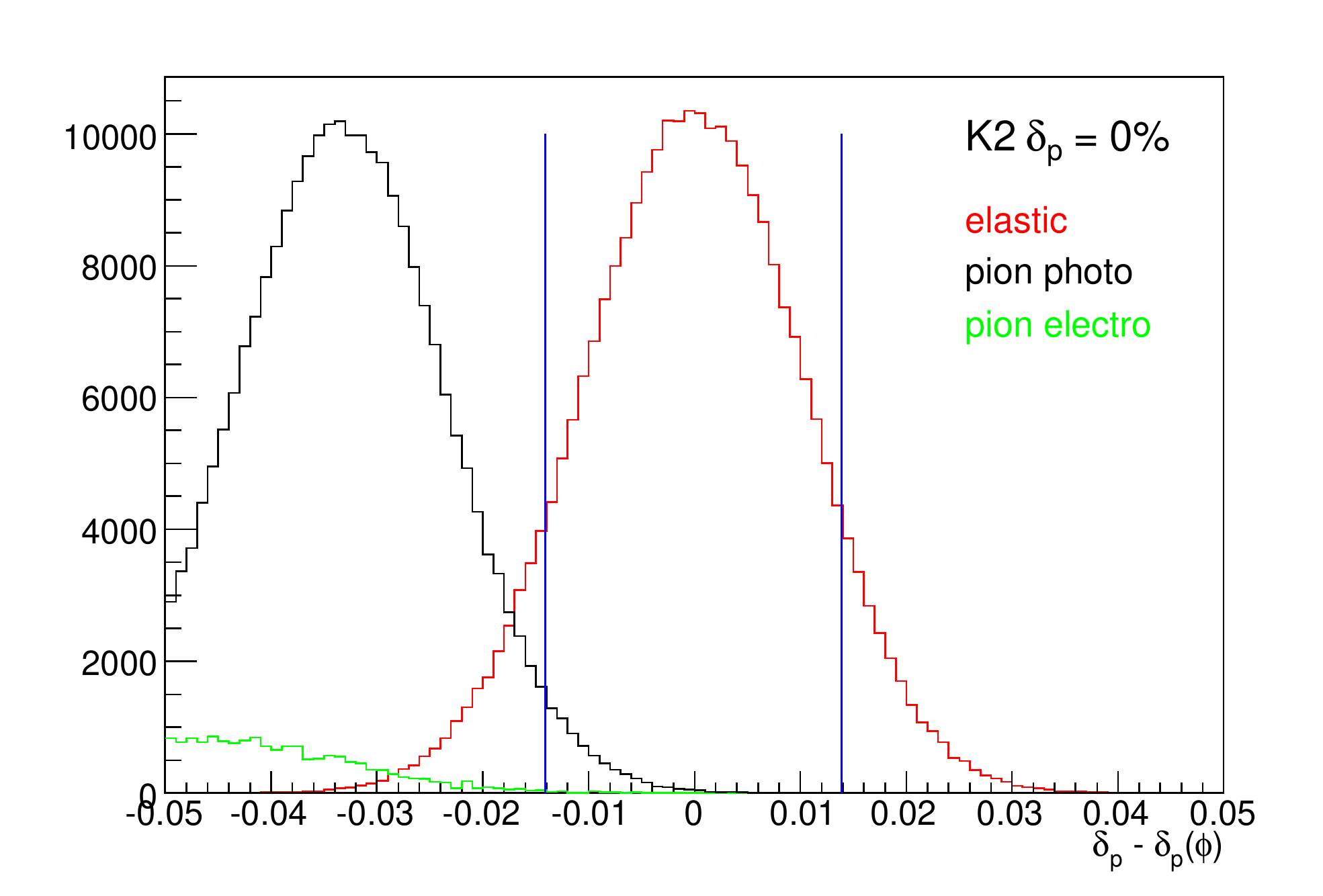}
    \caption{Phase space simulation for $ep$, $\pi^0 p$ and $ep\pi^0$ with E$_{\gamma}$ = 1190 MeV.}
    \label{fig:electro}
  \end{center}
\end{figure}

The electroproduction reaction $e+p\to e+p+\pi^0$ was checked as well. Although the virtual photon flux is $\sim 3$ times larger than the real photon, the phase space is much smaller (3-body) than the photoproduction, we expect the effect is even smaller. Fig.~\ref{fig:electro} shows the phase space simulation for elastic, pion photoproduction, and electroproduction at kinematics K2. With the elastic cuts, the electroproduction phase space is 50 times smaller compared to the photoproduction.
\section{Rate Estimation and Polarization corrections}
With all the information above, we can estimate the $\pi^0 p$ to $ep$ rate ratio assuming all the decayed photons were detected by:
\begin{equation}
r_0= N_{\pi^0 p}/N_{ep}= \sum_{E_{\gamma}}R_{phase}\times R_{XS}\times \Gamma.
\end{equation}
The results are listed in Table.~\ref{tab:rate}.
\begin{table}[t]
\begin{center}
\begin{tabular}{|c| c|}
\hline
 $E_{\gamma}$ range [MeV] & $r_0$   \\
\hline
1150-1160 & $0.5\times 10^{-6}$\\
1160-1170 & $0.22\times 10^{-5}$\\
1170-1180 & $0.12\times 10^{-4}$\\
1180-1185 & $0.23\times 10^{-4}$\\
1185-1190 & $0.18\times 10^{-4}$\\
1190-1192 & $0.1\times 10^{-4}$\\
\hline
$r_0$& $\sim 1\times 10^{-4}$\\
\hline
\end{tabular}
\caption{Estimated ratio of $\pi^0 p$ to $ep$ for kinematics K2.}
\label{tab:rate}
\end{center}
\end{table}
So the total $\pi^0 p$ to $ep$ ratio at K2 is $\sim 1\times 10^{-4}$ if all the decayed photons can be detected in the BigBite. Actually, during the experiment, only part of the BigBite shower counter was turned on, if the BigBite acceptance is taking into account, the $\pi^0 p$ rate will be further reduced.
\subsection{BigBite Acceptance}
For K2, a section of 3x3 shower blocks were on. The area of each shower block is $8.5$ cm $\times 8.5$ cm. The calorimeter was about 3 m away from the target. To estimate the upper limit, we naively assume that $\pi^0$ aimed at the center of the 9 shower blocks, so the in-plane and out-of-plane acceptance is about $\pm 2.4^{\circ}$. This corresponds to $\pm 35^{\circ}$ in the C. M. frame where the photons are uniformly distributed. The $\pi^0 p$ rate would be further suppressed by:
\begin{equation}
f_{BB}= (\cos 0^{\circ}-\cos 35^{\circ})/\cos 0^{\circ} = 0.18
\end{equation}
After multiplying the factor above, the $\pi^{0}p$ to $ep$ ratio is:
\begin{equation}
r = r_0\times f_{BB}= 3\times 10^{-4}\times 0.18 = 0.54 \times 10^{-4}.
\end{equation}
For different kinematics, the pion momentum is different as well as the acceptance of BigBite. $f_{BB}$ for each kinematics is listed in Table~\ref{tab:xs}. Together with the other information mentioned earlier, the $\pi^0 p$ to $ep$ ratio $r$ for each kinematics are obtained in Table~\ref{tab:xs}. This is a very conservative estimation, since the central angle of $\pi^0$ was actually $0.5 \sim 1.0$ degree off the center of the electrons, and the electrons actually were bent upwards by the BigBite magnet while the photons went straight through and hit at the lower region. Therefore, the effective acceptance for the photons is even smaller. The final ratio of the rates is $< 1\times 10^{-4}$ for all the kinematics we have taken.
\begin{table}[ht]
\begin{center}
\begin{tabular}{|c |c| c| c|}
\hline
Kine. & shower blocks & acceptance in C.M. [deg] & $f_{BB}$\\
\hline
K1 & 3x3 & $\pm 33$ &  0.16 \\
K2 & 3x3 & $\pm 35$  &  0.18\\
K3 & 3x3 & $\pm 31$  &  0.14\\
K4 & 3x3 & $\pm 32$  &  0.15\\
K5 & 3x3 & $\pm 30$  &  0.13\\
K6 & 4x4 & $\pm 40$ &  0.23\\
K7 & 5x5 & $\pm 47$ &  0.32\\
K8 & 5x5 & $\pm 44$ &  0.28\\
\hline
\end{tabular}
\caption{$ep$ and $\pi^0 p$ differential cross sections in the lab frame and the ratio $R_{XS}$ for different kinematics.}
\label{tab:xs}
\end{center}
\end{table}
\subsection{Hall C Inclusive Data}
\begin{figure}[hbt]
  \begin{center}
    \includegraphics[angle=0,width=.49\textwidth]{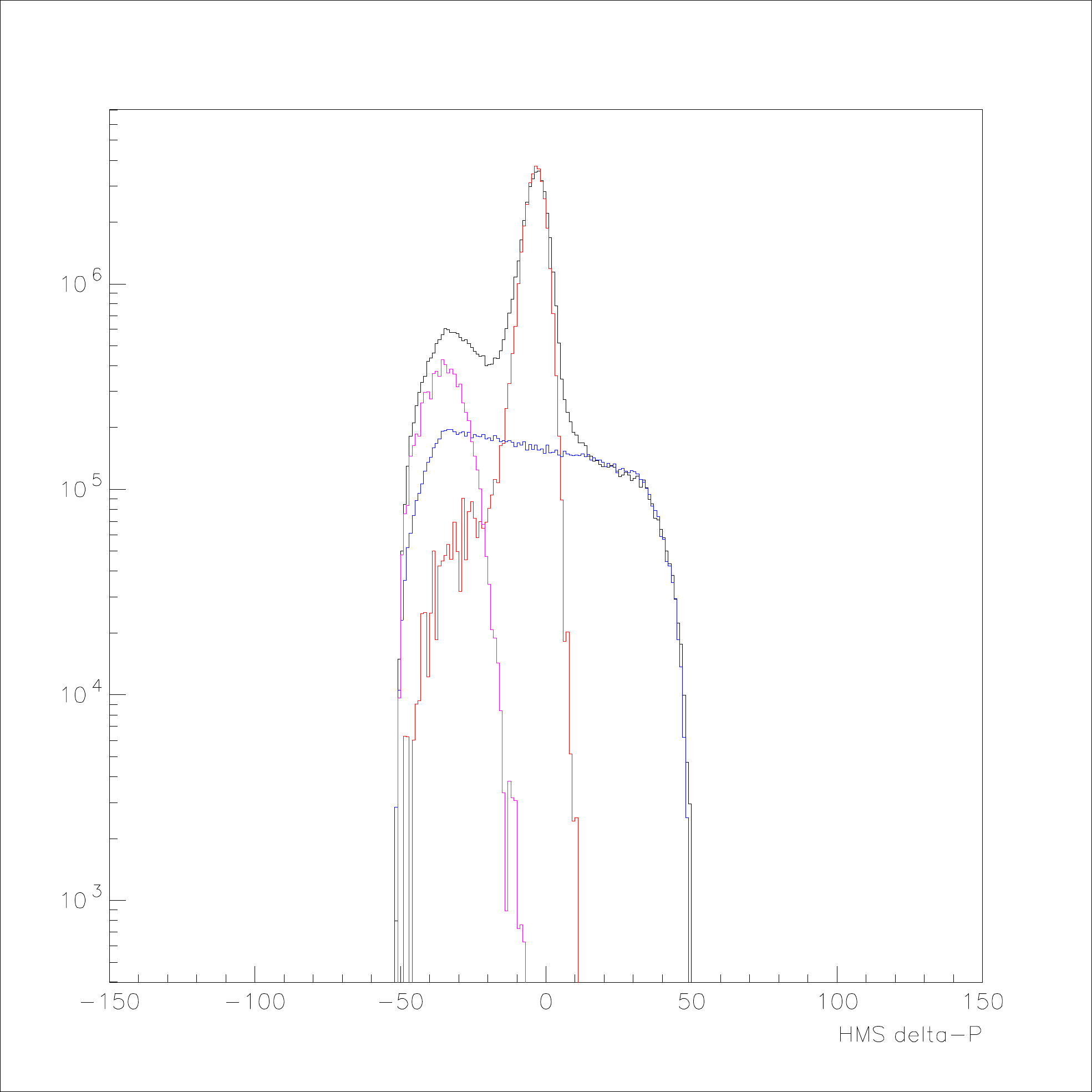}
    \includegraphics[angle=0,width=.49\textwidth]{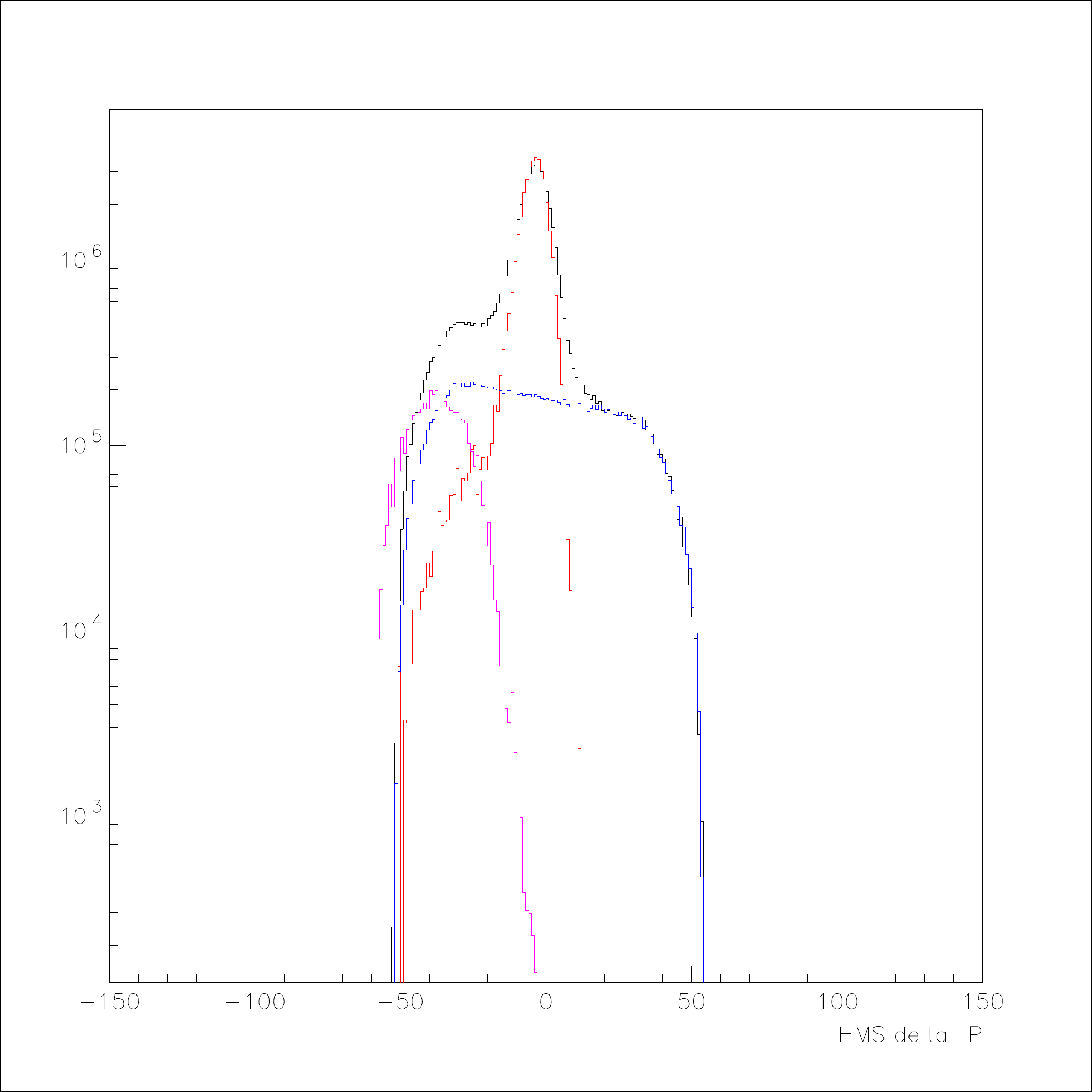}
    \caption{The proton singles spectra and the full background simulation from Hall C Super-Rosenbluth experiment with beam energy 849 MeV (left panel) and 985 MeV (right panel). The spectra in red in the proton elastic peak, and the one in magenta is the simulated pion production.}
    \label{fig:hallc}
  \end{center}
\end{figure}

The Hall C Super-Rosenbluth experiment~\cite{hallc_ep} took the singles elastic data at similar $Q^2$ with a bit lower beam energies. Fig.~\ref{fig:hallc} shows the full simulation of the proton singles spectra at 2 different beam energies. One can clearly see that the higher energy moves the pion production closer to the elastic peak, but the pion contamination is still much less than $1\%$ if tight elastic cuts are applied. In our experiment, the coincidence trigger and the limited BigBite acceptance greatly suppressed the inelastic background. In addition, compared to the HMS, the Hall A HRS has much better resolution which makes it much easier to cut out protons from the pion production. In another word, the smallness of the pion contamination in our data is also expected from the Hall C data and simulation.
\subsection{Corrected Proton Polarizations}
\begin{figure}[hbt]
  \begin{center}
    \includegraphics[angle=0,width=.45\textwidth]{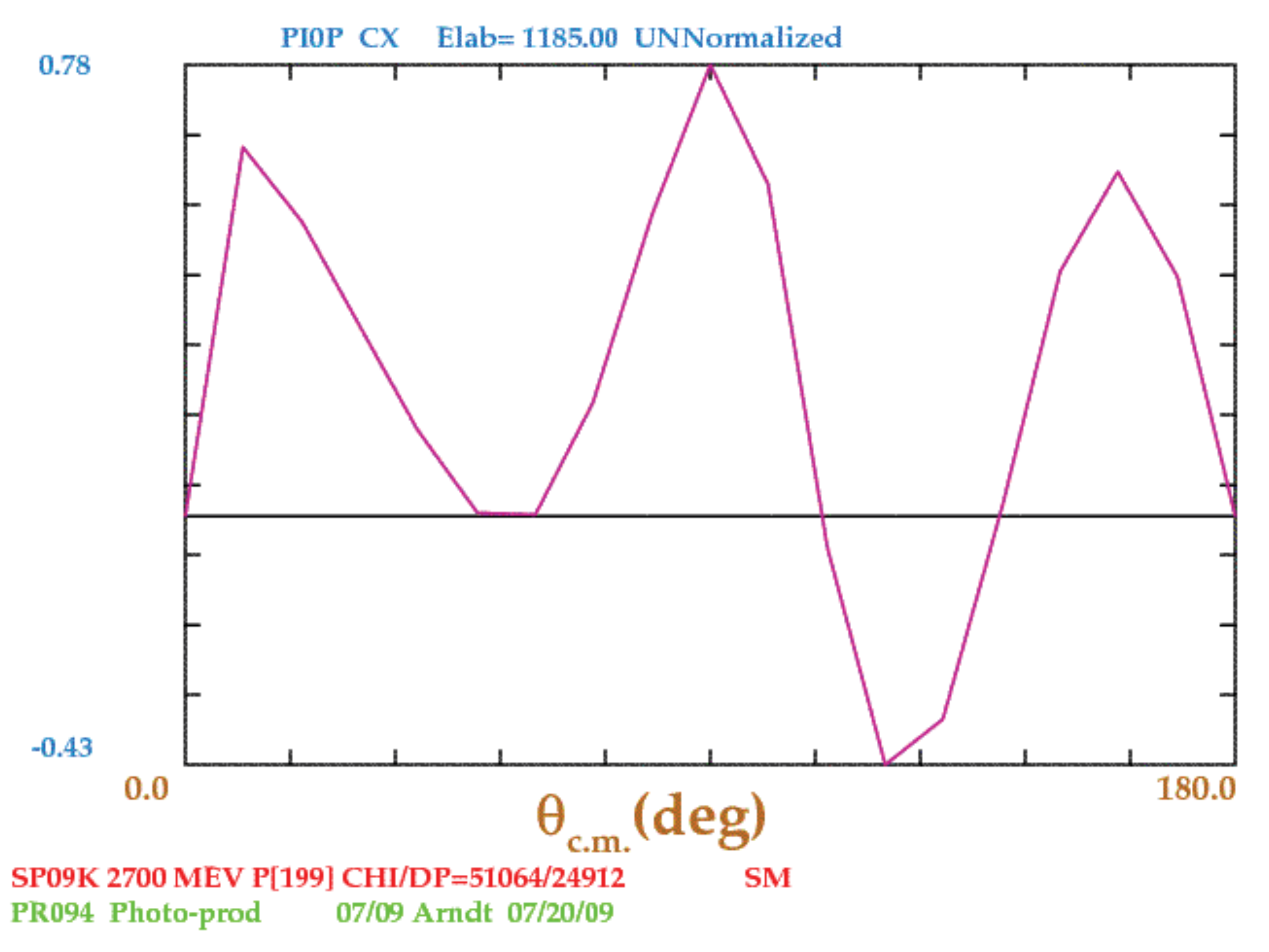}
    \includegraphics[angle=0,width=.45\textwidth]{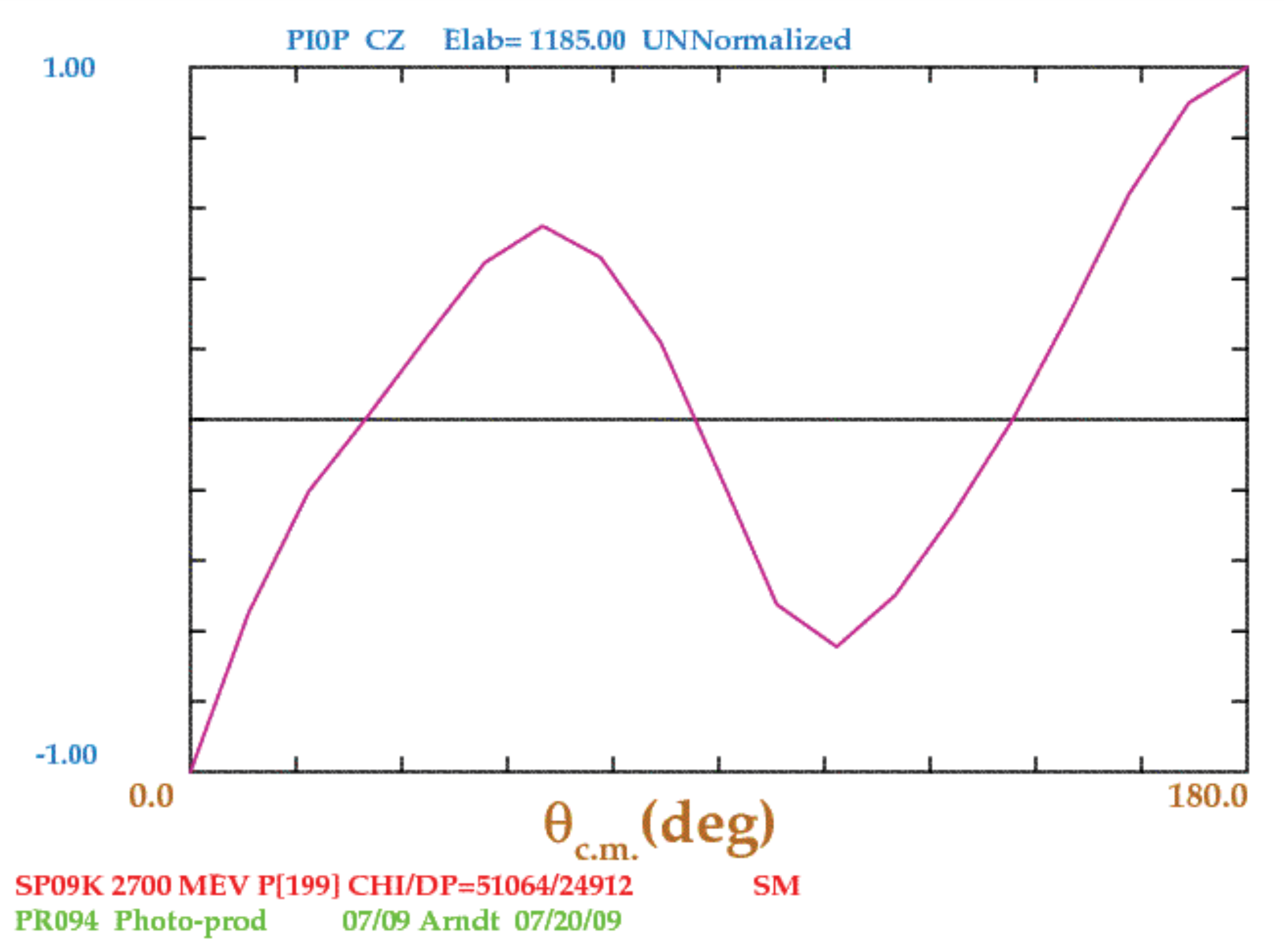}
    \caption{Calculations for the $\pi^0 p$ polarization observable at E$_{\gamma}$ = 1185 MeV.}
    \label{fig:pionc}
  \end{center}
\end{figure}

Now we can look at the the possible correction to the proton polarizations with $1\times 10^{-4}$ $\pi^0 p$ contamination after we applied the elastic cut. For kinematics K2, as shown in Fig.~\ref{fig:pionc}, the proton polarization for $\pi^0 p$ from~\cite{said} are listed in Table.~\ref{tab:pol}.
\begin{table}[hb]
\begin{center}
\begin{tabular}{|c |c |c|}
\hline
$E_{\gamma}$ (MeV) & $C_x$ & $C_z$\\
\hline
1180 & -0.0095 & 0.4456\\
\hline
1185 & 0.0053  & 0.4470\\
\hline
1190 & 0.0194  & 0.4475\\
\hline
elastic (K2) & -0.208 & 0.186\\
\hline
\end{tabular}
\caption{Polarization observable}
\label{tab:pol}
\end{center}
\end{table}
The corrected $C_x$ and $C_z$ for the elastic events are:
\begin{equation}
C_{ep} = \frac{C_{raw}-r\cdot C_{\pi^0 p}}{1-r}
\end{equation}
where $r=1\times 10^{-4}$ is the estimated $\pi^0 p$ to $ep$ ratio. The corrected form factor ratio would shift by $\sim 0.0003$. The results for the other kinematics are similar or even smaller.
\section{Summary}
 With the procedure presented above, we conservatively estimated the contribution from $\pi^0 p$ to be $< 10^{-4}$ level. The resulting correction to the proton polarization is also at $10^{-4}$ level which is negligible.
\clearpage
\newpage

%% file: app5.tex
\chapter{Cross Section Data}

\begin{table}[b]
 \begin{center}
 \begin{tabular}{|c|ccccccc|}
 \hline
 $Q^2$ & $E$ & $E_p$ & $\theta_e$ & $\varepsilon$ & $\sigma$ & $\delta_\sigma$ & Ref.\\
 $[(\mathrm{GeV}/c)^2]$ & [GeV] & [GeV] & [$^\circ$] & [1] & [nb/sr] & [nb/sr] & \\ 
 \hline\hline
0.2922 &  0.6240 & 0.4683 & 59.997 &0.58070  & 55.56   &    2.278     & \cite{janss} \\
0.2916 &  0.5280 & 0.3726 & 74.996 &0.43960  & 32.16   &    1.319     & \cite{janss} \\
0.2923 &  0.4680 & 0.3123 & 89.995 &0.31590  & 20.55   &    1.048     & \cite{janss} \\
0.2916 &  0.3990 & 0.2436 & 119.993&0.13340  & 10.50   &   0.5251     & \cite{janss} \\
0.2915 &  0.3800 & 0.2247 & 134.993&0.07340  & 8.812   &   0.3525     & \cite{janss} \\
\hline
0.3498  & 0.6920 & 0.5056 & 59.997 &0.57710  & 36.94   &    1.884       &\cite{janss}\\
0.3500  & 0.5880 & 0.4015 & 74.996 &0.43580  & 22.06   &   0.8823       &\cite{janss}\\
0.3503  & 0.4270 & 0.2403 &134.993 &0.07240  & 6.458   &   0.2583       &\cite{janss}\\
\hline
0.3894  & 0.9000 & 0.6925 & 46.557 &0.70860  & 59.11   &    2.896   &    \cite{janss}\\
0.3891  & 0.7360 & 0.5287 & 59.997 &0.57460  & 30.66   &    1.226   &    \cite{janss}\\
0.3897  & 0.6270 & 0.4193 & 74.996 &0.43330  & 17.65   &   0.7235   &    \cite{janss}\\
0.3894  & 0.5570 & 0.3495 & 89.995 &0.31050  & 11.82   &   0.5792   &    \cite{janss}\\
0.3898  & 0.4790 & 0.2713 &119.993 &0.13050  & 6.218   &   0.3109   &    \cite{janss}\\
0.3893  & 0.4570 & 0.2495 &134.993 &0.07170  & 5.123   &   0.2049   &    \cite{janss}\\
0.3894  & 0.4470 & 0.2395 &144.992 &0.04280  & 4.690   &   0.2345   &    \cite{janss}\\
0.3894  & 1.9035 & 1.6960 & 19.999 &0.93540  & 408.9   &    8.996   &    \cite{goitein}\\
0.3903  & 1.5370 & 1.3290 & 25.249 &0.89970  & 226.5   &    4.559   &    \cite{berger}\\
0.3891  & 1.2490 & 1.0416 & 31.738 &0.84780  & 131.6   &    2.577   &    \cite{berger}\\
0.3892  & 1.2310 & 1.0236 & 32.268 &0.84330  & 130.0   &    2.580   &    \cite{berger}\\
0.3892  & 1.1420 & 0.9346 & 35.148 &0.81780  & 107.4   &    2.188   &    \cite{berger}\\
0.3890  & 0.8480 & 0.6407 & 50.057 &0.67380  & 45.62   &   0.9317   &    \cite{berger}\\
0.3895  & 0.6960 & 0.4884 & 64.716 &0.52860  & 25.14   &   0.4075   &    \cite{berger}\\
0.3894  & 0.5560 & 0.3485 & 90.265 &0.30840  & 11.71   &   0.2287   &    \cite{berger}\\
\hline
0.4671  & 0.9500 & 0.7011  &49.507 &0.67490   &33.12   &    1.689    &   \cite{janss}\\
0.4672  & 0.9000 & 0.6510  &53.037 &0.63930   &27.62   &    1.381    &   \cite{janss}\\
0.4677  & 0.7000 & 0.4508  &74.996 &0.42850   &11.67   &   0.4668    &   \cite{janss}\\
0.4675  & 0.5150 & 0.2659 &134.993 &0.07040   &3.529   &   0.1765    &   \cite{janss}\\
0.4674  & 0.5040 & 0.2549 &144.992 &0.04200   &3.240   &   0.1620    &   \cite{janss}\\
\hline
0.5061  & 0.9500 & 0.6803 & 52.517 &0.64240   &24.26    &   1.189     &  \cite{janss}\\
0.5066  & 0.7350 & 0.4650 & 74.996 &0.42610   &9.320    &  0.3821     &  \cite{janss}\\
0.5064  & 0.5430 & 0.2732 &134.993 &0.06980   &2.882    &  0.1441     &  \cite{janss}\\
0.5072  & 1.7700 & 1.4997 & 25.249 &0.89700   &127.3    &   2.574     &  \cite{berger}\\
\hline
 \end{tabular}
 \end{center}
 \end{table}
 
  \begin{table}[t]
 \begin{center}
 \begin{tabular}{|c|ccccccc|}
  \hline
  $Q^2$ & $E$ & $E_p$ & $\theta_e$ & $\varepsilon$ & $\sigma$ & $\delta_\sigma$ & Ref.\\
 $[(\mathrm{GeV}/c)^2]$ & [GeV] & [GeV] & [$^\circ$] & [1] & [nb/sr] & [nb/sr] & \\
  \hline\hline
0.5451  & 0.9500 & 0.6595  &55.597 &0.60900  & 18.17    &  0.9084    &   \cite{janss}\\
0.5452  & 0.9000 & 0.6095  &59.797 &0.56700  & 14.20    &  0.7383    &   \cite{janss}\\
0.5453  & 0.7690 & 0.4784  &74.996 &0.42380  & 7.793    &  0.3195    &   \cite{janss}\\
0.5445  & 0.5700 & 0.2798 &134.993 &0.06920  & 2.454    &  0.1227    &   \cite{janss}\\
0.5456  & 0.5590 & 0.2683 &144.992 &0.04130  & 2.347    &  0.1173    &   \cite{janss}\\
\hline
0.5840  & 0.9500 & 0.6388 & 58.747 &0.57510   &13.27      &0.6504     &  \cite{janss}\\
0.5837  & 0.8020 & 0.4910 & 74.996 &0.42150   &6.573      &0.3352     &  \cite{janss}\\
0.5833  & 0.5970 & 0.2862 &134.993 &0.06860   &2.126      &0.1063     &  \cite{janss}\\
0.5841  & 2.3617 & 2.0505 & 19.999 &0.93240   &155.0      & 4.031     &  \cite{goitein}\\
0.5840  & 1.0720 & 0.7608 & 50.057 &0.66300   &17.69      &0.3563     &  \cite{berger}\\
0.5843  & 1.0420 & 0.7306 & 51.957 &0.64360   &16.64      &0.3373     &  \cite{berger}\\
0.5844  & 0.8920 & 0.5806 & 64.166 &0.52180   &9.945      &0.1983     &  \cite{berger}\\
0.5837  & 0.8860 & 0.5749 & 64.716 &0.51650   &9.656      &0.1985     &  \cite{berger}\\
0.5845  & 0.7180 & 0.4065 & 90.075 &0.29950   &4.517      &0.9927E-01 &  \cite{berger}\\
0.5844  & 0.7170 & 0.4056 & 90.265 &0.29820   &4.504      &0.8936E-01 &  \cite{berger}\\
0.5846  & 0.6470 & 0.3354 &110.294 &0.17210   &2.926      &0.6754E-01 &  \cite{berger}\\
0.5834  & 0.6450 & 0.3341 &110.714 &0.17000   &2.969      &0.6057E-01 &  \cite{berger}\\
0.5847  & 1.9120 & 1.6004 & 25.249 &0.89530   &89.17      & 1.781     &  \cite{berger}\\
0.5842  & 1.6290 & 1.3177 & 30.238 &0.85450   &60.86      & 1.190     &  \cite{berger}\\
0.5844  & 1.5400 & 1.2286 & 32.268 &0.83670   &51.12      &0.9907     &  \cite{berger}\\
0.5841  & 1.5220 & 1.2107 & 32.698 &0.83290   &48.94      &0.9816     &  \cite{berger}\\
0.5843  & 1.4310 & 1.1196 & 35.148 &0.81040   &42.28      &0.8441     &  \cite{berger}\\
\hline
0.7009  & 0.9500  &0.5765  &68.886 &0.46990   &5.588      &0.2738      & \cite{janss}\\
0.7005  & 0.8990  &0.5257  &74.996 &0.41460   &4.392      &0.2196      & \cite{janss}\\
0.7006  & 0.8640  &0.4907  &79.996 &0.37200   &3.609      &0.1732      & \cite{janss}\\
0.7012  & 0.6770  &0.3034 &134.993 &0.06680   &1.280      &0.6402E-01  & \cite{janss}\\
0.7013  & 0.6640  &0.2903 &144.992 &0.03980   &1.177      &0.5886E-01  & \cite{janss}\\
\hline
0.7790 &  1.7890 & 1.3739  &32.698 &0.82630   &23.13      &0.4753      & \cite{berger}\\
0.7784 &  1.6830 & 1.2682  &35.148 &0.80320   &19.65      &0.5852      & \cite{berger}\\
0.7791 &  1.3920 & 0.9768  &44.478 &0.71000   &11.25      &0.2274      & \cite{berger}\\
0.7791 &  1.0640 & 0.6488  &64.166 &0.51020   &4.457      &0.1089      & \cite{berger}\\
0.7792 &  0.8650 & 0.4498  &90.075 &0.28990   &2.118      &0.4660E-01  & \cite{berger}\\
0.7783 &  0.7840 & 0.3693 &110.124 &0.16650   &1.384      &0.2976E-01  & \cite{berger}\\
\hline
 \end{tabular}
 \end{center}
 \end{table}
 
\clearpage
\newpage

%% file: biblio.tex
\bibliography{main}
\bibliographystyle{unsrt}